\documentclass[12pt]{iopart}

\usepackage{graphicx}
\usepackage{subfigure}
\usepackage{pdfsync}
\usepackage{booktabs}
\usepackage{amssymb}
\usepackage{arydshln}
\usepackage{bm}
\usepackage{multirow}
\usepackage{cancel}
\usepackage[colorlinks,citecolor=blue,anchorcolor=red,menucolor=red,linkcolor=red,filecolor=red,runcolor=red,urlcolor=blue,frenchlinks=red]{hyperref}

\graphicspath{{Figures/}} %

\newcommand{\er}{$\pm$}
\newcommand{\tz}{$\to$}

\begin{document}

\title{An updated review of the new hadron states}

\author{Hua-Xing Chen$^1$\footnote{Corresponding author}, Wei Chen$^2$, Xiang Liu$^{3,4}$, Yan-Rui Liu$^5$ {\rm and} Shi-Lin Zhu$^{6}$\footnote{Corresponding author}}
\address{
$^1$School of Physics, Southeast University, Nanjing 210094, China\\
$^2$School of Physics, Sun Yat-Sen University, Guangzhou 510275, China\\
$^3$School of Physical Science and Technology, Lanzhou University, Lanzhou 730000, China \\
$^4$Research Center for Hadron and CSR Physics, Lanzhou University and Institute of Modern Physics of CAS, Lanzhou 730000, China \\
$^5$School of Physics, Shandong University, Jinan 250100, China \\
$^6$School of Physics and Center of High Energy Physics, Peking University, Beijing 100871, China}
\ead{hxchen@seu.edu.cn, chenwei29@mail.sysu.edu.cn, xiangliu@lzu.edu.cn, yrliu@sdu.edu.cn {\rm and} zhusl@pku.edu.cn}
\vspace{10pt}
\begin{indented}
\item[] April 2022
\end{indented}

\begin{abstract}
The past decades witnessed the golden era of hadron physics. Many excited open heavy flavor mesons and baryons have been observed since 2017. We shall provide an updated review of the recent experimental and theoretical progresses in this active field. Besides the conventional heavy hadrons, we shall also review the recently observed open heavy flavor tetraquark states $X(2900)$ and $T_{cc}^+(3875)$ as well as the hidden heavy flavor multiquark states $X(6900)$, $P_{cs}(4459)^0$, $Z_{cs}(3985)^-$, $Z_{cs}(4000)^+$, and $Z_{cs}(4220)^+$. We will also cover the recent progresses on the glueballs and light hybrid mesons, which are the direct manifestations of the non-Abelian $SU(3)$ gauge interaction of the Quantum Chromodynamics in the low-energy region.
\end{abstract}


\vspace{2pc}


\submitto{\RPP}
%
%
%
\tableofcontents

\section{Introduction}
\label{sec1}

In particle physics a hadron is a composite subatomic particle made of quarks and gluons held together by the strong interaction. The conventional hadrons are categorized into two broad families: mesons and baryons. According to the quark model proposed by Gell-Mann and Zweig sixty years ago~\cite{Gell-Mann:1964ewy,Zweig:1964ruk}, a conventional meson is composed of one quark and one antiquark, and a conventional baryon is composed of three quarks. Such a simple model has been very successful in explaining hadron properties. All the ground-state light mesons and baryons composed of $up/down/strange$ quarks are well established, and there have been a huge number of excited light mesons and baryons observed in experiments~\cite{pdg}.

Besides, a lot of open heavy flavor mesons and baryons were observed after the discoveries of the $charm$~\footnote{The charm quark, shortly denoted as the $c$ quark, was predicted in 1970 by the GIM mechanism~\cite{Glashow:1970gm}, and discovered in 1974 independently at the Brooklyn National Laboratory~\cite{E598:1974sol} and at the Stanford Linear Accelerator Center~\cite{SLAC-SP-017:1974ind}.} and $bottom$~\footnote{The bottom quark, also known as the beauty quark and shortly denoted as the $b$ quark, was predicted by Kobayashi and Maskawa to explain the $CP$ violation in 1973~\cite{Kobayashi:1973fv}, and discovered in 1977 at the Fermilab~\cite{Herb:1977ek}.} quarks. Some of them have been reviewed in our previous paper~\cite{Chen:2016spr}, and in this paper we shall provide an updated review on their recent experimental and theoretical progresses. We will review the open heavy flavor mesons in Sec.~\ref{sec2} and the open heavy flavor baryons in Sec.~\ref{sec3}. We also refer to the reviews~\cite{Korner:1994nh,Copley:1979wj,Ivanov:1998ms,Capstick:2000qj,Bali:2000gf,Manohar:2000dt,Aoki:2001ra,Bianco:2003vb,Colangelo:2004vu,Rosner:2006jz,Crede:2013kia,Karliner:2008sv,Klempt:2009pi,Cheng:2015iom,Cheng:2021qpd} for more discussions.

At the birth of the quark model~\cite{Gell-Mann:1964ewy,Zweig:1964ruk}, the multiquark states, such as the $qq\bar q\bar q$ tetraquark states and the $qqqq\bar q$ pentaquark states, were proposed together with the conventional $q\bar q$ mesons and $qqq$ baryons. There are hundreds of hadrons observed in particle experiments in the last century, but most of them can be well categorized into the conventional $q\bar q$ mesons and $qqq$ baryons~\cite{pdg}. There exist only a few multiquark candidates, {\it e.g.}, the light scalar mesons $\sigma/f_0(500)$, $\kappa/K_0^*(700)$, $a_0(980)$, and $f_0(980)$ are good candidates for the light tetraquark states, whose detailed discussions can be found in the reviews~\cite{pdg,Amsler:2018zkm,Pelaez:2015qba,Close:2002zu,Amsler:2004ps,Bugg:2004xu}.

In 2003 the interests in the multiquark states were revived by the discovery of several important candidates, including the $X(3872)/\chi_{c1}(3872)$ observed by Belle~\cite{Belle:2003nnu}, the $D_{s0}^*(2317)$ observed by BaBar~\cite{BaBar:2003oey}, the $D_{s1}(2460)$ observed by CLEO~\cite{CLEO:2003ggt}, the $\Theta^+(1540)$ observed by LEPS~\cite{LEPS:2003wug,Diakonov:1997mm} but not confirmed by the subsequent experiments~\cite{Liu:2014yva,MartinezTorres:2010zzb}, and the anomalously proton-antiproton mass threshold enhancement observed by BESII~\cite{BES:2003aic}. Since 2003 there have been significant progresses in this field, and many multiquark candidates were continually observed in the BaBar, Belle, BESII/BESIII, CDF, CMS, COMPASS, D0, and LHCb experiments~\cite{pdg}, etc. Some of them have been reviewed in our previous papers~\cite{Chen:2016qju,Liu:2019zoy}, and in this paper we shall provide an updated review on recent experimental and theoretical progresses of this active field in the past five years. We will discuss the open heavy flavor multiquark candidates $X_{0,1}(2900)$~\cite{LHCb:2020bls,LHCb:2020pxc} and $T_{cc}^+$~\cite{LHCb:2021vvq,LHCb:2021auc} in Sec.~\ref{sec4}, and the hidden heavy flavor multiquark candidates $X(6900)$~\cite{LHCb:2020bwg}, $P_{cs}(4459)^0$~\cite{LHCb:2020jpq}, $Z_{cs}(3985)^-$~\cite{BESIII:2020qkh}, $Z_{cs}(4000)^+$~\cite{LHCb:2021uow}, and $Z_{cs}(4220)^+$~\cite{LHCb:2021uow} in Sec.~\ref{sec5}. We also refer to the reviews~\cite{Jaffe:2004ph,Swanson:2006st,Eichten:2007qx,Zhu:2007wz,Voloshin:2007dx,Godfrey:2008nc,Nielsen:2009uh,Drenska:2010kg,Brambilla:2010cs,Druzhinin:2011qd,Liu:2013waa,Brambilla:2014jmp,Esposito:2014rxa,Olsen:2014qna,Briceno:2015rlt,Hosaka:2016pey,Richard:2016eis,Oset:2016lyh,Oset:2016nvf,Lebed:2016hpi,Esposito:2016noz,Guo:2017jvc,Ali:2017jda,Olsen:2017bmm,Karliner:2017qhf,Albuquerque:2018jkn,Guo:2019twa,Brambilla:2019esw,Yang:2020atz,Ortega:2020tng,Barabanov:2020jvn,Yuan:2021wpg,JPAC:2021rxu,Chen:2021ftn,Meng:2022ozq,Brambilla:2022ura,Mai:2022eur} for more discussions.

With the advent of Quantum Chromodynamics (QCD) as the theory of the strong interaction, there should exist the glueballs and hybrid mesons which contain gluons as the dynamical degree of freedom. A glueball is composed of two or more ``valence'' gluons, and a hybrid meson is composed of one valence quark and one valence antiquark together with one or more ``valence'' gluons. They are important for the understanding of non-perturbative QCD~\cite{Fritzsch:1973pi,Fritzsch:1975tx,Freund:1975pn}, which is asymptotically free~\cite{Gross:1973id,Politzer:1973fx} and at the same time color-confining~\cite{Wilson:1974sk}. There have been various experimental and theoretical investigations on the glueballs and hybrid mesons in the past fifty years, but the academic community has not formed a common understanding on their nature. Some of their discussions can be found in the reviews~\cite{Klempt:2007cp,Crede:2008vw,Mathieu:2008me,Meyer:2010ku,Meyer:2015eta,Ochs:2013gi,Sonnenschein:2016pim,Bass:2018xmz,Ketzer:2019wmd,Roberts:2021nhw,Fang:2021wes,Jin:2021vct}, and in this paper we shall briefly review their recent experimental and theoretical results in Sec.~\ref{sec6}.

The summary and outlook of the present review will be provided in Sec.~\ref{sec7}.

The past decades witnessed the golden era of hadron physics, and there were so many excellent works even just within the past five years. We wish to but we are unable to review all of them in this limited review. In this paper, we shall not cover most of the following topics:
\begin{itemize}

\item the recent experiments updating the previous experiments that were not performed in the last five years;

\item the recent theoretical studies on the hadrons that were not observed in the last five years;

\item the $top$ quark~\footnote{The top quark, shortly denoted as the $t$ quark, was predicted by Kobayashi and Maskawa in 1973 to explain the $CP$ violation together with the $bottom$ quark~\cite{Kobayashi:1973fv}, and discovered in 1995 at the Fermilab~\cite{CDF:1995wbb,D0:1995jca}.} and the possible $top$ hadrons;

\item the experimental and theoretical studies that are closely related to the weak interaction, $CP$ violation, heavy-ion collisions, and dark matter, etc.;

\item $\bullet$ $\bullet$ ~~~ $\bullet$ $\bullet$ $\bullet$

\end{itemize}

\subsection{Internal structure of hadrons}
\label{sec1.1}

\begin{figure}[hbtp]
\begin{center}
\subfigure[]{\includegraphics[width=0.3\textwidth]{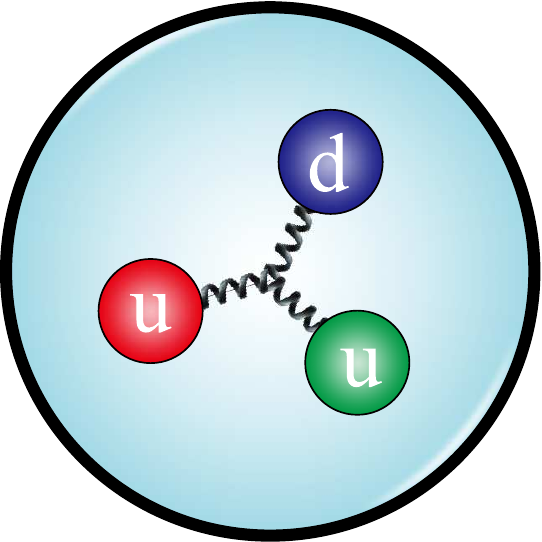}}
~~~~~~~~~~
\subfigure[]{\includegraphics[width=0.3\textwidth]{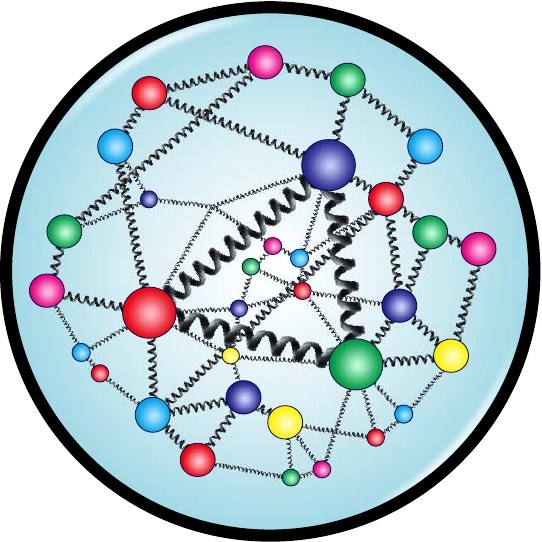}}
\end{center}
\caption{The internal structure of the proton from the viewpoints of (a) traditional quark model and (b) Quantum Chromodynamics (QCD).}
\label{fig:proton}
\end{figure}

Before reviewing the new hadron states observed in the past five years, we would like to discuss the internal structure of hadrons generally. Take the proton as an example. In the traditional quark model it is composed of two up quarks and one down quark, as shown in Fig.~\ref{fig:proton}(a). However, with the development of QCD as the theory of the strong interaction, we realize that its internal structure is not so simple at all. As shown in Fig.~\ref{fig:proton}(b), a proton actually contains two valence up quarks and one valence down quark, together with numberless sea quarks and gluons. Moreover, it is hard to tell which quark is ``valence'' and which quark is ``sea''. Hence, the proton is totally different from the hydrogen atom, which is exactly made of one proton and one electron bound by the electromagnetic interaction.

Then, why is the traditional quark model so simple but still very successful? One key reason is that the internal symmetry of hadrons is well described by the valence quarks and antiquarks through the group theory. There are four important internal symmetries among the quarks and antiquarks inside hadrons, which are associated with:
\begin{itemize}

\item the color degree of freedom,

\item the flavor degree of freedom,

\item the spin degree of freedom,

\item the orbital degree of freedom.

\end{itemize}
We shall discuss them separately in the following subsections.

\subsubsection{Color structure.}
\label{sec1.1.1}

\begin{figure}[hbtp]
\begin{center}
\includegraphics[width=1\textwidth]{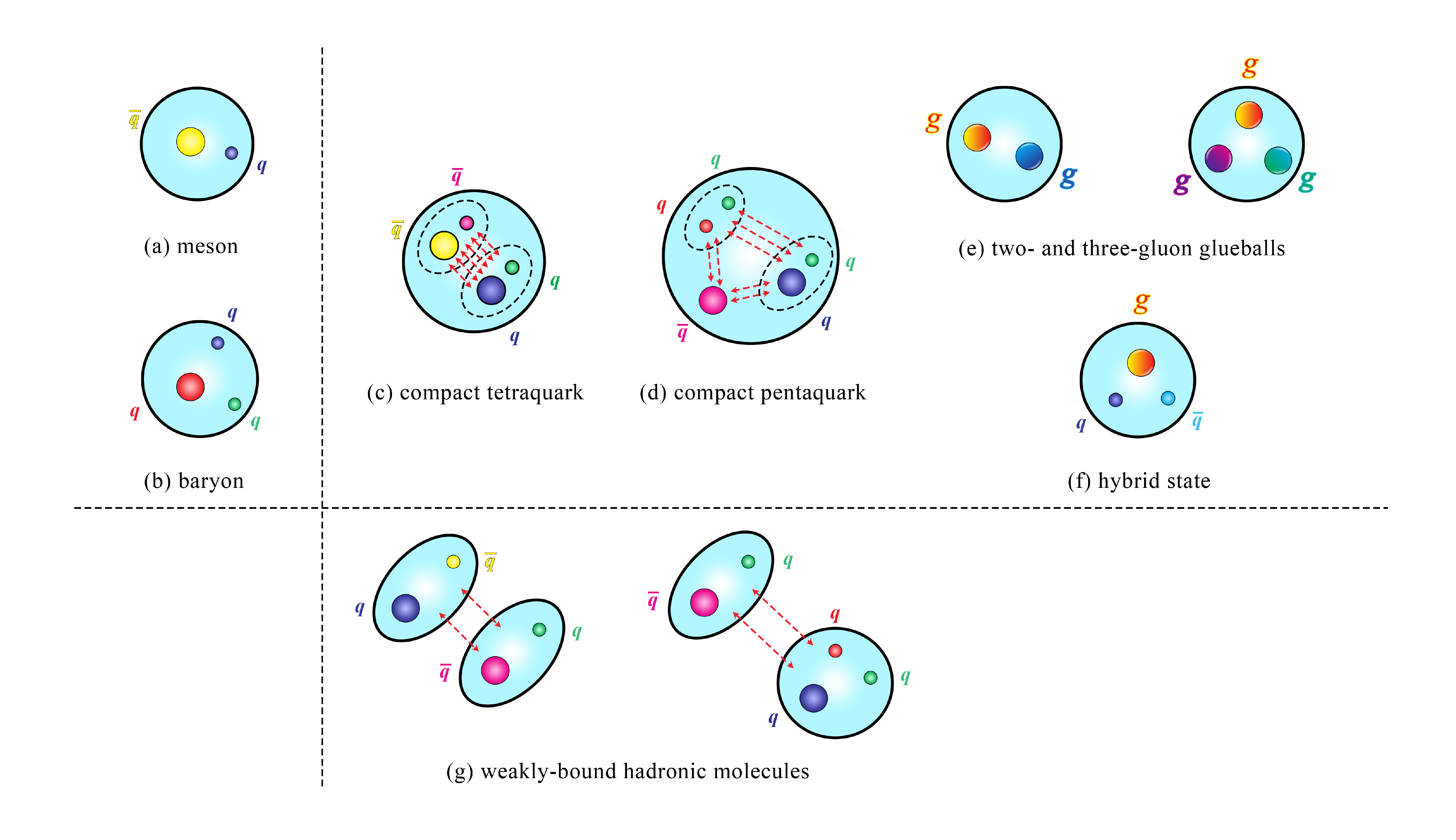}
\end{center}
\caption{Categorization of hadrons according to the color degree of freedom into the (a) $q \bar q$ meson, (b) $qqq$ baryon, (c) compact $qq\bar q\bar q$ tetraquark, (d) compact $qqqq\bar q$ pentaquark, (e) two- and three-gluon glueballs, (f) hybrid state, and (g) hadronic molecules. The hadrons in the left panel are conventional hadrons, and those in the right panel are exotic hadrons. The hadrons in the top panel are tightly bound by the strong interaction, and those in the bottom panel are weakly bound by the residual strong interaction.}
\label{fig:color}
\end{figure}

The color degree of freedom follows the color $SU(3)$ group, shortly denoted as the $SU(3)_C$ sometimes. As shown in Fig.~\ref{fig:color}, we can apply it to categorize hadrons into the conventional $q\bar q$ mesons and $qqq$ baryons, the compact $qq\bar q\bar q$ tetraquarks and $qqqq\bar q$ pentaquarks, as well as the weakly-bound hadronic molecules. If we further take into account the gluon degree of freedom, there may exist the glueballs and hybrid states. Let us further discuss them partly in Sec.~\ref{sec1.2} and partly here.

One quark and one antiquark can constitute a conventional meson with the color structure
\begin{eqnarray}
{\rm Meson} &:& \mathbf{3}_q \otimes \mathbf{\bar 3}_{\bar q} \rightarrow \mathbf{1}_{\bar q q} \, ,
\end{eqnarray}
and three quarks can constitute a conventional baryon with the totally antisymmetric color structure
\begin{eqnarray}
{\rm Baryon} &:& \mathbf{3}_q \otimes \mathbf{3}_q \otimes \mathbf{3}_q \rightarrow \mathbf{1}_{q q q} \, .
\end{eqnarray}
In contrast, the color structure of the $qq\bar q\bar q$ tetraquark state can be either the hadronic molecule type:
\begin{eqnarray}
{\rm Molecule} \,:\, \mathbf{3}_q \otimes \mathbf{3}_q \otimes \mathbf{\bar 3}_{\bar q} \otimes \mathbf{\bar 3}_{\bar q} &\rightarrow& \mathbf{1}_{[\bar q q]_\mathbf{1} [\bar q q]_\mathbf{1}} \, ,
\end{eqnarray}
or the compact tetraquark (diquark-antidiquark) type:
\begin{eqnarray}
{\rm Tetraquark} \,:\, \mathbf{3}_q \otimes \mathbf{3}_q \otimes \mathbf{\bar 3}_{\bar q} \otimes \mathbf{\bar 3}_{\bar q}
&\rightarrow& \mathbf{1}_{[q q]_\mathbf{\bar 3} [\bar q \bar q]_\mathbf{3}} \oplus \mathbf{1}_{[q q]_\mathbf{6} [\bar q \bar q]_\mathbf{\bar 6}} \, .
\end{eqnarray}
The color structure of the $qqq q\bar q$ pentaquark state can also be either the hadronic molecule type:
\begin{eqnarray}
{\rm Molecule} \,:\, \mathbf{3}_q \otimes \mathbf{3}_q \otimes \mathbf{3}_q \otimes \mathbf{3}_q \otimes \mathbf{\bar 3}_{\bar q} &\rightarrow& \mathbf{1}_{[q q q]_\mathbf{1} [\bar q q]_\mathbf{1}} \, ,
\end{eqnarray}
or the compact pentaquark (diquark-diquark-antiquark) type:
\begin{eqnarray}
{\rm Pentaquark} &:& \mathbf{3}_q \otimes \mathbf{3}_q \otimes \mathbf{3}_q \otimes \mathbf{3}_q \otimes \mathbf{\bar 3}_{\bar q}
\\ \nonumber &\rightarrow& \mathbf{1}_{[q q]_\mathbf{\bar 3} [q q]_\mathbf{\bar 3} [\bar q]_\mathbf{\bar 3}}
\oplus \mathbf{1}_{[q q]_\mathbf{\bar 3} [q q]_\mathbf{6} [\bar q]_\mathbf{\bar 3}}
\oplus \mathbf{1}_{[q q]_\mathbf{6} [q q]_\mathbf{\bar 3} [\bar q]_\mathbf{\bar 3}} \, .
\end{eqnarray}
Besides, there are some other possible color structures, such as the color-octet-color-octet configuration $\mathbf{1}_{[\bar q q]_\mathbf{8} [\bar q q]_\mathbf{8}}$ for the tetraquark state and the diquark-triquark configuration $\mathbf{1}_{[q q]_\mathbf{\bar 3} [q q \bar q]_\mathbf{3}}$ for the pentaquark state, etc.

Algebraically, there exist some relations between the color configurations of the compact multiquark states and those of the hadronic molecular states. For example, we can write down four color configurations for the tetraquark states:
\begin{eqnarray}
\nonumber \mathbf{1}_{[\bar q q]_\mathbf{1} [\bar q q]_\mathbf{1}} &\longrightarrow& \delta_{ab} \delta_{cd} \times \bar q^a q^b \bar q^c q^d \, ,
\\
\mathbf{1}_{[\bar q q]_\mathbf{8} [\bar q q]_\mathbf{8}} &\longrightarrow& \lambda^n_{ab} \lambda^n_{cd} \times \bar q^a q^b \bar q^c q^d \, ,
\\
\nonumber \mathbf{1}_{[q q]_\mathbf{\bar 3} [\bar q \bar q]_\mathbf{3}} &\longrightarrow& (\delta_{ab} \delta_{cd} - \delta_{ad} \delta_{cb}) \times \bar q^a q^b \bar q^c q^d \, ,
\\
\nonumber \mathbf{1}_{[q q]_\mathbf{6} [\bar q \bar q]_\mathbf{\bar 6}} &\longrightarrow& (\delta_{ab} \delta_{cd} + \delta_{ad} \delta_{cb}) \times \bar q^a q^b \bar q^c q^d \, ,
\end{eqnarray}
which can be related through the color rearrangements
\begin{eqnarray}
\delta_{ad} \delta_{bc} &=& {1\over3} \delta_{ab} \delta_{cd} + {1\over2}\lambda^n_{ab} \lambda^n_{cd} \, ,
\label{sec1:color}
\\ \nonumber \lambda^n_{ad} \lambda^n_{bc} &=& {16\over9} \delta_{ab} \delta_{cd} - {1\over3}\lambda^n_{ab} \lambda^n_{cd} \, ,
\end{eqnarray}
with $a \cdots d = 1 \cdots 3$ and $n = 1 \cdots 8$ the color indices. However, the compact multiquark states and the hadronic molecular states are totally different from each other. As shown in Fig.~\ref{fig:color}, the compact multiquark states are tightly bound by the strong interaction directly, while the hadronic molecular states are weakly bound by the residual strong interaction. In some sense they are similar to the atoms and chemical molecules.

There have been heated debates on whether the multiquark candidates observed in experiments are hadronic molecular states or compact multiquark states. Besides, some of these candidates were proposed to be caused by the kinematical effects, such as the triangle singularities~\cite{Guo:2019twa}. These debates have been significantly improving our understanding of their nature as well as the non-perturbative behaviors of the strong interaction in the low energy region, and hopefully, will continue to deepen our insights into the multiquark states until we finally understand them.

\subsubsection{Flavor structure.}
\label{sec1.1.2}

The flavor symmetry is an approximate symmetry, which follows the $SU(N_f)$ group, with $N_f$ the number of flavors.

\begin{figure}[hbtp]
\begin{center}
\subfigure[]{\includegraphics[width=0.4\textwidth]{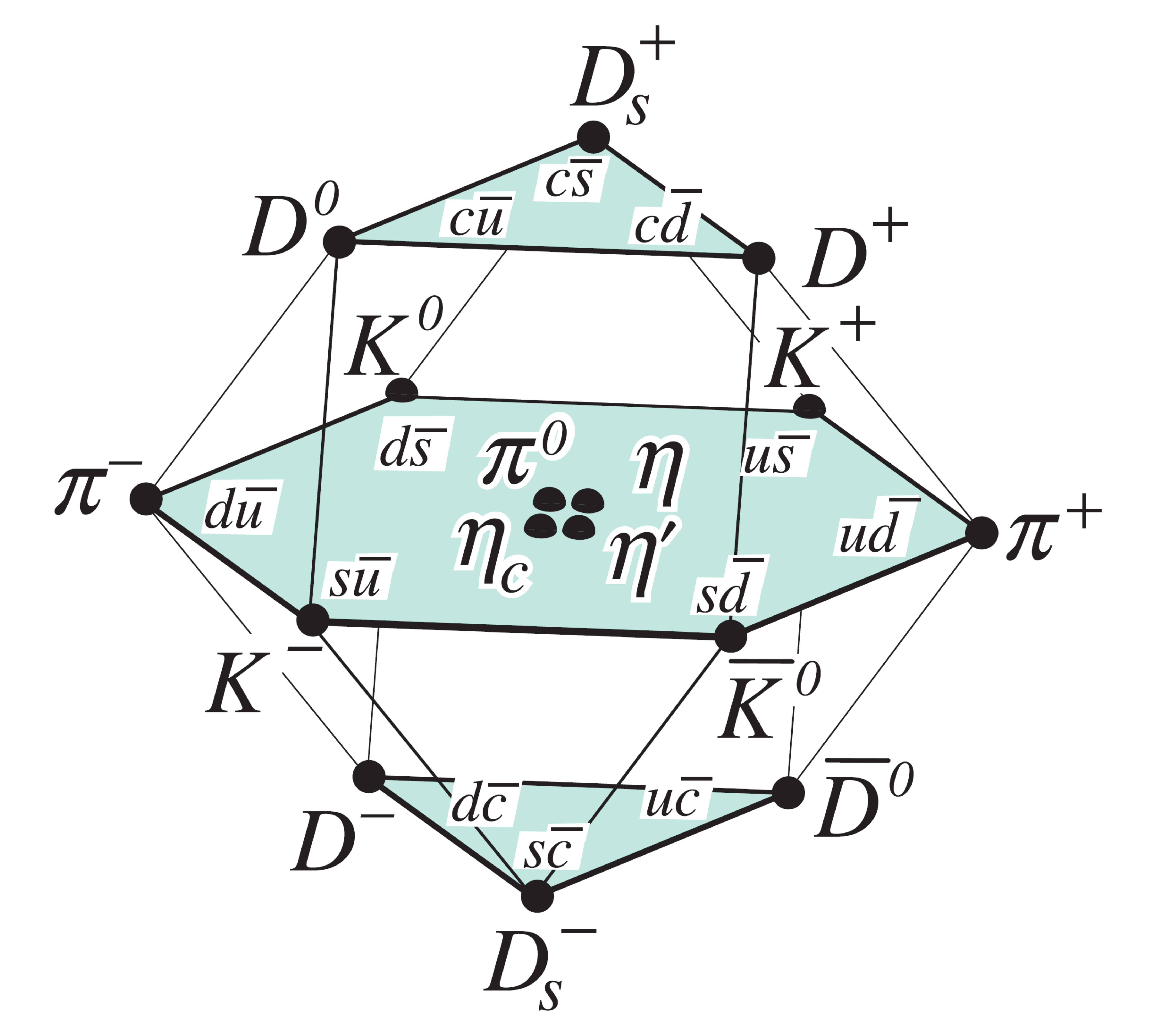}}
~~~~~
\subfigure[]{\includegraphics[width=0.4\textwidth]{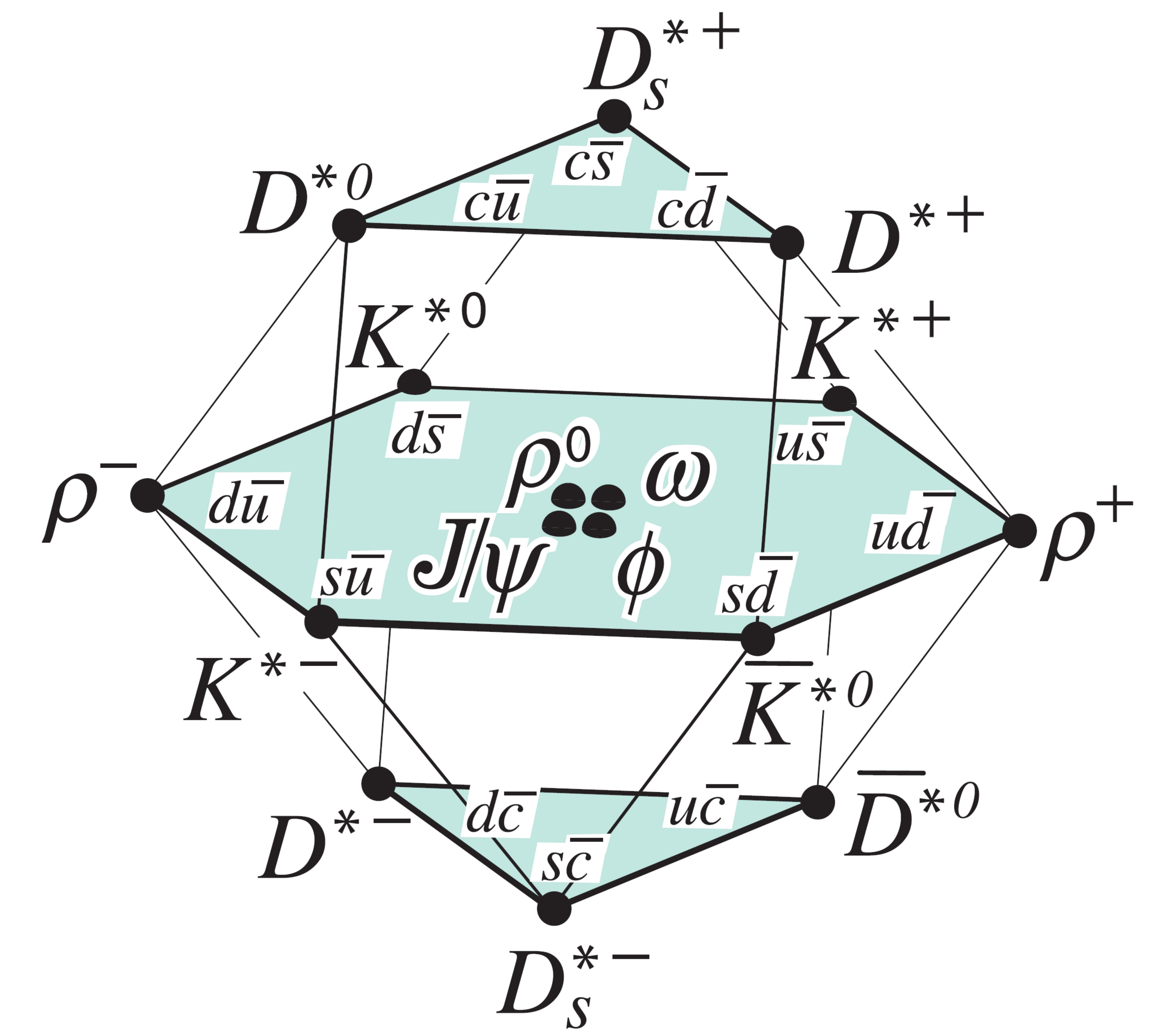}}
\\
\subfigure[]{\includegraphics[width=0.4\textwidth]{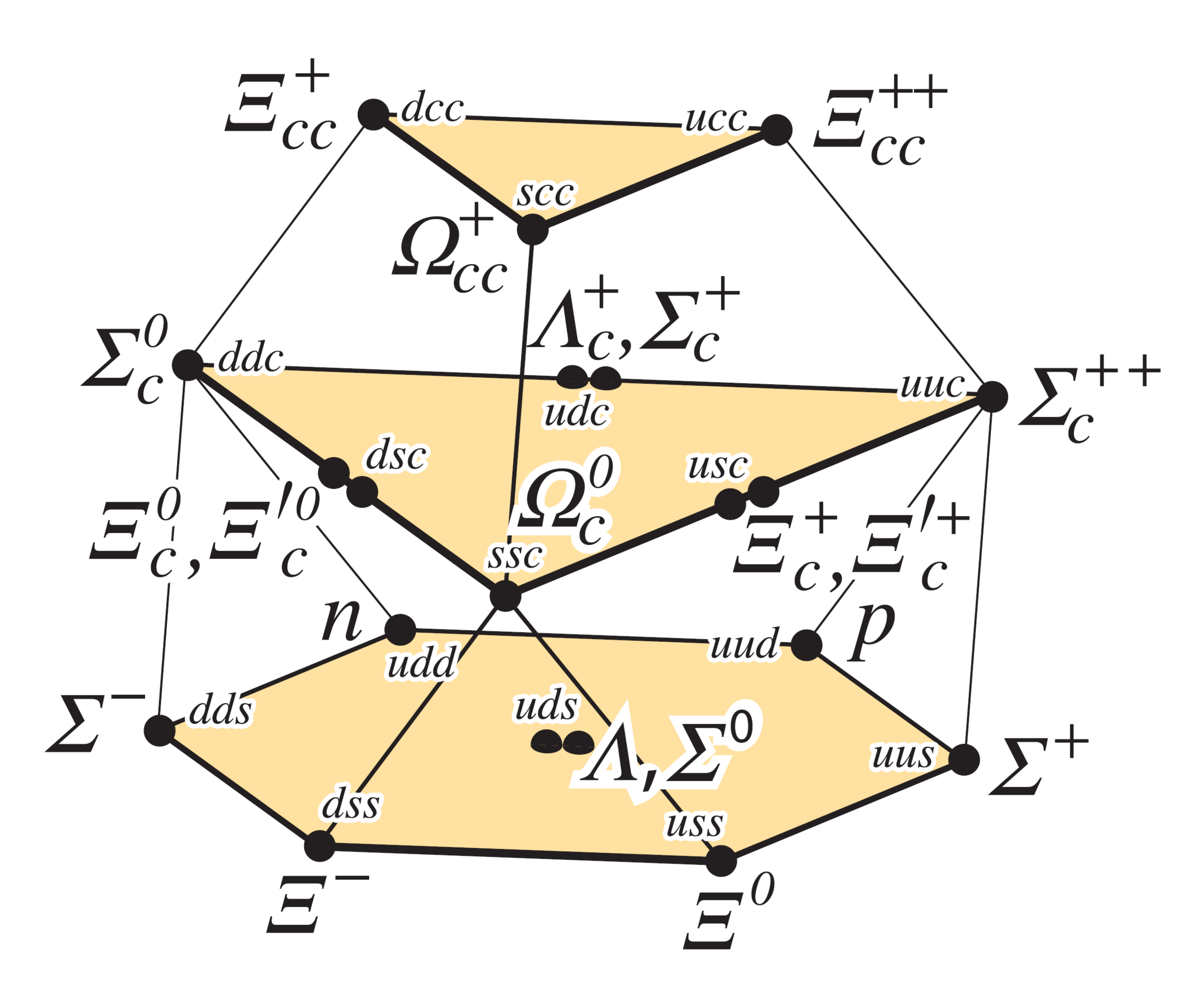}}
~~~~~
\subfigure[]{\includegraphics[width=0.4\textwidth]{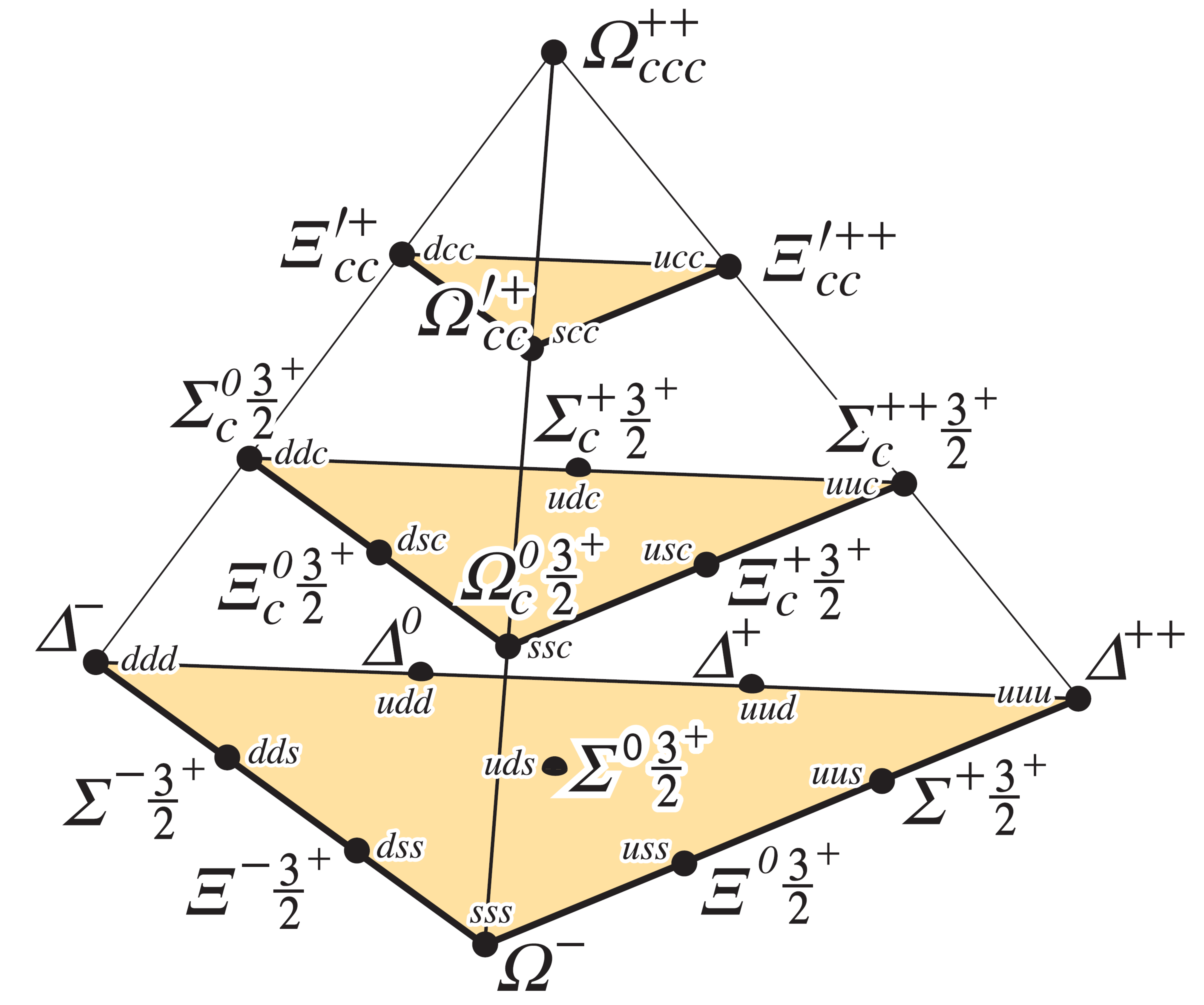}}
\end{center}
\caption{Flavor $SU(4)$ weight diagrams for the ground-state mesons and baryons made of $up/down/strange/charm$ quarks: (a) the 16-plet for the pseudoscalar mesons with an $SU(3)$ octet, (b) the 16-plet for the vector mesons with an $SU(3)$ octet, (c) the 20-plet for the spin-1/2 baryons with an $SU(3)$ octet, and (d) the 20-plet for the spin-3/2 baryons with an $SU(3)$ decuplet. Source: Ref.~\cite{ParticleDataGroup:2018ovx}.}
\label{fig:weight}
\end{figure}

If we only consider the light $up$ and $down$ quarks, the flavor symmetry becomes the isospin $SU(2)$ symmetry, which works quite well. The isospin representations for the $q\bar q$ mesons and $qqq$ baryons are
\begin{equation}
SU(2)_{\rm isospin}
\left\{\begin{array}{ccl}
{\rm Meson}  & : & \mathbf{2}_q \otimes \mathbf{\bar 2}_{\bar q} = \mathbf{1} \oplus \mathbf{3} \, ,
\\[3mm]
{\rm Baryon} & : & \mathbf{2}_q \otimes \mathbf{2}_{q} \otimes \mathbf{2}_{q} = \mathbf{2} \oplus \mathbf{2} \oplus \mathbf{4} \, .
\end{array}\right.
\end{equation}
If we further consider the light $strange$ quark, we have the flavor $SU(3)$ symmetry, which still works but not so well:
\begin{equation}
SU(3)_{\rm flavor}
\left\{\begin{array}{ccl}
{\rm Meson}  & : & \mathbf{3}_q \otimes \mathbf{\bar 3}_{\bar q} = \mathbf{1} \oplus \mathbf{8} \, ,
\\[3mm]
{\rm Baryon} & : & \mathbf{3}_q \otimes \mathbf{3}_{q} \otimes \mathbf{3}_{q} = \mathbf{1} \oplus \mathbf{8} \oplus \mathbf{8} \oplus \mathbf{10} \, .
\end{array}\right.
\end{equation}
We may take into account the heavy $charm$ or $bottom$ quark, and obtain the flavor $SU(4)$ symmetry:
\begin{equation}
SU(4)_{\rm flavor}
\left\{\begin{array}{ccl}
{\rm Meson}  & : & \mathbf{4}_q \otimes \mathbf{\bar 4}_{\bar q} = \mathbf{1} \oplus \mathbf{15} \, ,
\\[3mm]
{\rm Baryon} & : & \mathbf{4}_q \otimes \mathbf{4}_{q} \otimes \mathbf{4}_{q} = \mathbf{4} \oplus \mathbf{20} \oplus \mathbf{20} \oplus \mathbf{20} \, .
\end{array}\right.
\end{equation}
This symmetry is badly broken, but we can still use it to categorize hadrons, as shown in Fig.~\ref{fig:weight} for the ground-state mesons and baryons made of $up/down/strange/charm$ quarks. Technically, we can also separately deal with the heavy $charm/bottom$ quarks and the light $up/down/strange$ quarks, {\it e.g.}, we can still use the flavor $SU(3)$ symmetry to categorize the open heavy flavor mesons and baryons, which will be discussed in Sec.~\ref{sec1.3}.

The flavor structure of the multiquark states is quite complicated. For example, the flavor $SU(3)$ representations of the compact $qq\bar q\bar q$ tetraquark states are
\begin{eqnarray}
&& \mathbf{3}_q \otimes \mathbf{3}_q \otimes \mathbf{\bar 3}_{\bar q} \otimes \mathbf{\bar 3}_{\bar q}
\\ \nonumber &=&
\left(\mathbf{1}\oplus\mathbf{8}\right)_{[q q]_\mathbf{\bar 3} [\bar q \bar q]_\mathbf{3}}
\oplus \left(\mathbf{8}\oplus\mathbf{10}\right)_{[q q]_\mathbf{6} [\bar q \bar q]_\mathbf{3}}
\oplus \left(\mathbf{8}\oplus\overline{\mathbf{10}}\right)_{[q q]_\mathbf{\bar 3} [\bar q \bar q]_\mathbf{\bar 6}}
\oplus \left(\mathbf{1}\oplus\mathbf{8}\oplus\mathbf{27}\right)_{[q q]_\mathbf{6} [\bar q \bar q]_\mathbf{\bar 6}} \, .
\end{eqnarray}
Among these multiplets, the light nonet $\left(\mathbf{1}\oplus\mathbf{8}\right)_{[q q]_\mathbf{\bar 3} [\bar q \bar q]_\mathbf{3}}$ was suggested to lie below 1~GeV a long time ago~\cite{Jaffe:1976ig,Jaffe:1976ih}, which can be used to explain the nine light scalar mesons $\sigma/f_0(500)$, $\kappa/K_0^*(700)$, $a_0(980)$, and $f_0(980)$~\cite{pdg,Amsler:2018zkm,Pelaez:2015qba,Close:2002zu,Amsler:2004ps,Bugg:2004xu}.

Sometimes one needs to further consider the chiral symmetry. The spontaneous breaking of the chiral symmetry
\begin{equation}
SU(N_f)_L \otimes SU(N_f)_R \rightarrow SU(N_f)_V \, ,
\end{equation}
plays a dynamical role in the presence of the Nambu-Goldstone bosons. This mechanism also makes the QCD vacuum possess nontrivial structures, such as the nonzero quark condensates. Interested readers may consult Refs.~\cite{Chen:2008qv,Chen:2009sf,Chen:2010ba,Chen:2011rh,Dmitrasinovic:2016hup,Dmitrasinovic:2020wye,Dmitrasinovic:2011yf,Chen:2012vs,Chen:2013efa,Chen:2012ex,Chen:2012ut,Chen:2013jra,Chen:2013gnu,Chen:2020arr,Dong:2022otb} for a systematical study on chiral properties of the light baryon and tetraquark fields.

\subsubsection{Spin and orbital structures.}
\label{sec1.1.3}

Besides the color and flavor degrees of freedom, there are the spin and orbital degrees of freedom among quarks and antiquarks. One often needs to consider these four degrees of freedom as a whole. For example, the light $SU(3)$ flavor and spin are combined to be an approximate flavor-spin $SU(6)$ symmetry, which is quite useful when studying light baryons.

Within the conventional quark model, the spin of a hadron is contributed by its internal valence quarks and antiquarks. In 1987, the European Muon collaboration measured the fraction of the proton spin carried by quarks to be~\cite{EuropeanMuon:1987isl,EuropeanMuon:1989yki,Hughes:1983kf}:
\begin{equation}
\Delta \Sigma(Q^2 = 10.7{\rm~GeV}^2 ) = 0.060 \pm 0.047 \pm 0.069 \, ,
\end{equation}
which was consistent with zero. This result shocked the physics community and created the so-called ``proton spin crisis''~\cite{Cheng:1996jr,Filippone:2001ux,Bass:2004xa,Aidala:2012mv,Leader:2013jra,Ji:2016djn,Deur:2018roz,Ji:2020ena,Liu:2021lke}.

The experimental and theoretical studies on the proton spin show clearly that the proton as well as all the other hadrons have complicated internal structures that we still do not fully understand. Here we would like to note once more that the traditional quark model is simple but successful, (partly) because the internal symmetry of hadrons is well described by the valence quarks and antiquarks. Accordingly, we need to pay much attention to various symmetries in the study of hadron spectroscopy. One good example is the ground-state singly heavy baryons, which can be found at the end of this section.

A useful tool to study the hadron spectroscopy is the hadronic interpolating currents/fields/operators made of quark and gluon fields, which can describe the color, flavor, spin, and orbital degrees of freedom simultaneously. They are widely used in the lattice QCD and QCD sum rule studies. Interested readers may consult Refs.~\cite{Chen:2006hy,Chen:2006hyp,Chen:2007xr,Chen:2009gs,Chen:2008ej,Chen:2018kuu,Chen:2015fwa,Cui:2019roq,Dong:2020okt,Jiao:2009ra,Su:2020reg,Chen:2014vha,Chen:2021hxs} for systematical QCD sum rule studies on the light tetraquark and sexaquark/dibaryon currents.

Based on the Fierz and color rearrangements, one can derive some useful relations between the local currents of the compact multiquark color configurations and those of the hadronic molecular color configurations. However, this equivalence is just between the local currents, while the compact multiquark states and the hadronic molecular states have very different spatial configurations, as shown in Fig.~\ref{fig:color}. To describe them exactly, one probably needs non-local interpolating currents, but we are still incapable of using such currents to perform QCD sum rule analyses.

\subsection{Gluon degree of freedom}
\label{sec1.2}

With the advent of QCD as the theory of the strong interaction, the gluon degree of freedom naturally leads to the existence of the glueballs and hybrid mesons. However, their nature still remains elusive due to our poor understanding of the gluon degree of freedom. Experimentally, it is not easy to identify the glueballs and hybrid mesons unambiguously, and there is currently no definite experimental evidence on their existence. Theoretically, it is also not easy to define the constituent gluon precisely, although there have been some proposals to construct glueballs and hybrid mesons using constituent gluons~\cite{Horn:1977rq,Coyne:1980zd,Chanowitz:1980gu,Cho:2015rsa}.

The key question of the constituent gluon model concerns the mass of the valence gluon, and the pioneering works~\cite{Barnes:1981ac,Cornwall:1982zn} gave two discrepant answers. The authors of Ref.~\cite{Barnes:1981ac} argued that the valence gluon is still a massless particle, and it gains a dynamical mass that is the pole position of the dressed gluon propagator. In contrast, the authors of Ref.~\cite{Cornwall:1982zn} argued that the valence gluon has to be {\it a priori} considered as a massive particle, and the non-perturbative effects of QCD cause a mass term to appear in the gluon propagator. It is still controversial how to describe the valence gluon inside glueballs and hybrid mesons, so there is currently not a good and intuitive way to well categorize them.

Instead, we can always categorize the glueballs and hybrid mesons according to the color degree of freedom, as shown in Fig.~\ref{fig:color}:
\begin{itemize}

\item Two valence gluons can constitute a glueball with the color structure
\begin{eqnarray}
\mbox{Two-gluon glueball} &:& \mathbf{8}_g \otimes \mathbf{8}_g \rightarrow \mathbf{1}_{gg} \, .
\end{eqnarray}
We denote it as $G^nG_n$, with $G^n$ the valence gluon of the color index $n=1\cdots8$.

\item Three valence gluons can constitute a glueball with the color structure
\begin{eqnarray}
\mbox{Three-gluon glueball} &:& \mathbf{8}_g \otimes \mathbf{8}_g \otimes \mathbf{8}_g \rightarrow \mathbf{1}^S_{ggg} \oplus \mathbf{1}^A_{ggg} \, ,
\end{eqnarray}
where $\mathbf{1}^S_{ggg}$ denotes the symmetric color configuration $d_{nml}G^nG^mG^l$ and $\mathbf{1}^A_{ggg}$ denotes the antisymmetric color configuration $f_{nml}G^nG^mG^l$, with $d^{nml}$ and $f^{nml}$ the totally symmetric and antisymmetric $SU(3)$ structure constants, respectively.

\item One quark-antiquark pair together with one valence gluon can constitute a hybrid state with the color structure
\begin{eqnarray}
\mbox{One-gluon hybrid} &:& \mathbf{3}_q \otimes \mathbf{\bar 3}_{\bar q} \otimes \mathbf{8}_g \rightarrow \mathbf{1}_{\bar q q g} \, .
\end{eqnarray}
We denote it as $\bar q_a \lambda_n^{ab} q_b G^n$.

\item One quark-antiquark pair together with two valence gluons can constitute a hybrid state with the color structure
\begin{eqnarray}
\mbox{Two-gluon hybrid} &:& \mathbf{3}_q \otimes \mathbf{\bar 3}_{\bar q} \otimes \mathbf{8}_g \otimes \mathbf{8}_g \rightarrow \mathbf{1}^S_{\bar q q gg} \oplus \mathbf{1}^A_{\bar q q gg} \, ,
\end{eqnarray}
where $\mathbf{1}^S_{\bar q q gg}$ denotes the symmetric color configuration $d_{nml}\bar q^a \lambda^n_{ab} q^bG^mG^l$ and $\mathbf{1}^A_{\bar q q gg}$ denotes the antisymmetric color configuration $f_{nml}\bar q^a \lambda^n_{ab} q^bG^mG^l$.

\end{itemize}
Their recent experimental and theoretical progresses will be reviewed briefly in Sec.~\ref{sec6}.

\subsection{Heavy quark symmetry}
\label{sec1.3}

The heavy quark symmetry plays an important role in the study of heavy hadrons~\cite{Neubert:1993mb,Manohar:2000dt,Rosner:1985dx,Nussinov:1986hw,Voloshin:1986dir,Lepage:1987gg,Politzer:1988wp,Isgur:1989vq,Isgur:1991wq,Eichten:1989zv,Georgi:1990um,Grinstein:1990mj,Eichten:1993ub}. In this subsection we apply it to categorize the singly heavy mesons and baryons, based on which we shall review their recent experimental and theoretical progresses in Sec.~\ref{sec2} and Sec.~\ref{sec3}. We shall omit the word ``singly'' most of the time, which can largely simplify our discussions.

Let us investigate the heavy meson system first. In the heavy quark limit $m_Q\to\infty$, the one-gluon-exchange interaction between a heavy quark and a light antiquark is independent of the heavy quark mass, which results in the heavy quark flavor symmetry and heavy quark spin symmetry. Now the one-gluon-exchange potential contains only the Coulomb part and the spin-orbit part of the light antiquark, so the heavy quark looks like a static color source for the light antiquark. This picture is similar to the hydrogen system, and therefore, its hadron properties are mainly determined by the light degrees of freedom.

To be specific, we focus on the $Q\bar{q}$ system, with $Q$ the heavy quark and $\bar q$ the light antiquark. Its total angular momentum
\begin{equation}
{J}={L} \otimes ({s}_Q \otimes {s}_{\bar q})_{S} \, ,
\end{equation}
is reduced in the heavy quark limit to be
\begin{equation}
{J}={s}_Q \otimes ({L} \otimes {s}_{\bar q})_{j} \, ,
\end{equation}
where $S$ and $L$ are the total spin and orbital angular momenta, $s_Q$ and $s_{\bar q}$ are the heavy and light quark spin angular momenta, and $j = L \otimes s_{\bar q}$ is the angular momentum of the light degrees of freedom.

As a result, the two mesons with the same $(L,j)$ form a degenerate doublet. There is one lowest-lying $1S$ doublet $(J^P = 0^-,1^-)=(D,D^*)$, satisfying
\begin{equation}
{j} = {1 \over 2} \, , \, {J} = {s}_Q \otimes {j} = 0 \oplus 1 \, , \, P = - \, ;
\end{equation}
there are two excited $1P$ doublets $(0^+,1^+)=(D_0^*,D_1)$ and $(1^+,2^+)=(D_1^\prime,D_2^*)$, satisfying
\begin{equation}
{j} = {1 \over 2} \oplus {3 \over 2} \, , \, {J} = {s}_Q \otimes {j} = (0 \oplus 1) \oplus (1 \oplus 2) \, , \, P = + \, ;
\end{equation}
there are also two excited $1D$ doublets $(1^-,2^-)=(D_1^*,D_2)$ and $(2^-,3^-)=(D_2^\prime,D_3^*)$, satisfying
\begin{equation}
{j} = {3 \over 2} \oplus {5 \over 2} \, , \, {J} = {s}_Q \otimes {j} = (1 \oplus 2) \oplus (2 \oplus 3) \, , \, P = - \, .
\end{equation}
With the observation of more and more $Q\bar{q}$ mesons, one may also find the possible doublets with the radial excitations~\cite{Colangelo:2010te}.

\begin{figure}[hbtp]
\begin{center}
\includegraphics[width=0.4\textwidth]{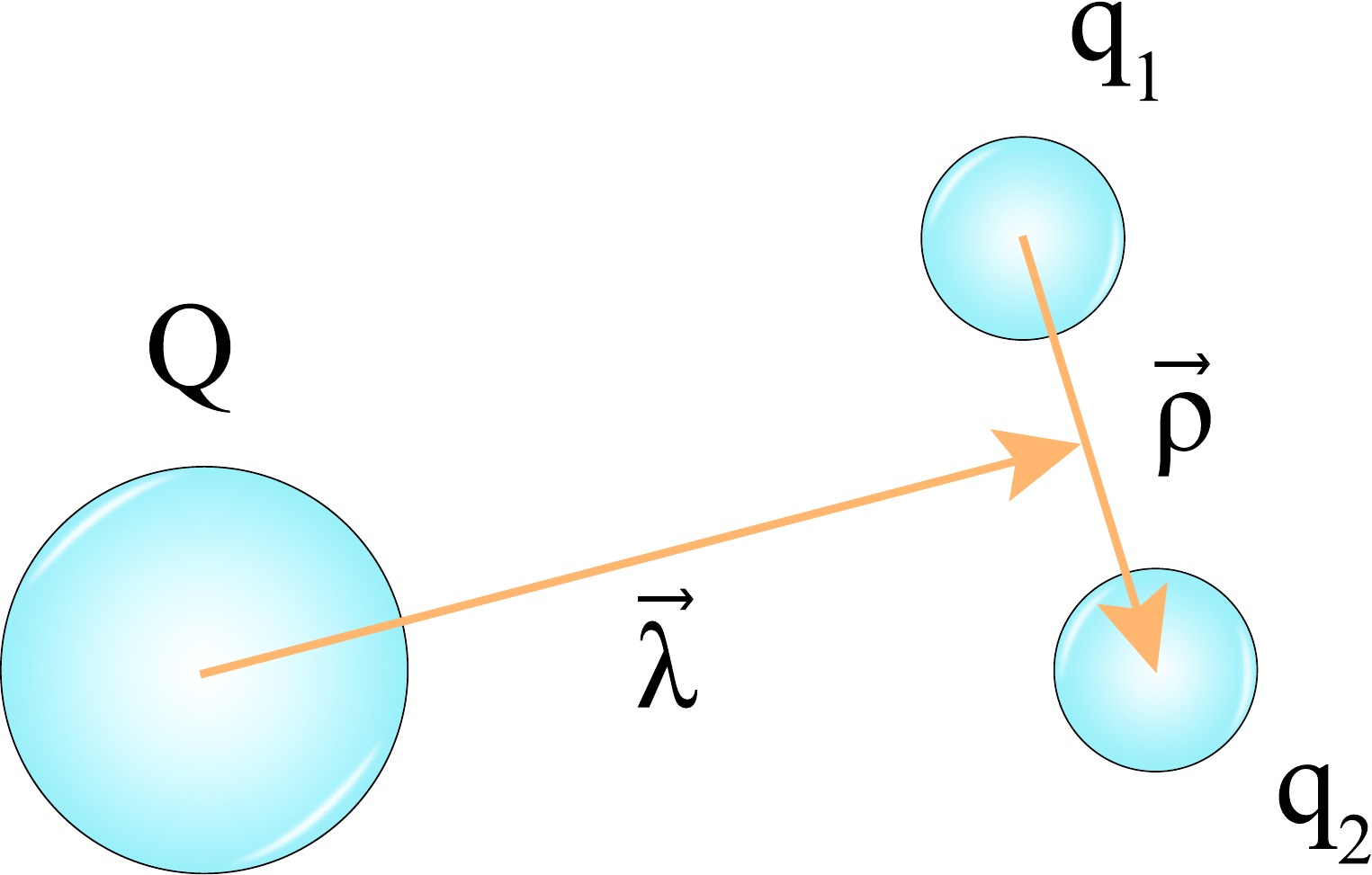}
\end{center}
\caption{Jacobi coordinates $\vec \lambda$ and $\vec \rho$ for the heavy baryon system.}
\label{fig:Jacobi}
\end{figure}

Similar to the heavy meson system, the heavy quark in the singly heavy baryon system also looks like a static color source for the two light quarks in the heavy quark limit $m_Q\to\infty$. This picture is similar to the atom system, and the heavy baryon properties are also mainly determined by the light degrees of freedom. However, its internal structure is much more complicated than that of the heavy meson.

As shown in Fig.~\ref{fig:Jacobi}, we focus on the $Qq_1q_2$ system, with $Q$ the heavy quark and $q_{1,2}$ the two light quarks:
\begin{itemize}

\item Its total orbital angular momentum can be separated into
\begin{equation}
L = l_\lambda \otimes l_\rho \, ,
\end{equation}
where the former $l_\lambda$ denotes the orbital angular momentum between the heavy quark and the two-light-quark system, and the latter $l_\rho$ denotes the orbital angular momentum between the two light quarks. We call them the $\lambda$- and $\rho$-mode excitations, respectively.

\item Its total spin angular momentum can be separated into
\begin{equation}
S = {s}_Q \otimes {s}_{q_1} \otimes {s}_{q_2} = {s}_Q \otimes s_l \, ,
\end{equation}
where $s_Q$ and $s_{q_{1,2}}$ are the heavy and light quark spin angular momenta, and $s_l = s_{qq} = {s}_{q_1} \otimes {s}_{q_2}$ is the total spin angular momentum of the light component.

\item Its total angular momentum
\begin{equation}
{J}=\left(l_\lambda \otimes l_\rho\right)_L \otimes \left({s}_Q \otimes {s}_{q_1} \otimes {s}_{q_2}\right)_{S} \, ,
\end{equation}
is reduced in the heavy quark limit to be
\begin{equation}
{J}={s}_Q \otimes \left(\left(l_\lambda \otimes l_\rho\right)_L \otimes \left({s}_{q_1} \otimes {s}_{q_2}\right)_{s_l}\right)_{j_l} \, ,
\end{equation}
where $j_l = L \otimes s_l = l_\lambda \otimes l_\rho \otimes {s}_{q_1} \otimes {s}_{q_2}$ is the total angular momentum of the light component. Besides, we use $j_{qq} = l_\rho \otimes s_l = l_\rho \otimes {s}_{q_1} \otimes {s}_{q_2}$ to denote the total angular momentum of the two light quarks.
\end{itemize}
As a result, sometimes two heavy baryons can have the same quantum numbers $(l_\lambda, l_\rho, L, s_l, j_l)$, and they form a degenerate doublet; sometimes there is only one heavy baryon with the quantum numbers $(l_\lambda, l_\rho, L, s_l, j_l)$, and it forms a singlet itself.

\begin{figure}[hbtp]
\begin{center}
\includegraphics[width=0.6\textwidth]{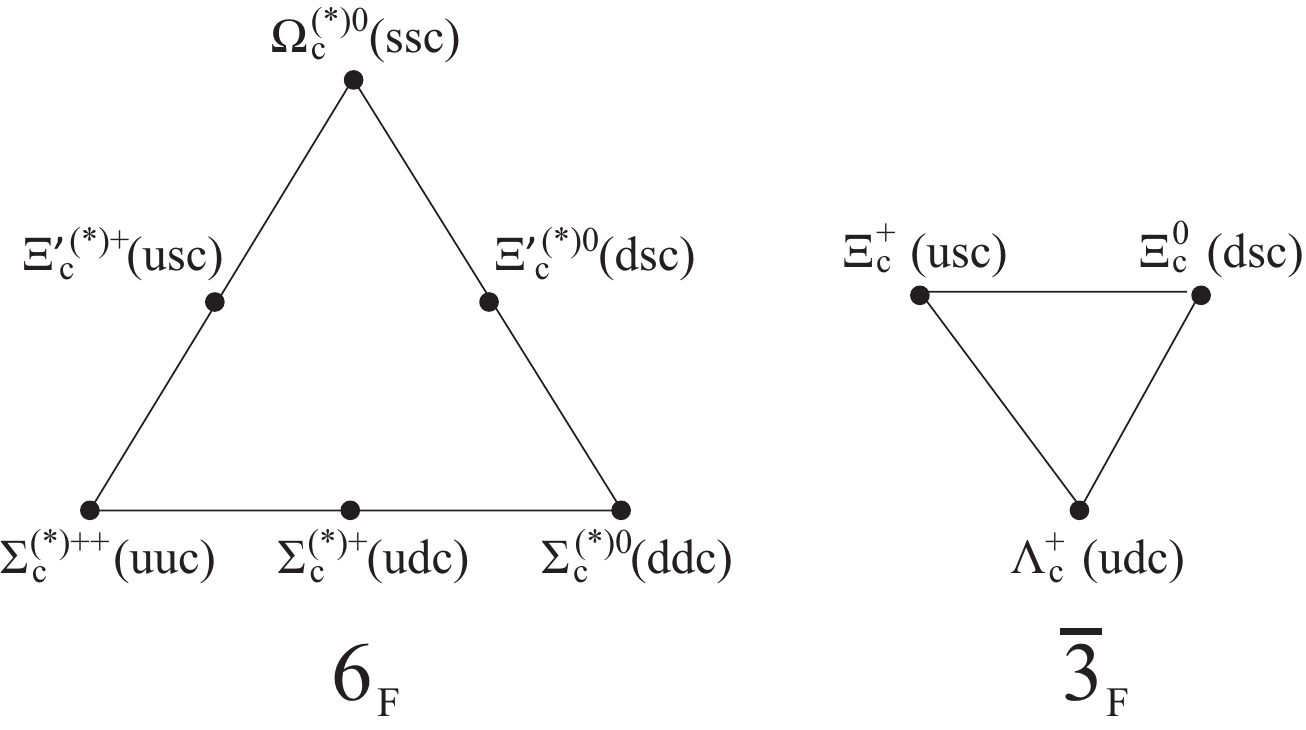}
\end{center}
\caption{Flavor $SU(3)$ flavor multiplets $\mathbf{6}_F$ and $\mathbf{\bar 3}_F$ of the ground-state charmed baryons. One uses the symbols $\Xi_c$ and $\Xi_c^\prime$ to denote the charmed-strange baryons ($usc$ and $dsc$) belonging to the flavor representations $\mathbf{\bar 3}_F$ and $\mathbf{6}_F$ respectively, but the superscript~$^\prime$ is usually omitted for experimental excited states since they can not be directly differentiated in experiments.}
\label{fig:heavybaryon}
\end{figure}

\begin{figure}[hbtp]
\begin{center}
\includegraphics[width=1.0\textwidth]{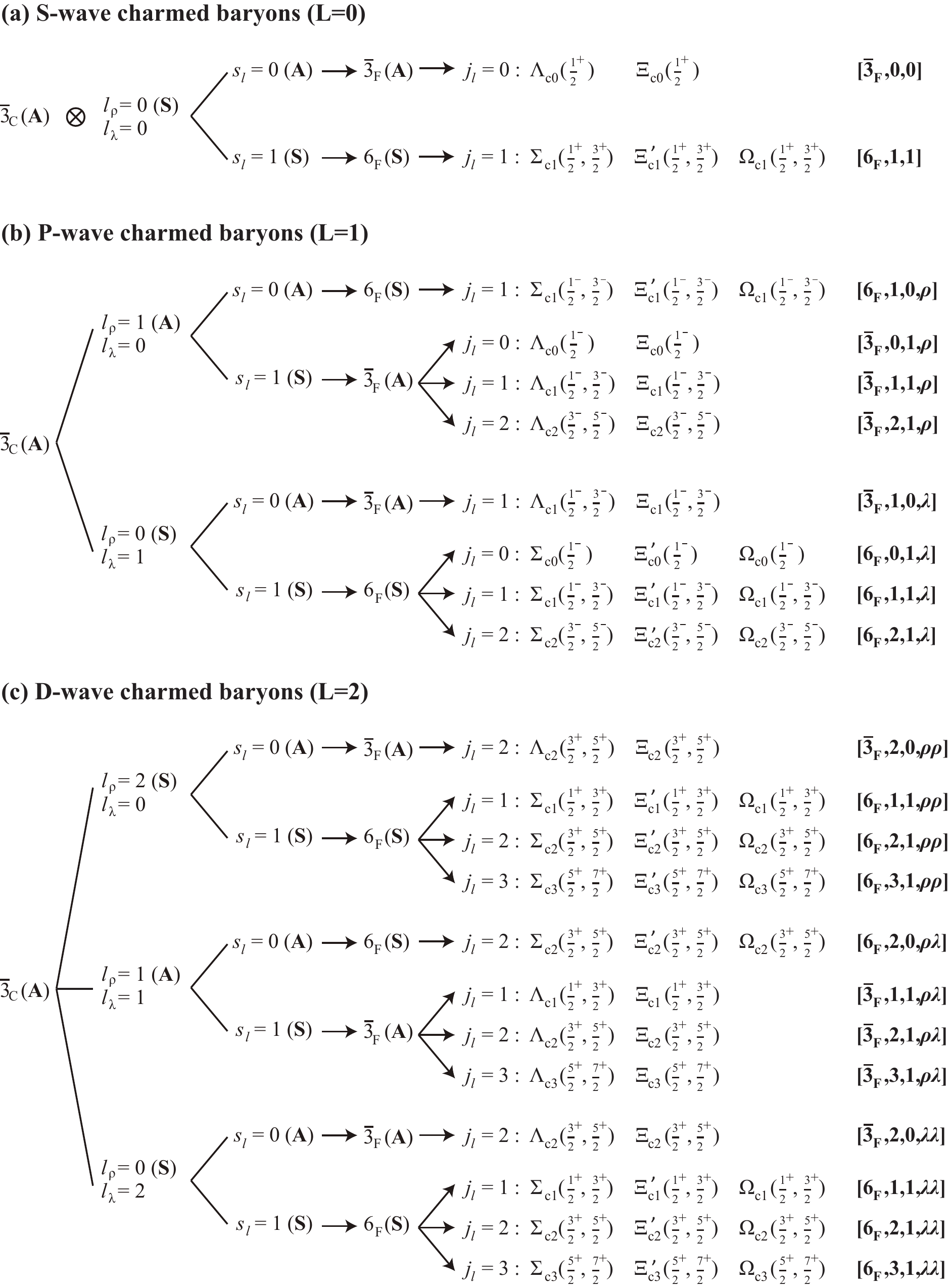}
\end{center}
\caption{Categorization of $S$-, $P$-, and $D$-wave charmed baryons.}
\label{fig:categorization}
\end{figure}

To understand the heavy baryon system well, one needs to carefully examine the four internal structures/symmetries between the two light quarks:
\begin{itemize}

\item The color structure of the two light quarks is always antisymmetric ($\mathbf{\bar 3}_C$).

\item The flavor structure of the two light quarks is either symmetric ($\mathbf{6}_F$) or antisymmetric ($\mathbf{\bar 3}_F$), {\it e.g.}, see Fig.~\ref{fig:heavybaryon} for the ground-state charmed baryons.

\item The spin structure of the two light quarks is either symmetric ($s_l = 1$) or antisymmetric ($s_l = 0$).

\item The orbital structure of the two light quarks is either symmetric ($l_\rho = 0/2/\cdots$) or antisymmetric ($l_\rho = 1/3/\cdots$).

\end{itemize}
According to the Pauli principle, the total wave function of the two light quarks is antisymmetric, and we can categorize the heavy baryons into several multiplets. There is one or two heavy baryons in each multiplet, with the total angular momenta $J = j_l \otimes s_Q = |j_l \pm 1/2|$. For examples, we can categorize the $S$-wave charmed baryons into two multiplets, the $P$-wave charmed baryons into eight multiplets, and the $D$-wave charmed baryons into twelve multiplets. We denote these multiplets as $[Flavor, j_l, s_l(, \rho/\lambda/\rho\rho/\rho\lambda/\lambda\lambda)]$, and show them systematically in Fig.~\ref{fig:categorization}.

Especially, the $S$-wave charmed baryons form one flavor $\mathbf{\bar 3}_F$ representation of $J^P = 1/2^+$, one flavor $\mathbf{6}_F$ representation of $J^P = 1/2^+$, and one flavor $\mathbf{6}_F$ representation of $J^P = 3/2^+$. The flavor $\mathbf{\bar 3}_F$ representation of $J^P = 1/2^+$ constitutes a charmed baryon multiplet itself, and the two flavor $\mathbf{6}_F$ representations of $J^P = 1/2^+$ and $3/2^+$ together constitute another charmed baryon multiplet. Experimentally, all these $S$-wave charmed and bottom baryons, except the $\Omega_b^*$ of $J^P = 3/2^+$, have been well established~\cite{pdg}, and they indeed form such two multiplets. This is a great success of the conventional quark model in the classification of heavy hadrons.

\section{Excited open heavy flavor mesons}
\label{sec2}

In this section we shall provide an updated review on the experimental and theoretical progresses on the open heavy flavor mesons since 2017. The heavy quark symmetry plays an important role, and we have applied it to categorize the open heavy flavor mesons in Sec.~\ref{sec1.3}, based on which we shall review them in this section. Besides, we need the following notations for the charmed mesons: the words ``natural parity states'', labeled as $D_J^*$, denote the charmed mesons of $P=(-1)^J$, such as $J^P = 0^+/1^-/2^+/\cdots$; the words ``unnatural parity states'', labeled as $D_J$, denote the charmed mesons of $P=(-1)^{J+1}$, such as $J^P = 0^-/1^+/2^-/\cdots$. The former $D_J^*$ can decay to $D\pi$, but the latter $D_J$ can not. Similar notations are used for the charmed-strange, bottom, and bottom-strange mesons, which will be separately reviewed in the following subsections.

\subsection{Charmed mesons}
\label{sec2.1}

The $1S$ charmed mesons $D$ and $D^*$ as well as the $1P$ ones $D_0^*(2300)$, $D_1(2420)$, $D_1(2430)$, and $D_2^*(2460)$ are well established in experiments~\cite{pdg}. They complete one $S$-wave doublet $(0^-, 1^-)$ as well as two $P$-wave doublets $(0^+, 1^+)$ and $(1^+, 2^+)$. There still exist some problems to fully understand the $D_0^*(2300)$ and $D_1(2420)$ possibly due to the chiral symmetry breaking effect, and we refer to the recent theoretical studies~\cite{Moir:2016srx,Du:2017zvv,Guo:2018tjx,Du:2019oki,Du:2020pui,Meissner:2020khl,Gayer:2021xzv} for detailed discussions.

\begin{figure}[hbtp]
\begin{center}
\subfigure[]{\includegraphics[width=0.3\textwidth]{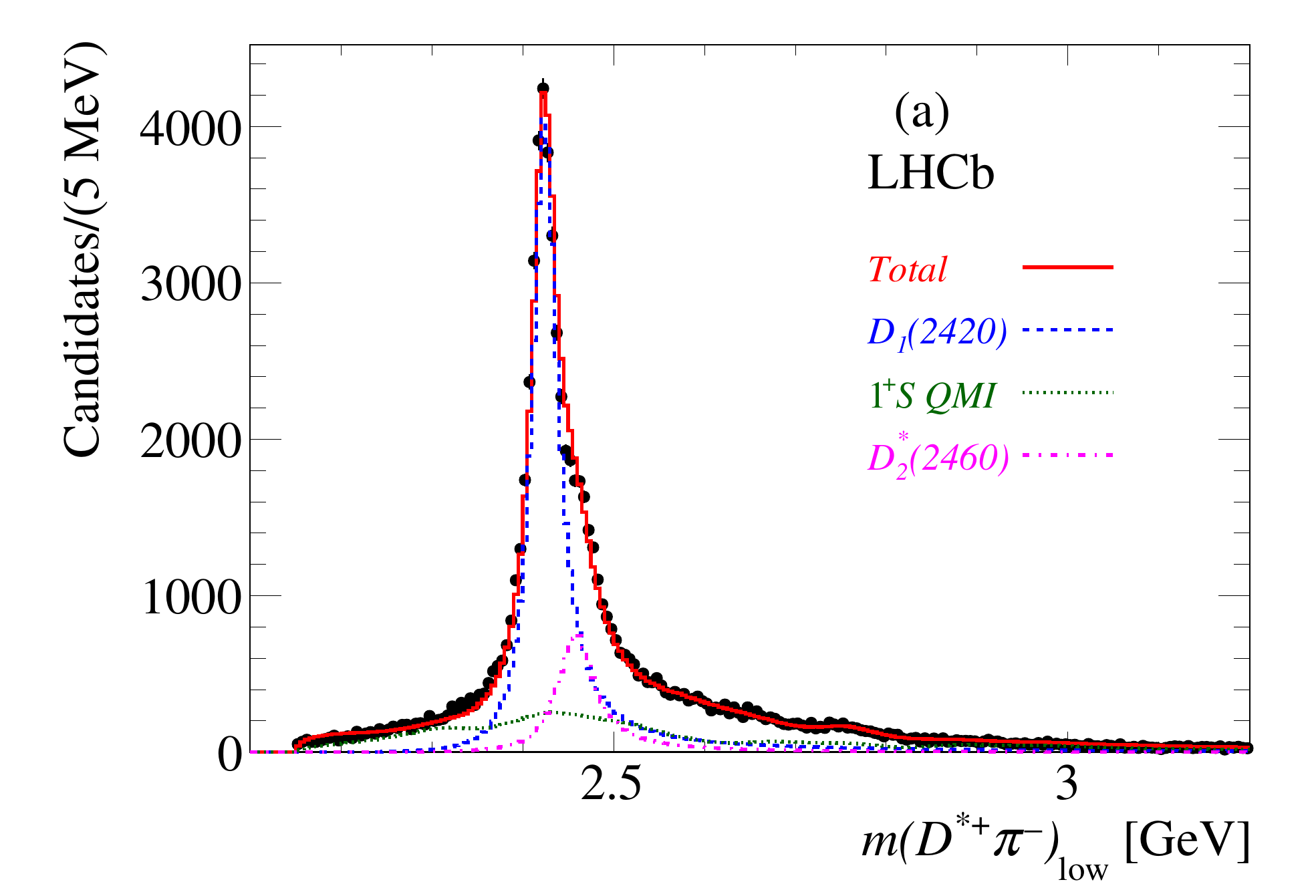}}
~~
\subfigure[]{\includegraphics[width=0.3\textwidth]{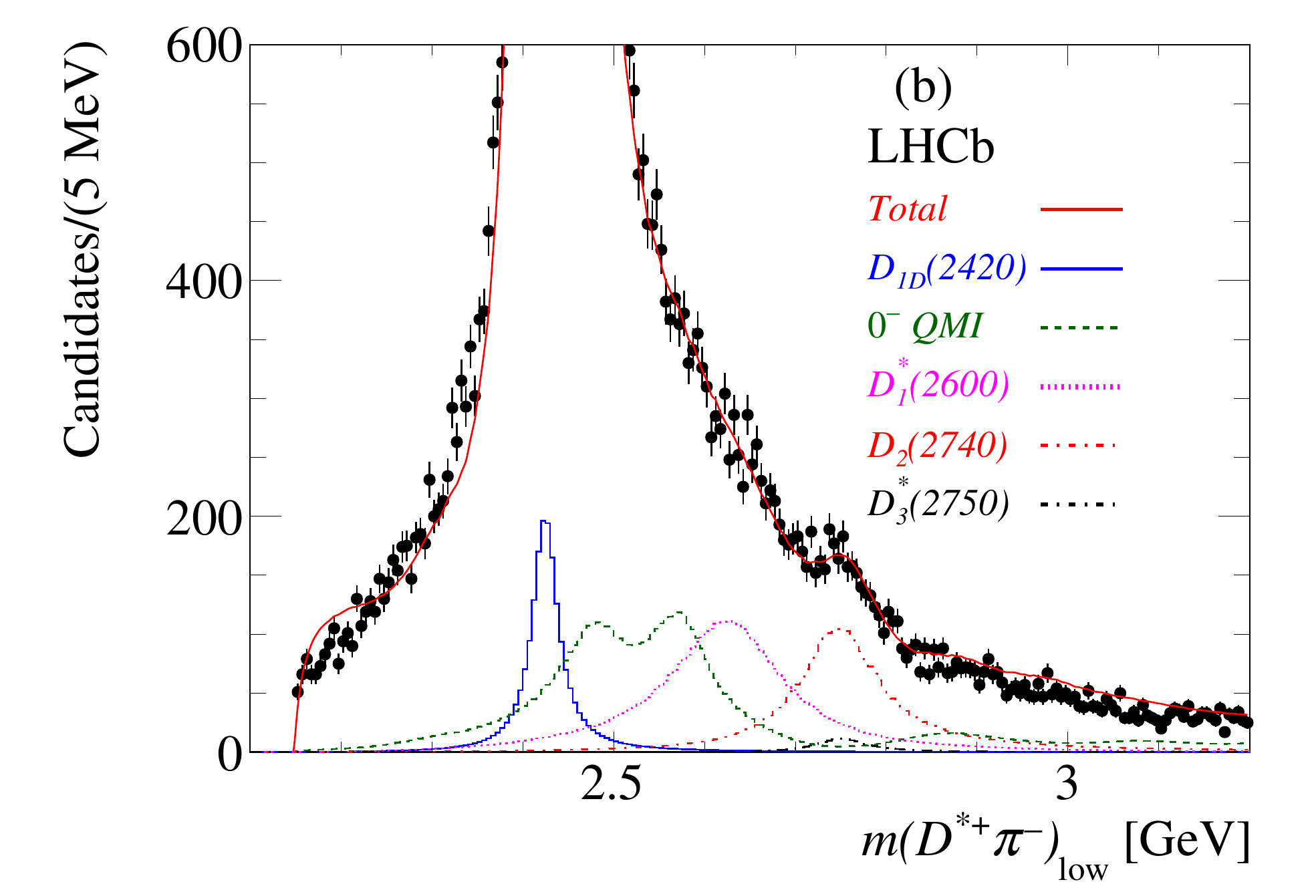}}
~~
\subfigure[]{\includegraphics[width=0.3\textwidth]{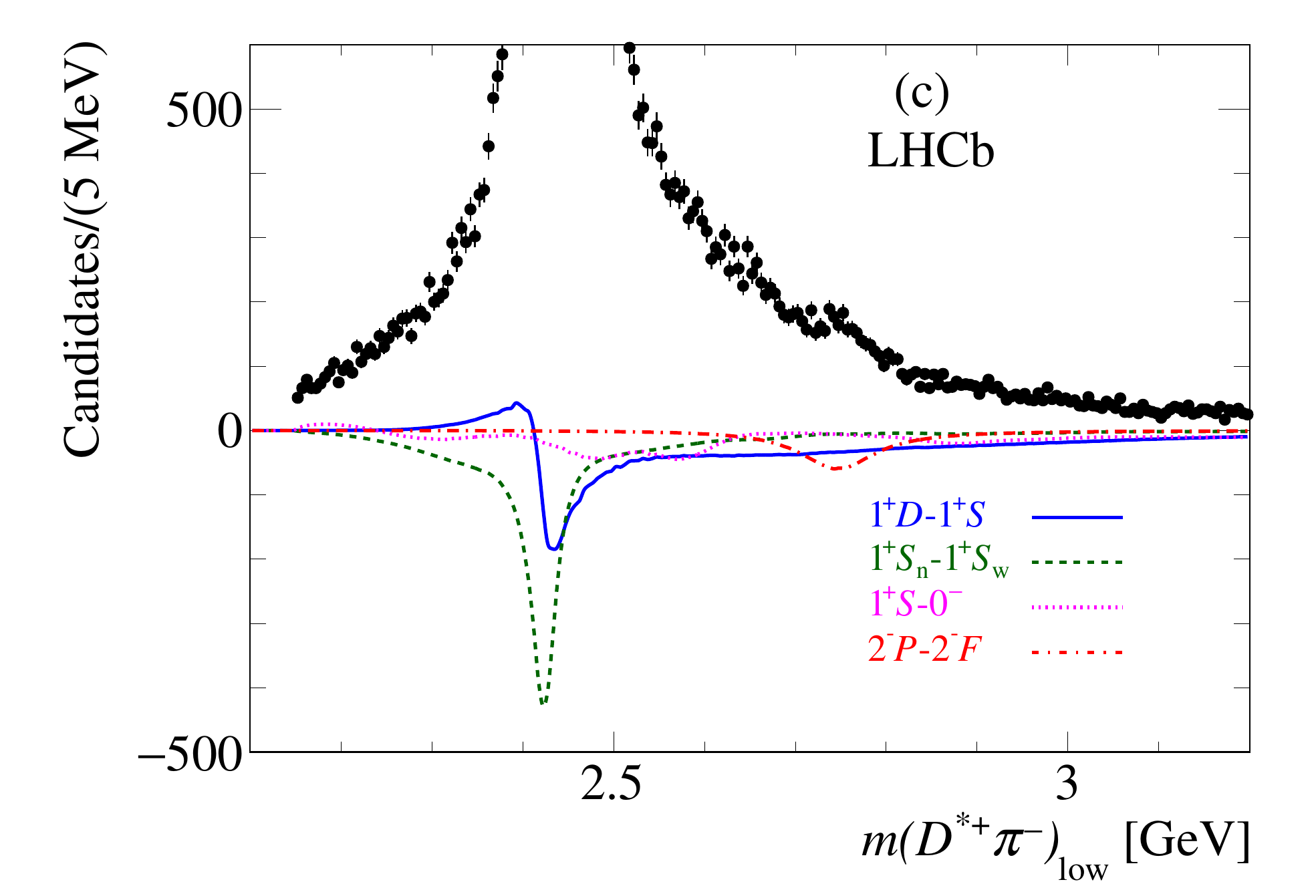}}
\end{center}
\caption{The $D^{*+}\pi^-$ invariant mass distribution of the $B^- \rightarrow D^{*+}\pi^-\pi^-$ decay, projected to the total dataset from the quasi-model-independent fitting model with (a,b) all amplitude contributions and (c) the significant interference terms. Source: Ref.~\cite{LHCb:2019juy}.}
\label{fig:D}
\end{figure}

Higher charmed mesons are still not understood well, including the $D_0(2550)^0$~\cite{BaBar:2010zpy}, $D_1^*(2600)^0$~\cite{BaBar:2010zpy}, $D_2(2740)^0$~\cite{BaBar:2010zpy}, $D^*_1(2760)^0$~\cite{BaBar:2010zpy,LHCb:2015eqv}, $D^*_3(2760)^{-,0}$~\cite{BaBar:2010zpy,LHCb:2015klp,LHCb:2016lxy}, $D_J(3000)^0$~\cite{LHCb:2013jjb}, $D_J^*(3000)^0$~\cite{LHCb:2013jjb}, and $D_2^*(3000)^0$~\cite{LHCb:2016lxy}, etc. All of them have been reviewed in our previous paper~\cite{Chen:2016spr}, so we do not review them here but just mention the comprehensive LHCb experiment performed in 2019~\cite{LHCb:2019juy}. As depicted in Fig.~\ref{fig:D}, LHCb performed a four-body amplitude analysis of the $B^- \rightarrow D^{*+}\pi^-\pi^-$ decay, and determined resonance parameters and quantum numbers of the $D_0(2550)^0$, $D_1^*(2600)^0$, $D_2(2740)^0$, and $D^*_3(2760)^0$ to be:
\begin{eqnarray}
D_0(2550)^0{\rm~of~}J^P = 0^-      &:& M = 2518 \pm 2 \pm 7{\rm~MeV} \, ,
\\ \nonumber                        && \Gamma = 199 \pm 5 \pm 17{\rm~MeV} \, ;
\\ D_1^*(2600)^0{\rm~of~}J^P = 1^- &:& M = 2641.9 \pm 1.8 \pm 4.5{\rm~MeV} \, ,
\\ \nonumber                        && \Gamma = 149 \pm 4 \pm 20{\rm~MeV} \, ;
\\ D_2(2740)^0{\rm~of~}J^P = 2^-   &:& M = 2751 \pm 3 \pm 7{\rm~MeV} \, ,
\\ \nonumber                        && \Gamma = 102 \pm 6 \pm 26{\rm~MeV} \, ;
\\ D^*_3(2760)^0{\rm~of~}J^P = 3^- &:& M = 2753 \pm 4 \pm 6{\rm~MeV} \, ,
\\ \nonumber                        && \Gamma = 66 \pm 10 \pm 14{\rm~MeV} \, .
\end{eqnarray}

In Table~\ref{sec2:charm}, we summarize the theoretical results of the charmed mesons obtained using the original Godfrey-Isgur (GI) relativized quark model~\cite{Godfrey:1985xj} updated by Godfrey and Moats (GI-Original)~\cite{Godfrey:2015dva}, the QCD motivated relativistic quark model based on the quasipotential approach (R.~Q.~M.)~\cite{Ebert:2009ua}, and the modified GI model taking into account the screening effect (GI-Screen)~\cite{Song:2015fha}. Possible experimental candidates are given for comparison. Especially, the three charmed mesons $D^*_1(2760)^0$, $D_2(2740)^0$, and $D^*_3(2760)^0$ probably take part in forming the two $D$-wave doublets $(1^-, 2^-)$ and $(2^-, 3^-)$. One may wonder whether the $D_2(2740)^0$ can be further separated into two substructures, which needs to be carefully examined in future experiments.

\begin{table*}[htbp]
\renewcommand{\arraystretch}{1.5}
\scriptsize
\caption{Theoretical results of the charmed mesons obtained using the original GI model updated by Godfrey and Moats (GI-Original)~\cite{Godfrey:2015dva}, the QCD motivated relativistic quark model (R.~Q.~M.)~\cite{Ebert:2009ua}, and the modified GI model taking into account the screening effect (GI-Screen)~\cite{Song:2015fha}. Possible experimental candidates are given for comparisons, with uncertainties the sum of their statistic and systematical uncertainties. The notation $L_J$ denotes the mixing state of $^1L_J$ and $^3L_J$. Masses are in units of MeV.}
\centering
\begin{tabular}{ c c c c c c c }
\toprule[1pt]
& $n \ ^{2S+1}L_J$ & Exp. Mass~\cite{pdg}  & Exp. Width~\cite{pdg} & GI-Original~\cite{Godfrey:2015dva} & R.~Q.~M.~\cite{Ebert:2009ua} & GI-Screen~\cite{Song:2015fha}
\\ \midrule[1pt]
$D^0$                   & $1 \ ^1S_0$      & $1864.83 \pm 0.05$                                & $\sim10^{-13}$~s                   & 1877  & 1871  & 1861 \\
$D^{\ast0}$             & $1 \ ^3S_1$      & $2006.85 \pm 0.05$                                & $<2.1$                             & 2041  & 2010  & 2020 \\
\hline
$D_{0}^\ast(2300)^0$    & $1 \ ^3P_0$      & $2300 \pm 19$                                     & $274 \pm 40$                       & 2399  & 2406  & 2365 \\
$D_{1}(2420)^0$         & $1 \ P_1$        & $2420.8 \pm 0.5$                                  & $31.7 \pm 2.5$                     & 2456  & 2426  & 2424 \\
$D_{1}(2430)^0$         & $1 \ P_1^\prime$ & $2427\pm36$                                       & $384 ^{+130}_{-105}$               & 2467  & 2469  & 2434 \\
$D_{2}^\ast(2460)^0$    & $1 \ ^3P_2$      & $2460.7 \pm 0.4$                                  & $47.5 \pm 1.1$                     & 2502  & 2460  & 2468 \\
\hline
$D_1^*(2760)^0$         & $1 \ ^3D_1$      & $2781 \pm 22$~\cite{LHCb:2015eqv}                 & $177 \pm 38$~\cite{LHCb:2015eqv}   & 2817  & 2788  & 2762 \\
\hdashline[2pt/2pt]
\multirow{2}{*}{$D_2(2740)^0$} & $1 \ D_2$ & \multirow{2}{*}{$2751 \pm 8$~\cite{LHCb:2019juy}}
                                           & \multirow{2}{*}{$102 \pm 27$~\cite{LHCb:2019juy}}                                      & 2816  & 2806  & --   \\
                        & $1 \ D_2^\prime$ &                                                   &                                    & 2845  & 2850  & 2789 \\
\hdashline[2pt/2pt]
$D_3^*(2760)^0$         & $1 \ ^3D_3$      & $2753 \pm 7$~\cite{LHCb:2019juy}                  & $66 \pm 17$~\cite{LHCb:2019juy}    & 2833  & 2863  & 2779 \\
\hline
--                      & $1 \ ^3F_2$      & --                                                & --                                 & 3132  & 3090  & 3053 \\
--                      & $1 \ F_3$        & --                                                & --                                 & 3108  & 3129  & --   \\
--                      & $1 \ F_3^\prime$ & --                                                & --                                 & 3143  & 3145  & --   \\
--                      & $1 \ ^3F_4$      & --                                                & --                                 & 3113  & 3187  & 3037 \\
\hline
$D_0(2550)^0$           & $2 \ ^1S_0$      & $2518 \pm 7$~\cite{LHCb:2019juy}                  & $199\pm18$~\cite{LHCb:2019juy}     & 2581  & 2581  & 2534 \\
$D_1^*(2600)^0$         & $2 \ ^3S_1$      & $2641.9 \pm 4.8$~\cite{LHCb:2019juy}              & $149\pm20$~\cite{LHCb:2019juy}     & 2643  & 2632  & 2593 \\
\hline
--                      & $2 \ ^3P_0$      & --                                                & --                                 & 2931  & 2919  & 2856 \\
--                      & $2 \ P_1$        & --                                                & --                                 & 2924  & 2932  & --   \\
--                      & $2 \ P_1^\prime$ & --                                                & --                                 & 2961  & 3021  & --   \\
--                      & $2 \ ^3P_2$      & --                                                & --                                 & 2957  & 3012  & 2884 \\
\hline
--                      & $2 \ ^3D_1$      & --                                                & --                                 & 3231  & 3228  & 3131 \\
--                      & $2 \ D_2$        & --                                                & --                                 & 3212  & 3259  & --   \\
--                      & $2 \ D_2^\prime$ & --                                                & --                                 & 3248  & 3307  & --   \\
--                      & $2 \ ^3D_3$      & --                                                & --                                 & 3226  & 3335  & 3129 \\
\hline
$D_J(3000)^0$           & $3 \ ^1S_0$      & $2971.8\pm8.7$~\cite{LHCb:2013jjb}                & $188.1\pm44.8$~\cite{LHCb:2013jjb} & 3068  & 3062  & 2976 \\
$D_J^*(3000)^0$         & $3 \ ^3S_1$      & $3008.1\pm4.0$~\cite{LHCb:2013jjb}                & $110.5\pm11.5$~\cite{LHCb:2013jjb} & 3110  & 3096  & 3015 \\
\hline
--                      & $3 \ ^3P_0$      & --                                                & --                                 & 3343  & 3346  & --   \\
--                      & $3 \ P_1$        & --                                                & --                                 & 3328  & 3365  & --   \\
--                      & $3 \ P_1^\prime$ & --                                                & --                                 & 3360  & 3461  & --   \\
$D^*_2(3000)$           & $3 \ ^3P_2$      & $ 3214 \pm 57$~\cite{LHCb:2016lxy}                & $186\pm81$~\cite{LHCb:2016lxy}     & 3353  & 3407  & --
\\ \bottomrule[1pt]
\end {tabular}
\label{sec2:charm}
\end{table*}

\subsection{Charmed-strange mesons and the $D_{s0}(2590)$}
\label{sec2.2}

The $1S$ charmed-strange mesons $D_s$ and $D_s^*$ as well as the $1P$ ones $D_{s0}^*(2317)$, $D_{s1}(2460)$, $D_{s1}(2536)$, and $D_{s2}^*(2573)$ are well established~\cite{pdg}. They complete one $S$-wave doublet $(0^-, 1^-)$ as well as two $P$-wave doublets $(0^+, 1^+)$ and $(1^+, 2^+)$. There still exist some problems to fully understand the $D_{s0}^*(2317)$ and $D_{s1}(2460)$, which are probably related to their nearby $DK/D^*K$ thresholds. These problems have attracted intensive discussions on the nature of the near-threshold bound and virtual states, {\it e.g.}, see Refs.~\cite{Matuschek:2020gqe,Yang:2021tvc,Albaladejo:2022sux} and references therein. We also refer interested readers to Sec.~\ref{sec5.1.2} of the present review, where we shall review some of the discussions on the nature of the exotic fully-charm tetraquark structures observed by LHCb in the di-$J/\psi$ invariant mass spectrum~\cite{LHCb:2020bwg}.

Higher charmed-strange mesons are still not well understood, including the $D_{s0}(2590)^+$~\cite{LHCb:2020gnv}, $D_{sJ}^\ast(2632)$~\cite{SELEX:2004drx}, $D_{s1}^\ast(2700)$~\cite{BaBar:2006gme}, $D_{s1}^\ast(2860)$~\cite{BaBar:2006gme,LHCb:2014ott,LHCb:2014ioa}, $D_{s3}^\ast(2860)$~\cite{BaBar:2006gme,LHCb:2014ott,LHCb:2014ioa}, and $D_{sJ}(3040)$~\cite{BaBar:2009rro}, etc. Most of them have been reviewed in our previous paper~\cite{Chen:2016spr}, except the $D_{s0}(2590)^+$ to be reviewed in this subsection. We summarize the theoretical results of the charmed-strange mesons~\cite{Godfrey:2015dva,Ebert:2009ua,Song:2015nia} in Table~\ref{sec2:cs}, with possible experimental candidates given for comparison. Especially, the two $D_{s1}^\ast(2860)$ and $D_{s3}^\ast(2860)$ mesons probably take part in forming the two $D$-wave doublets $(1^-, 2^-)$ and $(2^-, 3^-)$, so there may be two $D_{s2}$ mesons around 2860 MeV still waiting to be observed.

\begin{table*}[htbp]
\renewcommand{\arraystretch}{1.5}
\scriptsize
\caption{Theoretical results of the charmed-strange mesons obtained using the original GI model updated by Godfrey and Moats (GI-Original)~\cite{Godfrey:2015dva}, the QCD motivated relativistic quark model (R.~Q.~M.)~\cite{Ebert:2009ua}, and the modified GI model taking into account the screening effect (GI-Screen)~\cite{Song:2015nia}. Notations are the same as in Table~\ref{sec2:charm}.}
\centering
\begin{tabular}{ c c c c c c c }
\toprule[1pt]
& $n \ ^{2S+1}L_J$ & Exp. Mass~\cite{pdg} & Exp. Width~\cite{pdg} & GI-Original~\cite{Godfrey:2015dva} & R.~Q.~M.~\cite{Ebert:2009ua} & GI-Screen~\cite{Song:2015nia}
\\ \midrule[1pt]
$D_s$               & $1 \ ^1S_0$      & $1968.34 \pm 0.07$                                         & $\sim10^{-12}$~s                   & 1979  & 1969  & 1967 \\
$D_s^{\ast}$        & $1 \ ^3S_1$      & $2112.2 \pm 0.4$                                           & $<1.9$                             & 2129  & 2111  & 2115 \\
\hline
$D_{s0}^\ast(2317)$ & $1 \ ^3P_0$      & $2317.8 \pm 0.5$                                           & $<3.8$                             & 2484  & 2509  & 2463 \\
$D_{s1}(2460)$      & $1 \ P_1$        & $2459.5 \pm 0.6$                                           & $<3.5$                             & 2549  & 2536  & 2529 \\
$D_{s1}(2536)$      & $1 \ P_1^\prime$ & $2535.11 \pm 0.06$                                         & $0.92 \pm 0.05$                    & 2556  & 2574  & 2534 \\
$D_{s2}^\ast(2573)$ & $1 \ ^3P_2$      & $2569.1 \pm 0.8$                                           & $16.9 \pm 0.7$                     & 2592  & 2571  & 2571 \\
\hline
$D_{s1}^\ast(2860)$ & $1 \ ^3D_1$      & $2859\pm27$                                                & $159 \pm 80$                       & 2899  & 2913  & 2865 \\
--                  & $1 \ D_2$        & --                                                         & --                                 & 2900  & 2931  & --   \\
--                  & $1 \ D_2^\prime$ & --                                                         & --                                 & 2926  & 2961  & --   \\
$D_{s3}^\ast(2860)$ & $1 \ ^3D_3$      & $2860.5\pm7.0$                                             & $53 \pm 10$                        & 2917  & 2971  & 2883 \\
\hline
--                  & $1 \ ^3F_2$      & --                                                         & --                                 & 3208  & 3230  & 3159 \\
--                  & $1 \ F_3$        & --                                                         & --                                 & 3186  & 3254  & --   \\
--                  & $1 \ F_3^\prime$ & --                                                         & --                                 & 3218  & 3266  & --   \\
--                  & $1 \ ^3F_4$      & --                                                         & --                                 & 3190  & 3300  & 3143 \\
\hline
$D_{s0}(2590)$      & $2 \ ^1S_0$      & $2591 \pm 9$~\cite{LHCb:2020gnv}                           & $89 \pm 20$~\cite{LHCb:2020gnv}    & 2673  & 2688  & 2646 \\
$D_{s1}^\ast(2700)$ & $2 \ ^3S_1$      & $2708.3^{+4.0}_{-3.4}$                                     & $120 \pm 11$                       & 2732  & 2731  & 2704 \\
\hline
--                  & $2 \ ^3P_0$      & --                                                         & --                                 & 3005  & 3054  & 2960 \\
\hdashline[2pt/2pt]
\multirow{2}{*}{$D_{sJ}(3040)$}        & $2 \ P_1$        & \multirow{2}{*}{$3044^{+31}_{-~9}$}
                                                          & \multirow{2}{*}{$239^{+58}_{-55}$}                                           & 3018  & 3067  & --   \\
                    & $2 \ P_1^\prime$ &                                                            &                                    & 3038  & 3154  & 2992 \\
\hdashline[2pt/2pt]
--                  & $2 \ ^3P_2$      & --                                                         & --                                 & 3048  & 3142  & 3004 \\
\hline
--                  & $2 \ ^3D_1$      & --                                                         & --                                 & 3306  & 3383  & 3244 \\
--                  & $2 \ D_2$        & --                                                         & --                                 & 3298  & 3403  & --   \\
--                  & $2 \ D_2^\prime$ & --                                                         & --                                 & 3323  & 3456  & --   \\
--                  & $2 \ ^3D_3$      & --                                                         & --                                 & 3311  & 3469  & 3251 \\
\hline
--                  & $3 \ ^1S_0$      & --                                                         & --                                 & 3154  & 3219  & --   \\
--                  & $3 \ ^3S_1$      & --                                                         & --                                 & 3193  & 3242  & --   \\
\hline
--                  & $3 \ ^3P_0$      & --                                                         & --                                 & 3412  & 3513  & --   \\
--                  & $3 \ P_1$        & --                                                         & --                                 & 3416  & 3519  & --   \\
--                  & $3 \ P_1^\prime$ & --                                                         & --                                 & 3433  & 3618  & --   \\
--                  & $3 \ ^3P_2$      & --                                                         & --                                 & 3439  & 3580  & --
\\ \bottomrule[1pt]
\end {tabular}
\label{sec2:cs}
\end{table*}

\begin{figure}[hbtp]
\begin{center}
\subfigure[]{\includegraphics[width=0.4\textwidth]{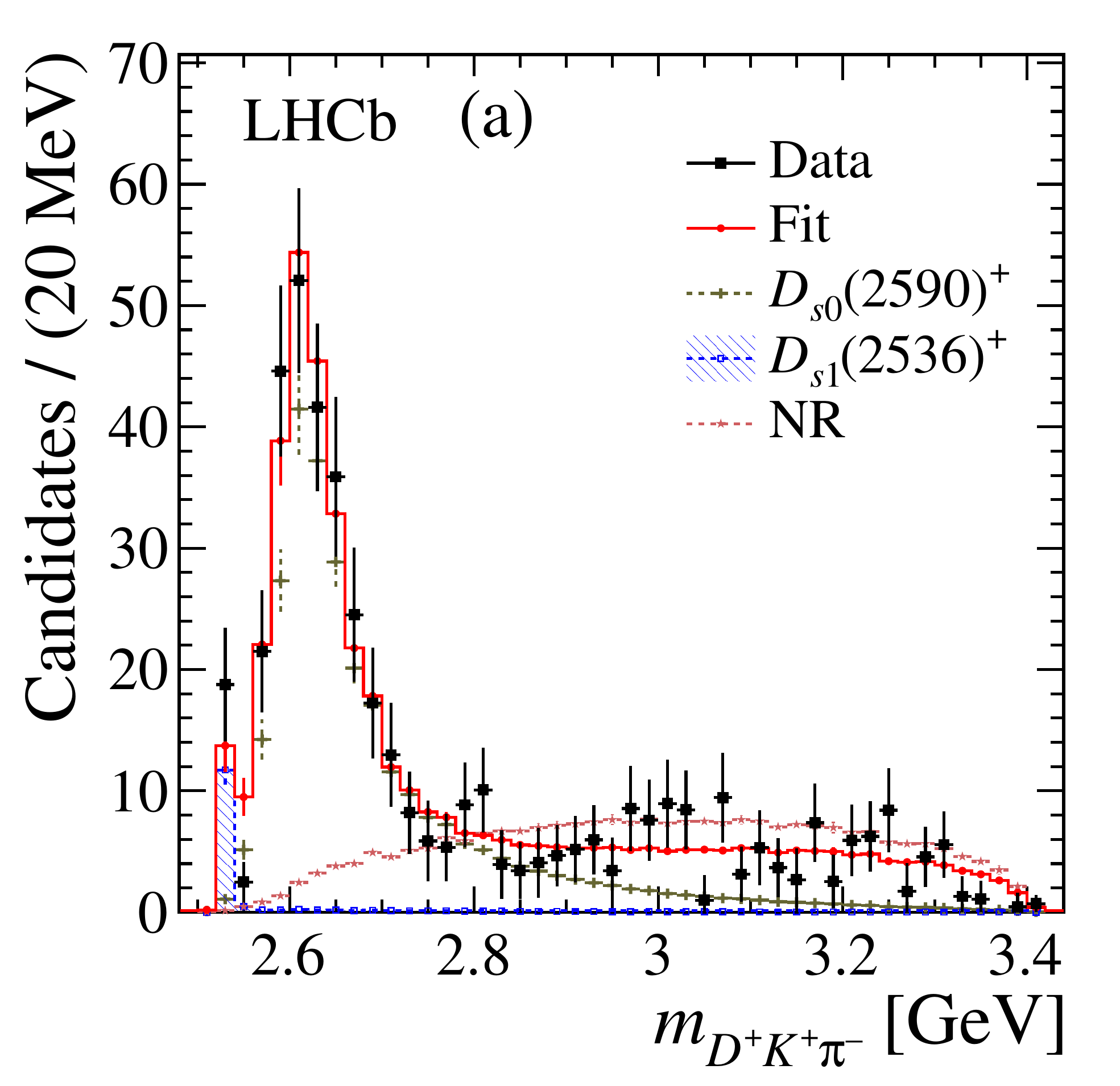}}
~~~~~~~~~~
\subfigure[]{\includegraphics[width=0.4\textwidth]{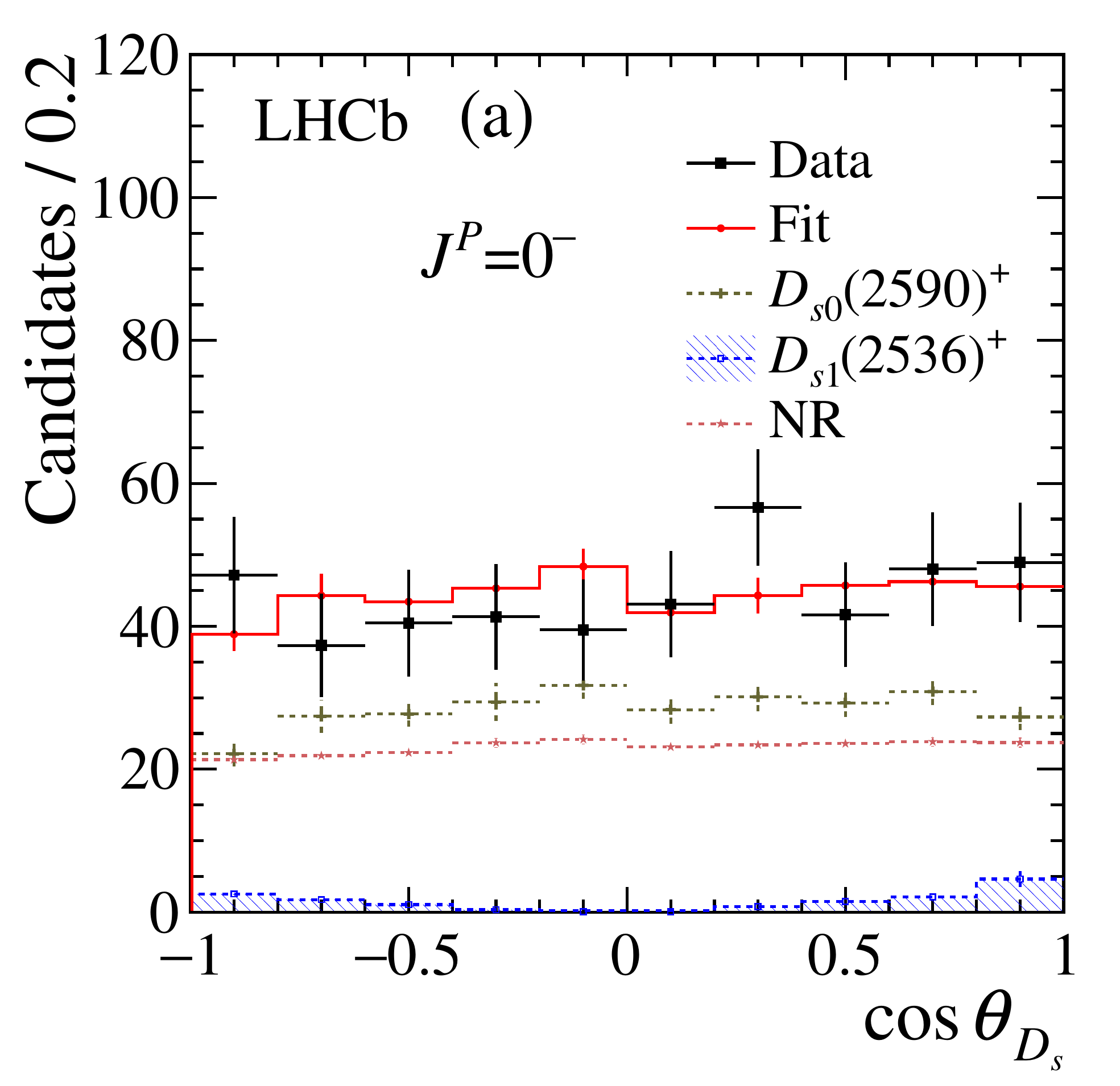}}
\end{center}
\caption{(a) The $D^+K^+\pi^-$ invariant mass distribution of the $B^0 \to D^-D^+K^+\pi^-$ decay, where the $D_{s0}(2590)^+$ was observed with large statistical significance. (b) Comparison of the $\cos\theta_{Ds}$ distribution for the spin-parity quantum number of $D_{s0}(2590)^+$ assumed to be $J^P = 0^-$. Source: Ref.~\cite{LHCb:2020gnv}.}
\label{fig:Ds2590}
\end{figure}

Recently, the LHCb collaboration studied the $B^0 \to D^-D^+K^+\pi^-$ decay, and observed a new excited charmed-strange meson decaying into the $D^+K^+\pi^-$ final state, which was named as $D_{s0}(2590)^+$~\cite{LHCb:2020gnv}. Its resonance parameters were measured to be
\begin{eqnarray}
D_{s0}(2590)^+ &:& M = 2591 \pm 6 \pm 7{\rm~MeV} \, ,
\\ \nonumber    && \Gamma = 89 \pm 16 \pm 12{\rm~MeV} \, .
\end{eqnarray}
As shown in Fig.~\ref{fig:Ds2590}(a), it was observed with quite large statistical significance. As depicted in Fig.~\ref{fig:Ds2590}(b), LHCb tested three possible spin-parity models ($J^P = 0^-/1^+/2^-$), and found that the $J^P = 0^-$ one leads to the best fit quality. The $D_{s0}(2590)^+$ was suggested by LHCb to be the $D_s(2^1S_0)$ meson, {\it i.e.}, the radial excitation of the ground-state $D_s$ meson. As shown in Table~\ref{sec2:cs}, the mass of $D_{s0}(2590)^+$ is lower than the quark model predictions~\cite{Godfrey:2015dva,Ebert:2009ua,Song:2015nia} by about 80 MeV.

In Ref.~\cite{Wang:2021orp} the authors assumed the $D_{s0}(2590)^+$ to be the $D_s(2^1S_0)$ meson, and calculated its width to be about $23.8$~MeV. This value is much smaller than its experimental width $\Gamma_{D_{s0}(2590)^+} = 89 \pm 16 \pm 12$~MeV. They further introduced a model independent parameter and a mass independent ratio, whose theoretical predictions were also much smaller than the experimental data. Accordingly, they concluded that the experimental mass and width of the $D_{s0}(2590)^+$ as the candidate of the $D_s(2^1S_0)$ do not match to each other.

In Ref.~\cite{Ni:2021pce} the authors studied the $D_{s0}(2590)^+$ with a semi-relativistic potential model. They obtained its mass to be about 60~MeV larger than the experimental data and its width to be about 19~MeV, five times smaller than the experimental data. Accordingly, they also concluded that more observations are needed in future experiments to establish the $D_s(2^1S_0)$ meson and uncover the nature of the $D_{s0}(2590)^+$.

In order to understand whether the $D_{s0}(2590)^+$ can be interpreted as the $D_s(2^1S_0)$ meson or not, the authors of Ref.~\cite{Xie:2021dwe} took into account the $D^*K$ contribution to the bare $D_s(2^1S_0)$ meson. The dressed mass of the $D_s(2^1S_0)$ was lowered by 88~MeV with only 10\% of the $D^*K$ component in its wave function, and therefore, consistent with the experimental mass of the $D_{s0}(2590)^+$. They proposed to carefully examine the two-body $D^*K$ partial decay of the $D_{s0}(2590)^+$ to verify their results.

Later in Ref.~\cite{Ortega:2021fem} the authors studied the $D_{s0}(2590)^+$ as the dressed $D_s(2^1S_0)$ meson in a coupled-channel framework. Their calculations were performed by including the $D^{(*)}K^{(*)}$, $D^{(*)}_s\omega$, and $D^{(*)}_s\eta$ channels. They analyzed the masses, widths, and production line shapes of the resulting state, and their results reinforced the assignment of the $D_{s0}(2590)^+$ as the $D_s(2^1S_0)$ meson with the nearby meson-meson thresholds taken into account.

In a recent study~\cite{Gao:2022bsb}, the authors took into account the screening effects, and studied the mass spectrum and strong decays of the excited charmed-strange mesons using the modified relativized quark model. Their calculated mass and width of the $D_{s0}(2590)^+$ are consistent with the experimental observations, indicating that it can be reasonably interpreted as the $D_s(2^1S_0)$ state.

In a short summary, we want to remind the readers that one should not take the deviation from the quark model predictions too seriously since the typical theoretical uncertainty might reach 100 MeV sometimes. In fact, the mass splitting between the $D_{s1}^\ast(2700)$ and $D_{s0}(2590)^+$ is almost the same as the mass splitting between the $D_{1}^\ast(2641)$ and $D_{0}(2518)^+$~\cite{LHCb:2019juy}, which strongly indicates that the $D_{s0}(2590)^+$ is very probably the $D_s(2^1S_0)$ state.

\subsection{Bottom mesons}
\label{sec2.3}

The $1S$ bottom mesons $B$ and $B^*$ complete one $S$-wave doublet $(0^-, 1^-)$. Higher bottom mesons are still not understood well, including the $B_1(5721)^{0,+}$~\cite{OPAL:1994hqv,D0:2007vzd,CDF:2013www}, $B^*_2(5747)^{0,+}$~\cite{OPAL:1994hqv,D0:2007vzd,CDF:2013www}, $B_J(5840)^{0,+}$~\cite{LHCb:2015aaf}, and $B_J(5970)^{0,+}$~\cite{CDF:2013www,LHCb:2015aaf}, etc. All of them have been reviewed in our previous paper~\cite{Chen:2016spr}, so we do not review them here.

We summarize the theoretical results of the bottom mesons~\cite{Godfrey:2016nwn,Ebert:2009ua,Sun:2014wea} in Table~\ref{sec2:bottom}, with possible experimental candidates given for comparison. Especially, the two bottom mesons $B_1(5721)^{0}$ and $B^*_2(5747)^{0}$ probably take part in forming the two $P$-wave doublets $(0^+, 1^+)$ and $(1^+, 2^+)$. Therefore, there may be one $B^*_0$ meson missing and the $B_1(5721)^{0}$ may be further separated into two substructures.

\begin{table*}[htbp]
\renewcommand{\arraystretch}{1.5}
\scriptsize
\caption{Theoretical results of the bottom mesons obtained using the original GI model updated by Godfrey, Moats, and Swanson (GI-Original)~\cite{Godfrey:2016nwn}, the QCD motivated relativistic quark model (R.~Q.~M.)~\cite{Ebert:2009ua}, and the modified GI model (GI-Modified)~\cite{Sun:2014wea}. Notations are the same as in Table~\ref{sec2:charm}.}
\centering
\begin{tabular}{ c c c c c c c }
\toprule[1pt]
& $n \ ^{2S+1}L_J$ & Exp. Mass~\cite{pdg} & Exp. Width~\cite{pdg} & GI-Original~\cite{Godfrey:2016nwn} & R.~Q.~M.~\cite{Ebert:2009ua} & GI-Modified~\cite{Sun:2014wea}
\\ \midrule[1pt]
$B^0$           & $1 \ ^1S_0$      & $5279.65 \pm 0.12$                                    & $\sim10^{-12}$~s                   & 5312  & 5280 & 5309 \\
$B^{*}$         & $1 \ ^3S_1$      & $5324.70 \pm 0.21$                                    & ?                                  & 5371  & 5326 & 5369 \\
\hline
--              & $1 \ ^3P_0$      & --                                                    & --                                 & 5756  & 5749 & 5756 \\
\hdashline[2pt/2pt]
\multirow{2}{*}{$B_1(5721)^0$}     & $1 \ P_1$ & \multirow{2}{*}{$5726.1\pm1.3$}
                                               & \multirow{2}{*}{$27.5 \pm 3.4$}                                                & 5777  & 5723 & 5779 \\
                & $1 \ P_1^\prime$ &                                                       &                                    & 5784  & 5774 & 5782 \\
\hdashline[2pt/2pt]
$B^*_2(5747)^0$ & $1 \ ^3P_2$      & $5739.5\pm0.7$                                        & $24.2 \pm 1.7$                     & 5797  & 5741 & 5796 \\
\hline
--              & $1 \ ^3D_1$      & --                                                    & --                                 & 6110  & 6119 & 6110 \\
--              & $1 \ D_2$        & --                                                    & --                                 & 6095  & 6103 & 6108 \\
--              & $1 \ D_2^\prime$ & --                                                    & --                                 & 6124  & 6121 & 6113 \\
--              & $1^3D_3$         & --                                                    & --                                 & 6106  & 6091 & 6105 \\
\hline
--              & $1 \ ^3F_2$      & --                                                    & --                                 & 6387  & 6412 & 6387 \\
--              & $1 \ F_3$        & --                                                    & --                                 & 6358  & 6391 & 6375 \\
--              & $1 \ F_3^\prime$ & --                                                    & --                                 & 6396  & 6420 & 6380 \\
--              & $1 \ ^3F_4$      & --                                                    & --                                 & 6364  & 6380 & 6364 \\
\hline
$B_J(5840)^0$   & $2 \ ^1S_0$      & $5863\pm9$                                            & $127 \pm 38$                       & 5904  & 5890 & 5904 \\
$B_J(5970)^0$   & $2 \ ^3S_1$      & $5971\pm5$                                            & $81 \pm 12$                        & 5933  & 5906 & 5934 \\
\hline
--              & $2 \ ^3P_0$      & --                                                    & --                                 & 6213  & 6221 & 6214 \\
--              & $2 \ P_1$        & --                                                    & --                                 & 6197  & 6209 & 6206 \\
--              & $2 \ P_1^\prime$ & --                                                    & --                                 & 6228  & 6281 & 6219 \\
--              & $2 \ ^3P_2$      & --                                                    & --                                 & 6213  & 6260 & 6213 \\
\hline
--              & $2 \ ^3D_1$      & --                                                    & --                                 & 6475  & 6534 & 6475 \\
--              & $2 \ D_2$        & --                                                    & --                                 & 6450  & 6528 & 6464 \\
--              & $2 \ D_2^\prime$ & --                                                    & --                                 & 6486  & 6554 & 6472 \\
--              & $2 \ ^3D_3$      & --                                                    & --                                 & 6460  & 6542 & 6459 \\
\hline
--              & $3 \ ^1S_0$      & --                                                    & --                                 & 6335  & 6379 & 6334 \\
--              & $3 \ ^3S_1$      & --                                                    & --                                 & 6355  & 6387 & 6355 \\
\hline
--              & $3 \ ^3P_0$      & --                                                    & --                                 & 6576  & 6629 & -- \\
--              & $3 \ P_1$        & --                                                    & --                                 & 6557  & 6650 & -- \\
--              & $3 \ P_1^\prime$ & --                                                    & --                                 & 6585  & 6685 & -- \\
--              & $3 \ ^3P_2$      & --                                                    & --                                 & 6570  & 6678 & --
\\ \bottomrule[1pt]
\end {tabular}
\label{sec2:bottom}
\end{table*}

\subsection{Bottom-strange mesons and the $B_{sJ}(6064)/B_{sJ}(6114)$}
\label{sec2.4}

The $1S$ bottom-strange mesons $B_s$ and $B_s^*$ complete the $S$-wave doublet $(0^-, 1^-)$. Higher bottom-strange mesons are still not understood well, including the $B_{s1}(5830)$~\cite{OPAL:1994hqv,CDF:2007avt}, $B^*_{s2}(5840)$~\cite{OPAL:1994hqv,CDF:2007avt,D0:2007die}, $B_{sJ}(6064)$~\cite{LHCb:2020pet}, and $B_{sJ}(6114)$~\cite{LHCb:2020pet}, etc. The former two have been reviewed in our previous paper~\cite{Chen:2016spr}, and the latter two will be reviewed in this subsection.

In Table~\ref{sec2:bs}, we summarize the theoretical results of the bottom-strange mesons~\cite{Godfrey:2016nwn,Ebert:2009ua,Sun:2014wea}, with possible experimental candidates for comparison. Especially, the two $B_{s1}(5830)$ and $B^*_{s2}(5840)$ mesons probably take part in forming the two $P$-wave doublets $(0^+, 1^+)$ and $(1^+, 2^+)$, so there may be one $B^*_{s0}$ meson missing and the $B_{s1}(5830)$ may be further separated into two substructures.

\begin{table*}[htbp]
\renewcommand{\arraystretch}{1.5}
\scriptsize
\caption{Theoretical results of the bottom-strange mesons obtained using the original GI model updated by Godfrey, Moats, and Swanson (GI-Original)~\cite{Godfrey:2016nwn}, the QCD motivated relativistic quark model (R.~Q.~M.)~\cite{Ebert:2009ua}, and the modified GI model (GI-Modified)~\cite{Sun:2014wea}. Notations are the same as in Table~\ref{sec2:charm}.}
\centering
\begin{tabular}{ c c c c c c c }
\toprule[1pt]
& $n \ ^{2S+1}L_J$ & Exp. Mass~\cite{pdg} & Exp. Width~\cite{pdg} & GI-Original~\cite{Godfrey:2016nwn} & R.~Q.~M.~\cite{Ebert:2009ua} & GI-Modified~\cite{Sun:2014wea}
\\ \midrule[1pt]
$B_s$           & $1 \ ^1S_0$      & $5366.88 \pm 0.14$                                    & $\sim10^{-12}$~s                   & 5394  & 5372 & 5390 \\
$B_s^{*}$       & $1 \ ^3S_1$      & $5415.4^{+1.8}_{-1.5}$                                & ?                                  & 5450  & 5414 & 5447 \\
\hline
--              & $1 \ ^3P_0$      & --                                                    & --                                 & 5831  & 5833 & 5830 \\
\hdashline[2pt/2pt]
\multirow{2}{*}{$B_{s1}(5830)$}    & $1 \ P_1$ & \multirow{2}{*}{$5828.70\pm0.20$}
                                               & \multirow{2}{*}{$0.5\pm0.4$}                                                   & 5857  & 5831 & 5858 \\
                & $1 \ P_1^\prime$ &                                                       &                                    & 5861  & 5865 & 5859 \\
\hdashline[2pt/2pt]
$B^*_{s2}(5840)$& $1 \ ^3P_2$      & $5839.86\pm0.12$                                      & $1.49 \pm 0.27$                    & 5876  & 5842 & 5875 \\
\hline
\multirow{4}{*}
{$\begin{array}{c}
B_{sJ}(6064)
\\ \\
B_{sJ}(6114)
\end{array}$}
                & $1 \ ^3D_1$      &
\multirow{4}{*}
{$\begin{array}{c}
6063.5 \pm 1.4~\cite{LHCb:2020pet}
\\ \\
6114 \pm 6~\cite{LHCb:2020pet}
\end{array}$}
&
\multirow{4}{*}
{$\begin{array}{c}
26 \pm 6~\cite{LHCb:2020pet}
\\ \\
66 \pm 28~\cite{LHCb:2020pet}
\end{array}$}
                                                                                           & 6182  & 6209 & 6181 \\
                & $1 \ D_2$        &                                                       &                                    & 6169  & 6189 & 6180 \\
                & $1 \ D_2^\prime$ &                                                       &                                    & 6196  & 6218 & 6185 \\
                & $1^3D_3$         &                                                       &                                    & 6179  & 6191 & 6178 \\
\hline
--              & $1 \ ^3F_2$      & --                                                    & --                                 & 6454  & 6501 & 6453 \\
--              & $1 \ F_3$        & --                                                    & --                                 & 6425  & 6468 & 6441 \\
--              & $1 \ F_3^\prime$ & --                                                    & --                                 & 6462  & 6515 & 6446 \\
--              & $1 \ ^3F_4$      & --                                                    & --                                 & 6432  & 6475 & 6431 \\
\hline
--              & $2 \ ^1S_0$      & --                                                    & --                                 & 5984  & 5976 & 5985 \\
--              & $2 \ ^3S_1$      & --                                                    & --                                 & 6012  & 5992 & 6013 \\
\hline
--              & $2 \ ^3P_0$      & --                                                    & --                                 & 6279  & 6318 & 6279 \\
--              & $2 \ P_1$        & --                                                    & --                                 & 6279  & 6321 & 6284 \\
--              & $2 \ P_1^\prime$ & --                                                    & --                                 & 6296  & 6345 & 6291 \\
--              & $2 \ ^3P_2$      & --                                                    & --                                 & 6295  & 6359 & 6295 \\
\hline
--              & $2 \ ^3D_1$      & --                                                    & --                                 & 6542  & 6629 & 6542 \\
--              & $2 \ D_2$        & --                                                    & --                                 & 6526  & 6625 & 6536 \\
--              & $2 \ D_2^\prime$ & --                                                    & --                                 & 6553  & 6651 & 6542 \\
--              & $2 \ ^3D_3$      & --                                                    & --                                 & 6535  & 6637 & 6534 \\
\hline
--              & $3 \ ^1S_0$      & --                                                    & --                                 & 6410  & 6467 & 6409 \\
--              & $3 \ ^3S_1$      & --                                                    & --                                 & 6429  & 6475 & 6429 \\
\hline
--              & $3 \ ^3P_0$      & --                                                    & --                                 & 6639  & 6731 & -- \\
--              & $3 \ P_1$        & --                                                    & --                                 & 6635  & 6761 & -- \\
--              & $3 \ P_1^\prime$ & --                                                    & --                                 & 6650  & 6768 & -- \\
--              & $3 \ ^3P_2$      & --                                                    & --                                 & 6648  & 6780 & --
\\ \bottomrule[1pt]
\end {tabular}
\label{sec2:bs}
\end{table*}

\begin{figure}[hbtp]
\begin{center}
\subfigure[]{\includegraphics[width=0.45\textwidth]{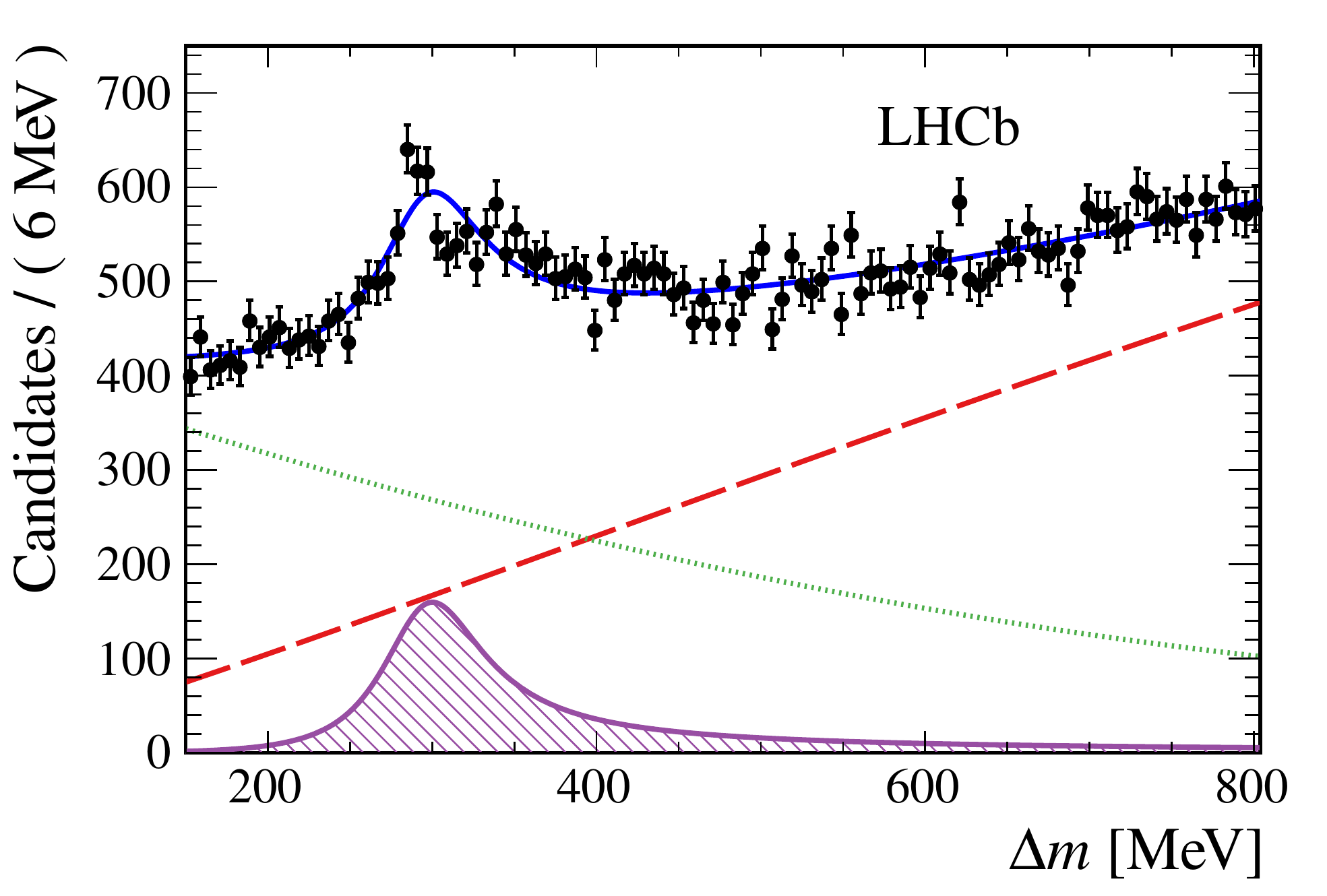}}
~~~~~
\subfigure[]{\includegraphics[width=0.45\textwidth]{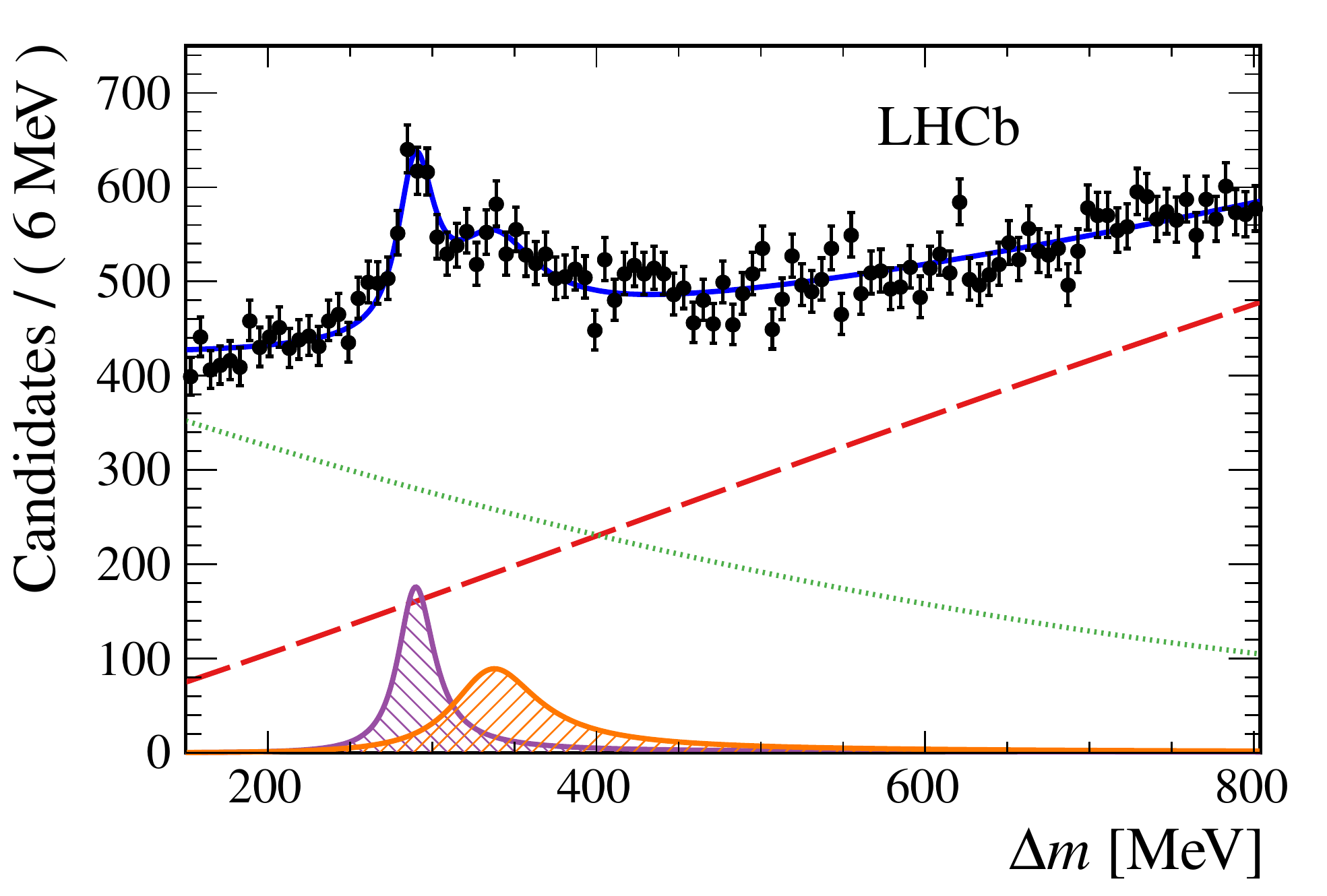}}
\end{center}
\caption{The $B^+ K^-$ invariant mass distributions with the prompt kaon $p_T> 2$~GeV and the fit models: (a) one-peak hypothesis and (b) two-peak hypothesis. Source: Ref.~\cite{LHCb:2020pet}.}
\label{fig:Bs6064}
\end{figure}

Recently, the LHCb collaboration observed a structure in the $B^\pm K^\mp$ mass spectrum in proton-proton collisions~\cite{LHCb:2020pet}. As depicted in Fig.~\ref{fig:Bs6064}, they found that this structure is not described well by a single resonance, and the significance for a two-peak structure relative to a one-peak hypothesis is greater than $7\sigma$. These two states were named as $B_{sJ}(6064)$ and $B_{sJ}(6114)$, whose masses and widths were measured to be
\begin{eqnarray}
B_{sJ}(6064) &:& M = 6063.5 \pm 1.2 \pm 0.8 {\rm~MeV} \, ,
\label{sec2:Bs6064}
\\ \nonumber && \Gamma = 26 \pm 4 \pm 4 {\rm~MeV} \, ;
\\ B_{sJ}(6114) &:& M = 6114 \pm 3 \pm 5 {\rm~MeV} \, ,
\label{sec2:Bs6114}
\\ \nonumber && \Gamma = 66 \pm 18 \pm 21 {\rm~MeV} \, .
\end{eqnarray}
These results were obtained under the assumption of a direct decay through $B^\pm K^\mp$. Besides, LHCb also considered the assumption of a decay through $B^{*\pm}K^\mp$ with a missing photon from the $B^{*\pm} \to B^\pm \gamma$ decay, and the above values were shifted to be
\begin{eqnarray}
B_{sJ}(6109) &:& M = 6108.8 \pm 1.1 \pm 0.7 {\rm~MeV} \, ,
\label{sec2:Bs6109}
\\ \nonumber && \Gamma = 22 \pm 5 \pm 4 {\rm~MeV} \, ;
\\ B_{sJ}(6158) &:& M = 6158 \pm 4 \pm 5 {\rm~MeV} \, ,
\\ \nonumber && \Gamma = 72 \pm 18 \pm 25 {\rm~MeV} \, .
\end{eqnarray}
The above structures were suggested by LHCb to be the $D$-wave bottom-strange mesons. Based on Eqs.~(\ref{sec2:Bs6064}-\ref{sec2:Bs6114}), the two $B_{sJ}(6064)$ and $B_{sJ}(6114)$ mesons may take part in forming the two $D$-wave doublets $(1^-, 2^-)$ and $(2^-, 3^-)$~\cite{Chen:2022fye}. If this is the case, their quantum numbers should be $J^P = 1^-$ and $3^-$, and there might be two $B_{s2}$ mesons still missing. However, as shown in Table~\ref{sec2:bs}, their masses are significantly lower than the quark model predictions~\cite{Godfrey:2016nwn,Ebert:2009ua,Sun:2014wea}.

\begin{figure}[hbtp]
\begin{center}
\includegraphics[width=0.5\textwidth]{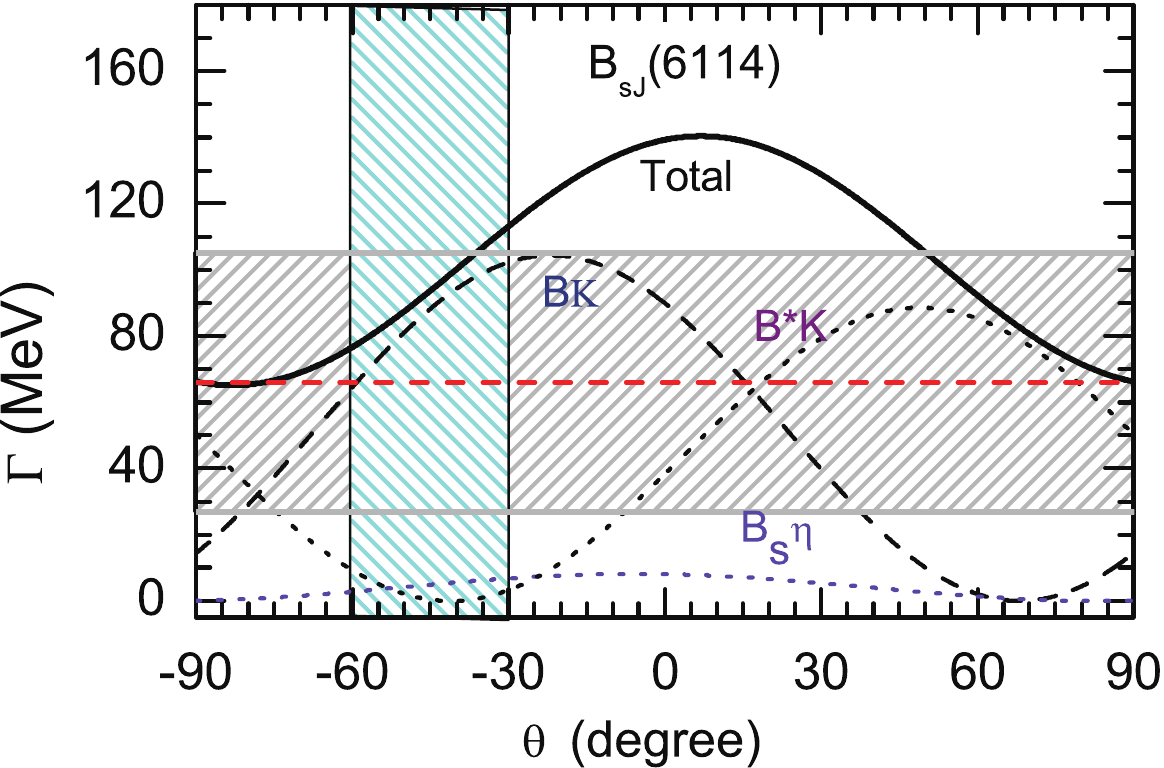}
\end{center}
\caption{Partial and total decay widths of the $B_{sJ}(6114)$ as the mixed state $B_s(|SD\rangle_H)$ via the $2^3S_1$-$1^3D_1$ mixing. Source: Ref.~\cite{Li:2021hss}.}
\label{fig:Bs6114}
\end{figure}

In Ref.~\cite{Li:2021hss} the authors investigated systematically the mass spectrum as well as the strong and radiative decays of the $1P$-, $1D$-, and $2S$-wave $B$ and $B_s$ states in the constituent quark model. Their results suggest that the $B_{sJ}(6114)$ can be interpreted as the mixed state $B_s(|SD\rangle_H)$ via the $2^3S_1$-$1^3D_1$ mixing. They considered the following mixing scheme:
\begin{equation}
\left(
  \begin{array}{c}
  |SD\rangle_L\\
  |SD\rangle_H
  \end{array}\right)=
  \left(
  \begin{array}{cc}
   \cos\theta &\sin\theta\\
  -\sin\theta &\cos\theta
  \end{array}
\right)
\left(
  \begin{array}{c}
  2^3S_1\\
  1^3D_1
  \end{array}\right) \, ,
\end{equation}
with the mixing angle determined in Ref.~\cite{Zhong:2009sk}, $\theta \approx -45^\circ \pm 16^\circ$. As shown in Fig.~\ref{fig:Bs6114}, the $B_s(|SD\rangle_H)$ state has the decay width around $95\pm15$~MeV and dominantly decays into the $B^\pm K^\mp$ channel with a branching fraction about 90\%. These results are consistent with the experimental data of the $B_{sJ}(6114)$. This assignment was supported by Ref.~\cite{Patel:2022hhl} based on the screening potential. On the other hand, the results of Ref.~\cite{Li:2021hss} also support the possible interpretation of the $B_{sJ}(6114)$ as a pure $B_s(1^3D_1)$ state if there is little mixing between $2^3S_1$-$1^3D_1$.

In Ref.~\cite{Li:2021hss} the authors also studied the $B_{sJ}(6064)$ and proposed two possible explanations. Firstly, this structure may be mainly caused by the $B_{sJ}(6109)$ resonance decaying into $B^{*+}K^-$, and the $B_{sJ}(6109)$ can be interpreted as the higher-mass mixed state $B_s(1D^\prime_2)$ of $J^P = 2^-$ via the $1^1D_2$-$1^3D_2$ mixing. Note that the $B_{sJ}(6064)$ from Eq.~(\ref{sec2:Bs6064}) and the $B_{sJ}(6109)$ from Eq.~(\ref{sec2:Bs6109}) were obtained by LHCb under different assumptions. Secondly, the $B_{sJ}(6064)$ may be interpreted as a pure $B_s(1^3D_3)$ state.

Later in Ref.~\cite{Chen:2021uou} the authors calculated the mass spectra of both the two-body $b \bar s$ system and the four-body $b \bar s q \bar q$ ($q = u/d$) system by considering the possible production of quark-antiquark pairs in the vacuum. They found it possible to explain the $B_{sJ}(6064)$ and $B_{sJ}(6114)$ as the conventional $B_s(2S)$ and/or $B_s(1D)$ mesons in the chiral quark model. They also found it possible to explain these states as the exotic $b \bar s q \bar q$ tetraquark states.

In Ref.~\cite{Kong:2021ohg} the authors systematically studied the $\bar B^{(*)}K^{(*)}$ hadronic molecular states in a quasipotential Bethe-Salpter equation approach together with the one-boson-exchange model. Their potential was achieved with the help of the hidden-gauge Lagrangians. They proposed that the ${\bar B}_{sJ}(6114)$, the antiparticle of $B_{sJ}(6114)$, is very close to the $\bar B K^*$ threshold, making it the best candidate of the $\bar B K^*$ molecular state of $(I)J^P = (0)1^+$.

At the end of this section, we want to emphasize that the LHCb collaboration made great contributions to the heavy meson spectrum with good accuracy. In order to pin down the underlying structure of these states, more efforts are needed to make the uncertainty of the theoretical predictions controllable. The motion of the light quark in the excited heavy mesons is highly relativistic. The relativistic correction has to be considered very carefully together with possible configuration mixings induced by various fine and hyperfine interactions. The spatial distribution of the higher excited heavy mesons is generally more extended than the ground states. Hence, the confinement potential plays a more important role in its competition with the one-gluon-exchange potential in the higher excited states than in the ground states. The excited meson spectrum will be very sensitive to the different forms of the confinement potential and different confinement mechanism. From the comparison of the experimental data and quark model prediction, we know that the commonly used linear potential works for the low-lying states. Whether it still works well for the very higher excited states needs to be tested carefully.

\section{Excited open heavy flavor baryons}
\label{sec3}

In this section we shall provide an updated review on the experimental and theoretical progresses on the open heavy flavor baryons since 2017. Similar to the open heavy flavor mesons, the heavy quark symmetry also plays an important role here, and we have applied it to categorize the singly heavy baryons in Sec.~\ref{sec1.3}. Note that the configuration mixing of the singly heavy baryons is more complicated than that of the heavy mesons. Because the heavy quark mass is actually not infinite, this framework works better for the singly bottom baryons than for the singly charmed baryons. So we shall first review the singly bottom baryons in Sec.~\ref{sec3.1}, and then review the singly charmed baryons in Sec.~\ref{sec3.2}. Besides, the doubly heavy baryons are also interesting, which will be shortly reviewed in Sec.~\ref{sec3.3}.

We shall omit the word ``singly'' for the singly heavy/charmed/bottom baryons most of the time, and we shall also omit their charges most of the time, which can largely simplify our notations. Besides, the symbols $\Xi_c$ and $\Xi_c^\prime$ are used to denote the charmed-strange baryons ($qsc$ with $q=u/d$) belonging to the flavor representations $\mathbf{\bar 3}_F$ and $\mathbf{6}_F$ respectively, but the superscript~$^\prime$ is usually omitted for the experimentally observed states since they can not be directly differentiated in experiments.

\subsection{Singly bottom baryons}
\label{sec3.1}

All the $1S$ bottom baryons are well established in experiments~\cite{pdg} except the $\Omega_b^{*-}$, {\it i.e.}, $(\Lambda_b^0, \Xi_b^{0}, \Xi_b^{-})$, $(\Sigma_b^{+}, \Sigma_b^{0}, \Sigma_b^{-}, \Xi_b^{\prime0}, \Xi_b^{\prime-}, \Omega_b^-)$, and $(\Sigma_b^{*+}, \Sigma_b^{*0}, \Sigma_b^{*-}, \Xi_b^{*0}, \Xi_b^{*-}, \xcancel{\Omega_b^{*-}})$. They almost complete the two $S$-wave bottom baryon multiplets $[\mathbf{\bar 3}_F, 0, 0]$ and $[\mathbf{6}_F, 1, 1]$, as shown in Fig.~\ref{fig:categorization}. Higher bottom baryons are much more complicated, including:
\begin{itemize}

\item $\Lambda_b(5912)$~\cite{LHCb:2012kxf}, $\Lambda_b(5920)$~\cite{LHCb:2012kxf,CDF:2013pvu}, $\Lambda_b(6072)$~\cite{CMS:2020zzv,LHCb:2020lzx}, $\Lambda_b(6146)$~\cite{LHCb:2019soc}, and $\Lambda_b(6152)$~\cite{LHCb:2019soc};

\item $\Sigma_b(6097)$~\cite{LHCb:2018haf};

\item $\Xi_b(6100)$~\cite{CMS:2021rvl}, $\Xi_b(6227)$~\cite{LHCb:2018vuc,LHCb:2020xpu}, $\Xi_b(6327)$~\cite{LHCb:2021ssn}, and $\Xi_b(6333)$~\cite{LHCb:2021ssn};

\item $\Omega_b(6316)$~\cite{LHCb:2020tqd}, $\Omega_b(6330)$~\cite{LHCb:2020tqd}, $\Omega_b(6340)$~\cite{LHCb:2020tqd}, and $\Omega_b(6350)$~\cite{LHCb:2020tqd}.

\end{itemize}
The two bottom baryons $\Lambda_b(5912)$ and $\Lambda_b(5920)$ have been reviewed in our previous paper~\cite{Chen:2016spr}. They are good candidates for the $P$-wave bottom baryons belonging to the $P$-wave bottom baryon doublet $[\mathbf{\bar 3}_F, 1, 0, \lambda]$. All the other twelve excited bottom baryons were observed by LHCb and CMS in the past five years~\cite{Chen:2016spr}, which will be reviewed in this subsection.

In Table~\ref{sec3:bottom}, we summarize the theoretical results of the bottom baryons obtained from various quark models~\cite{Ebert:2011kk,Roberts:2007ni,Yoshida:2015tia}, including the $1S$ and $2S$ states as well as the $1P$ and $1D$ states of the $\lambda$-mode. Possible experimental candidates are given for comparison. Especially, the $\Xi_b(6100)$ may be the partner state of the $\Lambda_b(5912)$ and $\Lambda_b(5920)$ belonging to the $P$-wave bottom baryon doublet $[\mathbf{\bar 3}_F, 1, 0, \lambda]$, which will be reviewed in Sec.~\ref{sec3.1.1}. The four bottom baryons $\Lambda_b(6146)$, $\Lambda_b(6152)$, $\Xi_b(6327)$, and $\Xi_b(6333)$ may complete the $D$-wave bottom baryon doublet $[\mathbf{\bar 3}_F, 2, 0, \lambda\lambda]$, which will be reviewed in Sec.~\ref{sec3.1.2} and Sec.~\ref{sec3.1.3}.

Many excited bottom baryons are good candidates for the $P$-wave bottom baryons belonging to the flavor $\mathbf{6}_F$ representation, which were studied in Refs.~\cite{Liu:2007fg,Mao:2015gya,Chen:2016phw,Mao:2017wbz,Cui:2019dzj,Yang:2019cvw,Chen:2020mpy,Mao:2020jln} using the methods of QCD sum rules~\cite{Shifman:1978bx,Reinders:1984sr} and light-cone sum rules~\cite{Balitsky:1989ry,Braun:1988qv,Chernyak:1990ag,Ball:1998je,Ball:1998kk,Ball:1998sk,Ball:1998ff,Ball:2004rg,Ball:2006wn,Ball:2007rt,Ball:2007zt} within the framework of heavy quark effective theory~\cite{Isgur:1989vq,Grinstein:1990mj,Georgi:1990um,Eichten:1989zv,Falk:1990yz}. In a recent study~\cite{Yang:2020zrh} the authors systematically investigated the four $P$-wave bottom baryon multiplets of the flavor $\mathbf{6}_F$, {\it i.e.}, $[\mathbf{6}_F, 1, 0, \rho]$, $[\mathbf{6}_F, 0, 1, \lambda]$, $[\mathbf{6}_F, 1, 1, \lambda]$, and $[\mathbf{6}_F, 2, 1, \lambda]$. Their masses, mass splittings within the same multiplets, and decay properties are summarized in Table~\ref{sec3:bottomsumrule}. These results can explain the $\Sigma_b(6097)$, $\Xi_b(6227)$, $\Omega_b(6316)$, $\Omega_b(6330)$, $\Omega_b(6340)$, and $\Omega_b(6350)$ as a whole, which will be reviewed in Sec.~\ref{sec3.1.4}, Sec.~\ref{sec3.1.5}, and Sec.~\ref{sec3.1.6}.

Especially, there exist many possible explanations for the $\Omega_b(6316)$, $\Omega_b(6330)$, $\Omega_b(6340)$, and $\Omega_b(6350)$, {\it e.g.}, those given in Table~\ref{sec3:bottom} are taken from Ref.~\cite{Liang:2020hbo} and those given in Table~\ref{sec3:bottomsumrule} are taken from Refs.~\cite{Chen:2020mpy,Yang:2020zrh}. See Sec.~\ref{sec3.1.6} for more discussions.

\begin{table*}[htb]
\scriptsize
\caption{Theoretical results of the bottom baryons obtained from various quark models~\cite{Ebert:2011kk,Roberts:2007ni,Yoshida:2015tia}, including the $1S$ and $2S$ states as well as the $1P$ and $1D$ states of the $\lambda$-mode. Possible experimental candidates are given for comparisons, with uncertainties the sum of their statistic and systematical uncertainties. Masses are in units of MeV.}
\centering
\begin{tabular}{ c c c c c c c }
\toprule[1pt]
& $J^P~(nL)$ & Exp. Mass~\cite{pdg} & Exp. Width~\cite{pdg} & RQM~\cite{Ebert:2011kk} & NQM~\cite{Roberts:2007ni} & NQM~\cite{Yoshida:2015tia}
\\ \midrule[1pt]
$\Lambda_b$             & $1/2^+~(1S)$     & $5619.60\pm0.17$                                  & $\sim10^{-12}$~s                   & 5620  & 5612  & 5618  \\
$\Xi_b$                 & $1/2^+~(1S)$     & $5797.0\pm0.6$                                    & $\sim10^{-12}$~s                   & 5803  & 5806  & --    \\
$\Sigma_b$              & $1/2^+~(1S)$     & $5810.56\pm0.25$                                  & $5.0 \pm 0.5$                      & 5808  & 5833  & 5823  \\
$\Sigma_b^{*}$          & $3/2^+~(1S)$     & $5830.32\pm0.27$                                  & $9.4 \pm 0.5$                      & 5834  & 5858  & 5845  \\
$\Xi_b^{\prime}$        & $1/2^+~(1S)$     & $5935.02\pm0.05$                                  & $<0.08$                            & 5936  & 5970  & --    \\
$\Xi_b^*$               & $3/2^+~(1S)$     & $5955.33\pm0.13$                                  & $1.65 \pm 0.33$                    & 5963  & 5980  & --    \\
$\Omega_b$              & $1/2^+~(1S)$     & $6046.1\pm1.7$                                    & $\sim10^{-12}$~s                   & 6064  & 6081  & 6076  \\
$\Omega_b^{*}$          & $3/2^+~(1S)$     & --                                                & --                                 & 6088  & 6102  & 6094  \\
\hline
$\Lambda_b$             & $1/2^-~(1P)$     & $\Lambda_b(5912):5912.20\pm0.21$                  & $<0.66$                            & 5930  & 5939  & 5938  \\
$\Lambda_b$             & $3/2^-~(1P)$     & $\Lambda_b(5920):5919.92\pm0.19$                  & $<0.63$                            & 5942  & 5941  & 5939  \\
$\Xi_b$                 & $1/2^-~(1P)$     & --                                                & --                                 & 6120  & 6090  & --    \\
$\Xi_b$                 & $3/2^-~(1P)$     & $\Xi_b(6100):6100.3\pm0.6$~\cite{CMS:2021rvl}     & $<1.9$~\cite{CMS:2021rvl}          & 6130  & 6093  & --    \\
$\Sigma_b$              & $1/2^-~(1P)$     & --                                                & --                                 & 6095  & 6099  & 6127  \\
$\Sigma_b$              & $1/2^-~(1P)$     & --                                                & --                                 & 6101  & 6106  & 6135  \\
$\Sigma_b$              & $3/2^-~(1P)$     & --                                                & --                                 & 6087  & 6101  & 6132  \\
$\Sigma_b$              & $3/2^-~(1P)$     & $\Sigma_b(6097):6095.8\pm1.7$~\cite{LHCb:2018haf} & $31.0 \pm 5.5$~\cite{LHCb:2018haf} & 6096  & 6105  & 6141  \\
$\Sigma_b$              & $5/2^-~(1P)$     & --                                                & --                                 & 6084  & 6172  & 6144  \\
$\Xi^\prime_b$          & $1/2^-~(1P)$     & --                                                & --                                 & 6227  & 6188  & --    \\
$\Xi^\prime_b$          & $1/2^-~(1P)$     & --                                                & --                                 & 6233  & --    & --    \\
$\Xi^\prime_b$          & $3/2^-~(1P)$     & --                                                & --                                 & 6224  & 6190  & --    \\
$\Xi^\prime_b$          & $3/2^-~(1P)$     & $\Xi_b(6227):6226.9\pm2.0$~\cite{LHCb:2018vuc}    & $18.1 \pm 5.7$~\cite{LHCb:2018vuc} & 6234  & --    & --    \\
$\Xi^\prime_b$          & $5/2^-~(1P)$     & --                                                & --                                 & 6226  & 6201  & --    \\
$\Omega_b$              & $1/2^-~(1P)$     & --                                                & --                                 & 6330  & 6301  & 6333  \\
$\Omega_b$              & $1/2^-~(1P)$     &$\Omega_b(6316):6315.64\pm0.59$~\cite{LHCb:2020tqd}& $<2.8$~\cite{LHCb:2020tqd}         & 6339  & 6312  & 6340  \\
$\Omega_b$              & $3/2^-~(1P)$     &$\Omega_b(6330):6330.30\pm0.58$~\cite{LHCb:2020tqd}& $<3.1$~\cite{LHCb:2020tqd}         & 6331  & 6304  & 6336  \\
$\Omega_b$              & $3/2^-~(1P)$     &$\Omega_b(6340):6339.71\pm0.57$~\cite{LHCb:2020tqd}& $<1.5$~\cite{LHCb:2020tqd}         & 6340  & 6311  & 6344  \\
$\Omega_b$              & $5/2^-~(1P)$     &$\Omega_b(6350):6349.88\pm0.61$~\cite{LHCb:2020tqd}& $1.4^{+1.0}_{-0.8}$~\cite{LHCb:2020tqd}  & 6334  & 6311  & 6345  \\
\hline
$\Lambda_b$             & $1/2^+~(2S)$     & $\Lambda_b(6072):6072.3\pm3.0$~\cite{LHCb:2020lzx}& $72 \pm 11$~\cite{LHCb:2020lzx}    & 6089  & 6107  & 6153  \\
$\Xi_b$                 & $1/2^+~(2S)$     & --                                                & --                                 & 6266  & --    & --    \\
$\Sigma_b$              & $1/2^+~(2S)$     & --                                                & --                                 & 6213  & 6294  & 6343  \\
$\Sigma_b$              & $3/2^+~(2S)$     & --                                                & --                                 & 6226  & 6308  & 6356  \\
$\Xi_b^{\prime}$        & $1/2^+~(2S)$     & --                                                & --                                 & 6329  & --    & --    \\
$\Xi_b^{\prime}$        & $3/2^+~(2S)$     & --                                                & --                                 & 6342  & 6311  & --    \\
$\Omega_b$              & $1/2^+~(2S)$     & --                                                & --                                 & 6450  & 6472  & 6517  \\
$\Omega_b$              & $3/2^+~(2S)$     & --                                                & --                                 & 6461  & 6478  & 6528  \\
\hline
$\Lambda_b$             & $3/2^+~(1D)$     & $\Lambda_b(6152):6152.51\pm0.37$~\cite{LHCb:2019soc}      & $2.1 \pm 0.9$~\cite{LHCb:2019soc}                    & 6190  & 6181  & 6211  \\
$\Lambda_b$             & $5/2^+~(1D)$     & $\Lambda_b(6146):6146.17\pm0.43$~\cite{LHCb:2019soc}      & $2.9 \pm 1.3$~\cite{LHCb:2019soc}                    & 6196  & 6183  & 6212  \\
$\Xi_b$                 & $3/2^+~(1D)$     & $\Xi_b(6327):6327.28^{+0.34}_{-0.33}$~\cite{LHCb:2021ssn} & $<2.20$~\cite{LHCb:2021ssn}                          & 6366  & --    & --    \\
$\Xi_b$                 & $5/2^+~(1D)$     & $\Xi_b(6333):6332.69^{+0.28}_{-0.29}$~\cite{LHCb:2021ssn} & $<1.55$~\cite{LHCb:2021ssn}                          & 6373  & 6300  & --    \\
$\Sigma_b$              & $1/2^+~(1D)$     & --                                                & --                                 & 6311  & --    & 6395  \\
$\Sigma_b$              & $3/2^+~(1D)$     & --                                                & --                                 & 6285  & --    & 6393  \\
$\Sigma_b$              & $3/2^+~(1D)$     & --                                                & --                                 & 6326  & --    & --    \\
$\Sigma_b$              & $5/2^+~(1D)$     & --                                                & --                                 & 6270  & 6325  & 6397  \\
$\Sigma_b$              & $5/2^+~(1D)$     & --                                                & --                                 & 6284  & 6328  & 6402  \\
$\Sigma_b$              & $7/2^+~(1D)$     & --                                                & --                                 & 6260  & 6333  & --    \\
$\Xi^\prime_b$          & $1/2^+~(1D)$     & --                                                & --                                 & 6447  & --    & --    \\
$\Xi^\prime_b$          & $3/2^+~(1D)$     & --                                                & --                                 & 6431  & --    & --    \\
$\Xi^\prime_b$          & $3/2^+~(1D)$     & --                                                & --                                 & 6459  & --    & --    \\
$\Xi^\prime_b$          & $5/2^+~(1D)$     & --                                                & --                                 & 6420  & 6393  & --    \\
$\Xi^\prime_b$          & $5/2^+~(1D)$     & --                                                & --                                 & 6432  & --    & --    \\
$\Xi^\prime_b$          & $7/2^+~(1D)$     & --                                                & --                                 & 6414  & 6395  & --    \\
$\Omega_b$              & $1/2^+~(1D)$     & --                                                & --                                 & 6540  & --    & 6561  \\
$\Omega_b$              & $3/2^+~(1D)$     & --                                                & --                                 & 6530  & --    & 6559  \\
$\Omega_b$              & $3/2^+~(1D)$     & --                                                & --                                 & 6549  & --    & --    \\
$\Omega_b$              & $5/2^+~(1D)$     & --                                                & --                                 & 6520  & 6492  & 6561  \\
$\Omega_b$              & $5/2^+~(1D)$     & --                                                & --                                 & 6529  & 6494  & 6566  \\
$\Omega_b$              & $7/2^+~(1D)$     & --                                                & --                                 & 6517  & 6497  & --
\\ \bottomrule[1pt]
\end{tabular}
\label{sec3:bottom}
\end{table*}

\begin{table*}[hbt]
\renewcommand{\arraystretch}{1.5}
\scriptsize
\caption{Mass spectrum and decay properties of the $P$-wave bottom baryons belonging to the $SU(3)$ flavor $\mathbf{6}_F$ representation, calculated in Ref.~\cite{Yang:2020zrh} using the methods of QCD sum rules and light-cone sum rules within the framework of heavy quark effective theory. There are four $P$-wave multiplets of the flavor $\mathbf{6}_F$, {\it i.e.}, $[\mathbf{6}_F, 1, 0, \rho]$, $[\mathbf{6}_F, 0, 1, \lambda]$, $[\mathbf{6}_F, 1, 1, \lambda]$, and $[\mathbf{6}_F, 2, 1, \lambda]$. Their possible experimental candidates are given in the last column for comparisons.}
\centering
\begin{tabular}{ c | c | c | c | c | c}
\hline\hline
\multirow{2}{*}{HQET state} & ~~~Mass~~~ & Difference & Main Decay channel & Total width  & \multirow{2}{*}{~Candidate~}
\\                          & ({GeV})    & ({MeV})    & ({MeV})            & ({MeV})      &
\\ \hline\hline
$[\Sigma_b({1\over2}^-), 1, 0, \rho]$                                                         & $6.05\pm0.12$ & \multirow{2}{*}{$3\pm1$} &
$\Gamma\left(\Sigma_b\pi\right)\approx710$                                                    & $710$ & --
\\ \cline{1-2}\cline{4-6}
$[\Sigma_b({3\over2}^-), 1, 0, \rho]$                                                         & $6.05\pm0.12$ & &
$\Gamma\left(\Sigma_b^*\pi\right)\approx410$                                                  & $410$ & --
\\ \hline
$[\Sigma_b({1\over2}^-), 0, 1, \lambda]$                                                      & $6.05\pm0.11$ & -- &
$\Gamma\left(\Lambda_b\pi\right)\approx1300$                                                  & $1300$ &--
\\ \hline
$[\Sigma_b({1\over2}^-), 1, 1, \lambda]$                                                      & $6.06\pm0.13$ & \multirow{4}{*}{$6\pm3$}       &
$\Gamma\left(\Sigma_b\pi\right)\approx14.1$                                                   & $14.3{^{+21.2}_{-10.9}}$ & --
\\ \cline{1-2}\cline{4-6}
$[\Sigma_b({3\over2}^-), 1, 1, \lambda]$                                                      & $6.07\pm0.13$ &                                &
$\begin{array}{c}
\Gamma\left(\Sigma_b\pi\right) \approx 0.55 \\
\Gamma\left(\Sigma_b^{*}\pi\right) \approx 4.0 \\
\Gamma\left(\Lambda_b\rho (\to\pi\pi) \right) \approx 0.23
\end{array}$                                                                                  & $4.8{^{+5.9}_{-2.9}}$    & --
\\ \hline
$[\Sigma_b({3\over2}^-),2,1,\lambda]$                                                         & $6.11\pm0.16$ & \multirow{4}{*}{$12\pm5$}      &
$\begin{array}{c}
\Gamma\left(\Lambda_b\pi\right)\approx49.6\\
\Gamma\left(\Sigma_b \pi\right)\approx1.6\\
\Gamma\left(\Sigma_b^{*}\pi\right)\approx0.25
\end{array}$                                                                                  & $51.4{^{+76.5}_{-32.9}}$ & $\Sigma_b(6097) $
\\ \cline{1-2}\cline{4-6}
$[\Sigma_b({5\over2}^-),2,1,\lambda]$                                                         & $6.12\pm0.15$ & &
$\begin{array}{c}
\Gamma\left(\Lambda_b\pi\right)\approx20.8\\
\Gamma\left(\Sigma_b \pi\right)\approx0.36\\
\Gamma\left(\Sigma_b^{*}\pi\right)\approx1.1
\end{array}$                                                                                  & $22.3{^{+23.8}_{-13.8}}$ & --
\\ \hline
$[\Xi^\prime_b({1\over2}^-),1,0,\rho]$                                                        & $6.18\pm0.12$ & \multirow{2}{*}{$3 \pm 1$} &
$\Gamma\left(\Xi_b^{\prime}\pi\right)\approx250$                                              & $250$ & --
\\ \cline{1-2}\cline{4-6}
$[\Xi^\prime_b({3\over2}^-),1,0,\rho]$                                                        & $6.19\pm0.11$ & &
$\Gamma\left(\Xi_b^{*}\pi\right)\approx160$                                                   & $160$ & --
\\ \hline
$[\Xi^\prime_b({1\over2}^-),0,1,\lambda]$                                                     & $6.20 \pm 0.11$ & -- &
$\begin{array}{c}
\Gamma\left(\Xi_b\pi\right)\approx990\\
\Gamma\left(\Lambda_b K\right)\approx910
\end{array}$                                                                                  & $1900$ & --
\\ \hline
$[\Xi^\prime_b({1\over2}^-),1,1,\lambda]$                                                     & $6.21\pm0.11$ & \multirow{3}{*}{$7 \pm 2$} &
$\begin{array}{c}
\Gamma\left(\Xi_b^{\prime}\pi\right)\approx4.5\\
\Gamma\left(\Xi_b^{*}\pi\right)\approx0.16
\end{array}$                                                                                  & $4.7{^{+5.8}_{-3.3}}$    & --
\\ \cline{1-2}\cline{4-6}
$[\Xi^\prime_b({3\over2}^-),1,1,\lambda]$                                                     & $6.22 \pm 0.11$                                & &
$\begin{array}{c}
\Gamma\left(\Xi_ b^{\prime}\pi\right)\approx0.34\\
\Gamma\left(\Xi_b^{*}\pi\right)\approx1.4
\end{array}$                                                                                  & $1.8{^{+1.1}_{-1.0}}$    & --
\\ \hline
$[\Xi_b^{\prime}({3\over2}^-),2,1,\lambda]$                                                   & $6.23\pm0.15$  & \multirow{5}{*}{$11\pm5$} &
$\begin{array}{c}
\Gamma\left(\Xi_b\pi\right)\approx19.0\\
\Gamma\left(\Lambda_b K\right)\approx7.4\\
\Gamma\left(\Xi_b^{\prime}\pi\right)\approx0.79\\
\Gamma\left(\Xi_b^{*}\pi\right)\approx0.13
\end{array}$                                                                                  & $27.3{^{+28.5}_{-14.2}}$  & $\Xi_b(6227)$
\\ \cline{1-2}\cline{4-6}
$[\Xi_b^{\prime}({5\over2}^-),2,1,\lambda]$                                                   & $6.24\pm0.14$                                  & &
$\begin{array}{c}
\Gamma\left(\Xi_b\pi\right)\approx8.1\\
\Gamma\left(\Lambda_b K\right)\approx3.4\\
\Gamma\left(\Xi_b^{\prime}\pi\right)\approx0.17\\
\Gamma\left(\Xi_b^{*}\pi\right)\approx0.58
\end{array}$                                                                                 & $12.3{^{+12.3}_{-~6.1}}$   & --
\\ \hline
$[\Omega_b({1\over2}^-),1,0,\rho]$                                                           & $6.32 \pm 0.11$                               & \multirow{2}{*}{$2 \pm 1$}  & -- & $\sim~0$ & \multirow{2}{*}{$\Omega_b(6316)$}
\\ \cline{1-2}\cline{4-5}
$[\Omega_b({3\over2}^-),1,0,\rho]$                                                           & $6.32 \pm 0.11$                               &                             & -- & $\sim~0$
\\ \hline
$[\Omega_b({1\over2}^-),0,1,\lambda]$                                                        & $6.34\pm0.11$                                 & -- &
$\Gamma\left(\Xi_b K\right)\approx2700$                                                      & $2700$                                        &--
\\ \hline
$[\Omega_b({1\over2}^-),1,1,\lambda]$                                                        & $6.34 \pm 0.10$                               & \multirow{2}{*}{$6 \pm 2$}  & -- & $\sim~0$ & $\Omega_b(6330)$
\\ \cline{1-2}\cline{4-6}
$[\Omega_b({3\over2}^-),1,1,\lambda]$                                                        & $6.34 \pm 0.09$                               &                             & -- & $\sim~0$ & $\Omega_b(6340)$
\\ \hline
$[\Omega_b({3\over2}^-),2,1,\lambda]$                                                        & $6.35\pm0.13$                                 & \multirow{2}{*}{$10\pm4$}   &
$\Gamma\left(\Xi_b K\right)\approx4.6$                                                       & $4.6{^{+3.3}_{-1.9}}$                         & $\Omega_b(6350)$
\\ \cline{1-2}\cline{4-6}
$[\Omega_b({5\over2}^-),2,1,\lambda]$                                                        & $6.36\pm0.12$                                 & &
$\Gamma\left(\Xi_b K\right)\approx2.5$                                                       & $2.5{^{+3.5}_{-1.6}}$                         & --
\\ \hline\hline
\end{tabular}
\label{sec3:bottomsumrule}
\end{table*}

\subsubsection{$\Xi_b(6100)$.}
\label{sec3.1.1}

\begin{figure}[hbtp]
\begin{center}
\subfigure[]{\includegraphics[width=0.4\textwidth]{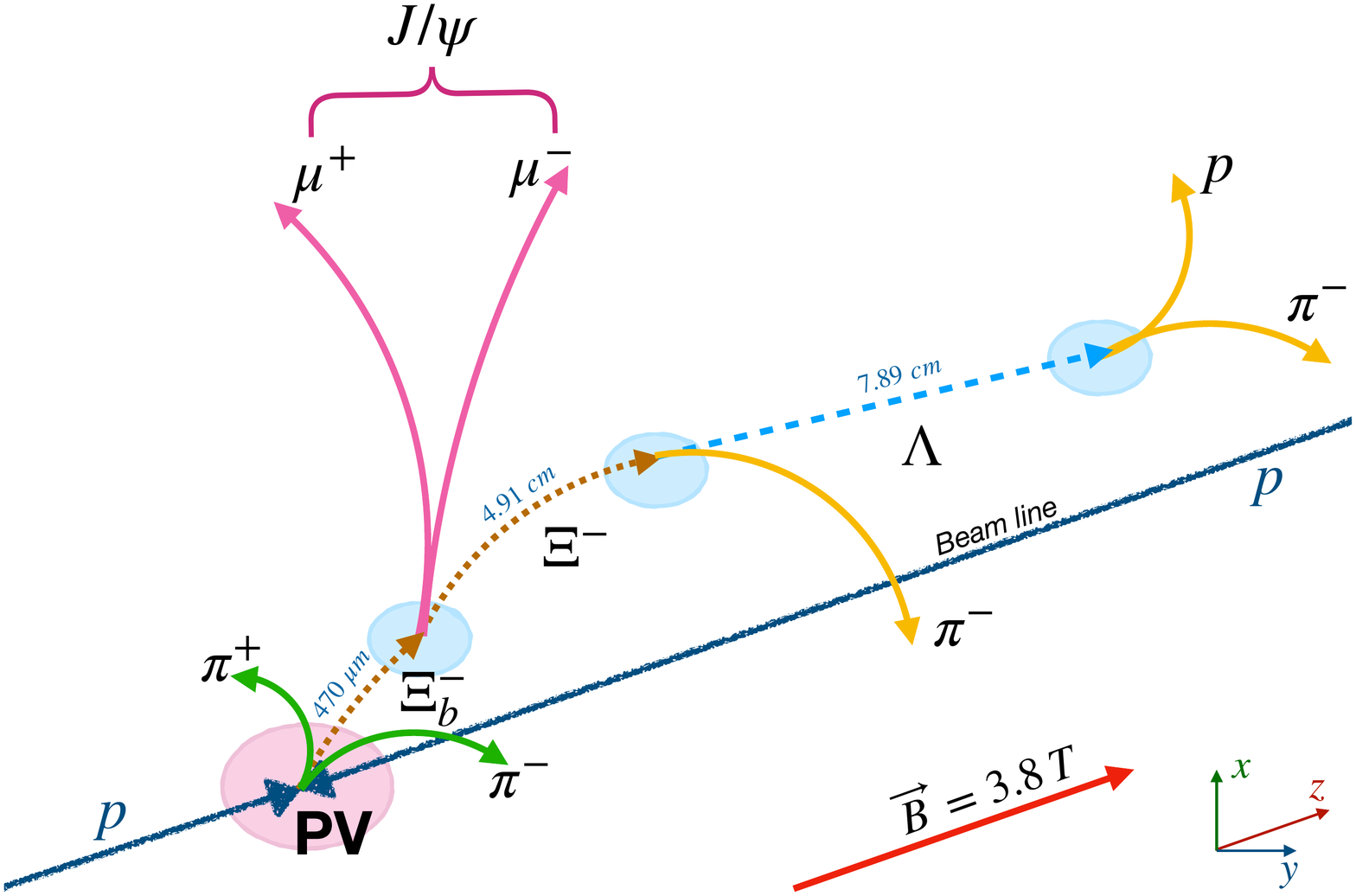}}
~~~~~~~~~~
\subfigure[]{\includegraphics[width=0.4\textwidth]{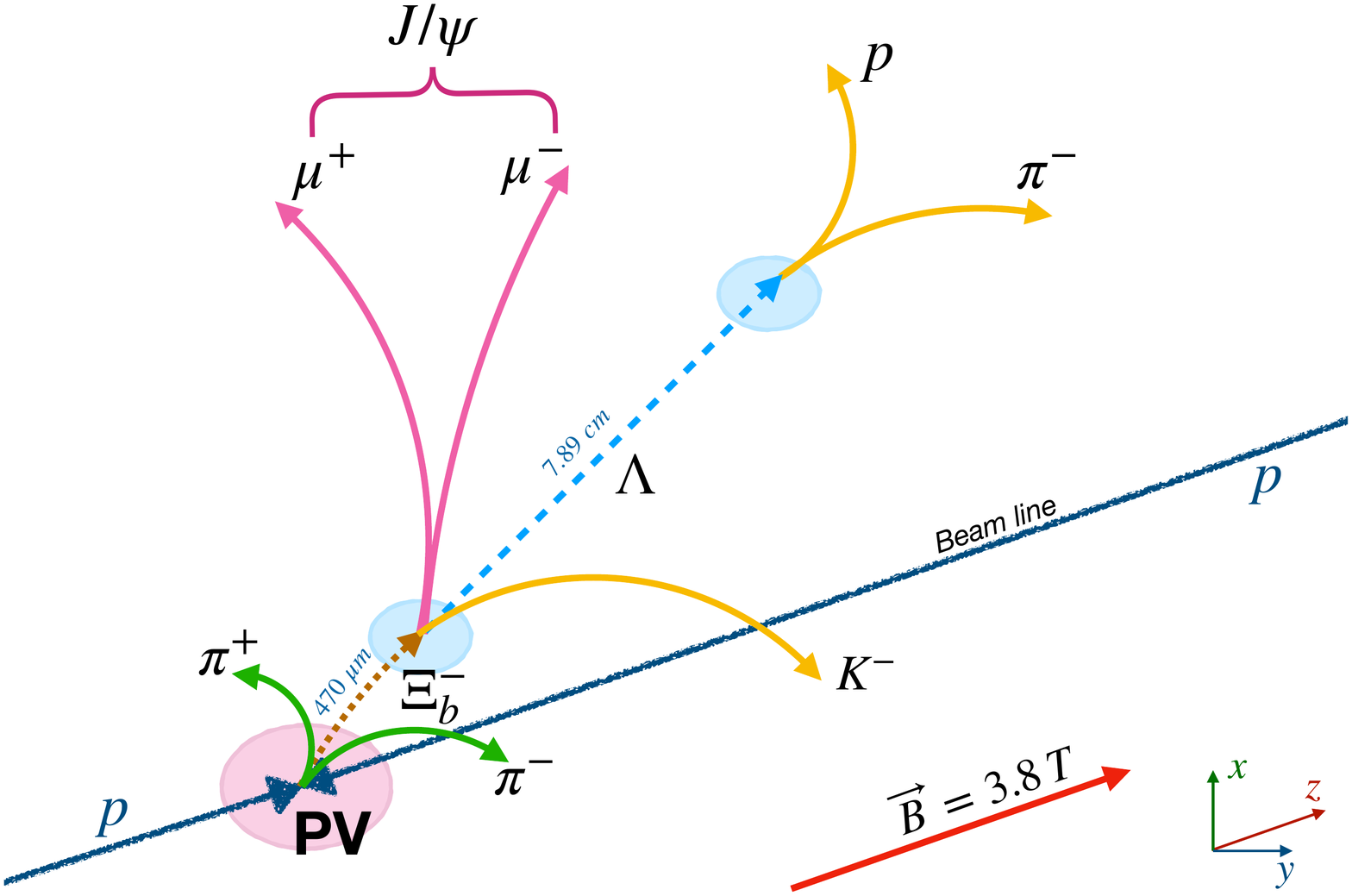}}
\end{center}
\caption{The $\Xi_b(6100)^- \to \Xi_b^- \pi^+\pi^-$ decay topology, where the $\Xi_b^-$ decays to (a) $J/\psi \Xi^-$ or to (b) $J/\psi \Lambda K^-$. Source: Ref.~\cite{CMS:2021rvl}.}
\label{fig:Xi6100}
\end{figure}

In 2021 the CMS collaboration observed a narrow $\Xi_b^-$ resonance in the $\Xi_b^- \pi^+\pi^-$ invariant mass spectrum~\cite{CMS:2021rvl}, with a nice decay topology depicted in Fig.~\ref{fig:Xi6100}. Its mass was measured to be
\begin{equation}
\Xi_b(6100)^-~:~M=6100.3 \pm 0.2 \pm 0.1 \pm 0.6{\rm~MeV}\, .
\end{equation}
Its natural width was determined to be compatible with zero, with an upper limit of 1.9~MeV at 95\% C.~L.

The $\Xi_b(6100)$ was suggested by CMS to be the $\Xi_b(1P)$ baryon of $J^P = 3/2^-$ belonging to the $[\mathbf{\bar 3}_F, 1, 0, \lambda]$ doublet~\cite{CMS:2021rvl,Wang:2017kfr,Kawakami:2019hpp,Chen:2018orb}. It may be the partner state of the $\Lambda_b(5912)$ and $\Lambda_b(5920)$, and therefore, there may be one excited $\Xi_b$ baryon of $J^P = 1/2^-$ still waiting to be observed. This assignment is supported by Ref.~\cite{Arifi:2021orx}, where the authors considered relativistic corrections up to the order $1/m_Q^2$ in the constituent quark model, and evaluated its width to be about $0.63$-$1.36$~MeV.

\begin{figure}[hbtp]
\begin{center}
\includegraphics[width=0.4\textwidth]{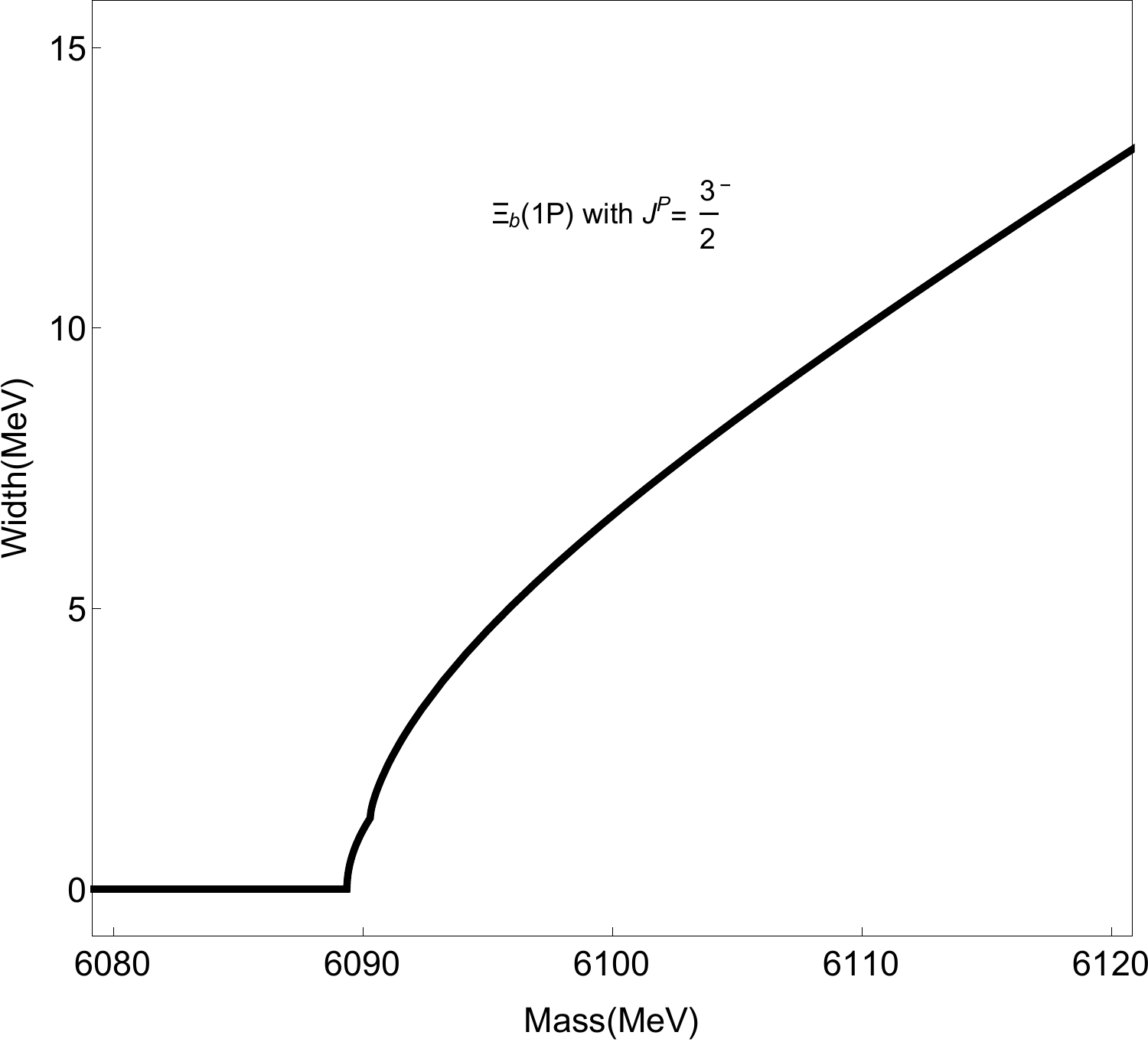}
\end{center}
\caption{Total decay width of the $J^P = 3/2^-$ $\Xi_b(1P)$ state versus its initial mass. Source: Ref.~\cite{He:2021xrh}.}
\label{fig:Ximassdependence}
\end{figure}

Later in Ref.~\cite{He:2021xrh} the authors studied strong decay behaviors of the $\lambda$-mode $\Xi_b$ and $\Xi^\prime_b$ baryons within the $^3P_0$ model. They obtained that the $\Xi_b(1P)$ state of $J^P = 1/2^-$ mainly decays into the $\Xi_b^\prime \pi$ channel with the partial decay width about 12~MeV, and the $\Xi_b(1P)$ state of $J^P = 3/2^-$ mainly decays into the $\Xi_b^* \pi$ channel with the partial decay width about 7~MeV. As shown in Fig.~\ref{fig:Ximassdependence}, these widths are sensitive to the masses of $\Xi_b(1P)$ states, so their results moderately support the assignment of the $\Xi_b(6100)$ as the $\Xi_b(1P)$ state of $J^P = 3/2^-$.

\subsubsection{$\Lambda_b(6072)$, $\Lambda_b(6146)$, and $\Lambda_b(6152)$.}
\label{sec3.1.2}

\begin{figure}[hbtp]
\begin{center}
\includegraphics[width=0.7\textwidth]{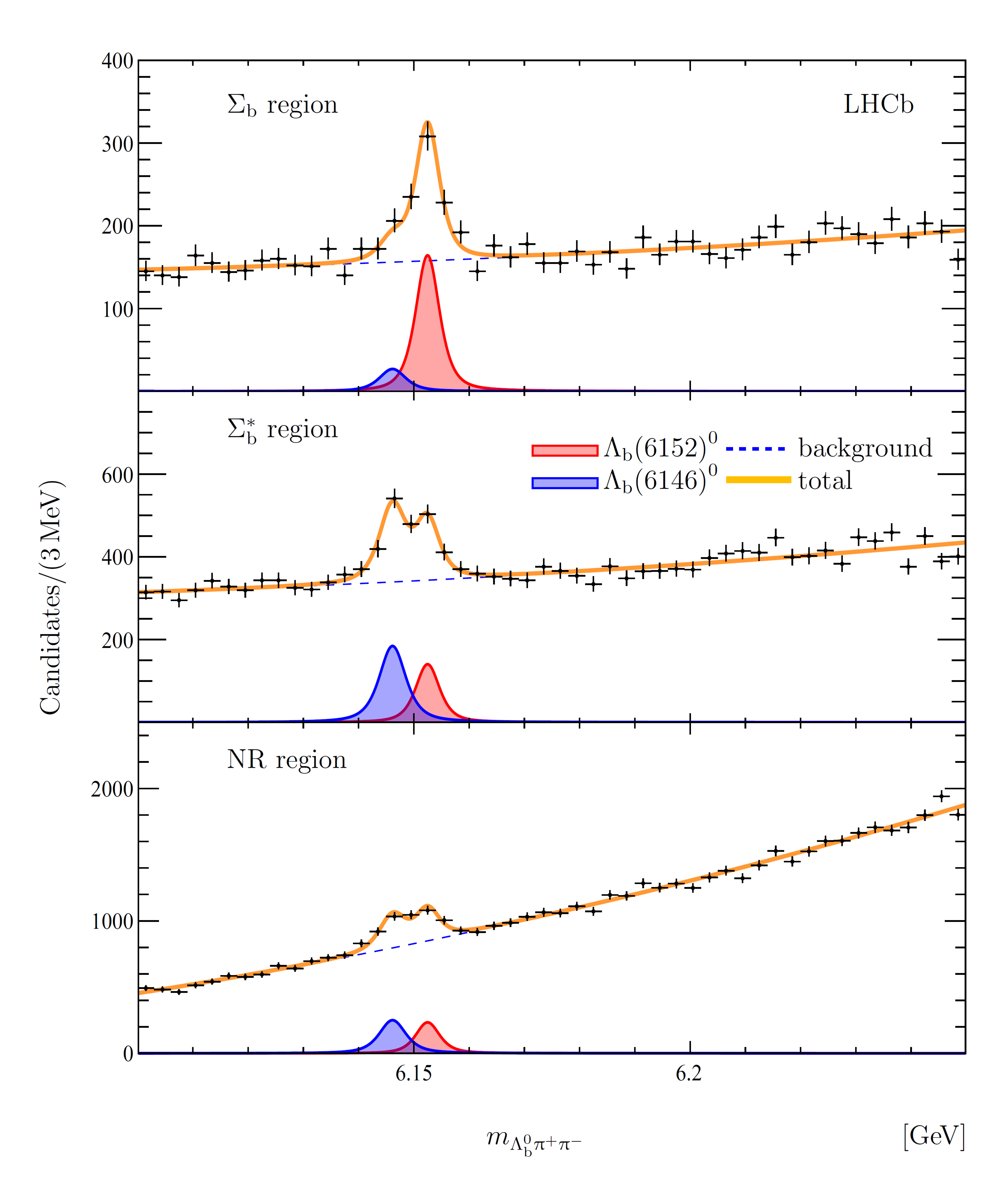}
\end{center}
\caption{The $\Lambda_b^0 \pi^+ \pi^-$ invariant mass distributions for the three regions in $\Lambda_b^0 \pi^\pm$ mass: (top) $\Sigma_b$, (middle) $\Sigma_b^*$, and (bottom) nonresonant region. Source: Ref.~\cite{LHCb:2019soc}.}
\label{fig:L6146a}
\end{figure}

\begin{figure}[hbtp]
\begin{center}
\includegraphics[width=0.5\textwidth]{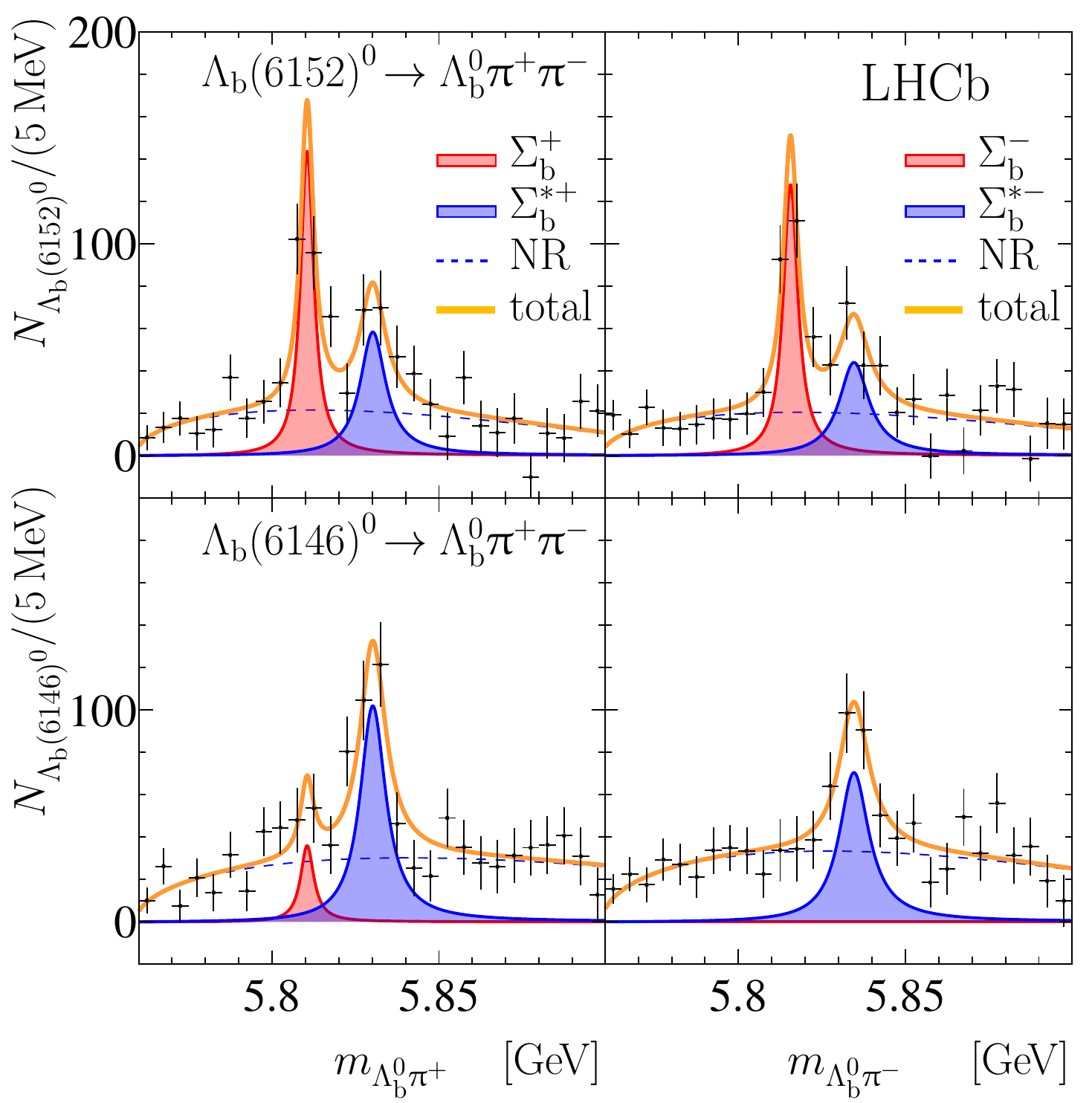}
\end{center}
\caption{Background-subtracted mass distribution of (left) $\Lambda^0_b\pi^+$ and (right) $\Lambda^0_b\pi^-$ combinations from (top) $\Lambda_b(6152)^0 \to \Lambda_b^0 \pi^+ \pi^-$ and (bottom) $\Lambda_b(6146)^0 \to \Lambda_b^0 \pi^+ \pi^-$ decays. Source: Ref.~\cite{LHCb:2019soc}.}
\label{fig:L6146b}
\end{figure}

In 2019 the LHCb collaboration reported their observation of a narrow structure in the $\Lambda_b^0 \pi^+ \pi^-$ invariant mass distribution~\cite{LHCb:2019soc}, as depicted in Fig.~\ref{fig:L6146a}. It was interpreted as two almost degenerate narrow states, denoted as the $\Lambda_b(6146)^0$ and $\Lambda_b(6152)^0$, whose masses and widths were measured to be:
\begin{eqnarray}
\Lambda_b(6146)^0 &:& M = 6146.17 \pm 0.33 \pm 0.22 \pm 0.16~{\rm MeV} \, ,
\\ \nonumber && \Gamma = 2.9 \pm 1.3 \pm 0.3~{\rm MeV} \, ;
\\ \Lambda_b(6152)^0 &:& M = 6152.51 \pm 0.26 \pm 0.22 \pm 0.16~{\rm MeV} \, ,
\\ \nonumber && \Gamma = 2.1 \pm 0.8 \pm 0.3~{\rm MeV} \, .
\end{eqnarray}
Interestingly, LHCb observed significant $\Lambda_b(6146)^0 \to \Sigma_b^{*\pm} \pi^\mp$, $\Lambda_b(6152)^0 \to \Sigma_b^\pm \pi^\mp$, and $\Lambda_b(6152)^0 \to \Sigma_b^{*\pm} \pi^\mp$ signals, but they did not observe significant $\Lambda_b(6146)^0 \to \Sigma_b^{\pm} \pi^\mp$ signal. We depict these decay behaviors in Fig.~\ref{fig:L6146b} to be discussed later.

\begin{figure}[hbtp]
\begin{center}
\includegraphics[width=0.5\textwidth]{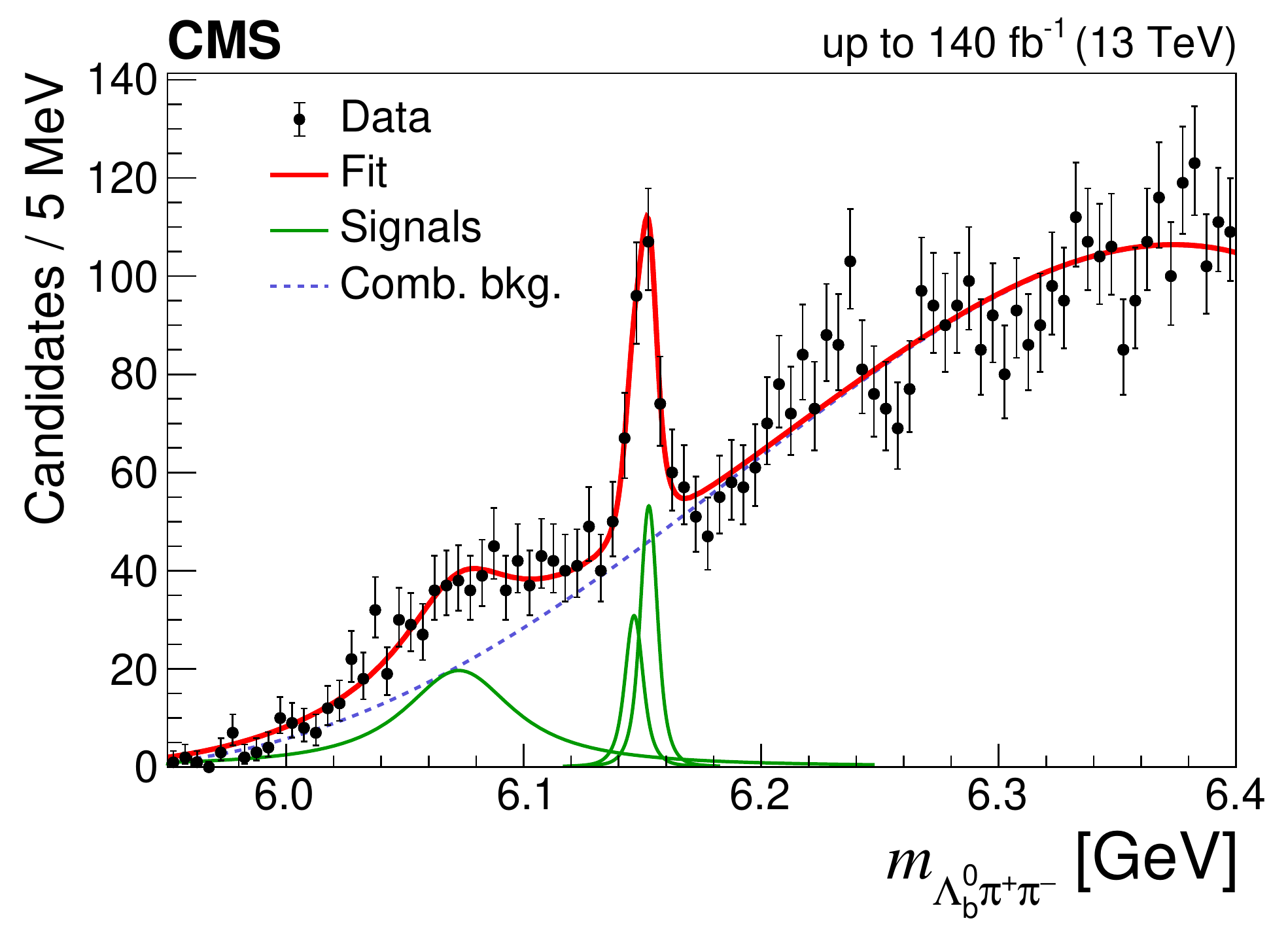}
\end{center}
\caption{The $\Lambda_b^0 \pi^+ \pi^-$ invariant mass distribution in the high mass region. Source: Ref.~\cite{CMS:2020zzv}.}
\label{fig:L6072}
\end{figure}

Later in 2020 the CMS collaboration investigated the $\Lambda_b^0 \pi^+ \pi^-$ invariant mass spectrum in the mass range up to 6400~MeV~\cite{CMS:2020zzv}. They observed a narrow peak with a significance over $5\sigma$. Its mass is close to 6150~MeV, which is consistent with the superposition of the $\Lambda_b(6146)^0$ and $\Lambda_b(6152)^0$ baryons. Their masses were measured to be:
\begin{eqnarray}
\Lambda_b(6146)^0 &:& M = 6146.5 \pm 1.9 \pm 0.8 \pm 0.2~{\rm MeV} \, ,
\\ \Lambda_b(6152)^0 &:& M = 6152.7 \pm 1.1 \pm 0.4 \pm 0.2~{\rm MeV} \, .
\end{eqnarray}
At the same time they observed a broad excess of events in the $\Lambda_b^0 \pi^+ \pi^-$ mass distribution in the region of $6040$-$6100$~MeV, as depicted in Fig.~\ref{fig:L6072}. They fitted it with a single Breit-Wigner function, and measured its mass and width to be $6073 \pm 5$~MeV and $55 \pm 11$~MeV. One month later, the LHCb collaboration confirmed the above signal, and measured its mass and width to be~\cite{LHCb:2020lzx}:
\begin{eqnarray}
\Lambda_b(6072)^0 &:& M = 6072.3 \pm 2.9 \pm 0.6 \pm 0.2~{\rm MeV} \, ,
\\ \nonumber && \Gamma = 72 \pm 11 \pm 2~{\rm MeV} \, .
\end{eqnarray}

The $\Lambda_b(6146)$ and $\Lambda_b(6152)$ were suggested by LHCb to form the $\Lambda_b(1D)$ doublet $[\mathbf{\bar 3}_F, 2, 0, \lambda\lambda]$ with the quantum numbers $J^P = 3/2^+$ and $5/2^+$~\cite{LHCb:2019soc,Capstick:1985xss,Chen:2014nyo}. This is supported by Ref.~\cite{Jia:2019bkr} through the Regge approach, where the masses of the $\Lambda_b(1D)$ doublet were calculated to be $6144.8$~MeV and $6152.2$~MeV. Various theoretical studies in Refs.~\cite{Faustov:2020gun,Azizi:2020tgh,Oudichhya:2021yln,Kim:2021ywp} also support this explanation.

Later in Ref.~\cite{Liang:2019aag} the authors studied the strong decays of the $\Lambda_b(2S)$, $\Lambda_b(1D)$, $\Sigma_b(2S)$, and $\Sigma_b(1P)$ bottom baryons within the quark pair creation model. Their results suggest that the $\Lambda_b(6146)$ and $\Lambda_b(6152)$ can be reasonably classified into the $\Lambda_b(1D)$ doublet, while other assignments are disfavored. They evaluated the partial branching ratio of the $\Sigma_b\pi$ and $\Sigma_b^*\pi$ channels to be about 6 for the $J^P = 3/2^+$ state, and to be about $6 \times 10^{-3}$ for the $J^P=5/2^+$ state.

Together with the experimental decay modes observed by LHCb~\cite{LHCb:2019soc} and depicted in Fig.~\ref{fig:L6146b}, the results of Ref.~\cite{Liang:2019aag} favor the interpretations of the $\Lambda_b(6146)$ and $\Lambda_b(6152)$ as the $\Lambda_b(1D)$ states of $J^P = 5/2^+$ and $J^P = 3/2^+$, respectively. However, such two assignments face a serious problem of the mass reverse that the $\Lambda_b(1D)$ state of $J^P=5/2^+$ probably has a mass larger than its $J^P=3/2^+$ partner, as shown in Table~\ref{sec3:bottom}. This problem was further investigated in Refs.~\cite{Wang:2019uaj,Chen:2019ywy} using two similar quark models, and the same conclusion was obtained, which needs to be clarified in the future.

\begin{figure}[hbtp]
\begin{center}
\includegraphics[width=0.5\textwidth]{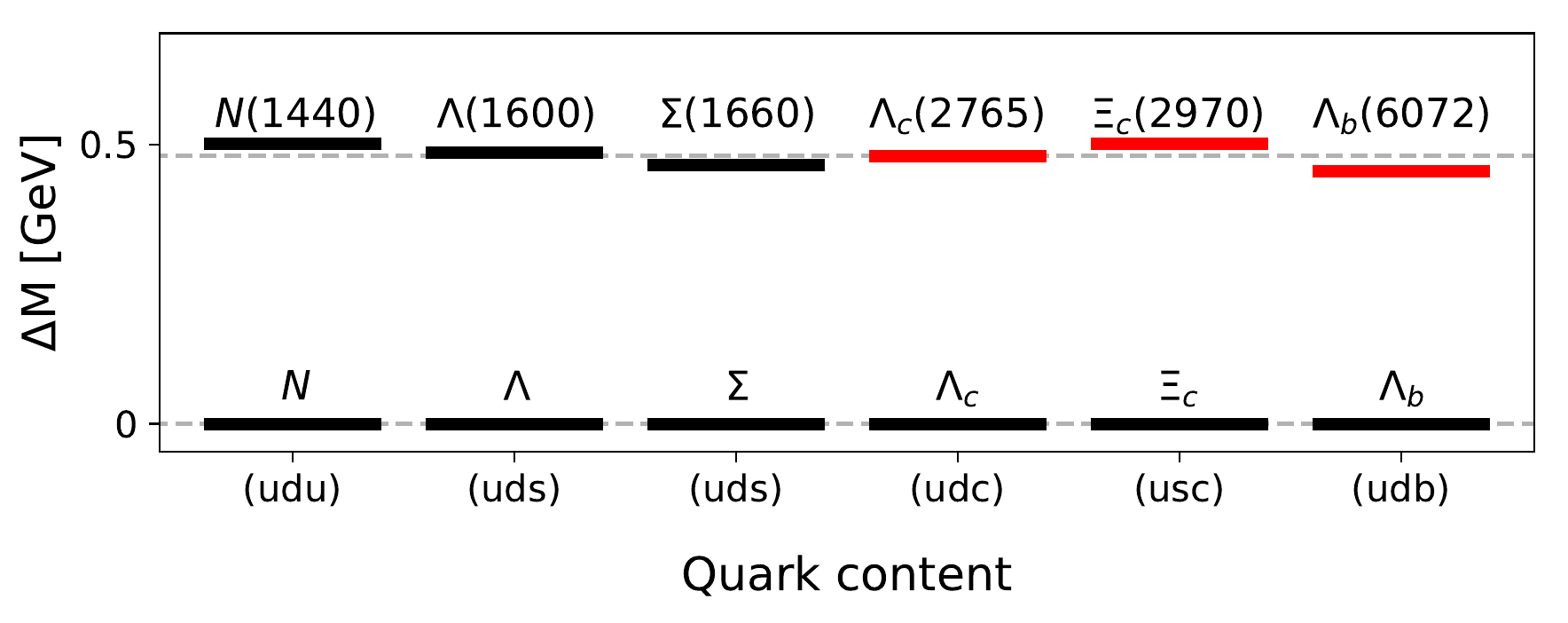}
\end{center}
\caption{Excitation energies of several $2S$ baryons and possible candidates with various flavor contents. The states with red bars are with undetermined spin and parity. Source: Ref.~\cite{Arifi:2020yfp}.}
\label{fig:roper}
\end{figure}

The $\Lambda_b(6072)$ was suggested by LHCb to be the $\Lambda_b(2S)$ state of $J^P = 1/2^+$~\cite{LHCb:2020lzx,Capstick:1985xss,Roberts:2007ni,Ebert:2011kk,Yamaguchi:2014era,Chen:2018vuc}. This assignment is supported by Refs.~\cite{Arifi:2020yfp,Suenaga:2022ajn}, where the authors studied its sequential three-body decay process into $\Lambda_b^0 \pi^+ \pi^-$ through the $\Sigma_b^{(*)-}$ and $\Sigma_b^{(*)+}$. The authors further proposed several common features of the $2S$ baryons, {\it e.g.}, their excitation energies are all around 500 MeV, which are rather flavor independent as shown in Fig.~\ref{fig:roper}. We refer to Ref.~\cite{Chen:2021eyk} for more discussions on the mass gaps existing in heavy baryons. Theoretical studies of Ref.~\cite{Azizi:2020ljx} based on QCD sum rules support this assignment.

Later in Ref.~\cite{Liang:2020kvn} the authors tentatively assigned the $\Lambda_b(6072)$ as the $\Lambda_b(2S)$, $\Sigma_b(1P)$, and $\Lambda_b(1P)$ states, and studied its decay behavior within the $^3P_0$ model. Their results support the assignment of the $\Lambda_b(6072)$ as the $\Lambda_b(1P)$ state with the $\rho$-mode excitation. The masses of the $P$-wave $\Sigma_b$ states are expected to be close to the measured mass of $\Lambda_b(6072)$, as shown in Tables~\ref{sec3:bottom} and \ref{sec3:bottomsumrule}. This possibility was discussed in Ref.~\cite{Xiao:2020gjo}, where the authors studied the strong decays of the $\Sigma_b(1P)$ states using the chiral quark model. Their results suggest that the broad structure ``$\Lambda_b(6072)$'' may arise from the overlapping of the two $P$-wave $\Sigma_b$ states of the $\lambda$-mode, with the quantum numbers $J^P=1/2^-$ and $3/2^-$.

\subsubsection{$\Xi_b(6327)$ and $\Xi_b(6333)$.}
\label{sec3.1.3}

\begin{figure}[hbtp]
\begin{center}
\includegraphics[width=0.5\textwidth]{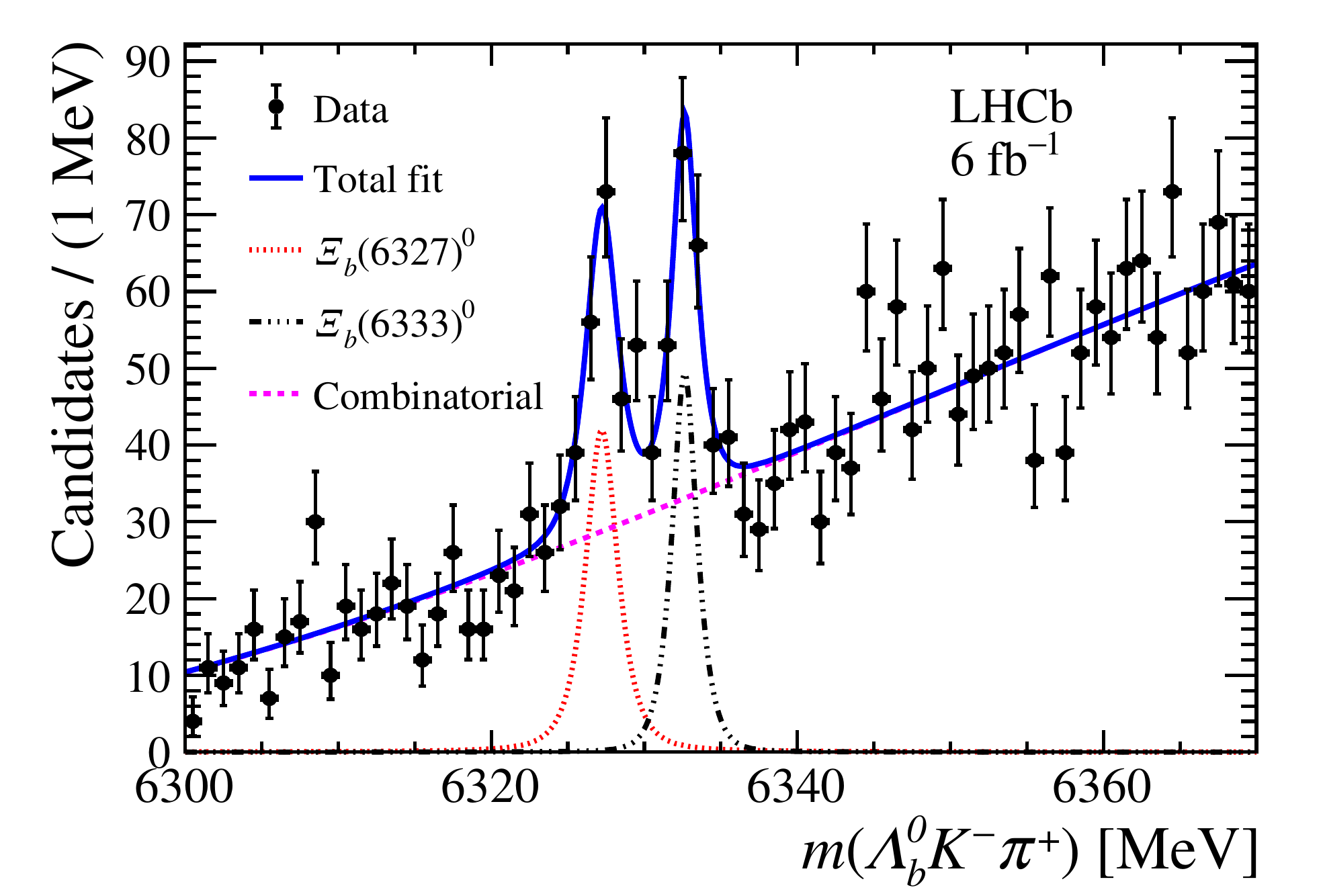}
\end{center}
\caption{The $\Lambda^0_b K^-\pi^+$ invariant mass spectrum. Source: Ref.~\cite{LHCb:2021ssn}.}
\label{fig:Xi6327}
\end{figure}

Very recently the LHCb collaboration reported their observation of two narrow $\Xi^0_b$ baryons in the $\Lambda^0_b K^-\pi^+$ mass spectrum~\cite{LHCb:2021ssn}, as depicted in Fig.~\ref{fig:Xi6327}. Their masses were measured to be
\begin{eqnarray}
\Xi_b(6327)^0 &:& M = 6327.28 ^{+0.23}_{-0.21} \pm 0.08 \pm 0.24{\rm~MeV} \, ,
\\
\Xi_b(6333)^0 &:& M = 6332.69 ^{+0.17}_{-0.18} \pm 0.03 \pm 0.22{\rm~MeV} \, ,
\end{eqnarray}
with the mass splitting
\begin{equation}
\Delta M = 5.41 ^{+0.26}_{-0.27} \pm 0.06{\rm~MeV}\, .
\end{equation}
Their corresponding widths were observed to be consistent with zero, and upper limits at 90\% (95\%) C.~L. were set to be
\begin{eqnarray}
\Xi_b(6327)^0 &:& \Gamma < 2.20~(2.56){\rm~MeV} \, ,
\\
\Xi_b(6333)^0 &:& \Gamma < 1.55~(1.85){\rm~MeV} \, .
\end{eqnarray}

Based on the theoretical studies of Refs.~\cite{Chen:2019ywy,Yao:2018jmc}, the $\Xi_b(6327)$ and $\Xi_b(6333)$ were suggested by LHCb to be the $\Xi_b(1D)$ states of $J^P = 3/2^+$ and $5/2^+$, belonging to the $[\mathbf{\bar 3}_F, 2, 0, \lambda\lambda]$ doublet. Hence, they may be the partner states of the $\Lambda_b(6146)$ and $\Lambda_b(6152)$. This explanation is supported by Refs.~\cite{Bijker:2020tns,Wang:2022zqv} within the chiral quark model and the $^3P_0$ model.

\subsubsection{$\Sigma_b(6097)$.}
\label{sec3.1.4}

\begin{figure}[hbtp]
\begin{center}
\subfigure[]{\includegraphics[width=0.4\textwidth]{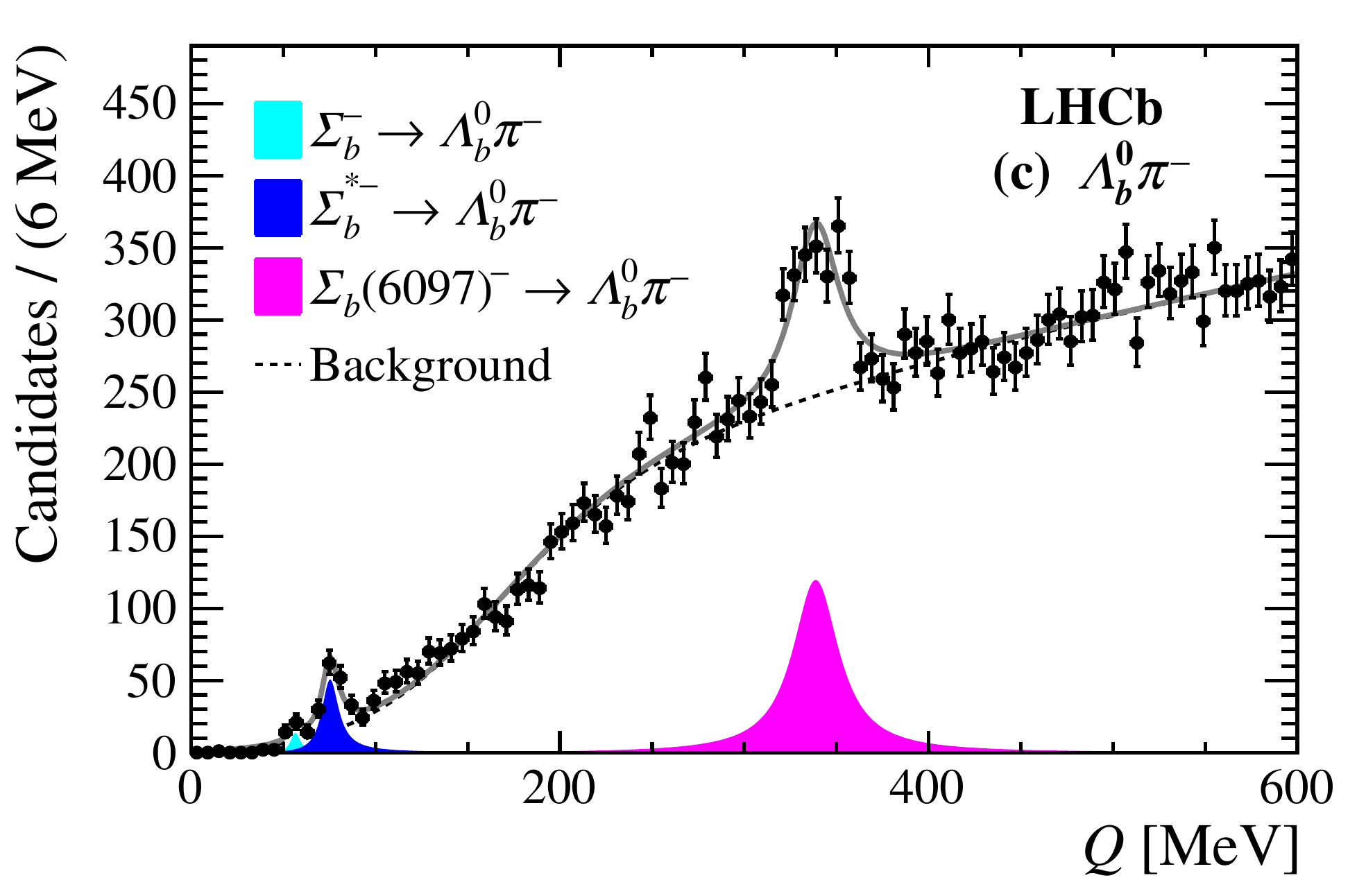}}
~~~~~
\subfigure[]{\includegraphics[width=0.4\textwidth]{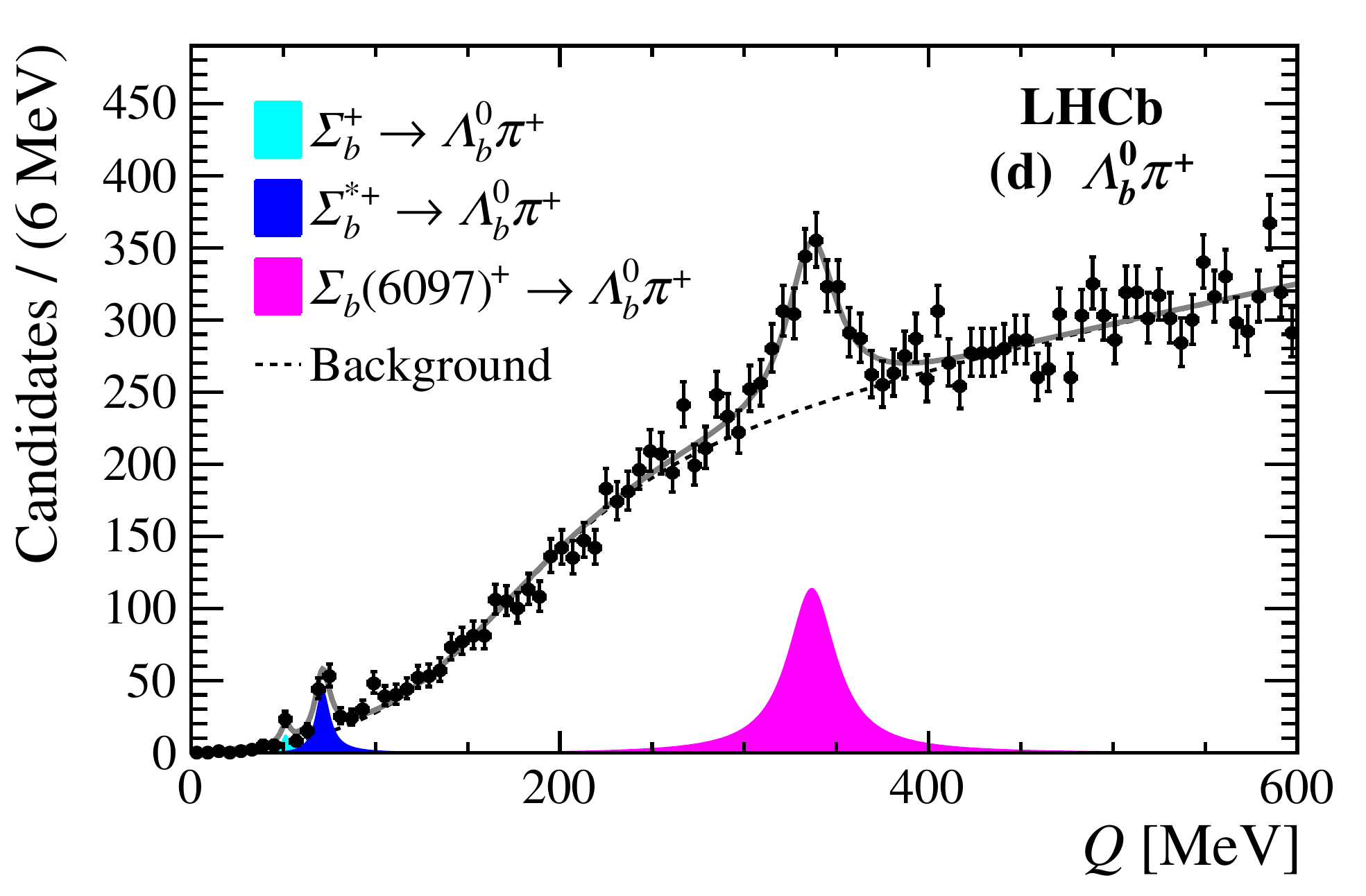}}
\end{center}
\caption{Mass distribution for selected (a) $\Lambda^0_b \pi^+$ and (b) $\Lambda^0_b \pi^-$ candidates. Source: Ref.~\cite{LHCb:2018haf}.}
\label{fig:S6097}
\end{figure}

In 2018 the LHCb collaboration reported their observation of two excited $\Sigma_b^\pm$ resonances in the $\Lambda^0_b \pi^+$ and $\Lambda^0_b \pi^-$ systems~\cite{LHCb:2018haf}, as depicted in Fig.~\ref{fig:S6097}. Their parameters were determined to be
\begin{eqnarray}
\Sigma_b(6097)^+ &:& M = 6095.8 \pm 1.7 \pm 0.4{\rm~MeV} \, ,
\\ \nonumber && \Gamma = 31.0 \pm 5.5 \pm 0.7{\rm~MeV} \, ,
\\ \Sigma_b(6097)^- &:& M = 6098.0 \pm 1.7 \pm 0.5{\rm~MeV} \, ,
\\ \nonumber && \Gamma = 28.9 \pm 4.2 \pm 0.9{\rm~MeV} \, ,
\end{eqnarray}
with the mass splitting
\begin{equation}
\Delta M = - 2.2 \pm 2.4 \pm 0.3{\rm~MeV}\, .
\end{equation}

\begin{figure}[hbtp]
\begin{center}
\includegraphics[width=0.5\textwidth]{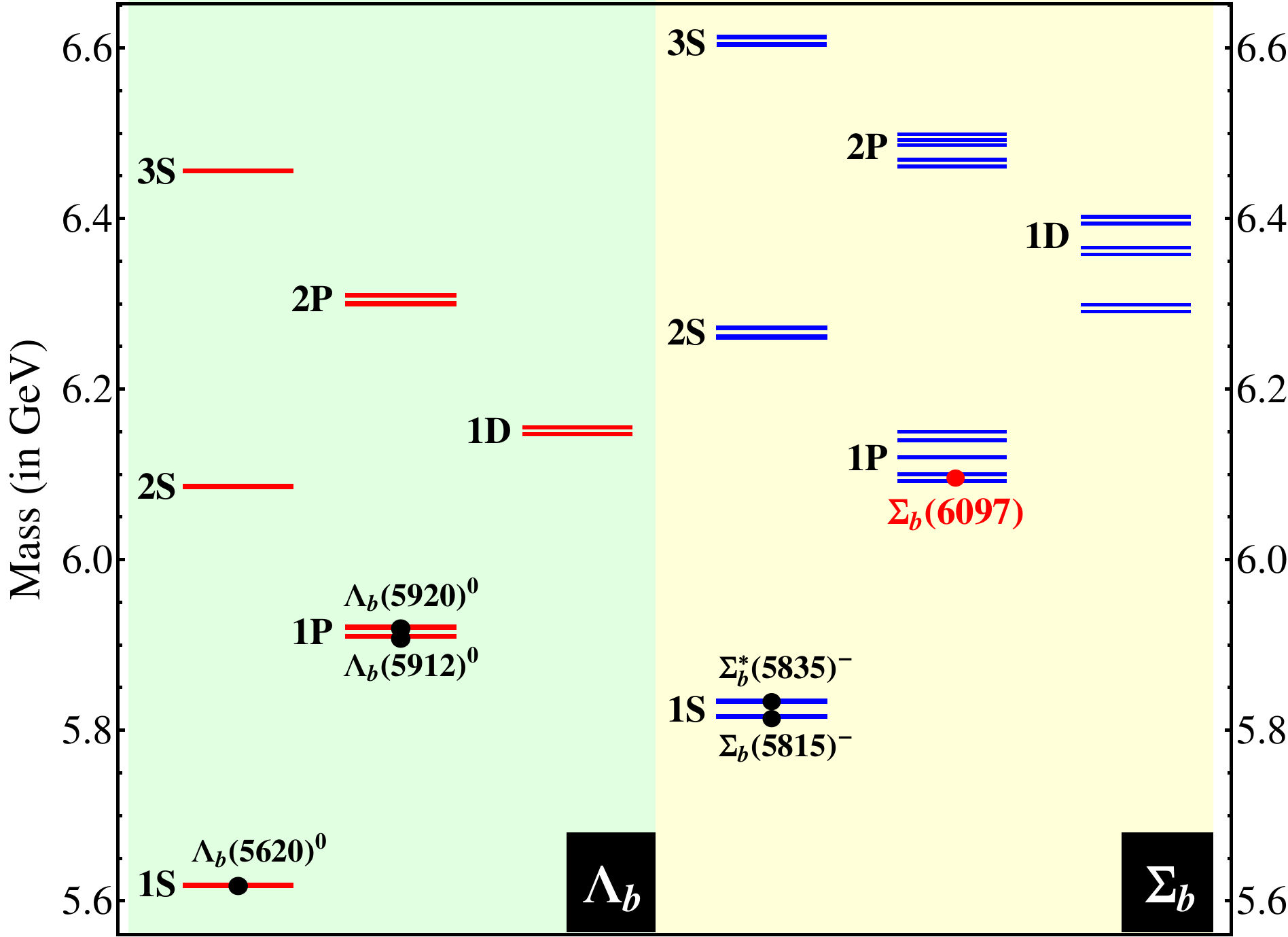}
\end{center}
\caption{Mass spectrum of the $\Lambda_b$ (left) and $\Sigma_b$ baryons (right) calculated in the nonrelativistic quark potential model. Source: Ref.~\cite{Chen:2018vuc}.}
\label{fig:Sigmamass}
\end{figure}

In Ref.~\cite{Chen:2018vuc} the authors studied the mass spectrum and strong decay properties of the $2S$ and $1P$ bottom baryons. Their masses were calculated in the nonrelativistic quark potential model, as shown in Fig.~\ref{fig:Sigmamass}. Their strong decay properties were calculated within the $^3P_0$ model. The obtained results support the assignment of the $\Sigma_b(6097)$ as the $\Sigma_b(1P)$ state with either $J^P = 3/2^-$ or $5/2^-$. This explanation is supported by Refs.~\cite{Wang:2018fjm,Yang:2018lzg,Wang:2021bmz} based on the chiral quark model and the $^3P_0$ model.

In Refs.~\cite{Aliev:2018lcs,Azizi:2020azq} the authors applied the method of QCD sum rules to study the strong decay properties of the $2S$ and $1P$ bottom baryons. Their results support the interpretation of the $\Sigma_b(6097)$ as the $\Sigma_b(1P)$ state with the quantum number $J^P = 3/2^-$. This assignment is supported by Refs.~\cite{Cui:2019dzj,Yang:2020zrh} based on the method of light-cone sum rules.

\subsubsection{$\Xi_b(6227)$.}
\label{sec3.1.5}

\begin{figure}[hbtp]
\begin{center}
\subfigure[]{\includegraphics[width=0.3\textwidth]{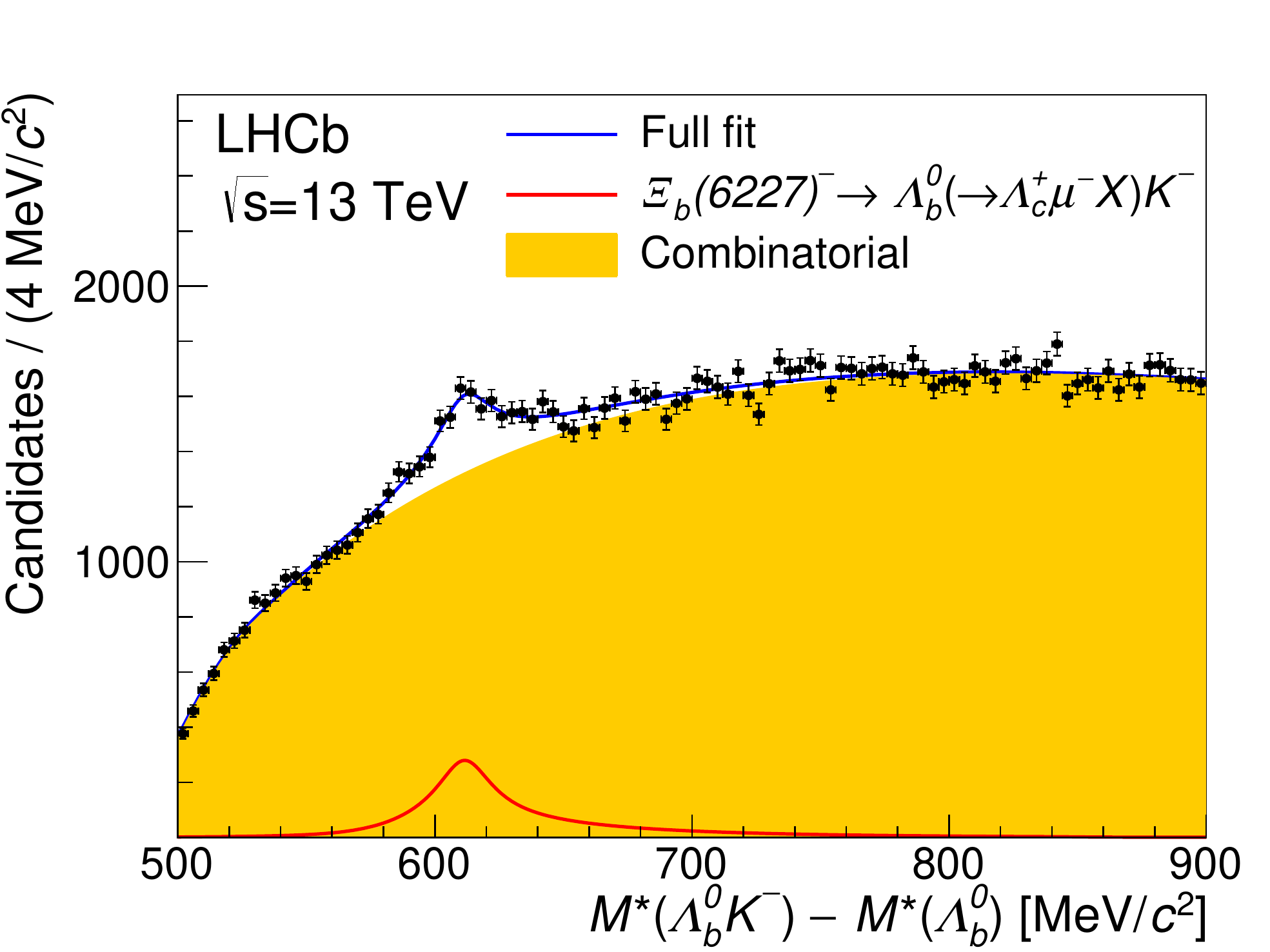}}
~~
\subfigure[]{\includegraphics[width=0.3\textwidth]{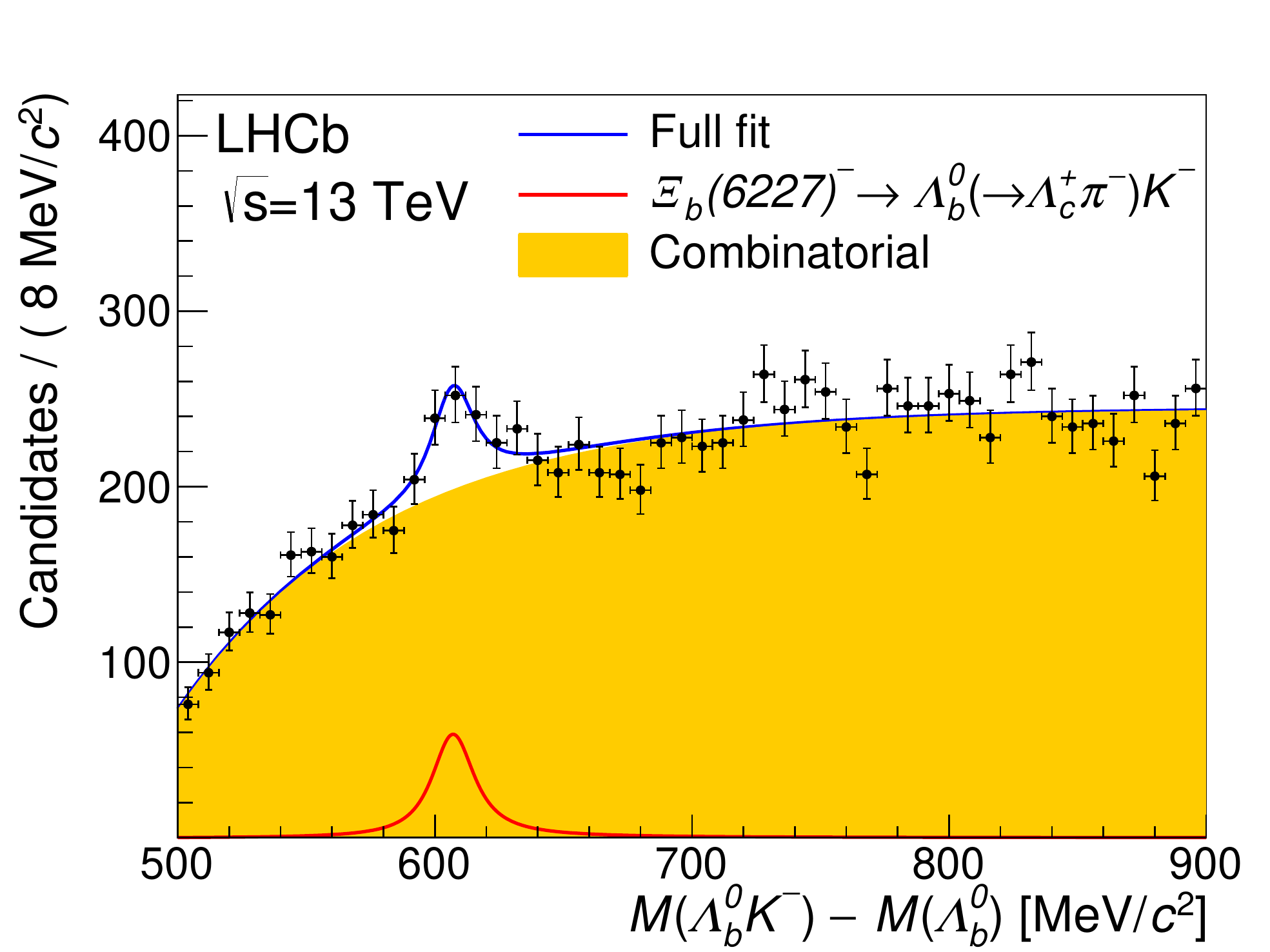}}
~~
\subfigure[]{\includegraphics[width=0.3\textwidth]{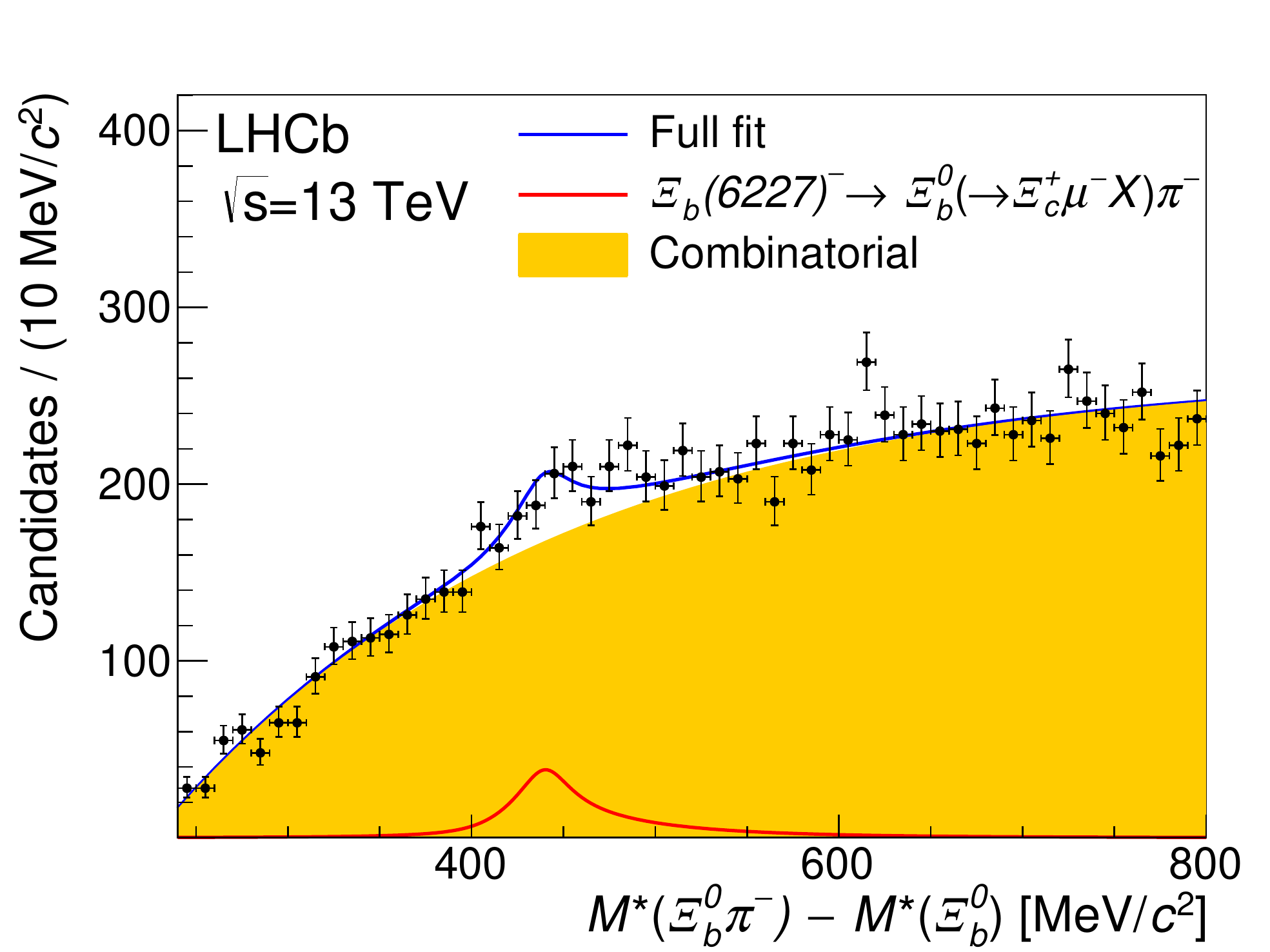}}
\end{center}
\caption{Spectra of mass differences for the $\Xi_b(6227)^-$ candidates, reconstructed in the final states (a) $\Lambda^0_b K^-$ with $\Lambda^0_b \to \Lambda^+_c \pi^-$, (b) $\Lambda^0_b K^-$ with $\Lambda^0_b \to \Lambda^+_c \mu^- X$, and (c) $\Xi^0_b \pi^-$ with $\Xi^0_b \to \Xi^+_c \mu^- X$. Source: Ref.~\cite{LHCb:2018vuc}.}
\label{fig:Xi6227}
\end{figure}

In 2018 the LHCb collaboration reported the observation of a peak in both the $\Lambda^0_b K^-$ and $\Xi^0_b \pi^-$ invariant mass spectra~\cite{LHCb:2018vuc}, as depicted in Fig.~\ref{fig:Xi6227}. It was named as $\Xi_b(6227)^-$, whose mass and width were measured to be
\begin{eqnarray}
\Xi_b(6227)^- &:& M = 6226.9 \pm 2.0 \pm 0.3 \pm 0.2{\rm~MeV} \, ,
\\ \nonumber && \Gamma = 18.1 \pm 5.4 \pm 1.8{\rm~MeV} \, .
\end{eqnarray}
LHCb also measured its relative production rates to the $\Lambda^0_b K^-$ and $\Xi^0_b \pi^-$ final states. Using the relation between the two decay constants $f_{\Xi^0_b} \approx 0.1 f_{\Lambda^0_b}$~\cite{Voloshin:2015xxa,Hsiao:2015txa,Jiang:2018iqa}, they further derived the ratio
\begin{equation}
{\mathcal{B}\left( \Xi_b(6227)^- \to \Lambda^0_b K^- \right) \over \mathcal{B}\left( \Xi_b(6227)^- \to \Xi^0_b \pi^- \right)} \approx 1.0 \pm 0.5 \, ,
\end{equation}
with sizable uncertainty due to various theoretical assumptions and experimental inputs.

Later in 2021 the LHCb collaboration reported the observation of an excited $\Xi^0_b$ resonance decaying to the $\Xi^-_b \pi^+$ final state~\cite{LHCb:2020xpu}. It is the partner state of the $\Xi_b(6227)^-$, and so labeled as $\Xi_b(6227)^0$. Its mass and width were measured to be
\begin{eqnarray}
\Xi_b(6227)^0 &:& M = 6227.1 ^{+1.4}_{-1.5} \pm 0.5{\rm~MeV} \, ,
\\ \nonumber && \Gamma = 18.6 ^{+5.0}_{-4.1} \pm 1.4{\rm~MeV} \, .
\end{eqnarray}

\begin{figure}[hbtp]
\begin{center}
\includegraphics[width=0.5\textwidth]{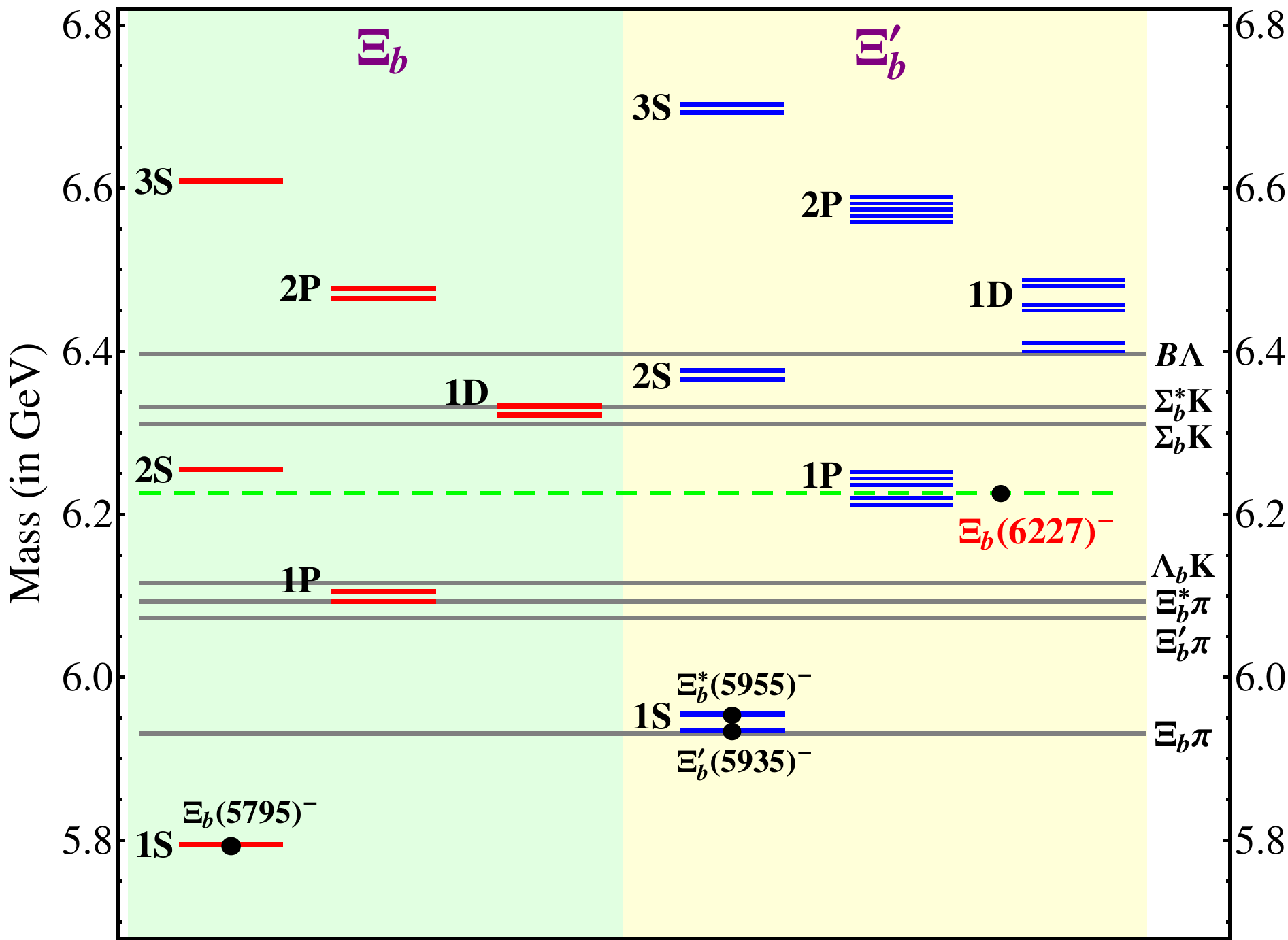}
\end{center}
\caption{Mass spectrum of bottom-strange baryons calculated in the nonrelativistic quark potential model. The red solid lines (left) correspond to the predicted masses of the flavor $\mathbf{\bar 3}_F$ $\Xi_b$ states, and the blue solid lines (right) correspond to the flavor $\mathbf{6}_F$ $\Xi^\prime_b$ states. Source: Ref.~\cite{Chen:2018orb}.}
\label{fig:Ximass}
\end{figure}

In Ref.~\cite{Chen:2018orb} the authors carried out a phenomenological analysis of the $2S$ and $1P$ bottom-strange baryons. They calculated their masses, as shown in Fig.~\ref{fig:Ximass}. They also studied their two-body OZI-allowed strong decay behaviors. Their results support the interpretation of the $\Xi_b(6227)$ as the flavor $\mathbf{6}_F$ $\Xi_b^\prime(1P)$ state with either $J^P = 3/2^-$ or $5/2^-$. Hence, it may be the partner state of the $\Sigma_b(6097)$. This assignment is supported by Ref.~\cite{Wang:2018fjm} based on the chiral quark model, but not supported by Ref.~\cite{Kawakami:2019hpp} within a three-flavor chiral model.

In Refs.~\cite{Aliev:2018lcs,Azizi:2020azq} the authors employed the QCD sum rule technique to study masses and decay properties of the $2S$ and $1P$ bottom-strange baryons. Their results favor the interpretation of the $\Xi_b(6227)$ as the flavor $\mathbf{6}_F$ $\Xi_b^\prime(1P)$ state with the quantum number $J^P = 3/2^-$. This assignment is supported by Refs.~\cite{Cui:2019dzj,Yang:2020zrh} based on the light-cone sum rule method.

Besides the above conventional explanations, there exist some exotic explanations for the $\Xi_b(6227)$. In Ref.~\cite{Huang:2018bed} the authors assumed the $\Xi_b(6227)$ to be a pure $\Sigma_b\bar K$ molecular state, and studied its strong decays into the $\Lambda_b\bar K$, $\Xi_b\pi$, and $\Xi^\prime_b\pi$ channels with different spin-parity assignments. They found that both the total decay width and the ratio of the partial decay widths into $\Lambda_b\bar K$ and $\Xi_b\pi$ can be reproduced with the assumption that the $\Xi_b(6227)$ is an $S$-wave $\Sigma_b\bar K$ bound state of $J^P = 1/2^-$. This assumption is supported by Ref.~\cite{Nieves:2019jhp} using a similar coupled-channel unitarized model.

In Ref.~\cite{Yu:2018yxl} the authors investigated the $\Xi_b(6227)$ as a dynamically generated state from the meson-baryon interaction in coupled channels. They applied an extension of the local hidden-gauge approach in the Bethe-Salpeter equation to obtain two poles at 6220.30 MeV with a width 25.20 MeV and at 6240.21 MeV with a width 29.30 MeV, in the $J^P=1/2^-$ and $3/2^-$ sectors respectively. These two poles can both explain the $\Xi_b(6227)$ as a dynamically generated molecular state.

\subsubsection{$\Omega_b(6316)$, $\Omega_b(6330)$, $\Omega_b(6340)$, and $\Omega_b(6350)$.}
\label{sec3.1.6}

\begin{figure}[hbtp]
\begin{center}
\subfigure[]{\includegraphics[width=0.8\textwidth]{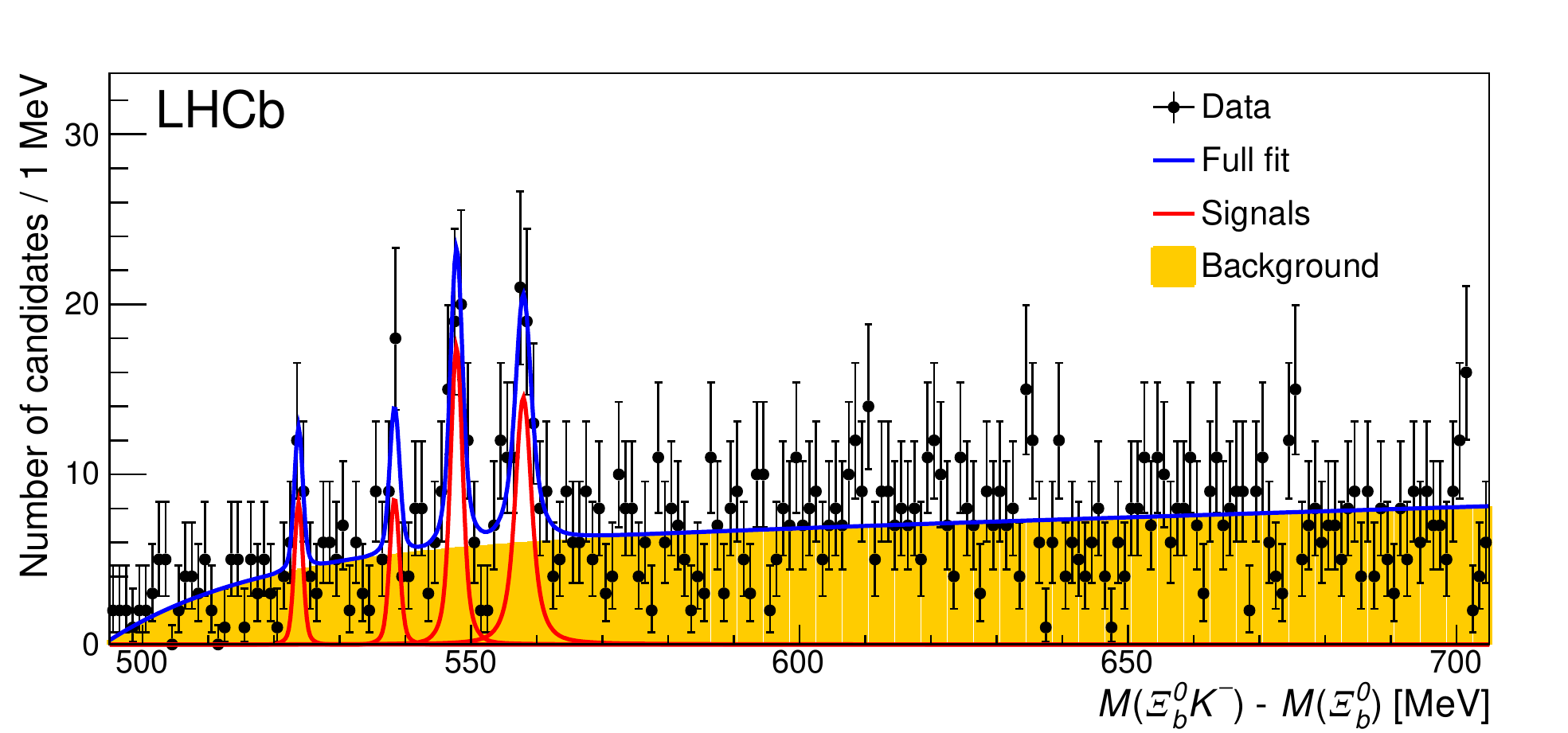}}
\subfigure[]{\includegraphics[width=0.8\textwidth]{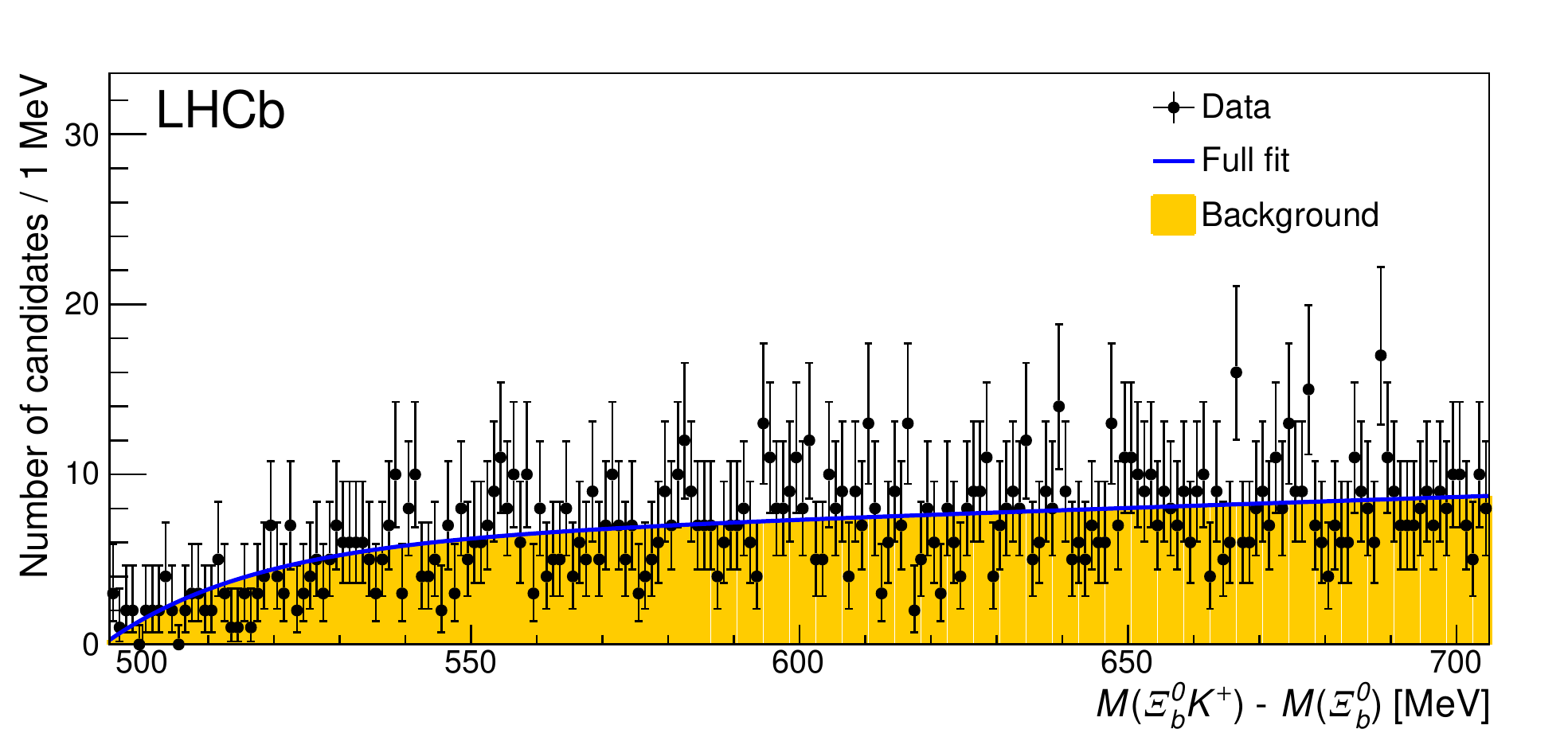}}
\end{center}
\caption{Distribution of the mass difference for (a) right-sign $\Xi^0_b K^-$ candidates and (b) wrong-sign $\Xi^0_b K^+$ candidates. Source: Ref.~\cite{LHCb:2020tqd}.}
\label{fig:O6316}
\end{figure}

In 2020 the LHCb collaboration reported the observation of four narrow peaks in the $\Xi^0_b K^-$ invariant mass spectrum~\cite{LHCb:2020tqd}. As depicted in Fig.~\ref{fig:O6316}, these four peaks were seen in the right-sign spectrum of the $\Xi^0_b K^-$ candidates, whereas no significant peak was seen in the corresponding wrong-sign $\Xi^0_b K^+$ distribution. Their masses and widths (or 90\% C.~L. upper limits on the natural widths) were measured to be:
\begin{eqnarray}
\Omega_b(6316)^- &:& M = 6315.64 \pm 0.31 \pm 0.07 \pm 0.50{\rm~MeV} \, ,
\\ \nonumber      && \Gamma < 2.8{\rm~MeV} \, ;
\\ \Omega_b(6330)^- &:& M = 6330.30 \pm 0.28 \pm 0.07 \pm 0.50{\rm~MeV} \, ,
\\ \nonumber      && \Gamma < 3.1{\rm~MeV} \, ;
\\ \Omega_b(6340)^- &:& M = 6339.71 \pm 0.26 \pm 0.05 \pm 0.50{\rm~MeV} \, ,
\\ \nonumber      && \Gamma < 1.5{\rm~MeV} \, ;
\\ \Omega_b(6350)^- &:& M = 6349.88 \pm 0.35 \pm 0.05 \pm 0.50{\rm~MeV} \, ,
\\ \nonumber      && \Gamma = 1.4 ^{+1.0}_{-0.8} \pm 0.1 {\rm~MeV} < 2.8{\rm~MeV} \, .
\end{eqnarray}

The four peaks observed by LHCb are consistent with the expectations for the excited $\Omega_b$ resonances. However, there do exist many possible interpretations for them~\cite{Xiao:2020oif,Mutuk:2020rzm,Karliner:2020fqe,Xu:2020ofp,Liang:2020dxr,Kakadiya:2022zvy}, since there can be as many as five $\lambda$-mode and two $\rho$-mode $\Omega_b(1P)$ states, as shown in Fig.~\ref{fig:categorization}. We briefly review two possible explanations as follows.

In Ref.~\cite{Liang:2020hbo} the authors investigated the strong decays of the low-lying $\Omega_b$ states within the $^3P_0$ model, and their results support the interpretations of the $\Omega_b(6316)$, $\Omega_b(6330)$, $\Omega_b(6340)$, and $\Omega_b(6350)$ as the four $\lambda$-mode $\Omega_b(1P)$ states of $J^P = 1/2^-$, $3/2^-$, $3/2^-$, and $5/2^-$. There are altogether five $\lambda$-mode excitations, and the remaining one of $J^P = 1/2^-$ was predicted to have a width too broad to be observed in experiments. The authors also investigated the $\Omega_b(2S)$ and $\Omega_b(1D)$ states, which were found to have relatively narrow total widths and mainly decay into the $\Xi_b \bar K$, $\Xi^\prime_b\bar K$, and $\Xi^*_b \bar K$ final states.

In Refs.~\cite{Chen:2020mpy,Yang:2020zrh} the authors further took the two $\rho$-mode excitations into account, and studied the mass spectrum and decay properties of the $P$-wave $\Omega_b$ baryons using the methods of QCD sum rules and light-cone sum rules within the framework of heavy quark effective theory. The results are summarized in Table~\ref{sec3:bottomsumrule}: the $\Omega_b(6350)$ is a $P$-wave $\Omega_b$ baryon of $J^P = 3/2^-$ with the $\lambda$-mode excitation, which has a $J^P = 5/2^-$ partner with the mass $10.0^{+4.6}_{-3.8}$~MeV larger; the $\Omega_b(6330)$ and $\Omega_b(6340)$ are the partner states both with the $\lambda$-mode excitation, which have $J^P = 1/2^-$ and $3/2^-$ respectively; the $\Omega_b(6316)$ is a $P$-wave $\Omega_b$ baryon of either $J^P = 1/2^-$ or $3/2^-$ with the $\rho$-mode excitation, which can be further separated into two states with the mass splitting $2.3^{+1.0}_{-0.9}$~MeV.

Especially, it is straightforward to understand within the framework of heavy quark effective theory (HQET) why the $\Omega_b(6316)$, $\Omega_b(6330)$, and $\Omega_b(6340)$ have natural widths consistent with zero but they can still be observed in the $\Xi_b^0 K^-$ mass spectrum~\cite{LHCb:2020tqd,Chen:2020mpy} (see Table~\ref{sec3:bottomsumrule}). The HQET is an effective theory, so the three $J^P=1/2^-$ $\Omega_b(1P)$ states can mix together and the three $J^P=3/2^-$ $\Omega_b(1P)$ states can also mix together, making it possible to observe all of them in the $\Xi_b^0 K^-$ mass spectrum. Since the HQET works quite well for the bottom system, such a mixing is not large and some of them still have very narrow widths.

\subsection{Singly charmed baryons}
\label{sec3.2}

Because the HQET works better for the singly bottom baryons than for the singly charmed baryons, we shall pay more attention to the mixing effect when reviewing the singly charmed baryons in this subsection.

All the $1S$ charmed baryons are well established in experiments~\cite{pdg}, including $(\Lambda_c^+, \Xi_c^{+}, \Xi_c^{0})$, $(\Sigma_c^{++}, \Sigma_c^{+}, \Sigma_c^{0}, \Xi_c^{\prime+}, \Xi_c^{\prime0}, \Omega_c^0)$, and $(\Sigma_c^{*++}, \Sigma_c^{*+}, \Sigma_c^{*0}, \Xi_c^{*+}, \Xi_c^{*0}, \Omega_c^{*0})$. They complete the two $S$-wave charmed baryon multiplets $[\mathbf{\bar 3}_F, 0, 0]$ and $[\mathbf{6}_F, 1, 1]$, as shown in Fig.~\ref{fig:categorization}. Higher charmed baryons are much more complicated, including:
\begin{itemize}

\item $\Lambda_c(2595)$~\cite{CLEO:1994oxm}, $\Lambda_c(2625)$~\cite{ARGUS:1993vtm}, $\Lambda_c(2765)$~\cite{CLEO:2000mbh}, $\Lambda_c(2860)$~\cite{LHCb:2017jym}, $\Lambda_c(2880)$~\cite{CLEO:2000mbh}, and $\Lambda_c(2940)$~\cite{BaBar:2006itc};

\item $\Sigma_c(2800)$~\cite{Belle:2004zjl};

\item $\Xi_c(2790)$~\cite{CLEO:2000ibb}, $\Xi_c(2815)$~\cite{CLEO:1999msf}, $\Xi_c(2923)$~\cite{BaBar:2007xtc,LHCb:2020iby}, $\Xi_c(2939)$~\cite{BaBar:2007xtc,LHCb:2020iby}, $\Xi_c(2965)$~\cite{LHCb:2020iby}, $\Xi_c(2970)$~\cite{Belle:2006edu,Belle:2020tom}, $\Xi_c(3055)$~\cite{BaBar:2007zjt}, $\Xi_c(3080)$~\cite{Belle:2006edu}, and $\Xi_c(3123)$~\cite{BaBar:2007zjt};

\item $\Omega_c(3000)$~\cite{LHCb:2017uwr}, $\Omega_c(3050)$~\cite{LHCb:2017uwr}, $\Omega_c(3066)$~\cite{LHCb:2017uwr}, $\Omega_c(3090)$~\cite{LHCb:2017uwr}, and $\Omega_c(3119)$~\cite{LHCb:2017uwr}.

\end{itemize}
We refer to the recent comprehensive review of the charmed baryons~\cite{Cheng:2021qpd} for more discussions. Some of them were reviewed in our previous paper~\cite{Chen:2016spr}, and in this subsection we shall review the rest of them, including the $\Lambda_c(2860)$, $\Xi_c(2923)$, $\Xi_c(2939)$, $\Xi_c(2965)$, $\Omega_c(3000)$, $\Omega_c(3050)$, $\Omega_c(3066)$, $\Omega_c(3090)$, and $\Omega_c(3119)$.

\begin{table*}[htb]
\tiny
\caption{Theoretical results of the charmed baryons obtained from various quark models~\cite{Ebert:2011kk,Roberts:2007ni,Yoshida:2015tia,Chen:2016iyi}, including the $1S$ and $2S$ states as well as the $1P$ and $1D$ states of the $\lambda$-mode. Possible experimental candidates are given for comparison, with uncertainties the sum of their statistic and systematical uncertainties. Masses are in units of MeV.}
\centering
\begin{tabular}{ c c c c c c c c }
\toprule[1pt]
& $J^P~(nL)$ & Exp. Mass~\cite{pdg} & Exp. Width~\cite{pdg} & RQM~\cite{Ebert:2011kk} & NQM~\cite{Roberts:2007ni} & NQM~\cite{Yoshida:2015tia} & NQM~\cite{Chen:2016iyi}
\\ \midrule[1pt]
$\Lambda_c$             & $1/2^+~(1S)$     & $2286.46\pm0.14$                                  & $\sim10^{-13}$~s                   & 2286  & 2268  & 2285   & 2286  \\
$\Xi_c$                 & $1/2^+~(1S)$     & $2467.94^{+0.17}_{-0.20}$                         & $\sim10^{-13}$~s                   & 2476  & 2466  & --     & 2470  \\
$\Sigma_c$              & $1/2^+~(1S)$     & $2452.9\pm0.4$                                    & $<4.6$                             & 2443  & 2455  & 2460   & 2456  \\
$\Sigma_c^{*}$          & $3/2^+~(1S)$     & $2517.5\pm2.3$                                    & $<17$                              & 2519  & 2519  & 2523   & 2515  \\
$\Xi_c^{\prime}$        & $1/2^+~(1S)$     & $2578.4\pm0.5$                                    & ?                                  & 2579  & 2594  & --     & 2579  \\
$\Xi_c^{*}$             & $3/2^+~(1S)$     & $2645.56^{+0.24}_{-0.30}$                         & $2.14 \pm 0.19$                    & 2649  & 2649  & --     & 2649  \\
$\Omega_c$              & $1/2^+~(1S)$     & $2695.2\pm1.7$                                    & $\sim10^{-13}$~s                   & 2698  & 2718  & 2731   & --    \\
$\Omega_c^{*}$          & $3/2^+~(1S)$     & $2765.9\pm2.0$                                    & ?                                  & 2768  & 2776  & 2779   & --    \\
\hline
$\Lambda_c$             & $1/2^-~(1P)$     & $\Lambda_c(2595):2592.25\pm0.28$                  & $2.59 \pm 0.56$                    & 2598  & 2625  & 2628   & 2614  \\
$\Lambda_c$             & $3/2^-~(1P)$     & $\Lambda_c(2625):2628.11\pm0.19$                  & $<0.97$                            & 2627  & 2636  & 2630   & 2639  \\
$\Xi_c$                 & $1/2^-~(1P)$     & $\Xi_c(2790):2792.4\pm0.5$                        & $8.9 \pm 1.0$                      & 2792  & 2773  & --     & 2793  \\
$\Xi_c$                 & $3/2^-~(1P)$     & $\Xi_c(2815):2816.74^{+0.20}_{-0.23}$             & $2.43 \pm 0.26$                    & 2819  & 2783  & --     & 2820  \\
$\Sigma_c$              & $1/2^-~(1P)$     & --                                                & --                                 & 2713  & 2748  & 2802   & 2702  \\
\hdashline[2pt/2pt]
$\Sigma_c$              & $1/2^-~(1P)$     &\multirow{4}{*}{$\Sigma_c(2800):2792^{+14}_{-~5}$} &\multirow{4}{*}{$62^{+64}_{-44}$}   & 2799  & 2768  & 2826   & 2765  \\
$\Sigma_c$              & $3/2^-~(1P)$     &                                                   &                                    & 2773  & 2763  & 2807   & 2785  \\
$\Sigma_c$              & $3/2^-~(1P)$     &                                                   &                                    & 2798  & 2776  & 2837   & 2798  \\
$\Sigma_c$              & $5/2^-~(1P)$     &                                                   &                                    & 2789  & 2790  & 2839   & 2790  \\
\hdashline[2pt/2pt]
$\Xi^\prime_c$          & $1/2^-~(1P)$     & --                                                & --                                 & 2854  & 2855  & --     & 2839  \\
$\Xi^\prime_c$          & $1/2^-~(1P)$     & $\Xi_c(2923):2923.04\pm0.35$~\cite{LHCb:2020iby}  & $7.1 \pm 2.0$~\cite{LHCb:2020iby}  & 2936  & --    & --     & 2900  \\
$\Xi^\prime_c$          & $3/2^-~(1P)$     & $\Xi_c(2939):2938.55\pm0.30$~\cite{LHCb:2020iby}  & $10.2 \pm 1.4$~\cite{LHCb:2020iby} & 2912  & 2866  & --     & 2921  \\
$\Xi^\prime_c$          & $3/2^-~(1P)$     & $\Xi_c(2965):2964.88\pm0.33$~\cite{LHCb:2020iby}  & $14.1 \pm 1.6$~\cite{LHCb:2020iby} & 2935  & --    & --     & 2932  \\
$\Xi^\prime_c$          & $5/2^-~(1P)$     & --                                                & --                                 & 2929  & 2895  & --     & 2927  \\
$\Omega_c$              & $1/2^-~(1P)$     & --                                                & --                                 & 2966  & 2977  & 3030   & --    \\
$\Omega_c$              & $1/2^-~(1P)$     & $\Omega_c(3000):3000.41\pm0.22$                   & $4.5 \pm 0.7$                      & 3055  & 2990  & 3048   & --    \\
$\Omega_c$              & $3/2^-~(1P)$     & $\Omega_c(3050):3050.20\pm0.13$                   & $<1.2$                             & 3029  & 2986  & 3033   & --    \\
$\Omega_c$              & $3/2^-~(1P)$     & $\Omega_c(3066):3065.46\pm0.28$                   & $3.5 \pm 0.4$                      & 3054  & 2994  & 3056   & --    \\
$\Omega_c$              & $5/2^-~(1P)$     & $\Omega_c(3090):3090.0\pm0.5$                     & $8.7 \pm 1.3$                      & 3051  & 3014  & 3057   & --    \\
\hline
$\Lambda_c$             & $1/2^+~(2S)$     & $\Lambda_c(2765):2766.6\pm2.4$                    & $\sim50$                           & 2769  & 2791  & 2857   & 2772  \\
$\Xi_c$                 & $1/2^+~(2S)$     & $\Xi_c(2970):2966.34^{+0.17}_{-1.00}$             & $20.9^{+2.4}_{-3.5}$               & 2959  & --    & --     & 2940  \\
$\Sigma_c$              & $1/2^+~(2S)$     & --                                                & --                                 & 2901  & 2958  & 3029   & 2850  \\
$\Sigma_c$              & $3/2^+~(2S)$     & --                                                & --                                 & 2936  & 2995  & 3065   & 2876  \\
$\Xi_c^{\prime}$        & $1/2^+~(2S)$     & --                                                & --                                 & 2983  & --    & --     & 2977  \\
$\Xi_c^{\prime}$        & $3/2^+~(2S)$     & --                                                & --                                 & 3026  & 3012  & --     & 3007  \\
\hdashline[2pt/2pt]
$\Omega_c$              & $1/2^+~(2S)$     & \multirow{2}{*}{$\Omega_c(3119):3119.1\pm1.0$}    & \multirow{2}{*}{$<2.6$}            & 3088  & 3152  & 3227   & --    \\
$\Omega_c$              & $3/2^+~(2S)$     &                                                   &                                    & 3123  & 3190  & 3257   & --    \\
\hline
$\Lambda_c$             & $3/2^+~(1D)$     & $\Lambda_c(2860):2856.1^{+2.3}_{-5.9}$            & $67.6^{+11.8}_{-21.6}$             & 2874  & 2887  & 2920   & 2843  \\
$\Lambda_c$             & $5/2^+~(1D)$     & $\Lambda_c(2880):2881.63\pm0.24$                  & $5.6^{+0.8}_{-0.6}$                & 2880  & 2887  & 2922   & 2851  \\
$\Xi_c$                 & $3/2^+~(1D)$     & $\Xi_c(3055):3055.9\pm0.4$                        & $7.8 \pm 1.9$                      & 3059  & --    & --     & 3033  \\
$\Xi_c$                 & $5/2^+~(1D)$     & $\Xi_c(3080):3077.2\pm0.4$                        & $3.6 \pm 1.1$                      & 3076  & 3004  & --     & 3040  \\
$\Sigma_c$              & $1/2^+~(1D)$     & --                                                & --                                 & 3041  & --    & 3103   & 2949  \\
$\Sigma_c$              & $3/2^+~(1D)$     & --                                                & --                                 & 3040  & --    & 3094   & 2952  \\
$\Sigma_c$              & $3/2^+~(1D)$     & --                                                & --                                 & 3043  & --    & --     & 2964  \\
$\Sigma_c$              & $5/2^+~(1D)$     & --                                                & --                                 & 3023  & 3003  & 3099   & 2942  \\
$\Sigma_c$              & $5/2^+~(1D)$     & --                                                & --                                 & 3038  & 3010  & 3114   & 2962  \\
$\Sigma_c$              & $7/2^+~(1D)$     & --                                                & --                                 & 3013  & 3015  & --     & 2943  \\
$\Xi^\prime_c$          & $1/2^+~(1D)$     & --                                                & --                                 & 3163  & --    & --     & 3075  \\
$\Xi^\prime_c$          & $3/2^+~(1D)$     & --                                                & --                                 & 3160  & --    & --     & 3081  \\
$\Xi^\prime_c$          & $3/2^+~(1D)$     & --                                                & --                                 & 3167  & --    & --     & 3089  \\
$\Xi^\prime_c$          & $5/2^+~(1D)$     & --                                                & --                                 & 3153  & 3080  & --     & 3077  \\
$\Xi^\prime_c$          & $5/2^+~(1D)$     & --                                                & --                                 & 3166  & --    & --     & 3091  \\
$\Xi^\prime_c$          & $7/2^+~(1D)$     & --                                                & --                                 & 3147  & 3094  & --     & 3078  \\
$\Omega_c$              & $1/2^+~(1D)$     & --                                                & --                                 & 3287  & --    & 3292   & --    \\
$\Omega_c$              & $3/2^+~(1D)$     & --                                                & --                                 & 3282  & --    & 3285   & --    \\
$\Omega_c$              & $3/2^+~(1D)$     & --                                                & --                                 & 3298  & --    & --     & --    \\
$\Omega_c$              & $5/2^+~(1D)$     & --                                                & --                                 & 3286  & 3196  & 3288   & --    \\
$\Omega_c$              & $5/2^+~(1D)$     & --                                                & --                                 & 3297  & 3203  & 3299   & --    \\
$\Omega_c$              & $7/2^+~(1D)$     & --                                                & --                                 & 3283  & 3206  & --     & --
\\ \bottomrule[1pt]
\end{tabular}
\label{sec3:charm}
\end{table*}

In Table~\ref{sec3:charm}, we summarize the theoretical results of the charmed baryons obtained from various quark models~\cite{Ebert:2011kk,Roberts:2007ni,Yoshida:2015tia,Chen:2016iyi}, including the $1S$ and $2S$ states as well as the $1P$ and $1D$ states of the $\lambda$-mode. Possible experimental candidates are given for comparison. Especially, the four charmed baryons $\Lambda_c(2595)$, $\Lambda_c(2625)$, $\Xi_c(2790)$ and $\Xi_c(2815)$ may complete the $P$-wave charmed baryon doublet $[\mathbf{\bar 3}_F, 1, 0, \lambda]$. The four charmed baryons $\Lambda_c(2860)$, $\Lambda_c(2880)$, $\Xi_c(3055)$, and $\Xi_c(3080)$ may complete the $D$-wave charmed baryon doublet $[\mathbf{\bar 3}_F, 2, 0, \lambda\lambda]$, the first of which will be reviewed in Sec.~\ref{sec3.2.1}.

Some of the excited charmed baryons are good candidates for the $P$-wave charmed baryons belonging to the flavor $\mathbf{6}_F$ representation, which were studied in Refs.~\cite{Chen:2015kpa,Chen:2016phw,Chen:2017sci,Mao:2017wbz,Yang:2020zjl} using the methods of QCD sum rules and light-cone sum rules within the framework of heavy quark effective theory. In a recent study~\cite{Yang:2021lce} the authors systematically investigated the four $P$-wave charmed baryon multiplets of the flavor $\mathbf{6}_F$, {\it i.e.}, $[\mathbf{6}_F, 1, 0, \rho]$, $[\mathbf{6}_F, 0, 1, \lambda]$, $[\mathbf{6}_F, 1, 1, \lambda]$, and $[\mathbf{6}_F, 2, 1, \lambda]$. Their masses, mass splittings within the same multiplets, and decay properties are summarized in Table~\ref{sec3:charmsumrule}, where the mixing effect has been taken into account. These results can explain the $\Sigma_c(2800)$, $\Xi_c(2923)$, $\Xi_c(2939)$, $\Xi_c(2965)$, $\Omega_c(3000)$, $\Omega_c(3050)$, $\Omega_c(3066)$, $\Omega_c(3090)$, and $\Omega_c(3119)$ as a whole, which will be reviewed in Sec.~\ref{sec3.2.2} and Sec.~\ref{sec3.2.3}. Especially, the $\Sigma_c(2800)$ may be further separated into several substructures, which needs to be carefully examined in future experiments.

There exist many possible explanations for the $\Omega_c(3000)$, $\Omega_c(3050)$, $\Omega_c(3066)$, $\Omega_c(3090)$, and $\Omega_c(3119)$, {\it e.g.}, those given in Table~\ref{sec3:charm} are taken from Refs.~\cite{Wang:2017hej,Wang:2017kfr} and those given in Table~\ref{sec3:charmsumrule} are taken from Ref.~\cite{Yang:2021lce}. See Sec.~\ref{sec3.2.3} for more discussions.

\begin{table*}[hbtp]
\renewcommand{\arraystretch}{1.4}
\scriptsize
\caption{Mass spectrum and decay properties of the $P$-wave charmed baryons belonging to the $SU(3)$ flavor $\mathbf{6}_F$ representation, calculated in Ref.~\cite{Yang:2021lce} using the methods of QCD sum rules and light-cone sum rules within the framework of heavy quark effective theory. There are four $P$-wave multiplets of the flavor $\mathbf{6}_F$, {\it i.e.}, $[\mathbf{6}_F, 1, 0, \rho]$, $[\mathbf{6}_F, 0, 1, \lambda]$, $[\mathbf{6}_F, 1, 1, \lambda]$, and $[\mathbf{6}_F, 2, 1, \lambda]$. Their mixing effect has been taken into account, and their possible experimental candidates are given in the last column for comparisons. The mixing angle $\theta_1^\prime$ is between $[\Omega_c({1/2}^-),1,0,\rho]$ and $[\Omega_c({1/2}^-),0,1,\lambda]$, and the mixing angle $\theta_2^\prime$ is between $[\Omega_c({3/2}^-),1,0,\rho]$ and $[\Omega_c({3/2}^-),2,1,\lambda]$.}
\centering
\begin{tabular}{ c | c | c | c | c | c | c | c }
\hline\hline
\multirow{2}{*}{HQET state} & \multirow{2}{*}{Mixing} & \multirow{2}{*}{Mixed state} & Mass & Difference & Main decay channel & Width & \multirow{2}{*}{Candidate}
\\                                                                                 &&&({GeV})& ({MeV})   & ({MeV})            &({MeV}) &
\\ \hline\hline
$[\Sigma_c({1\over2}^-),1,0,\rho]$        & --                                         & $[\Sigma_c({1\over2}^-),1,0,\rho]$    & $2.77^{+0.16}_{-0.12}$ & \multirow{2}{*}{$15^{+6}_{-5}$} &
$\Gamma\left(\Sigma_c\pi\right) \approx 380$                                           & $390^{+630}_{-270}$ & --
\\ \cline{1-4} \cline{6-8}
$[\Sigma_c({3\over2}^-),1,0,\rho]$        & --                                         & $[\Sigma_c({3\over2}^-),1,0,\rho]$    & $2.79^{+0.16}_{-0.12}$ & &
$\Gamma\left(\Sigma_c^{*} \pi\right) \approx 220$                                      & $220^{+360}_{-150}$ & --
\\ \hline
$[\Sigma_c({1\over2}^-),0,1,\lambda]$     & \multirow{5}{*}{$\theta_1\simeq0^\circ$}   & $[\Sigma_c({1\over2}^-),0,1,\lambda]$ & $2.83^{+0.06}_{-0.04}$ & -- &
$\Gamma\left(\Lambda_c \pi\right) \approx 610$                                         & $610^{+860}_{-410}$ & --
\\ \cline{1-1}\cline{3-8}
$[\Sigma_c({1\over2}^-),1,1,\lambda]$     &                                            & $[\Sigma_c({1\over2}^-),1,1,\lambda]$ & $2.73^{+0.17}_{-0.18}$ & \multirow{5}{*}{$23^{+19}_{-43}$}&
$\begin{array}{c}
\Gamma\left(\Lambda_c \pi\right) \simeq 0 \\
\Gamma\left(\Sigma_c\pi\right) \approx 37\\
\Gamma\left(\Lambda_c(\rho\to\pi\pi)\right)\approx9.2\\
\Gamma\left(\Sigma_c\rho(\to\pi\pi)\right)\approx1.2
\end{array}$                                                                           & $48^{+70}_{-29}$    & \multirow{10}{*}{$\Sigma_c(2800)$}
\\ \cline{1-4} \cline{6-7}
$[\Sigma_c({3\over2}^-),1,1,\lambda]$     & \multirow{3}{*}{$\theta_2\approx37^\circ$} & $|\Sigma_c({3\over2}^-)\rangle_1$     & $2.75^{+0.17}_{-0.17}$ & &
$\begin{array}{c}
\Gamma\left(\Lambda_c\pi\right)\approx13\\
\Gamma\left(\Sigma_c\pi\right)\approx3.3\\
\Gamma\left(\Sigma_c^{*}\pi\right)\approx6.4
\end{array}$                                                                           & $24^{+23}_{-10}$    &
\\ \cline{1-1} \cline{3-7}
$[\Sigma_c({3\over2}^-),2,1,\lambda]$     &                                            & $|\Sigma_c({3\over2}^-)\rangle_2$     & $2.80^{+0.14}_{-0.12}$ & \multirow{4}{*}{$68^{+51}_{-51}$}&
$\begin{array}{c}
\Gamma\left(\Lambda_c\pi\right)\approx23\\
\Gamma\left(\Sigma_c^{*}\pi\right)\approx3.5
\end{array}$                                                                           & $28^{+36}_{-16}$    &
\\ \cline{1-4} \cline{6-7}
$[\Sigma_c({5\over2}^-),2,1,\lambda]$     & --                                         & $[\Sigma_c({5\over2}^-),2,1,\lambda]$ & $2.87^{+0.12}_{-0.11}$ & &
$\begin{array}{c}
\Gamma\left(\Lambda_c\pi\right)\approx12\\
\Gamma\left(\Sigma_c\pi\right)\approx0.39\\
\Gamma\left(\Sigma_c^{*}\pi\right)\approx0.61
\end{array}$                                                                           & $13^{+18}_{-~8}$    &
\\ \hline
$[\Xi^\prime_c({1\over2}^-),1,0,\rho]$    & --                                         & $[\Xi^\prime_c({1\over2}^-),1,0,\rho]$    & $2.88^{+0.15}_{-0.13}$ & \multirow{2}{*}{$13^{+6}_{-5}$} &
$\Gamma\left(\Xi_c^{\prime}\pi\right)\approx110$                                       & $110^{+170}_{-~80}$ & --
\\ \cline{1-4} \cline{6-8}
$[\Xi^\prime_c({3\over2}^-),1,0,\rho]$    & --                                         & $[\Xi^\prime_c({3\over2}^-),1,0,\rho]$    & $2.89^{+0.15}_{-0.13}$ & &
$\Gamma\left(\Xi_c^{*}\pi\right)\approx58$                                             & $59^{+88}_{-39}$    & --
\\ \hline
$[\Xi_c^\prime({1\over2}^-),0,1,\lambda]$ & \multirow{6}{*}{$\theta_1\simeq0^\circ$}   & $[\Xi_c^\prime({1\over2}^-),0,1,\lambda]$ & $2.90^{+0.13}_{-0.12}$ & -- &
$\begin{array}{c}
\Gamma\left(\Lambda_c \bar K\right)\approx400\\
\Gamma\left(\Xi_c \pi\right)\approx360
\end{array}$                                                                           & $760^{+820}_{-370}$ & --
\\ \cline{1-1}\cline{3-8}
$[\Xi_c^\prime({1\over2}^-),1,1,\lambda]$ &                                            & $[\Xi_c^\prime({1\over2}^-),1,1,\lambda]$ & $2.91^{+0.13}_{-0.12}$ & \multirow{5}{*}{$27^{+16}_{-27}$} &
$\begin{array}{c}
\Gamma\left(\Lambda_c \bar K\right)\simeq0 \\
\Gamma\left(\Xi_c \pi\right)\simeq0 \\
\Gamma\left(\Xi_c^{\prime}\pi\right)\approx12\\
\Gamma\left(\Xi_c\rho(\to\pi\pi)\right)\approx1.7
\end{array}$                                                                           & $14^{+17}_{-~8}$    & $\Xi_c(2923)$
\\ \cline{1-4}\cline{6-8}
$[\Xi_c^\prime({3\over2}^-),1,1,\lambda]$ & \multirow{4}{*}{$\theta_2\approx37^\circ$} & $|\Xi_c^\prime({3\over2}^-)\rangle_1$     & $2.94^{+0.12}_{-0.11}$ & &
$\begin{array}{c}
\Gamma\left(\Lambda_c \bar K\right)\approx2.3\\
\Gamma\left(\Xi_c\pi\right)\approx4.6\\
\Gamma\left(\Xi_ c^{\prime}\pi\right)\approx2.0\\
\Gamma\left(\Xi_c^{*}\pi\right)\approx2.1
\end{array}$                                                                           & $12^{+10}_{-~4}$    & $\Xi_c(2939)$
\\ \cline{1-1}\cline{3-8}
$[\Xi_c^\prime({3\over2}^-),2,1,\lambda]$ &                                            & $|\Xi_c^\prime({3\over2}^-)\rangle_2$     & $2.97^{+0.24}_{-0.15}$ & \multirow{4}{*}{$56^{+30}_{-35}$} &
$\begin{array}{c}
\Gamma\left(\Lambda_c \bar K\right)\approx6.3\\
\Gamma\left(\Xi_c\pi\right)\approx11\\
\Gamma\left(\Xi_c^{*}\pi\right)\approx1.3
\end{array}$                                                                           & $19^{+22}_{-~9}$    & $\Xi_c(2965)$
\\ \cline{1-4} \cline{6-8}
$[\Xi^\prime_c({5\over2}^-),2,1,\lambda]$ & --                                         & $[\Xi^\prime_c({5\over2}^-),2,1,\lambda]$ & $3.02^{+0.23}_{-0.14}$ & &
$\begin{array}{c}
\Gamma\left(\Lambda_c \bar K\right)\approx6.3\\
\Gamma\left(\Xi_c \pi\right)\approx9.6\\
\Gamma\left(\Xi_c^{*} \pi\right)\approx1.5
\end{array}$                                                                           & $18^{+20}_{-~8}$    & --
\\ \hline
$[\Omega_c({1\over2}^-),1,0,\rho]$        & $\theta_1^\prime\simeq0^\circ$             & $[\Omega_c({1\over2}^-),1,0,\rho]$    & $2.99^{+0.15}_{-0.15}$ & \multirow{2}{*}{$12^{+5}_{-5}$} &
$\Gamma\left(\Xi_c \bar K\right) \simeq 0$                                             & $\simeq0$                & \multirow{2}{*}{$\Omega_c(3000)$}
\\ \cline{1-4} \cline{6-7}
$[\Omega_c({3\over2}^-),1,0,\rho]$        & $\theta_2^\prime\simeq0^\circ$             & $[\Omega_c({3\over2}^-),1,0,\rho]$    & $3.00^{+0.15}_{-0.15}$ & &
$\Gamma\left(\Xi_c \bar K\right) \simeq 0$                                             & $\simeq0$                &
\\ \hline
$[\Omega_c({1\over2}^-),0,1,\lambda]$     & \multirow{2}{*}{$\theta_1\simeq0^\circ$}   & $[\Omega_c({1\over2}^-),0,1,\lambda]$ & $3.03^{+0.18}_{-0.19}$ & -- &
$\Gamma\left(\Xi_c \bar K\right)\approx980$                                            & $980^{+1530}_{-~670}$    & --
\\ \cline{1-1} \cline{3-8}
$[\Omega_c({1\over2}^-),1,1,\lambda]$     &                                            & $[\Omega_c({1\over2}^-),1,1,\lambda]$ & $3.04^{+0.11}_{-0.09}$ & \multirow{2}{*}{$27^{+15}_{-23}$} &
$\Gamma\left(\Xi_c \bar K\right)\simeq0$                                               & $\simeq0$                & $\Omega_c(3050)$
\\ \cline{1-4} \cline{6-8}
$[\Omega_c({3\over2}^-),1,1,\lambda]$     & \multirow{2}{*}{$\theta_2\approx37^\circ$} & $|\Omega_c({3\over2}^-)\rangle_1$     & $3.06^{+0.10}_{-0.09}$ & &
$\Gamma\left(\Xi_c \bar K\right)\approx2.0$                                            & $2.0^{+3.5}_{-1.5}$      & $\Omega_c(3066)$
\\ \cline{1-1} \cline{3-8}
$[\Omega_c({3\over2}^-),2,1,\lambda]$     &                                            & $|\Omega_c({3\over2}^-)\rangle_2$     & $3.09^{+0.22}_{-0.17}$ & \multirow{2}{*}{$51^{+26}_{-29}$} &
$\Gamma\left(\Xi_c \bar K\right)\approx6.3$                                            & $6.4^{+11.2}_{-~4.8}$    & $\Omega_c(3090)$
\\ \cline{1-4} \cline{6-8}
$[\Omega_c({5\over2}^-),2,1,\lambda]$     & --                                         & $[\Omega_c({5\over2}^-),2,1,\lambda]$ & $3.14^{+0.21}_{-0.15}$ & &
$\Gamma\left(\Xi_c \bar K\right)\approx5.5$                                            & $5.5^{+9.6}_{-4.1}$      & $\Omega_c(3119)$
\\ \hline\hline
\end{tabular}
\label{sec3:charmsumrule}
\end{table*}

\subsubsection{$\Lambda_c(2860)$.}
\label{sec3.2.1}

\begin{figure}[hbtp]
\begin{center}
\subfigure[]{\includegraphics[width=0.32\textwidth]{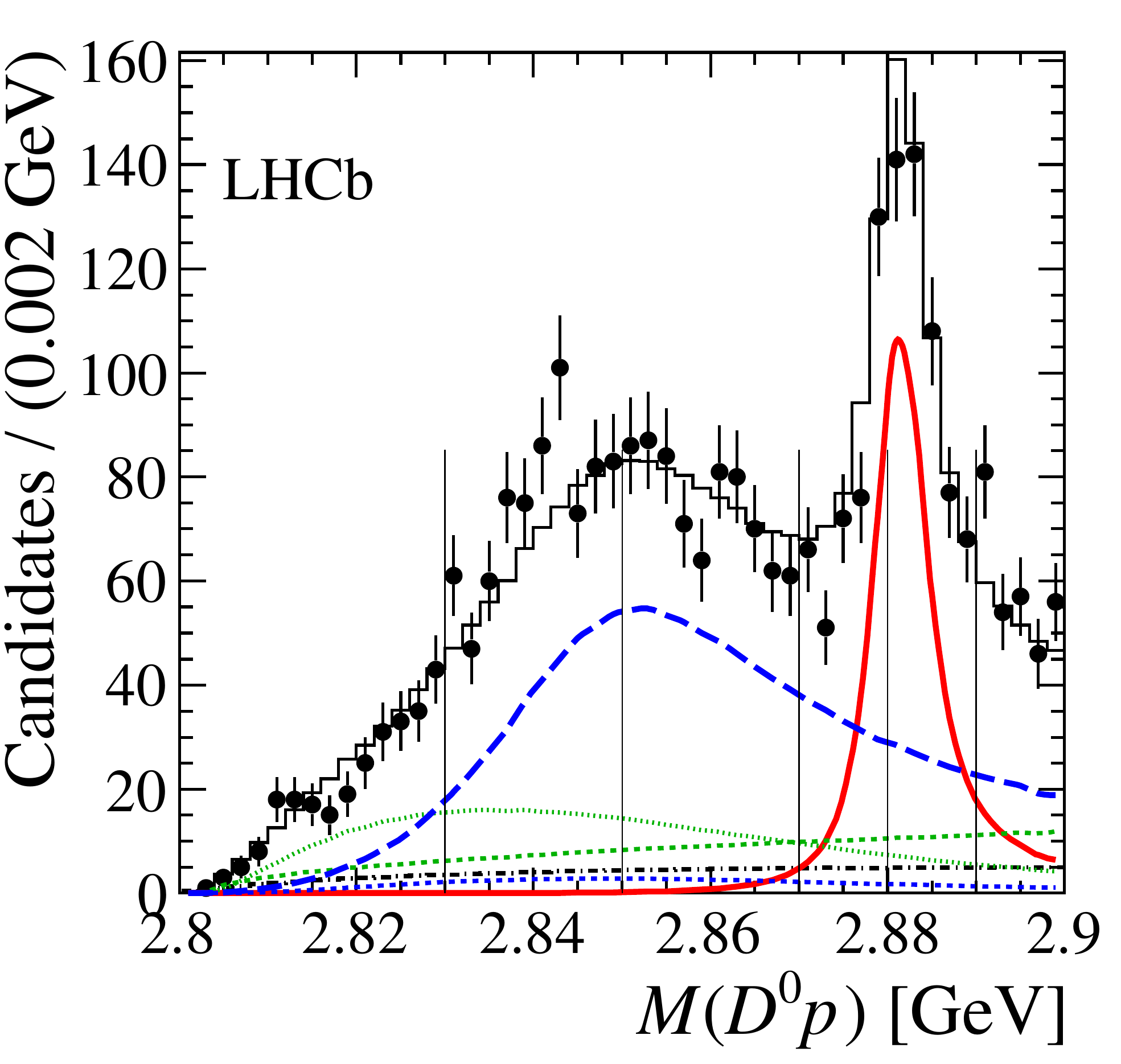}}
~~
\subfigure[]{\includegraphics[width=0.32\textwidth]{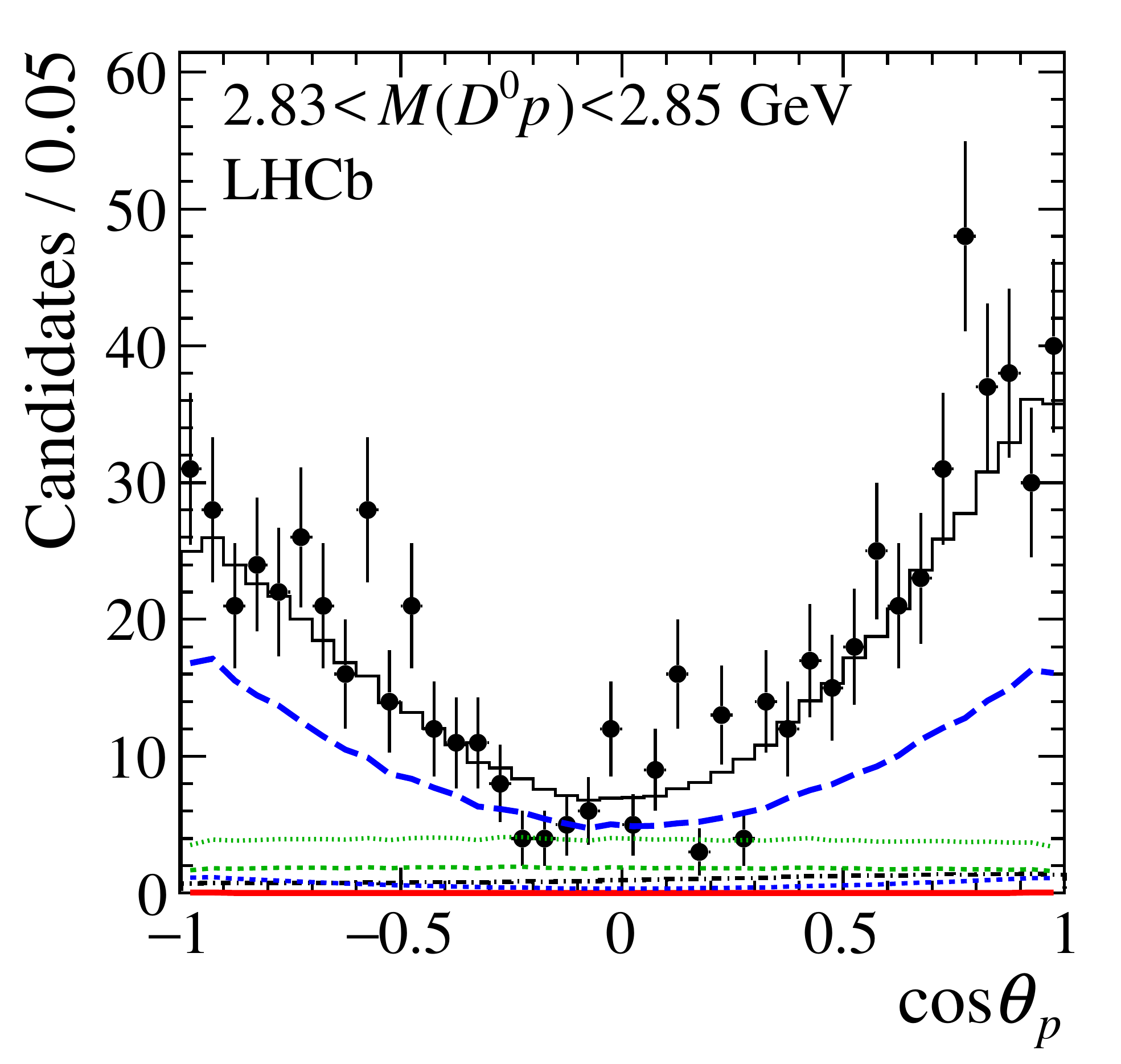}}
\includegraphics[width=0.3\textwidth]{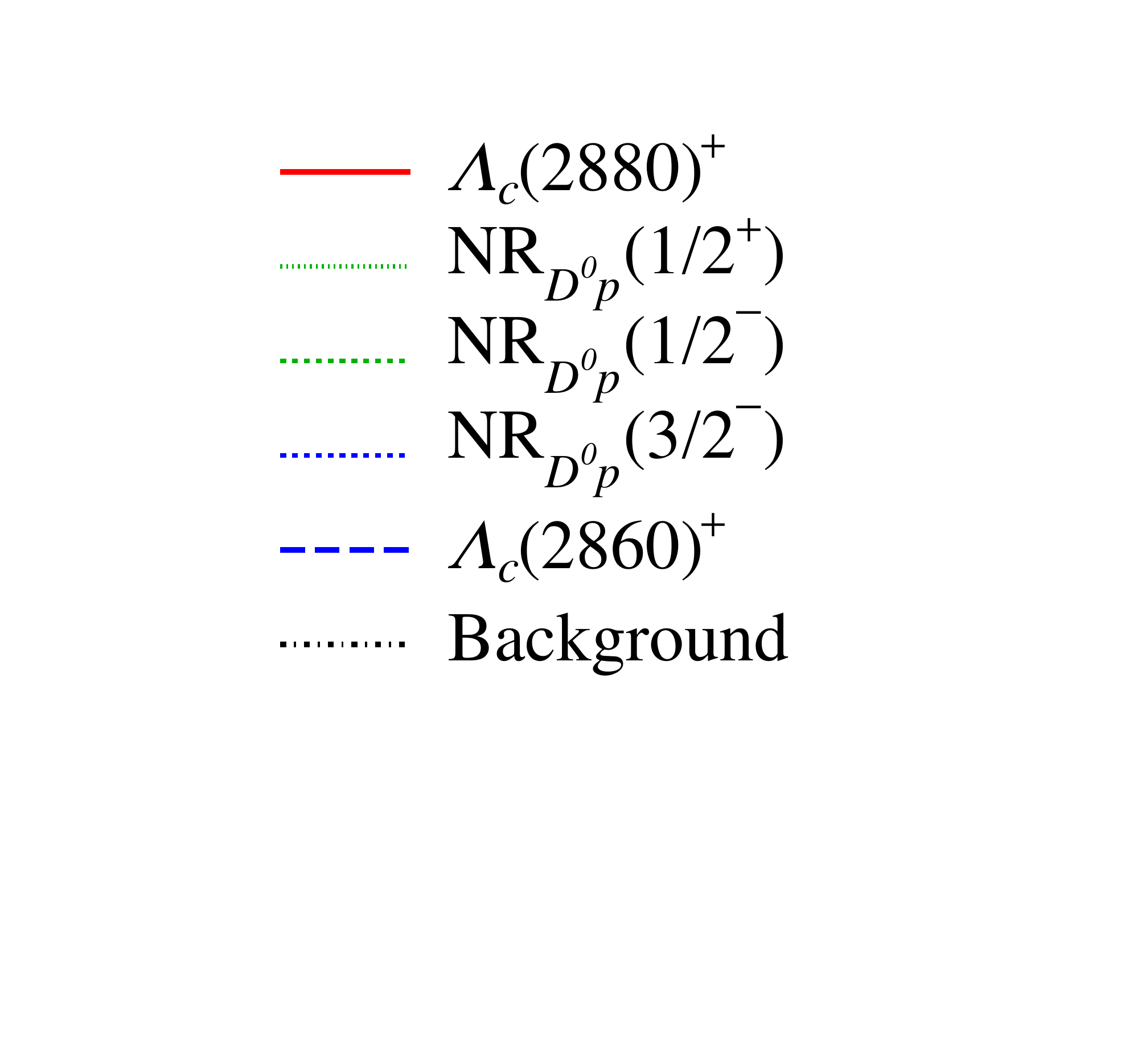}
\end{center}
\caption{Results for the fit of the $\Lambda^0_b \to D^0 p \pi^-$ Dalitz plot distribution in the near-threshold $D^0p$ mass region: (a) $M(D^0p)$ projection and (b) $\cos\theta_p$ projection. Source: Ref.~\cite{LHCb:2017jym}.}
\label{fig:L2860}
\end{figure}

In 2017 the LHCb collaboration performed a comprehensive amplitude analysis of the $\Lambda^0_b \to D^0 p \pi^-$ decay in the region of the phase space containing $D^0p$ resonant contributions~\cite{LHCb:2017jym}. They found the preferred spin of the $\Lambda_c(2880)^+$ to be $J = 5/2$, and measured its mass and width to be
\begin{eqnarray}
\Lambda_c(2880)^+ &:& M = 2881.75 \pm 0.29 \pm 0.07 ^{+0.14}_{-0.20}{\rm~MeV} \, ,
\\ \nonumber      && \Gamma = 5.43 ^{+0.77}_{-0.71} \pm 0.29 ^{+0.75}_{-0.00}{\rm~MeV} \, .
\end{eqnarray}
They found the most likely spin-parity quantum number of the $\Lambda_c(2940)^+$ to be $J^P = 3/2^-$, and measured its mass and width to be
\begin{eqnarray}
\Lambda_c(2940)^+ &:& M = 2944.8 ^{+3.5}_{-2.5} \pm 0.4 ^{+0.1}_{-4.6}{\rm~MeV} \, ,
\\ \nonumber      && \Gamma = 27.7 ^{+8.2}_{-6.0} \pm 0.9 ^{+~5.2}_{-10.4}{\rm~MeV} \, .
\end{eqnarray}
Besides, they observed a near-threshold enhancement in the $D^0 p$ amplitude, as depicted in Fig.~\ref{fig:L2860}. It is consistent with a resonant state of $J^P = 3/2^+$, with the mass and width measured to be
\begin{eqnarray}
\Lambda_c(2860)^+ &:& M = 2856.1 ^{+2.0}_{-1.7} \pm 0.5 ^{+1.1}_{-5.6}{\rm~MeV} \, ,
\\ \nonumber      && \Gamma = 67.6 ^{+10.1}_{-~8.1} \pm 1.4 ^{+~5.9}_{-20.0}{\rm~MeV} \, .
\end{eqnarray}

The existence of the $\Lambda_c(2860)$ had been predicted in Refs.~\cite{Chen:2016iyi,Lu:2016ctt,Chen:2016phw} before the LHCb experiment~\cite{LHCb:2017jym}. Its mass is consistent with the predictions for the $\Lambda_c(1D)$ state of $J^P = 3/2^+$ based on the nonrelativistic quark potential model~\cite{Chen:2016iyi} and from QCD sum rules within the framework of heavy quark effective theory~\cite{Chen:2016phw}. Accordingly, the four charmed baryons $\Lambda_c(2860)$, $\Lambda_c(2880)$, $\Xi_c(3055)$, and $\Xi_c(3080)$ may complete the $D$-wave charmed baryon doublet $[\mathbf{\bar 3}_F, 2, 0, \lambda\lambda]$. This picture is supported by Refs.~\cite{Yao:2018jmc} through the constituent quark model. Based on this assignment, the production of the $\Lambda_c(2860)$ at ${\rm \bar P}$ANDA was studied in Ref.~\cite{Lin:2021wrb}, and they found that the total cross section of the $p\bar p \to \Lambda^-_c \Lambda_c(2860)^+ \to \Lambda^-_c p D^0$ process may reach up to $10\mu$b.

In Ref.~\cite{Chen:2017aqm} the authors carried out an analysis of the mass spectra of the $\lambda$-mode charmed baryons, and their results suggest that the four charmed baryons $\Lambda_c(2860)$, $\Lambda_c(2880)$, $\Xi_c(3055)$, and $\Xi_c(3080)$ together form the $D$-wave charmed baryon doublet $[\mathbf{\bar 3}_F, 2, 0, \lambda\lambda]$. However, they further studied their two-body OZI-allowed decays, and found it difficult to explain the $\Lambda_c(2880)$ and $\Xi_c(3080)$ as the $D$-wave charmed baryons of $J^P = 5/2^+$ with the $\lambda$-mode excitation. Hence, the relation between $\Lambda_c(2860)$ and $\Lambda_c(2880)$ still needs to be carefully examined in future experimental and theoretical studies~\cite{Gong:2021jkb,Guo:2019ytq}.

\subsubsection{$\Xi_c(2923)$, $\Xi_c(2939)$, and $\Xi_c(2965)$.}
\label{sec3.2.2}

\begin{figure}[hbtp]
\begin{center}
\includegraphics[width=0.5\textwidth]{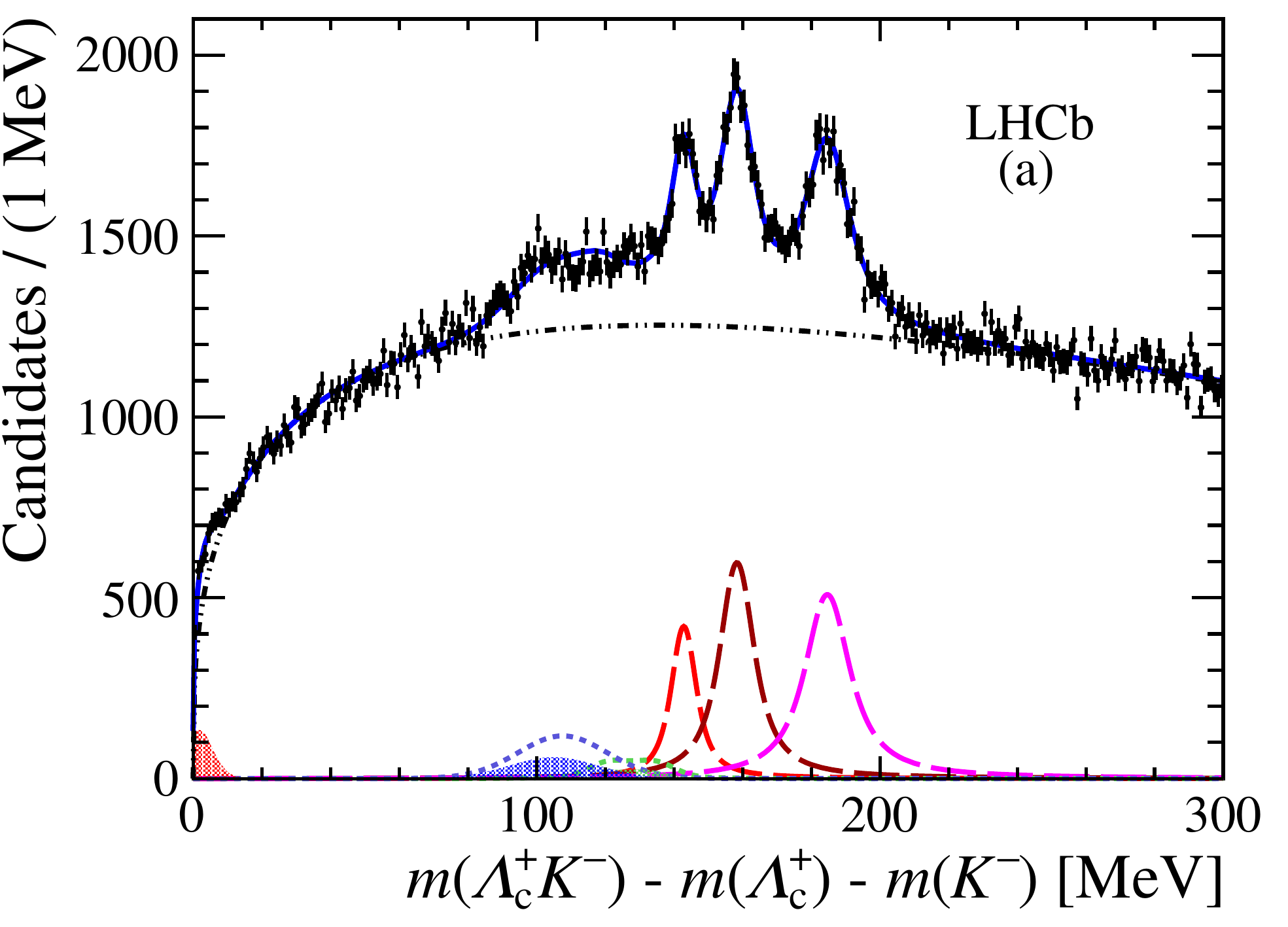}
\includegraphics[width=0.3\textwidth]{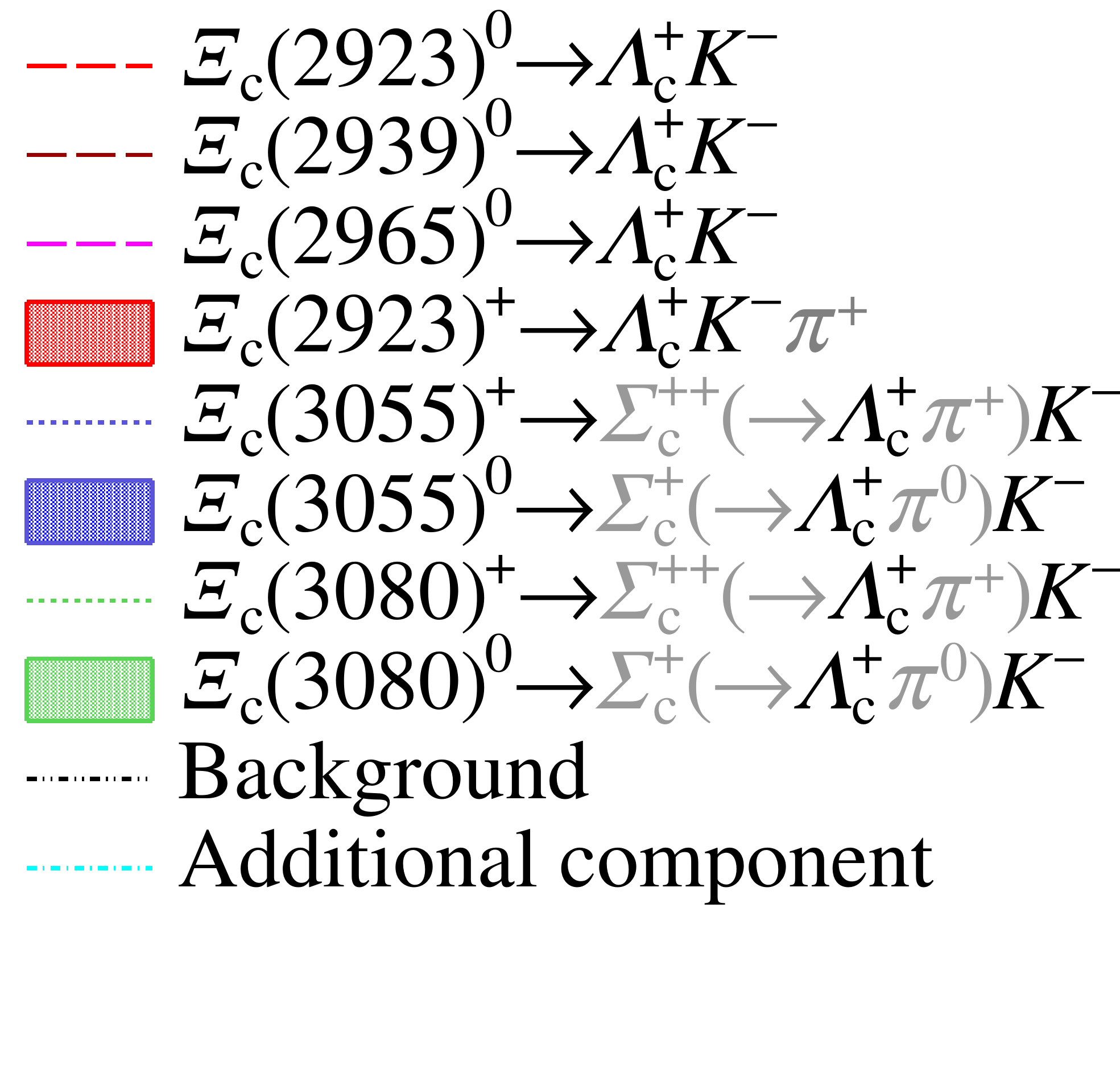}
\end{center}
\caption{Distributions of the reconstructed invariant mass difference $\Delta m = m(\Lambda^+_c K^-) - m(\Lambda^+_c) - m(K^-)$ for all candidates passing the selection requirements. Source: Ref.~\cite{LHCb:2020iby}.}
\label{fig:Xi2923}
\end{figure}

In 2007 the $\Xi_c(2930)^0$ state was observed in the $\Lambda_c^+K^-$ invariant mass spectrum of the $B^-\to \Lambda_c^+\bar \Lambda_c^-K^-$ decay by the BaBar collaboration~\cite{BaBar:2007xtc}, and later confirmed by the Belle collaboration~\cite{Belle:2017jrt,Belle:2018yob}. In 2020 the LHCb collaboration reinvestigated the $\Lambda^+_c K^-$ invariant mass spectrum, and observed three excited $\Xi^0_c$ states with a large significance~\cite{LHCb:2020iby}, as depicted in Fig.~\ref{fig:Xi2923}. Their masses and widths were measured to be
\begin{eqnarray}
\Xi_c(2923)^0 &:& M = 2923.04 \pm 0.25 \pm 0.20 \pm 0.14{\rm~MeV} \, ,
\\ \nonumber   && \Gamma = 7.1 \pm 0.8 \pm 1.8{\rm~MeV} \, ;
\\ \Xi_c(2939)^0 &:& M = 2938.55 \pm 0.21 \pm 0.17 \pm 0.14{\rm~MeV} \, ,
\\ \nonumber      && \Gamma = 10.2 \pm 0.8 \pm 1.1{\rm~MeV} \, ;
\\ \Xi_c(2965)^0 &:& M = 2964.88 \pm 0.26 \pm 0.14 \pm 0.14{\rm~MeV} \, ,
\\ \nonumber      && \Gamma = 14.1 \pm 0.9 \pm 1.3{\rm~MeV} \, .
\end{eqnarray}
As suggested by LHCb, the $\Xi_c(2930)^0$ signal may be due to the overlap of the two narrow $\Xi_c(2923)^0$ and $\Xi_c(2939)^0$ baryons, so we have not included the $\Xi_c(2930)^0$ in Tables~\ref{sec3:charm} and \ref{sec3:charmsumrule}. Although the $\Xi_c(2965)^0$ is in the vicinity of the known $\Xi_c(2970)^0$ baryon observed by Belle in the $\Lambda^+_c K^- \pi^+$ final state~\cite{Belle:2006edu}, its mass and width differ significantly from those of the $\Xi_c(2970)^0$, {\it i.e.}, $M_{\Xi_c(2970)^0} = 2970.9^{+0.4}_{-0.6}$~MeV and $\Gamma_{\Xi_c(2970)^0} = 28.1^{+3.4}_{-4.0}$~MeV~\cite{pdg}. Therefore, the $\Xi_c(2965)^0$ and $\Xi_c(2970)^0$ states may be two different baryons~\cite{Cheng:2021qpd}.

We briefly review several possible explanations for the $\Xi_c(2923)$, $\Xi_c(2939)$, and $\Xi_c(2965)$ as follows, and more possible explanations can be found in Refs.~\cite{Bahtiyar:2020uuj,Bijker:2020tns,Hu:2020zwc}. We shall notice that all these explanations are not (exactly) the same, partly because the heavy quark effective theory does not work quite well for these charmed baryons. Therefore, one needs to properly take the mixing effect into account, and their explanations become not so ``unique''. Similar situations also occur in other frameworks/models.

In Ref.~\cite{Lu:2020ivo} the author studied the $\Xi_c(2923)$, $\Xi_c(2939)$, and $\Xi_c(2965)$ using the constituent quark model. His results indicate that the $\Xi_c(2923)$ and $\Xi_c(2939)$ are the $\lambda$-mode $\Xi_c^\prime(1P)$ states of $J^P = 3/2^-$ and $5/2^-$ respectively, and the $\Xi_c(2965)$ can be assigned as the $\Xi_c^\prime(2S)$ state of $J^P = 1/2^+$. In Ref.~\cite{Wang:2020gkn} the authors studied them within the chiral quark model. Their results suggest that the $\Xi_c(2923)$ and $\Xi_c(2939)$ are most likely to be explained as the two $\lambda$-mode $\Xi_c^\prime(1P)$ states of $J^P = 3/2^-$, and the $\Xi_c(2965)$ may correspond to the $\lambda$-mode $\Xi_c^\prime(1P)$ state of $J^P = 5/2^-$.

In Refs.~\cite{Yang:2020zjl,Yang:2021lce} the authors studied the $\Xi_c(2923)$, $\Xi_c(2939)$, and $\Xi_c(2965)$ using the methods of QCD sum rules and light-cone sum rules within the framework of heavy quark effective theory. Their results are summarized in Table~\ref{sec3:charmsumrule}: the $\Xi_c(2923)$ and $\Xi_c(2939)$ can be interpreted as the $P$-wave $\Xi_c^\prime(1P)$ baryons of $J^P = 1/2^-$ and $3/2^-$ respectively, both of which belong to the $[{\bf 6}_F, 1, 1, \lambda]$ doublet; the $\Xi_c(2965)$ can be interpreted as the $P$-wave $\Xi_c^\prime(1P)$ baryon of $J^P = 3/2^-$, belonging to the $[{\bf 6}_F, 2, 1, \lambda]$ doublet; accordingly, the $\Xi_c(2965)$ has a partner state still missing, that is the $\Xi_c^\prime(1P)$ of $J^P = 5/2^-$, whose mass is about $56^{+30}_{-35}$~MeV higher and width about $18.1^{+19.7}_{-~8.3}$~MeV. In Ref.~\cite{Agaev:2020fut} the authors studied them also using the QCD sum rule method. Their results support the interpretation of the $\Xi_c(2923)$ and $\Xi_c(2939)$ as the $P$-wave $\Xi_c^\prime$ baryons of $J^P = 1/2^-$ and $3/2^-$, respectively. But they interpreted the $\Xi_c(2965)$ as the $\Xi_c(2S)$ or $\Xi_c^\prime(2S)$ state of $J^P = 1/2^+$.

Besides the above conventional explanations, there exist some exotic explanations for the $\Xi_c(2923)$, $\Xi_c(2939)$, and $\Xi_c(2965)$. In Ref.~\cite{Zhu:2020jke} the authors studied the strong decay modes of the $D\bar \Lambda$-$D\bar \Sigma$ molecular state. Their results support the interpretation of the $\Xi_c(2923)$ as a molecule, while the decay widths of $\Xi_c(2939)$ and $\Xi_c(2965)$ can not be reproduced in the molecular picture. Based on this interpretation, the authors of Ref.~\cite{Yang:2021rmp} evaluated the widths of the $\Xi_c(2923) \to \gamma \Xi_c$ and $\Xi_c(2923) \to \gamma \Xi^\prime_c$ decays to be approximately 1.23-11.66~keV and 0.30-3.71~keV, respectively.

\subsubsection{$\Omega_c(3000)$, $\Omega_c(3050)$, $\Omega_c(3066)$, $\Omega_c(3090)$, and $\Omega_c(3119)$.}
\label{sec3.2.3}

\begin{figure}[hbtp]
\begin{center}
\subfigure[]{\includegraphics[width=0.4\textwidth]{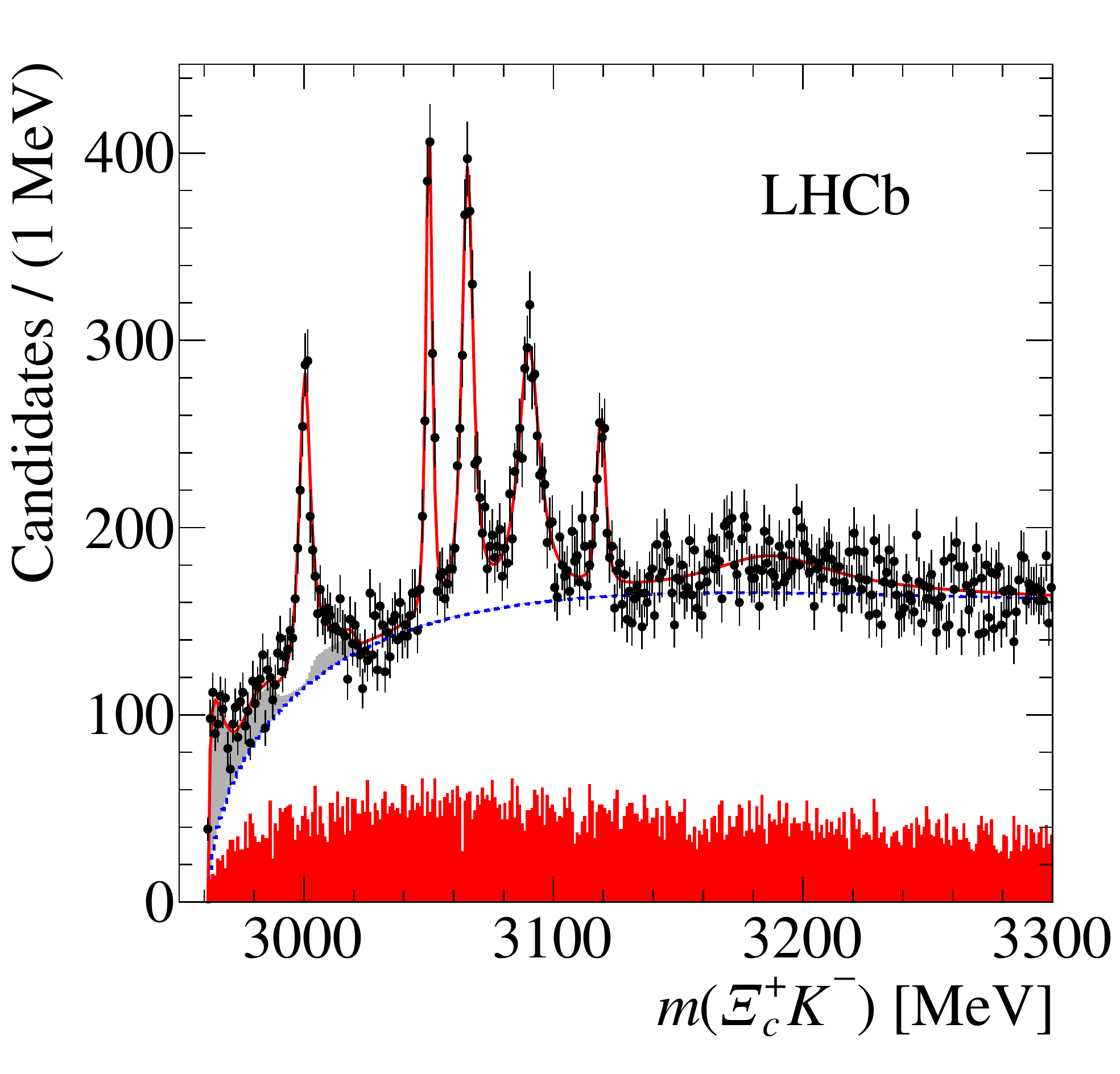}}
~~~~~
\subfigure[]{\includegraphics[width=0.51\textwidth]{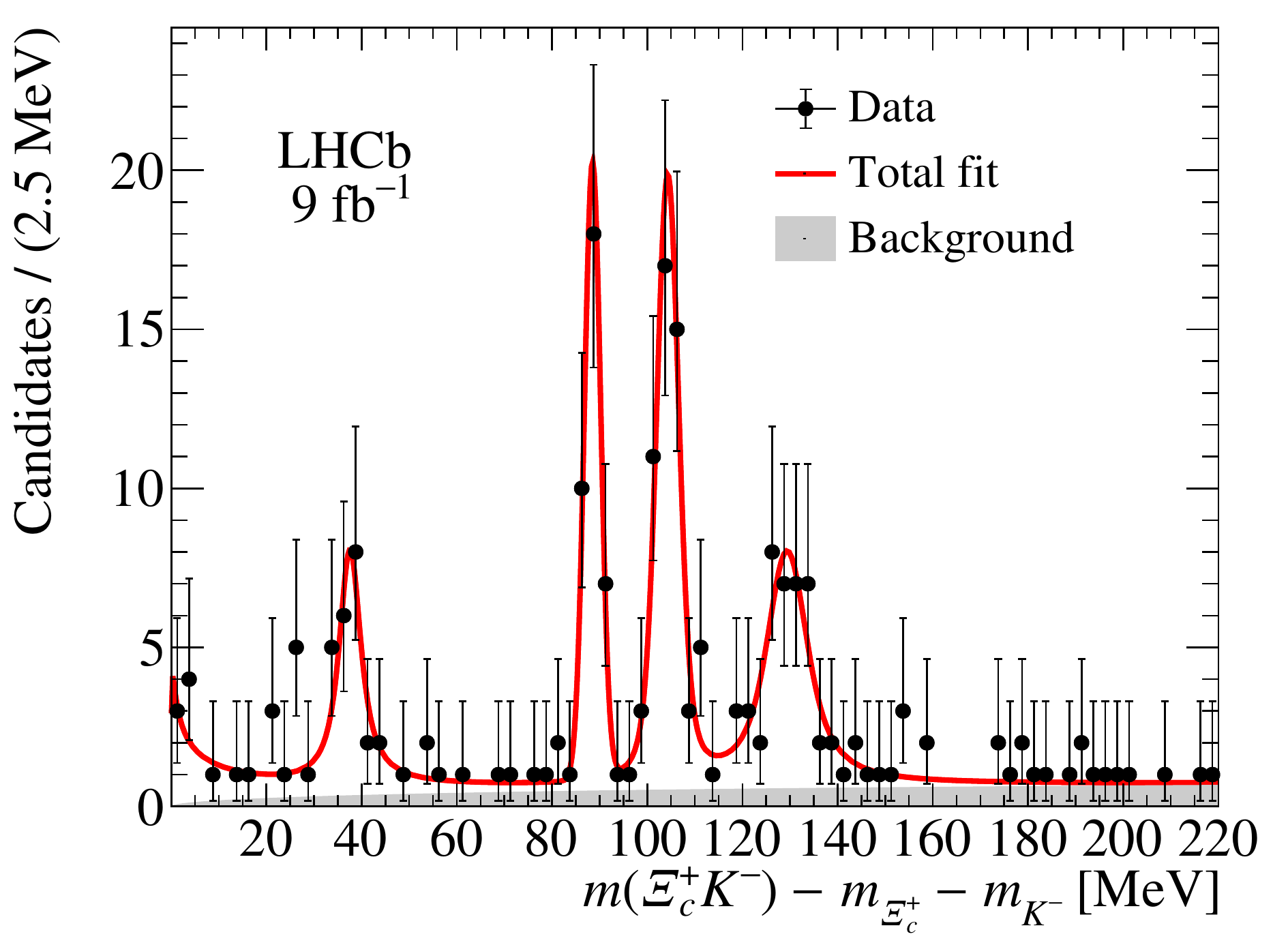}}
\end{center}
\caption{(a) The $\Xi^+_c K^-$ invariant mass spectrum in $pp$ collisions, measured by LHCb in 2017. Source: Ref.~\cite{LHCb:2017uwr}. (b) Distributions of the reconstructed invariant mass difference $\Delta m = m(\Xi^+_c K^-) - m(\Xi^+_c) - m(K^-)$ of the $\Omega_b^- \to  \Xi^+_c K^- \pi^-$ decay, measured by LHCb in 2021. Source: Ref.~\cite{LHCb:2021ptx}.}
\label{fig:O3000ac}
\end{figure}

\begin{figure}[hbtp]
\begin{center}
\includegraphics[width=0.6\textwidth]{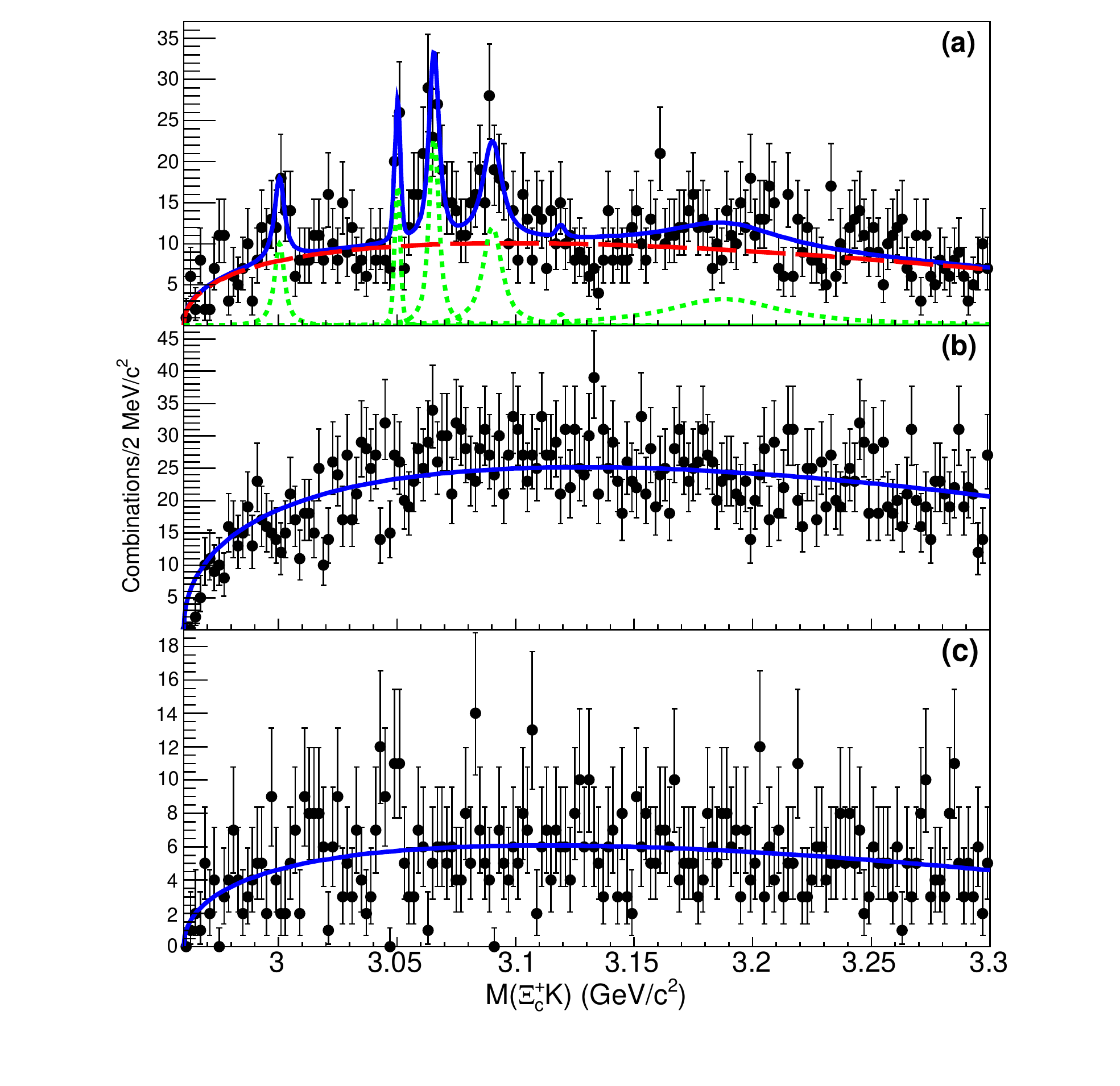}
\end{center}
\caption{(a) The $\Xi^+_c K^-$ invariant mass distribution in $e^+e^-$ collisions, measured by Belle in 2017. (b) A threshold function fit to the wrong-sign $\Xi^+_c K^+$ invariant mass distribution. (c) A threshold function fit to the invariant mass distribution for sidebands to the $\Xi^+_c$ candidates in combination with $K^-$ candidates. Source: Ref.~\cite{Belle:2017ext}.}
\label{fig:O3000b}
\end{figure}

In 2017 the LHCb collaboration studied the $\Xi^+_c K^-$ invariant mass spectrum in $pp$ collisions, and observed as many as five narrow excited $\Omega^0_c$ states~\cite{LHCb:2017uwr}. As depicted in Fig.~\ref{fig:O3000ac}(a), they are labeled as $\Omega_c(3000)^{0}$, $\Omega_c(3050)^{0}$, $\Omega_c(3066)^{0}$, $\Omega_c(3090)^{0}$, and $\Omega_c(3119)^{0}$, whose masses and widths were measured to be:
\begin{eqnarray}
\Omega_c(3000)^{0}   &:& M = 3000.4 \pm 0.2 \pm 0.1^{+0.3}_{-0.5} ~{\rm MeV} \, ,
\\ \nonumber && \Gamma = 4.5 \pm 0.6 \pm 0.3 ~{\rm MeV} \, ;
\\ \Omega_c(3050)^{0}&:& M = 3050.2 \pm 0.1 \pm 0.1^{+0.3}_{-0.5} ~{\rm MeV} \, ,
\\ \nonumber && \Gamma = 0.8 \pm 0.2 \pm 0.1 ~{\rm MeV} \, ;
\\ \Omega_c(3066)^{0}&:& M = 3065.6 \pm 0.1 \pm 0.3^{+0.3}_{-0.5} ~{\rm MeV} \, ,
\\ \nonumber && \Gamma = 3.5 \pm 0.4 \pm 0.2 ~{\rm MeV} \, ;
\\ \Omega_c(3090)^{0}&:& M = 3090.2 \pm 0.3 \pm 0.5^{+0.3}_{-0.5} ~{\rm MeV} \, ,
\\ \nonumber && \Gamma = 8.7 \pm 1.0 \pm 0.8 ~{\rm MeV} \, ;
\\ \Omega_c(3119)^{0}&:& M = 3119.1 \pm 0.3 \pm 0.9^{+0.3}_{-0.5} ~{\rm MeV} \, ,
\\ \nonumber && \Gamma = 1.1 \pm 0.8 \pm 0.4 ~{\rm MeV} \, .
\end{eqnarray}
The former four states $\Omega_c(3000)^{0}$, $\Omega_c(3050)^{0}$, $\Omega_c(3066)^{0}$, and $\Omega_c(3090)^{0}$ were confirmed in the later LHCb and Belle experiments~\cite{LHCb:2021ptx,Belle:2017ext}, as depicted in Fig.~\ref{fig:O3000ac}(b) and Fig.~\ref{fig:O3000b}. Their masses, natural widths, and relative production rates were also measured in Ref.~\cite{LHCb:2021ptx}, together with a test of spin hypotheses.

Similar to the $\Omega_b(6316)$, $\Omega_b(6330)$, $\Omega_b(6340)$, and $\Omega_b(6350)$ states reviewed in Sec.~\ref{sec3.1.6}, there exist many possible interpretations for the $\Omega_c(3000)$, $\Omega_c(3050)$, $\Omega_c(3066)$, $\Omega_c(3090)$, and $\Omega_c(3119)$. We briefly review some possible explanations as follows, and more possible explanations can be found in Refs.~\cite{Cheng:2021qpd,Agaev:2017jyt,Wang:2017vnc,Zhao:2017fov,Chen:2017gnu,Liu:2017xzo,Santopinto:2018ljf,Duan:2020xcc}. Note that there can be five $\lambda$-mode and two $\rho$-mode $\Omega_c(1P)$ states as well as two $\Omega_c(2S)$ states, all of which may be used to explain these five $\Omega_c$ baryons.

Some theoretical studies used only the five $\lambda$-mode excitations to explain the five $\Omega_c$ baryons observed by LHCb. In Refs.~\cite{Padmanath:2017lng,Padmanath:2013bla} the authors studied the ground- and excited-state spectra of the $\Omega_c$ baryons with spin up to $7/2$ from lattice QCD with dynamical quark fields. They predicted the quantum numbers of the five $\Omega_c$ baryons to be: the $\Omega_c(3000)$ and $\Omega_c(3050)$ both have $J^P = 1/2^-$, the $\Omega_c(3066)$ and $\Omega_c(3090)$ both have $J^P = 3/2^-$, and the $\Omega_c(3119)$ is possibly a $J^P=5/2^-$ state. This is supported by Ref.~\cite{Karliner:2017kfm}, where the authors interpreted the five $\Omega_c$ baryons as bound states of a charm quark and a $P$-wave $ss$ diquark. This is also supported by Ref.~\cite{Wang:2017zjw} through the QCD sum rule approach. However, this picture is not favored by the recent LHCb experiment~\cite{LHCb:2021ptx} with a 3.5$\sigma$ significance.

Some theoretical studies further considered the two $\Omega_c(2S)$ excitations. For example, an alternative interpretation was proposed in Ref.~\cite{Karliner:2017kfm}: the $\Omega_c(3090)$ and $\Omega_c(3119)$ are $\Omega_c(2S)$ states with $J^P=1/2^+$ and $3/2^+$; the $\Omega_c(3000)$, $\Omega_c(3050)$, and $\Omega_c(3066)$ are $\Omega_c(1P)$ states with $J^P=3/2^-$, $3/2^-$, and $5/2^-$; therefore, there are two $\Omega_c(1P)$ states of $J^P = 1/2^-$ still missing. This is partly supported by Ref.~\cite{Cheng:2017ove} through the Regge approach. A similar interpretation was proposed in Refs.~\cite{Wang:2017hej,Wang:2017kfr} based on the constituent quark model: the $\Omega_c(3000)$, $\Omega_c(3050)$, $\Omega_c(3066)$, and $\Omega_c(3090)$ are $\Omega_c(1P)$ states with $J^P=1/2^-$, $3/2^-$, $3/2^-$, and $5/2^-$ respectively, and the $\Omega_c(3119)$ may correspond to one of the two $\Omega_c(2S)$ states. This picture is favored by the recent LHCb experiment~\cite{LHCb:2021ptx}, and partly supported by Refs.~\cite{Aliev:2017led,Agaev:2017lip,Wang:2017xam} based on the QCD sum rule method.

In Ref.~\cite{Yang:2021lce} the authors further took the two $\rho$-mode excitations into account, and studied the mass spectrum and decay properties of the $P$-wave $\Omega_c$ baryons using the methods of QCD sum rules and light-cone sum rules within the framework of heavy quark effective theory. Their results are summarized in Table~\ref{sec3:charmsumrule}: the $\Omega_c(3050)$ and $\Omega_c(3066)$ are the $P$-wave $\Omega_c$ baryons of $J^P=1/2^-$ and $3/2^-$ respectively, which mainly belong to the $[\mathbf{6}_F,1,1,\lambda]$ doublet; the $\Omega_c(3090)$ and $\Omega_c(3119)$ are the $P$-wave $\Omega_c$ baryons of $J^P=3/2^-$ and $5/2^-$ respectively, which mainly belong to the $[\mathbf{6}_F,2,1,\lambda]$ doublet; the $\Omega_c(3000)$ is a $P$-wave $\Omega_c$ baryon of either $J^P = 1/2^-$ or $3/2^-$ with the $\rho$-mode excitation, which can be further separated into two states with the mass splitting $12\pm5$~MeV. This picture is also favored by the recent LHCb experiment~\cite{LHCb:2021ptx} but with the ordering reversed.

The above explanations are all based on the conventional quark model, and there also exist some exotic explanations for the five $\Omega_c$ baryons observed by LHCb~\cite{Wang:2017smo,Chen:2017xat,Nieves:2017jjx,Huang:2018wgr,Debastiani:2018adr,Liu:2018bkx,Liang:2017ejq,Huang:2017dwn,Xu:2019kkt}. In Ref.~\cite{Debastiani:2017ewu} the authors investigated them as meson-baryon molecular states, which are dynamically generated from the meson-baryon interactions in the charm sector. They considered seven $J^P = 1/2^-$ channels ($\Xi_c \bar K$, $\Xi^\prime_c \bar K$, $\Xi D$, $\Omega_c \eta$, $\Xi D^*$, $\Xi_c \bar K^*$, and $\Xi_c^\prime \bar K^*$) and six $J^P = 3/2^-$ channels ($\Xi_c^* \bar K$, $\Omega_c^* \eta$, $\Xi D^*$, $\Xi_c \bar K^*$, $\Xi^* D$ and $\Xi_c^\prime \bar K^*$). They obtained two states with $J^P = 1/2^-$ to explain the $\Omega_c(3050)$ and $\Omega_c(3090)$. They also obtained one state with $J^P = 3/2^-$ to explain the $\Omega_c(3119)$. A similar study was done in Ref.~\cite{Montana:2017kjw}, which employed effective Lagrangians to reproduce the masses and widths of $\Omega_c(3050)$ and $\Omega_c(3090)$.

Besides the meson-baryon molecular picture, the compact pentaquark picture has also been investigated~\cite{Ali:2017wsf,Anisovich:2017aqa,Yang:2017rpg}. In Ref.~\cite{An:2017lwg} the authors employed the constituent quark model and found four $sscq \bar q$ ($q=u/d$) configurations with $J^P = 1/2^-$ or $3/2^-$, which lie at energies very close to the five $\Omega_c$ baryons. According to Ref.~\cite{An:2017lwg}, the pentaquark configurations might form sizable components of these $\Omega_c$ baryons. Theoretical studies based on the chiral quark-soliton model can be found in Refs.~\cite{Kim:2017jpx,Kim:2017khv,Kim:2018cku,Polyakov:2022eub}, and these results partly support the $\Omega_c(3050)$ and $\Omega_c(3119)$ as exotic heavy pentaquark baryons.

To end this subsection, we would like to note that the above explanations are just possible explanations, and there exist many other possibilities for the five excited $\Omega_c$ baryons as well as the four excited $\Omega_b$ baryons, all of which were first observed by LHCb~\cite{LHCb:2017uwr,LHCb:2020iby,LHCb:2020tqd} in recent years. Although further experimental and theoretical studies are still demanded to fully understand them, their beautiful fine structure is in any case directly related to the rich internal structure of the excited singly heavy baryons. Recalling that the development of quantum physics is closely related to the better understanding of the gross, fine, and hyperfine structures of atomic spectra, one naturally guesses that the currently undergoing studies on the singly heavy baryons would not only improve our understandings of their internal structures, but also enrich our knowledge of the quantum physics~\cite{Chen:2016spr,Chen:2020mpy}.

\subsection{Doubly charmed baryons}
\label{sec3.3}

The doubly heavy baryons, as an ideal platform to study the heavy quark symmetry, have been investigated in various experimental and theoretical studies during the past fifty years~\cite{pdg}. In 2002 the SELEX collaboration reported their observation of the doubly charmed baryon $\Xi_{cc}^+(3519)$ in the $\Xi_{cc}^+\to \Lambda_c^+ K^- \pi^+$ decay process~\cite{SELEX:2002wqn}. Its mass was measured to be $3519 \pm 1$~MeV, and its lifetime was measured to be less than $33$~fs at 90\%~C.~L. However, this state was not confirmed in the later experiments~\cite{Ratti:2003ez,BaBar:2006bab,Belle:2006edu,Li:2021iwf}.

\begin{figure}[hbtp]
\begin{center}
\includegraphics[width=0.5\textwidth]{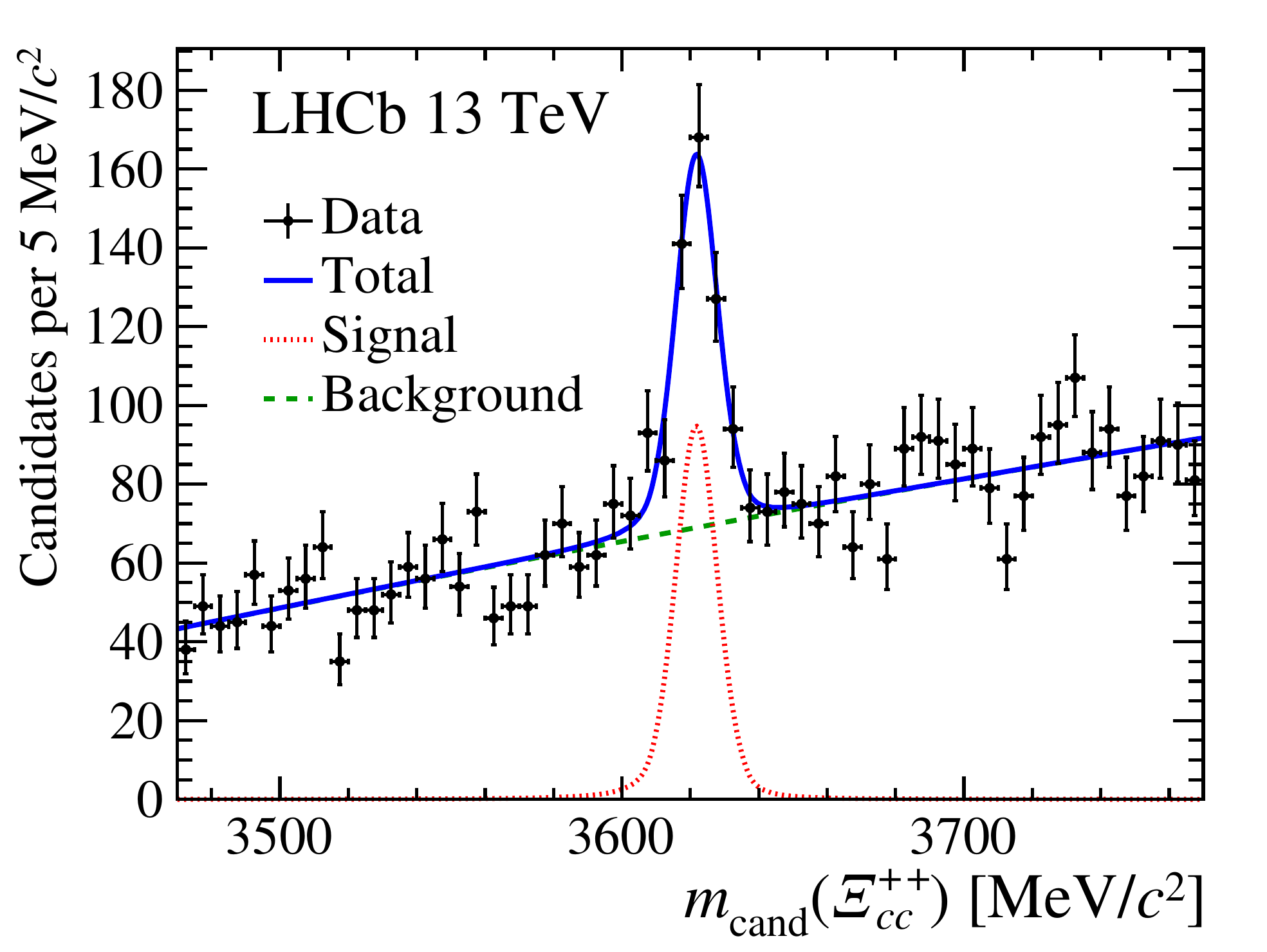}
\end{center}
\caption{Invariant mass distribution of $\Lambda^+_c K^- \pi^+ \pi^+$ candidates with fit projections overlaid. Source: Ref.~\cite{LHCb:2017iph}.}
\label{fig:Xicc}
\end{figure}

In 2017 the LHCb collaboration reported their discovery of the doubly charmed baryon $\Xi^{++}_{cc}(3621)$ in the $\Lambda^+_c K^- \pi^+ \pi^+$ mass spectrum~\cite{LHCb:2017iph}, as shown in Fig.~\ref{fig:Xicc}. Its mass and lifetime were measured in the later LHCb experiments~\cite{LHCb:2019epo,LHCb:2018zpl} to be
\begin{eqnarray}
\Xi^{++}_{cc}(3621)   &:& M = 3621.55 \pm 0.23 \pm 0.30~{\rm MeV} \, ,
\\ \nonumber && \tau = 0.256 ^{+0.024}_{-0.022} \pm 0.014 ~{\rm ps} \, .
\end{eqnarray}
This mass value is significantly larger than the mass of $\Xi_{cc}^+(3519)$ observed by SELEX~\cite{SELEX:2002wqn}. Because the $\Xi_{cc}^{++}$ and $\Xi_{cc}^{+}$ are isospin partners whose mass difference should be only a few MeV, the LHCb experiment~\cite{LHCb:2017iph} did not confirm the SELEX experiment~\cite{SELEX:2002wqn} neither.

The four-body decay channel $\Xi^{++}_{cc} \to \Lambda^+_c K^- \pi^+ \pi^+$ was previously suggested in Ref.~\cite{Yu:2017zst} as the most favorable decay mode to search for the doubly charmed baryon $\Xi^{++}_{cc}$. Besides, the decay process $\Xi^{++}_{cc} \to \Xi^+_c \pi^+$ was suggested in Refs.~\cite{Yu:2017zst,Sharma:2017txj} to have a sizable branching fraction, where the LHCb collaboration confirmed the existence of the $\Xi^{++}_{cc}(3621)$~\cite{LHCb:2018pcs}. The decay process $\Xi^{++}_{cc}(3621) \to \Xi^{\prime+}_c \pi^+$ was recently reported in another LHCb paper~\cite{LHCb:2022rpd}.

The discovery of the $\Xi^{++}_{cc}(3621)$ attracted much attention from the particle physics community, and many theoretical methods and models were applied to study it, {\it e.g.,} see Refs.~\cite{Mathur:2018rwu,Bahtiyar:2018vub} for some relevant lattice QCD studies. Since the $\Xi^{++}_{cc}(3621)$ is the ground-state baryon whose strong decay channels are all kinematically forbidden, many of its theoretical studies are closely related to the weak and electromagnetic interactions, {\it e.g.,} see Refs.~\cite{Wang:2017mqp,Wang:2017azm,Shi:2020qde,Pan:2020qqo,Geng:2017mxn,Cheng:2020wmk,Gutsche:2017hux,Gutsche:2019iac,HillerBlin:2018gjw,Li:2017pxa,Xiao:2017udy,Cui:2017udv,Ke:2019lcf,Li:2020ggh,Li:2020uok,Li:2020qrh,Li:2021rfj,Zhang:2022jst,Chen:2018koh,Berezhnoy:2018bde,Hu:2021gdg}. Since the present review is mainly concerned with the strong interaction, we shall not review the doubly charmed baryon $\Xi^{++}_{cc}(3621)$ in details.

Very recently, the doubly heavy baryon $\Xi^{+}_{bc}$ was investigated by the LHCb collaboration~\cite{LHCb:2022fbu}. They performed the first search for the $\Xi^+_{bc} \to J/\psi \Xi^+_c$ decay, and observed two peaking structures with a local (global) significance of 4.3 (2.8) and 4.1 (2.4) standard deviations at masses of $6571$~MeV and $6694$~MeV, respectively. We shall not review these structures in details.

\section{Open heavy flavor multiquark states}
\label{sec4}

At the birth of the quark model, the multiquark states, such as the $qq\bar q\bar q$ tetraquark states and the $qqqq\bar q$ pentaquark states, were proposed together with the conventional $q\bar q$ mesons and $qqq$ baryons~\cite{Gell-Mann:1964ewy,Zweig:1964ruk}. However, there were only a few multiquark candidates in the last century~\cite{pdg}, {\it e.g.}, the light scalar mesons $\sigma/f_0(500)$, $\kappa/K_0^*(700)$, $a_0(980)$, and $f_0(980)$ are good candidates for the light tetraquark states~\cite{pdg,Pelaez:2015qba,Close:2002zu,Amsler:2004ps,Bugg:2004xu}. Since 2003 there have been significant progresses in this field. Many multiquark candidates were observed in the BaBar, Belle, BESII/BESIII, CDF, CMS, COMPASS, D0, and LHCb experiments~\cite{pdg}, etc. Some of them have been reviewed in our previous papers~\cite{Chen:2016qju,Liu:2019zoy}, and in this section we shall review the following open heavy flavor multiquark candidates observed in the past five years:
\begin{itemize}

\item the open-charm tetraquark states $X_0(2900)$ and $X_1(2900)$ observed by LHCb in 2020~\cite{LHCb:2020bls,LHCb:2020pxc} will be reviewed in Sec.~\ref{sec4.1};

\item the doubly charmed tetraquark state $T_{cc}^+(3875)$ observed by LHCb in 2021~\cite{LHCb:2021vvq,LHCb:2021auc} will be reviewed in Sec.~\ref{sec4.2}.

\end{itemize}
Before doing this, we would like to mention two other open heavy flavor multiquark candidates, $T^a_{c\bar s 0}(2900)^{++}$ and $T^a_{c \bar s 0}(2900)^0$, which were observed by LHCb very recently in the $D^+_s \pi^\pm$ invariant mass spectrum of the $B^+ \to D^- D^+_s \pi^+$ and $B^0 \to \bar D^0 D^+_s \pi^-$ decays~\cite{LHCbTcs1,LHCbTcs2}. Their masses and widths were measured to be:
\begin{eqnarray}
T^a_{c\bar s 0}(2900)^{++} &:&    M= 2921 \pm 17 \pm 19  {\rm~MeV} \, ,
\\ \nonumber            &&    \Gamma=    137 \pm 32 \pm 14 {\rm~MeV} \, ;
\\   T^a_{c \bar s 0}(2900)^0 &:&    M= 2892 \pm 14 \pm 15  {\rm~MeV} \, ,
\\ \nonumber            &&    \Gamma=    119 \pm 26 \pm 12 {\rm~MeV} \, .
\end{eqnarray}
Assuming that they belong to the same isospin triplet, the LHCb experiment determined their averaged values to be:
\begin{eqnarray}
T^a_{c\bar s 0}(2900) &:&    M= 2908 \pm 11 \pm 20  {\rm~MeV} \, ,
\\ \nonumber            &&    \Gamma=    136 \pm 23 \pm 11 {\rm~MeV} \, .
\end{eqnarray}
The quark contents of $T^a_{c\bar s 0}(2900)^{++}$ and $T^a_{c \bar s 0}(2900)^0$ are $c\bar s u \bar d$ and $c\bar s u\bar d$ respectively, and their spin-parity quantum numbers were determined to be $J^P= 0^+$ with a high significance.

The $T^a_{c\bar s 0}(2900)$ had been investigated in Refs.~\cite{Azizi:2018mte,Lu:2020qmp,He:2020jna,Cheng:2020nho,Albuquerque:2020ugi,Guo:2021mja,Chen:2017rhl} before this LHCb measurement~\cite{LHCbTcs1,LHCbTcs2}. Especially, the authors of Ref.~\cite{Chen:2017rhl} applied the QCD sum rule method to study the $sq \bar q \bar c$ ($q=u/d$) tetraquark state of $J^P = 0^+$ through the tetraquark current built of the light vector diquark $s^T C \gamma_\mu q$ and the heavy vector antidiquark $\bar q \gamma_\mu C \bar c^T$. They calculated its mass to be $2.91 \pm 0.14$~GeV, and proposed to search in the $D_s^{(*)-} \pi^-$ channel for the exotic doubly-charged tetraquark state with the quark content $sd \bar u \bar c$. After the LHCb measurement~\cite{LHCbTcs1,LHCbTcs2}, the $T^a_{c\bar s 0}(2900)^{++}$ and $T^a_{c \bar s 0}(2900)^0$ have attracted much attention and there have been some theoretical studies~\cite{Agaev:2022duz,An:2022vtg,Ge:2022dsp,Chen:2022svh}, which we shall not discuss any further.

\subsection{Open-charm tetraquark states $X_0(2900)$ and $X_1(2900)$}
\label{sec4.1}

\begin{figure}[hbtp]
\begin{center}
\subfigure[]{\includegraphics[width=0.4\textwidth]{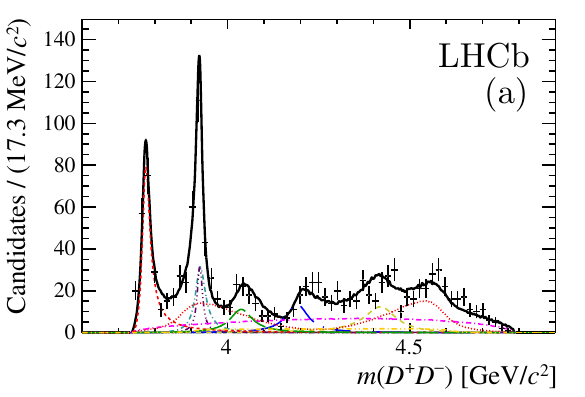}}
~~
\subfigure[]{\includegraphics[width=0.4\textwidth]{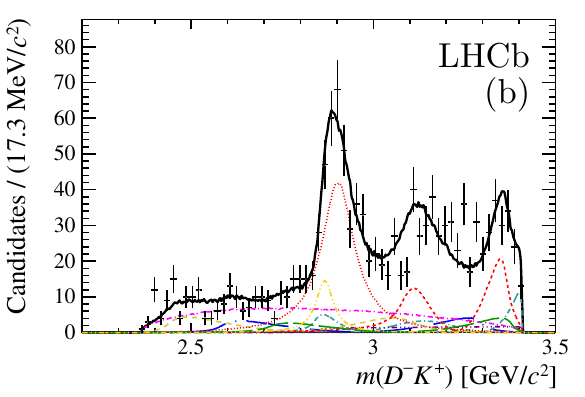}}
\\
\subfigure[]{\includegraphics[width=0.4\textwidth]{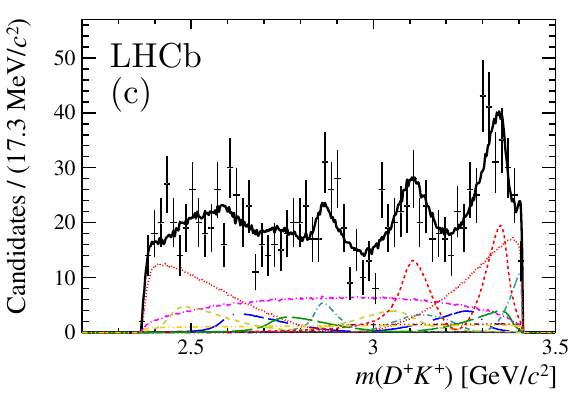}}
~~
\subfigure[]{\includegraphics[width=0.4\textwidth]{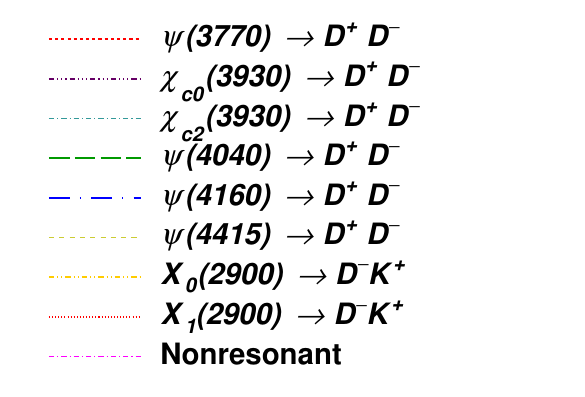}}
\end{center}
\caption{Comparisons of the invariant mass distributions of $B^+ \to D^+ D^- K^+$ candidates in data to the fit projection of the baseline model, in the (a) $D^+D^-$, (b) $D^-K^+$, and (c) $D^+K^+$ channels. Source: Ref.~\cite{LHCb:2020pxc}.}
\label{fig:X2900}
\end{figure}

In 2020 the LHCb collaboration studied the $B^+ \to D^+ D^- K^+$ decay, and observed an exotic peak in the $D^- K^+$ invariant mass spectrum~\cite{LHCb:2020bls,LHCb:2020pxc}, as depicted in Fig.~\ref{fig:X2900}. They found it necessary to include new spin-0 and spin-1 resonances in the $D^- K^+$ channel in order to obtain a good description of the data, whose masses and widths were determined to be
\begin{eqnarray}
     X_0(2900) &:&    M= 2866 \pm 7 \pm 2  {\rm~MeV} \, ,
\label{sec4:X02900}
\\ \nonumber            &&    \Gamma=    57 \pm 12 \pm 4 {\rm~MeV} \, ;
\\   X_1(2900) &:&    M= 2904 \pm 5 \pm 1  {\rm~MeV} \, ,
\label{sec4:X12900}
\\ \nonumber            &&    \Gamma=    110 \pm 11 \pm 4 {\rm~MeV} \, .
\end{eqnarray}
These two resonances were both observed in the $D^- K^+$ final state, implying: a) the spin-parity quantum numbers of $X_0(2900)$ and $X_1(2900)$ are $J^P = 0^+$ and $1^-$, respectively; b) their quark contents are $u d \bar s \bar c$, so they are exotic tetraquark states with valence quarks of four different flavors. Actually, this is the first time observing such exotic hadrons with open heavy flavor, given that the $X(5568)$ observed by D0~\cite{D0:2016mwd,D0:2017qqm} was not confirmed by the subsequent ATLAS, CDF, CMS, and LHCb experiments~\cite{ATLAS:2018udc,CDF:2017dwr,CMS:2017hfy,LHCb:2016dxl}.

There are several possible explanations for the $X_0(2900)$ and $X_1(2900)$, such as the hadronic molecular states and compact tetraquark states, or they may also be caused by the triangle singularities. We shall separately discuss these scenarios as follows, but note that it is still difficult to determine which explanation is more suitable at this moment, so further experimental and theoretical investigations are crucially needed to better understand them.

\subsubsection{Hadronic molecular picture.}
\label{sec4.1.1}

Many theoretical studies support the interpretation of the $X_0(2900)$ as the $\bar D^* K^*$ molecular state of $(I)J^P = (0)0^{+}$, and a few theoretical studies support the interpretation of the $X_1(2900)$ as the $\bar D_1 K$ molecular state of $(I)J^P = (0)1^{-}$.

In 2010 before the above LHCb experiment~\cite{LHCb:2020bls,LHCb:2020pxc}, the authors of Ref.~\cite{Molina:2010tx} studied the vector-meson-vector-meson interaction in a coupled-channel unitary approach, and predicted a $\bar D^* K^*$ bound state of $(I)J^P=(0)0^+$ with the mass 2848~MeV and width 59~MeV. It can be used to describe the $X_0(2900)$. Recently in Ref.~\cite{Molina:2020hde} the authors updated their previous calculations, and reported two more states stemming from the same interaction, one with $(I)J^P = (0)1^+$ and the other with $(I)J^P = (0)2^+$. Their masses were evaluated to be $2861$~MeV and $2775$~MeV, and they can be observed in the $\bar D^* K$ and $\bar D K$ channels, respectively. Based on the existence of the $X_0(2900)$ as the $\bar D^* K^*$ molecular state of $(I)J^P = (0)0^{+}$, the authors of Ref.~\cite{Hu:2020mxp} extracted the whole heavy-quark-symmetry multiplets formed by the $(\bar D, \bar D^*)$ doublet and the $K^*$ meson. Their results support the existence of the $\bar D K^*$ hadronic molecule with $(I)J^P = (0)1^{+}$ as well as the $\bar D^* K^*$ hadronic molecules with $(I)J^P = (0)1^{+}$ and $(0)2^{+}$. They further proposed to search for these partner states in the $\bar B^0 \to D^{*+} \bar D^{*0} K^-$ and $\bar B^0 \to D^{(*)+} K^- K^{(*)0}$ reactions~\cite{Dai:2022htx,Dai:2022qwh}.

\begin{figure}[hbtp]
\begin{center}
\includegraphics[width=0.8\textwidth]{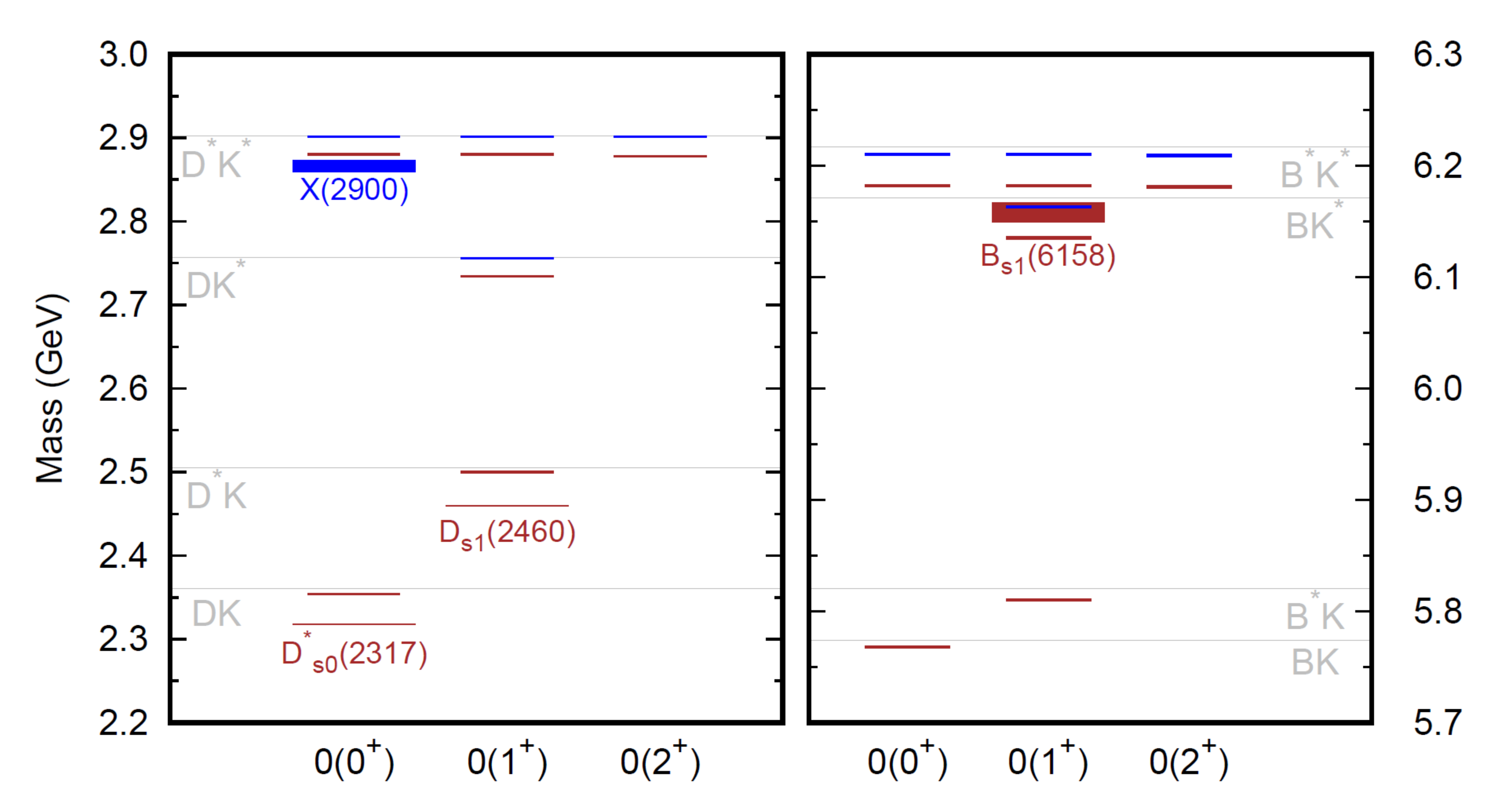}
\end{center}
\caption{Bound states from the interactions $D^{(*)}K^{(*)}/\bar{B}^{(*)}K^{(*)}$ (brown lines) and $\bar{D}^{(*)}K^{(*)}/B^{(*)}K^{(*)}$ (blue lines) at cutoff 1.4 and 2.0~GeV, respectively. Source: Ref.~\cite{Kong:2021ohg}.}
\label{fig:X2900BS}
\end{figure}

In Ref.~\cite{Liu:2020nil} the authors studied the $\bar D^{(*)}$ and $K^{(*)}$ interaction using the one-boson-exchange model, and their results suggest that a) the $X_0(2900)$ can be reasonably understood as the $\bar D^* K^*$ molecular state of $(I)J^P = (0)0^{+}$, or at least it has a large molecular component, and b) the $X_1(2900)$ can not be interpreted as a molecular state. Later in Ref.~\cite{Kong:2021ohg} the authors investigated the heavy-strange meson systems $D^{(*)}K^{(*)}$ and $\bar D^{(*)}K^{(*)}$ through a quasipotenial Bethe-Salpter equation approach based on the one-boson-exchange model. As shown in Fig.~\ref{fig:X2900BS}, they obtained a broad scalar state near the $\bar D^{(*)}K^{(*)}$ threshold after including the coupled-channel effect. This state is consistent with the experimentally observed $X_0(2900)$, although its binding energy is a little smaller than the experimental value.

\begin{figure}[hbtp]
\begin{center}
\includegraphics[width=0.5\textwidth]{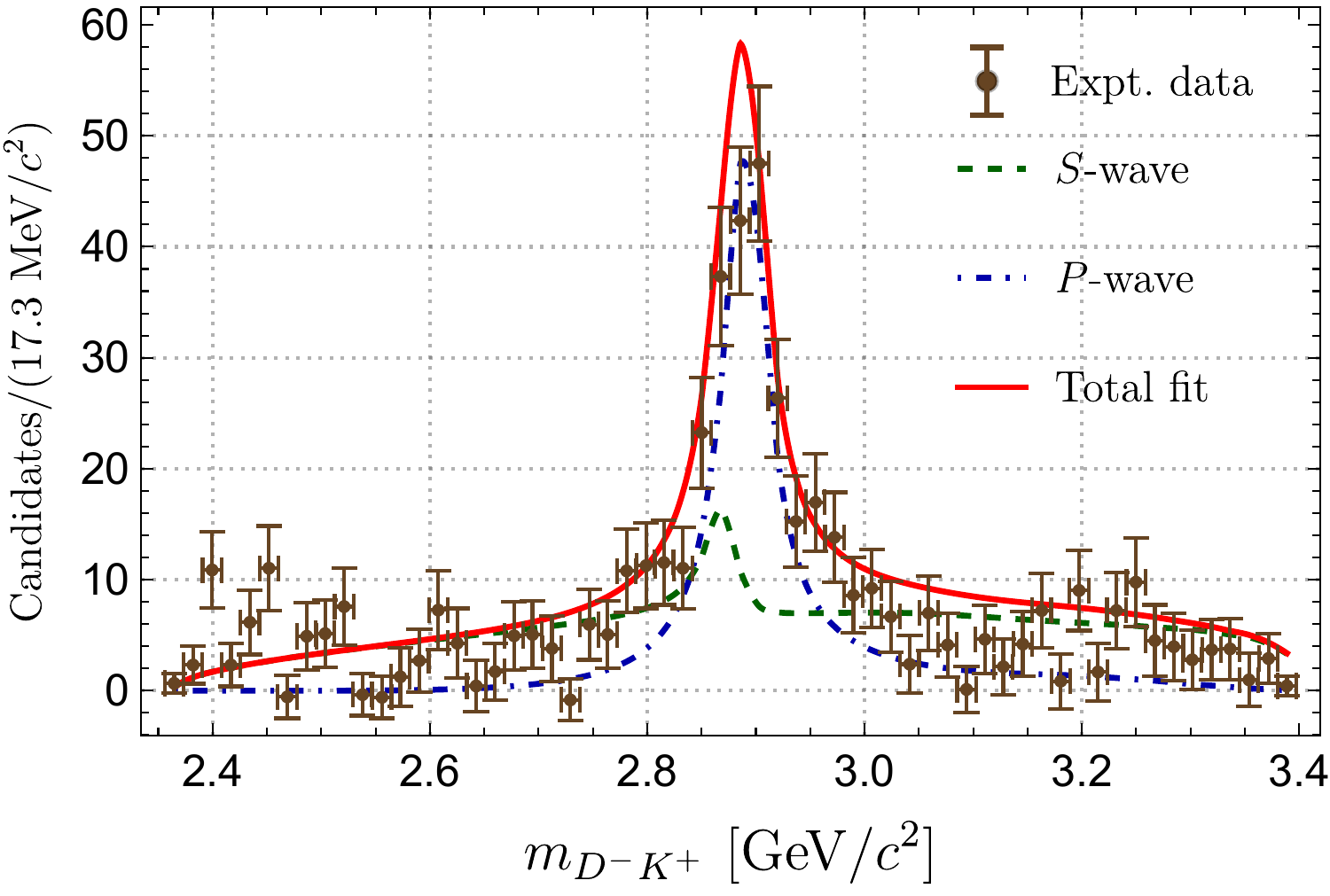}
\end{center}
\caption{The fitted $D^-K^+$ invariant mass distributions for the $B^+ \to D^+ D^- K^+$ decay. The experimental data are extracted from the LHCb experiment~\cite{LHCb:2020pxc} with the reflection contributions from charmonia subtracted. Source: Ref.~\cite{Wang:2021lwy}.}
\label{fig:X2900fit}
\end{figure}

In a recent study~\cite{Wang:2021lwy} the authors studied the $S$- and $P$-wave $\bar D^* K^*$ interactions in a coupled-channel formalism, in order to describe the $D^- K^+$ invariant mass spectrum of the $B^+ \to D^+ D^- K^+$ decay measured by LHCb~\cite{LHCb:2020bls,LHCb:2020pxc}. As shown in Fig.~\ref{fig:X2900fit}, they obtained two sharp peaks corresponding to the $\bar D^* K^*$ molecular states of $(I)J^P=(0)0^+$ and $(0)1^-$, whose masses are in good agreement with those of the $X_0(2900)$ and $X_1(2900)$, respectively; however, the width of the $0^+$ state is larger than that of the $1^-$ one, which is different from those given in Eqs.~(\ref{sec4:X02900}-\ref{sec4:X12900}).

\begin{figure}[hbtp]
\begin{center}
\subfigure[]{\includegraphics[width=0.3\textwidth]{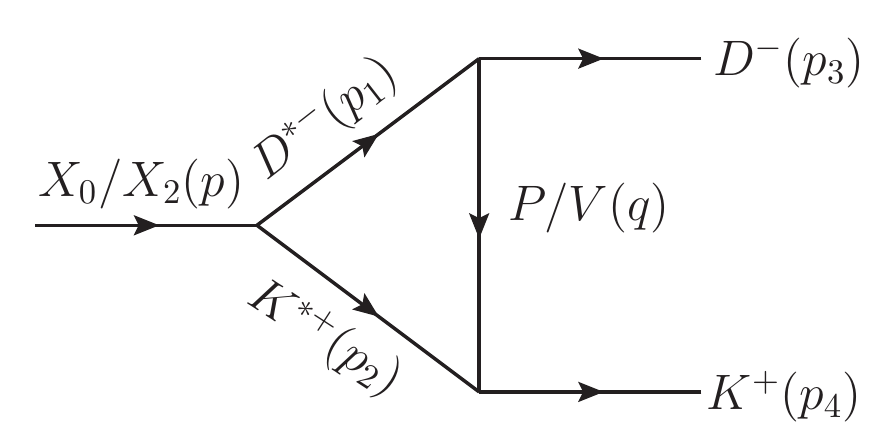}}
~~
\subfigure[]{\includegraphics[width=0.3\textwidth]{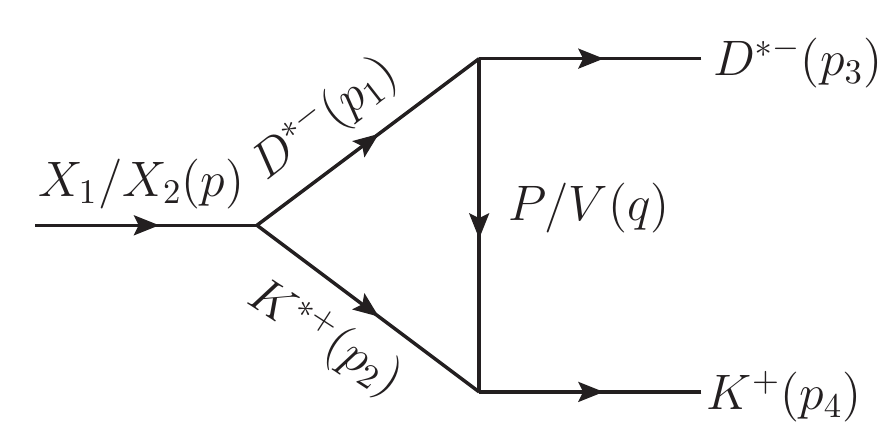}}
~~
\subfigure[]{\includegraphics[width=0.3\textwidth]{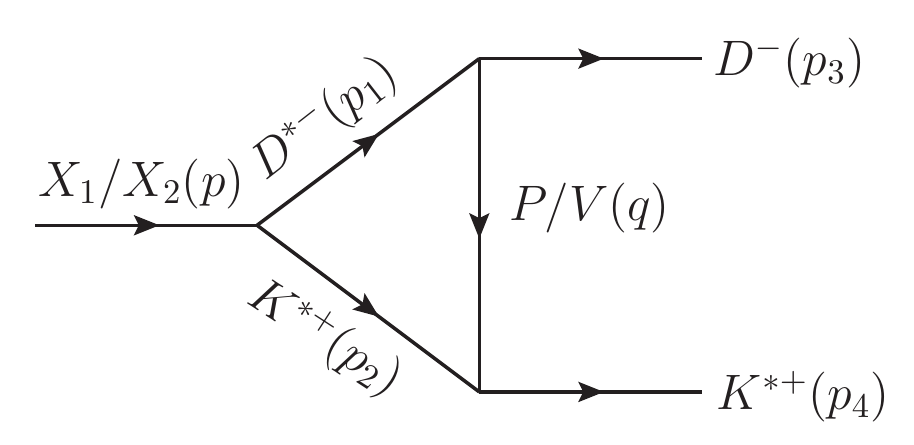}}
\\
\subfigure[]{\includegraphics[width=0.3\textwidth]{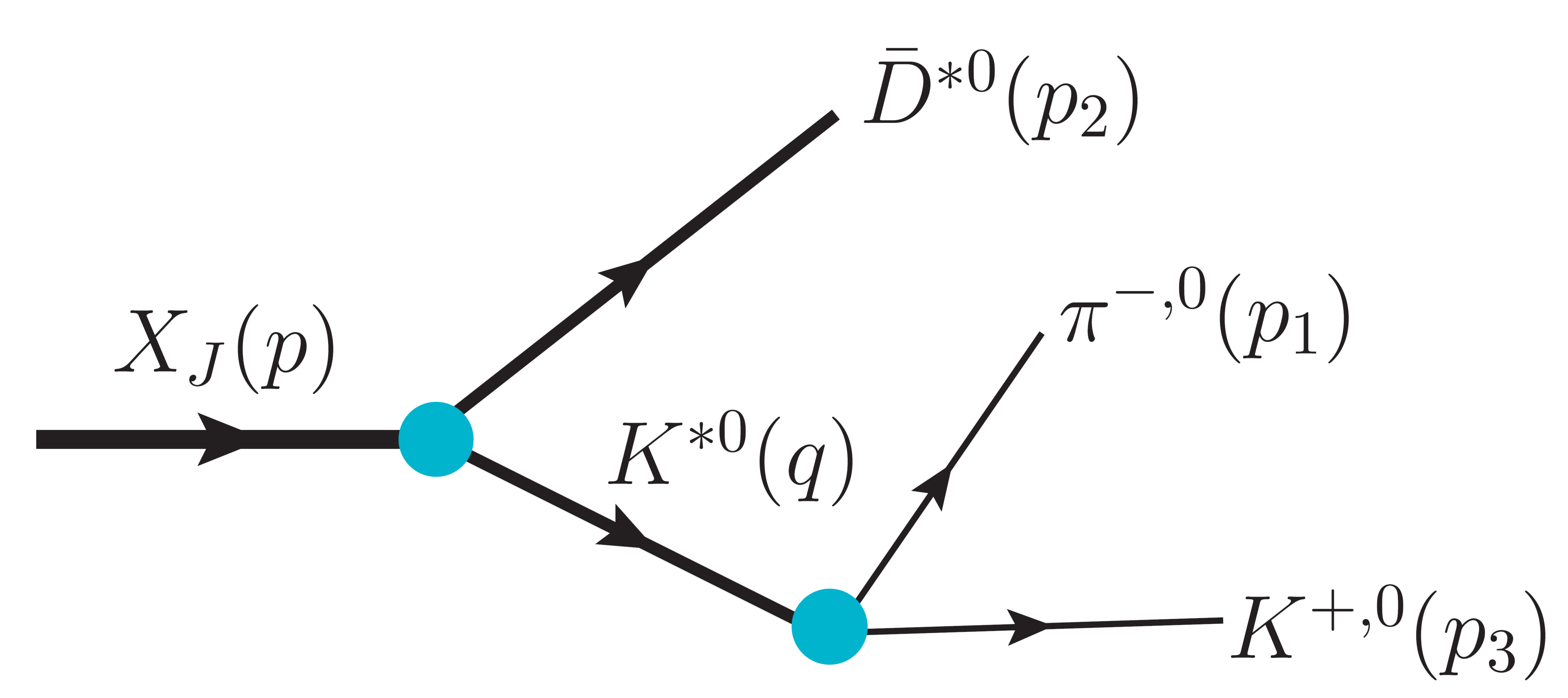}}
~~
\subfigure[]{\includegraphics[width=0.3\textwidth]{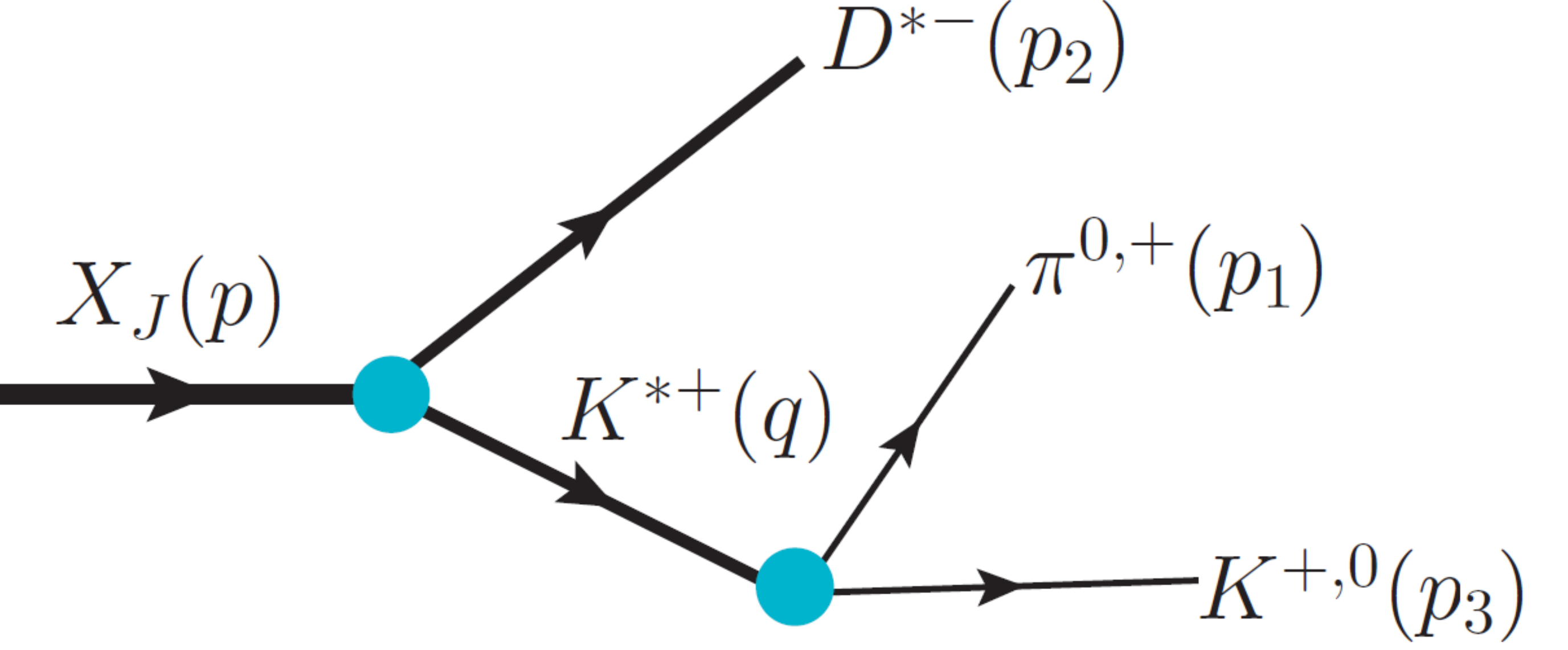}}
\end{center}
\caption{The triangle diagrams contributing to the decay processes of the $X_{J(J=0,1,2)}$ into the (a) $D^-K^+$, (b) $D^{*-}K^+$, and (c) $D^-K^{*+}$ channels, and the tree diagrams contributing to the decay processes of the $X_{J(J=0,1,2)}$ into the (d) $\bar D^{*0} K^{*0}(\rightarrow K \pi)$ and (e) $D^{*-} K^{*+}(\rightarrow K \pi)$ channels. Source: Refs.~\cite{Xiao:2020ltm,Huang:2020ptc}.}
\label{fig:X2900decay}
\end{figure}

The authors of Ref.~\cite{Xiao:2020ltm} assigned the $X_0(2900)$ to be an $S$-wave isoscalar $\bar D^* K^*$ hadronic molecule, and applied the effective Lagrangian approach to calculate its partial decay widths. They estimated the contributions from several triangle diagrams, as shown in Fig.~\ref{fig:X2900decay}(a,b,c), and their results are in agreement with the experimental data when using the cut-off parameters around 1~GeV. Within these constrained parameters, they further studied the other two $S$-wave isoscalar $\bar D^* K^*$ hadronic molecules $X_{J(J=1,2)}$ and calculated their partial decay widths. A similar study was done in Ref.~\cite{Huang:2020ptc}, where the authors studied the $\bar D^* K^*$ hadronic molecules through their two-body decays into $D^-K^+$ via triangle diagrams. They also considered the three-body decays into the $\bar D^* K \pi$ final state, as shown in Fig.~\ref{fig:X2900decay}(d,e). Their results suggest that the $X_0(2900)$ has a large $\bar D^* K^*$ component but with a non-negligible compact tetraquark component, while the $X_1(2900)$ can not be explained in the hadronic molecular picture.

The above LHCb experiment~\cite{LHCb:2020bls,LHCb:2020pxc} has an interesting feature that the higher resonance $X_1(2900)$ of $J^P = 1^-$ has a width significantly larger than the lower one $X_0(2900)$ of $J^P = 0^+$, although the $X_1(2900) \to D^-K^+$ is a $P$-wave decay and the $X_0(2900) \to D^-K^+$ is an $S$-wave decay. Paying attention to this feature, their possible interpretations were studied in Ref.~\cite{Chen:2020aos} using the method of QCD sum rules. Their results suggest that the $X_0(2900)$ can be interpreted as the $S$-wave $D^{*-}K^{*+}$ molecule state of $J^P = 0^+$, and the $X_1(2900)$ can be interpreted as the $P$-wave $u d \bar s \bar c$ compact tetraquark state of $J^P = 1^-$. In Ref.~\cite{Agaev:2020nrc} the authors investigated the $X_0(2900)$ and applied the QCD sum rule method to calculate its mass and width. Their results support it to be the $\bar D^{*0}K^{*0}$ molecule state of $J^P = 0^+$.

Later in Ref.~\cite{Chen:2021erj} the author further took into account the isospin quantum number, and the updated QCD sum rule analysis supports the interpretation of the $X_0(2900)$ as the $\bar D^* K^*$ molecular state of $(I)J^P = (0)0^{+}$. At the same time its decay properties as well as its productions in $B$ and $B^*$ decays were investigated through the current algebra. Recently in Ref.~\cite{Chen:2021xlu}, the author improved the QCD sum rule method, and proposed a possible binding mechanism induced by the exchanged light quarks. This mechanism is similar to the covalent bond in the chemical molecules induced by the shared electrons. The obtained results support the interpretation of the $X_0(2900)$ as the $\bar D^* K^*$ covalent molecular state of $(I)J^P = (0)0^{+}$ bound by the shared $ud$ quarks.

Since the $X_1(2900)$ lies just below the $\bar D_1 K$ threshold, it is natural to investigate whether it can be interpreted as the $\bar D_1 K$ molecular state or not~\cite{He:2020btl}. This assignment was supported by Ref.~\cite{Qi:2021iyv}, where the authors investigated the $X_1(2900)$ as an $S$-wave $\bar D_1 K$ molecular state through the Bethe-Salpeter equation approach. Later in Ref.~\cite{Chen:2021tad} the authors analyzed the LHCb data~\cite{LHCb:2020bls,LHCb:2020pxc} with the couple-channel $K$-matrix approaches as well as the Flatt{\'e}-like parameterization. Their results favor the interpretation of the $X_1(2900)$ as the $\bar D_1 K$ molecular state of $(I)J^P = (0)1^-$. However, the authors of Ref.~\cite{Dong:2020rgs} applied the vector-meson-exchange model to study this possibility and found that the potential between $\bar D_1$ and $K$ is attractive but too weak to form any bound state.

\subsubsection{Compact tetraquark picture.}
\label{sec4.1.2}

Some theoretical studies support the interpretation of the $X_0(2900)$ and $X_1(2900)$ as the $ud \bar s \bar c$ compact tetraquark states, but some do not.

\begin{figure}[hbtp]
\begin{center}
\includegraphics[width=0.6\textwidth]{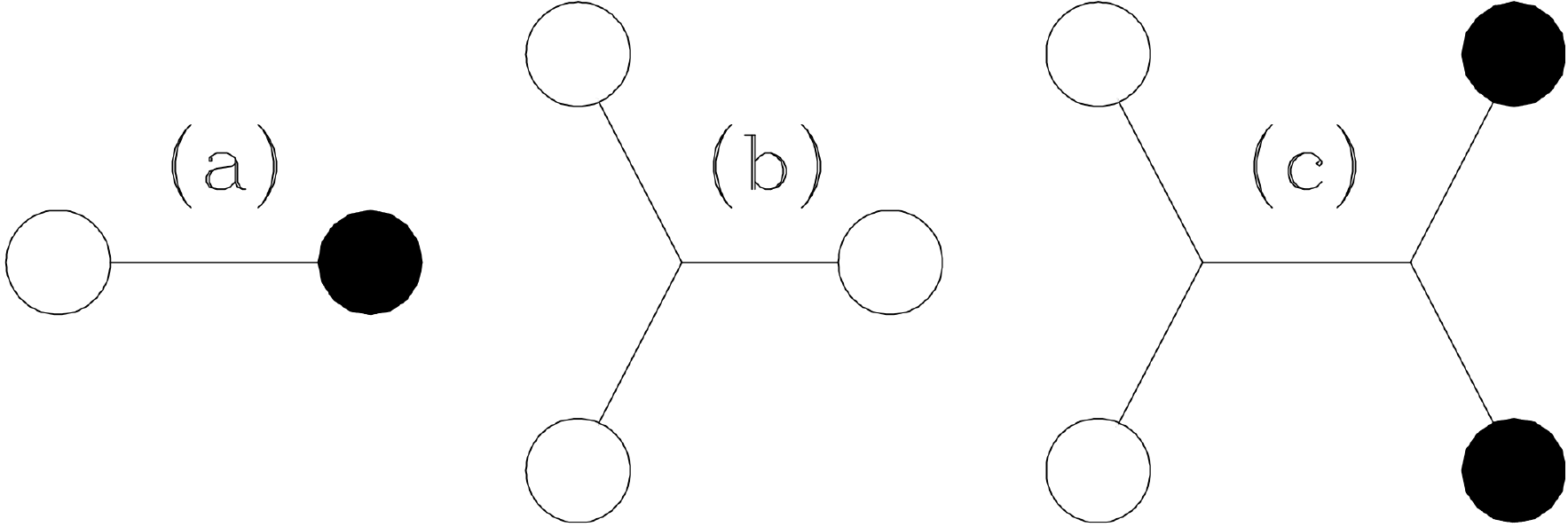}
\end{center}
\caption{QCD strings connecting quarks (open circles) and antiquarks (filled circles) within the string-junction picture. Source: Ref.~\cite{Karliner:2020vsi}.}
\label{fig:X2900string}
\end{figure}

In Ref.~\cite{Karliner:2020vsi} the authors investigated the string-junction picture with universal quark masses for the mesons and baryons, as shown in Fig~\ref{fig:X2900string}. They interpreted the $X_0(2900)$ as a $ud \bar s \bar c$ isosinglet compact tetraquark state, whose mass was calculated to be $2863\pm12$~MeV. They also predicted its analogous $ud \bar s \bar b$ tetraquark state to be at $6213\pm12$~MeV. Later in Ref.~\cite{He:2020jna} the authors applied the two-body chromomagnetic interactions to explain the $X_0(2900)$ and $X_1(2900)$ as the radially excited $ud \bar s \bar c$ tetraquark state of $J^P = 0^+$ and the orbitally excited one of $J^P = 1^-$, respectively. They further used the flavor $SU(3)$ symmetry to investigate their partners made of $d s \bar u \bar c$ and $us \bar d \bar c$, and proposed to observe them through the decays $X_{d s \bar u \bar c} \to D^-K^-/D^-_s \pi^-$ and $X_{us \bar d \bar c} \to D^-_s \pi^+$. The chromomagnetic interaction model was applied in Refs.~\cite{Cheng:2020nho,Guo:2021mja} to support the assignment of the $X_0(2900)$ as a compact $ud \bar s \bar c$ tetraquark state of $J^P = 0^+$.

The method of QCD sum rules were used in Refs.~\cite{Zhang:2020oze,Wang:2020xyc,Agaev:2021knl,Ozdem:2022ydv} to investigate the $X_0(2900)$ and $X_1(2900)$ as the $ud \bar s \bar c$ compact tetraquark states. In Ref.~\cite{Zhang:2020oze} the author studied the $ud \bar c \bar s$ tetraquark state using four configurations of interpolating currents with $J^P = 0^+$, and his result supports the interpretation of the $X_0(2900)$ as a $ud \bar c \bar s$ tetraquark state of $J^P = 0^+$. Later in Ref.~\cite{Wang:2020xyc} the author applied the QCD sum rule method to study the $ud \bar c \bar s$ tetraquark states with the $J^P = 0^+$ interpolating current of the axialvector-diquark-axialvector-antidiquark type. Its mass was calculated to be $2.91\pm0.12$~GeV, supporting the assignment of the $X_0(2900)$ as a scalar tetraquark state. In a recent QCD sum rule study~\cite{Agaev:2021knl} the authors modeled the $X_1(2900)$ as an exotic vector state built of the light diquark $u^T C \gamma_5 d$ and heavy antidiquark $\bar c \gamma_\mu \gamma_5 C \bar s^T$. Its mass and width were calculated to be $M = 2890 \pm 122$~MeV and $\Gamma = 93 \pm 13$~MeV respectively, consistent with the experimental parameters of $X_1(2900)$.

However, some theoretical studies do not support the compact tetraquark picture in explaining the $X_0(2900)$. In Ref.~\cite{Lu:2020qmp} the authors systematically calculated the mass spectrum of the open heavy flavor tetraquark states $qq \bar q \bar Q$ within an extended relativized quark model. They applied the variational method to solve a four-body relativized Hamiltonian, including the Coulomb potential, confining potential, spin-spin interactions, and relativistic corrections. They calculated the masses of four scalar $ud \bar s \bar c$ states to be 2765, 3065, 3152, and 3396~MeV, disfavoring the assignment of the $X_0(2900)$ as a compact tetraquark state. A similar study was done in Ref.~\cite{Wang:2020prk}, where the authors studied the $S$-wave $qq\bar s\bar c$ states in the compact tetraquark scenario with the quark model, considering the Coulomb, linear confinement, and hyperfine interactions. They calculated the mass and width of the $(I)J^P=(0)0^+$ state to be $M = 2649$~MeV and $\Gamma = 48.1$~MeV respectively, which can not be used to explain the $X_0(2900)$; while they calculated the mass and width of the $(I)J^P=(1)0^+$ state to be $M = 2941$~MeV and $\Gamma = 26.6$~MeV respectively, which may be a good candidate for the $X_0(2900)$.

\begin{figure}[hbtp]
\begin{center}
\includegraphics[width=0.4\textwidth]{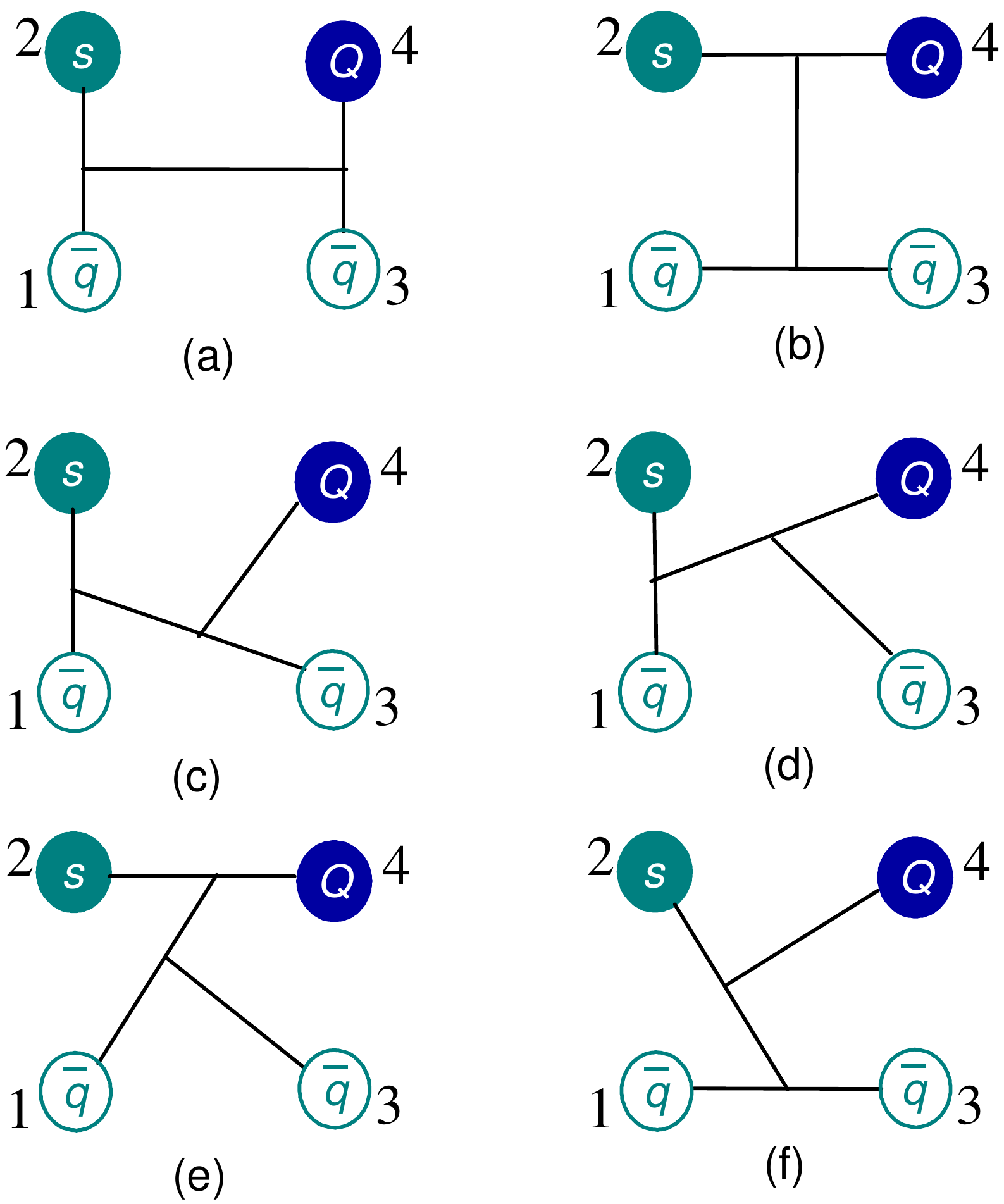}
\end{center}
\caption{Possible configurations for the $qq \bar s\bar Q$ ($q=u/d$, $Q=c/b$) tetraquark states: (a) meson-meson, (b) diquark-antidiquark, and (c-f) $K$-type arrangements of quarks. Source: Ref.~\cite{Yang:2021izl}.}
\label{fig:X2900config}
\end{figure}

Later in Ref.~\cite{Yang:2021izl} the authors investigated the low-lying $qq \bar s\bar Q$ tetraquark states of $(I)J^P = (0/1)0^+/1^+/2^+$ using a chiral quark model. They took into account many possible tetraquark configurations, including the meson-meson and diquark-antidiquark configurations as well as the $K$-type arrangements of quarks, as shown in Fig.~\ref{fig:X2900config}. They applied the variational method to solve the four-body Schr\"odinger-like equation, but the obtained $X_{0,1}(2900)$ signals are unstable in the single-channel calculation. They only found one candidate of $(I)J^P=(0)1^+$ in the coupled-channel calculation with the mass 2.94~GeV, which may be a good candidate for the $X_1(2900)$.

\subsubsection{More theoretical studies.}
\label{sec4.1.3}

In Refs.~\cite{Albuquerque:2020ugi,Narison:2021vfl,Albuquerque:2022weq} the authors performed a comprehensive QCD sum rule analysis of the $J^P = 0^+$ and $1^-$ $ud\bar s \bar c$ tetraquark states using the interpolating currents given in Table~\ref{sec4:X2900current}. They considered the stability criteria by including the factorised perturbative NLO corrections and the contributions of quark and gluon condensates up to dimension-6 in the OPE. They found three almost degenerate states of $J^P = 0^+$ through the interpolating currents ${\cal O}^0_{SS}$, ${\cal O}^0_{VV}$, and ${\cal O}^0_{D^*K^*}$, whose masses were calculated to be:
\begin{eqnarray}
\nonumber M_{SS} &=& 2736\pm21~{\rm MeV} \, ,
\\ M_{AA} &=& 2675\pm65~{\rm MeV} \, ,
\\ \nonumber M_{D^*K^*} &=& 2808\pm41~{\rm MeV} \, .
\end{eqnarray}
Then they assumed the physical state to be a superposition of these hypothetical states. This physical state is called ``Tetramole'', whose mass was calculated to be
\begin{equation}
M_{Tetramole;0^+} = 2743\pm18~{\rm MeV} \, .
\end{equation}
Therefore, it may be a good candidate for the $X_0(2900)$ though its mass is lighter. This discrepancy was reduced by further considering the mixing of the Tetramole state with the radially excited $\bar D K$ molecule. Another Tetramole state of $J^P = 1^-$ was similarly constructed in Ref.~\cite{Albuquerque:2020ugi} to explain the $X_1(2900)$. More theoretical investigations considering the hadronic molecular and compact tetraquark pictures simultaneously can be found in Refs.~\cite{Xue:2020vtq,Mutuk:2020igv}.

\begin{table*}[hbtp]
\renewcommand{\arraystretch}{1.5}
\scriptsize
\caption{Interpolating currents describing the scalar $(0^+)$ and vector $(1^-)$ hadronic molecules and compact tetraquarks. Source: Ref.~\cite{Albuquerque:2020ugi}.}
\centering
\begin{tabular}{ll}
&
\\ \hline \hline
Scalar states ($0^+$) & Vector states ($1^-$) \\
\hline
  Tetraquarks \\
 $ {\cal O}^0_{SS} = \epsilon_{i j k} \:\epsilon_{m n k} \left(
  u_i^T\, C \gamma_5 \,d_j \right) \left( \bar{c}_m\,
  \gamma_5 C \,\bar{s}_n^T\right) $ &  ${\cal O}^1_{AP} = \epsilon_{m n k}\: \epsilon_{i j k}\left( \bar{c}_m\, \gamma_\mu C \,
  \bar{s}_n^T\right) \left(
  u_i^T\, C \,d_j \right) $ \\

$   {\cal O}^0_{PP} = \epsilon_{i j k} \:\epsilon_{m n k} \left(
  u_i^T\, C \,d_j \right) \left( \bar{c}_m\,
  C \,\bar{s}_n^T\right) $ & ${\cal O}^1_{PA} = \epsilon_{m n k}\: \epsilon_{i j k}\left( \bar{c}_m\,  C \,
  \bar{s}_n^T\right) \left(
  u_i^T\, C\gamma_\mu \,d_j \right) $  \\

 $  {\cal O}^0_{VV} = \epsilon_{i j k} \:\epsilon_{m n k} \left(
  u_i^T\, C \gamma_5 \gamma_\mu \,d_j \right) \left( \bar{c}_m\,
  \gamma^\mu \gamma_5 C \,\bar{s}_n^T\right) $ & $ {\cal O}^1_{SV} = \epsilon_{i j k} \:\epsilon_{m n k} \left(
  u_i^T\, C \gamma_5 \,d_j \right) \left( \bar{c}_m\,
  \gamma_\mu \gamma_5 C \,
  \bar{s}_n^T\right) $ \\

 $  {\cal O}^0_{AA} = \epsilon_{i j k} \:\epsilon_{m n k} \left(
  u_i^T\, C \gamma_\mu \,d_j \right) \left( \bar{c}_m\,
  \gamma^\mu C \,\bar{s}_n^T\right) $ & $ {\cal O}^1_{VS} =\epsilon_{i j k} \:\epsilon_{m n k} \left(
  u_i^T\, C \gamma_5 \gamma_\mu \,d_j \right) \left( \bar{c}_m\,
  \gamma_5 C \,
  \bar{s}_n^T\right) $\\
  Molecules \\
$  {\cal O}^0_{DK}=(\bar c\gamma_5 d)(\bar s\gamma_5 u)$& ${\cal O}^1_{D_1 K} =\left(\bar{c} \gamma_\mu \gamma_5 d \right)
  \left( \bar{s} \gamma_5\,u\right) $  \\
${\cal O}^0_{D^*K^*}=(\bar c\gamma^\mu d)(\bar s\gamma_\mu u)$&$ {\cal O}^1_{D K_1} = \left(\bar{c} \gamma_5 d \right)
  \left( \bar{s}\, \gamma_\mu \gamma_5 \,u\right) $\\
$ {\cal O}^0_{D_1K_1}=(\bar c\gamma^\mu\gamma_5 d)(\bar s\gamma_
  \mu\gamma_5 u)$ & $ {\cal O}^1_{D^* K^*_0} = \left(\bar{c} \gamma_\mu  d \right)
  \left( \bar{s}\,u\right) $ \\
 $  {\cal O}^0_{D^*_0K^*_0}=(\bar c d)(\bar s u)$ &
$ {\cal O}^1_{D^*_0 K^*} =   \left(\bar{c} \,d \right)
  \left( \bar{s} \,\gamma_\mu \,u\right)$ \\
\\ \hline\hline
\end{tabular}
\label{sec4:X2900current}
\end{table*}

\begin{figure}[hbtp]
\begin{center}
\includegraphics[width=0.6\textwidth]{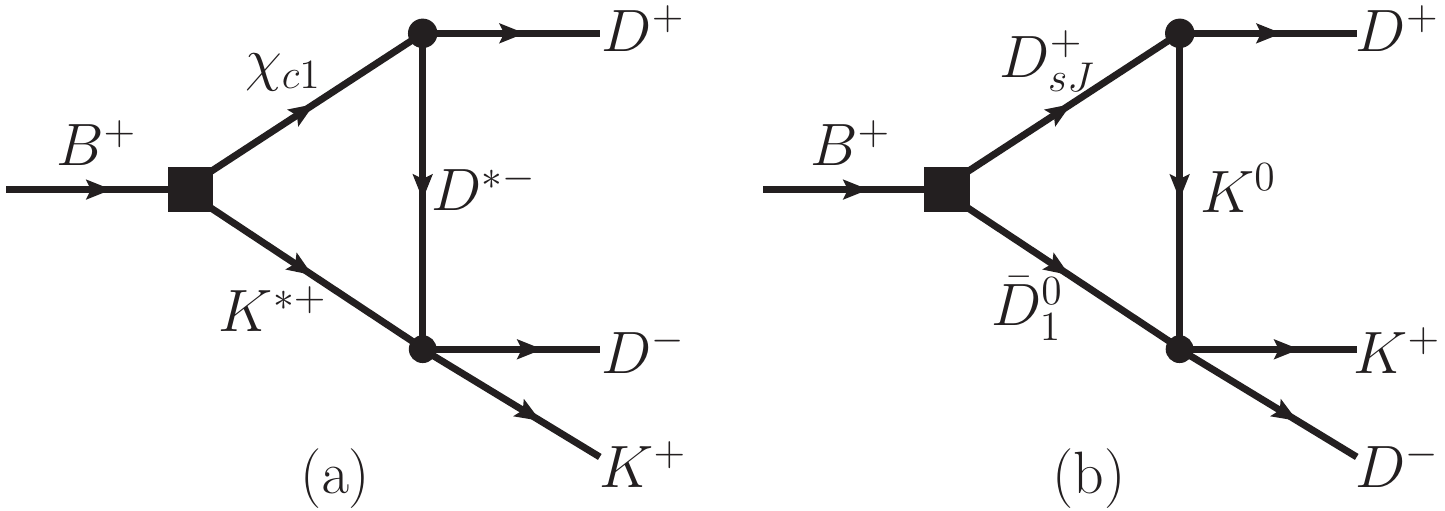}
\end{center}
\caption{The $B^+ \to D^+D^-K^+$ decay process via the triangle rescattering diagrams. Source: Ref.~\cite{Liu:2020orv}.}
\label{fig:X2900triangle}
\end{figure}

Besides the hadronic molecular and compact tetraquark pictures, the triangle singularity was used in Ref.~\cite{Liu:2020orv} to explain the $X_{0,1}(2900)$, where the authors examined the rescattering processes that may contribute to the $B^+ \to D^+D^-K^+$ decay, as depicted in Fig.~\ref{fig:X2900triangle}. Two resonance-like peaks around the $D^{*-}K^{*+}$ and $\bar D^0_1K^0$ thresholds were obtained in the $D^-K^+$ invariant mass spectrum, which can be used to simulate the $X_0(2900)$ and $X_1(2900)$ states without introducing any genuine exotic state.

\begin{figure}[hbtp]
\begin{center}
\includegraphics[width=1\textwidth]{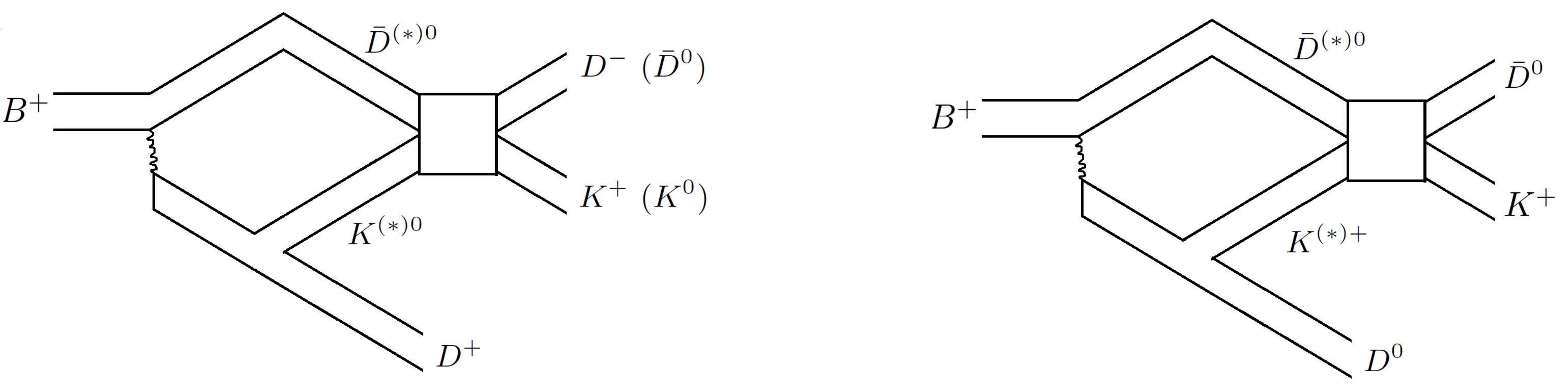}
\end{center}
\caption{Production of the $X_{0,1}(2900)$ states (left) and their possible charged partner states (right) in $B^+$ decays. Source: Ref.~\cite{Burns:2020xne}.}
\label{fig:X2900production}
\end{figure}

In order to discriminate among possible interpretations of the $X_{0,1}(2900)$, the authors of Refs.~\cite{Burns:2020xne,Burns:2020epm} performed an exhaustive analysis on their decay behaviors as well as their productions in $B^0$ and $B^+$ decays, as shown in Fig.~\ref{fig:X2900production}. They considered a number of competing models, including triangle diagrams mediated by quark exchange or pion exchange, and resonance scenarios including hadronic molecules and compact tetraquarks. They found characteristic differences in the predictions of these different models. We refer interested readers to these references for detailed discussions, some of which are listed as follows:
\begin{itemize}

\item The triangle scenario with the quark exchange is characterized by the striking prediction that in $B^+ \to D^+ X$, the $X(2900)$ states are seen in $D^-K^+$ but not in $\bar D^0K^0$, whereas in $B^0 \to D^0X$ decay the pattern reverses, with the states seen in $\bar D^0K^0$ but not in $D^-K^+$. The modes which are forbidden in this scenario are allowed in the alternative triangle scenario where the interactions are based on the one pion exchange. Their fit fractions, however, may be smaller than those in the resonance scenario.

\item In the resonance scenario, the neutral $X(2900)$ states have the same fit fractions regardless of whether they are $I = 0$ or $I = 1$. The two possibilities, hadronic molecules and compact tetraquarks, would instead be distinguished by the existence of a charged partner in the latter case, which has enormous fit fractions even exceeding that of the observed $X_1(2900)$ in its discovery mode.

\end{itemize}

Branching fractions of the $B^- \to D^- X_{0,1}(2900)$ decays were calculated in Ref.~\cite{Chen:2020eyu} using the rescattering mechanism. Productions of the $X_{0,1}(2900)$ states in the $\Lambda_b$ and $\Xi_b$ decays were studied in Ref.~\cite{Hsiao:2021tyq}, where the authors also investigated their partners $X^\prime_{0,1}$ with the quark content $us\bar d \bar c$. They estimated their branching fractions $\mathcal{B}(\Lambda_b^0 \to \Sigma^0_c X^\prime_{0,1}, X^\prime_{0,1} \to D^-_s \pi^+/ \bar D^0 \bar K^0)$ to be at the level of $10^{-4}$, which can be accessible in future experiments.

\subsection{Doubly charmed tetraquark state $T_{cc}^+(3875)$}
\label{sec4.2}

\begin{figure}[hbtp]
\begin{center}
\includegraphics[width=0.6\textwidth]{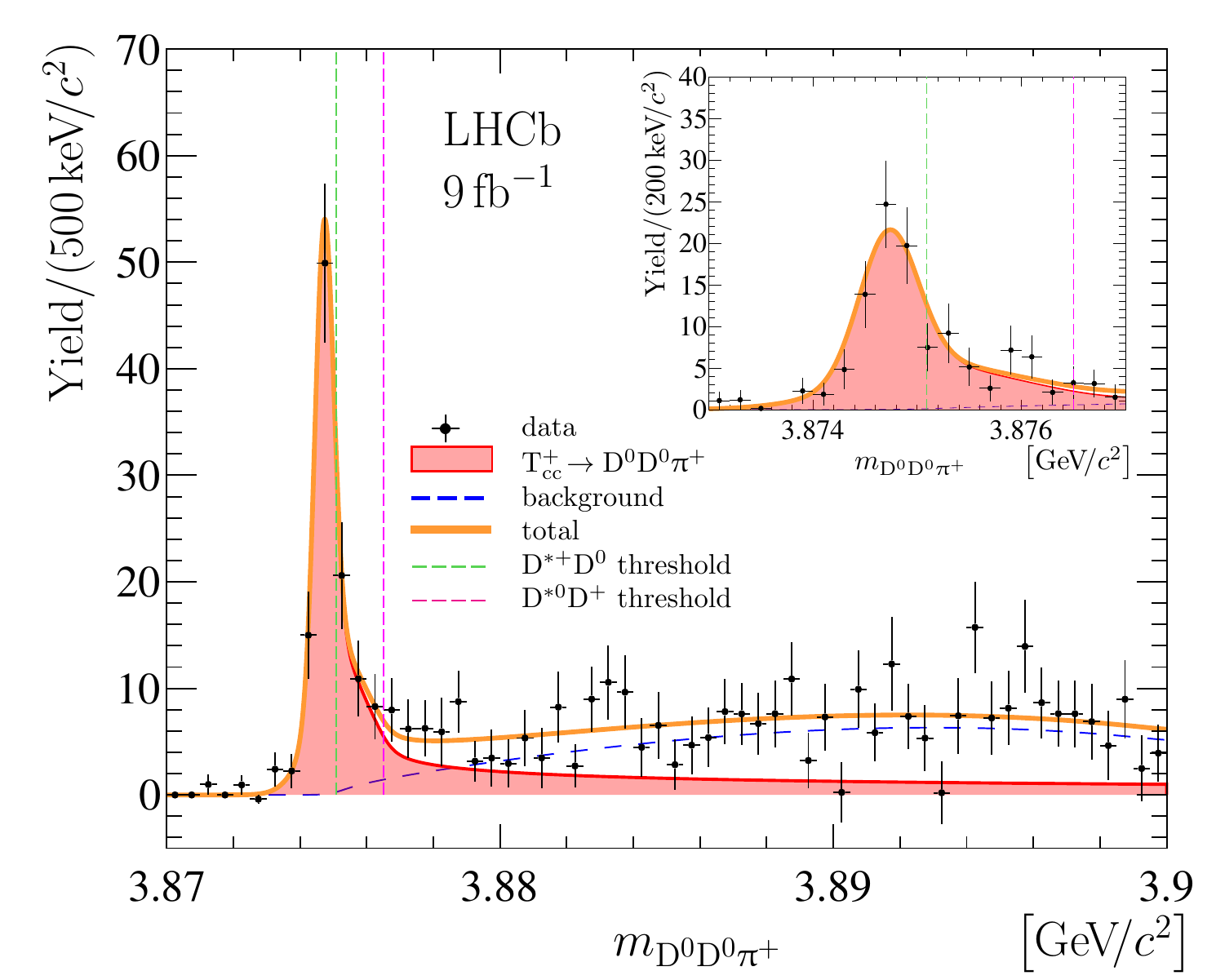}
\end{center}
\caption{The $D^0D^0\pi^+$ mass distribution. Source: Ref.~\cite{LHCb:2021auc}.}
\label{fig:TccDDpi}
\end{figure}

In 2003 the famous charmonium-like state $X(3872)/\chi_{c1}(3872)$ was discovered by the Belle collaboration in the $\pi^+ \pi^- J/\psi$ mass spectrum of the exclusive $B^\pm \to K^\pm \pi^+ \pi^- J/\psi$ decays~\cite{Belle:2003nnu}. Its quark content is $c \bar c q \bar q$ ($q=u/d$), and its partner state with the quark content $c c \bar q \bar q$ may also exist. With the development of experimental techniques in the past twenty years, the LHCb collaboration recently reported the observation of the first doubly charmed tetraquark state $T^+_{cc}$ in the $D^0D^0\pi^+$ mass spectrum just below the $D^{*+}D^0$ mass threshold~\cite{LHCb:2021vvq,LHCb:2021auc}, as depicted in Fig.~\ref{fig:TccDDpi}. This state has the quark content $cc\bar u\bar d$ and the spin-parity quantum number $J^P = 1^+$. Moreover, the LHCb experiment favors an isoscalar state. Using the Breit-Wigner parametrisation, its mass and width were measured to be:
\begin{eqnarray}
T^+_{cc} &:&  \Delta M_{\rm BW} = M_{\rm BW} - ( M_{D^{*+}} + M_{D^0} ) = -273 \pm 61 \pm 5 ^{+11}_{-14} {\rm~keV} \, ,
\\ \nonumber && \Gamma_{\rm BW} = 410 \pm 165 \pm 43 ^{+18}_{-38} {\rm~keV} \, .
\end{eqnarray}
Alternatively, LHCb analyzed their data with a resonance profile more suitable to the closeness of the $T^+_{cc}$ to the $D^{*+}D^0$ threshold, where the mass and width of $T^+_{cc}$ were measured to be:
\begin{eqnarray}
T^+_{cc} &:&  \Delta M_{\rm pole} = M_{\rm pole} - ( M_{D^{*+}} + M_{D^0} ) = -360 \pm 40 ^{+4}_{-0} {\rm~keV} \, ,
\\ \nonumber && \Gamma_{\rm pole} = 48 \pm 2 ^{+0}_{-14} {\rm~keV} \, .
\end{eqnarray}

\begin{figure}[hbtp]
\begin{center}
\includegraphics[width=1\textwidth]{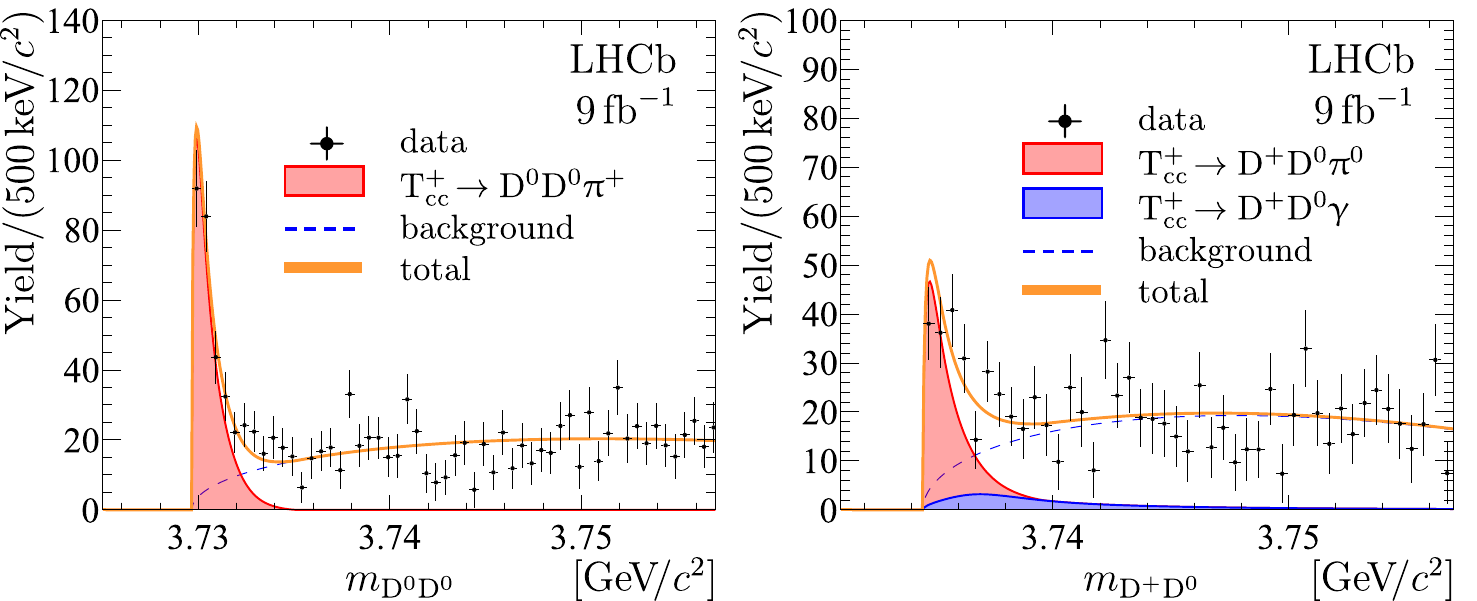}
\end{center}
\caption{Mass distributions for selected $D^0D^0$ (left) and $D^+D^0$ (right) combinations with the non-$D$ background subtracted. Source: Ref.~\cite{LHCb:2021auc}.}
\label{fig:TccDD}
\end{figure}

Besides, the LHCb collaboration examined several important parameters that can be useful to reveal the nature of the $T^+_{cc}$~\cite{LHCb:2021vvq,LHCb:2021auc}:
\begin{itemize}

\item The lower limit on the absolute value of the coupling constant of the $T^+_{cc}$ to the $D^*D$ system was determined to be:
\begin{equation}
|g| > 5.1 (4.3) {\rm~GeV~at~} 90~(95)\% {\rm~CL} \, .
\end{equation}

\item LHCb examined the low-energy limit of the amplitude, and estimated the scattering length $a$, the effective range $r$, and the compositeness $Z$ to be:
\begin{eqnarray}
\nonumber    a &=& \left[ -(7.16 \pm 0.51) + i (1.85 \pm 0.28) \right] {\rm~fm} \, ,
\\          -r &<& 11.9~(16.9){\rm~fm~at~} 90~(95)\% {\rm~CL} \, ,
\\ \nonumber Z &<& 0.52~(0.58){\rm~at~} 90~(95)\% {\rm~CL} \, .
\end{eqnarray}

\item Using $R_{\Delta E}$ to denote the characteristic size calculated from the binding energy and $R_a$ to denote the one calculated from the scattering length, LHCb determined these two parameters to be:
\begin{eqnarray}
R_{\Delta E} &=& 7.49 \pm 0.42 {\rm~fm} \, ,
\\ \nonumber R_a &=& 7.16 \pm 0.51 {\rm~fm} \, .
\end{eqnarray}
They are consistent with each other, and both correspond to a spatial extension significantly exceeding the typical scale for the heavy-flavor hadrons.

\item LHCb examined the $D^0D^0$ mass distribution of the $T^+_{cc} \to D^0D^0 \pi^+$ decay, and observed a narrow peak just above the $D^0D^0$ mass threshold, as shown in the left panel of Fig.~\ref{fig:TccDD}. This peak is (probably) not a genuine exotic state, but due to the proximity of the $T^+_{cc}$ to the $D^{*+}D^0$ mass threshold and the small energy released in the $D^{*+} \to D^0 \pi^+$ decay. A similar peak just above the $D^+D^0$ mass threshold was also observed, as shown in the right panel of Fig.~\ref{fig:TccDD}.

\end{itemize}

\subsubsection{Earlier theoretical studies.}
\label{sec4.2.1}

The doubly heavy tetraquark states $QQ\bar q\bar q$ ($Q=c/b$ and $q=u/d$) with two heavy quarks and two light antiquarks have been widely discussed since forty years ago, based on both the compact tetraquark and hadronic molecular pictures.

\begin{figure}[hbtp]
\begin{center}
\subfigure[]{\includegraphics[width=0.3\textwidth]{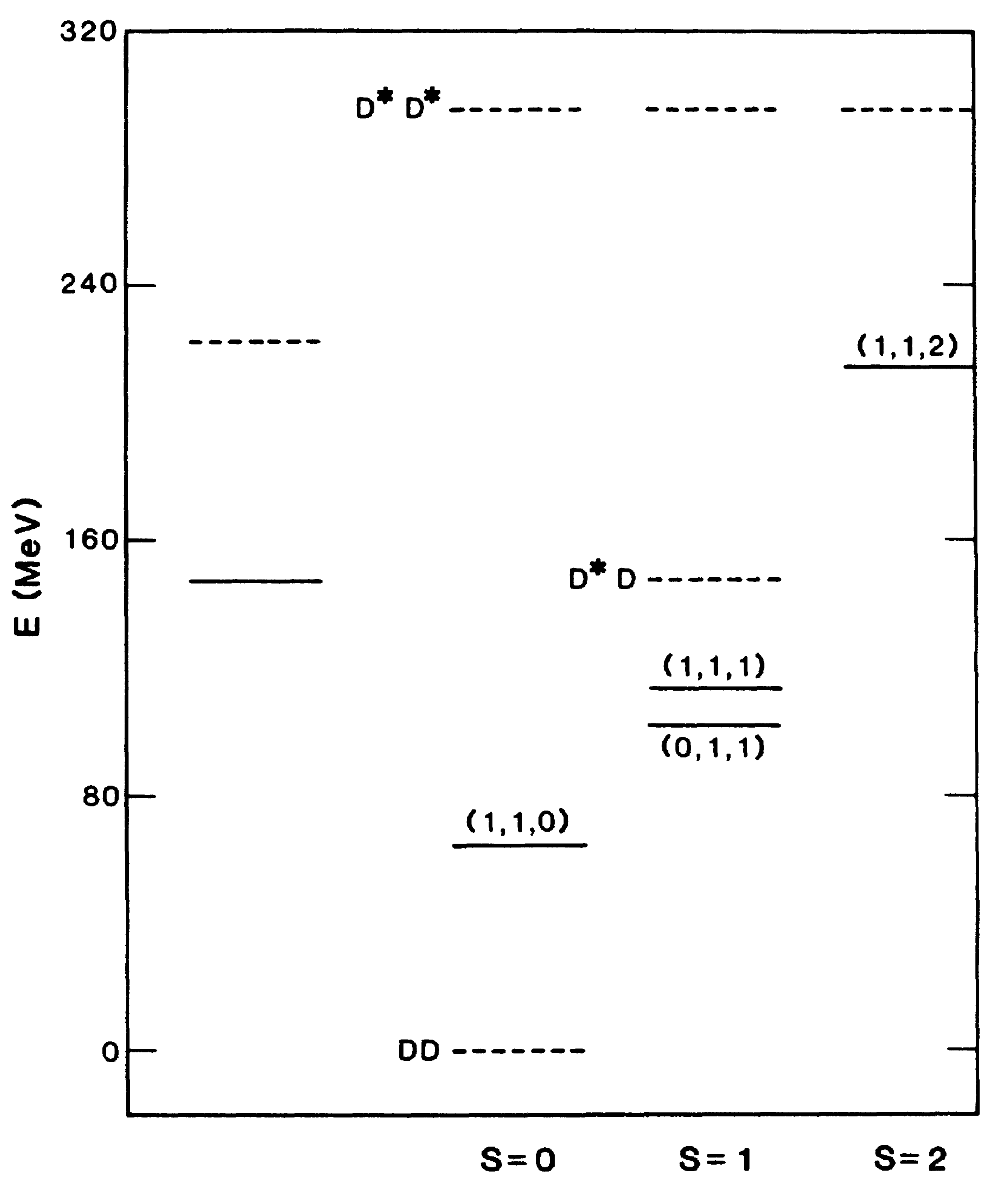}}
~~
\subfigure[]{\includegraphics[width=0.3\textwidth]{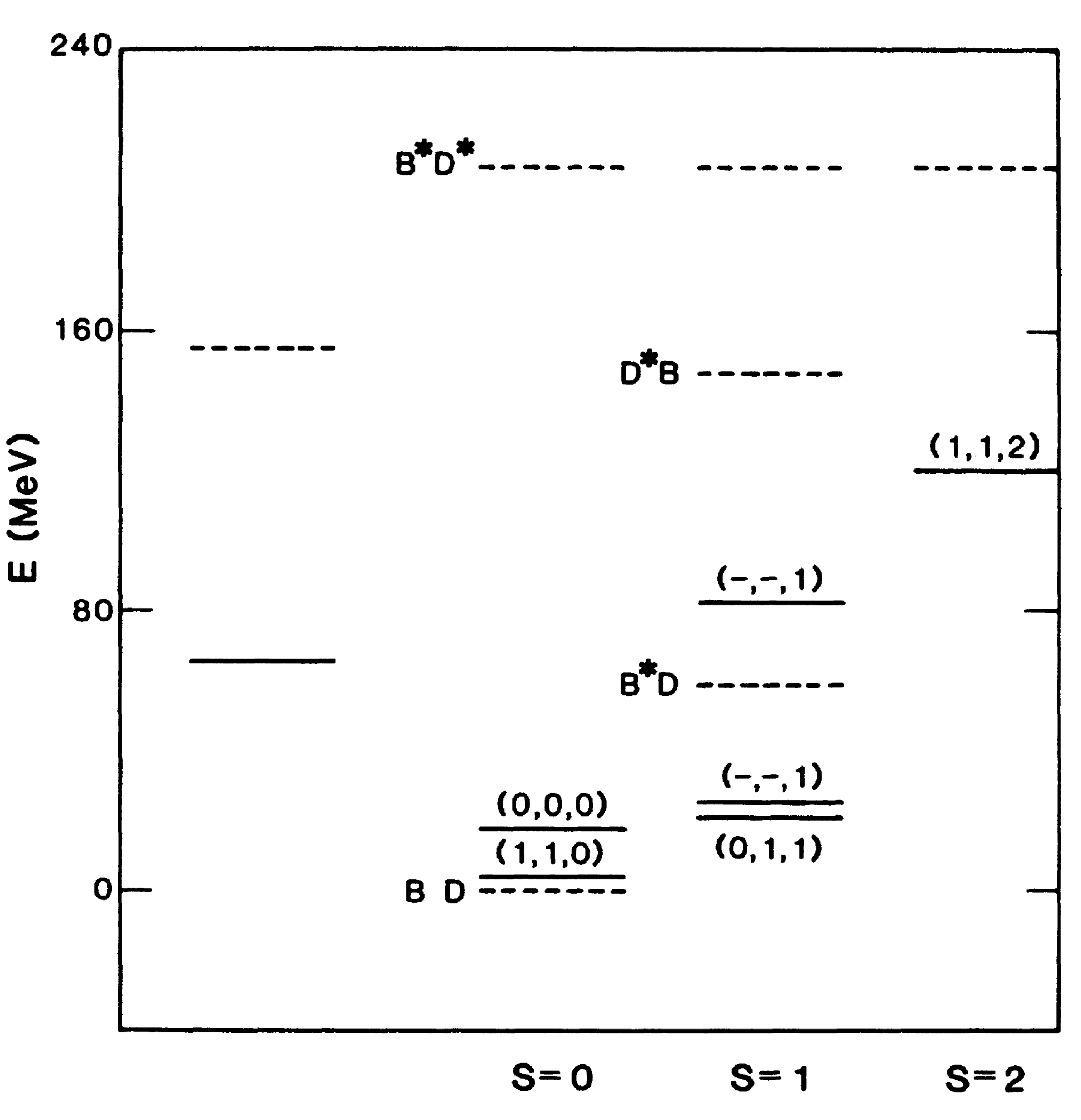}}
~~
\subfigure[]{\includegraphics[width=0.3\textwidth]{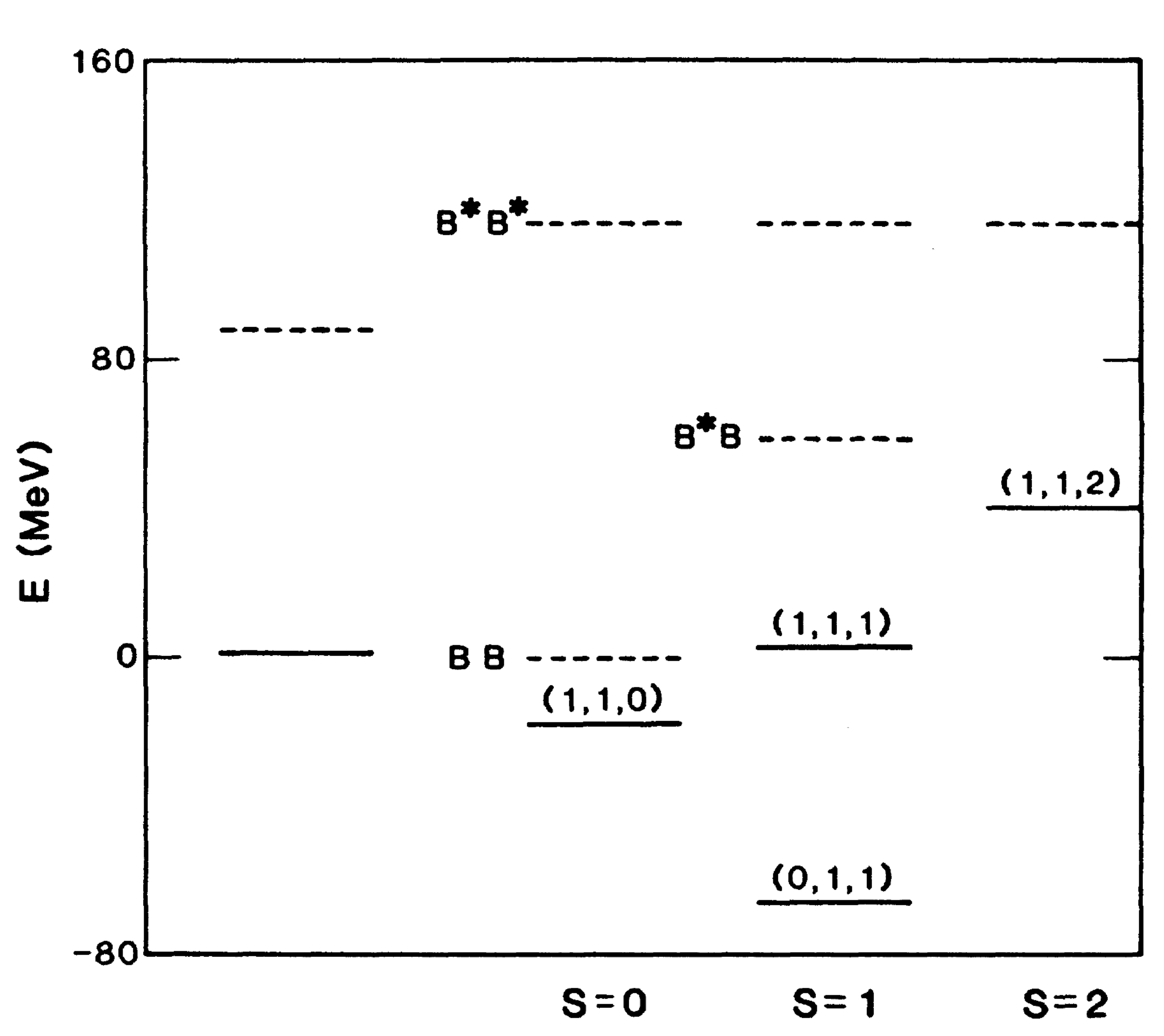}}
\end{center}
\caption{Hyperfine structures of the (a) $cc\bar q\bar q$, (b) $bc\bar q\bar q$, and (c) $bb\bar q\bar q$ systems, labeled by $(S_{\bar q\bar q}, S_{QQ}, S)$. Source: Ref.~\cite{Carlson:1987hh}.}
\label{fig:TccMIT}
\end{figure}

In the 1980's the authors of Refs.~\cite{Ballot:1983iv,Zouzou:1986qh} applied the nonrelativistic quark model to study the $QQ\bar q\bar q$ system. They used the standard two-body potentials, and found the exotic $QQ\bar q\bar q$ bound states with the diquark-antiquark-antiquark structure $[QQ]\bar q\bar q$, analogous to the heavy antibaryons $\bar Q^\prime \bar q \bar q$. One can apply the Pauli principle to the two identical heavy quarks, and naturally arrive at the diquark-antidiquark structure $[QQ][\bar q\bar q]$, which is widely used in the study of the doubly heavy tetraquark states. As discussed at the beginning of this review, there can be two different color configurations for this structure:
\begin{equation}
\mathbf{1}_{[Q Q]_\mathbf{\bar 3} [\bar q \bar q]_\mathbf{3}} {\rm~~and~~} \mathbf{1}_{[Q Q]_\mathbf{6} [\bar q \bar q]_\mathbf{\bar 6}} \, .
\end{equation}
The former antisymmetric configuration may lie lower and is stabler than the latter symmetric one. Especially, for the $S$-wave $cc\bar u\bar d$ state of $(I)J^P = (0)1^+$ with the antisymmetric color configuration $\mathbf{1}_{[cc]_\mathbf{\bar 3} [\bar q \bar q]_\mathbf{3}}$, the two charm quarks have the antisymmetric color, symmetric flavor, and symmetric orbital structures, so they must have the symmetric spin structure $S_{cc} = 1$; the two light quarks have the antisymmetric color, antisymmetric flavor, and symmetric orbital structures, so they must have the antisymmetric spin structure $S_{\bar q \bar q} = 0$. In Ref.~\cite{Carlson:1987hh} the authors labeled this state as $(S_{\bar q\bar q}, S_{QQ}, S) = (0,1,1)$ with $S$ the total spin angular momentum. They studied the $cc\bar q\bar q$, $bc\bar q\bar q$, and $bb\bar q\bar q$ systems using the many-body confining interaction from the MIT bag model. Their results are summarized in Fig.~\ref{fig:TccMIT} together with the relevant two-meson thresholds, where the $(I)J^P = (0)1^+$ $cc\bar u\bar d$ tetraquark state is below the $DD^*$ threshold by about 50~MeV.

Many theoretical studies of the doubly heavy tetraquark states are based on the compact diquark-antidiquark picture through various theoretical methods and models, such as various quark models~\cite{Heller:1986bt,Yan:2018gik,Richard:2018yrm,Qin:2020zlg,Meng:2020knc,Meng:2021yjr}, flux-tube model~\cite{Deng:2018kly}, QCD sum rules~\cite{Navarra:2007yw,Dias:2011mi,Du:2012wp,Agaev:2019qqn}, Bethe-Salpeter equations~\cite{Feng:2013kea}, heavy quark symmetry~\cite{Eichten:2017ffp,Braaten:2020nwp,Cheng:2020wxa}, etc. Theoretical studies based on other compact tetraquark structures that are slightly different from the diquark-antidiquark structure can be found in Refs.~\cite{Lipkin:1986dw,Moinester:1995fk}.

Besides the compact tetraquark picture, the hadronic molecular picture was also used to investigate the doubly heavy tetraquark states~\cite{Ding:2009vj,Molina:2010tx,Xu:2017tsr,Tang:2019nwv}. In Ref.~\cite{Manohar:1992nd} the authors investigated the one pion exchange between the ground-state heavy mesons, and calculated its corresponding long range binding potential using chiral perturbation theory. Their results suggest the existence of the stable doubly heavy tetraquark states $QQ\bar q\bar q$ in the limit where the heavy quark mass $m_Q$ goes to infinity, and especially, there may be the $bb \bar q \bar q$ bound states made of two bottom mesons. However, they also investigated the $D^{(*)}\bar B^{(*)}$ and $D^{(*)}D^{(*)}$ systems, which were found to be unbound. Similar conclusions were obtained in Ref.~\cite{Tornqvist:1993ng} considering the pion exchanges alone. Later the authors of Ref.~\cite{Janc:2004qn} argued that this method may miss some weakly bound states, and they indeed found a weakly bound $DD^{*}$ molecular state in the $(I)J^P=(0)1^+$ channel with the binding energy about $0.6$-$2.7$~MeV.

\begin{figure}[hbtp]
\begin{center}
\subfigure[]{\includegraphics[width=0.45\textwidth]{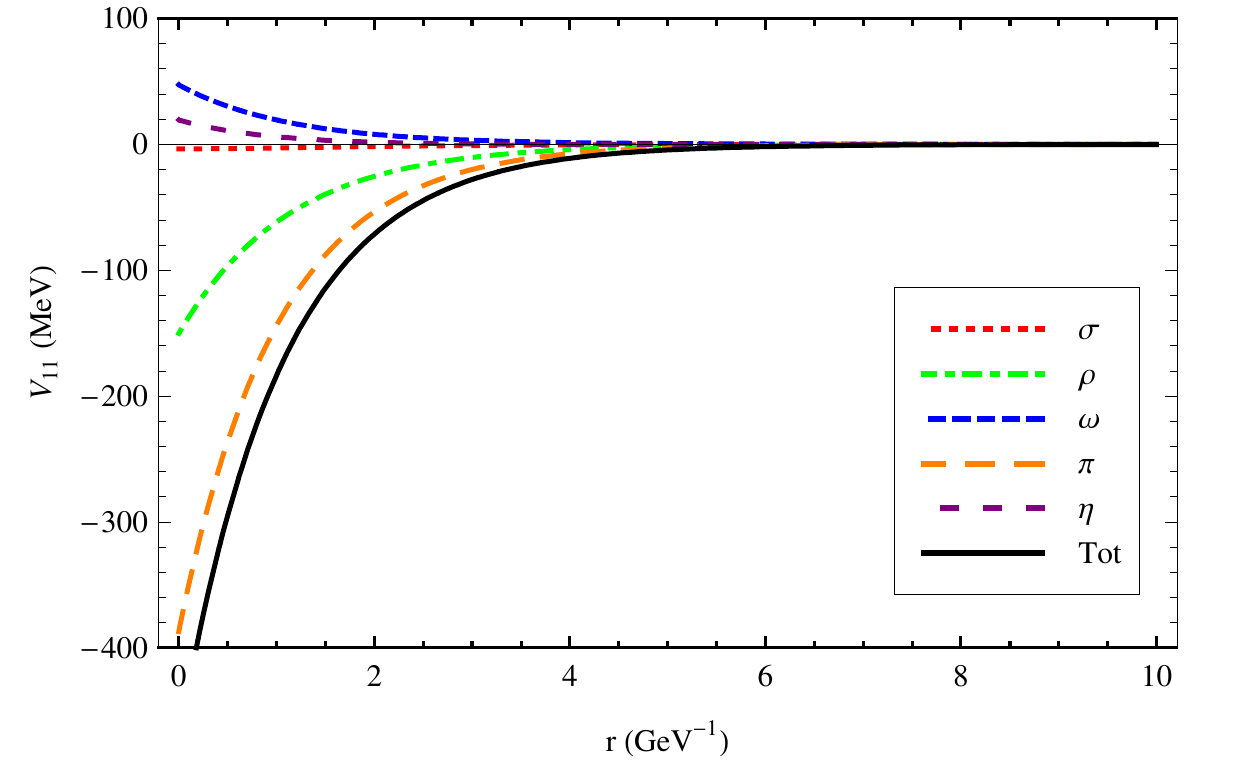}}
~~~
\subfigure[]{\includegraphics[width=0.45\textwidth]{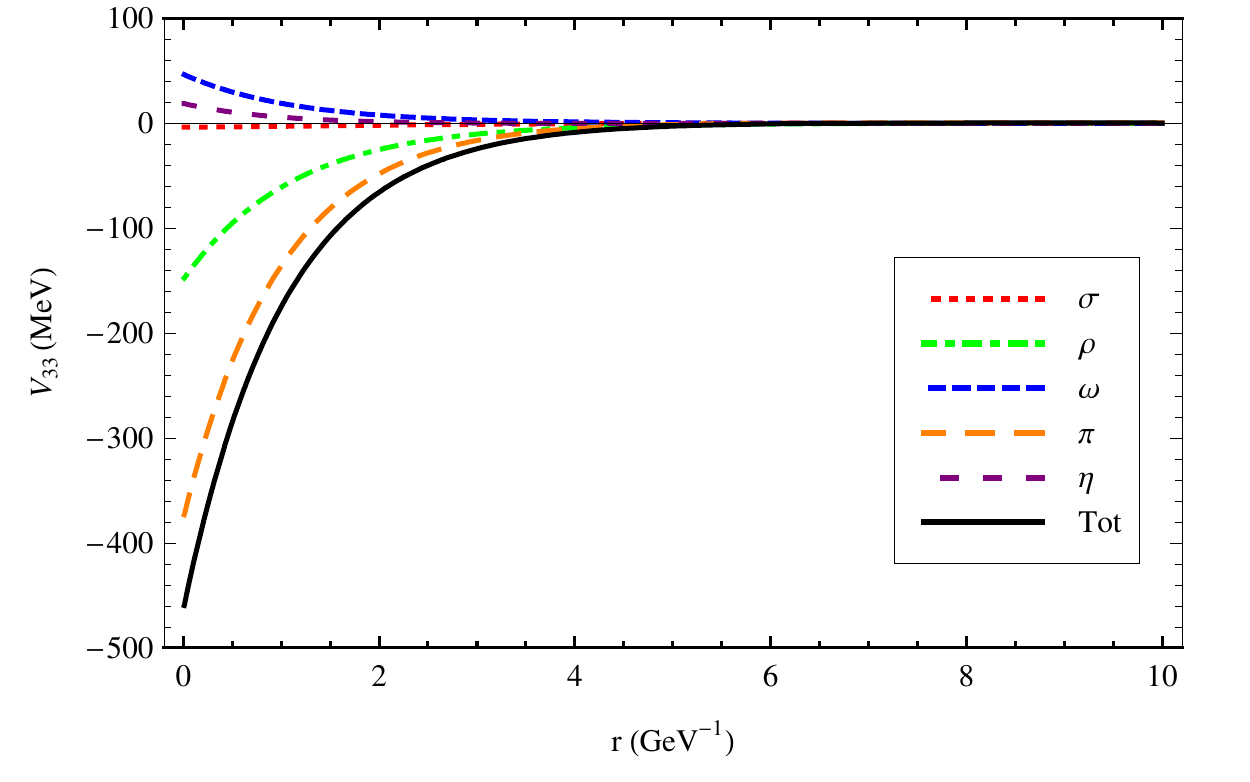}}
\end{center}
\caption{The effective potentials for the (a) $DD^*$ and (b) $D^*D^*$ systems in the $^3S_1$ channel with the cutoff $\Lambda = 1.00$~GeV. Source: Ref.~\cite{Li:2012ss}.}
\label{fig:TccOBE}
\end{figure}

The authors of Ref.~\cite{Ericson:1993wy} realized that some other mesons can also be exchanged when forming hadronic molecules. In Ref.~\cite{Li:2012ss} the long-range pion exchange, the medium-range $\eta/\sigma$ exchanges, and the short-range $\rho/\omega$ exchanges were systematically examined in the $D^{(*)}D^{(*)}$ systems. Their induced potentials are shown in Fig.~\ref{fig:TccOBE} for the $DD^*$ and $D^*D^*$ systems in the $^3S_1$ channel. Their results suggest the existence of the $DD^*$ and $D^*D^*$ hadronic molecules with $(I)J^P = (0)1^+$; the binding energy of the $DD^*$ one is 0.47~MeV when using the one-boson-exchange potential with the cutoff $\Lambda = 0.95$~GeV.

\begin{figure}[hbtp]
\begin{center}
\includegraphics[width=0.8\textwidth]{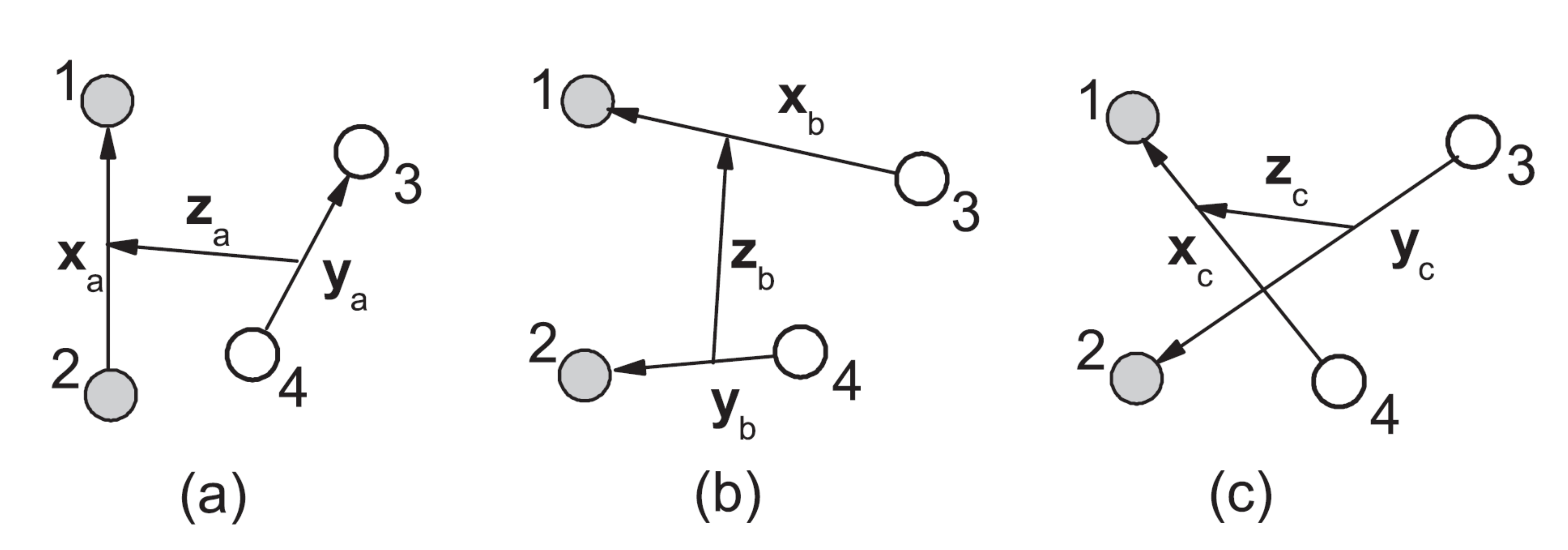}
\end{center}
\caption{Three independent relative coordinates for the $QQ\bar q\bar q$ system in the (a) diquark-antidiquark as well as the (b) direct meson-meson and (c) exchange meson-meson channels. The heavy and light quarks are denoted using the filled and open circles, respectively. Source: Ref.~\cite{Yang:2009zzp}.}
\label{fig:Tccquark}
\end{figure}

Some theoretical studies investigated the compact tetraquark and hadronic molecular structures simultaneously~\cite{Vijande:2009kj,Yang:2019itm,Tan:2020ldi}. In Ref.~\cite{Yang:2009zzp} the authors performed a systematic study on the $S$-wave $QQ\bar q\bar q$ states using the constituent quark model, considering both the meson-meson and diquark-antidiquark structures, as depicted in Fig.~\ref{fig:Tccquark}. Their results, obtained using the quark model proposed in Ref.~\cite{Bhaduri:1981pn}, suggest that the $cc\bar q\bar q$ state of $(I)J^P = (0)1^+$ is bound in a meson-meson structure with the binding energy about 1.8~MeV, but unbound in a diquark-antidiquark structure. They also investigated the mixing of the meson-meson and diquark-antidiquark structures, which enhances the binding energy to be 23.7~MeV.

\begin{figure}[hbtp]
\begin{center}
\includegraphics[width=0.4\textwidth]{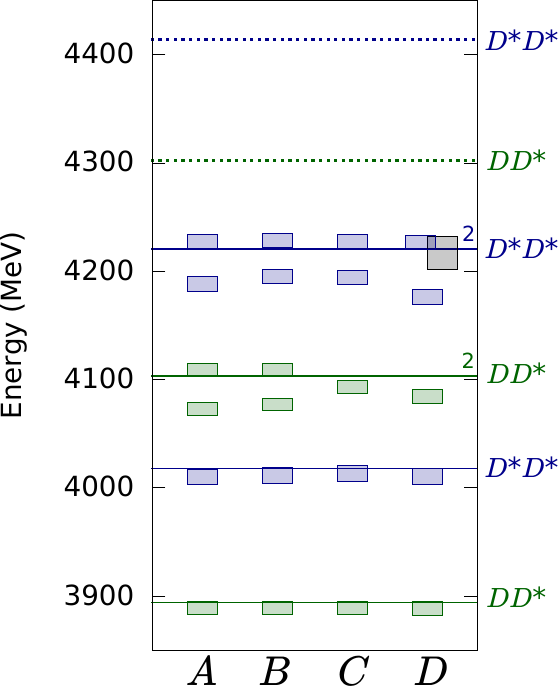}
\end{center}
\caption{Spectrum of the doubly charmed tetraquark states of $(I)J^P=(0)1^+$ with different bases of operators: (A) full basis of meson-meson and tetraquark operators, (B) only meson-meson operators, (C) only meson-meson operators excluding one $DD^*$ operator, and (D) the operators as in C supplemented with the tetraquark operators. Source: Ref.~\cite{Cheung:2017tnt}.}
\label{fig:Tcclattice}
\end{figure}

There have been a number of lattice QCD calculations on the doubly heavy tetraquark states~\cite{Ikeda:2013vwa,Pflaumer:2021ong}. In the bottom sector, there were several lattice QCD studies supporting the existence of the $bb\bar u\bar d$ bound state with $(I)J^P=(0)1^+$~\cite{Brown:2012tm,Bicudo:2012qt,Bicudo:2015kna,Bicudo:2015vta,Bicudo:2016ooe,Bicudo:2021qxj}, and the existence of the $bb\bar u\bar d$ resonance with $(I)J^P=(0)1^-$ was suggested in Ref.~\cite{Bicudo:2017szl}. More rigorous full lattice QCD studies supported the existence of the $bb\bar u\bar d$ bound states, and further predicted the existence of the $bb\bar u\bar s$ bound states~\cite{Francis:2016hui,Francis:2018jyb,Leskovec:2019ioa,Hudspith:2020tdf}. Simulations in the charm sector are more difficult than in the bottom sector. In Ref.~\cite{Cheung:2017tnt} the authors studied the doubly charmed tetraquark states by computing correlation functions involving large bases of the meson-meson and diquark-antidiquark operators. As shown in Fig.~\ref{fig:Tcclattice}, their results in the $(I)J^P=(0)1^+$ channel was not inconsistent with there being an attractive interaction, although there were no obvious signs of a bound $cc\bar u\bar d$ state in this channel. A similar study was done in Ref.~\cite{Junnarkar:2018twb}, where the authors performed a detailed lattice QCD study on the doubly charmed tetraquark states in both the spin-0 and spin-1 sectors. They found the lowest-lying $cc\bar u\bar d$ state of $(I)J^P=(0)1^+$ lies below its threshold by about $23\pm11$~MeV.

\begin{figure}[hbtp]
\begin{center}
\includegraphics[width=1\textwidth]{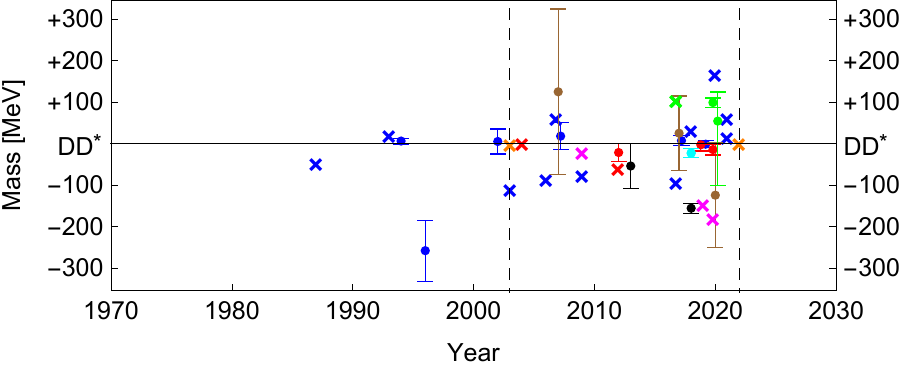}
\end{center}
\caption{Theoretical predictions on the mass of the doubly charmed tetraquark state $cc\bar u\bar d$ with $(I)J^P = (0)1^+$, with uncertainties (error bars) and without uncertainties (crosses), calculated based on the compact tetraquark picture through various quark models~\cite{Carlson:1987hh,Silvestre-Brac:1993zem,Semay:1994ht,Pepin:1996id,Gelman:2002wf,Vijande:2003ki,Cui:2006mp,Vijande:2007rf,Ebert:2007rn,Lee:2009rt,Luo:2017eub,Karliner:2017qjm,Park:2018wjk,Maiani:2019lpu,Lu:2020rog,Noh:2021lqs,Faustov:2021hjs} (blue), QCD sum rules~\cite{Navarra:2007yw,Wang:2017uld,Gao:2020ogo} (brown), heavy quark symmetry~\cite{Eichten:2017ffp,Braaten:2020nwp,Cheng:2020wxa} (green), and others~\cite{Deng:2018kly,Feng:2013kea} (black), as well as those calculated through the hadronic molecular picture~\cite{Janc:2004qn,Li:2012ss,Ohkoda:2012hv,Liu:2019stu,Ding:2020dio} (red), the quark model considering the mixture of the meson-meson and diquark-antidiquark structures~\cite{Yang:2009zzp,Yang:2019itm,Tan:2020ldi} (magenta), and lattice QCD~\cite{Junnarkar:2018twb} (cyan). The two dashed lines with orange crosses denote the $\chi_{c1}(3872)$ ($X(3872)$) first observed by Belle in 2003~\cite{Belle:2003nnu} and the $T^+_{cc}$ recently observed by LHCb in 2021~\cite{LHCb:2021vvq,LHCb:2021auc}.}
\label{fig:Tcc}
\end{figure}

To end this subsection, we collect as many theoretical predictions on the mass of the doubly charmed tetraquark state $cc\bar u\bar d$ with $(I)J^P = (0)1^+$ as we can, and summarize them in Fig.~\ref{fig:Tcc}. The average value of these mass predictions is
\begin{equation}
M_{| cc\bar u\bar d ; (I)J^P = (0)1^+ \rangle} \sim 3857{\rm~MeV} \, ,
\end{equation}
which differs from the experimental mass of the $T^+_{cc}$ by 18~MeV only.

\subsubsection{Hadronic molecular picture.}
\label{sec4.2.2}

The closeness of the $T^+_{cc}$ to the $D^{*+}D^0$ threshold renders it to be an excellent candidate for the $DD^*$ molecular state of $(I)J^P = (0)1^+$, whose existence had been predicted in some earlier theoretical studies~\cite{Janc:2004qn,Yang:2009zzp,Li:2012ss,Ohkoda:2012hv,Xu:2017tsr,Liu:2019stu,Ding:2020dio} before the LHCb experiment~\cite{LHCb:2021vvq,LHCb:2021auc}. After the LHCb experiment~\cite{LHCb:2021vvq,LHCb:2021auc}, some of these theoretical studies were updated together with new predictions.

\begin{figure}[hbtp]
\begin{center}
\includegraphics[width=0.6\textwidth]{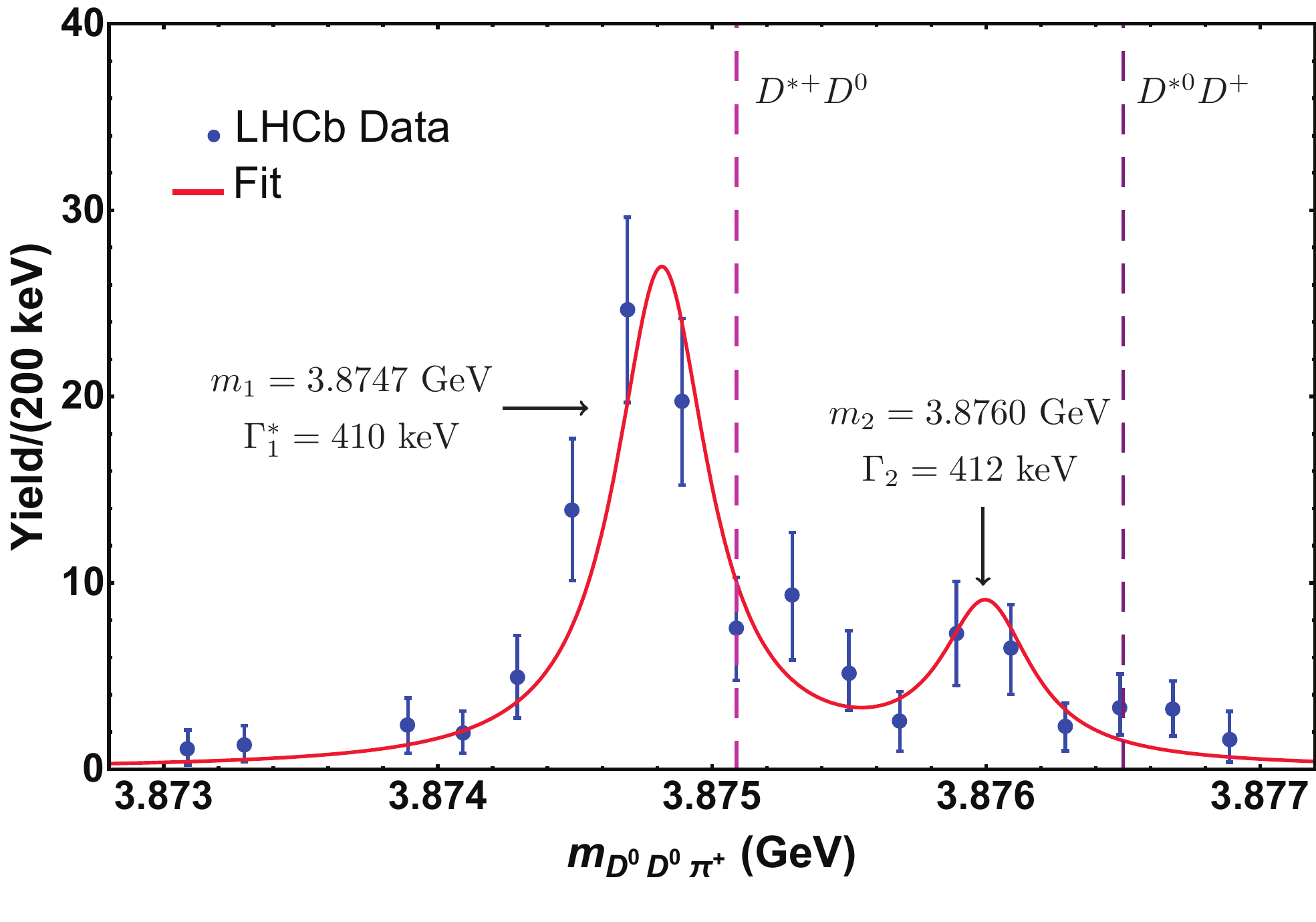}
\end{center}
\caption{Fit to the experimental data of the $D^0D^0\pi^+$ mass spectrum. The two peaks correspond to the $T^+_{cc}$ and $T^{\prime+}_{cc}$, and the two dash lines label the $D^{*+}D^0$ and $D^{*0}D^+$ mass thresholds. Source: Ref.~\cite{Chen:2021vhg}.}
\label{fig:Tccprime}
\end{figure}

In Ref.~\cite{Chen:2021vhg} the authors studied the $D^{*+}D^0$ and $D^{*0}D^+$ interactions with one-boson-exchange effective potentials. They examined the isospin breaking effect carefully, and reproduced the $T^+_{cc}$ mass in the doubly charmed molecular scenario. Besides, they predicted another doubly charmed tetraquark state $T^{\prime+}_{cc}$ with the mass $3876$~MeV and width $412$~keV, as depicted in Fig.~\ref{fig:Tccprime}, which decays into the $D^+D^0\gamma$ and $D^+D^0\pi^0$ channels. However, this $T^{\prime+}_{cc}$ state was not supported by Ref.~\cite{Feijoo:2021ppq} based on the chiral unitary model, where the authors proposed that the bump below the $D^{*0}D^+$ threshold is not a physical state, which is associated with the decay channel $D^0D^0\pi^+$ and its phase space. It is important to carefully examine this structure in future experiments.

In Ref.~\cite{Wang:2021yld} the authors investigated the $S$-wave interactions between the charmed meson doublets $(D, D^*)$ and $(D_1, D_2^*)$ through the one-boson-exchange model. They took into account the $S$- and $D$-wave mixing effect as well as the coupled-channel effect. Their results suggest that the $S$-wave $DD_1$ state of $(I)J^P = (0)1^-$, the $S$-wave $DD_2^*$ state of $(I)J^P = (0)2^-$, and the $S$-wave $D^*D_2^*$ state of $(I)J^P = (0)2^-$ are possible molecular candidates. Later in Ref.~\cite{Wang:2021ajy} the authors applied the same procedures to investigate the $S$-wave interactions between a pair of charmed mesons $(D_1, D_2^*)$. Their results suggest that the $S$-wave $D_1D_1$ state of $(I)J^P = (1)2^+$, the $S$-wave $D_1D^*_2$ state of $(I)J^P = (1)3^+$, and the $S$-wave $D^*_2D^*_2$ state of $(I)J^P = (1)4^+$ are possible molecular candidates.

In Ref.~\cite{Ren:2021dsi} the authors investigated the masses and decay properties of the $DD^*$ molecular states through some effective Lagrangians. Their results indicate that the $T^+_{cc}$ can be understood as a molecular state of $(I)J^P = (0)1^+$. They proposed to observe its beauty-partner $T^-_{bb}$, whose mass was calculated to be $10598^{+2}_{-3}$~MeV, in the $T^-_{bb} \to \bar B^0B^- \gamma$ decay channel. Later in Ref.~\cite{Deng:2021gnb} the authors performed a systematical investigation of the doubly heavy tetraquark states with the molecular configuration using the nonrelativistic quark model. They took into account a color screening confinement potential, meson-exchange interactions, and one-gluon-exchange interactions. Their results suggest the $T^+_{cc}$ to be a loosely-bound deuteron-like state with the binding energy around 0.34~MeV and a huge size of 4.32~fm, where both the meson exchange force and the coupled-channel effect play pivotal roles. Especially, they found that the $T^+_{cc}$ would not exist without the meson exchange force, while its beauty-partner $T^-_{bb}$ would still exist.

A coupled-channel analysis of the $D^{*+}D^0$ and $D^{*0}D^+$ system was performed in Ref.~\cite{Albaladejo:2021vln}, where the $T^+_{cc}$ was found to have a large molecular component. Assuming it to be an isoscalar state, the author applied the heavy quark spin symmetry to predict the existence of its partner, the $D^*D^*$ hadronic molecular state of $(I)J^P = (0)1^+$, with the mass $1\sim2$~MeV below the $D^{*+}D^{*0}$ threshold. A similar conclusion was obtained in Ref.~\cite{Dai:2021vgf}, where this binding energy was estimated to be about $2\sim3$~MeV. The $BB^*$ and $B^*B^*$ hadronic molecular states both of $(I)J^P = (0)1^+$ were similarly investigated in Ref.~\cite{Dai:2022ulk}, whose binding energies were calculated to be about 21~MeV and 19~MeV, respectively. A coupled-channel analysis including three-body effects was performed in Ref.~\cite{Du:2021zzh}, where the compositeness parameter of the $T^+_{cc}$ was estimated to be close to unity, implying it to be a hadronic molecule generated by the interactions in the $D^{*+}D^0$ and $D^{*0}D^+$ channels.

The decay properties of the $T^+_{cc}$ were studied in Refs.~\cite{Meng:2021jnw,Ling:2021bir,Yan:2021wdl,Fleming:2021wmk} as the $DD^*$ molecular state of $(I)J^P = (0)1^+$. In Ref.~\cite{Meng:2021jnw} the authors investigated the $T^+_{cc}$ within the hadronic molecular picture, and studied its strong decays $T^+_{cc} \to D^0 D^0 \pi^+$ and $T^+_{cc} \to D^+ D^0 \pi^0$ as well as its radiative decay $T^+_{cc} \to D^+ D^0 \gamma$. Their results suggest the partial width of the $T^+_{cc} \to D^0 D^0 \pi^+$ decay to be dominant, while the radiative decay width is tiny and so less likely to be detected. Similar conclusions were obtained in Ref.~\cite{Ling:2021bir} through the effective Lagrangian approach. The productions of the $T^+_{cc}$ in the $\gamma p$ reaction, proton-proton collisions, and heavy ion collisions were studied in Refs.~\cite{Huang:2021urd,Braaten:2022elw,Hu:2021gdg,Abreu:2022lfy} as the $DD^*$ molecular state of $(I)J^P = (0)1^+$, and its quasi-fission phenomenon induced by the nucleons was studied in Ref.~\cite{He:2022rta}.

The interpretation of the $T^+_{cc}$ as the $DD^*$ molecular state with $(I)J^P = (0)1^+$ was supported by Ref.~\cite{Xin:2021wcr} within the QCD sum rule method, but not supported by Ref.~\cite{Agaev:2022ast} through the same approach. Besides, the Bethe-Salpeter equation was applied in Refs.~\cite{Zhao:2021cvg,Ke:2021rxd} to explain the $T^+_{cc}$ as the $DD^*$ molecular state of $(I)J^P = (0)1^+$. The binding energy of its beauty-partner was calculated in Ref.~\cite{Ke:2021rxd} to be 17.38~MeV.

\subsubsection{Compact tetraquark picture.}
\label{sec4.2.3}

There are a few theoretical studies on the $T^+_{cc}$ within the compact tetraquark picture after the LHCb experiment~\cite{LHCb:2021vvq,LHCb:2021auc}.

In Ref.~\cite{Guo:2021yws} the authors studied the mass spectrum of the $S$-wave doubly heavy tetraquark states $QQ \bar q \bar q$ with the quantum numbers $J^P = 0^+/1^+/2^+$ based on the improved chromomagnetic interaction model. For the doubly charmed system, they found a stable tetraquark state $cc \bar u \bar d$ of $(I)J^P = (0)1^+$ below the $D^{*+}D^0$ threshold, which can be used to explain the $T^+_{cc}$. A similar study was done in Ref.~\cite{Weng:2021hje}, where the authors calculated the masses of the doubly heavy tetraquarks using an extended chromomagnetic model. They found three stable states lying below the thresholds of two pseudoscalar mesons, {\it i.e.}, the $(I)J^P = (0)1^+$ $bb\bar q \bar q$ tetraquark, the $(I)J^P = (0)0^+$ $c b \bar q \bar q$ tetraquark, and the $J^P = 1^+$ $b b \bar q \bar s$ tetraquark. In a recent study~\cite{Kim:2022mpa}, the authors applied the potential chiral-diquark model to find several $bb\bar q \bar q$ bound states, but they did not find the $cc\bar q \bar q$ and $cb\bar q \bar q$ (deep) bound states.

The interpretation of the $T^+_{cc}$ as the compact $cc\bar u \bar d$ state of $(I)J^P = (0)1^+$ was supported by Ref.~\cite{Agaev:2021vur} within the QCD sum rule method. Assuming the $T^+_{cc}$ to be a compact diquark-antidiquark state, the mass of its strange-partner with the quark content $cc \bar q \bar s$ was calculated in Ref.~\cite{Karliner:2021wju} to be $4106 \pm 12$~MeV.

\subsubsection{More theoretical studies.}
\label{sec4.2.4}

Similar to the quark model study of Ref.~\cite{Yang:2009zzp}, the authors studied the doubly heavy tetraquark states within the framework of chiral quark model in Ref.~\cite{Chen:2021tnn}, by considering the meson-meson and diquark-antidiquark structures as well as their coupling. Using the meson-meson structure, they found a weakly bound $DD^*$ molecular state, which is about 1.8~MeV below the $D^0D^{*+}$ threshold and so a good candidate for the $T^+_{cc}$. Moreover, they used the diquark-antidiquark structure, and found a deeper bound state with the mass 3700.9~MeV; their mixing can further decrease the energy of the system to 3660.7~MeV. A two-body Bethe-Salpeter equation was applied in Ref.~\cite{Santowsky:2021bhy} to study the internal competition between the meson-meson and diquark-antidiquark components in the wave functions of the doubly heavy tetraquark states. Their results indicate that the $T^+_{cc}$ is dominated by an internal $DD^*$ component and its diquark-antidiquark component is negligible. Productions of the $T^+_{cc}$ were studied in Ref.~\cite{Jin:2021cxj} within both the compact tetraquark and hadronic molecular pictures.

In Ref.~\cite{Baru:2021ldu} the authors discussed some general features of the effective range expansion, and estimated the finite-range corrections and the compositeness of the $T^+_{cc}$. Their results are not in contradiction with the molecular nature. A similar study was done in Ref.~\cite{Kinugawa:2021ykv}, where the authors discussed the role of the effective range in the weak-binding relation. In a recent study~\cite{Mikhasenko:2022rrl}, the author studied the effective-range expansion for the $D^*D$ scattering in LHCb model of $T^+_{cc}$ by taking into account three-body nature of the problem. The scattering amplitudes of the $D^0D^0\pi^+$ and $D^{*+}D^0$ coupled channels were studied in Ref.~\cite{Dai:2021wxi} based on the $K$-matrix method, and their pole analysis suggests that the $T^+_{cc}$ may originate from a $D^{*+}D^0$ virtual state formed by the attractive interaction between $D^0$ and $D^{*+}$ together with the coupling to the $D^0D^0\pi^+$ channel.

Besides the above theoretical studies, the $T^+_{cc}$ was studied in Refs.~\cite{Padmanath:2021qje,Padmanath:2022cvl,Ozdem:2021hmk,Azizi:2021aib,Aliev:2021dgx,Azizi:2021aib,Abreu:2021jwm,Andreev:2021eyj,Liu:2022uex} using the lattice QCD, QCD sum rules, AdS/QCD, and the machine learning method, which we shall not discuss any more.

\section{Hidden heavy flavor multiquark states}
\label{sec5}

In this section we shall review the following hidden heavy flavor multiquark candidates observed in recent years:
\begin{itemize}

\item the fully-charm tetraquark state $X(6900)$ observed by LHCb in 2020~\cite{LHCb:2020bwg} will be reviewed in Sec.~\ref{sec5.1};

\item the hidden-charm pentaquark state with strangeness, $P_{cs}(4459)^0$ observed by LHCb in 2020~\cite{LHCb:2020jpq}, will be reviewed in Sec.~\ref{sec5.2};

\item the three hidden-charm tetraquark states with strangeness, $Z_{cs}(3985)^-$ observed by BESIII in 2020~\cite{BESIII:2020qkh} as well as the $Z_{cs}(4000)^+$ and $Z_{cs}(4220)^+$ observed by LHCb in 2021~\cite{LHCb:2021uow}, will be reviewed in Sec.~\ref{sec5.3}.

\end{itemize}

\subsection{Fully-charm tetraquark state $X(6900)$}
\label{sec5.1}

\begin{figure*}[hbtp]
\begin{center}
\subfigure[]{\includegraphics[width=0.31\textwidth]{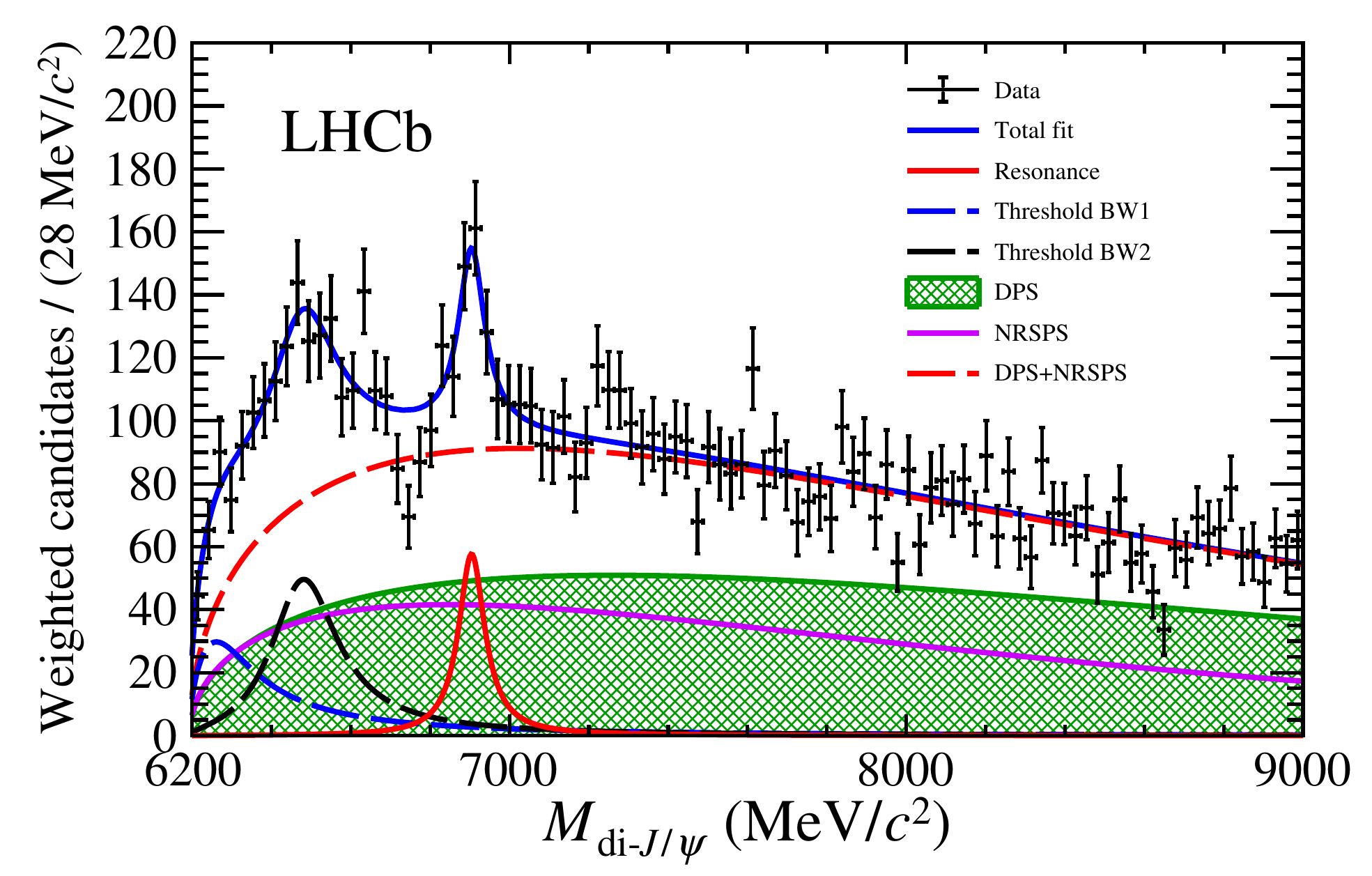}}
\subfigure[]{\includegraphics[width=0.31\textwidth]{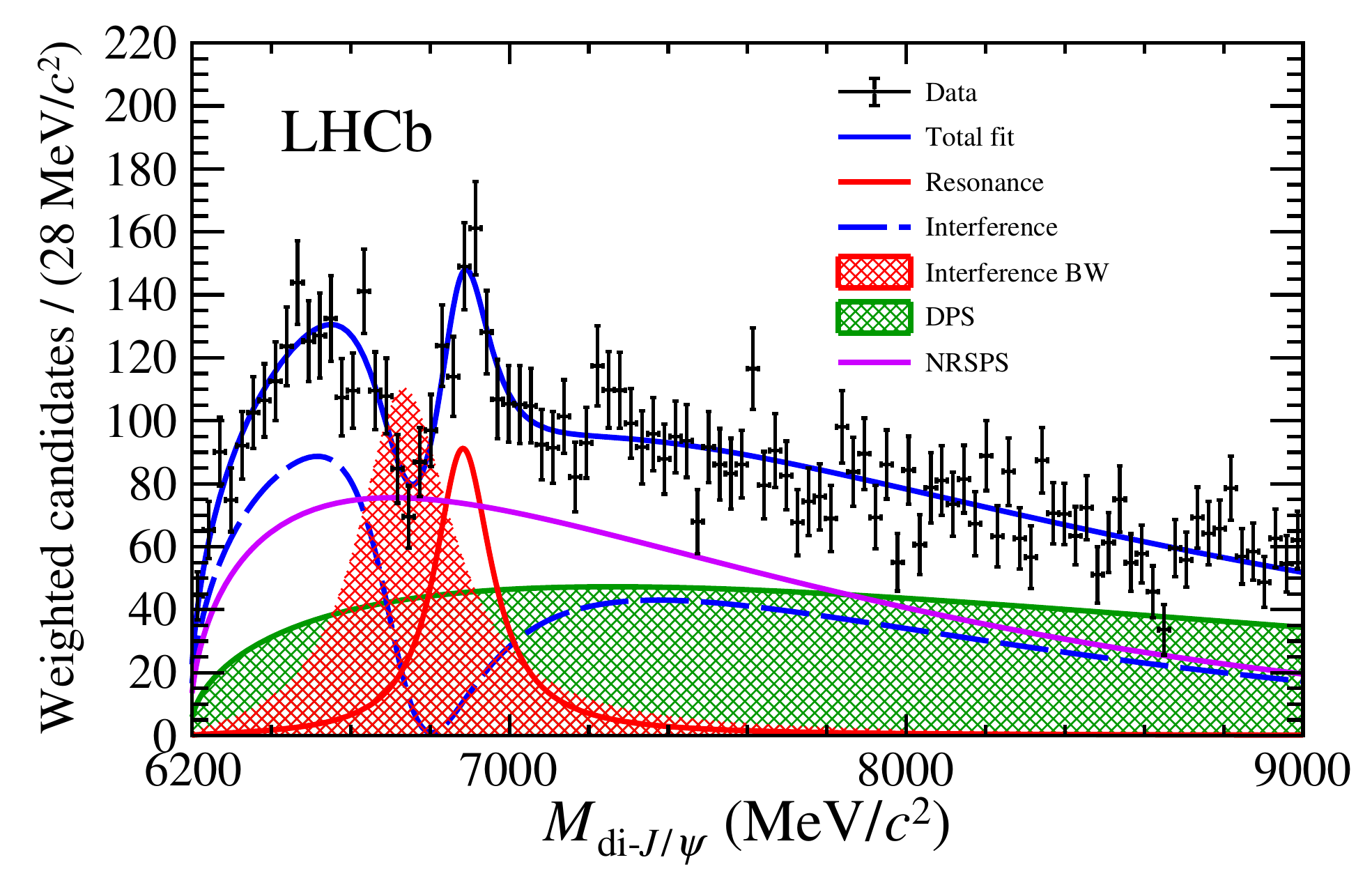}}
\subfigure[]{\includegraphics[width=0.31\textwidth]{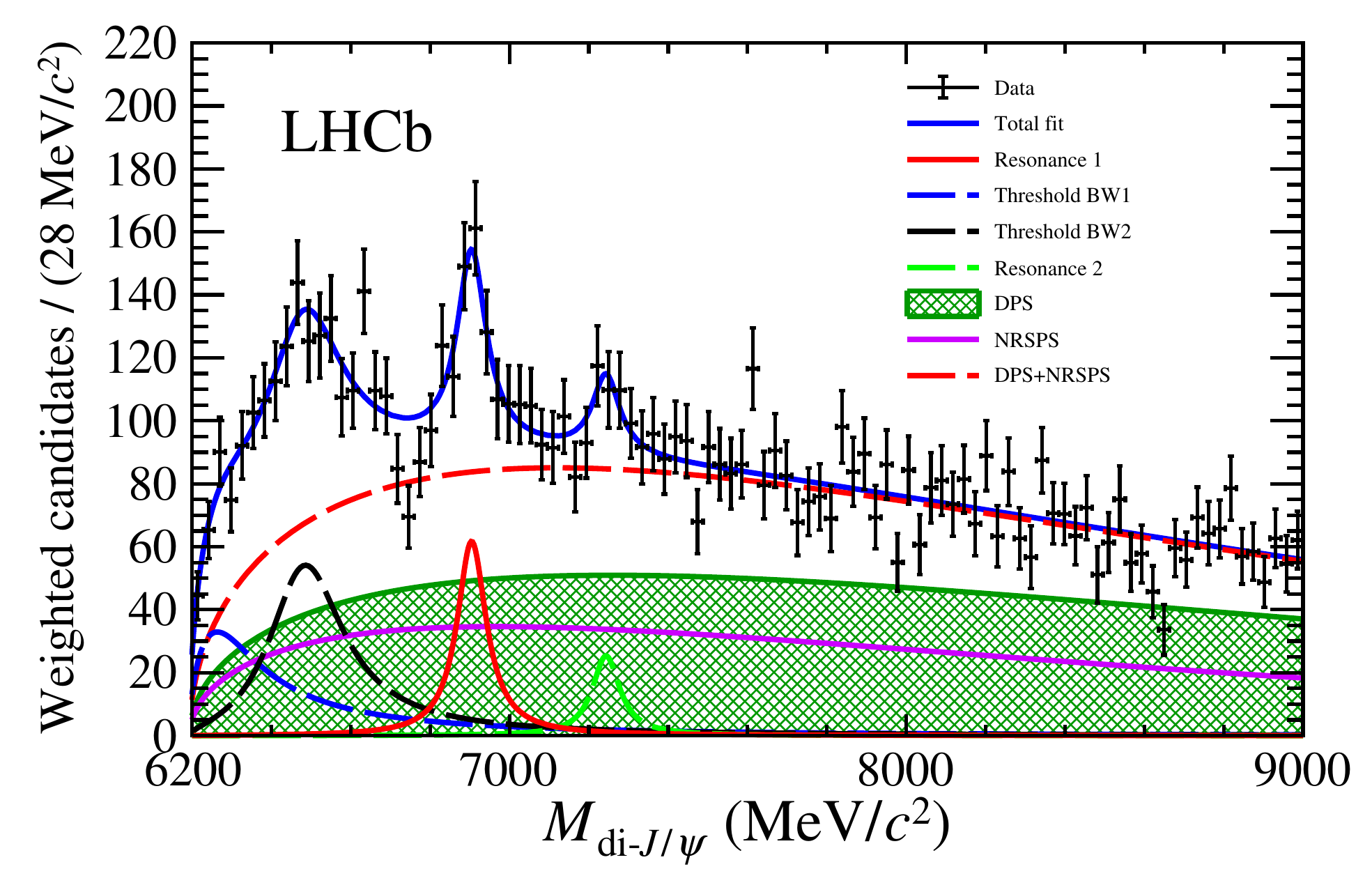}}
\end{center}
\caption{Invariant mass spectra of weighted di-$J/\psi$ candidates fitted (a) without and (b) with interferences, as well as (c) with an additional Breit-Wigner function introduced to describe the structure at around 7.2~GeV. Source: Ref.~\cite{LHCb:2020bwg}.}
\label{fig:X6900}
\end{figure*}

In 2020 the LHCb collaboration reported the observation of two exotic structures in the di-$J/\psi$ invariant mass spectrum~\cite{LHCb:2020bwg}. As shown in Fig.~\ref{fig:X6900}(a,b), they observed a broad structure ranging from 6.2 to 6.8~GeV, and a narrow structure around 6.9~GeV with a global significance larger than $5\sigma$. They described the latter as a resonance with the Breit-Wigner lineshape, and measured its mass and width to be
\begin{eqnarray}
X(6900) &:& M = 6905 \pm 11 \pm 7 {\rm~MeV} \, ,
\\ \nonumber && \Gamma = 80 \pm 19 \pm 33 {\rm~MeV} \, .
\end{eqnarray}
These values were obtained under the assumption that no interference with the nonresonant single-parton scattering continuum is present. Assuming the continuum interferes with the broad structure, the above parameters were shifted to be
\begin{eqnarray}
X(6900) &:& M = 6886 \pm 11 \pm 11 {\rm~MeV} \, ,
\\ \nonumber && \Gamma = 168 \pm 33 \pm 69 {\rm~MeV} \, .
\end{eqnarray}
Besides the above two structures, there exists a hint of another structure around 7.2~GeV. LHCb also performed a fit with an additional Breit-Wigner function introduced to describe this structure, as shown in Fig.~\ref{fig:X6900}(c). These exotic structures are good candidates for the fully-charm tetraquark states.

This LHCb measurement is in a remarkable coincidence with the QCD sum rule results~\cite{Chen:2016jxd} obtained in 2016. As partly shown in Table~\ref{tab:mass}, eighteen diquark-antidiquark $\left([cc] [\bar c \bar c]\right)$ currents were used in Ref.~\cite{Chen:2016jxd} through the QCD sum rule method to calculate the mass spectra of the $cc \bar c \bar c$ and $bb \bar b \bar b$ tetraquark states. Twelve currents have the positive parity, four of which correspond to the $S$-wave $cc \bar c \bar c$ tetraquark states within the quark model picture. Their masses were calculated to be about 6.5~GeV, suggesting that the broad structure around $6.2$-$6.8$~GeV can be interpreted as an $S$-wave $cc\bar c \bar c$ tetraquark state. The other seven currents have the negative parity, and their masses were calculated to be about 6.9~GeV, suggesting that the narrow structure around 6.9~GeV can be interpreted as a $P$-wave $cc\bar c \bar c$ tetraquark state. Later in Refs.~\cite{Chen:2020xwe,Chen:2022sbf} these currents were used to study decay properties of the $cc\bar c \bar c$ and $bb \bar b \bar b$ tetraquark states through the Firez transformation, and some similar currents were used in Refs.~\cite{Yang:2021zrc,Wang:2021taf,Wang:2021mma} to study the $bc\bar b \bar c$ and $bb \bar c \bar c$ tetraquark states.

\begin{table*}
\scriptsize
\begin{center}
\renewcommand{\arraystretch}{1.5}
\caption{Mass spectra of the $cc\bar c\bar c$ and $bb\bar b\bar b$ tetraquark states with the quantum numbers $J^{PC} = 0^{++}/0^{-+}/0^{--}/1^{++}/1^{+-}/1^{-+}/1^{--}/2^{++}$, calculated in Ref.~\cite{Chen:2016jxd} through the QCD sum rule method. Source: Ref.~\cite{Chen:2020xwe}.}
\label{tab:mass}
\begin{tabular}{c c c c}
\hline
~~~~$J^{PC}$ ~~~~& Currents &~~~~ $m_{c c\bar c \bar c}$ \mbox{(GeV)} ~~~~&~~~~ $m_{b b \bar b \bar b}$ \mbox{(GeV)} ~~~~
\\ \hline
$0^{++}$      & $J^{0^{++}}_1$              &  $6.44\pm0.15$              & $18.45\pm0.15$ \\
              & $J^{0^{++}}_2$              &  $6.46\pm0.16$              & $18.46\pm0.14$
\vspace{4pt} \\
$1^{+-}$      & $J^{1^{+-}}_{3\alpha}$      &  $6.51\pm0.15$              & $18.54\pm0.15$
\vspace{4pt} \\
$2^{++}$      & $J^{2^{++}}_{4\alpha\beta}$ &  $6.51\pm0.15$              & $18.53\pm0.15$
\vspace{4pt} \\
$0^{-+}$      & $J^{0^{-+}}_5$              &  $6.84\pm0.18$              & $18.77\pm0.18$ \\
              & $J^{0^{-+}}_6$              &  $6.85\pm0.18$              & $18.79\pm0.18$
\vspace{4pt} \\
$0^{--}$      & $J^{0^{--}}_7$              &  $6.84\pm0.18$              & $18.77\pm0.18$
\vspace{4pt} \\
$1^{-+}$      & $J^{1^{-+}}_{8\alpha}$      &  $6.84\pm0.18$              & $18.80\pm0.18$ \\
              & $J^{1^{-+}}_{9\alpha}$      &  $6.88\pm0.18$              & $18.83\pm0.18$
\vspace{4pt} \\
$1^{--}$      & $J^{1^{--}}_{10\alpha}$     &  $6.84\pm0.18$              & $18.77\pm0.18$ \\
              & $J^{1^{--}}_{11\alpha}$     &  $6.83\pm0.18$              & $18.77\pm0.16$ \\
\hline
\end{tabular}
\end{center}
\end{table*}

The observation of the $X(6900)$ immediately attracted much attention from the particle physics community. Especially, some theorists reanalysed the LHCb data on the di-$J/\psi$ mass spectrum~\cite{LHCb:2020bwg}, and proposed the existence of more exotic structures, {\it e.g.}, in Ref.~\cite{Wang:2020wrp} the authors reproduced three peaking structures at near 6.5, 6.9, and 7.3~GeV, and in Ref.~\cite{Dong:2020nwy} the authors proposed the existence of a near-threshold state at around 6.2~GeV. We shall review these theoretical studies in Sec.~\ref{sec5.1.2} in details.

Very recently, the CMS and ATLAS collaborations also examined the di-$J/\psi$ mass spectrum, and both of them confirmed the existence of the $X(6900)$~\cite{CMS,ATLAS}. Besides, the CMS collaboration observed two new structures in the di-$J/\psi$ mass spectrum, labeled as $X(6600)$ and $X(7200)$. Their masses and widths were measured to be~\cite{CMS}:
\begin{eqnarray}
X(6600) &:& M = 6552 \pm 10 \pm 12 {\rm~MeV} \, ,
\\ \nonumber && \Gamma = 124 \pm 29 \pm 34 {\rm~MeV} \, ;
\\ X(6900) &:& M = 6927 \pm 9 \pm 5 {\rm~MeV} \, ,
\\ \nonumber && \Gamma = 122 \pm 22 \pm 19 {\rm~MeV} \, ;
\\ X(7200) &:& M = 7287 \pm 19 \pm 5 {\rm~MeV} \, ,
\\ \nonumber && \Gamma = 95 \pm 46 \pm 20 {\rm~MeV} \, .
\end{eqnarray}
The ATLAS collaboration examined the di-$J/\psi$ mass spectrum, and their best fit was performed with three interfering resonances, whose masses and widths were measured to be~\cite{ATLAS}:
\begin{eqnarray}
X(6200) &:& M = 6.22 \pm 0.05 ^{+0.04}_{-0.05} {\rm~GeV} \, ,
\\ \nonumber && \Gamma = 0.31 \pm 0.12 ^{+0.07}_{-0.08} {\rm~GeV} \, ;
\\ X(6600) &:& M = 6.62 \pm 0.03 ^{+0.02}_{-0.01} {\rm~GeV} \, ,
\\ \nonumber && \Gamma = 0.31 \pm 0.09 ^{+0.06}_{-0.11} {\rm~GeV} \, ;
\\ X(6900) &:& M = 6.87 \pm 0.03 ^{+0.06}_{-0.01} {\rm~GeV} \, ,
\\ \nonumber && \Gamma = 0.12 \pm 0.04 ^{+0.03}_{-0.01} {\rm~GeV} \, .
\end{eqnarray}
The ATLAS collaboration also examined the $J/\psi \psi(2S)$ mass spectrum, and reported the evidence for an enhancement at 6.9~GeV and a resonance at 7.2~GeV. Their masses and widths were measured to be~\cite{ATLAS}:
\begin{eqnarray}
X(6900) &:& M = 6.78 \pm 0.36 ^{+0.35}_{-0.54} {\rm~GeV} \, ,
\\ \nonumber && \Gamma = 0.39 \pm 11 ^{+0.11}_{-0.07} {\rm~GeV} \, ;
\\ X(7200) &:& M = 7.22 \pm 0.03 ^{+0.02}_{-0.03} {\rm~GeV} \, ,
\\ \nonumber && \Gamma = 0.10 {^{+0.13}_{-0.07}} {^{+0.06}_{-0.05}} {\rm~GeV} \, .
\end{eqnarray}

Actually, the fully-heavy tetraquark states were already studied in the 1980's~\cite{Iwasaki:1975pv,Chao:1980dv,Ader:1981db,Li:1983ru,Heller:1985cb,Badalian:1985es}, but they did not receive much attention due to the absence of experimental data. With the running of LHC at 13 TeV, searching for these exotic tetraquark states has become an important experimental issue. Besides the above LHCb measurement~\cite{LHCb:2020bwg}, in 2017 the CMS collaboration found an excess in the $\Upsilon(1S)l^+l^-$ ($l=e,\mu$) mass distribution near 18.5~GeV~\cite{Durgut:2018lmn,Yi:2018fxo}. Its mass was measured to be $18.4 \pm 0.1 \pm 0.2$~GeV with a global significance of 3.6$\sigma$. This structure is a possible fully-bottom tetraquark state, but it was not confirmed in the later LHCb and CMS experiments~\cite{LHCb:2018uwm,CMS:2020qwa}.

Since the interaction between two bottomonia/charmonia may not be strong enough to form a tightly bound state, many theoretical studies on the fully-heavy tetraquark states adopt the compact tetraquark picture. The compact diquark-antidiquark structure $[QQ][\bar Q \bar Q]$ is the most popular one, but the mass predictions are quite model dependent. For example, the $b b \bar b \bar b$ tetraquark states may lie below the di-bottomonium threshold according to Refs.~\cite{Chen:2016jxd,Karliner:2016zzc,Bai:2016int,Wang:2017jtz,Debastiani:2017msn,Anwar:2017toa,Esposito:2018cwh}, while they may lie above this threshold according to Refs.~\cite{Wu:2016vtq,Hughes:2017xie,Wang:2019rdo,Chen:2019dvd,Liu:2019zuc,Deng:2020iqw}. Due to these controversial issues, more experimental and theoretical investigations are crucial to understand them~\cite{Maiani:2020pur,Chao:2020dml,Richard:2020hdw}.

We have reviewed many of the theoretical studies performed before the LHCb measurement~\cite{LHCb:2020bwg} in our previous paper~\cite{Liu:2019zoy}, so we shall only review those works after this measurement in this limited paper.

\subsubsection{Compact tetraquark picture.}
\label{sec5.1.1}

Besides the QCD sum rule study of Ref.~\cite{Chen:2016jxd}, many theoretical studies support the interpretation of the broad structure observed by LHCb ranging from 6.2 to 6.8~GeV~\cite{LHCb:2020bwg} as an $S$-wave $cc\bar c \bar c$ tetraquark state, and the narrow structure $X(6900)$ as a $P$-wave $cc\bar c \bar c$ tetraquark state.

\begin{figure*}[hbtp]
\begin{center}
\subfigure[]{\includegraphics[width=0.48\textwidth]{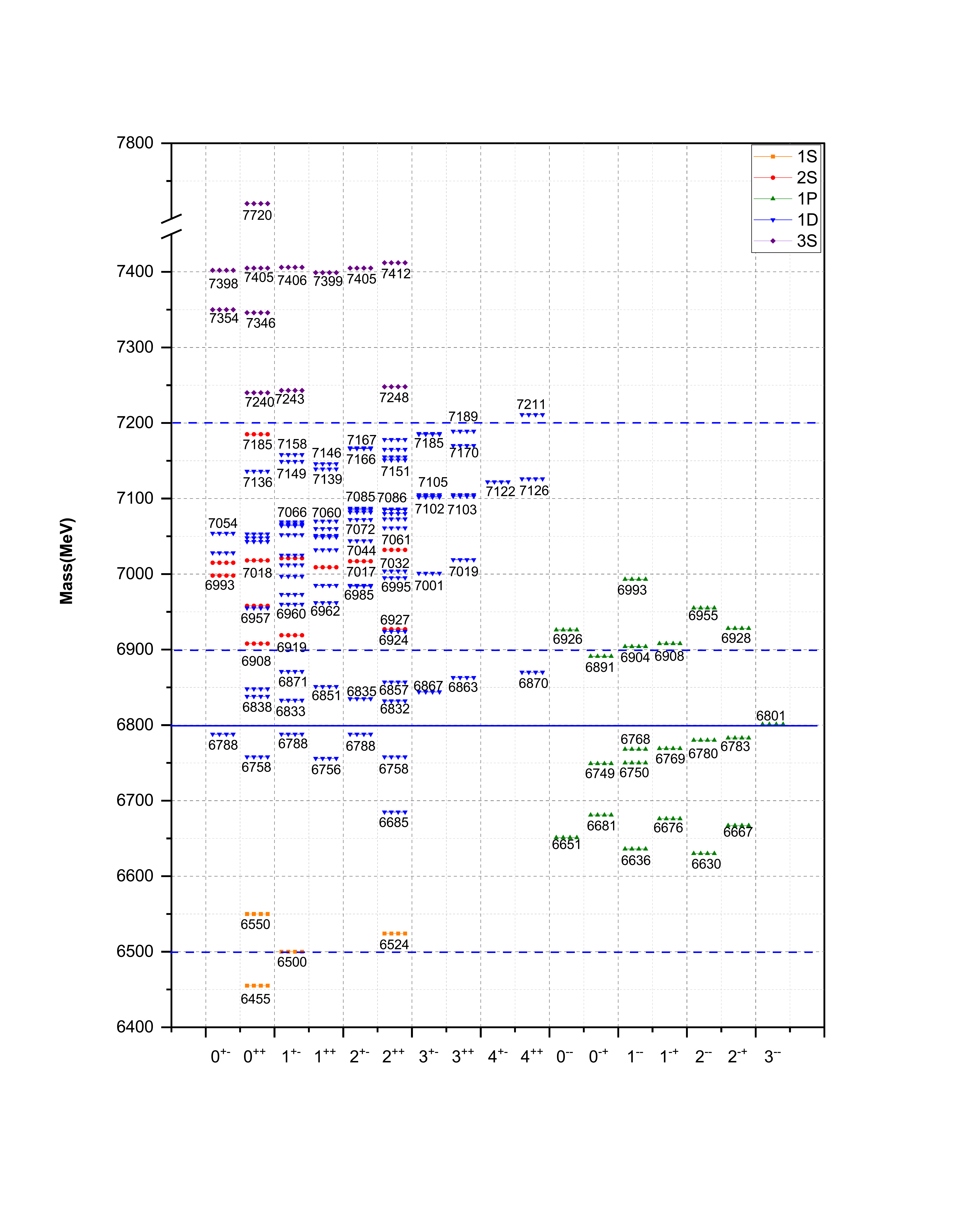}}
\subfigure[]{\includegraphics[width=0.48\textwidth]{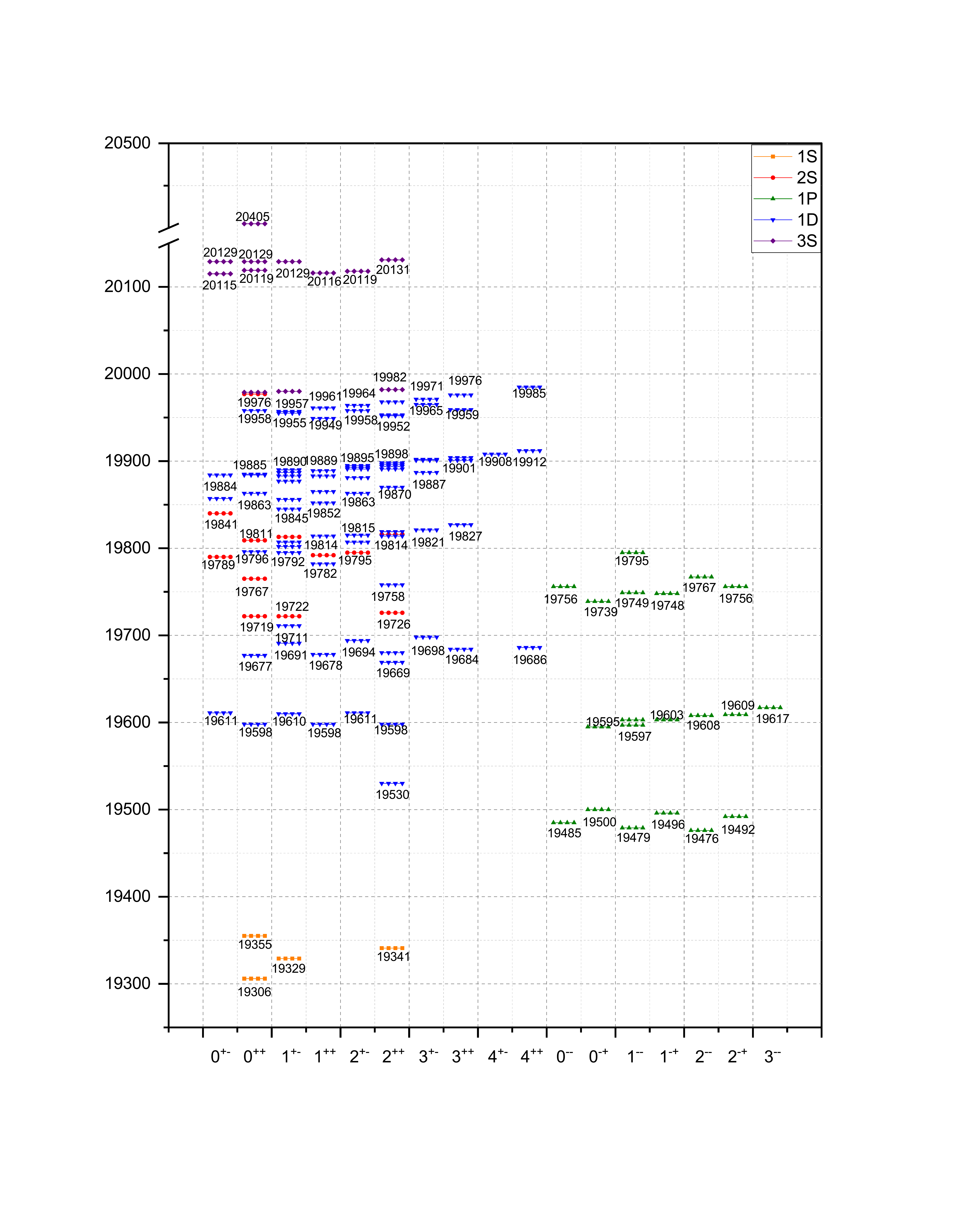}}
\end{center}
\caption{Mass spectra of the (a) $cc \bar c \bar c$ and (b) $b b \bar b \bar b$ tetraquark states up to the second radial excitations, calculated using the nonrelativistic potential quark model. Source: Ref.~\cite{Liu:2021rtn}.}
\label{fig:X6900NRPQM}
\end{figure*}

In Ref.~\cite{Liu:2020eha} the authors calculated the mass spectra of the $cc \bar c \bar c$ and $b b \bar b \bar b$ tetraquark states using a nonrelativistic potential quark model. Their Hamiltonian contains a linear confinement potential and a parameterized one-gluon-exchange potential. They labeled these states as $T_{(cc\bar c \bar c/bb\bar b \bar b)J^{PC}}$. As shown in Fig.~\ref{fig:X6900NRPQM}, their results suggest that the narrow structure $X(6900)$ may be explained as the $P$-wave states $T_{(cc\bar c \bar c)0^{-+}}(6891)$, $T_{(cc\bar c \bar c)1^{-+}}(6908)$, and $T_{(cc\bar c \bar c)2^{-+}}(6928)$, and the broad structure may be explained as the $S$-wave states $T_{(cc\bar c \bar c)0^{++}}(6455)$, $T_{(cc\bar c \bar c)0^{++}}(6550)$, and $T_{(cc\bar c \bar c)2^{++}}(6524)$. The interpretation of the narrow structure $X(6900)$ as a $P$-wave $cc\bar c \bar c$ tetraquark state was supported by Ref.~\cite{Tiwari:2021tmz} through the constituent quark model, and the interpretation of the broad structure as an $S$-wave $cc\bar c \bar c$ tetraquark state was supported by Refs.~\cite{Lu:2020cns,Faustov:2020qfm,Zhang:2020xtb,Li:2021ygk} based on the relativized quark model as well as QCD sum rules and Bethe-Salpeter equation.

The above nonrelativistic potential quark model study of Ref.~\cite{Liu:2020eha} was improved in Ref.~\cite{Liu:2021rtn} by including the higher mass spectra of the $2S$- and $1D$-wave states. As shown in Fig.~\ref{fig:X6900NRPQM}, these results suggest that the narrow structure $X(6900)$ may be caused by the $1P$-, $2S$-, and $1D$-wave $cc \bar c \bar c$ states, and the vague structure $X(7200)$ may be caused by the highest $2S$ state $T_{(cc\bar c \bar c)0^{++}}(7185)$, the two low-lying $3S$ states $T_{(cc\bar c \bar c)0^{++}}(7240)$ and $T_{(cc\bar c \bar c)2^{++}}(7248)$, and the high-lying $1D$ states with masses around 7.2 GeV.

The interpretation of the narrow structure $X(6900)$ as a $2S$ $cc\bar c \bar c$ tetraquark state was supported by many theoretical studies. In Ref.~\cite{Giron:2020wpx} the authors studied the mass spectrum of the $cc \bar c \bar c$ tetraquark states using the dynamical diquark model. They investigated the most minimal form of this model, where each diquark appears only in the color $\mathbf{\bar{3}}_c$ representation and the non-orbital spin couplings connect only quarks within each diquark. After neglecting the tensor coupling, they found three degenerate $S$-wave states and seven $P$-wave states satisfying an equal-spacing rule. Some of their masses were evaluated to be
\begin{eqnarray}
\nonumber 1S &:& M = 6264.0-6266.1 {\rm~MeV} \, ,
\\ \nonumber 1P &:& M = 6611.4 {\rm~MeV} \, ,
\\ 1D &:& M = 6860.5-6862.4 {\rm~MeV} \, ,
\\ \nonumber 2S &:& M = 6771.8 {\rm~MeV} \, ,
\\ \nonumber 2P &:& M = 7010.8-7013.0 {\rm~MeV} \, ,
\\ \nonumber 2D &:& M = 7213.3-7216.7 {\rm~MeV} \, .
\end{eqnarray}
Accordingly, the $X(6900)$ was interpreted in Ref.~\cite{Giron:2020wpx} as the $2S$ $cc\bar c \bar c$ tetraquark state, while it may also be interpreted as the $1D$ state. The $2S$ assignment was supported by Ref.~\cite{Karliner:2020dta}, where the authors calculated the mass of the $0^{++}(2S)$ $cc\bar c \bar c$ tetraquark state within the string-junction picture to be $6871\pm25$~MeV. Besides, the $X(6900)$ was interpreted in Refs.~\cite{Wang:2020ols,Ke:2021iyh,Zhu:2020xni} as the $3S$ $cc\bar c \bar c$ tetraquark state.

The full heavy tetraquarks states were systematically investigated in Ref.~\cite{Jin:2020jfc} within the chiral quark model, by considering both the meson-meson and diquark-antidiquark structures. They did not predict bound states in the meson-meson configuration, while in the diquark-antidiquark configuration they predicted several resonances ranging between $6.3\sim7.4$~GeV. Later in Ref.~\cite{Yang:2021hrb} a similar quark model calculation was done based on lattice QCD studies of the two-body $Q \bar Q$ interaction. They considered a more complete set of four-body configurations, including meson-meson, diquark-antidiquark, and $K$-type configurations, similar to those shown in Fig.~\ref{fig:X2900config} of Sec.~\ref{sec4.1.2}. They predicted several narrow resonances with decay widths less than 30 MeV, which were used to explain the structures observed by LHCb in the di-$J/\psi$ mass spectrum~\cite{LHCb:2020bwg}. More theoretical studies considering both the meson-meson and diquark-antidiquark configurations can be found in Refs.~\cite{Albuquerque:2020hio,Albuquerque:2021erv,Wu:2022qwd,Asadi:2021ids}, and there also exist some theoretical studies~\cite{Yang:2020wkh,Huang:2020dci} taking the two heavy quark-antiquark pairs as colored clusters.

\subsubsection{Discussions on the di-$J/\psi$ threshold.}
\label{sec5.1.2}

As shown in Fig.~\ref{fig:X6900}, the broad structure ranging from 6.2 to 6.8~GeV is just above the $J/\psi J/\psi$ threshold, and many theoretical studies were performed to study the relevant threshold effects. Actually, this is an old and complicated problem dating back to the study of the deuteron by S.~Weinberg in the 1960's~\cite{Weinberg:1962hj,Weinberg:1963zza,Weinberg:1965zz}. We refer interested readers to Refs.~\cite{Guo:2019twa,Baru:2003qq,Matuschek:2020gqe,Baru:2021ldu,Sekihara:2014kya,Kamiya:2016oao,Gamermann:2009uq,Guo:2015daa,Xiao:2016dsx,Du:2020bqj,Du:2021fmf,Esposito:2021vhu,Mai:2022eur} and references therein for detailed discussions. Here we follow Refs.~\cite{Baru:2003qq,Matuschek:2020gqe} to briefly introduce the derivation of the Weinberg compositeness criterion.

We assume that the physical bound state $| \Psi \rangle$ is composed of a bare compact state $| \psi_0 \rangle$ and a two-hadron channel $| h_1 h_2 \rangle_{\vec{p}}$:
\begin{equation}
| \Psi \rangle =
\left(\begin{array}{c}
\lambda \: | \psi_0 \rangle
\\
\chi(\vec{p}) \: | h_1 h_2 \rangle_{\vec{p}}
\end{array}\right) \, ,
\end{equation}
with $\vec{p}$ the relative momentum of the two hadrons, and the interaction governed by
\begin{equation}
\hat{\mathbf{H}} =
\left(\begin{array}{cc}
\hat{H}_c & \hat{V}
\\
\hat{V} & \hat{H}_{hh}^0
\end{array}\right) \, .
\end{equation}
After employing the Schr\"odinger equation $\hat{\mathbf{H}} | \Psi \rangle = E | \Psi \rangle$, we can derive
\begin{equation}
\chi(\vec{p}) = \lambda \frac{f(p^2)}{E-{p^2}/{(2\mu)}}.
\end{equation}
After employing the normalization condition $\langle \Psi | \Psi \rangle = 1$, we can further derive
\begin{equation}
1 = \lambda^2 \left( 1 + \int  \!\! \frac{d^3p}{(2\pi)^3} \frac{f^2(p^2)}{(E_B + {p^2}/{(2\mu)})^2} \right) \, ,
\end{equation}
where $f(p^2) = \langle \psi_0 | \hat{V} | h_1 h_2 \rangle_{\vec{p}}$, $E_B=m_1+m_2-M$ is the binding energy, and $\mu=m_1 m_2/(m_1+m_2)$ is the reduced mass of $h_1$ and $h_2$.

The field renormalization factor $Z$ is defined through
\begin{equation}
G(E) = \frac{1}{E-E_0-\Sigma(E)} = \frac{Z}{E+E_B}+ {\cal O}\left((E+E_B)^2\right) \, ,
\end{equation}
where
\begin{eqnarray}
-E_B &=& E_0 + \Sigma(-E_B) \, ,
\\ \nonumber
\Sigma(E) &=& \int \!\! \frac{d^3p}{(2\pi)^3} \frac{f^2(p^2)}{E-{p^2}/{(2\mu)+i 0}} \, .
\end{eqnarray}
Accordingly, we can derive
\begin{eqnarray}
Z &=& \frac{1}{1-\left.\partial\Sigma/\partial E\right|_{E=-E_B}}
\\ \nonumber &=& \left(1 + \int  \!\! \frac{d^3p}{(2\pi)^3} \frac{f^2(p^2)}{(E_B+{p^2}/{(2\mu)})^2}\right)^{-1}
\\ \nonumber &=& \lambda^2 \, .
\end{eqnarray}
Since $\lambda^2 = \left| \langle \psi_0 | \Psi \rangle \right|^2$ is just the probability of finding the bare state $| \psi_0 \rangle$ in the physical state $| \Psi \rangle$, the compositeness of $| \Psi \rangle$ is given by $1-\lambda^2 = 1-Z$.

We can employ the effective range expansion of $T$-matrix
\begin{equation}
T(E) = -\frac{2\pi}{\mu}\frac{1}{1/a + (r/2)k^2-i k} \, ,
\label{sec5:Tmatrix}
\end{equation}
and further derive
\begin{eqnarray}
a &= -2 \: \frac{1-Z}{2-Z} \frac1\gamma + \mathcal{O}(1/\beta) \, ,
\\ \nonumber
r &= - \frac{Z}{1-Z}  \frac1\gamma + \mathcal{O}(1/\beta) \, ,
\end{eqnarray}
where $E=k^2/(2\mu)$; $a$ is the scattering length and $r$ is the effective range; $\gamma=\sqrt{2\mu E_B}$ is the binding momentum and $\beta$ is the closest non-analyticity of the system not related to the threshold under investigation.

\begin{figure*}[hbtp]
\begin{center}
\subfigure[]{\includegraphics[width=0.3\textwidth]{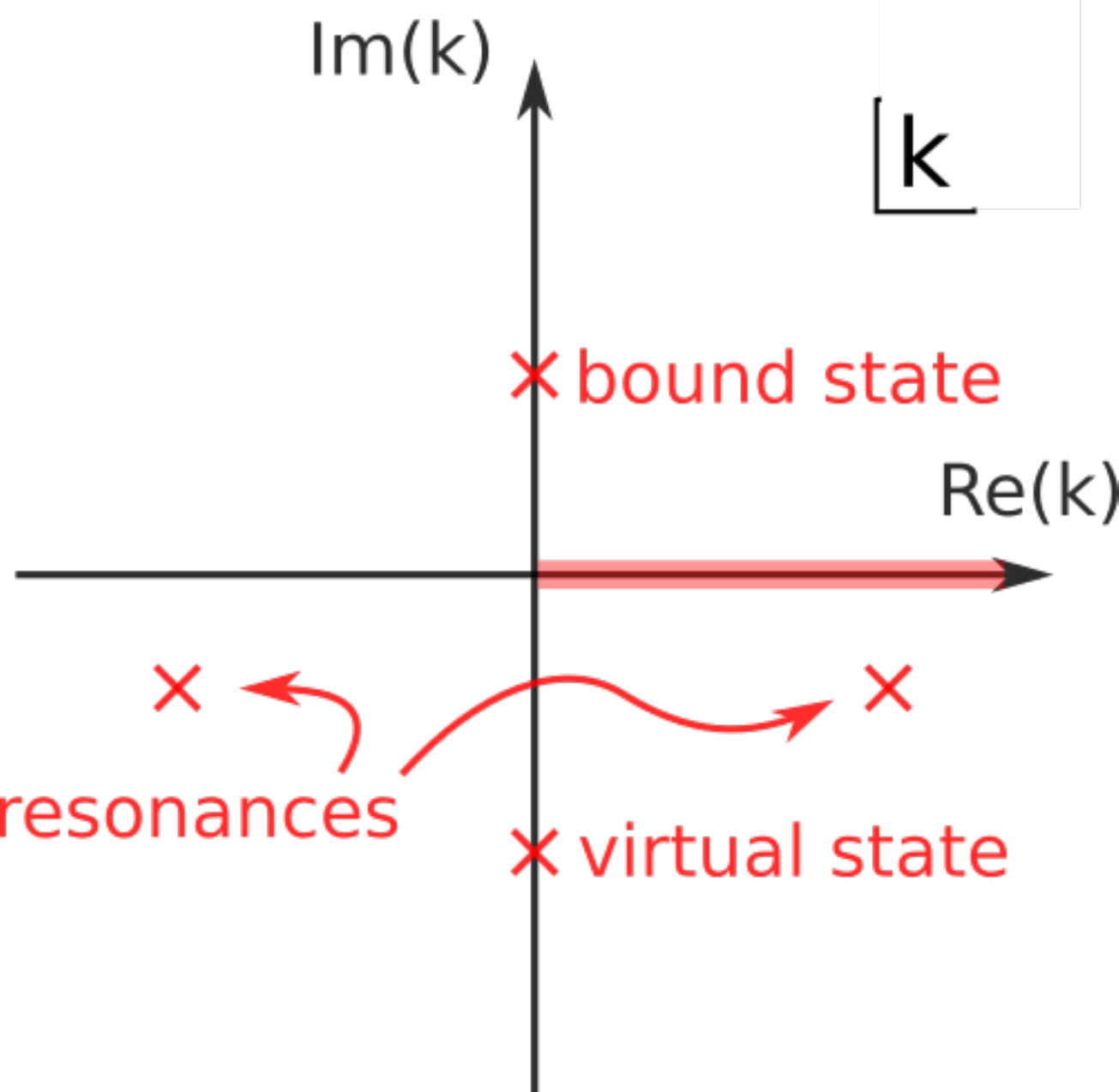}}
~~~~~~~~~~
\subfigure[]{\includegraphics[width=0.3\textwidth]{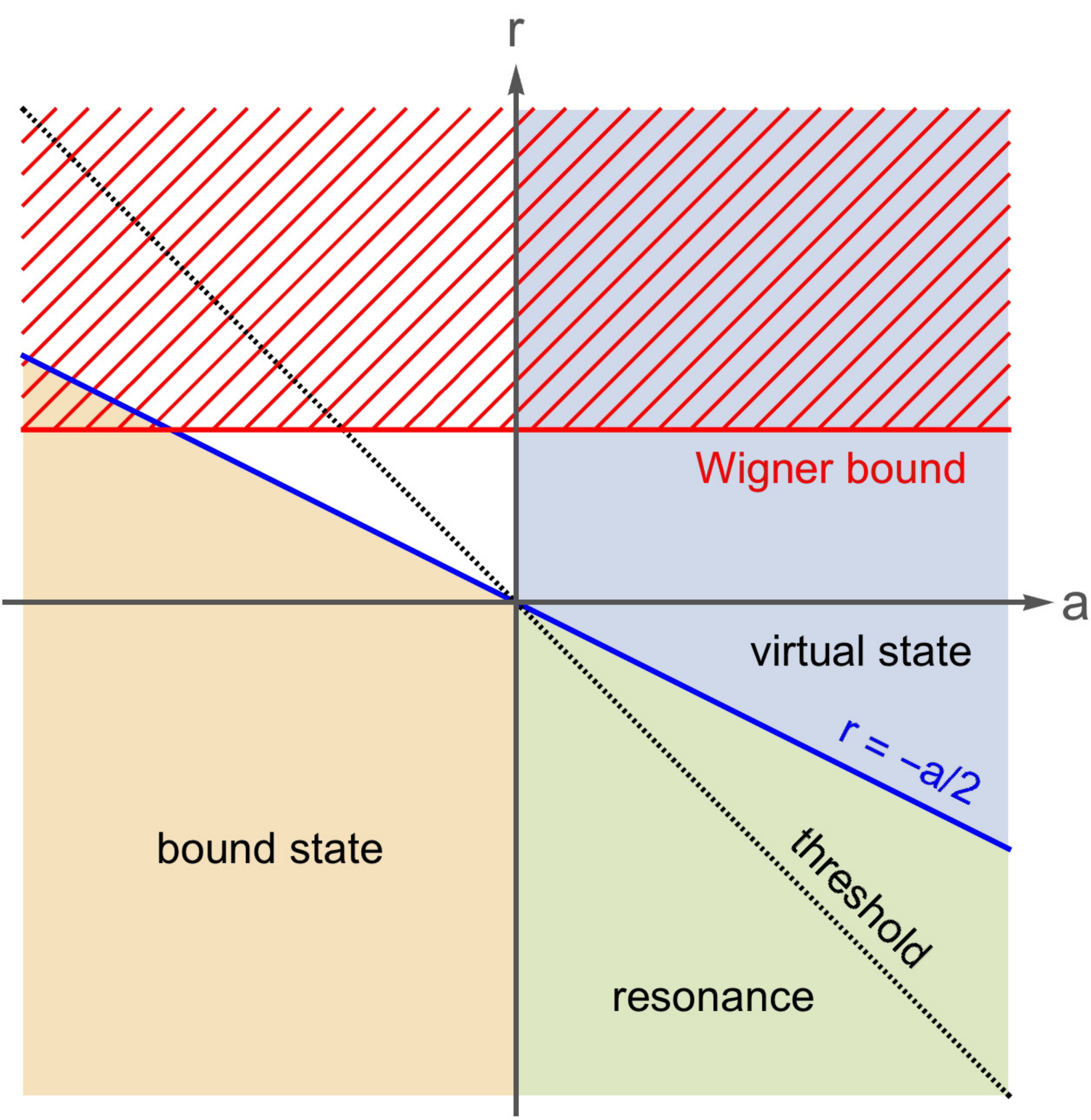}}
\end{center}
\caption{Types of poles in the (a) $k$-plane and (b) $r$-$a$ plane. Source: Ref.~\cite{Matuschek:2020gqe}.}
\label{fig:pole}
\end{figure*}

Eq.~(\ref{sec5:Tmatrix}) has two poles locating at
\begin{equation}
k = \frac{i}{r} \left( 1 \pm \sqrt{1+\frac{2r}{a}} \right) \, .
\end{equation}
As shown in the $k$-plane of Fig.~\ref{fig:pole}(a), one usually uses ``bound states'' to denote the states on the positive imaginary momentum axis, ``virtual states'' to denote those on the negative imaginary momentum axis, and ``resonances'' to denote all the other physically allowed states. These states are also shown in the $r$-$a$ plane of Fig.~\ref{fig:pole}(b). We note that it is not trivial at all to extend the notion of compositeness to states other than the bound states, which will not be discussed in this review any more.

In many theoretical studies one needs to consider more than one two-hadron channel, so the coupled-channel analyses are necessary. Moreover, in many theoretical studies one needs to investigate the experimental data with the near-threshold structures, for which it is possible to employ an effective Lagrangian without explicitly involving the bare states.

\begin{figure*}[hbtp]
\begin{center}
\includegraphics[width=0.6\textwidth]{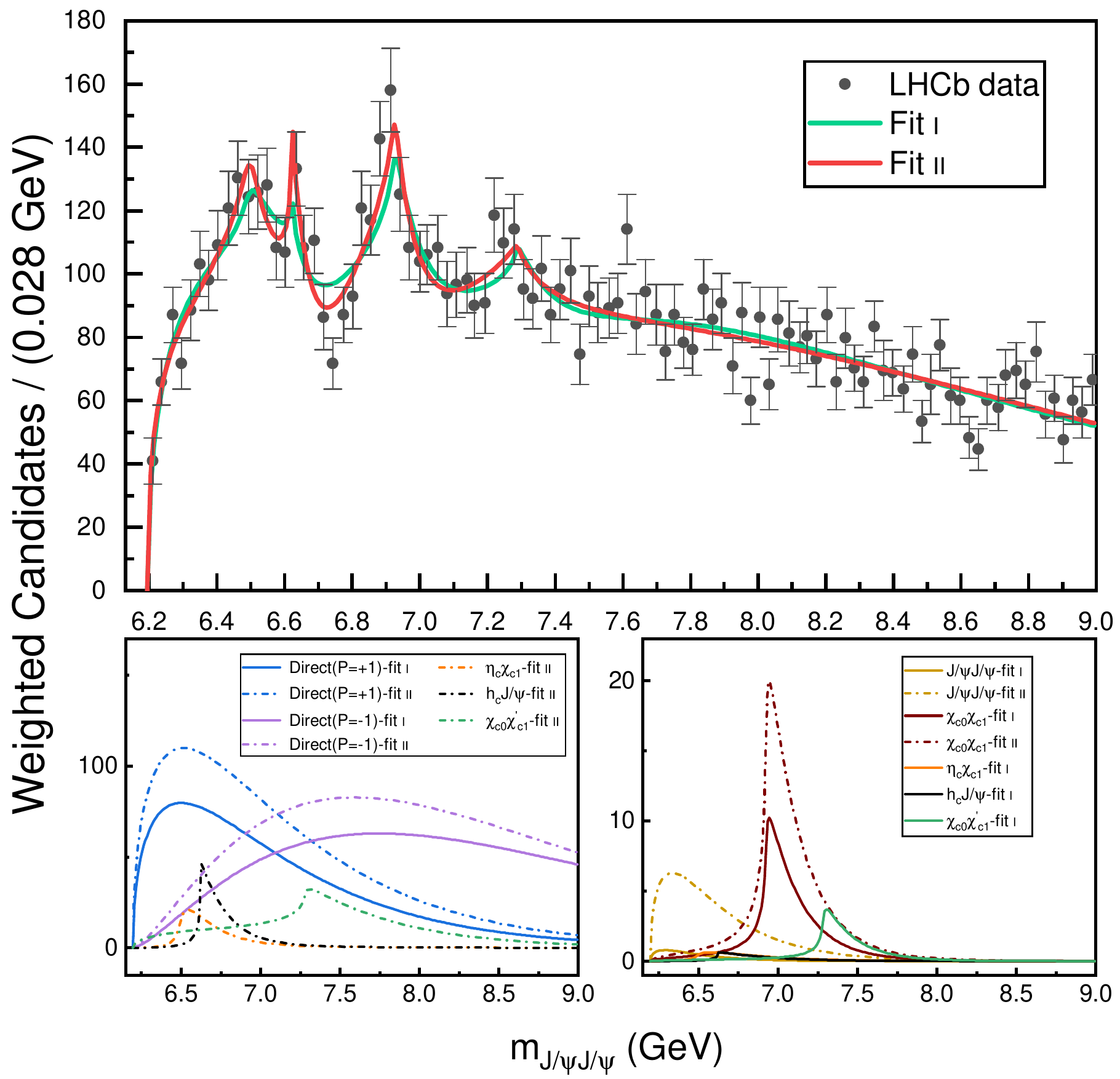}
\end{center}
\caption{Fits to the LHCb data of the di-$J/\psi$ invariant mass distribution based on a dynamical rescattering mechanism, considering both the direct production of $J/\psi J/\psi$ (bottom-left) and the transitions from all the possible double-charmonia into $J/\psi J/\psi$ (bottom-right). Fit I and Fit II were performed by using the same and independent cutoff parameters for intermediate charmonium states, respectively. Source: Ref.~\cite{Wang:2020wrp}.}
\label{fig:X6900dynamical}
\end{figure*}

In Ref.~\cite{Wang:2020wrp} the authors proposed a non-resonant dynamical mechanism to mimic the experimental data of a di-$J/\psi$ mass spectrum given by LHCb~\cite{LHCb:2020bwg}. As depicted in Fig.~\ref{fig:X6900dynamical}, three obvious peak structures near $6.5$, $6.9$, and $7.3$~GeV were reproduced, which naturally corresponded to three rescattering channels $\eta_c \chi_{c1}$, $\chi_{c0} \chi_{c1}$, and $\chi_{c0} \chi_{c1}^\prime$. Besides, the lineshape of other possible fully-charm tetraquark states in the $J/\psi \psi(3686)$ invariant mass spectrum were studied in Ref.~\cite{Wang:2020tpt} through a similar dynamical rescattering mechanism.

\begin{figure*}[hbtp]
\begin{center}
\subfigure[]{\includegraphics[width=0.43\textwidth]{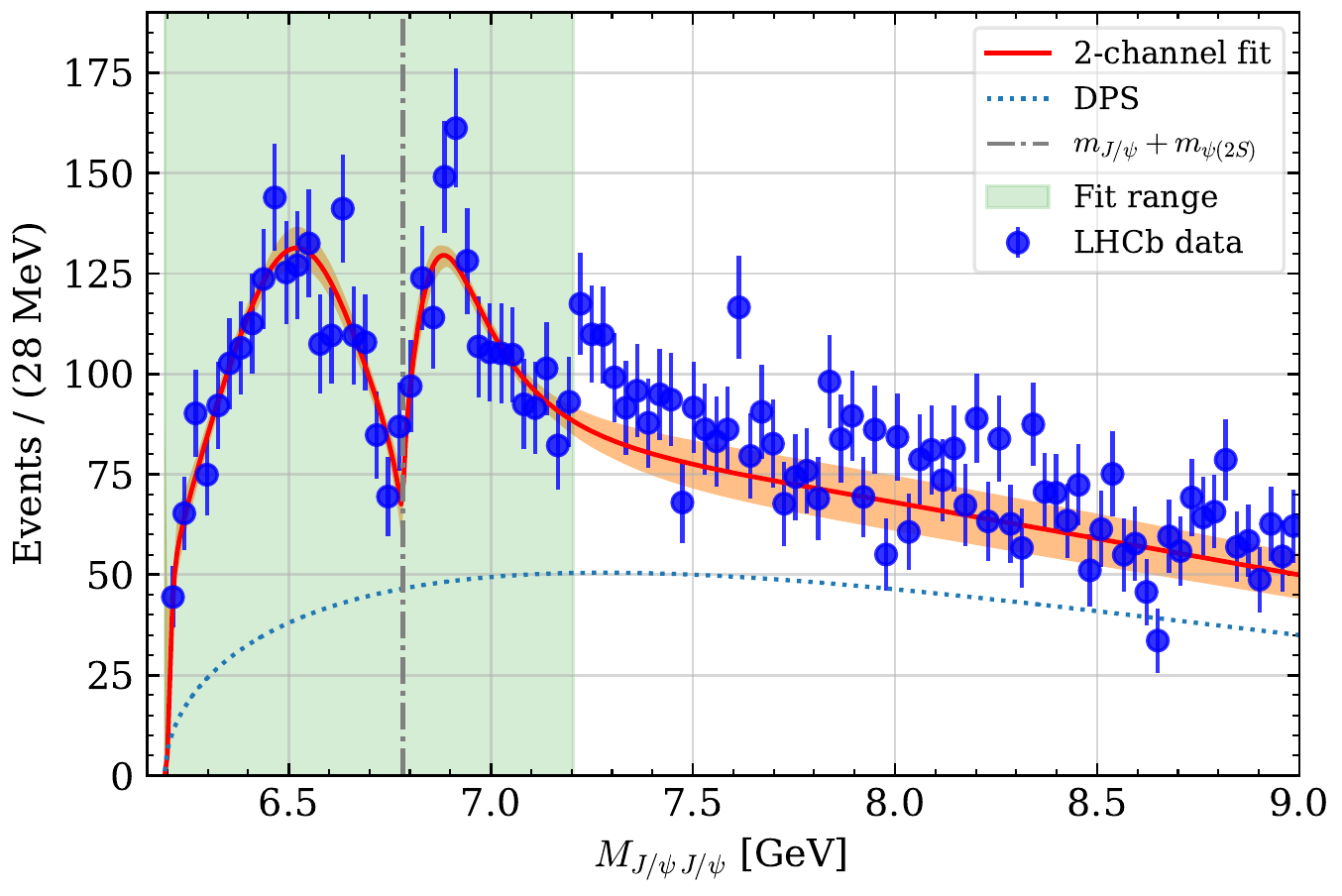}}
~~~
\subfigure[]{\includegraphics[width=0.463\textwidth]{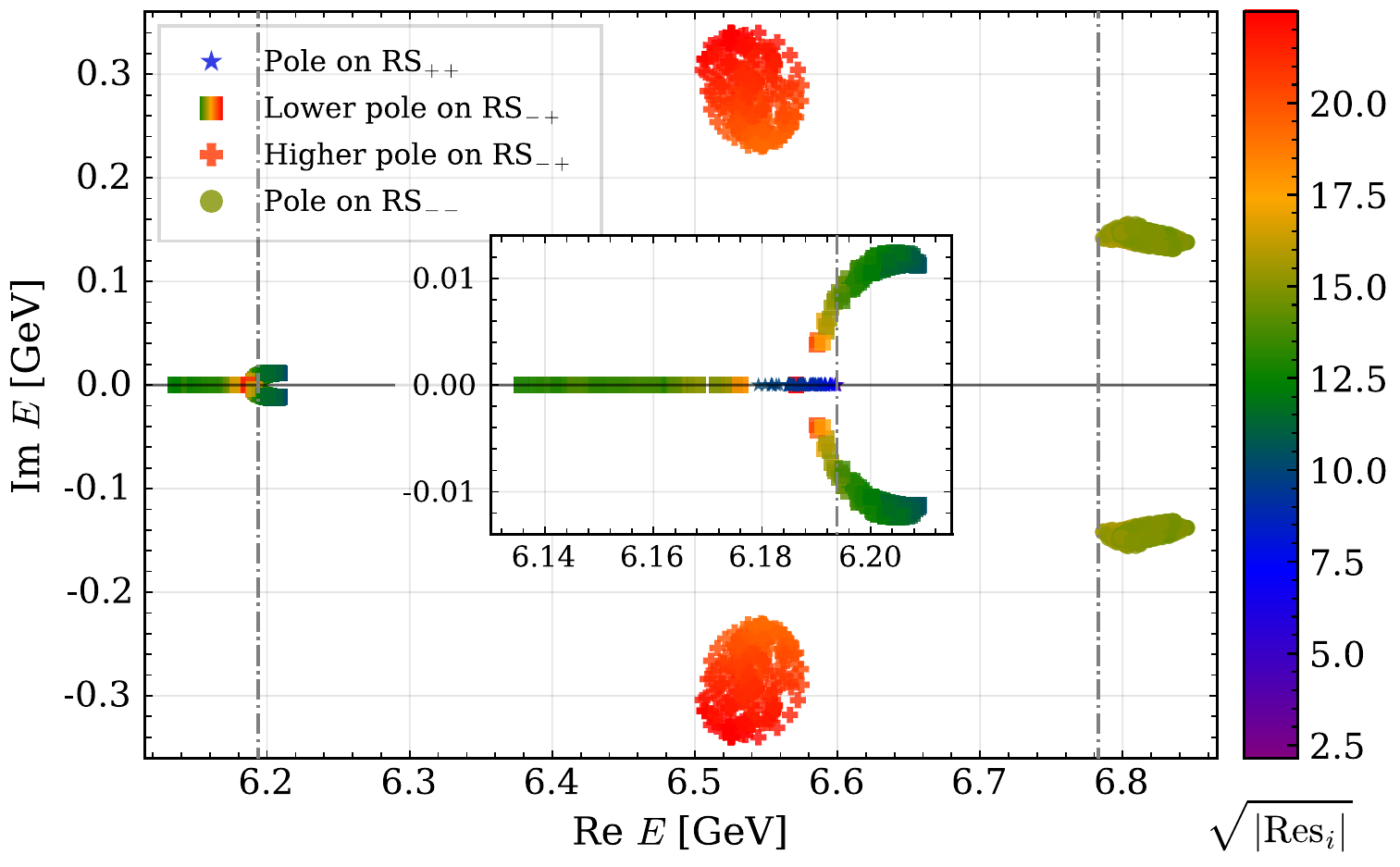}}
\end{center}
\caption{(a) Fit to the LHCb data of the di-$J/\psi$ invariant mass distribution, and (b) poles of the $T$-matrix from this fit with the subplot zooming in the poles around 6.2~GeV. Source: Ref.~\cite{Dong:2020nwy}.}
\label{fig:X6200}
\end{figure*}

The authors of Ref.~\cite{Dong:2020nwy} described the LHCb data on the di-$J/\psi$ spectrum~\cite{LHCb:2020bwg} through a coupled-channel approach without introducing any pre-existing bare pole. As shown in Fig.~\ref{fig:X6200}, their results suggest the existence of a near-threshold state in the $J/\psi J/\psi$ system with the quantum number $J^{PC} = 0^{++}$ or $2^{++}$. This state was labeled as $X(6200)$. A more general discussion can be found in Ref.~\cite{Dong:2020hxe}, where the authors proposed that there should be a threshold cusp at any $S$-wave threshold, showing up as a peak only for the channels with attractive interactions, with the width inversely proportional to the reduced mass relevant for the threshold.

\begin{figure*}[hbtp]
\begin{center}
\subfigure[]{\includegraphics[width=0.45\textwidth]{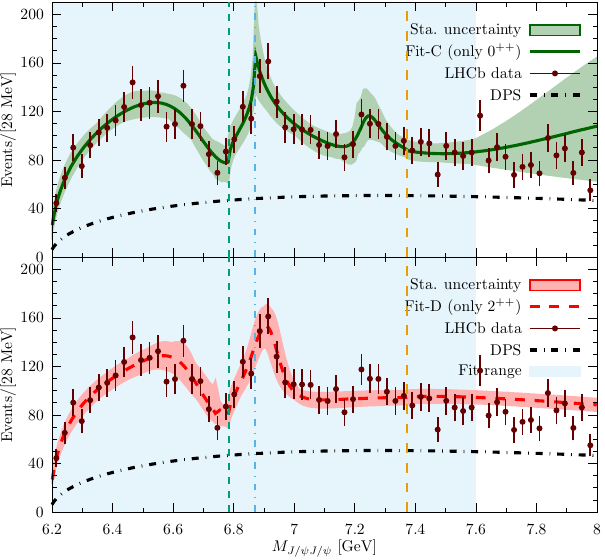}}
~~~
\subfigure[]{\includegraphics[width=0.45\textwidth]{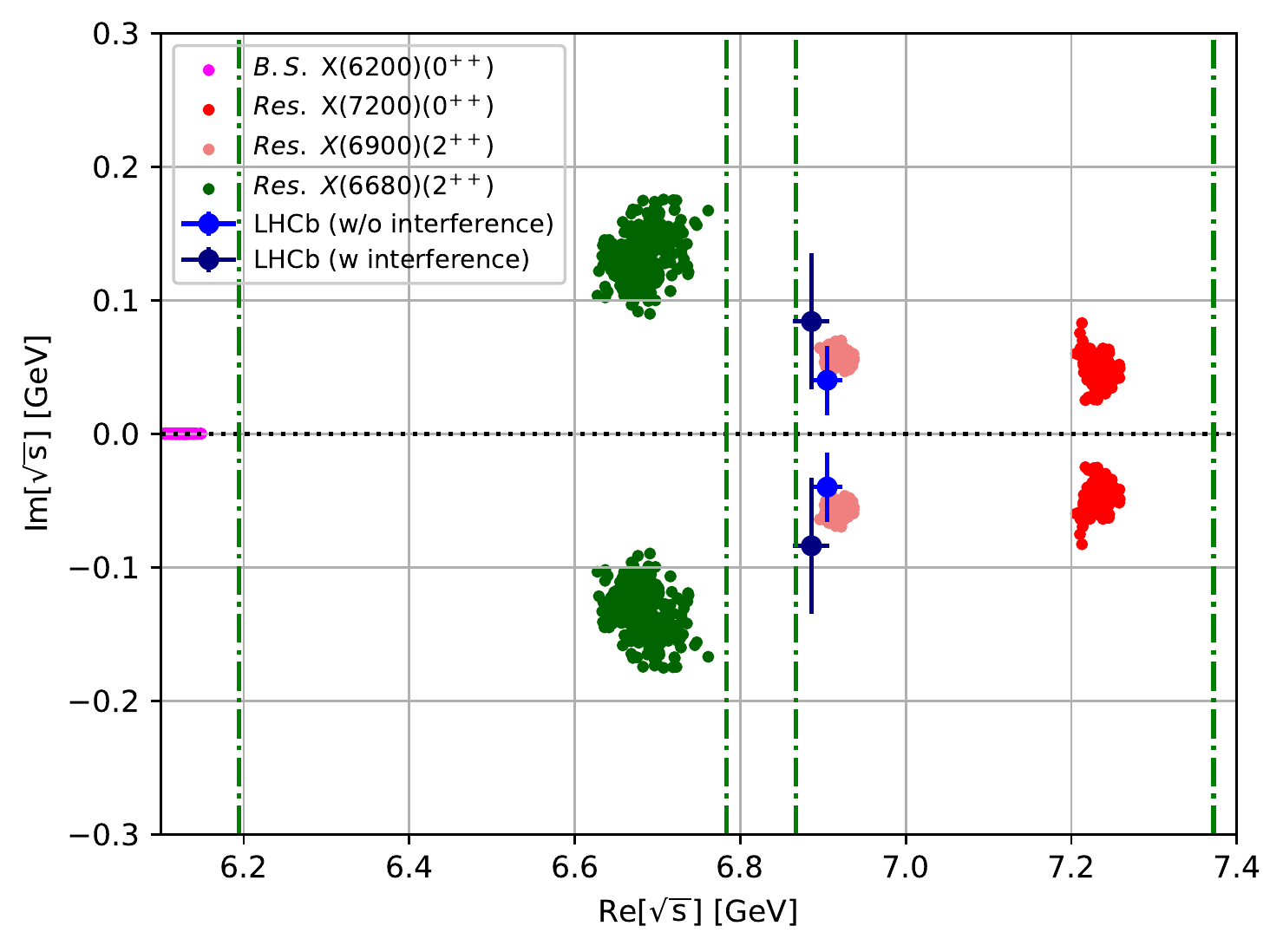}}
\end{center}
\caption{(a) Four-channel fit to the LHCb data of the di-$J/\psi$ invariant mass distribution and (b) poles of the $T$-matrix from this fit. Source: Ref.~\cite{Liang:2021fzr}.}
\label{fig:X6900fit}
\end{figure*}

Later in Ref.~\cite{Liang:2021fzr} the authors used the effective potentials as dynamical inputs, and performed a partial wave analysis to describe the LHCb data on the di-$J/\psi$ spectrum. They found a dynamically generated pole structure, which can be identified as the $X(6900)$ with the quantum number $J^{PC} = 2^{++}$. As shown in Fig.~\ref{fig:X6900fit}, their results suggest the existence of three other states: a bound state $X(6200)$ of $J^{PC} = 0^{++}$, a broad resonant state $X(6680)$ of $J^{PC} = 2^{++}$, and a narrow resonant state $X(7200)$ of $J^{PC} = 0^{++}$. The $X(6200)$ was further studied in Ref.~\cite{Nefediev:2021pww} within the QCD string model, and its quantum number $J^{PC} = 0^{++}$ was also favored.

In Ref.~\cite{Gong:2020bmg} the authors studied the Pomeron exchanges in the vector charmonium scattering. They found that the $X(6900)$, as a dynamically generated resonance pole, can arise from the Pomeron exchanges and coupled-channel effects between the $J/\psi J/\psi$ and $J/\psi \psi(2S)$ scattering. They also found a pole structure near the $J/\psi J/\psi$ threshold. Later in Ref.~\cite{Dong:2021lkh} the authors investigated the interaction between two charmonia through the exchange of gluons hadronizing into two pions at the longest distance. They argued that the exchange of correlated light mesons is able to provide a sizeable attraction to the $J/\psi J/\psi$ system, so it is possible for two $J/\psi$ mesons to form a bound state.

The $X(6900)$ was analyzed in Ref.~\cite{Guo:2020pvt} within the framework of the effective-range expansion, compositeness relation and width saturation, and a coupled-multichannel dynamical study. The authors considered the $J/\psi J/\psi$, $\chi_{c0}\chi_{c0}$, and $\chi_{c1}\chi_{c1}$ channels. Their results suggest that the two-meson components do not play a dominant role in the $X(6900)$. However, their coupled-channel analyses found another pole corresponding to a narrow resonance $X(6825)$ of the molecular origin, sitting below the $\chi_{c0}\chi_{c0}$ threshold. Later in Ref.~\cite{Cao:2020gul} a momentum-dependent Flatt{\'e}-like parameterization was applied to study the $X(6900)$. They found that the lowest state in the $J/\psi J/\psi$ mass spectrum, with the same quantum numbers as the $X(6900)$, is essential to describe the extremely deep dip below 6800 MeV by destructive interference. Their results suggest that the $X(6900)$ is probably not a $J/\psi \psi(2S)$ molecular state.

\subsubsection{Productions and decay properties.}
\label{sec5.1.3}

Productions of the fully-charm tetraquark state $X(6900)$ have attracted much attention from particle theorists, since this is an important issue closely related to the productions of the (double) charmonia~\cite{Brambilla:2010cs,Chapon:2020heu}.

In Ref.~\cite{Feng:2020riv} the authors presented a model-independent theoretical framework to study the $X(6900)$ at large $p_T$ in hadron collision experiments. They proposed that the non-perturbative yet universal gluon-to-$X(6900)$ fragmentation function can be decomposed into the product of the perturbatively calculable short-distance coefficients and the long-distance non-relativistic QCD (NRQCD) matrix elements. They further adopted the diquark ansatz to roughly estimate these NRQCD matrix elements, based on which they calculated the cross section for the $X(6900)$ inclusive production at LHC. This cross section, with $15$~GeV $\leq p_T \leq 60$~GeV, was calculated to be $33^{+4}_{-4}$~pb for the $J^{PC} = 0^{++}$ state, and $424^{+13}_{-21}$~pb for the $2^{++}$ state.

The same NRQCD approach was applied in Refs.~\cite{Feng:2020qee,Huang:2021vtb} to analyze productions of the $X(6900)$ as well as its $C$-odd partners at $B$ factories. A similar method was applied in Ref.~\cite{Ma:2020kwb} to explore the nature of $X(6900)$, where the authors proposed another state also locating around 6.9 GeV. They calculated the ratio of the production cross section for the $X(6900)$ to this undiscovered state, and found it sensitive to the nature of $X(6900)$, {\it e.g.}, if the cross section for the undiscovered state is larger than half of that for the $X(6900)$, the $X(6900)$ should be a tetraquark state.

\begin{figure*}[hbtp]
\begin{center}
\subfigure[]{\includegraphics[width=0.3\textwidth]{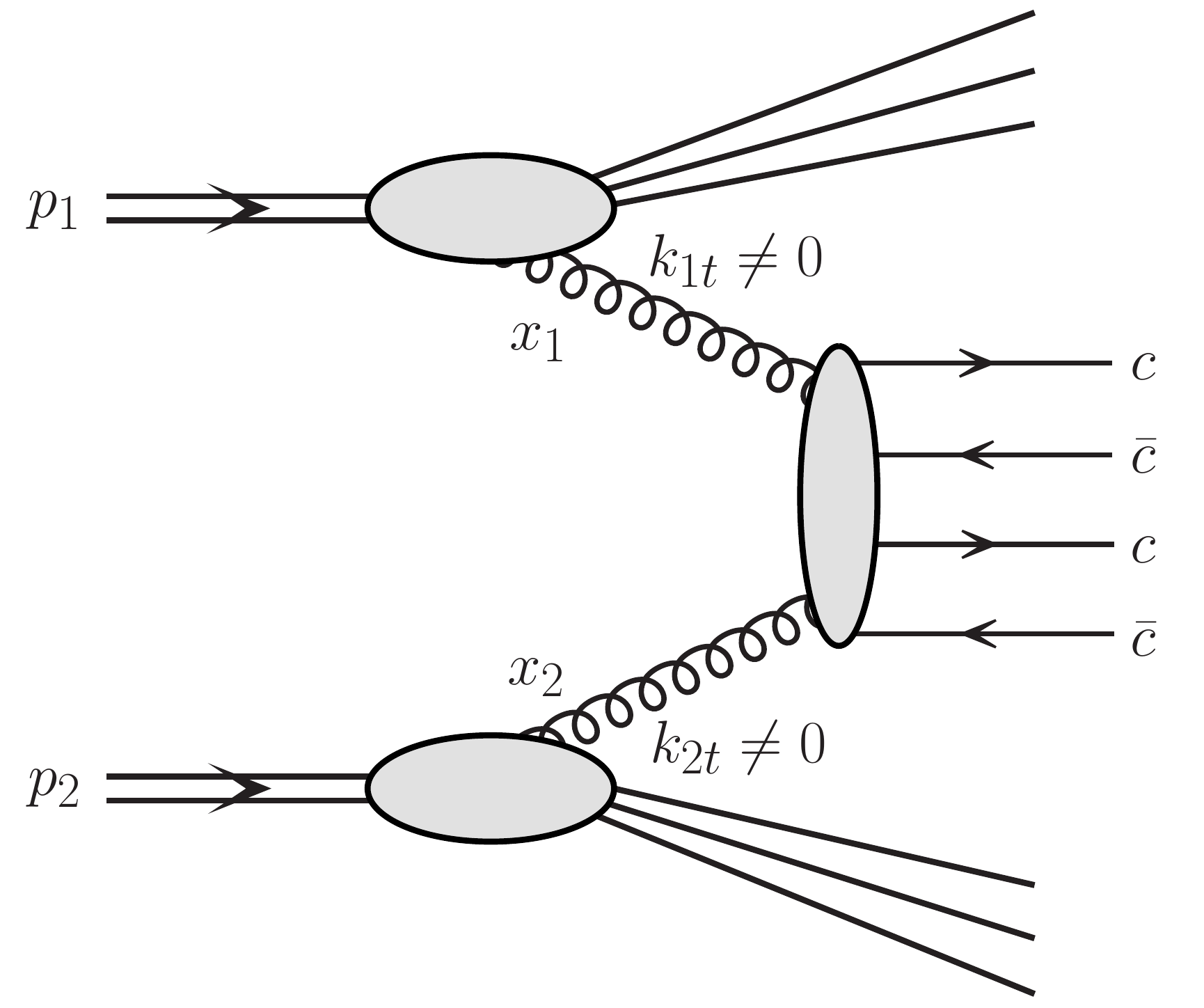}}
~
\subfigure[]{\includegraphics[width=0.3\textwidth]{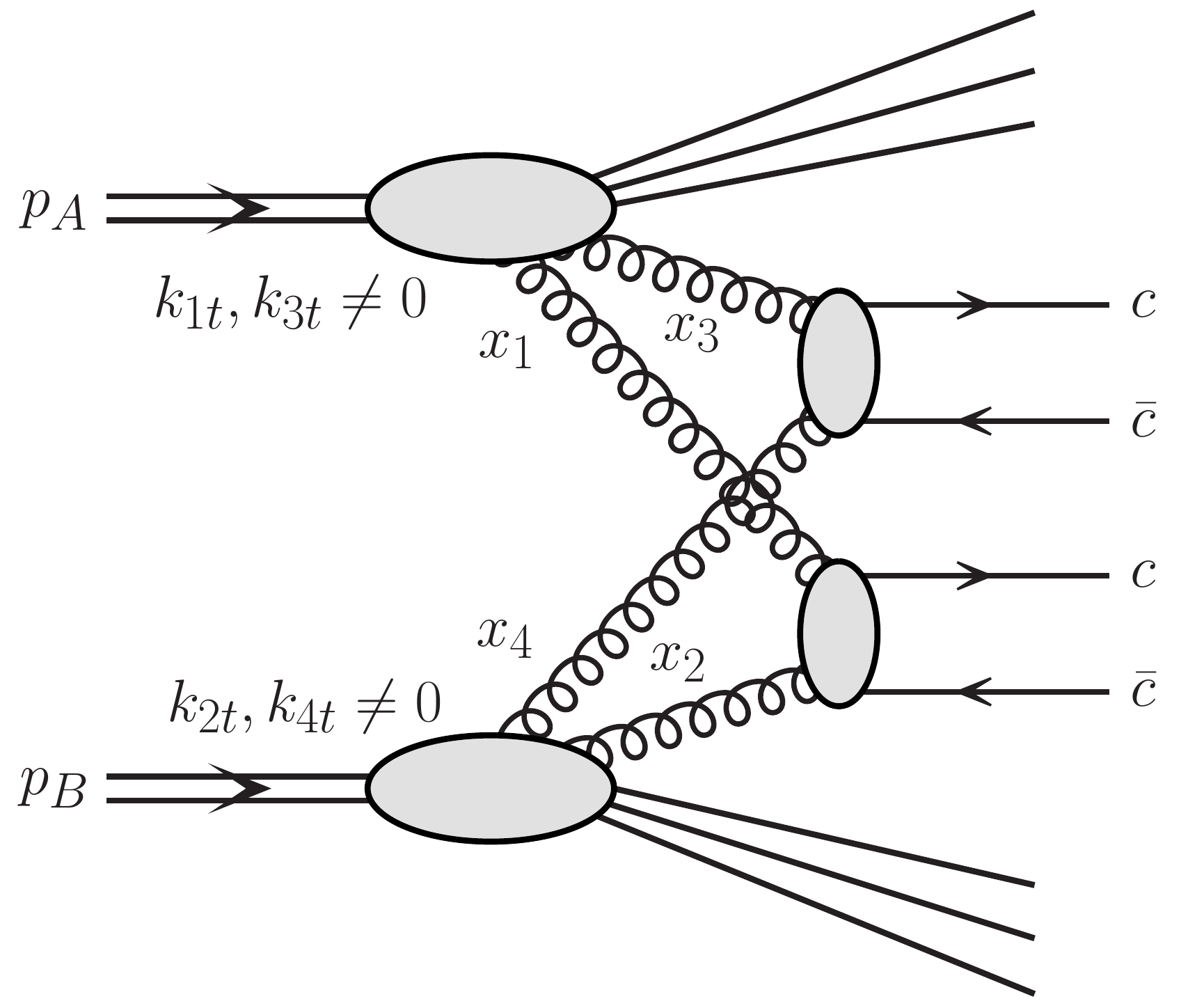}}
~
\subfigure[]{\includegraphics[width=0.3\textwidth]{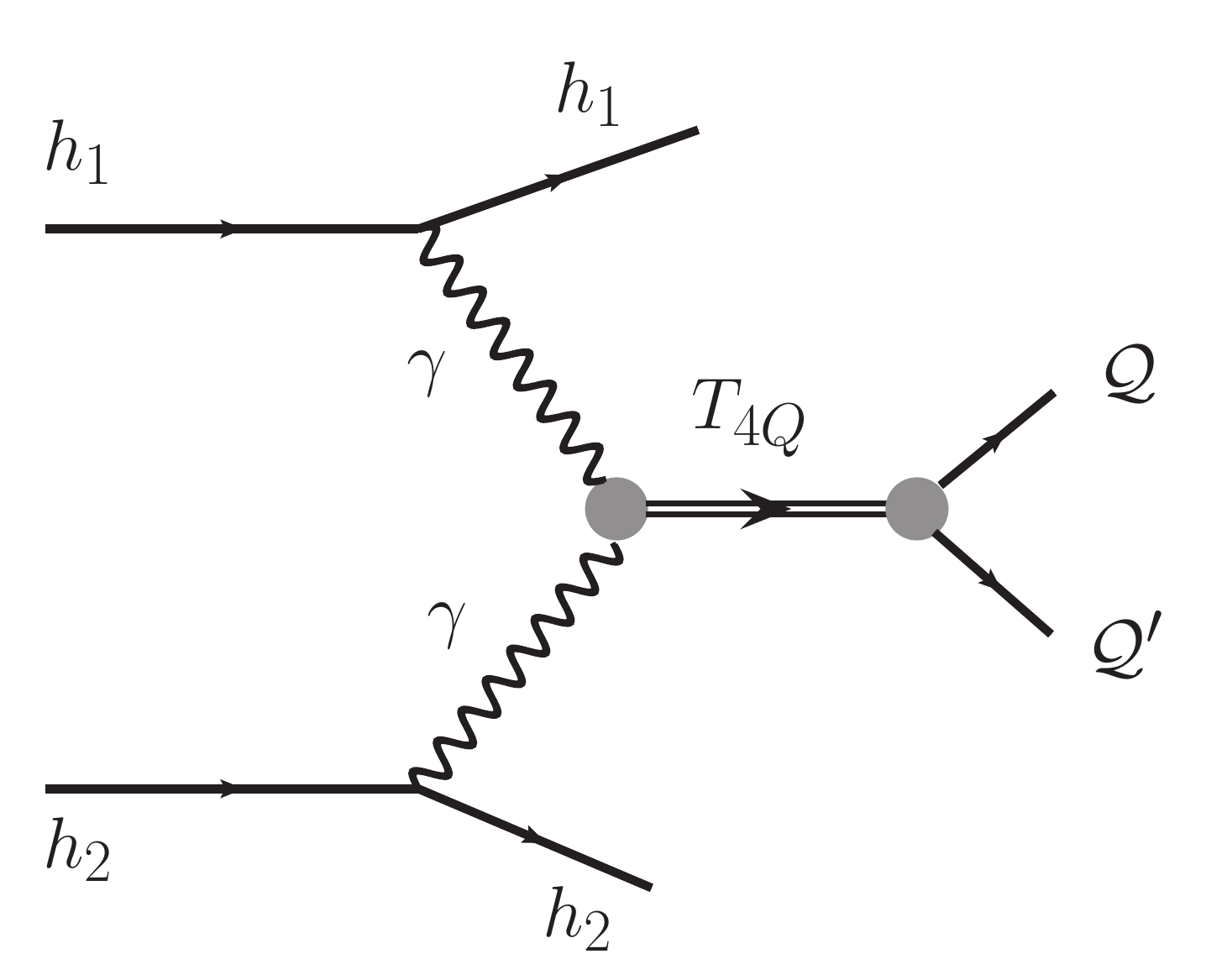}}
\end{center}
\caption{Typical diagrams for the $X(6900)$ productions through the (a) SPS and (b) DPS $gg \to X(6900)$ fusion mechanisms as well as through (c) the $\gamma \gamma$ interactions, with $h_i=p,\,Pb$ and ${\cal{Q}} = J/\psi$. Source: Refs.~\cite{Maciula:2020wri,Goncalves:2021ytq}.}
\label{fig:X6900production}
\end{figure*}

In Ref.~\cite{Maciula:2020wri} the authors studied the production mechanism of the $X(6900)$ in the $gg \to X(6900) \to J/\psi J/\psi$ process. As shown in Fig.~\ref{fig:X6900production}(a,b), they separately investigated the single parton scattering (SPS) and double parton scattering (DPS) mechanisms. They found that the contribution of the DPS mechanism is almost two orders of magnitude larger than that of the SPS one. They further considered two different spin scenarios $J^P=0^+$ and $J^P=0^-$, and found the $0^+$ assignment is preferred over the $0^-$ one. Later in Ref.~\cite{Goncalves:2021ytq} the authors investigated the production of the $X(6900)$ in the $\gamma \gamma$ interactions at LHC. As depicted in Fig.~\ref{fig:X6900production}(c), the cross sections for the $X(6900) (\to J/\psi J/\psi)$ production by $\gamma \gamma$ interactions in $pp$, $pPb$ and $PbPb$ collisions were estimated to be around 30~fb, 80~pb, and 200~nb, respectively.

The production of $X(6900)$ in the $\bar pp \to J/\psi J/\psi$ reaction was studied in Ref.~\cite{Wang:2020gmd} within an effective Lagrangian approach, where the authors found that the $X(6900)$ is only some cusp effect from the $J/\psi \psi(3770)$ channel. Their results suggest that the cross section for the $X(6900)$ production can reach dozens of pb at the best energy window $W = 6.9$~GeV, and its signal can be clearly distinguished from the background. They further estimated that dozens of $X(6900)$ can be detected with a data sample of $10.4$~fb$^{-1}$ collected with the D0 detector. Accordingly, they proposed to confirm the $X(6900)$ via the antiproton-proton scattering at D0. Later in Ref.~\cite{Esposito:2021ptx} the authors studied the production of $X(6900)$ in the ultra-peripheral heavy ion collisions, and estimated its production cross section to be around $250\sim1150$~nb, suggesting the ultra-peripheral heavy ion collisions as an ideal setup for the exotic searches.

Besides the above theoretical studies, the $X(6900)$ was also investigated in Refs.~\cite{Bedolla:2019zwg,Zhao:2020zjh,Mutuk:2021hmi,Wang:2021kfv,Weng:2020jao,Mikhasenko:2020qor,Sonnenschein:2020nwn,Eichmann:2020oqt,Zhu:2020snb,Wan:2020fsk,Gordillo:2020sgc,Liu:2020tqy,Majarshin:2021hex,Sombillo:2021rxv,Kuang:2022vdy,Zhuang:2021pci,Zhao:2020nwy,Lansberg:2020ejc,Andrade:2022rbn,Becchi:2020mjz,Becchi:2020uvq} through various theoretical methods and models, which we shall not discuss any more.

\subsection{Hidden-charm pentaquark state $P_{cs}(4459)$}
\label{sec5.2}

The LHCb collaboration discovered the famous hidden-charm pentaquark states $P_c(4312)$, $P_c(4337)$, $P_c(4380)$, $P_c(4440)$, and $P_c(4457)$ in 2015~\cite{LHCb:2015yax}, 2019~\cite{LHCb:2019kea}, and 2021~\cite{LHCb:2021chn}. These $P_c$ states contain at least five quarks $\bar c c uud$, so they are good candidates for the hidden-charm pentaquark states. It is natural to conjecture whether the hidden-charm pentaquark state with strangeness exists or not. Such a state is usually denoted as the ``$P_{cs}$'', whose quark content is $\bar c c s q q$ ($q=u/d$). It was suggested in Refs.~\cite{Chen:2015sxa,Cheng:2015cca,Santopinto:2016pkp,Shen:2020gpw} to search for the $P_{cs}$ state in the $J/\psi \Lambda$ invariant mass spectrum of the $\Xi_b^- \to J/\psi K^- \Lambda$ decay, and especially, this state was investigated in Ref.~\cite{Chen:2015sxa} as a molecular state.

\begin{figure*}[hbtp]
\begin{center}
\subfigure[]{\includegraphics[width=0.45\textwidth]{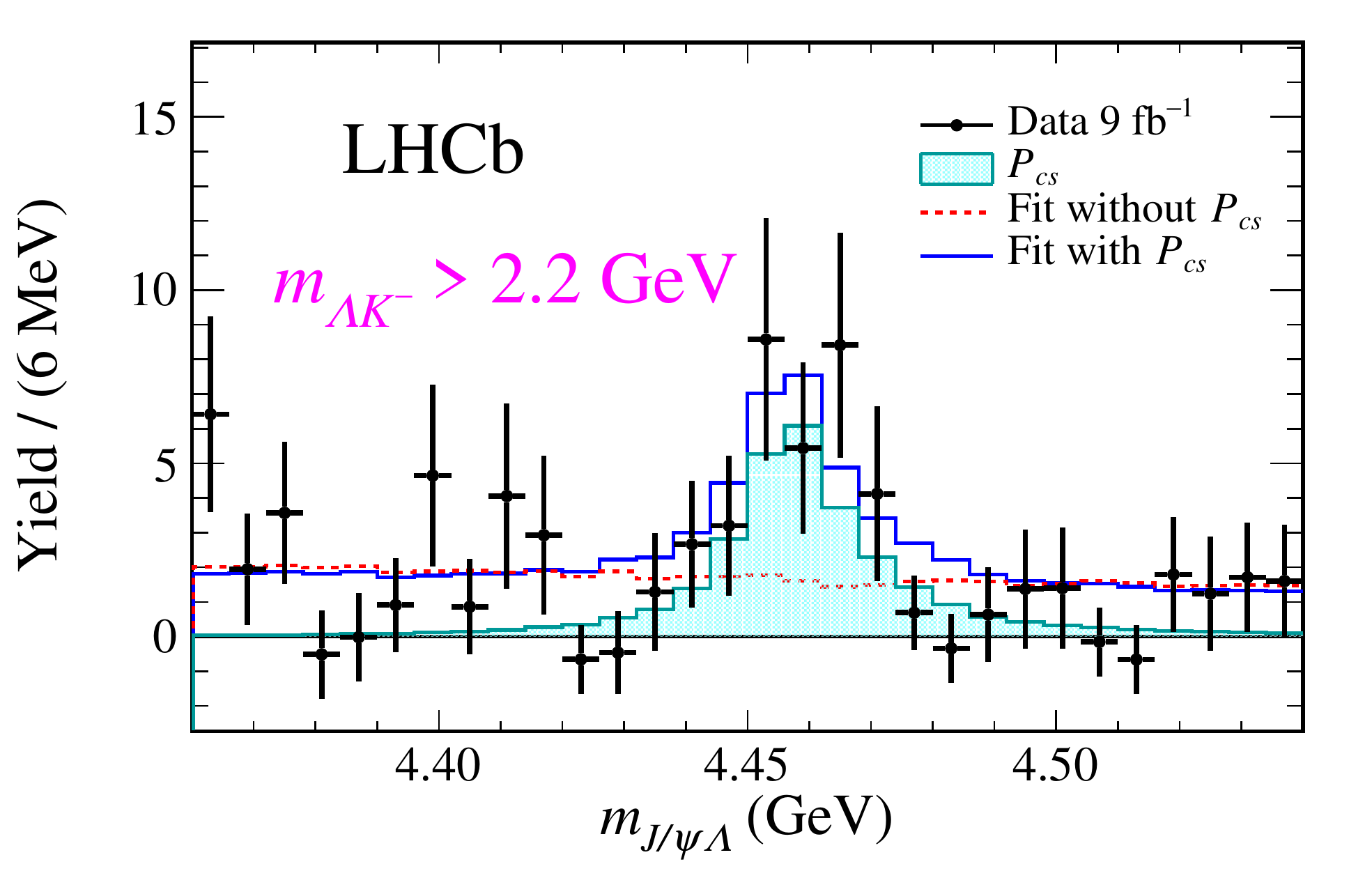}}
~~~
\subfigure[]{\includegraphics[width=0.45\textwidth]{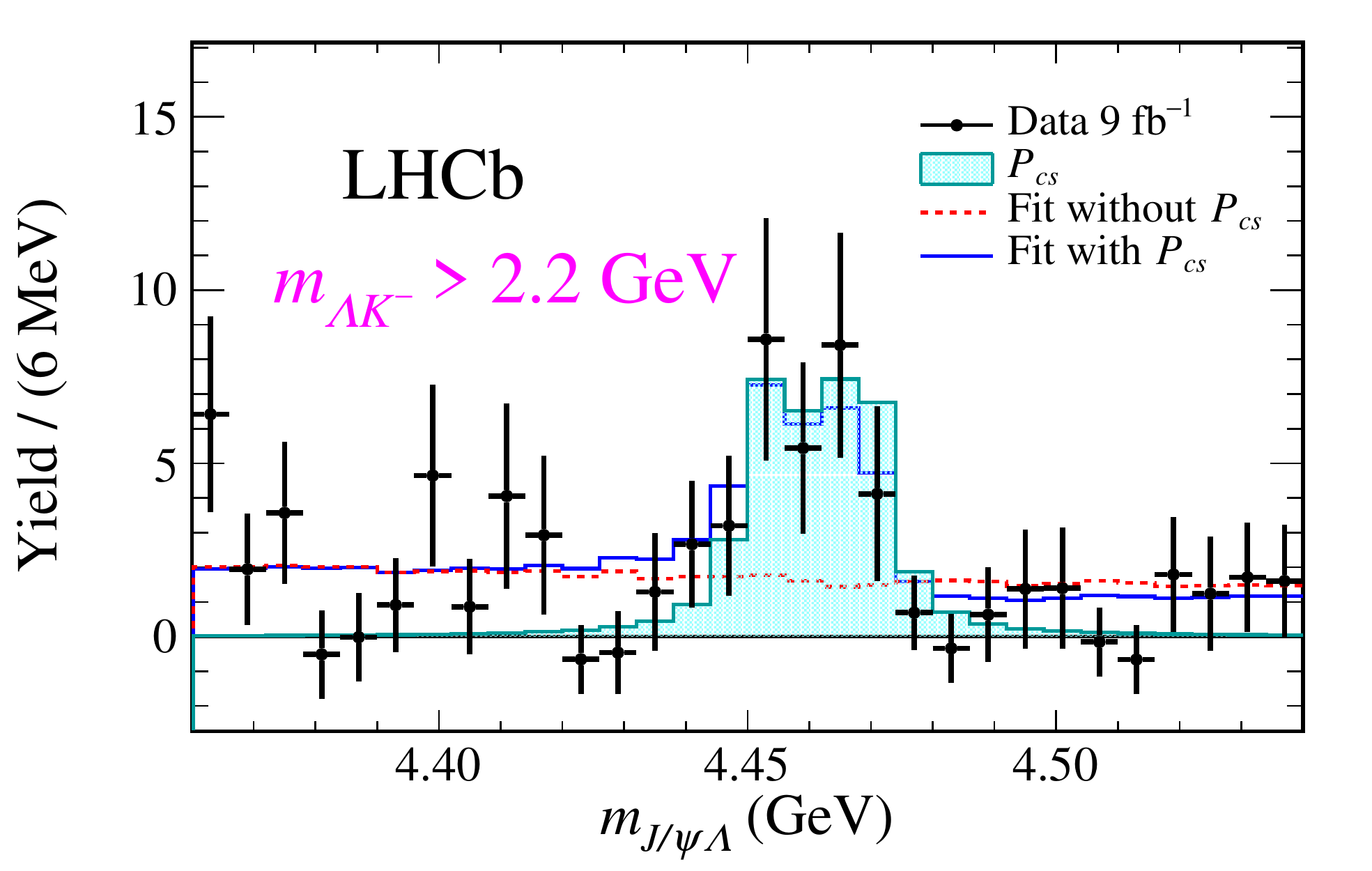}}
\end{center}
\caption{Projection of $m_{J/\psi \Lambda}$ with a $m_{\Lambda K^-} > 2.2$~GeV requirement, overlaid by the fits using (a) only one resonance and (b) two almost degenerate resonances to model the peak region. Source: Ref.~\cite{LHCb:2020jpq}.}
\label{fig:Pcs}
\end{figure*}

In 2020 the LHCb Collaboration reported the first evidence of a hidden-charm pentaquark state with strangeness, labeled as $P_{cs}(4459)^0$, in the $J/\psi \Lambda$ invariant mass spectrum of the $\Xi_b^- \to J/\psi K^- \Lambda$ decay~\cite{LHCb:2020jpq}, as shown in Fig.~\ref{fig:Pcs}. Its mass and width were measured to be:
\begin{eqnarray}
P_{cs}(4459)^0 &:& M = 4458.8 \pm  2.9^{+4.7}_{-1.1}{\rm~MeV} \, ,
\\ \nonumber && \Gamma = 17.3 \pm  6.5^{+8.0}_{-5.7}{\rm~MeV} \, .
\end{eqnarray}
Its spin-parity quantum number was not determined since the statistic is not enough.

The $P_{cs}(4459)^0$ is about 19~MeV below the $\bar D^{*0} \Xi_c^0$ threshold, so it is natural to interpret it as a $\bar D^{*} \Xi_c$ molecular state with the spin-parity quantum number $J^P = 1/2^-$ or $3/2^-$~\cite{Zou:2021sha,Karliner:2021xnq}. As suggested in Ref.~\cite{Wang:2019nvm}, there may exist two almost degenerate molecular states below the $\bar D^{*} \Xi_c$ threshold, similar to the two $P_c(4440)$ and $P_c(4457)$ states below the $\bar D^*\Sigma_c$ threshold. Accordingly, LHCb also performed a fit using two resonances to model the peak region, as shown in Fig.~\ref{fig:Pcs}(b). Their masses and widths were measured to be $4454.9 \pm 2.7$~MeV and $7.5 \pm 9.7$~MeV as well as $4467.8 \pm 3.7$~MeV and $5.2 \pm 5.3$~MeV, respectively. These mass values are in a remarkable coincidence with the masses of the two $\bar D^{*} \Xi_c$ molecular states calculated in Ref.~\cite{Wang:2019nvm} using the chiral effective field theory:
\begin{eqnarray}
\bar D^{*} \Xi_c {\rm~of~}J^P = 1/2^- &:& M = 4456.9^{+3.2}_{-3.3}{\rm~MeV} \, ,
\\ \nonumber \bar D^{*} \Xi_c {\rm~of~}J^P = 3/2^- &:& M = 4463.0^{+2.8}_{-3.0}{\rm~MeV} \, .
\end{eqnarray}
Besides, the existence of the $P_{cs}$ states had been predicted in Refs.~\cite{Wu:2010jy,Chen:2016ryt,Shen:2019evi,Xiao:2019gjd} before the LHCb experiment~\cite{LHCb:2020jpq}, and more theoretical studies at that time can be found in Refs.~\cite{Anisovich:2015zqa,Wang:2015wsa,Feijoo:2015kts,Lu:2016roh,Cheng:2019obk,Zhang:2020cdi,Liu:2020hcv}.

In Ref.~\cite{Peng:2020hql} the authors applied the effective field theory to study the $P_{cs}(4459)$ as a $\bar D^{*} \Xi_c$ molecular state. Based on the heavy quark spin symmetry, they found two almost degenerate states, and the $J=3/2$ one is a bit lighter than the $J=1/2$ one due to their coupling to the nearby $\bar D \Xi^\prime_c$ and $\bar D \Xi^*_c$ channels. They argued that the spectroscopy and the $J/\psi \Lambda$ decay mode might suggest a marginal preference for $J=3/2$ over $J=1/2$. Later in Ref.~\cite{Zhu:2021lhd} the authors performed a coupled-channel analysis of the $\bar D^* \Xi_c^*$, $\bar D^* \Xi^\prime_c$, $\bar D \Xi^*_c$, $\bar D^* \Xi_c$, $\bar D \Xi^\prime_c$, and $\bar D \Xi_c$ interactions in the quasipotential Bethe-Salpeter equation approach. They found two $\bar D^* \Xi_c$ molecular states of $J^P=1/2^-$ and $3/2^-$. Again, they found the $3/2^-$ state to be the lower one, which was related to the $P_{cs}(4459)$.

The $\bar D^* \Xi_c$ molecular state of $J^P=1/2^-$ was used in Ref.~\cite{Hu:2021nvs} to explain the $P_{cs}(4459)$, although the mass of the $J^P=3/2^-$ one was still calculated to be lighter, {\it i.e.}, the masses of the $J^P=1/2^-$ and $3/2^-$ states were calculated to be about $4461$~MeV and $4443$~MeV, respectively. However, the results of Ref.~\cite{Du:2021bgb} favor the $J^P=1/2^-$ state to be the lighter one and the $J^P=3/2^-$ state to be the heavier one, both of which are located in the energy region of the $P_{cs}(4459)$. The magnetic moments of the $\bar D^* \Xi_c$ molecular states were calculated in Ref.~\cite{Li:2021ryu} to be $\mu_{P_{cs}} = -0.062 \mu_N$ for the $J^P = 1/2^-$ state, and $\mu_{P_{cs}} = 0.465 \mu_N$ for the $J^P = 3/2^-$ state. More calculations within the meson-baryon, diquark-diquark-antiquark, and diquark-triquark pictures can be found in Refs.~\cite{Ozdem:2021ugy,Gao:2021hmv}. These results may be useful in determining the quantum numbers and the internal structure of the $P_{cs}(4459)$.

In Ref.~\cite{Xiao:2021rgp} the authors studied the $P_{cs}(4459)$ through the coupled-channel unitary approach combined with the heavy quark spin and local hidden-gauge symmetries. Their results suggest that the $P_{cs}(4459)$ can be explained as the $\bar D^* \Xi_c$ molecular state of either $J^P=1/2^-$ or $3/2^-$. Besides, they found another pole in the $J^P=1/2^-$ sector located at around $4310$~MeV, corresponding to a deep bound state of $\bar D \Xi_c$. The interpretation of the $P_{cs}(4459)$ as a $\bar D^* \Xi_c$ molecular state was supported by Ref.~\cite{Chen:2020uif} through the QCD sum rule method. More QCD sum rule studies on the hidden-charm pentaquark states can be found in Refs.~\cite{Chen:2015moa,Chen:2016otp,Xiang:2017byz,Chen:2019bip,Chen:2020pac,Chen:2020opr}.

\begin{figure*}[hbtp]
\begin{center}
\includegraphics[width=0.8\textwidth]{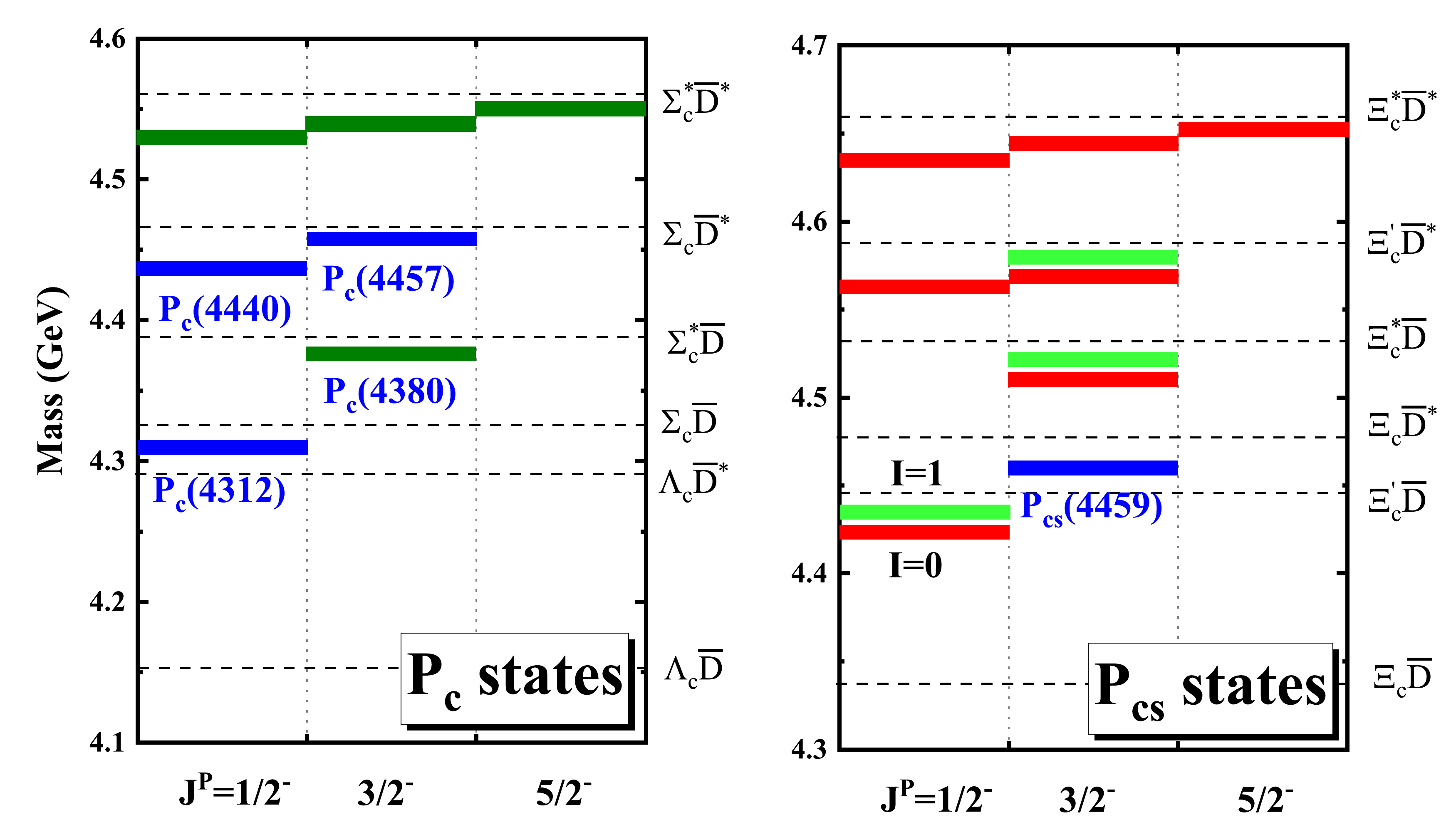}
\end{center}
\caption{Mass spectra of the $P_c$ and $P_{cs}$ states in the meson-baryon molecular scenario, calculated in Refs.~\cite{Chen:2022onm,Chen:2019asm} using the one-boson-exchange model. The $P_c$ states in the left panel all have $I=1/2$. The red and green lines in the right panel label the $P_{cs}$ states with $I = 0$ and $I = 1$, respectively. Source: Ref.~\cite{Chen:2022onm}.}
\label{fig:Pcsobe}
\end{figure*}

In Ref.~\cite{Chen:2020kco} the author performed a single-channel $\bar D^* \Xi_c$ and a four-channel $\bar D^* \Xi_c/\bar D \Xi^*_c/\bar D \Xi^\prime_c/\bar D^* \Xi^*_c$ analyses to study the $P_{cs}(4459)$ using the one-boson-exchange model. His results suggest that the $P_{cs}(4459)$ cannot be interpreted as a pure $\bar D^* \Xi_c$ molecular state, while it can be explained as a bound state of $(I)J^P = (0)3/2^-$ in the coupled channels $\bar D^* \Xi_c/\bar D \Xi^*_c/\bar D \Xi^\prime_c/\bar D^* \Xi^*_c$, with the $\bar D^* \Xi_c$ and $\bar D \Xi^*_c$ components being dominant. Based on this interpretation, the author further investigated the two-body strong decay behaviors of the $P_{cs}(4459)$ in Ref.~\cite{Chen:2021tip}. He calculated the partial decay width of the $D_s^* \Lambda_c$ channel to be around 0.6 to 2.0 MeV, while the width of the $J/\psi \Lambda$ channel is ten times smaller. In a recent study~\cite{Chen:2022onm}, the authors performed a coupled-channel analysis to systematically study the interactions between $\Xi_c^{(\prime,*)}$ and $\bar D^{(*)}$. As summarized in Fig.~\ref{fig:Pcsobe}, they predicted the existence of several hidden-charm molecular pentaquarks with strangeness, including the $\bar D \Xi_c^\prime$ molecules of $(I)J^P = (0/1){1\over2}^-$, the $\bar D \Xi_c^*$ molecules of $(0/1){3\over2}^-$, the $\bar D^* \Xi_c^\prime$ molecules of $(0){1\over2}^-$ and $(0/1){3\over2}^-$, and the $\bar D^* \Xi_c^*$ molecules of $(0){1\over2}^-/{3\over2}^-/{5\over2}^-$.

\begin{figure*}[hbtp]
\begin{center}
\includegraphics[width=1\textwidth]{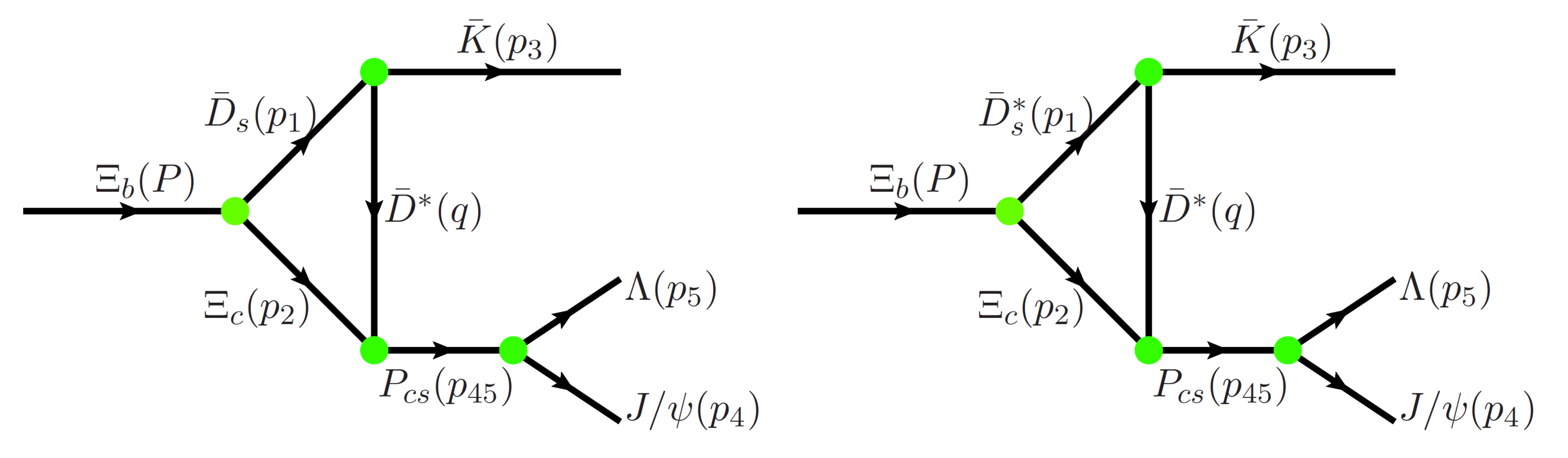}
\end{center}
\caption{Triangle diagrams for the $\Xi^-_b \to K^- P_{cs} (\to J/\psi\Lambda)$ decay process. Source: Ref.~\cite{Lu:2021irg}.}
\label{fig:Pcstriangle}
\end{figure*}

Based on the interpretation of the $P_{cs}(4459)$ as a $\bar D^{*} \Xi_c$ molecular state, its production in the $\Xi_b$ decay was studied in Ref.~\cite{Wu:2021caw} through an effective Lagrangian approach. The magnitude of the branching fraction for the $\Xi_b \to P_{cs}(4459) K$ decay was estimated to be of the order of $10^{-4}$. Later in Ref.~\cite{Lu:2021irg}, the $\Xi_b^- \to J/\psi K^- \Lambda$ decay was investigated via the triangle diagrams $\Xi_b \to \bar D^{(*)}_s \Xi_c \to (\bar D^{(*)} \bar K )\Xi_c \to P_{cs} \bar K \to (J/\psi \Lambda) \bar K$, as shown in Fig.~\ref{fig:Pcstriangle}. They calculated the production yield of the $\bar{D}^*\Xi_c$ molecular state with $J^P=3/2^-$ to be approximately one order of magnitude larger than that of the $J^P=1/2^-$ state, due to the interference with the $\bar{D}_s\Xi_c$ and $\bar{D}_s^*\Xi_c$ intermediate states.

Strong decays of the $P_{cs}(4459)$ into the $J/\psi \Lambda$, $\bar D_s \Lambda_c$, and $\bar D \Xi^{(\prime)}_c$ final states were investigated in Ref.~\cite{Yang:2021pio} through hadronic loops. Their results support its interpretation as the $\bar D^* \Xi_c$ molecular state of $J^P = 1/2^-$, with the $\bar D \Xi^\prime_c$ decay mode having the largest branching ratio. More theoretical studies can be found in Refs.~\cite{Cheng:2021gca,Clymton:2021thh,Liu:2020ajv}, which investigated the $P_{cs}(4459)$ in the $\gamma p \to K^+ P_{cs}(4459)$, $K^- p \to J/\psi \Lambda$, and $\Lambda_b \to J/\psi \Lambda \phi$ reactions.

\begin{figure*}[hbtp]
\begin{center}
\includegraphics[width=1\textwidth]{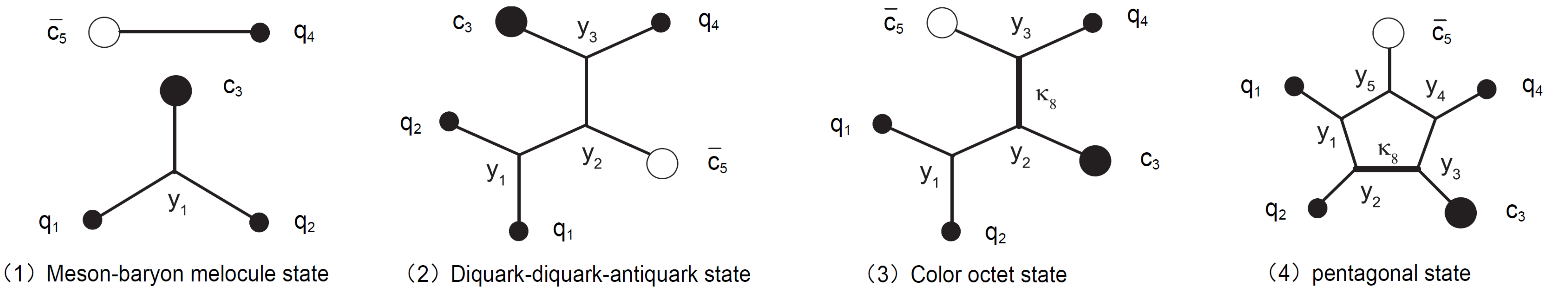}
\end{center}
\caption{Possible color structures of the hidden-charm pentaquark states $qqqc \bar c$ ($q = u/d/s$). Source: Ref.~\cite{Deng:2022vkv}.}
\label{fig:Pcsfluxtude}
\end{figure*}

Besides the $\bar D^{*} \Xi_c$ molecular picture, the compact pentaquark picture was also applied to investigate the $P_{cs}(4459)$. In Ref.~\cite{Deng:2022vkv} the author made an exhaustive investigation on the pentaquark states $qqqc \bar c$ ($q = u/d/s$) through a multiquark color flux-tube model. As shown in Fig.~\ref{fig:Pcsfluxtude}, he considered altogether four possible color structures, including the meson-baryon, diquark-diquark-antiquark, color-octet (diquark-triquark), and pentagonal configurations. His results suggest that the $P_{cs}(4459)$ has the pentagon structure with the spin-parity quantum number $J^P=1/2^-$.

The diquark-diquark-antiquark configuration was used in Refs.~\cite{Shi:2021wyt,Wang:2020eep,Azizi:2021utt} to investigate the $P_{cs}(4459)$. In Ref.~\cite{Shi:2021wyt} the authors studied the $P_{cs}(4459)$ through an effective Lagrangian approach, and their results support its assignment as a compact diquark-diquark-antiquark state of $(I)J^P = (0)1/2^-$ or $(0)3/2^-$. In Ref.~\cite{Wang:2020eep} the author studied the scalar-diquark-scalar-diquark-antiquark type pentaquark state using the QCD sum rule method, and his results support the interpretation of the $P_{cs}(4459)$ as the $udsc \bar c$ state of $J^P=1/2^-$. The authors of Ref.~\cite{Azizi:2021utt} studied the $P_{cs}(4459) \to J/\psi \Lambda$ decay also within the QCD sum rule framework, and calculated its partial decay width to be $15.87 \pm 3.11$~MeV.

Besides the $P_{cs}(4459)^0$, another hidden-charm pentaquark state with strangeness was reported by the LHCb collaboration very recently~\cite{LHCbTcs1}. It was named as $P^\Lambda_{\psi s}(4338)$ according to the new convention proposed by LHCb~\cite{Gershon:2022xnn}, or $P_{cs}(4338)^0$ according to the previous widely-used convention. This exotic structure was observed in the $J/\psi \Lambda$ mass distribution of the $B^- \to J/\psi \Lambda \bar p$ decay. Its mass and width were measure to be
\begin{eqnarray}
P^\Lambda_{\psi s}(4338)/P_{cs}(4338)^0 &:& M = 4338.2 \pm 0.7 \pm 0.4 {\rm~MeV} \, ,
\\ \nonumber && \Gamma = 7.0 \pm 1.2 \pm 1.3 {\rm~MeV} \, ,
\end{eqnarray}
and its spin-parity quantum numbers were determined to be $J^P= 1/2^-$ with a high significance.  After this LHCb measurement~\cite{LHCbTcs1}, the $P^\Lambda_{\psi s}(4338)/P_{cs}(4338)^0$ has attracted much attention and there have been some theoretical studies~\cite{Wang:2022mxy,Yan:2022wuz,Meng:2022wgl,Burns:2022uha,Ozdem:2022kei,Wang:2022tib,Nakamura:2022jpd}. We shall not review these studies, but just note that the $P^\Lambda_{\psi s}(4338)/P_{cs}(4338)^0$ is a good candidate for the $\bar D \Xi_c$ hadronic molecule of $J^P= 1/2^-$.

\subsection{Hidden-charm tetraquark states $Z_{cs}(3985)/Z_{cs}(4000)/Z_{cs}(4220)$}
\label{sec5.3}

\begin{figure*}[hbtp]
\begin{center}
\includegraphics[width=0.6\textwidth]{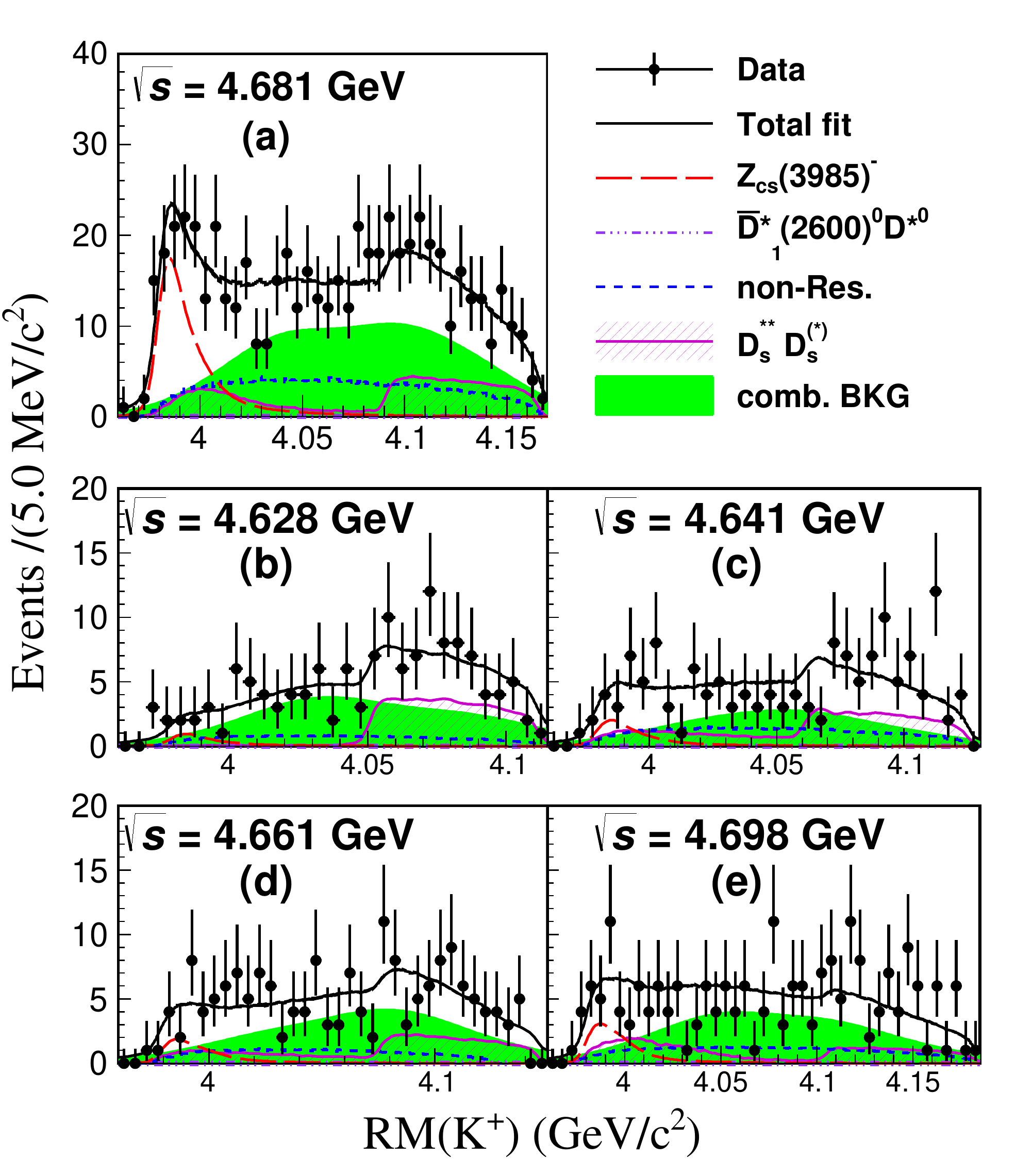}
\end{center}
\caption{Simultaneous unbinned maximum likelihood fit to the $K^+$ recoil-mass spectra in data at $\sqrt{s}$=4.628, 4.641, 4.661, 4.681 and 4.698~GeV. Source: Ref.~\cite{BESIII:2020qkh}.}
\label{fig:Zcs3985}
\end{figure*}

In 2020 the BESIII collaboration studied the $e^+e^-\to K^+ D_s^- D^{*0}$ and $K^+ D^{*-}_s D^0$ reactions based on 3.7~fb$^{-1}$ of data collected at $\sqrt{s}=$4.628, 4.641, 4.661, 4.681, and 4.698~GeV~\cite{BESIII:2020qkh}. They observed an excess of events near the $D_s^- D^{*0}$ and $D^{*-}_s D^0$ mass thresholds in the $K^+$ recoil-mass spectrum for events collected at $\sqrt{s}=4.681$~GeV, as shown in Fig.~\ref{fig:Zcs3985}. Its pole mass and width were measured to be
\begin{eqnarray}
Z_{cs}(3985)^- &:& M = 3982.5^{+1.8}_{-2.6}\pm2.1 {\rm~MeV} \, ,
\label{eq:Zcs3985}
\\ \nonumber && \Gamma = 12.8^{+5.3}_{-4.4}\pm3.0 {\rm~MeV} \, .
\end{eqnarray}
The significance of the resonance hypothesis was estimated to be $5.3\sigma$ over the contributions only from the conventional charmed mesons. A polarization analysis was recently performed in Ref.~\cite{Chen:2022yev} with the motivation of measuring its spin quantum number in the future. The $Z_{cs}(3985)^-$ is the first candidate for the charged hidden-charm tetraquark with strangeness.

\begin{figure*}[hbtp]
\centering
\includegraphics[width=1.0\textwidth]{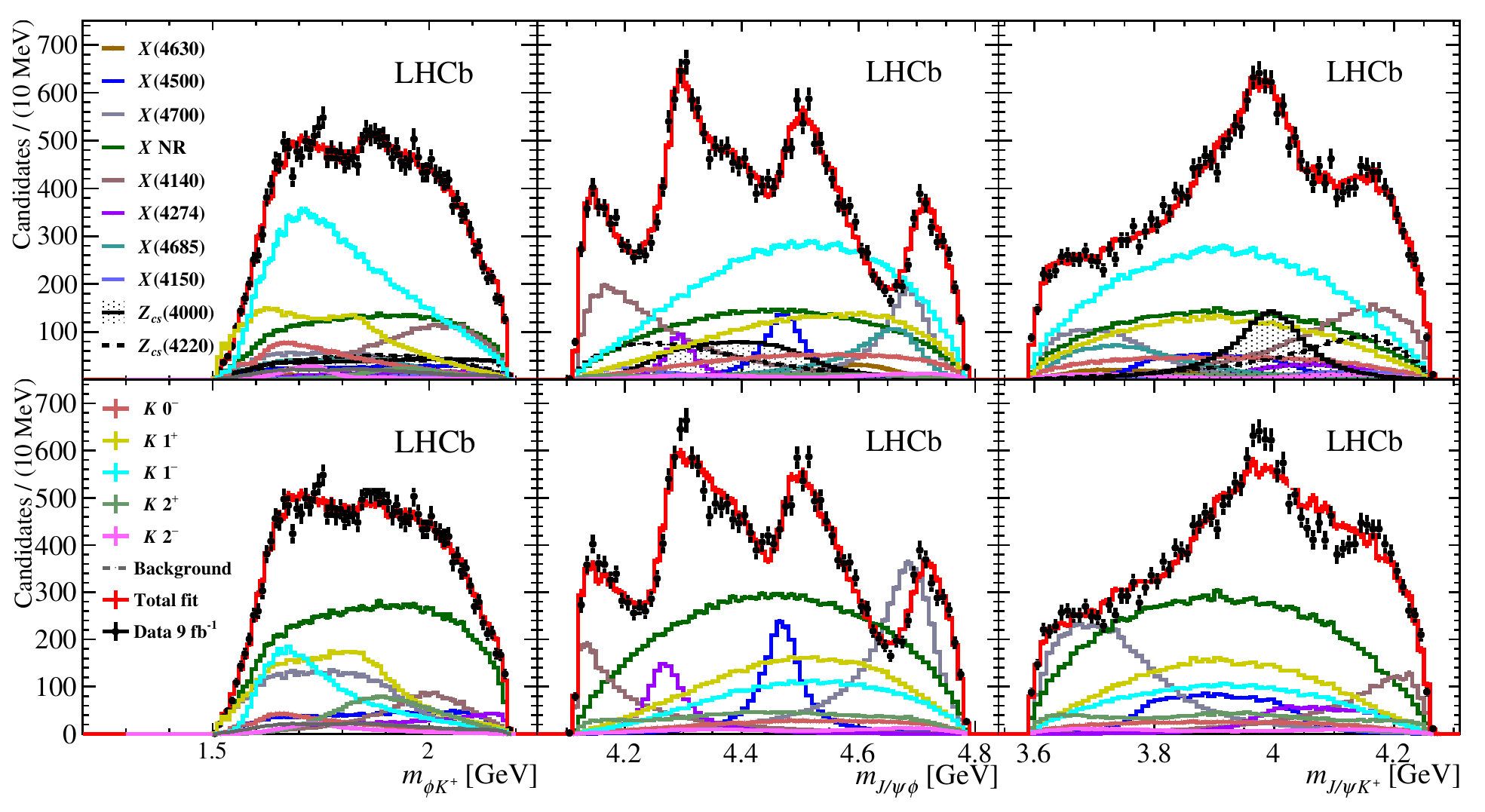}
\caption{Distributions of the $\phi K^+$ (left), $J/\psi \phi$ (middle), and $J/\psi K^+$ (right) invariant masses for the $B^+ \to J/\psi \phi K^+$ candidates (black data points) compared with the fit results (red solid lines) of the default model (top row) and the Run 1 model (bottom row). Source: Ref.~\cite{LHCb:2021uow}.}
\label{fig:Zcs4000}
\end{figure*}

Later in 2021 the LHCb collaboration reported their observation of two exotic structures in the $J/\psi K^+$ invariant mass spectrum of the $B^+ \to J/\psi \phi K^+$ decay~\cite{LHCb:2021uow}, as shown in Fig.~\ref{fig:Zcs4000}. They observed a relatively narrow $Z_{cs}(4000)^+$ state, whose mass and width were measured to be
\begin{eqnarray}
Z_{cs}(4000)^+ &:& M = 4003 \pm 6 ^{+~4}_{-14} {\rm~MeV} \, ,
\label{eq:Zcs4000}
\\ \nonumber && \Gamma = 131 \pm 15 \pm 26 {\rm~MeV} \, .
\end{eqnarray}
Its spin-parity quantum number was determined to be $J^P = 1^+$ with a high significance. They also observed a broader $Z_{cs}(4220)^+$ state with a significance of $5.9\sigma$. Its mass and width were measured to be
\begin{eqnarray}
Z_{cs}(4220)^+ &:& M = 4216 \pm 24 ^{+43}_{-30} {\rm~MeV} \, ,
\\ \nonumber && \Gamma = 233 \pm 52 ^{+97}_{-73} {\rm~MeV} \, ,
\end{eqnarray}
and its spin-parity quantum number was determined to be either $J^P = 1^+$ or $1^-$. Besides the $Z_{cs}(4000)^+$ and $Z_{cs}(4220)^+$, LHCb also observed two additional states in the $J/\psi \phi$ invariant mass spectrum, {\it i.e.}, the $X(4685)$ of $J^P = 1^+$ and the $X(4630)$ of $J^P = 1^-$ or $2^-$, which will not be discussed in the present review.

The $Z_{cs}(3985)^-$ state observed by BESIII has the quark content $c \bar c s \bar u$. The $Z_{cs}(4000)^+$ and $Z_{cs}(4220)^+$ states observed by LHCb have the quark content $c \bar c u \bar s$, while their antiparticles $Z_{cs}(4000)^-$ and $Z_{cs}(4220)^-$ have the quark content $c \bar c s \bar u$. However, one may wonder whether the $Z_{cs}(3985)^-$ and the $Z_{cs}(4000)^-$ are the same state or not. To simplify our notations, we shall omit their charges.

The $Z_{cs}(3985)$, $Z_{cs}(4000)$, and $Z_{cs}(4220)$ may be the possible strangeness partners of the $X(3872)$~\cite{Belle:2003nnu}, $Z_c(3900)$~\cite{BESIII:2013ris,Belle:2013yex}, and $Z_c(4020)$~\cite{BESIII:2013ouc}, respectively. Their existence had been predicted in various theoretical models before the BESIII and LHCb experiments~\cite{BESIII:2020qkh,LHCb:2021uow}, based on the $D \bar D_s^*/D^* \bar D_s$ hadronic molecular picture~\cite{Lee:2008uy}, the compact tetraquark picture~\cite{Ebert:2005nc,Dias:2013qga,Ferretti:2020ewe}, the hadro-quarkonium picture~\cite{Voloshin:2019ilw,Ferretti:2020ewe}, and the initial-single-chiral-particle-emission mechanism~\cite{Chen:2013wca}. We shall separately review these theoretical studies as follows, together with those performed after the BESIII and LHCb experiments.

\subsubsection{Hadronic molecular picture.}
\label{sec5.3.1}

The existence of a $D \bar D_s^*/D^* \bar D_s$ molecular state with the quantum number $J^P = 1^+$ was predicted in Ref.~\cite{Lee:2008uy} through the QCD sum rule method. They considered the antisymmetrical current
\begin{equation}
J_\mu = {i\over\sqrt{2}}\left[(\bar{d}_a\gamma_5 c_a)(\bar{c}_b\gamma_\mu s_b)-(\bar{d}_a\gamma_\mu c_a)(\bar{c}_b\gamma_5 s_b)\right] \, ,
\end{equation}
in order to keep a close relation with the $X(3872)$ (see the current $\eta_2^\alpha$ given in Eqs.~(\ref{sec5:X3872current})). Its mass was calculated to be $3.97 \pm 0.08$~GeV, so it can be used to explain the $Z_{cs}(3985)$ or $Z_{cs}(4000)$ as a loosely bound molecular state.

In a recent study~\cite{Chen:2021erj}, the author systematically studied the $D^{(*)} \bar D^{(*)}$ and $D^{(*)} \bar D_s^{(*)}$ hadronic molecular states through the QCD sum rule method. There are altogether six $S$-wave $D^{(*)} \bar D^{(*)}$ hadronic molecular states:
\begin{eqnarray}
\nonumber
| D \bar D; 0^{++} \rangle &=& | D \bar D \rangle_{J=0} \, ,
\\ \nonumber
| D \bar D^*; 1^{++} \rangle &=& \left( | D \bar D^* \rangle_{J=1} + | D^* \bar D \rangle_{J=1} \right) / \sqrt2 \, ,
\\
| D \bar D^*; 1^{+-} \rangle &=& \left( | D \bar D^* \rangle_{J=1} - | D^* \bar D \rangle_{J=1} \right) / \sqrt2 \, ,
\\ \nonumber
| D^* \bar D^*; 0^{++} \rangle &=& | D^* \bar D^* \rangle_{J=0} \, ,
\\ \nonumber
| D^* \bar D^*; 1^{+-} \rangle &=& | D^* \bar D^* \rangle_{J=1} \, ,
\\ \nonumber
| D^* \bar D^*; 2^{++} \rangle &=& | D^* \bar D^* \rangle_{J=2} \, .
\end{eqnarray}
Their corresponding interpolating currents are:
\begin{eqnarray}
\nonumber
\eta_1(x)              &=& \bar q_a(x) \gamma_5 c_a(x) ~ \bar c_b(x) \gamma_5 q_b(x) \, ,
\\ \nonumber
\eta_2^\alpha(x)       &=& \bar q_a(x) \gamma_5 c_a(x) ~ \bar c_b(x) \gamma^\alpha q_b(x) - \{ \gamma_5 \leftrightarrow \gamma^\alpha \} \, ,
\\
\eta_3^\alpha(x)       &=& \bar q_a(x) \gamma_5 c_a(x) ~ \bar c_b(x) \gamma^\alpha q_b(x) + \{ \gamma_5 \leftrightarrow \gamma^\alpha \} \, ,
\label{sec5:X3872current}
\\ \nonumber
\eta_4(x)              &=& \bar q_a(x) \gamma^\mu c_a(x) ~ \bar c_b(x) \gamma_\mu q_b(x) \, ,
\\ \nonumber
\eta_5^\alpha(x)       &=& \bar q_a(x) \gamma_\mu c_a(x) ~ \bar c_b(x)\sigma^{\alpha\mu}\gamma_5 q_b(x) - \{ \gamma_\mu \leftrightarrow \sigma^{\alpha\mu}\gamma_5 \} \, ,
\\ \nonumber
\eta_6^{\alpha\beta}(x) &=& P^{\alpha\beta,\mu\nu} \bar q_a(x) \gamma_\mu c_a(x) \bar c_b(x) \gamma_\nu q_b(x) \, .
\end{eqnarray}
These currents were applied to perform QCD sum rule analyses in Ref.~\cite{Chen:2021erj}, and their results support the interpretations of the $X(3872)$, $Z_c(3900)$, and $Z_c(4020)$ as the hadronic molecular states $| D \bar D^*; 1^{++} \rangle$, $| D \bar D^*; 1^{+-} \rangle$, and $| D^* \bar D^*; 0^{++} \rangle$, respectively.

Similarly, there are six $S$-wave $D^{(*)} \bar D_s^{(*)}$ hadronic molecular states:
\begin{eqnarray}
\nonumber
| D \bar D_s; 0^{+} \rangle &=& | D \bar D_s \rangle_{J=0} \, ,
\\ \nonumber
| D \bar D_s^{*}; 1^{++} \rangle &=& \left( | D \bar D_s^{*} \rangle_{J=1} + | D^* \bar D_s \rangle_{J=1} \right) / \sqrt2 \, ,
\\
| D \bar D_s^{*}; 1^{+-} \rangle &=& \left( | D \bar D_s^{*} \rangle_{J=1} - | D^* \bar D_s \rangle_{J=1} \right) / \sqrt2 \, ,
\\ \nonumber
| D^* \bar D_s^{*}; 0^{+} \rangle &=& | D^* \bar D_s^{*} \rangle_{J=0} \, ,
\\ \nonumber
| D^* \bar D_s^{*}; 1^{+} \rangle &=& | D^* \bar D_s^{*} \rangle_{J=1} \, ,
\\ \nonumber
| D^* \bar D_s^{*}; 2^{+} \rangle &=& | D^* \bar D_s^{*} \rangle_{J=2} \, .
\end{eqnarray}
Although the two states $| D \bar D_s^{*}; 1^{++} \rangle$ and $| D \bar D_s^{*}; 1^{+-} \rangle$ do not have the $C$-parity, we still denote their quantum numbers as $J^{PC} = 1^{+\pm}$, since they are the strangeness partners of $| D \bar D^{*}; 1^{++} \rangle$ and $| D \bar D^{*}; 1^{+-} \rangle$, respectively.

The interpolating currents corresponding to Eqs.~(\ref{sec5:X3872current}) were systematically constructed and applied to perform QCD sum rule analyses in Ref.~\cite{Chen:2021erj}. Their results suggest that the $Z_{cs}(3985)$, $Z_{cs}(4000)$, and $Z_{cs}(4220)$ can be interpreted as the hadronic molecular states $| D \bar D_s^{*}; 1^{++} \rangle$, $| D \bar D_s^{*}; 1^{+-} \rangle$, and $| D^* \bar D_s^{*}; 1^{+} \rangle$, respectively, and therefore, they may be the strangeness partners of the $X(3872)$, $Z_c(3900)$, and $Z_c(4020)$, respectively. This relationship is quite similar to that obtained in Ref.~\cite{Maiani:2021tri} within the diquark-antidiquark picture, which will be discussed later in Sec.~\ref{sec5.3.2}. To verify this relationship, the author of Ref.~\cite{Chen:2021erj} proposed to confirm the $Z_{cs}(3985)$ in the $J/\psi \pi K$ decay channel, considering that the $Z_c(3900)$ and $Z_{cs}(4000)$ were separately observed in the $J/\psi \pi$ and $J/\psi K$ channels, while the $X(3872)$ was observed in the $J/\psi \pi\pi$ channel. Moreover, productions of these charmonium-like states in $B$ meson decays as well as their decay properties were systematically studied Refs.~\cite{Chen:2021erj,Chen:2019wjd,Chen:2019eeq} through the Fierz transformation.

In Ref.~\cite{Meng:2020ihj} the authors studied the $Z_{cs}(3985)$ as a resonance through a coupled-channel analysis within the flavor $SU(3)$ symmetry and heavy quark spin symmetry. Their results suggest that the $Z_{cs}(3985)$, as the strangeness partner of the $Z_c(3900)$, can be explained as a $D \bar D_s^*/D^* \bar D_s$ molecular state. Besides, they predicted the strangeness partner of the $Z_c(4020)$ to have the mass around $4130$~MeV and the width around $30$~MeV. A similar result was obtained in Ref.~\cite{Yang:2020nrt}, where the authors applied the flavor $SU(3)$ symmetry to investigate the $Z_c(3900)$ and $Z_c(4020)$. The resonant poles of their strangeness partners were calculated to be around $3996 - i30$~MeV and $4138 - i 28$~MeV, respectively. They proposed to use the latter pole to explain the dip of the $J/\psi K$ mass distribution around 4.1 GeV, as shown in Fig.~\ref{fig:Zcs4000}. Many other theoretical studies~\cite{Du:2020vwb,Wang:2020htx,Cao:2020cfx,Yan:2021tcp,Ortega:2021enc,Baru:2021ddn,Cao:2021ton,Du:2022jjv} also suggest the existence of some exotic structure near the $D^* \bar D_s^*$ threshold around 4.1 GeV, so further experimental and theoretical studies are demanded to examine whether the $Z_{cs}(4220)$ is an effect of an event dip around the $D^* \bar D_s^*$ threshold.

Besides the above theoretical studies, the interpretation of the $Z_{cs}(3985)$ and/or $Z_{cs}(4000)$ states as the $D \bar D_s^*/D^* \bar D_s$ hadronic molecular state(s) was supported by some other theoretical studies~\cite{Sun:2020hjw,Ding:2021igr}. However, the authors of Ref.~\cite{Chen:2020yvq} studied the $Z_{cs}(3985)$ through the one-boson-exchange model considering coupled-channel effects, and their results exclude its assignment as a $D^{*0} D^-_s / D^0 D^{*-}_s / D^{*0} D^{*-}_s$ resonance with $(I)J^P = (1/2)1^+/0^-/1^-/2^-$. Besides, the elastic effective-range-expansion study of Ref.~\cite{Guo:2020vmu} suggest that both the two-meson molecular components and other degrees of freedom, such as the compact four-quark cores or heavier hadronic components, may play a role in the physical $Z_c(3900)$, $Z_c(4020)$, and $Z_{cs}(3985)$ resonances.

\subsubsection{Compact tetraquark picture.}
\label{sec5.3.2}

In Refs.~\cite{Maiani:2004vq,Maiani:2014aja} the authors proposed the ``type-I/II'' diquark-antidiquark models, where the $S$-wave tetraquark states are written in the spin basis as $|s_{cq}, \bar s_{\bar c \bar q} \rangle_J$, with $s_{cq}$ and $\bar s_{\bar c \bar q}$ the diquark and antidiquark spins, respectively. They used the ``good'' diquark of $s_{cq}=0$ and the ``bad'' diquark of $s_{cq}=1$~\cite{Jaffe:2004ph} to construct six $[cq][\bar c \bar q]$ ($q=u/d$) states, including two $J^{PC} = 0^{++}$ states, one $1^{++}$ state, two $1^{+-}$ states, and one $2^{++}$ state:
\begin{eqnarray}
\nonumber |0^{++}\rangle &=& |0_{cq}, 0_{\bar c \bar q}\rangle_{J=0} \, ,
\\ \nonumber |0^{++\prime}\rangle &=& |1_{cq}, 1_{\bar c \bar q}\rangle_{J=0} \, ,
\\ |1^{++}\rangle &=& \left(|0_{cq}, 1_{\bar c \bar q}\rangle_{J=1} + |1_{cq}, 0_{\bar c \bar q}\rangle_{J=1}\right)/\sqrt2 \, ,
\\ \nonumber |1^{+-}\rangle &=& \left(|0_{cq}, 1_{\bar c \bar q}\rangle_{J=1} - |1_{cq}, 0_{\bar c \bar q}\rangle_{J=1}\right)/\sqrt2 \, ,
\\ \nonumber |1^{+-\prime}\rangle &=& |1_{cq}, 1_{\bar c \bar q}\rangle_{J=1} \, ,
\\ \nonumber |2^{++}\rangle &=& |1_{cq}, 1_{\bar c \bar q}\rangle_{J=2} \, .
\end{eqnarray}
The three charmonium-like states $X(3872)$, $Z_c(3900)$, and $Z_c(4020)$ can be identified as the compact diquark-antidiquark states $|1^{++}\rangle$, $|1^{+-}\rangle$, and $|1^{+-\prime}\rangle$, respectively~\cite{Maiani:2014aja}.

Within the diquark-antidiquark picture, the authors of Ref.~\cite{Ebert:2005nc} systematically calculated the mass spectrum of the hidden-charm tetraquark states using the relativistic quark model. As summarized in Table~\ref{sec5:Zcsmass}, their results suggest that the $Z_{cs}(3985)$ and $Z_{cs}(4000)$, as possible strangeness partners of the $X(3872)$ and $Z_c(3900)$, can both be interpreted as the compact diquark-antidiquark states of $J^P = 1^+$. See Refs.~\cite{Wu:2018xdi,Faustov:2021hjs,Shi:2021jyr,Giron:2021sla} for more quark model calculations within the diquark-antidiquark picture, and there also exist some quark model calculations~\cite{Jin:2020yjn,Chen:2021uou,Yang:2021zhe,Han:2022yst} considering both the meson-meson and diquark-antidiquark structures.

Note that the states with the quark content $[cs][\bar c \bar q]$ ($q=u/d$) do not have the $C$-parity, but we still denote them as $|J^{PC}\rangle$ for simplicity. For example, we use the notation $|1^{+-}\rangle$ with the quark content $[cs][\bar c \bar q]$ to denote the strangeness partner of the neutral $|1^{+-}\rangle$ state with the quark content $[cq][\bar c \bar q]$. This state, $|1^{+-}\rangle$ with $[cs][\bar c \bar q]$, was further studied in Ref.~\cite{Dias:2013qga} through the QCD sum rule method. Its decay widths into the $J/\psi K$, $\eta_c K^*$, and $D \bar D_s^*/D^* \bar D_s$ channels were calculated to be $11.2$~MeV, $10.8$~MeV, and $2.9$~MeV, respectively. Its total width was calculated to be $24.9 \pm 12.6$~MeV, so it can be used to explain the $Z_{cs}(3985)$ as a compact diquark-antidiquark state. It may also be used to explain the $Z_{cs}(4000)$.

\begin{table*}[hbtp]
\caption{Masses of the hidden-charm tetraquark states (in MeV), calculated in Ref.~\cite{Ebert:2005nc} based on the compact diquark-antidiquark picture. $S$ and $A$ denote the ``good'' diquark of $s=0$ and the ``bad'' diquark of $s=1$, respectively.}
\label{sec5:Zcsmass}
\centering
\begin{tabular}{ccccc}
\toprule[1pt]
State& Diquark & \multicolumn{3}{c}{Mass}
\\ \cline{3-5}
$J^{PC}$   & content                         & $cq\bar c\bar q$ & $cs\bar c\bar s$ & $cs\bar c\bar q$
\\ \midrule[1pt]
$0^{++}$   & $S\bar S$                       & 3812             & 4051             & 3922
\\
$1^{+\pm}$ & $(S\bar A\pm \bar S A)/\sqrt2$  & 3871             & 4113             & 3982
\\
$0^{++}$   & $A\bar A$                       & 3852             & 4110             & 3967
\\
$1^{+-}$   & $A\bar A$                       & 3890             & 4143             & 4004
\\
$2^{++}$   & $A\bar A$                       & 3968             & 4209             & 4080
\\ \bottomrule[1pt]
\end{tabular}
\end{table*}

\begin{figure*}[hbtp]
\centering
\includegraphics[width=1.0\textwidth]{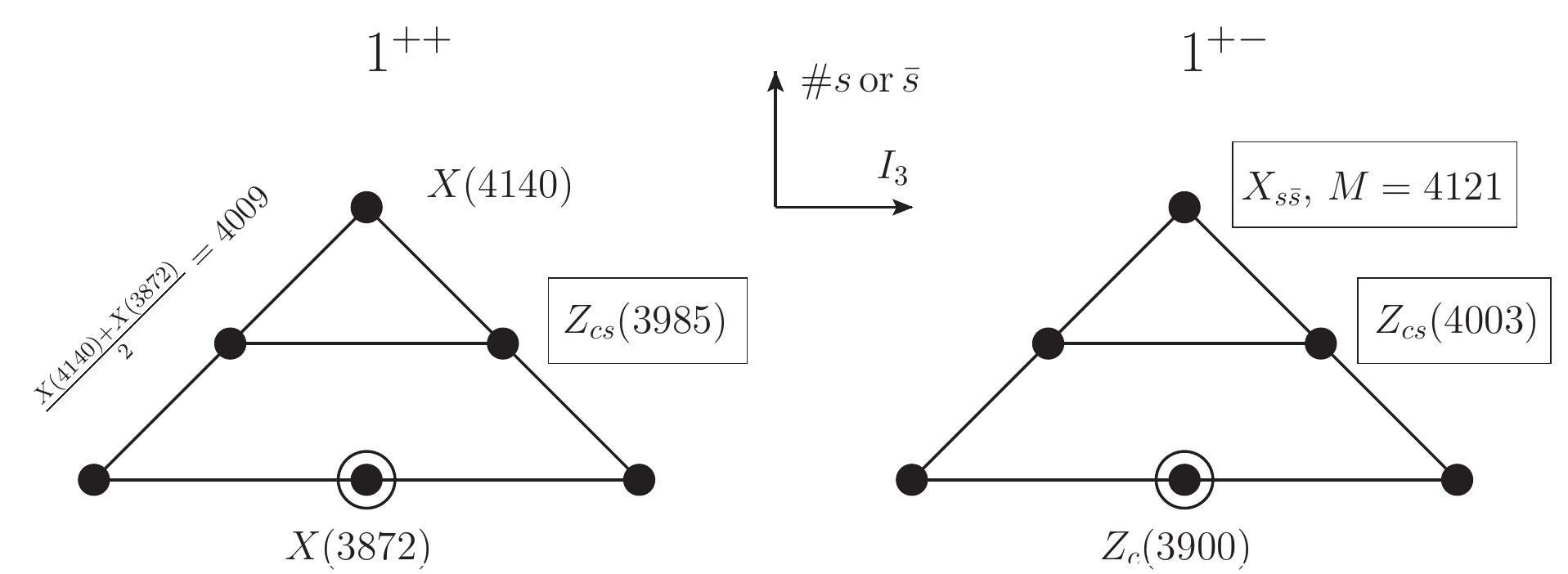}
\caption{Two flavor $SU(3)$ nonets with $J^P = 1^+$, derived from the diquark-antidiquark picture. Each of them contains four $Z_{cs}$ states with the quark contents $c u \bar c \bar s$, $c d \bar c \bar s$, $c s \bar c \bar u$, and $c s \bar c \bar d$. Source: Ref.~\cite{Maiani:2021tri}.}
\label{fig:Zcsdiquark}
\end{figure*}

The above diquark-antidiquark picture was recently improved in Ref.~\cite{Maiani:2021tri}. As shown in Fig.~\ref{fig:Zcsdiquark}, the authors proposed that the $Z_{cs}(3985)$ and $Z_{cs}(4000)$ states as well as the $X(3872)$ and $Z_c(3900)$ states neatly fit into two flavor $SU(3)$ nonets with $J^P = 1^+$ and opposite charge-conjugation, together with the $X(4140)$ interpreted as a $[cs] [\bar c \bar s]$ tetraquark state. Besides, they also expected a third nonet associated to the $Z_c(4020)$ and $Z_{cs}(4220)$. This relationship between $X(3872)/Z_c(3900)/Z_c(4020)$ and $Z_{cs}(3985)/Z_{cs}(4000)/Z_{cs}(4220)$ is quite similar to that obtained in Ref.~\cite{Chen:2021erj} within the hadronic molecular picture, which has been discussed in Sec.~\ref{sec5.3.1}.

\subsubsection{More theoretical studies.}
\label{sec5.3.3}

\begin{figure*}[hbtp]
\begin{center}
\subfigure[]{\includegraphics[width=0.3\textwidth]{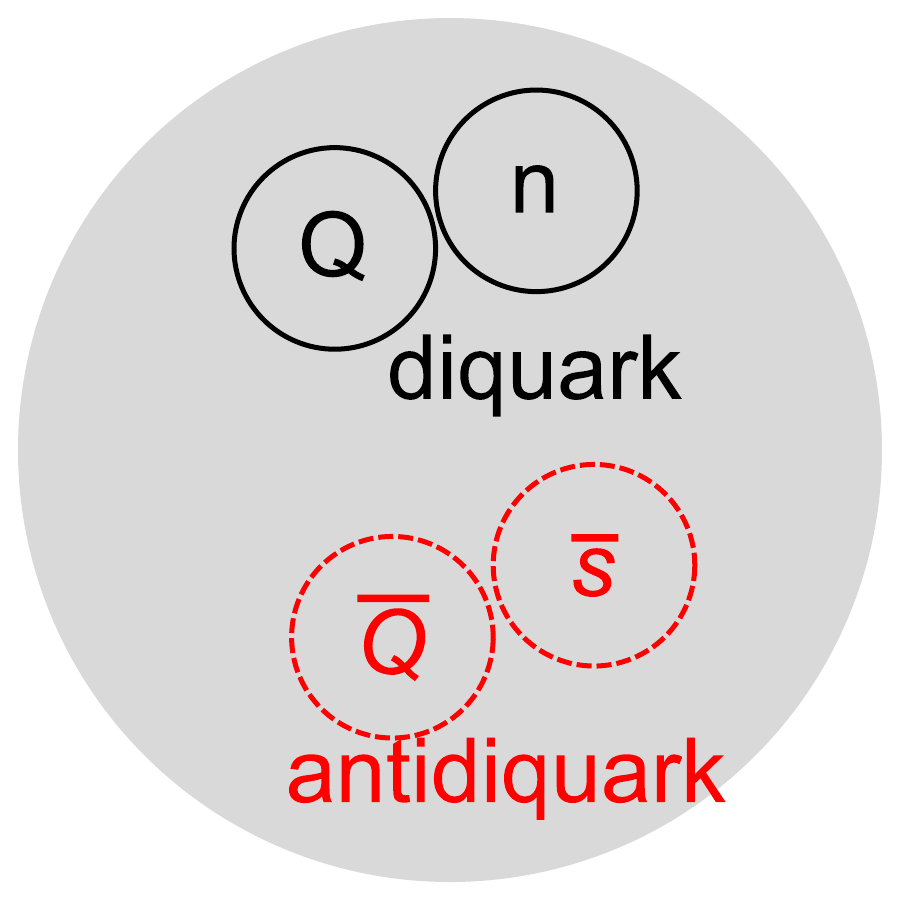}}
~~~~~~~~~~
\subfigure[]{\includegraphics[width=0.3\textwidth]{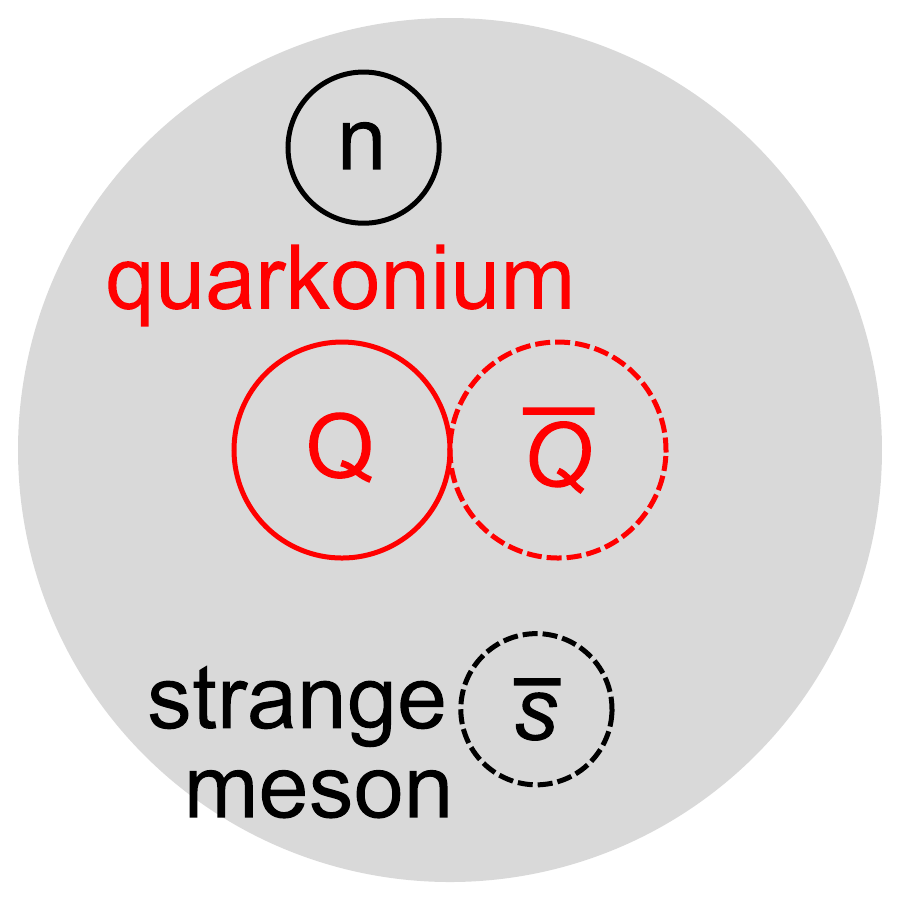}}
\end{center}
\caption{Schematic representation of (a) the compact diquark-antidiquark state and (b) the hadro-quarkonium state. Source: Ref.~\cite{Ferretti:2020ewe}.}
\label{fig:Zcspicture}
\end{figure*}

As depicted in Fig.~\ref{fig:Zcspicture}(b), the ``hadro-quarkonium'' picture was proposed in Refs.~\cite{Dubynskiy:2007tj,Voloshin:2007dx} to describe the resonance with a compact quarkonium state embedded into an excited light meson by a QCD analog of the van der Waals force. Later in Ref.~\cite{Voloshin:2018vym} the charged charmonium-like states $Z_c(4100)$~\cite{LHCb:2018oeg} and $Z_c(4200)$~\cite{Belle:2014nuw} were explained as two hadro-charmonium states related by the heavy quark spin symmetry. The existence of their strangeness partners, decaying to $\eta_c K$ and $J/\psi K$, was suggested in Ref.~\cite{Voloshin:2019ilw}. Accordingly, the $Z_{cs}(4000)$ and $Z_{cs}(4220)$, observed by LHCb in the $J/\psi K$ decay mode, may be explained as two hadro-charmonium states. Later in Ref.~\cite{Ferretti:2020ewe}, the authors studied the $\psi(2S) \otimes K$ bound state of $J^P = 1^+$, with the $\psi(2S)$ embedded into the $K$ meson. Its mass was calculated to be 3996~MeV, further supporting the interpretation of the $Z_{cs}(4000)$ as a hadro-charmonium state.

\begin{figure*}[hbtp]
\begin{center}
\includegraphics[width=1\textwidth]{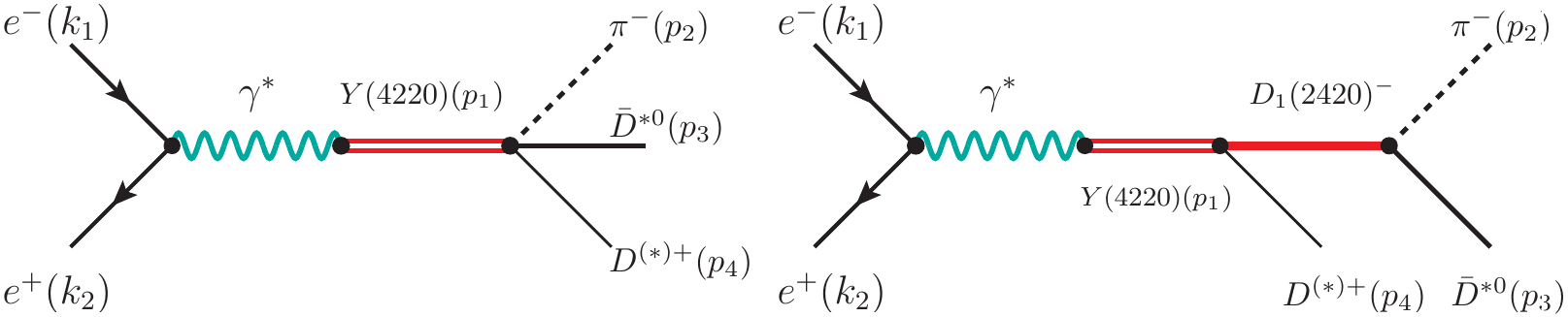}
\end{center}
\caption{Schematic diagrams contributing to the $e^{+}e^{-} \to Y(4220) \to D^{(*)+} \bar D^{*0} \pi^-$ process. Source: Ref.~\cite{Wang:2020axi}.}
\label{fig:Zcsreflection}
\end{figure*}

The initial-single-chiral-particle-emission mechanism was proposed in Ref.~\cite{Chen:2013wca}, through which the authors predicted the existence of several charged charmonium-like structures in the hidden-charm and open-strange channels near the $D \bar D_s^*/D^*\bar D_s$ and $D^* \bar D_s^*$ thresholds. One of these structures can be used to explain the $Z_{cs}(3985)$ observed by BESIII near the $D \bar D_s^*/D^*\bar D_s$ mass thresholds. This mechanism was further developed in Ref.~\cite{Wang:2020axi}, where the authors proposed a reflection mechanism to explain the $Z_c(3990)$ and $Z_c(4025)$, based on their line shapes observed by the BESIII experiments~\cite{BESIII:2013qmu,BESIII:2013mhi} in the $D^{(*)}\bar{D}^{*}$ mass spectra of the $e^{+}e^{-} \to \pi D^{(*)}\bar{D}^{*}$ processes. They described these line shapes by the reflection peak due to the intermediate resonance $\bar D_1(2420)$ involved in the $e^+e^- \to Y(4220) \to D^{(*)} \bar D_1(2420) \to D^{(*)} \bar D^* \pi$ process, as depicted in Fig.~\ref{fig:Zcsreflection}. The same reflection mechanism was used in a recent study~\cite{Wang:2020kej} to explain the $Z_{cs}(3985)$ as a reflection structure due to the intermediate resonance $D_{s2}(2573)$ involved in the $e^+e^- \to Y(4660) \to D_s^{*-} D_{s2}^*(2573)^+ \to D_s^{*-} D^0 K^+$ process. This mechanism was applied in Ref.~\cite{Wang:2020dmv} to study the three-body open-charm process at the $e^+e^-$ collisions, where the authors suggested a series of optimal center-of-mass energy points to search for new charmonium-like structures, as summarized in Table~\ref{sec5:reflection}.

\begin{table}
\caption{Proposal for inducing charmonium-like structures through the Critical Energy induced Enhancement mechanism. The last column shows the corresponding pole position of a reflective charmonium-like peak in the invariant mass of $D_{(s)}^{(*)}D_{(s)}^{(*)}$. Source: Ref.~\cite{Wang:2020dmv}.}
\label{sec5:reflection}
\setlength{\tabcolsep}{2.0mm}{
\begin{tabular}{cllccccc}
\bottomrule[1.2pt]
$\sqrt{s_{\textrm{critical}}}$  & Recommended  process   & Involved state & $ m_{D_{(s)}^{(*)}D_{(s)}^{(*)}}^{pole} $
\\ \bottomrule[1.2pt]
	    4286 & $e^+e^- \to DD^*\pi$  & $D_{1}(2420)$ ($1^+$) & 3890 \\
	    4429 & $e^+e^- \to D_sD_s^*\pi$  & $D_{s1}(2460)$ ($1^+$) & 4091   \\
	    4430 & $e^+e^- \to D_s^*D_s\pi$  & $D_{s0}(2317)$ ($0^+$) & 4092 \\
	    4431 & $e^+e^- \to D^*D^*\pi$  & $D_{1}(2420)$ ($1^+$) & 4035  \\
	    4471 & $e^+e^- \to D^*D\pi$  & $D_{2}^*(2460)$ ($2^+$) & 3910 \\
	    4474 & $e^+e^- \to DD^*\pi$ & $D^*(2600)$ ($1^-$) & 3906 \\
	    4504 & $e^+e^- \to D_sD^*K$  & $D_{s1}(2536)$ ($1^+$) &  3982 \\
	    4572 & $e^+e^- \to D_s^*D_s^*\pi$  & $D_{s1}(2460)$ ($1^+$) & 4234   \\
	    4615 & $e^+e^- \to D_sD^*K$ &  $D_s(2646)$ ($0^-$) &  3994 \\
	    4617 & $e^+e^- \to DD^*\pi$ & $D(2750)$ ($2^-$) & 3921 \\
	    4619 & $e^+e^- \to D^*D^*\pi$ & $D^*(2600)$ ($1^-$) &  4052 \\
	    4647 & $e^+e^- \to D_s^*D^*K$  & $D_{s1}(2536)$ ($1^+$) &  4125  \\
	    4677 & $e^+e^- \to D_sDK$ & $D^*_{s1}(2700)$  ($1^-$) &  3878\\
	    4750 & $e^+e^- \to DD^*\rho$ & $D(2885)$ ($1^+$)  & 3889   \\
	    4758 & $e^+e^- \to D_s^*D^*K$ &  $D_s(2646)$ ($0^-$)  & 4138  \\
	    4762 & $e^+e^- \to D^*D^*\pi$ & $D(2750)$  ($2^-$) & 4068\\
	    4820 & $e^+e^- \to D_s^*DK$ & $D^*_{s1}(2700)$  ($1^-$) &  4023\\
	    4894 & $e^+e^- \to D^*D\rho$  & $D(2884)$ ($2^+$)  & 3914  \\
	    4895 & $e^+e^- \to D^*D^*\rho$ & $D(2885)$ ($1^+$)  & 4034   \\
	    5013 & $e^+e^- \to DD\rho$ & $D(3148)$ ($2^-$)  & 3812 \\
	    5116 & $e^+e^- \to D_s^*D^*K^*$  & $D_{s}( 3004)$ ($2^+$)  & 4138    \\
	    5158 & $e^+e^- \to D^*D\rho$ & $D(3148)$ ($2^-$)  & 3960  \\
	    5229 & $e^+e^- \to D_sDK^*$ & $D_s(3260)$ ($2^-$)  & 3924  \\
	    5372 & $e^+e^- \to D_s^*DK^*$ & $D_s(3260)$ ($2^-$)  & 4070  \\
\bottomrule[1.2pt]
\end{tabular}}
\end{table}

Besides the above reflection mechanism, many other non-resonance models were applied to explain the $Z_{cs}(3985)$, $Z_{cs}(4000)$, and $Z_{cs}(4220)$, such as the recoupling mechanism~\cite{Simonov:2020ozp}, the threshold effects from the $D \bar D_s^*/D^* \bar D_s/D^* \bar D_s^*$ interactions~\cite{Ikeno:2020mra,Ikeno:2021mcb,Nakamura:2021bvs}, the $J/\psi K^{*+}$ and $\psi(2S) K^+$ threshold cusps enhanced by some nearby triangle singularities~\cite{Ge:2021sdq}, etc. The triangle loop mechanism was also applied in Refs.~\cite{Wu:2021ezz,Wu:2021cyc} to study their productions and decay properties.

At the end of this subsection, we emphasize that there have been quite a lot of experimental information on the $Z_{cs}(3985)$, $Z_{cs}(4000)$, and $Z_{cs}(4220)$ as well as the $X(3872)$, $Z_c(3900)$, and $Z_c(4020)$, all of which have been well studied in the literature to some extents, {\it e.g.}, see Refs.~\cite{Karliner:2021enr,Karliner:2021qok,Wan:2020oxt,Xu:2020evn,Albuquerque:2021tqd,Ozdem:2021yvo,Ozdem:2021hka,Wang:2020rcx,Wang:2020iqt,Azizi:2020zyq,Sungu:2020zvk,Chen:2010ze,Liu:2021ojf,Yang:2021jof,Zhu:2021vtd} that are not reviewed in this paper. Although we still do not fully understand them yet, these exotic structures have significantly improved our understanding on the nature of the exotic hadrons, and will probably give us more surprises in the near future.

\section{Glueballs and light hybrid mesons}
\label{sec6}

In this section we shall review experimental and theoretical progresses on the glueballs and light hybrid mesons briefly, and we shall mainly concentrate on recent theoretical results obtained using lattice QCD and QCD sum rules. Note that there have been various investigations in the past fifty years, but the academic community has not formed a common understanding on their nature.

Two major obstacles to study the glueballs and hybrid mesons are both related to the gluon degree of freedom, which is crucial to understand non-perturbative behaviors of the strong interaction at low energy region. Firstly, it is very difficult to experimentally identify glueballs and hybrid mesons unambiguously, and there is currently no definite experimental evidence on their existence. This is partly due to the difficulty in differentiating them from conventional $\bar q q$ mesons and exotic tetraquark states, and there is still not any good method to overcome this difficulty. Moreover, there can be strong mixing among these states, making their isolation even more difficult. On the other hand, theoretically, one can always investigate the mathematics of mixing in a multi-state model through some non-diagonal Hamiltonian, which will be discussed using the scalar glueball as an example in Sec.~\ref{sec6.1.2}.

Secondly, it is very difficult to define the gluon degree of freedom theoretically, {\it e.g.}, there have been some proposals to construct glueballs and hybrid mesons using constituent gluons, but a precise definition of the constituent gluon is still lacking~\cite{Coyne:1980zd,Chanowitz:1980gu,Cho:2015rsa}. Fortunately, one does not need to directly confront this difficulty when applying the methods of lattice QCD and QCD sum rules. Unfortunately, one may at the same time lose some insights into the nature and formation of the glueballs and hybrid mesons. In order to reach a more general picture, we shall first review some of the constituent gluon models~\cite{Barnes:1981ac,Cornwall:1982zn} at the beginning of the next subsection, from which we can feel the changes in this field.

\subsection{Glueballs}
\label{sec6.1}

Glueballs, composed of the valence gluons, are important for the understanding of non-perturbative QCD~\cite{Fritzsch:1973pi,Fritzsch:1975tx,Freund:1975pn}, which is asymptotically free~\cite{Gross:1973id,Politzer:1973fx} and at the same time confining~\cite{Wilson:1974sk}. There have been tremendous theoretical studies on them in the past fifty years using various methods and models, such as the MIT bag model~\cite{Jaffe:1975fd,DeGrand:1975cf,Carlson:1982er,Chanowitz:1982qj}, flux-tube model~\cite{Isgur:1983wj,Isgur:1984bm}, constituent gluon model~\cite{Szczepaniak:1995cw,Llanes-Estrada:2005bii,Mathieu:2008bf,Boulanger:2008aj}, AdS/QCD model~\cite{deTeramond:2005su,Boschi-Filho:2005xct,Karch:2006pv,Forkel:2007ru,Li:2013oda,Brunner:2015yha,Sonnenschein:2015zaa,Brunner:2018wbv,Domokos:2022djc}, lattice QCD~\cite{Michael:1988jr,Richards:2010ck,Yamanaka:2019yek,Morningstar:1999rf,Chen:2005mg,Meyer:2004gx,Gregory:2012hu,Athenodorou:2020ani}, QCD sum rules~\cite{Novikov:1979va,Kataev:1981aw,Kataev:1981gr,Shuryak:1982dp,Kaminski:2009qg,Narison:2021xhc,Chen:2021cjr,Chen:2021bck,Pimikov:2022brd}, and others~\cite{Chao:2005si,Ye:2012gu,Lu:2013jj}.

\begin{figure}[hbtp]
\begin{center}
\includegraphics[width=0.6\textwidth]{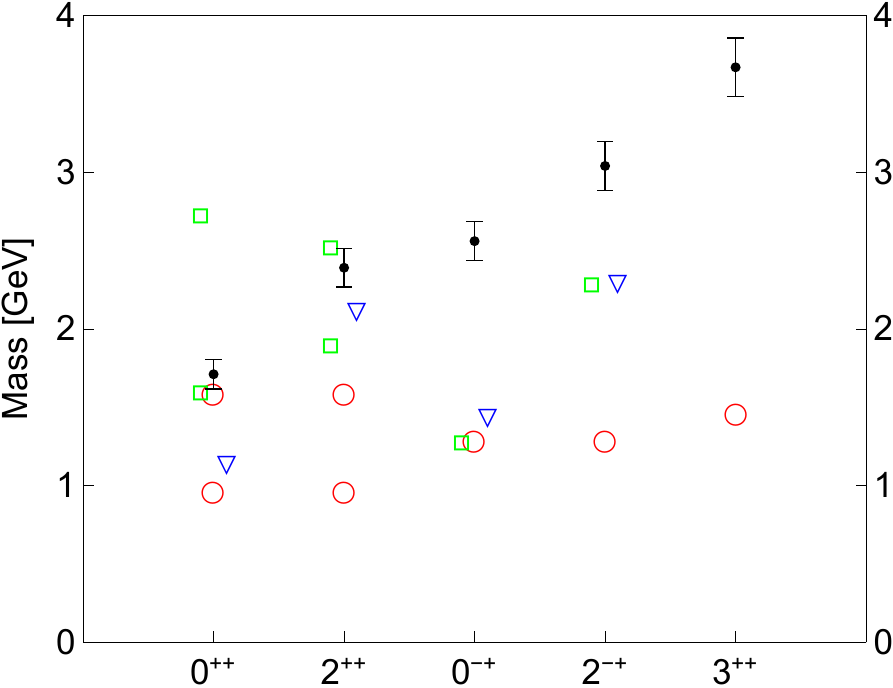}
\end{center}
\caption{Mass spectra of glueballs calculated using the MIT bag model from Ref.~\cite{Jaffe:1975fd} (red circles), Ref.~\cite{Carlson:1982er} (green squares), and Ref.~\cite{Chanowitz:1982qj} (blue triangles). Lattice QCD calculations from Ref.~\cite{Chen:2005mg} (black error bars) are given for comparisons.}
\label{fig:MITbag}
\end{figure}

Within the MIT bag model proposed in Ref.~\cite{Chodos:1974je}, the quarks and gluons are confined in a bag with a boundary condition and a constant energy density $B$. Later in Ref.~\cite{Jaffe:1975fd} the authors applied this method to the glueballs and calculated their masses. This model was further developed in Refs.~\cite{Carlson:1982er,Chanowitz:1982qj}, where the former one took into account the spin splitting induced by the one-gluon-exchange interaction, and the latter one assumed a non-constant energy density $B$. Their results were obtained more than forty years ago, which are generally lower than those of lattice QCD calculations~\cite{Morningstar:1999rf,Chen:2005mg,Meyer:2004gx,Gregory:2012hu,Athenodorou:2020ani}, as shown in Fig.~\ref{fig:MITbag}.

\begin{figure}[hbtp]
\begin{center}
\subfigure[]{\includegraphics[width=0.45\textwidth]{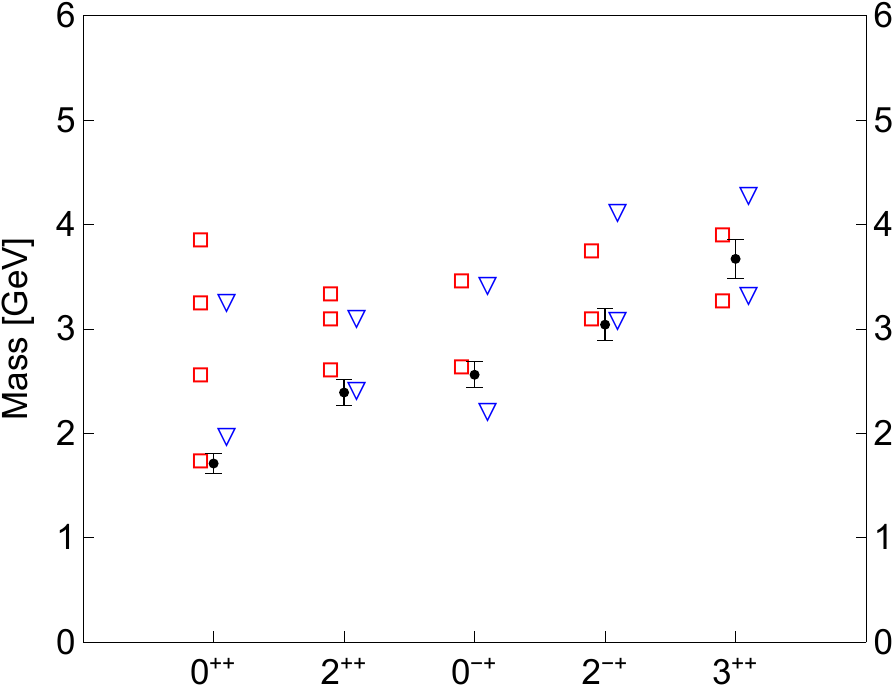}}
~~~~~
\subfigure[]{\includegraphics[width=0.45\textwidth]{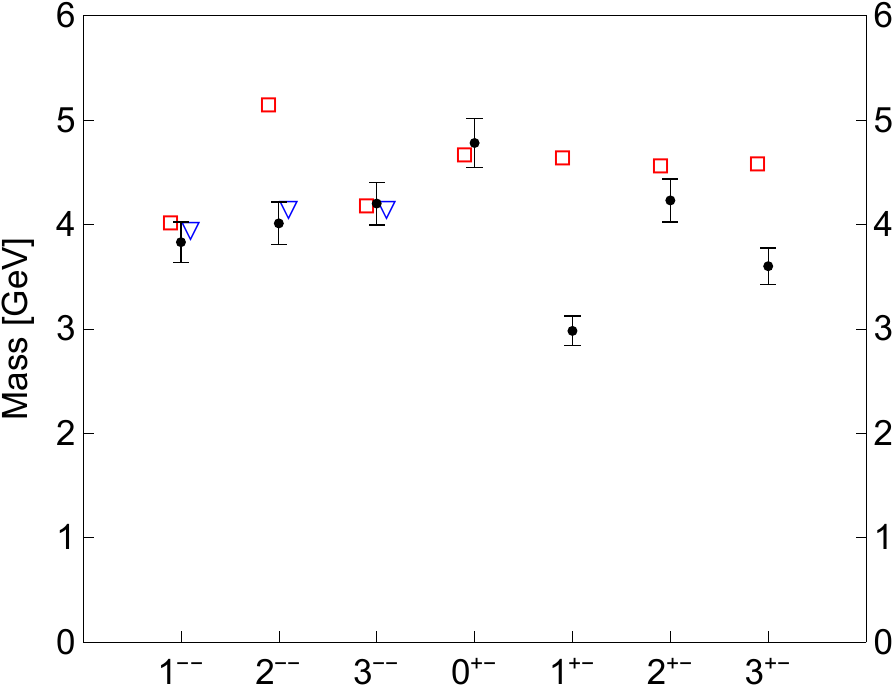}}
\end{center}
\caption{Mass spectra of (a) two-gluon glueballs from Ref.~\cite{Mathieu:2008bf} (red squares) and Ref.~\cite{Szczepaniak:2003mr} (blue triangles) as well as (b) three-gluon glueballs from Ref.~\cite{Mathieu:2008pb} (red squares) and Ref.~\cite{Llanes-Estrada:2005bii} (blue triangles). Lattice QCD calculations from Ref.~\cite{Chen:2005mg} (black error bars) are given for comparisons.}
\label{fig:constituentgluon}
\end{figure}

The key question of the constituent gluon model is concerning the mass of the constituent gluon, given that the gluon itself remains massless to all orders in the perturbation theory. The pioneering works~\cite{Barnes:1981ac,Cornwall:1982zn} gave two discrepant answers. Some theoretical studies originally from Ref.~\cite{Barnes:1981ac} argued that the valence gluon is still a massless particle, and it gains a dynamical mass that is the pole position of the dressed gluon propagator. Some other theoretical studies originally from Ref.~\cite{Cornwall:1982zn} argued that the valence gluon has to be {\it a priori} considered as a massive particle, and the non-perturbative effects of QCD cause a mass term to appear in the gluon propagator, which is typically around $m_{\rm gluon} = 0.5 \pm 0.2$~GeV~\cite{Bernard:1981pg,Bonnet:2000kw,Aguilar:2001zy,Borot:2013qs}. It is still controversial how to describe the constituent gluon inside the glueballs and hybrid mesons, but we shall not go further on this viewpoint. We just show some results obtained using the constituent gluon model in Fig.~\ref{fig:constituentgluon}. These results were obtained in the past twenty years, which are generally consistent with lattice QCD calculations~\cite{Morningstar:1999rf,Chen:2005mg,Meyer:2004gx,Gregory:2012hu,Athenodorou:2020ani}.

\subsubsection{Lattice QCD and QCD sum rule calculations.}
\label{sec6.1.1}

In both lattice QCD and QCD sum rules one does not need to directly confront the difficulty in defining the constituent gluon, because it is the well-defined gluon field that is used in these two methods. One can use the gluon field strength tensor $G_{\mu\nu} = {\lambda_a \over 2}G^a_{\mu\nu}$~($\mu,\nu=0\cdots3$) and its dual field $\tilde G_{\mu\nu} = G^{\rho\sigma} \times \epsilon_{\mu\nu\rho\sigma}/2$ to construct relativistic glueball currents, which are often used in QCD sum rule studies. Their corresponding non-relativistic operators are composed of the chromoelectric and chromomagnetic gluon fields ($i,j=1\cdots3$):
\begin{equation}
E_i = G_{i0} ~~~ {\rm and} ~~~ B_i = -{1\over2} \epsilon_{ijk} G^{jk} \, ,
\end{equation}
some of which have been successfully applied in lattice QCD studies.

The low-lying two- and three-gluon glueballs were systematically classified in Ref.~\cite{Jaffe:1985qp}, and their corresponding non-relativistic low-dimension operators were constructed at the same time. Their corresponding relativistic low-dimension currents were systematically constructed later in Refs.~\cite{Chen:2021cjr,Chen:2021bck}. We briefly summarize some of their results as follows.

The authors of Ref.~\cite{Jaffe:1985qp} classified representations of the Lorentz group using the isomorphism of its algebra to the algebra of $SU(2) \times SU(2)$, with each irreducible representation labeled by a pair $(k, l)$. The gluon field strength tensor $G^a_{\mu\nu}$ transforms as $[(1,0) \oplus (0, 1)]$, and the two-gluon glueball operator ${\rm Tr}[G_{\mu\nu}G_{\rho\sigma}] = G^a_{\mu\nu}G_{a,\rho\sigma}$ transforms as the symmetric part of $[(1,0)\oplus(0,1)]^2$:
\begin{eqnarray}
[(1,0)\oplus(0,1)]^2_{\bf S} &=& [(1,0)^2]^{(0,0)} \oplus [(0,1)^2]^{(0,0)}
\\ \nonumber && \oplus [(1,0)\otimes(0,1)]^{(1,1)}
\\ \nonumber && \oplus [(1,0)^2]^{(2,0)} \oplus [(0,1)^2]^{(0,2)} \, .
\end{eqnarray}
It corresponds to seven non-relativistic operators:
\begin{equation*}
\renewcommand{\arraystretch}{1.2}
\begin{array}{l l l}
\hline
J^{PC}~~~~~ &   {\rm Lorentz~representation}~~~~~  &   {\rm Operator}
\\ \hline
0^{++}   &   [(1,0)^2]^{(0,0)}+[(0,1)^2]^{(0,0)}   &   \vec E_a^2 - \vec B_a^2
\\
0^{++}   &   [(1,0)\otimes(0,1)]^{(1,1)}           &   \vec E_a^2 + \vec B_a^2
\\
0^{-+}   &   [(1,0)^2]^{(0,0)}-[(0,1)^2]^{(0,0)}   &   \vec E_a \cdot \vec B_a
\\
1^{-+}   &   [(1,0)\otimes(0,1)]^{(1,1)}           &   \vec E_a \times \vec B_a
\\
2^{++}   &   [(1,0)^2]^{(2,0)}+[(0,1)^2]^{(0,2)}   &   \mathcal{S}[ E_a^i E_a^j - B_a^i B_a^j ]
\\
2^{++}   &   [(1,0)\otimes(0,1)]^{(1,1)}           &   \mathcal{S}[ E_a^i E_a^j + B_a^i B_a^j ]
\\
2^{-+}   &   [(1,0)^2]^{(2,0)}-[(0,1)^2]^{(0,2)}   &   \mathcal{S}[ E_a^i B_a^j - B_a^i E_a^j ]
\\ \hline
\end{array}
\end{equation*}
One can further construct their semi-corresponding relativistic currents:
\begin{eqnarray}
\nonumber J_0 &=& g_s^2 G_a^{\mu\nu} G^{a}_{\mu\nu} \, ,
\\ \nonumber \tilde J_0 &=& g_s^2 G_a^{\mu\nu} \tilde G^{a}_{\mu\nu} \, ,
\\ J_1^{\alpha\beta} &=& g_s^2 G_a^{\alpha\mu} \tilde G^{a,\beta}_{\mu} - \{ \alpha \leftrightarrow \beta \} \, ,
\label{sec6:twogluon}
\\ \nonumber J_2^{\alpha_1\alpha_2,\beta_1\beta_2} &=& \mathcal{S}^\prime[ g_s^2 G_a^{\alpha_1\beta_1} G^{a,\alpha_2\beta_2} ] \, ,
\\ \nonumber \tilde J_2^{\alpha_1\alpha_2,\beta_1\beta_2} &=& \mathcal{S}^\prime[ g_s^2 G_a^{\alpha_1\beta_1} \tilde G^{a,\alpha_2\beta_2} ] \, .
\end{eqnarray}
In the above expressions $\mathcal{S}$ denotes symmetrization and subtracting trace terms in the set $\{ij\}$, and $\mathcal{S}^\prime$ denotes symmetrization and subtracting trace terms in the two sets $\{\alpha_1 \alpha_2\}$ and $\{\beta_1 \beta_2\}$ simultaneously.

Similarly, one can investigate the $C=+1$ three-gluon glueball operator ${\rm Tr}G_{\alpha\beta}[G_{\mu\nu},G_{\rho\sigma}]_- = f_{abc} G^a_{\alpha\beta} G^b_{\mu\nu}G^c_{\rho\sigma}$, and decompose it into:
\begin{eqnarray}
[(1, 0)\oplus(0, 1)]^3_{\bf A} &=& [(1, 0)^3]^{(0,0)} \oplus [(0, 1)^3]^{(0,0)}
\\ \nonumber && \oplus [[(1, 0)^2]^{(1,0)} \otimes (0,1)]^{(1,1)} \oplus [(1,0) \otimes [(0, 1)^2]^{(0,1)}]^{(1,1)} \, .
\end{eqnarray}
One can also investigate the $C=-1$ three-gluon glueball operator ${\rm Tr}G_{\alpha\beta}[G_{\mu\nu},G_{\rho\sigma}]_+ = d_{abc} G^a_{\alpha\beta} G^b_{\mu\nu}G^c_{\rho\sigma}$, and decompose it into:
\begin{eqnarray}
[(1, 0)\oplus(0, 1)]^3_{\bf S} &=& [(1, 0)^3]^{(1,0)} \oplus [(0, 1)^3]^{(0,1)}
\\ \nonumber && \oplus [[(1, 0)^2]^{(0,0)} \otimes (0,1)]^{(0,1)} \oplus [(1,0) \otimes [(0, 1)^2]^{(0,0)}]^{(1,0)}
\\ \nonumber && \oplus [[(1, 0)^2]^{(2,0)} \otimes (0,1)]^{(2,1)} \oplus [(1,0) \otimes [(0, 1)^2]^{(0,2)}]^{(1,2)}
\\ \nonumber && \oplus [(1, 0)^3]^{(3,0)} \oplus [(0, 1)^3]^{(0,3)} \, .
\end{eqnarray}
Some of their corresponding non-relativistic and relativistic operators were constructed in Refs.~\cite{Jaffe:1985qp,Chen:2021cjr,Chen:2021bck}. Interestingly, it was explicitly proven in Ref.~\cite{Chen:2021bck} that the current $J_1^{\alpha\beta}$ defined in Eq.~(\ref{sec6:twogluon}) vanishes, suggesting that the low-lying two-gluon glueball of $J^{PC} = 1^{-+}$ does not exist within the relativistic framework. Besides, all the three-gluon glueball currents of $J^{PC} = 1^{\pm+}$ vanish, suggesting that the low-lying three-gluon glueballs of $J^{PC} = 1^{\pm+}$ do not exist within the relativistic framework neither.

\begin{figure}[hbtp]
\begin{center}
\includegraphics[width=0.8\textwidth]{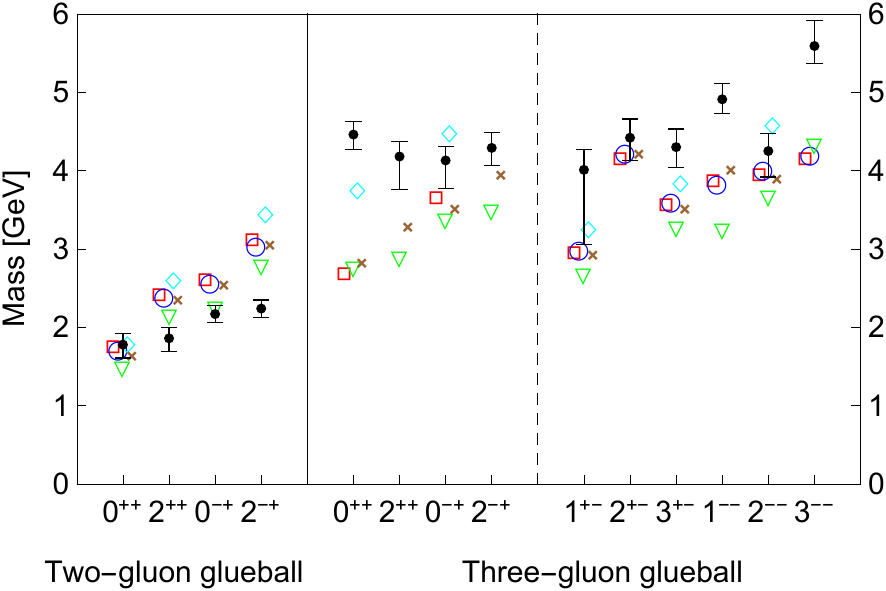}
\end{center}
\caption{Mass spectra of two- and three-gluon glueballs calculated using lattice QCD from Ref.~\cite{Morningstar:1999rf} (red squares), Ref.~\cite{Chen:2005mg} (blue circles), Ref.~\cite{Meyer:2004gx} (green triangles), Ref.~\cite{Gregory:2012hu} (cyan diamonds), and Ref.~\cite{Athenodorou:2020ani} (brown crosses). QCD sum rule results from Ref.~\cite{Chen:2021bck} (black error bars) are given for comparisons.}
\label{fig:gluonsumrule}
\end{figure}

The above non-relativistic and relativistic two- and three-gluon glueball operators have been successfully used in lattice QCD and QCD sum rule analyses. In Table~\ref{sec6:comparison}, we summarize the lattice QCD results from Refs.~\cite{Morningstar:1999rf,Chen:2005mg,Meyer:2004gx,Gregory:2012hu,Athenodorou:2020ani} and the QCD sum rule results from Refs.~\cite{Chen:2021cjr,Chen:2021bck}. These results were obtained in the past twenty years, which are generally consistent with each other, as shown in Fig.~\ref{fig:gluonsumrule}. Based on these results, we shall separately review experimental and theoretical progresses on the scalar, tensor, and pseudoscalar two-gluon glueballs as well as the three-gluon glueballs in the following subsections.

\begin{table*}[hbt]
\renewcommand{\arraystretch}{1.5}
\scriptsize
\caption{Masses of two- and three-gluon glueballs, in units of MeV. Lattice QCD results from Refs.~\cite{Morningstar:1999rf,Chen:2005mg,Meyer:2004gx,Gregory:2012hu,Athenodorou:2020ani} are listed in the 2nd-6th columns. QCD sum rule results from Ref.~\cite{Chen:2021bck} are listed in the 7th column.}
\centering
\begin{tabular}{c | c | c | c | c | c | c}
\hline\hline
Glueball  & ~~~~~Ref.~\cite{Morningstar:1999rf}~~~~~  & ~~~~~Ref.~\cite{Chen:2005mg}~~~~~ &  ~~~~~Ref.~\cite{Meyer:2004gx}~~~~~  & ~Ref.~\cite{Gregory:2012hu}~ & Ref.~\cite{Athenodorou:2020ani} & QSR~\cite{Chen:2021bck}
\\ \hline \hline
$|{\rm GG};0^{++}\rangle$       & $1730\pm50\pm80$    & $1710\pm50\pm80$                  & $1475\pm30\pm65$                     & $1795\pm60$                  & $1653\pm26$                     & $1780^{+140}_{-170}$
\\
$|{\rm GG};2^{++}\rangle$       & $2400\pm25\pm120$   & $2390\pm30\pm120$                 & $2150\pm30\pm100$                    & $2620\pm50$                  & $2376\pm32$                     & $1860^{+140}_{-170}$
\\
$|{\rm GG};0^{-+}\rangle$       & $2590\pm40\pm130$   & $2560\pm35\pm120$                 & $2250\pm60\pm100$                    & --                           & $2561\pm40$                     & $2170^{+110}_{-110}$
\\
$|{\rm GG};2^{-+}\rangle$       & $3100\pm30\pm150$   & $3040\pm40\pm150$                 & $2780\pm50\pm130$                    & $3460\pm320$                 & $3070\pm60$                     & $2240^{+110}_{-110}$
\\ \hline
$|{\rm GGG};0^{++}\rangle$      & $2670\pm180\pm130$  & --                                & $2755\pm70\pm120$                    & $3760\pm240$                 & $2842\pm40$                     & $4460^{+170}_{-190}$
\\
$|{\rm GGG};2^{++}\rangle$      & --                  & --                                & $2880\pm100\pm130$                   & --                           & $3300\pm50$                     & $4180^{+190}_{-420}$
\\
$|{\rm GGG};0^{-+}\rangle$      & $3640\pm60\pm180$   & --                                & $3370\pm150\pm150$                   & $4490\pm590$                 & $3540\pm80$                     & $4130^{+180}_{-360}$
\\
$|{\rm GGG};2^{-+}\rangle$      & --                  & --                                & $3480\pm140\pm160$                   & --                           & $3970\pm70$                     & $4290^{+200}_{-220}$
\\ \hline
$|{\rm GGG};1^{+-}\rangle$      & $2940\pm30\pm140$   & $2980\pm30\pm140$                 & $2670\pm65\pm120$                    & $3270\pm340$                 & $2944\pm42$                     & $4010^{+260}_{-950}$
\\
$|{\rm GGG};2^{+-}\rangle$      & $4140\pm50\pm200$   & $4230\pm50\pm200$                 & --                                   & --                           & $4240\pm80$                     & $4420^{+240}_{-290}$
\\
$|{\rm GGG};3^{+-}\rangle$      & $3550\pm40\pm170$   & $3600\pm40\pm170$                 & $3270\pm90\pm150$                    & $3850\pm350$                 & $3530\pm80$                     & $4300^{+230}_{-260}$
\\
$|{\rm GGG};1^{--}\rangle$      & $3850\pm50\pm190$   & $3830\pm40\pm190$                 & $3240\pm330\pm150$                   & --                           & $4030\pm70$                     & $4910^{+200}_{-180}$
\\
$|{\rm GGG};2^{--}\rangle$      & $3930\pm40\pm190$   & $4010\pm45\pm200$                 & $3660\pm130\pm170$                   & $4590\pm740$                 & $3920\pm90$                     & $4250^{+220}_{-330}$
\\
$|{\rm GGG};3^{--}\rangle$      & $4130\pm90\pm200$   & $4200\pm45\pm200$                 & $4330\pm260\pm200$                   & --                           & --                              & $5590^{+330}_{-220}$
\\ \hline \hline
\end{tabular}
\label{sec6:comparison}
\end{table*}

\subsubsection{Scalar glueballs and the $f_0(1500)/f_0(1710)$.}
\label{sec6.1.2}

\begin{figure}[hbtp]
\begin{center}
\subfigure[]{\includegraphics[width=0.5\textwidth]{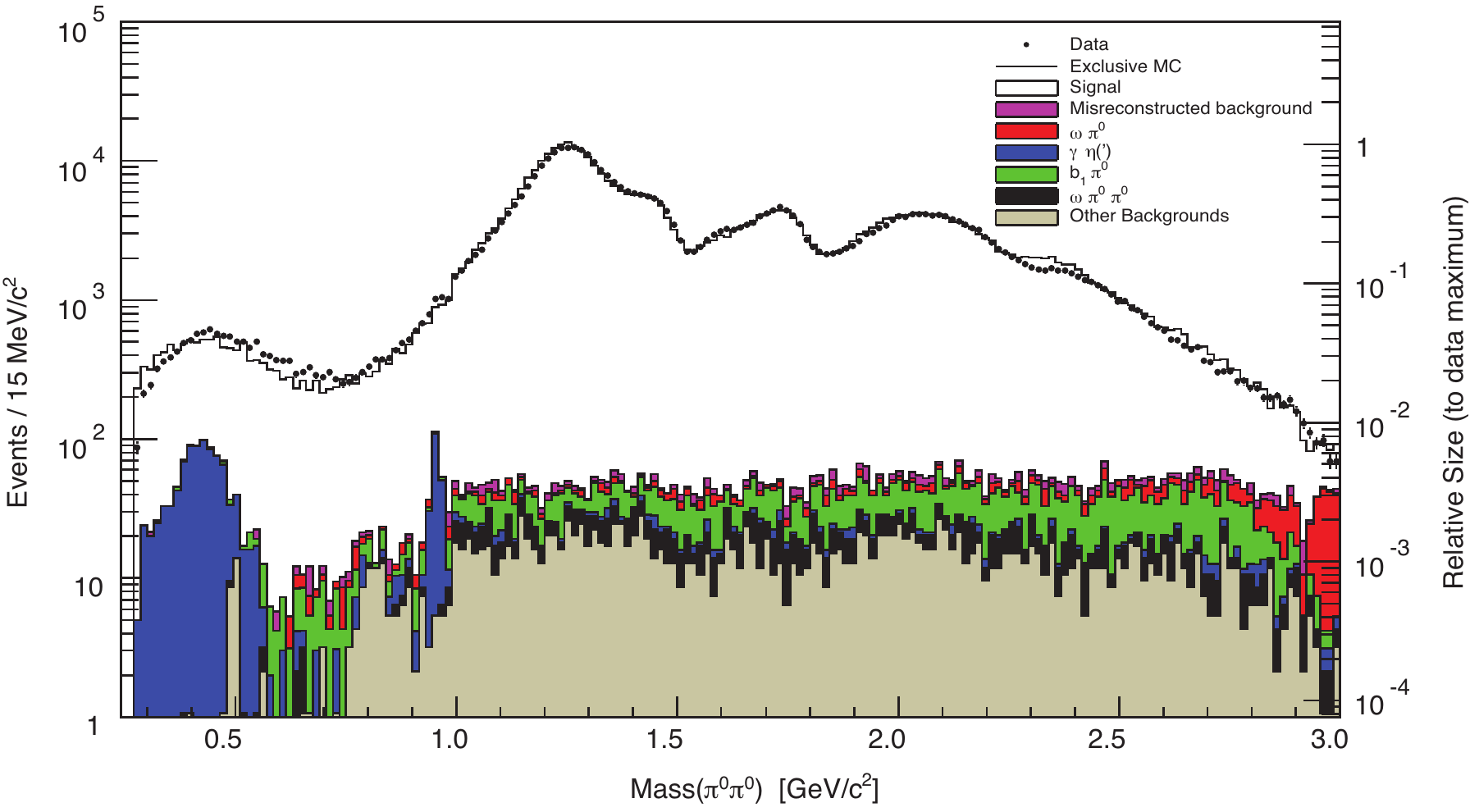}}
~~
\subfigure[]{\includegraphics[width=0.43\textwidth]{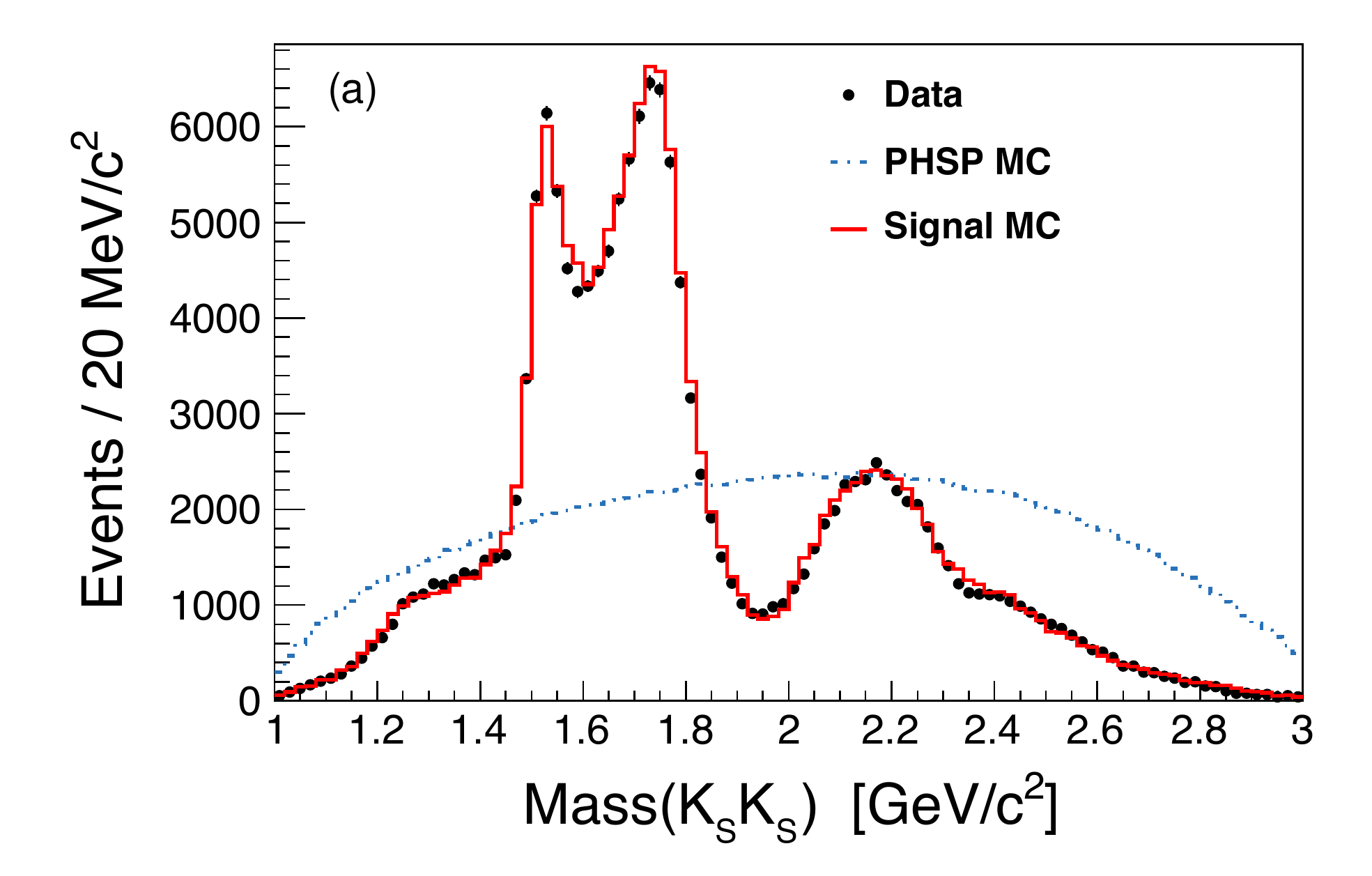}}
\end{center}
\caption{The (a) $\pi^0 \pi^0$ and (b) $K^0_S K^0_S$ invariant mass spectra of the $J/\psi \to \gamma \pi^0 \pi^0$ and $J/\psi \to \gamma K^0_S K^0_S$ decays, respectively. Source: Refs.~\cite{BESIII:2015rug,BESIII:2018ubj}.}
\label{fig:f0}
\end{figure}

There are as many as nine light isoscalar scalar mesons listed in PDG2020, {\it i.e.}, $f_0(500)$, $f_0(980)$, $f_0(1370)$, $f_0(1500)$, $f_0(1710)$, $f_0(2020)$, $f_0(2100)$, $f_0(2200)$, and $f_0(2330)$~\cite{pdg}. A tremendous number of experimental and theoretical analyses have been performed to study them. Especially, these resonances were intensively studied in the radiative $J/\psi$ decays, and the data with the highest statistics nowadays stem from the BESIII experiments~\cite{BES:1999dmf,BES:2006nqh,Ablikim:2006db,BESIII:2012rtd,BESIII:2013qqz,BESIII:2015rug,BESIII:2016qzq,BESIII:2018ubj,BESIII:2020nme,Bugg:2009ch,Fang:2021wes}. For example, we depict the $\pi^0 \pi^0$ and $K^0_S K^0_S$ invariant mass spectra of the $J/\psi \to \gamma \pi^0 \pi^0$ and $J/\psi \to \gamma K^0_S K^0_S$ decays in Fig.~\ref{fig:f0}.

\begin{figure}[hbtp]
\begin{center}
\subfigure[]{\includegraphics[width=0.45\textwidth]{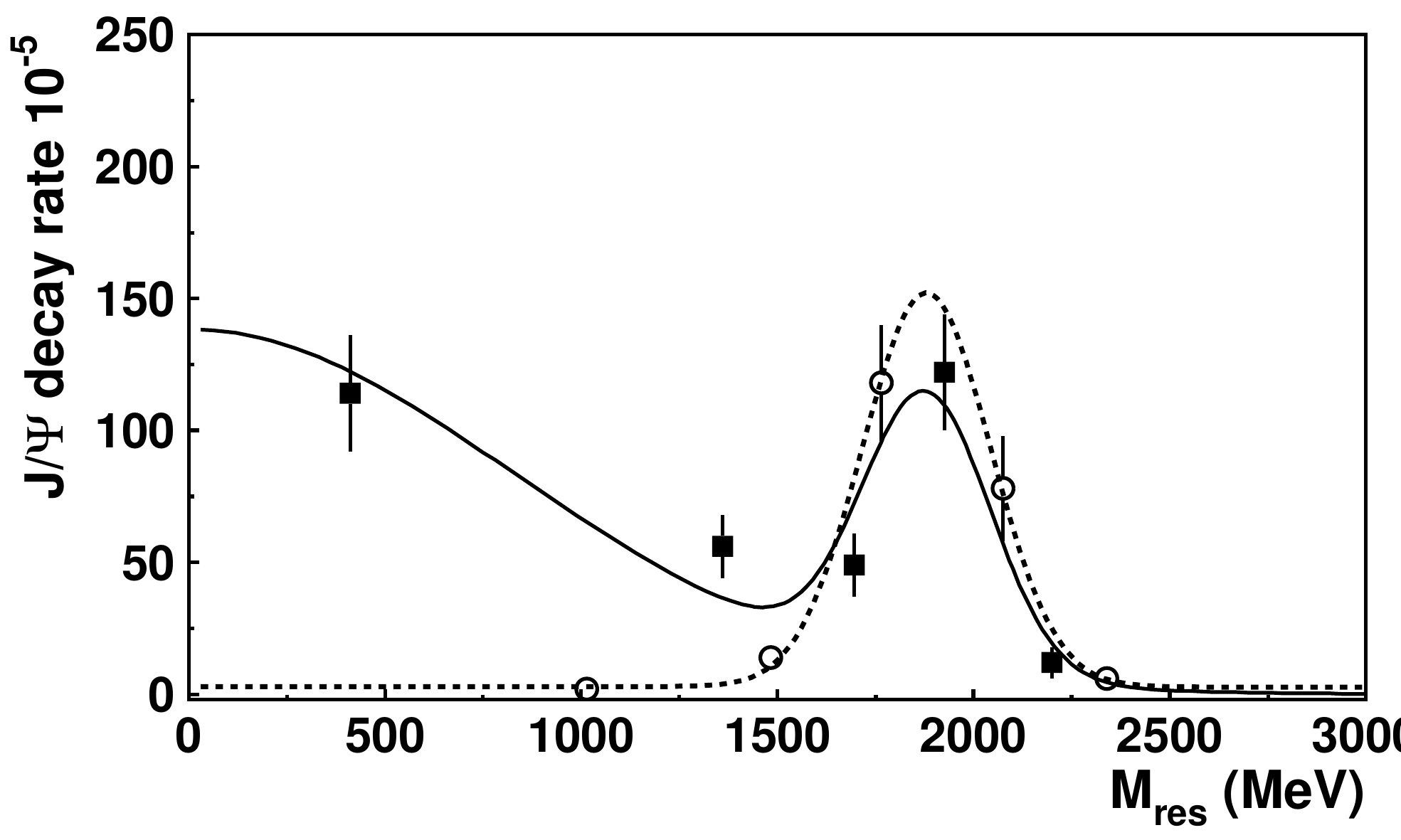}}
~~~
\subfigure[]{\includegraphics[width=0.45\textwidth]{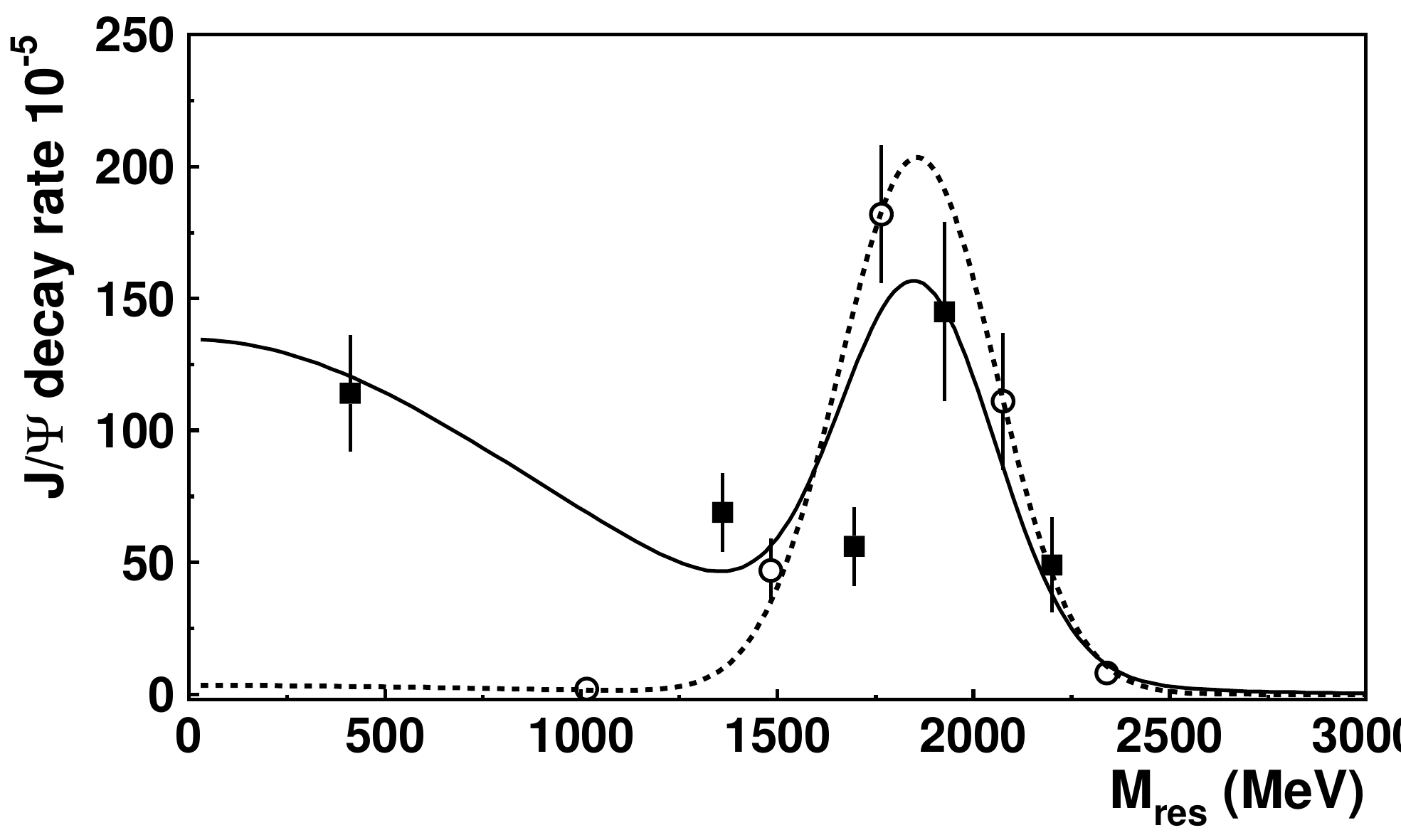}}
\end{center}
\caption{Yield of the isoscalar scalar $SU(3)$ flavor singlet (squares) and octet (circles) mesons produced in the radiative $J/\psi$ decays: a) for the $\pi\pi$, $K \bar K$, $\eta \eta$, and $\omega \phi$ decays, and b) when the $4\pi$ decays and $\omega\omega$ are included. The data points are extracted from a coupled-channel analysis of BESIII data on radiative $J/\psi$ decays constrained by a large number of further data. Source: Ref.~\cite{Sarantsev:2021ein}.}
\label{fig:f0yield}
\end{figure}

In a recent study~\cite{Sarantsev:2021ein} the authors performed a coupled-channel analysis of the above BESIII data and found as many as ten isoscalar scalar mesons. Nine of them have been listed in PDG2020~\cite{pdg}, while the $f_0(1770)$ is not, and it was suggested to be different from the $f_0(1710)$~\cite{Bugg:2004xu}. Their masses, widths, and decay modes were determined and partly summarized in Table~\ref{sec6:scalarmeson}. Moreover, they found that the masses of the eight states above 1~GeV fall onto two regular Regge trajectories, so they concluded that these mesons can be interpreted as mainly $SU(3)$ flavor singlet and octet states. However, their yield in the radiative $J/\psi$ decays peaks, as depicted in Fig.~\ref{fig:f0yield}, and this peak was proposed to be a scalar glueball. Its mass and width were determined to be $M = 1865 \pm 25 ^{+10}_{-30}$~MeV and $\Gamma = 370 \pm 50 ^{+30}_{-20}$~MeV, and its yield in the radiative $J/\psi$ decays was determined to be $(5.8 \pm 1.0) \times 10^{-3}$.

\begin{table*}[hbtp]
\renewcommand{\arraystretch}{1.5}
\caption{Pole masses and widths of ten isoscalar scalar mesons taken from Ref.~\cite{Sarantsev:2021ein}. The PDG2020 values~\cite{pdg} are listed as small numbers for comparison. Source: Ref.~\cite{Sarantsev:2021ein}.}
\label{sec6:scalarmeson}
\centering
\begin{tabular}{cccccc}
\hline\hline
Name              & $f_0(500)$             & $f_0(1370)$              & $f_0(1710)$           & $f_0(2020)$                    & $f_0(2200)$
\\ \hline
$M$  [MeV]        & 410\er 20              & 1370\er 40               & 1700\er 18            & 1925\er 25                     & 2200\er 25
\\[-1.5ex]
                  & \scriptsize 400\tz 550 & \scriptsize 1200\tz 1500 & \scriptsize 1704\er12 & \scriptsize 1992\er 16         & \scriptsize 2187\er 14
\\
$\Gamma$ [MeV]    & 480\er30               &  390\er 40               & 255\er 25             & 320\er 35                      & 150\er 30
\\[-1.5ex]
                  & \scriptsize 400\tz 700 & \scriptsize 100\tz 500   & \scriptsize 123\er 18 & \scriptsize 442\er60           & \scriptsize $\sim 200$
\\ \hline\hline
Name              & $f_0(980)$             & $f_0(1500)$              & $f_0(1770)$           & $f_0(2100)$                    & $f_0(2330)$
\\ \hline
$M$  [MeV]        & 1014\er 8              & 1483\,\er\,15            & 1765\er 15            & 2075\er 20                     & 2340\er 20
\\[-1.5ex]
                  & \scriptsize 990\er 20  & \scriptsize 1506\,\er\,6 &                       & \scriptsize 2086$^{+20}_{-24}$ & \scriptsize$\sim$2330
\\
$\Gamma$ [MeV]    & 71\er10                & 116\er 12                & 180\er 20             & 260\er 25                      & 165\er 25
\\[-1.5ex]
                  & \scriptsize 10\tz 100  & \scriptsize 112\er 9     &                       & \scriptsize 284$^{+60}_{-32}$  & \scriptsize 250\er 20
\\ \hline\hline
\vspace{-2mm}
\end{tabular}
\end{table*}

In Ref.~\cite{Rodas:2021tyb} the JPAC collaboration performed a systematic analysis on the $J/\psi \to \gamma\pi^0\pi^0$ and $J/\psi \to \gamma K^0_S K^0_S$ decays measured by BESIII~\cite{BESIII:2015rug,BESIII:2018ubj}. They investigated seven scalar and tensor resonances in the 1.0-2.5~GeV mass range, and their results suggest that the $f_0(1710)$ appears in $J/\psi \to \gamma f_0$ more strongly than the $f_0(1500)$. Accordingly, they proposed that this affinity of the $f_0(1710)$ to the gluon-rich initial state, together with a coupling to $K \bar K$ larger by one order of magnitude, is the hint for a sizeable glueball component. The BESIII collaboration quantitatively estimated the $f_0(1500)$ and $f_0(1710)$ production rates in the radiative $J/\psi$ decays to be~\cite{Jin:2021vct}:
\begin{eqnarray}
\mathcal{B}( J/\psi \to \gamma f_0(1500)) &\sim& 0.29 \times 10^{-3} \, ,
\\ \nonumber \mathcal{B}( J/\psi \to \gamma f_0(1710)) &\sim& 2.2 \times 10^{-3} \, ,
\end{eqnarray}
and they arrived at the same conclusion after comparing with the lattice QCD prediction on the production rate of the scalar glueball in the radiative $J/\psi$ decays~\cite{Gui:2012gx}:
\begin{equation}
\mathcal{B}( J/\psi \to \gamma |gg; 0^{++}\rangle) = (3.8\pm0.9) \times 10^{-3} \, .
\end{equation}

\begin{figure}[hbtp]
\begin{center}
\includegraphics[width=1\textwidth]{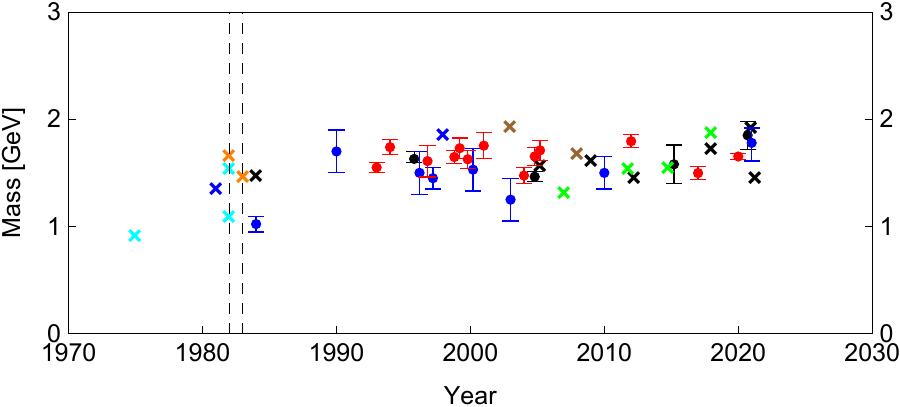}
\end{center}
\caption{Theoretical predictions of the scalar glueball mass with uncertainties (error bars) and without uncertainties (crosses), calculated using the MIT bag model~\cite{Jaffe:1975fd,Carlson:1982er,Chanowitz:1982qj} (cyan), constituent gluon model~\cite{Szczepaniak:2003mr,Mathieu:2008bf} (brown), AdS/QCD model~\cite{Colangelo:2007pt,Boschi-Filho:2012ijd,Chen:2015zhh,Rinaldi:2018yhf,Rinaldi:2021dxh} (green), lattice QCD~\cite{Bali:1993fb,Chen:1994uw,Teper:1997am,Morningstar:1999rf,Vaccarino:1999ku,Niedermayer:2000yx,Liu:2001je,Meyer:2004gx,Chen:2005mg,Loan:2005ff,Gregory:2012hu,Sun:2017ipk,Athenodorou:2020ani} (red), QCD sum rules~\cite{Novikov:1981xi,Narison:1984hu,Bagan:1990sy,Narison:1996fm,Kisslinger:1997gs,Huang:1998wj,Forkel:2000fd,Forkel:2003mk,Wen:2010as,Chen:2021bck} (blue), and others~\cite{Isgur:1984bm,Anisovich:1996zj,Close:2005vf,Kaidalov:2005kz,Ganbold:2009ak,Tsue:2012kf,Fariborz:2015dou,Fariborz:2018unf,Sarantsev:2021ein,Huber:2021yfy,Li:2021gsx} (black). The two dashed lines with orange crosses denote the $f_0(1500)$ first observed in 1983~\cite{Gray:1983cw,Serpukhov-Brussels-AnnecyLAPP:1983xdr} and $f_0(1710)$ first observed in 1982~\cite{Etkin:1982se,Burke:1982am,Edwards:1981ex}, whose masses were measured to be $1506 \pm 6$~MeV and $1704 \pm 12$~MeV, respectively~\cite{pdg}.}
\label{fig:scalarglueball}
\end{figure}

The scalar mesons $f_0(1500)$ first observed in 1983~\cite{Gray:1983cw,Serpukhov-Brussels-AnnecyLAPP:1983xdr} and $f_0(1710)$ first observed in 1982~\cite{Etkin:1982se,Burke:1982am,Edwards:1981ex} were both speculated to be good candidates for the lightest scalar glueball, because they are copiously produced in the gluon-rich processes and both have masses close to most of the theoretical predictions. Their masses and widths were measured to be~\cite{pdg}:
\begin{eqnarray}
f_0(1500) &:& M = 1506 \pm 6 {\rm~MeV} \, ,
\\ \nonumber && \Gamma = 112 \pm 9 {\rm~MeV} \, ;
\\ f_0(1710) &:& M = 1704 \pm 12 {\rm~MeV} \, ,
\\ \nonumber && \Gamma = 123 \pm 18 {\rm~MeV} \, .
\end{eqnarray}
We collect as many theoretical predictions on the scalar glueball mass as we can, and summarize them in Fig.~\ref{fig:scalarglueball}. The average value of the mass predictions obtained after the year 1990 is
\begin{equation}
M_{| {gg} ; 0^{++} \rangle} \sim 1650{\rm~MeV} \, .
\end{equation}
Note that there probably exists strong mixing between the scalar glueball and scalar $\bar q q$ mesons, which has not been taken into account in many theoretical calculations, and therefore, the above value can only and mainly describe the theoretical mass of a ``pure'' scalar glueball.

In Refs.~\cite{Amsler:1995tu,Amsler:1995td} the authors proposed to study the mixing of the scalar glueball with the scalar $\bar q q$ mesons. This approach was followed by a large number of theoretical studies, and we briefly introduce it based on Refs.~\cite{Close:2000yk,Close:2005vf}. In the basis of $|gg\rangle$, $|s \bar s\rangle$, and $|n \bar n\rangle\equiv|u \bar u + d \bar d\rangle/\sqrt2$, the mass matrix for their mixing is expressed as
\begin{equation}
{\mathcal M} =
\left(\begin{array}{ccc}
M_{gg}     & f            & \sqrt{2} f  \\
f          & M_{s\bar{s}} & 0           \\
\sqrt{2} f & 0            & M_{n\bar{n}}
\end{array}\right) \, ,
\end{equation}
where $M_{gg}$, $M_{s\bar{s}}$, and $M_{n\bar{n}}$ represent the masses of the pure states $|gg\rangle$, $|{s\bar{s}}\rangle$, and $|{n\bar{n}}\rangle$, respectively. The parameter $f$ is defined as $f\equiv \langle s\bar{s} | \mathcal{V} | gg\rangle =\langle n\bar{n} | \mathcal{V} | gg \rangle /\sqrt{2}$, with $\mathcal{V}$ the potential causing such a mixing. One usually treats the three scalar mesons $f_0(1710)$, $f_0(1500)$, and $f_0(1370)$ as the eigenstates of $\mathcal M$, and many useful information can be extracted after solving this mass matrix. For example, in Ref.~\cite{Close:2005vf} the authors obtained the branching ratios $\mathcal{B}(J/\psi\to \phi |gg\rangle) \simeq \mathcal{B}(J/\psi\to \omega |gg\rangle)/2 \sim 10^{-4}$ with the pure glueball mass around 1.5~GeV.

The above mixing scheme seems necessary to understand the scalar glueball. In a recent study~\cite{Klempt:2021wpg} the authors studied the mixing between the scalar glueball and six scalar mesons from their decays into two pseudoscalar mesons. These six mesons are the $f_0(1370)$, $f_0(1500)$, $f_0(1710)$, $f_0(1770)$, $f_0(2020)$, and $f_0(2100)$, whose parameters have been listed in Table~\ref{sec6:scalarmeson}. To make this study possible, they first investigated the $f_0(1370)$ and $f_0(1500)$, and studied their mixing in the quark basis
\begin{equation}
\left(\begin{array}{c}
H\\
L
\end{array}\right)
=
\left(\begin{array}{cc}
\cos\varphi  & -\sin\varphi  \\
\sin\varphi  & \cos\varphi
\end{array}\right)
\left(\begin{array}{c}
|n\bar n\rangle \\
|s\bar s\rangle
\end{array}\right) \, ,
\end{equation}
where $\varphi$ is the quarkonia mixing angle; $H$ and $L$ represent the higher- and lower-mass states, respectively. Then they took into account the scalar glueball $|gg\rangle$ to obtain
\begin{eqnarray}
{H^\prime} &=& \left( |n\bar n\rangle \cos\varphi^{\prime} - |s\bar s\rangle \sin\varphi^{\prime} \right) \cos\phi^H + |gg\rangle \sin\phi^H \, ,
\\ \nonumber
{L^\prime} &=& \left( |n\bar n\rangle \sin\varphi^{\prime} + |s\bar s\rangle \cos\varphi^{\prime} \right) \cos\phi^L + |gg\rangle \sin\phi^L \, ,
\end{eqnarray}
where $\varphi^\prime$ is the modified quarkonia mixing angle; $\phi^H$ and $\phi^L$ are the quarkonia-glueball mixing angles of the modified states $H^\prime$ and $L^\prime$, respectively.

\begin{figure}[hbtp]
\begin{center}
\includegraphics[width=0.6\textwidth]{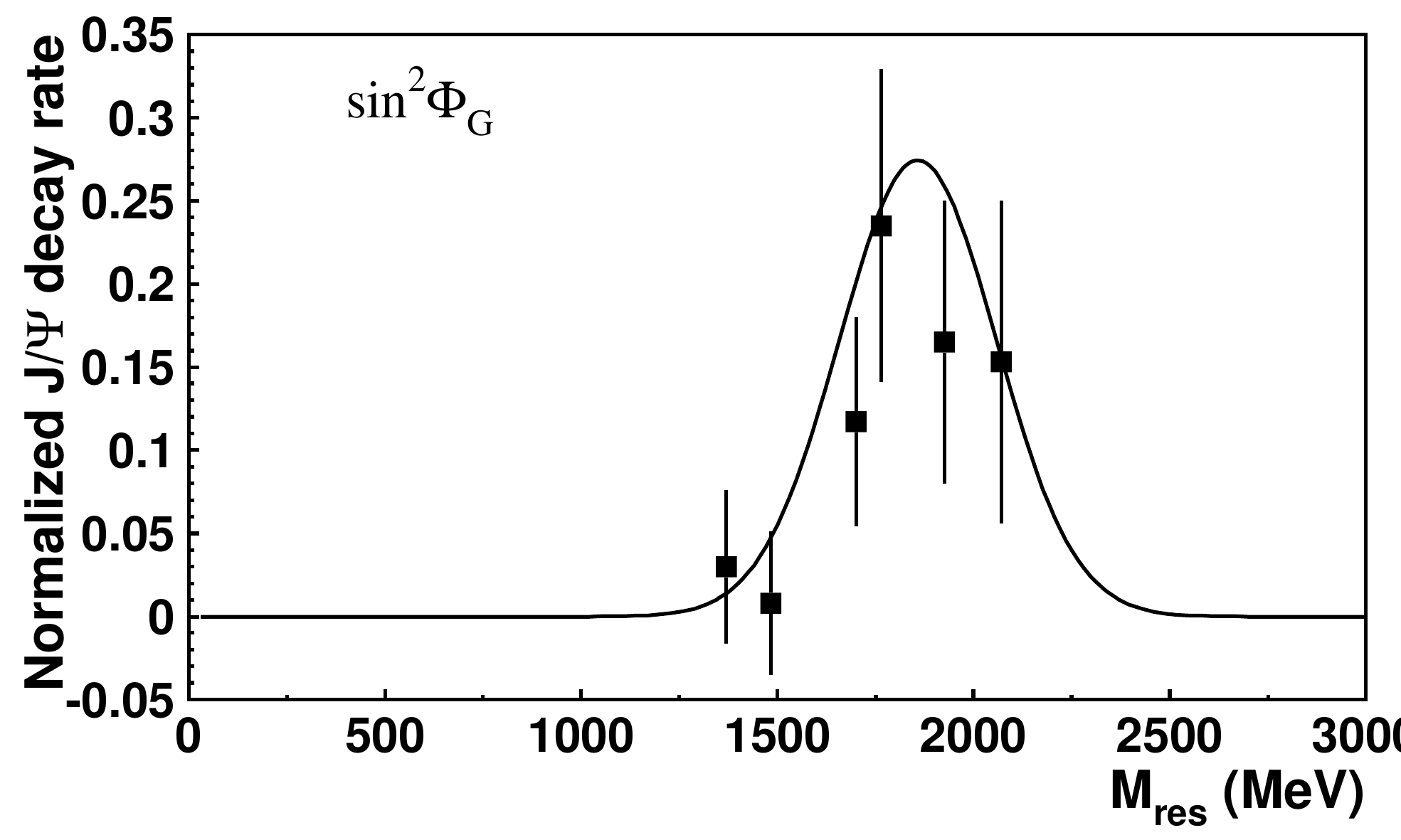}
\end{center}
\caption{The glueball contents inside the six scalar mesons $f_0(1370)$, $f_0(1500)$, $f_0(1710)$, $f_0(1770)$, $f_0(2020)$, and $f_0(2100)$. The data points are extracted from the experiments studying their decays into two pseudoscalar mesons. Source: Ref.~\cite{Klempt:2021wpg}.}
\label{fig:gluonfraction}
\end{figure}

The $f_0(1710)$ and $f_0(1770)$ as well as the $f_0(2020)$ and $f_0(2100)$ can be similarly investigated. Using the experimental decay data as inputs, the authors of Ref.~\cite{Klempt:2021wpg} determined the six quarkonia-glueball mixing angles, and extracted the fractional glueball contents of these resonances to be:
\begin{equation*}
\begin{array}{cccccc}
f_0(1370)    & f_0(1500)   & f_0(1710)     & f_0(1770)     & f_0(2020)    & f_0(2100)
\\
(5\pm4)\%    & <5\%        & (12\pm6)\%    & (25\pm10)\%   & (16\pm9)\%   & (17\pm8)\%
\end{array}
\end{equation*}
The sum of these fractions is $(78\pm18)\%$. As depicted in Fig.~\ref{fig:gluonfraction}, these fractions were used to further propose the existence of a distributed scalar glueball, whose mass and width were determined to be $M\approx1865$~MeV and $\Gamma \approx 370$~MeV, respectively.

\subsubsection{Tensor glueballs and the $f_2(2340)$.}
\label{sec6.1.3}

Besides the scalar glueball, the tensor glueball can also be investigated in the radiative $J/\psi$ decays~\cite{BES:1999dmf,BES:2006nqh,BESIII:2012rtd,BESIII:2013qqz,BESIII:2015rug,BESIII:2016qzq,BESIII:2018ubj,Bugg:2009ch,BESIII:2020nme,Fang:2021wes}. There are as many as twelve light isoscalar tensor mesons listed in PDG2020, {\it i.e.}, $f_2(1270)$, $f_2(1430)$, $f_2(1525)$, $f_2(1565)$, $f_2(1640)$, $f_2(1810)$, $f_2(1910)$, $f_2(1950)$, $f_2(2010)$, $f_2(2150)$, $f_2(2300)$, and $f_2(2340)$; besides, the $f_J(2200)$ may also have the quantum number $I^GJ^{PC} = 0^+ 2^{++}$~\cite{pdg}. This number is more than that of the isoscalar scalar mesons, making the investigation of the tensor glueball even more complicated.

\begin{figure}[hbtp]
\begin{center}
\includegraphics[width=1\textwidth]{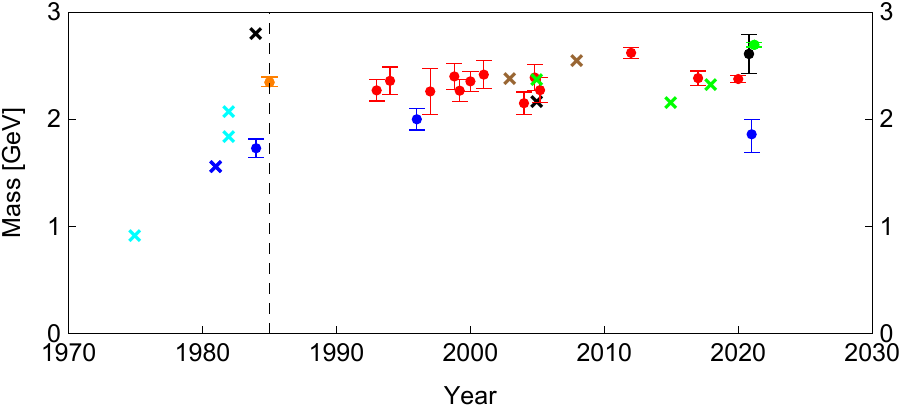}
\end{center}
\caption{Theoretical predictions on the tensor glueball mass with uncertainties (error bars) and without uncertainties (crosses), calculated using the MIT bag model~\cite{Jaffe:1975fd,Carlson:1982er,Chanowitz:1982qj} (cyan), constituent gluon model~\cite{Szczepaniak:2003mr,Mathieu:2008bf} (brown), AdS/QCD model~\cite{Boschi-Filho:2005xct,Chen:2015zhh,Rinaldi:2018yhf,Rinaldi:2021dxh} (green), lattice QCD~\cite{Bali:1993fb,Chen:1994uw,Teper:1997am,Morningstar:1999rf,Vaccarino:1999ku,Niedermayer:2000yx,Liu:2001je,Meyer:2004gx,Chen:2005mg,Loan:2005ff,Gregory:2012hu,Sun:2017ipk,Athenodorou:2020ani} (red), QCD sum rules~\cite{Novikov:1981xi,Narison:1984hu,Narison:1996fm,Chen:2021bck} (blue), and others~\cite{Isgur:1984bm,Kaidalov:2005kz,Huber:2021yfy} (black). The dashed line with a orange error bar denotes the $f_2(2340)$ first observed in 1985~\cite{Lindenbaum:1984wz,Etkin:1985se,Booth:1985kv}, whose mass was measured to be $2345^{+50}_{-40}$~MeV~\cite{pdg}.}
\label{fig:tensorglueball}
\end{figure}

We collect as many theoretical predictions on the tensor glueball mass as we can, and summarize them in Fig.~\ref{fig:tensorglueball}. The average value of the mass predictions obtained after 1990 is
\begin{equation}
M_{| {gg} ; 2^{++} \rangle} \sim 2340{\rm~MeV} \, .
\end{equation}
Accordingly, the tensor meson $f_2(2340)$ first observed in 1985~\cite{Lindenbaum:1984wz,Etkin:1985se,Booth:1985kv} becomes a possible candidate for the lightest tensor glueball, whose mass and width were measured to be~\cite{pdg}:
\begin{eqnarray}
f_2(2340) &:& M = 2345^{+50}_{-40} {\rm~MeV} \, ,
\\ \nonumber && \Gamma = 322^{+70}_{-60} {\rm~MeV} \, .
\end{eqnarray}
However, its production rates in radiative $J/\psi$ decays were measured by BESIII to be~\cite{BESIII:2013qqz,BESIII:2016qzq,BESIII:2018ubj}:
\begin{eqnarray}
\nonumber \mathcal{B}( J/\psi \to \gamma f_2(2340) \to \gamma K \bar K) &=& (5.54{^{+0.34}_{-0.40}}{^{+3.82}_{-1.49}}) \times 10^{-5} \, ,
\\ \mathcal{B}( J/\psi \to \gamma f_2(2340) \to \gamma \eta \eta) &=& (5.60{^{+0.62}_{-0.65}}{^{+2.37}_{-2.07}}) \times 10^{-5} \, ,
\\ \nonumber \mathcal{B}( J/\psi \to \gamma f_2(2340) \to \gamma \phi \phi) &=& (1.91\pm0.14{^{+0.72}_{-0.73}}) \times 10^{-5} \, ,
\end{eqnarray}
which appear to be substantially lower than the lattice QCD prediction on the production rate of the tensor glueball in radiative $J/\psi$ decays~\cite{Yang:2013xba}:
\begin{equation}
\mathcal{B}( J/\psi \to \gamma |gg; 2^{++}\rangle) = (1.1\pm0.2\pm0.1) \times 10^{-2} \, .
\end{equation}

\subsubsection{Pseudoscalar glueballs and the $X(2370)$.}
\label{sec6.1.4}

There are seven light isoscalar pseudoscalar mesons listed in PDG2020, {\it i.e.}, $\eta(548)$, $\eta^\prime(958)$, $\eta(1295)$, $\eta(1405)$, $\eta(1475)$, $\eta(1760)$, and $\eta(2225)$. Besides, the $X(1835)$, $X(2120)$, $X(2370)$, and $X(2500)$ observed by BESII/BESIII may also have the quantum number $I^GJ^{PC} = 0^+ 0^{-+}$~\cite{pdg}.

\begin{figure}[hbtp]
\begin{center}
\subfigure[]{\includegraphics[width=0.30\textwidth]{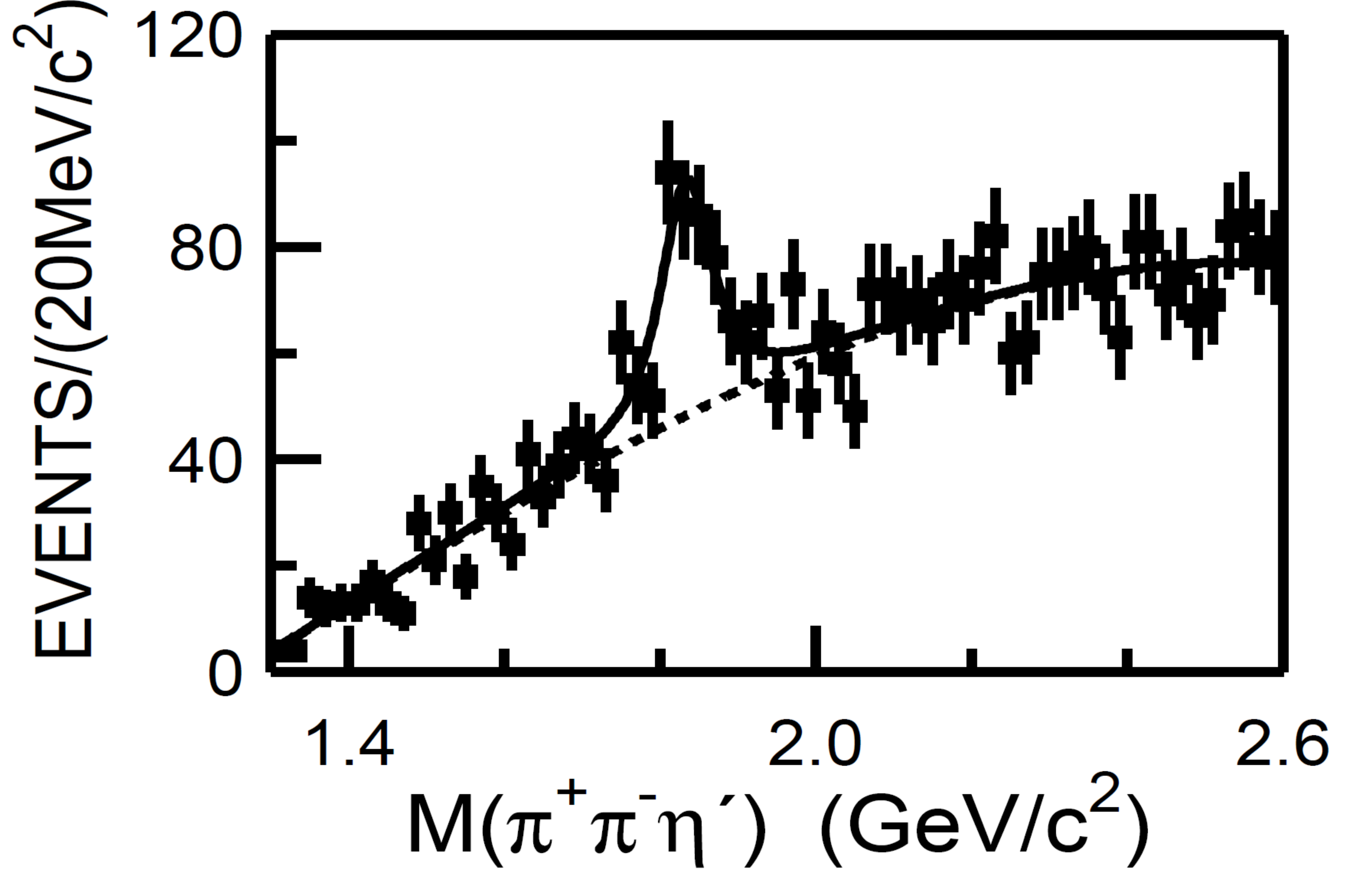}}
\subfigure[]{\includegraphics[width=0.32\textwidth]{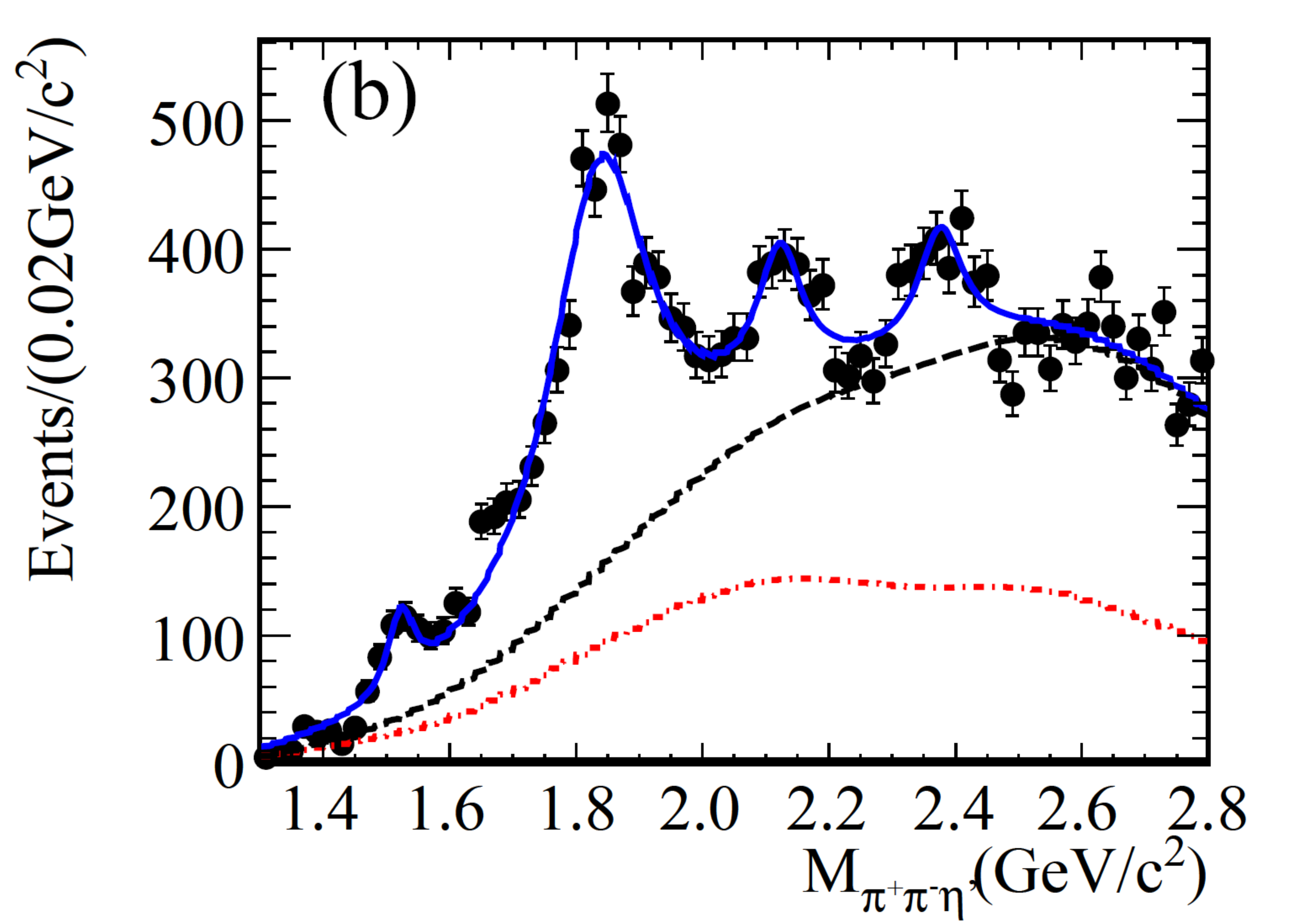}}
\subfigure[]{\includegraphics[width=0.32\textwidth]{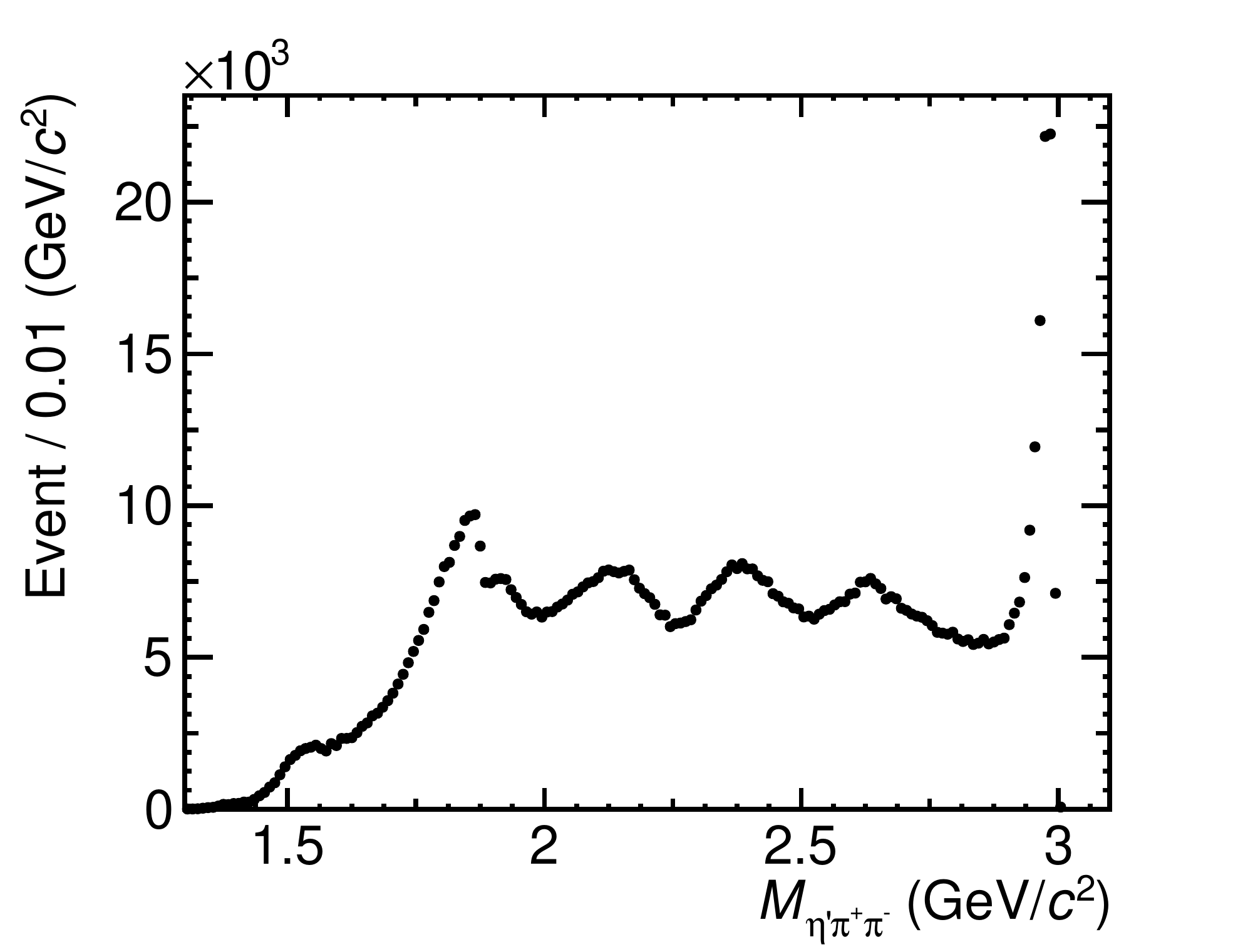}}
\end{center}
\caption{The $\pi^+ \pi^- \eta^\prime$ invariant mass distribution of the $J/\psi \to \gamma \pi^+ \pi^- \eta^\prime$ decay, investigated by (a) BESII in 2005, (b) BESIII in 2010, and (c) BESIII in 2022. Source: Refs.~\cite{BES:2005ega,BESIII:2010gmv,BESIII:2022iiq}.}
\label{fig:X1835}
\end{figure}

\begin{figure}[hbtp]
\begin{center}
\subfigure[]{\includegraphics[width=0.4\textwidth]{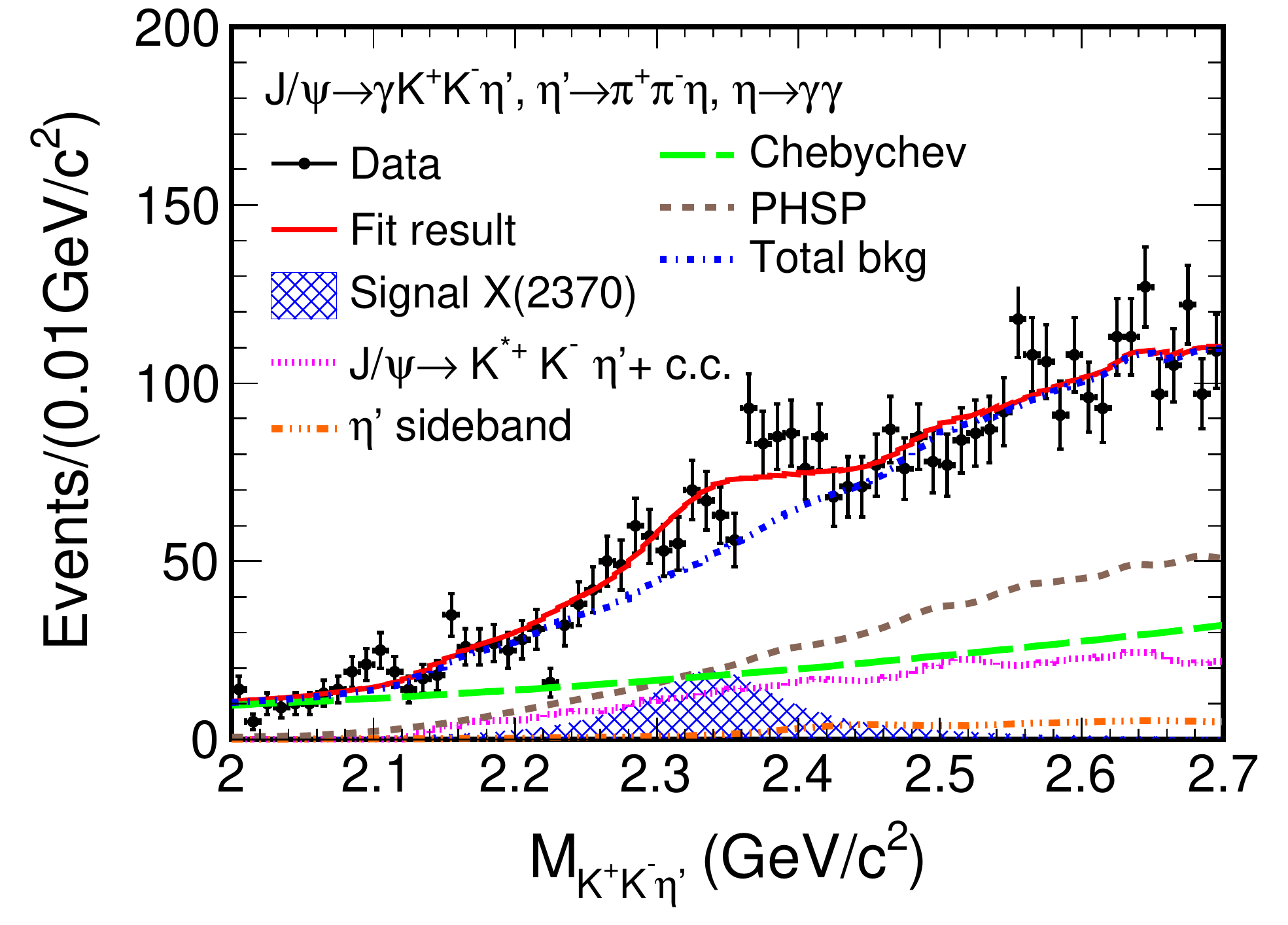}}
~
\subfigure[]{\includegraphics[width=0.4\textwidth]{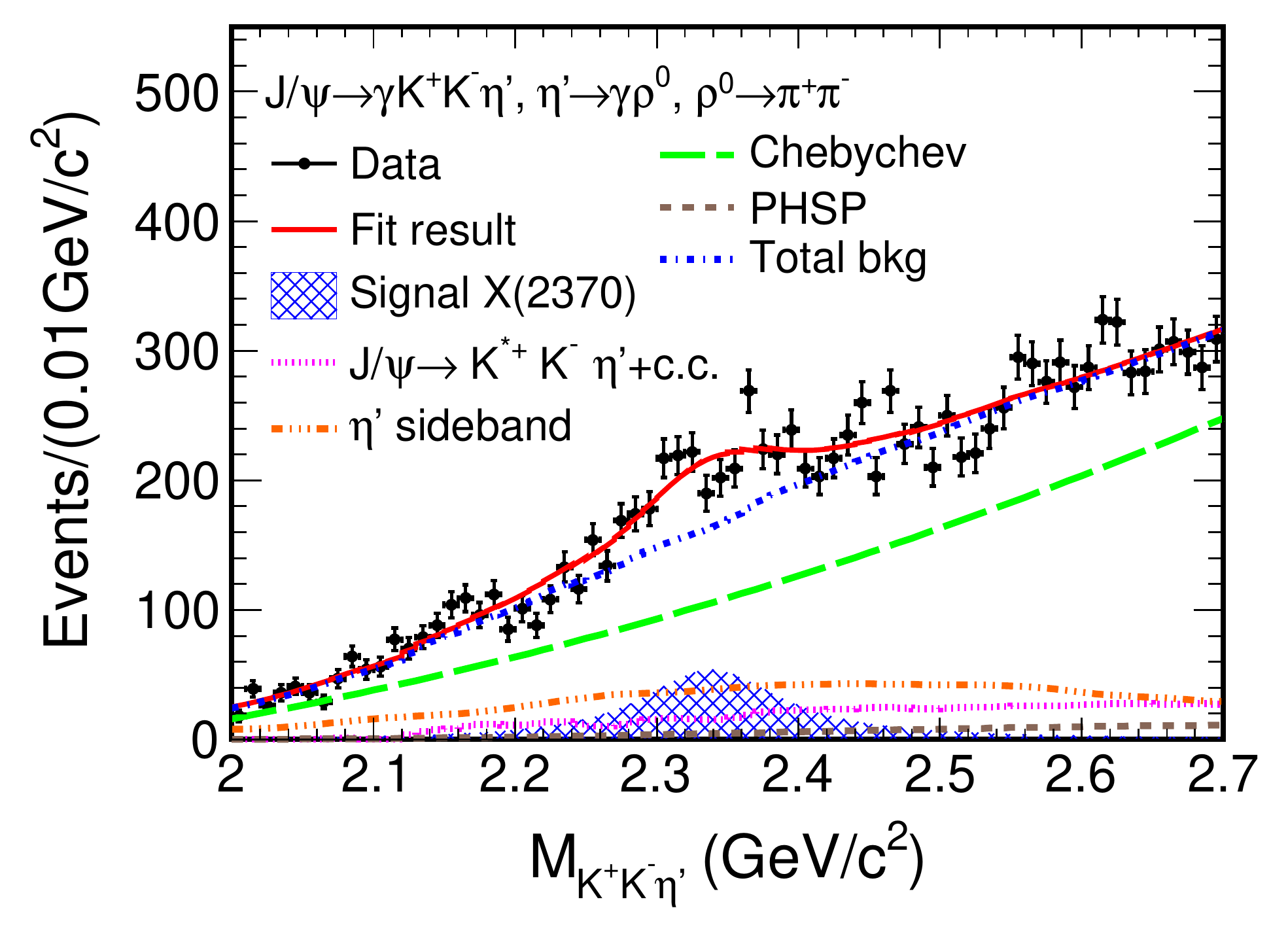}}
\\
\subfigure[]{\includegraphics[width=0.4\textwidth]{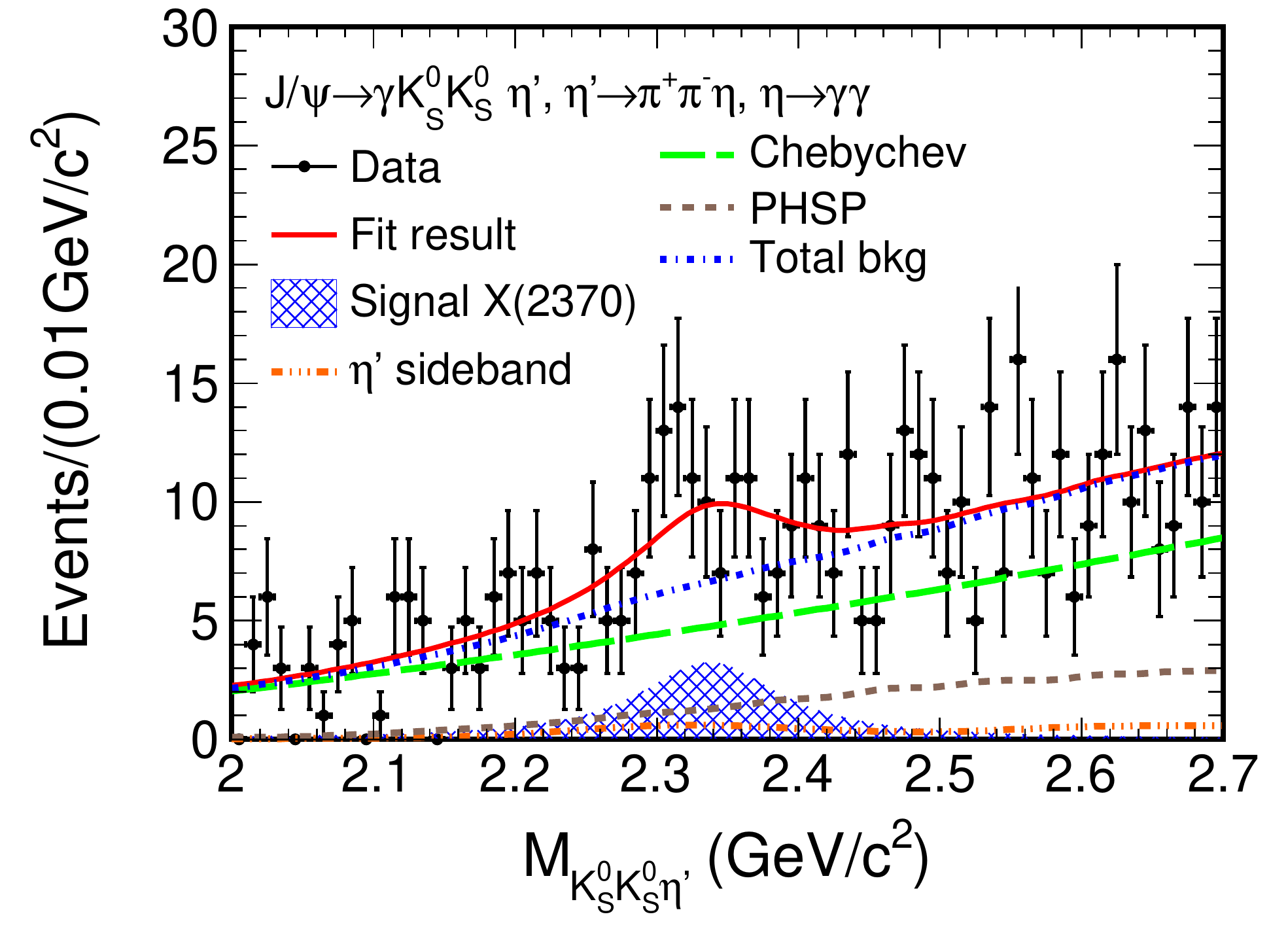}}
~
\subfigure[]{\includegraphics[width=0.4\textwidth]{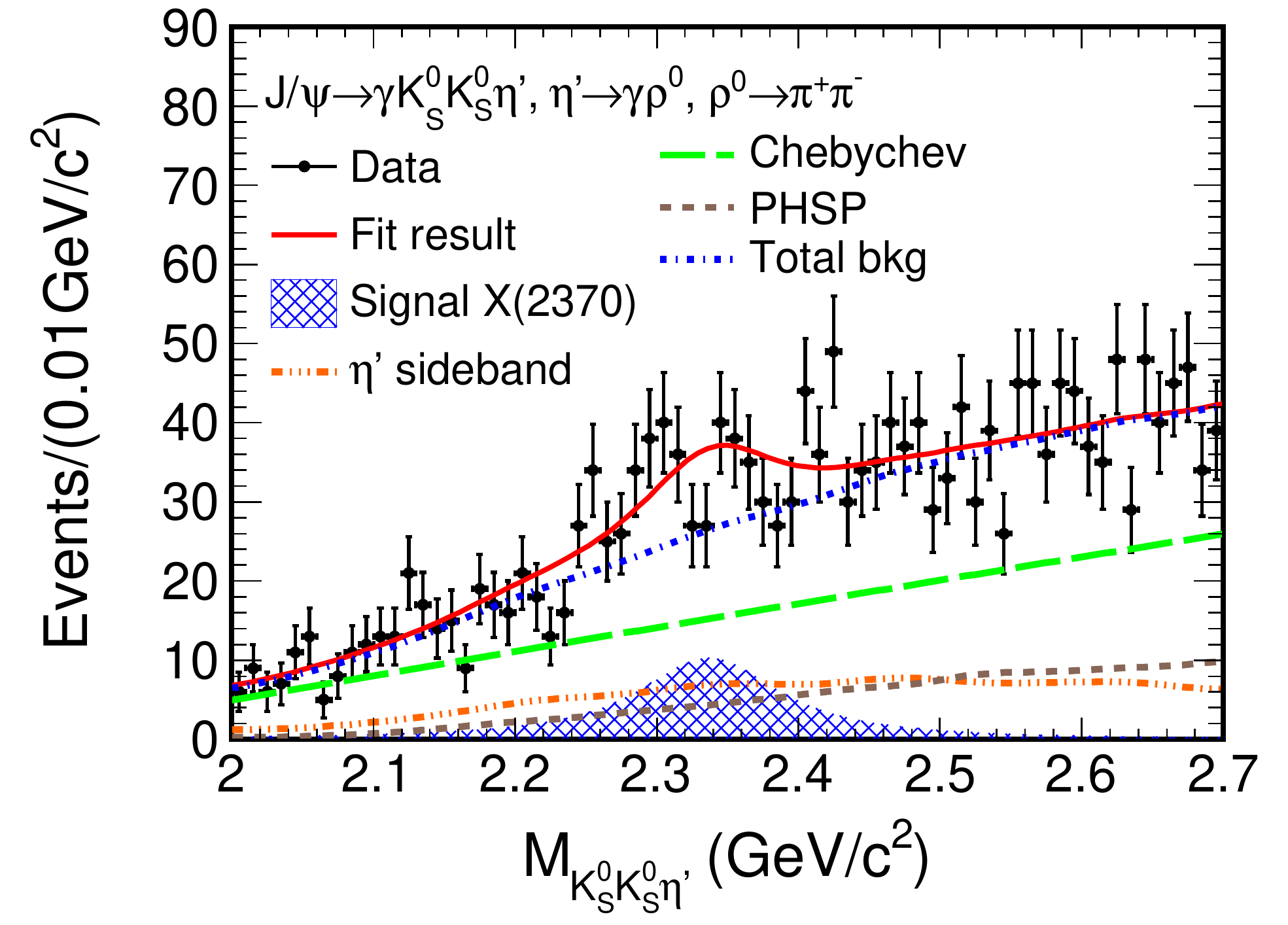}}
\end{center}
\caption{Fit results of the $X(2370)$ in the $\eta^\prime K \bar K$ invariant mass distribution of the decays: (a) $J/\psi \to \gamma X(2370) (\to K^+K^-\eta^\prime)$ with $\eta^\prime \to \pi^+ \pi^- \eta$, (b) $J/\psi \to \gamma X(2370) (\to K^+K^-\eta^\prime)$ with $\eta^\prime \to \gamma \rho^{0}$, (c) $J/\psi \to \gamma X(2370) (\to K^0_S K^0_S \eta^\prime)$ with $\eta^\prime \to \pi^+ \pi^- \eta$, and (d) $J/\psi \to \gamma X(2370) (\to K^0_S K^0_S \eta^\prime)$ with $\eta^\prime \to \gamma \rho^{0}$. Source: Ref.~\cite{BESIII:2019wkp}.}
\label{fig:X2370}
\end{figure}

In 2003 the BESII collaboration observed an anomalous proton-antiproton mass threshold enhancement in the $J/\psi \to \gamma p \bar p$ decays~\cite{BES:2003aic}, which was later confirmed by BESIII~\cite{BESIII:2010vwa,BESIII:2011aa} and CLEO~\cite{CLEO:2010fre}. In 2005 the BESII collaboration studied the $J/\psi \to \gamma \pi^+ \pi^- \eta^\prime$ decay and first observed the $X(1835)$ in the $\pi^+ \pi^- \eta^\prime$ invariant mass distribution~\cite{BES:2005ega}, as shown in Fig.~\ref{fig:X1835}(a). Later the BESIII collaboration confirmed this state, and observed two additional states $X(2120)$ and $X(2370)$~\cite{BESIII:2010gmv,BESIII:2016fbr,BESIII:2021xoh}, as shown in Fig.~\ref{fig:X1835}(b). The $X(1835)$ was confirmed in the $\eta K^0_S K^0_S$ channel of the $J/\psi \to \gamma \eta K^0_S K^0_S$ decay~\cite{BESIII:2015xco}. The $X(2370)$ was confirmed in the $\eta^\prime K \bar K$ channel of the $J/\psi \to \gamma \eta^\prime K \bar K$ decay~\cite{BESIII:2019wkp}, as depicted in Fig.~\ref{fig:X2370}, but not observed in the $\eta^\prime \eta \eta$ channel of the $J/\psi \to \gamma \eta^\prime \eta \eta$ decay~\cite{BESIII:2020ymv}. Besides, the $X(2500)$ was observed in the $\phi\phi$ mass spectrum of the $J/\psi \to \gamma\phi\phi$ decay~\cite{BESIII:2016qzq}.

Recently, the BESIII collaboration studied the $J/\psi \to \gamma \pi^+ \pi^- \eta^\prime$ process and observed a new resonance $X(2600)$ in the $\pi^+ \pi^- \eta^\prime$ invariant mass spectrum with a statistical significance larger than $20\sigma$~\cite{BESIII:2022iiq}, as shown in Fig.~\ref{fig:X1835}(c). Its mass and width were measured to be
\begin{eqnarray}
X(2600) &:& M = 2617.8 \pm 2.1 ^{+18.2}_{-~1.9} {\rm~MeV} \, ,
\\ \nonumber && \Gamma = 200 \pm 8 ^{+20}_{-17} {\rm~MeV} \, ,
\end{eqnarray}
and its spin-parity quantum number was determined to be $J^{PC} = 0^{-+}$ or $2^{-+}$. Interestingly, BESIII found its strong correlation to the scalar meson $f_0(1500)$:
\begin{eqnarray}
\nonumber && \mathcal{B}(J/\psi \to \gamma X(2600)) \cdot \mathcal{B}(X(2600) \to f_0(1500) \eta^\prime) \cdot \mathcal{B}(f_0(1500) \to \pi^+ \pi^-)
\\ &=& (3.39 \pm 0.18 ^{+0.91}_{-0.66} ) \times 10^{-5} \, ,
\\
\nonumber && \mathcal{B}(J/\psi \to \gamma X(2600)) \cdot \mathcal{B}(X(2600) \to f_2^\prime(1525) \eta^\prime) \cdot \mathcal{B}( f_2^\prime(1525) \to \pi^+ \pi^-)
\\ &=& (2.43 \pm 0.13 ^{+0.31}_{-1.11}) \times 10^{-5} \, .
\end{eqnarray}
Note the lattice QCD simulations strongly indicate the pseudoscalar glueball lies around 2.6~GeV. The $X(2600)$ is closely related to the $f_0(1500)$ and $f_2^\prime(1525)$. If the $f_0(1500)$ were the scalar glueball, the $X(2600)$ could be the pseudoscalar glueball. Otherwise, if either the $f_0(1500)$ or $f_2^\prime(1525)$ were dominated by the $\bar q q$ component, the $X(2600)$ could be an excited $\bar q q$ state. More investigations are necessary to understand their nature.

There are many possible explanations for the above pseudoscalar mesons, and some of them are candidates for the pseudoscalar glueball. For example, the $\eta(1405)/\eta(1475)$ was suggested to be a possible pseudoscalar glueball in Refs.~\cite{Donoghue:1980hw,Chanowitz:1980gu,Ishikawa:1980xv,Lacaze:1981sc,Faddeev:2003aw,Li:2007ky}, due to its copious production in the gluon-rich processes. Besides, the anomalously large isospin violation in the $\eta(1405)/\eta(1475) \to \pi\pi\pi$ decay~\cite{BESIII:2012aa} was investigated in Ref.~\cite{Wu:2011yx} through the triangle singularity mechanism. The $X(1835)$ was also interpreted as a candidate of the pseudoscalar glueball in Refs.~\cite{Kochelev:2005vd,He:2005nm,Li:2005vd}.

\begin{figure}[hbtp]
\begin{center}
\includegraphics[width=1\textwidth]{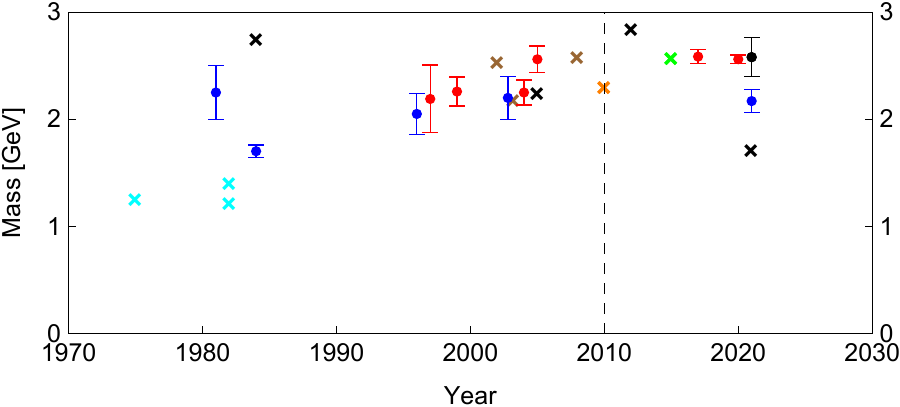}
\end{center}
\caption{Theoretical predictions on the pseudoscalar glueball mass with uncertainties (error bars) and without uncertainties (crosses), calculated using the MIT bag model~\cite{Jaffe:1975fd,Carlson:1982er,Chanowitz:1982qj} (cyan), constituent gluon model~\cite{Hou:2002jv,Szczepaniak:2003mr,Mathieu:2008bf} (brown), AdS/QCD model~\cite{Chen:2015zhh} (green), lattice QCD~\cite{Teper:1997am,Morningstar:1999rf,Meyer:2004gx,Chen:2005mg,Sun:2017ipk,Athenodorou:2020ani} (red), QCD sum rules~\cite{Novikov:1981xi,Narison:1984hu,Narison:1996fm,Forkel:2003mk,Chen:2021bck} (blue), and others~\cite{Isgur:1984bm,Kaidalov:2005kz,Tsue:2012kf,Huber:2021yfy,Li:2021gsx} (black). The dashed line with a orange error bar denotes the $X(2370)$ first observed in 2010~\cite{BESIII:2010gmv}, whose mass was measured to be $2341.6 \pm 6.5 \pm 5.7$~MeV~\cite{BESIII:2019wkp}.}
\label{fig:pseudoglueball}
\end{figure}

We collect as many theoretical predictions on the pseudoscalar glueball mass as we can, and summarize them in Fig.~\ref{fig:pseudoglueball}. The average value of the mass predictions obtained after the year 1990 is
\begin{equation}
M_{| {gg} ; 0^{-+} \rangle} \sim 2360{\rm~MeV} \, .
\end{equation}
Accordingly, the resonance $X(2370)$ first observed in 2010~\cite{BESIII:2010gmv} becomes a possible candidate for the low-lying pseudoscalar glueball, whose mass and width were measured to be~\cite{BESIII:2019wkp}:
\begin{eqnarray}
X(2370) &:& M = 2341.6 \pm 6.5 \pm 5.7 {\rm~MeV} \, ,
\\ \nonumber && \Gamma =  117 \pm 10 \pm 8 {\rm~MeV} \, .
\end{eqnarray}
Its production rates in the radiative $J/\psi$ decays were measured by BESIII to be~\cite{BESIII:2019wkp}:
\begin{eqnarray}
\mathcal{B}( J/\psi \to \gamma X(2370) \to \gamma K^+K^- \eta^\prime) &=& (1.79\pm0.23\pm0.65) \times 10^{-5} \, ,
\\ \nonumber \mathcal{B}( J/\psi \to \gamma X(2370) \to \gamma K^0_S K^0_S \eta^\prime) &=& (1.18\pm0.32\pm0.39) \times 10^{-5} \, .
\end{eqnarray}
Besides, the $X(2500)$ might be the same state as the $X(2370)$, and its production rate in the radiative $J/\psi$ decays was measured by BESIII to be~\cite{BESIII:2016qzq}:
\begin{equation}
\mathcal{B}( J/\psi \to \gamma X(2500) \to \gamma \phi \phi) = (1.7\pm0.2^{+0.2}_{-0.8}) \times 10^{-5} \, .
\end{equation}
The above production rates are also smaller than the lattice QCD prediction on the production rate of the pseudoscalar glueball in the radiative $J/\psi$ decays~\cite{Gui:2019dtm}:
\begin{equation}
\mathcal{B}( J/\psi \to \gamma |gg; 0^{-+}\rangle) = (2.31\pm0.80) \times 10^{-4} \, .
\end{equation}

In Ref.~\cite{Zhang:2021xvl} the authors applied the lattice QCD to study the mixing of the pseudoscalar glueball and the pseudoscalar charmonium $\eta_c$. By assuming the $X(2370)$ to be predominantly a pseudoscalar glueball, they determined the mixing angle to be $4.3^\circ\pm0.5^\circ$, which raises the $\eta_c$ mass by about $3.9$~MeV and its total width by about 7~MeV. A more complicated mixing scheme was examined in Ref.~\cite{Qin:2017qes} based on Refs.~\cite{Cheng:2008ss,Tsai:2011dp}, where the authors studied the mixing among the pseudoscalar glueball and the pseudoscalar quarkonia $\eta$, $\eta^\prime$, and $\eta_c$. Their results suggest that the mass of the pseudoscalar glueball can not be lower than 1.9~GeV.

The interpretation of the $X(2370)$ as the lowest-lying pseudoscalar glueball is supported by Ref.~\cite{Xian:2014jpa} based on Gaussian sum rules within the instanton vacuum model. However, this assignment is not favored in Ref.~\cite{Sun:2021kka}, where the authors investigated the radiative $J/\psi$ decays through chiral effective Lagrangians. See Refs.~\cite{Janowski:2011gt,Janowski:2014ppa,Eshraim:2012jv,Eshraim:2016mds} for more studies on decay behaviors of the pseudoscalar glueball through chiral Lagrangians.

\subsubsection{Three-gluon glueballs.}
\label{sec6.1.5}

The investigations of the three-gluon glueballs are closely related to the pomeron and odderon. One feature of the strong interaction is the dominance at high energies of the $C$-even exchange over the $C$-odd one. In the Regge language the large energy asymptotic of the $C$-even amplitude is due to the pomeron exchange~\cite{Kuraev:1977fs,Balitsky:1978ic,Low:1975sv,Nussinov:1975mw}, and that of the $C$-odd amplitude is due to the odderon exchange~\cite{Lukaszuk:1973nt,Bartels:1980pe,Kwiecinski:1980wb,Janik:1998xj}, both of which correspond to the leading singularities of the relevant amplitudes in the complex angular momentum plane.

With the advent of QCD as the theory of the strong interaction, a family of colorless $C$-even states, beginning with a $t$-channel exchange of two gluons, was demonstrated to play the role of the pomeron. Correspondingly, a family of colorless $C$-odd states, beginning with a $t$-channel exchange of three gluons, was predicted by QCD to play the role of the odderon, whose experimental existence remains inconclusive in spite of many assertions~\cite{Bouquet:1974cs,Nussinov:1975qb,Joynson:1975az,Bartels:1980pe,Jaroszewicz:1980mq,Erhan:1984mv,Breakstone:1985pe,Gauron:1992zc,Chen:1995pa,Dosch:2002ai,Khoze:2017swe,Martynov:2017zjz,Csorgo:2019ewn}. We refer to the reviews~\cite{Levin:1990gg,Braun:1998fs,Ewerz:2005rg,Block:2006hy} for their detailed discussions.

\begin{figure}[hbtp]
\begin{center}
\includegraphics[width=0.6\textwidth]{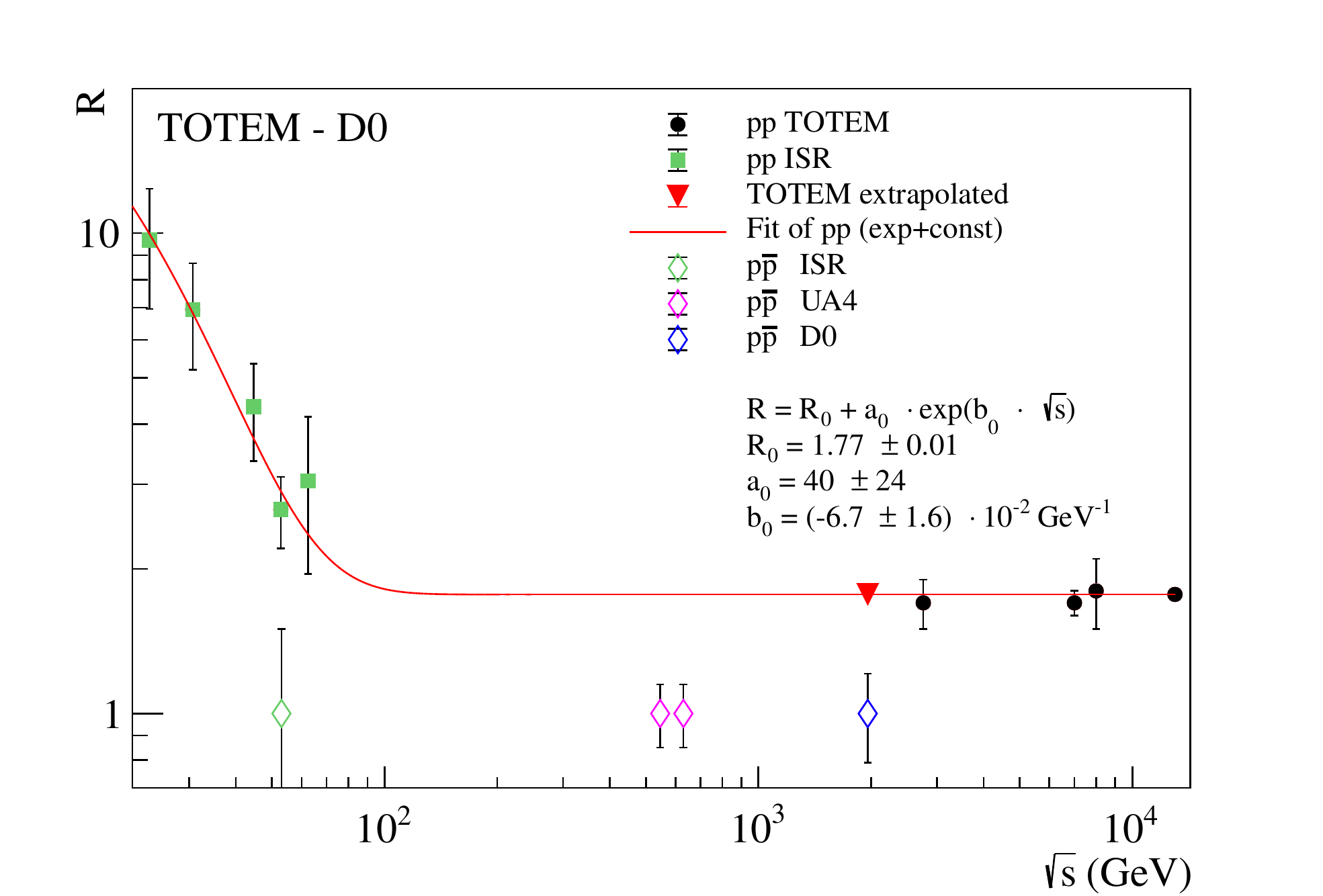}
\end{center}
\caption{The ratio $R$ of the differential cross sections measured at the bump and dip locations for ISR~\cite{Breakstone:1985pe,Nagy:1978iw}, S$p\bar p$S~\cite{UA4:1985oqn,UA4:1986cgb}, Tevatron~\cite{D0:2012erd}, and LHC~\cite{TOTEM:2011vxg,TOTEM:2015oop,TOTEM:2018hki,TOTEM:2018psk} $pp$ and $p \bar p$ elastic cross section data. Source: Ref.~\cite{TOTEM:2020zzr}.}
\label{fig:odderon}
\end{figure}

In 2020 the D0 and TOTEM collaborations studied their $pp$ and $p\bar p$~\cite{D0:2012erd} elastic cross section data, and found they differed with a significance of $3.4\sigma$~\cite{TOTEM:2020zzr}, as depicted in Fig.~\ref{fig:odderon}. This leads to the evidence of a $t$-channel exchanged odderon~\cite{COMPETE:2002jcr,Martynov:2018sga,Khoze:2018kna}, {\it i.e.}, a possible $C$-odd three-gluon glueball. They further combined their previous data~\cite{TOTEM:2017sdy} and increased the significance to be $5.2\sigma$-$5.7\sigma$. Accordingly, the D0 and TOTEM collaborations claimed that they accomplished the first experimental observation of the odderon. There were heated discussions and debates on this topic~\cite{Khoze:2018bus,Csorgo:2018uyp,Goncalves:2018nsp,Harland-Lang:2018ytk,Xie:2019soz,Szanyi:2019kkn,Troshin:2020krx,Petrov:2020mwj,Aushev:2021ofv,Sjostrand:2021dal,Bence:2021uji,Godizov:2021ksd,Capossoli:2021ope,TOTEM:2021imi}, which are mainly on the contribution of the odderon exchange in certain collisions. It is also important to directly study the odderon as well as the three-gluon glueball themselves.

\begin{figure}[hbtp]
\begin{center}
\includegraphics[width=0.6\textwidth]{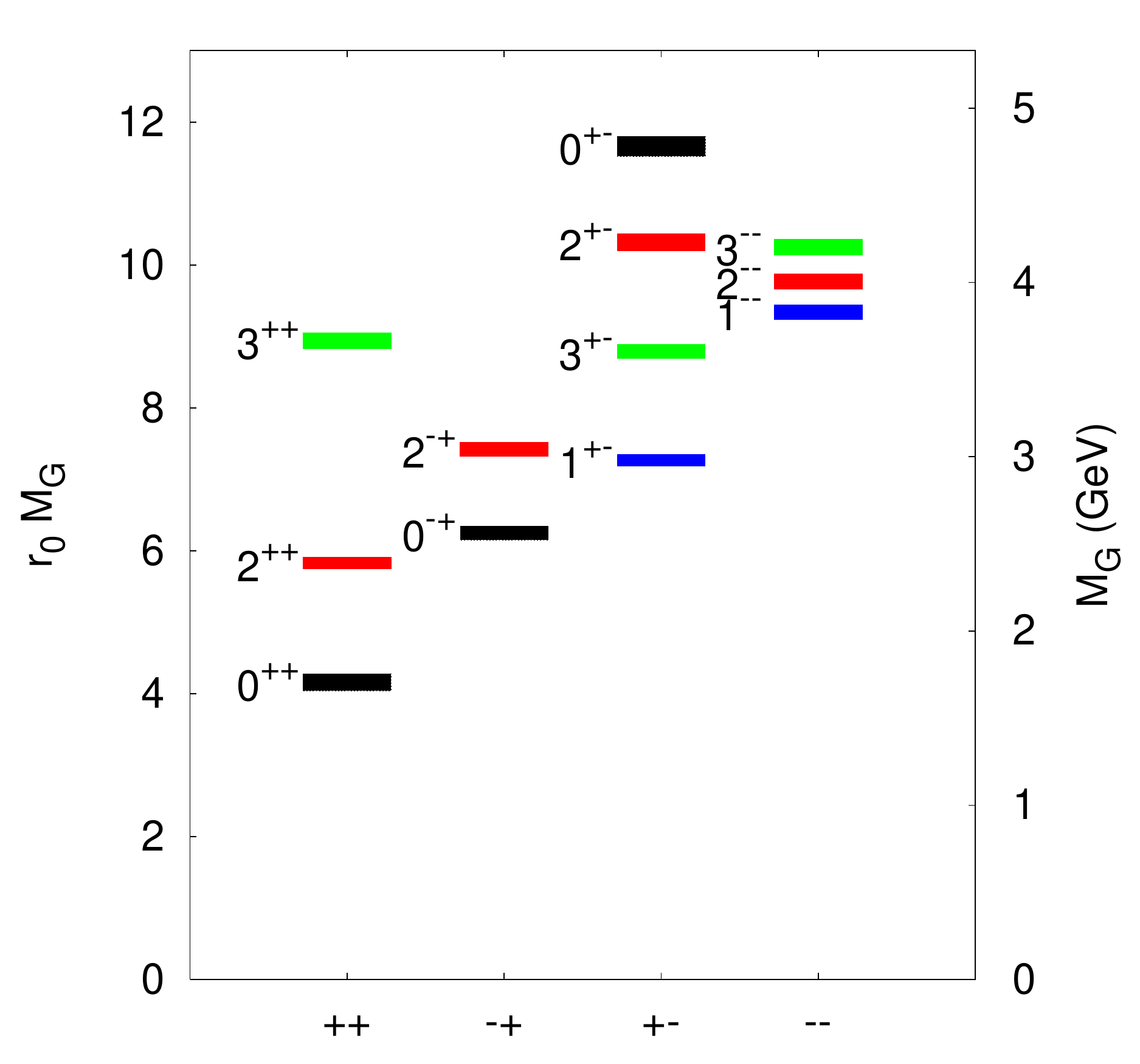}
\end{center}
\caption{Mass spectrum of glueballs calculated using lattice QCD in the pure $SU(3)$ gauge theory, given in terms of $r_0$ ($r^{-1}_0 = 410$~MeV) and in GeV. Source: Ref.~\cite{Chen:2005mg}.}
\label{fig:latticeglueball}
\end{figure}

Through lattice QCD and QCD sum rules, the three-gluon glueballs of both $C=+1$ and $C=-1$ have been systematically studied in Refs.~\cite{Morningstar:1999rf,Chen:2005mg,Meyer:2004gx,Gregory:2012hu,Athenodorou:2020ani,Latorre:1987wt,Hao:2005hu,Chen:2021bck}. Their mass predictions of the low-lying $J^{PC} = 0^{\pm+}/1^{\pm-}/2^{\pm\pm}/3^{\pm-}$ three-gluon glueballs are generally consistent with each other, as already shown in Table~\ref{sec6:comparison} and Fig.~\ref{fig:gluonsumrule}. Besides, the $J^{PC} = 0^{\pm-}$ three-gluon glueballs have been investigated in Refs.~\cite{Morningstar:1999rf,Chen:2005mg,Gregory:2012hu,Qiao:2014vva,Tang:2015twt,Pimikov:2016pag,Pimikov:2017bkk,Pimikov:2017xap,Qiao:2017jxc} through lattice QCD and QCD sum rules. These two glueballs are probably not low-lying any more, as depicted in Fig.~\ref{fig:latticeglueball}. We summarize their masses calculated in these studies as follows (in unit of MeV):
\begin{equation*}
\begin{scriptsize}
\begin{array}{ccccccc}
          & {\rm Ref.~\cite{Morningstar:1999rf}} & {\rm Ref.~\cite{Chen:2005mg}} & {\rm Ref.~\cite{Gregory:2012hu}} & {\rm Refs.~\cite{Qiao:2014vva,Tang:2015twt}} & {\rm Refs.~\cite{Pimikov:2016pag,Pimikov:2017bkk}}
\\
0^{+-}    & 4740\pm70\pm230                      & 4780\pm60\pm230               & 5450\pm830                       & 4570\pm130                                   & 9200^{+1300}_{-1400}
\\
0^{--}                                                                         &&& 5166\pm1000                      & 3810\pm120                                   & 6800^{+1100}_{-1200}
\end{array}
\end{scriptsize}
\end{equation*}
Especially, the three mass predictions of the $J^{PC} = 0^{--}$ three-gluon glueball differ significantly from each other. Moreover, in Ref.~\cite{Szanyi:2019kkn} the authors studied the glueball trajectories, and they argued that the odderon is essentially an object in the Regge pole model, and any derivation of its finite mass from the massless gluons meets conceptual difficulties in QCD, which prevents reliable calculations of its spectrum. All these theoretical controversies and divergences reflect that the academic community has not formed a common understanding on the nature of the glueballs, and further experimental and theoretical studies are crucially demanded to understand them better.

To end this subsection, we briefly discuss possible decay patterns of two- and three-gluon glueballs following Refs.~\cite{Chen:2021cjr,Chen:2021bck}. A pure two-gluon glueball, if it exists, can decay through exciting two color-octet quark-antiquark pairs from two gluons, and recombining into two color-singlet mesons. However, it is rather difficult to differentiate it from conventional $\bar q q$ mesons and exotic tetraquark states, as discussed at the beginning of this section.

\begin{figure}[hbtp]
\begin{center}
\subfigure[]{\includegraphics[width=0.45\textwidth]{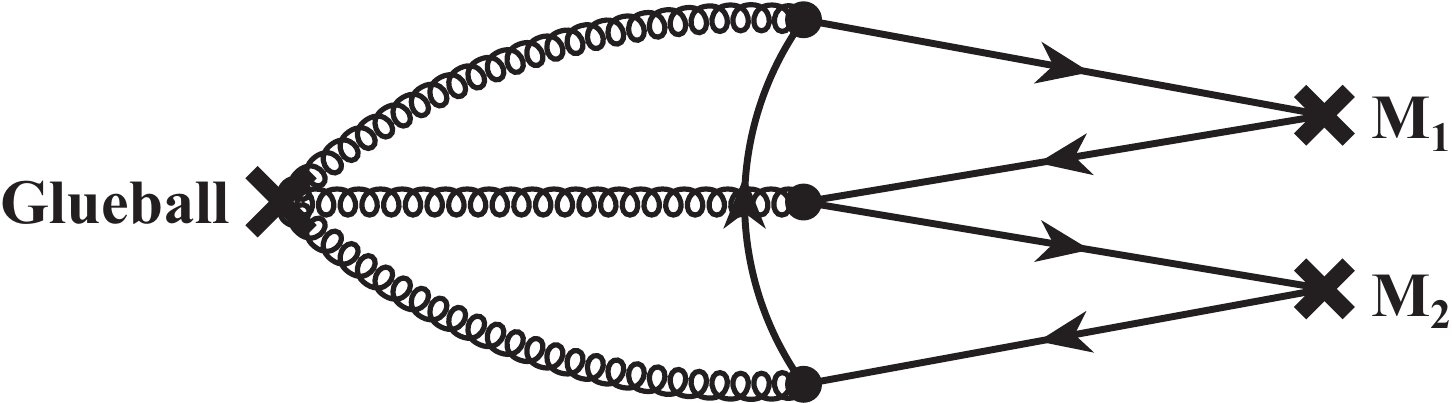}}
~~~~~
\subfigure[]{\includegraphics[width=0.45\textwidth]{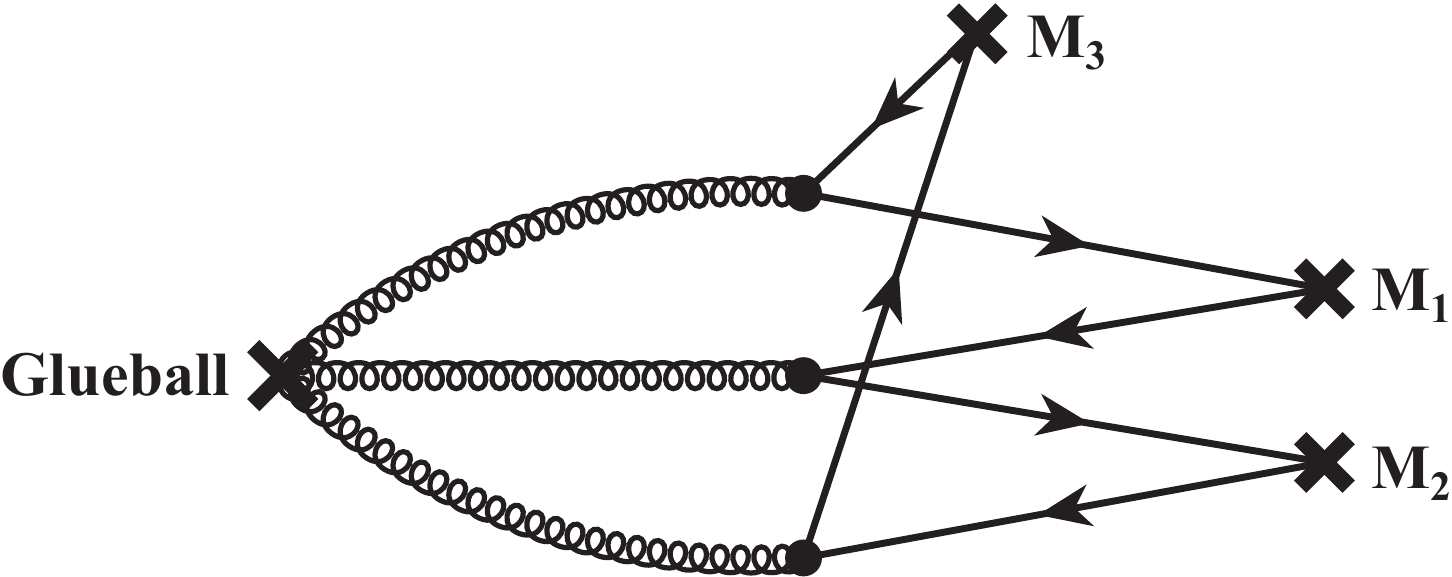}}
\end{center}
\caption{Two possible decay processes of the three-gluon glueball, which decays through exciting three color-octet quark-antiquark pairs from three gluons, and recombining into (a) two color-singlet mesons or (b) three color-singlet mesons.}
\label{fig:decayglueball}
\end{figure}

A pure three-gluon glueball, if it exists, can decay through exciting three color-octet quark-antiquark pairs from three gluons, and recombining into two color-singlet mesons or three mesons, as depicted in Fig.~\ref{fig:decayglueball}. The amplitudes of these two possible decay processes are both at the $\mathcal{O}(\alpha_s^{3/2})$ order, so the three-meson decay patterns are generally not suppressed severely compared to the two-meson decay patterns, or even enhanced due to the quark-antiquark annihilation during the two-meson decay process. This behavior can be useful in identifying its nature. We use $P$ and $V$ to denote light vector and pseudoscalar mesons respectively, and explicitly write down its possible decay pattern as follows:
\begin{eqnarray}
\nonumber       &0^{-} \to& ~~~~\,PPP,VVP,VVV~~~~~(S\mbox{-wave})         ~~\&~~  ~~~VP,VV~~~~(P\mbox{-wave}) \, ,
\\
\nonumber       &0^{+} \to& ~~~~\,VPP,VVP,VVV~~~~~(P\mbox{-wave})         ~~\&~~  ~~~PP,VV\,~~~(S\mbox{-wave}) \, ,
\\
\nonumber       &1^{-} \to& ~~~~\,VPP,VVP,VVV~~~~~(S\mbox{-wave})         ~~\&~~  PP,VP,VV~(P\mbox{-wave}) \, ,
\\
\nonumber       &1^{+} \to& PPP,VPP,VVP,VVV~(P\mbox{-wave})               ~~\&~~  ~~~VP,VV~~~~(S\mbox{-wave}) \, ,
\\
\nonumber       &2^{-} \to& ~~~~~~~~\,VVP,VVV~~~~~~~~~(S\mbox{-wave})     ~~\&~~  ~~~VP,VV~~~~(P\mbox{-wave}) \, ,
\\
\nonumber       &2^{+} \to& ~~~~VPP,VVP,VVV~~~~~(P\mbox{-wave})           ~~\&~~  ~~~~~~VV~~~~~~~(S\mbox{-wave}) \, ,
\\
\nonumber       &3^{-} \to& ~~~~~~~~~~~~\,VVV~~~~~~~~~~~~~(S\mbox{-wave}) ~~\&~~  ~~~~~~VV~~~~~~~(P\mbox{-wave}) \, ,
\\
\nonumber       &3^{+} \to& ~~~~~~~~VVP,VVV~~~~~~~~~(P\mbox{-wave})       ~~\&~~  ~~~VP,VV~~~~(D\mbox{-wave}) \, .
\end{eqnarray}
Considering their limited decay patterns and suppressed two-meson decay patterns, the spin-3 three-gluon glueballs may have relatively smaller widths. Accordingly, we propose to search for them in their three-meson decay patterns $VVV$ and $VVP$ in future particle experiments~\cite{Chen:2021cjr,Chen:2021bck}.

\subsection{Light hybrid mesons}
\label{sec6.2}

A hybrid meson is composed of one valence quark and one valence antiquark together with a few valence gluons. Its experimental confirmation is a direct test of QCD in the low energy sector. Especially, the hybrid mesons with $J^{PC} =0^{--}/0^{+-}/1^{-+}/2^{+-}/3^{-+}/4^{+-}/\cdots$ are of particular interests, since these exotic quantum numbers can not be accessed by conventional $\bar q q$ mesons and may arise from the gluon degree of freedom. However, it is still difficult to differentiate them from the exotic tetraquark states, and there is no good method to overcome this difficulty yet.

In the past half century there have been a lot of experimental and theoretical studies on hybrid mesons, but their nature still remains elusive, partly due to the above difficulty. In this paper we only briefly review recent progresses on the light hybrid mesons, which consist of valence gluons together with the light $up/down/strange$ quarks. We shall separately discuss single- and double-gluon hybrid mesons as follows.

\subsubsection{$\pi_1(1400)/\pi_1(1600)/\pi_1(2015)$ of $I^GJ^{PC} = 1^-1^{-+}$.}
\label{sec6.2.1}

Up to now there are three candidates of hybrid mesons observed in experiments with the exotic quantum number $I^GJ^{PC} = 1^-1^{-+}$, {\it i.e.}, the $\pi_1(1400)$~\cite{IHEP-Brussels-LosAlamos-AnnecyLAPP:1988iqi}, $\pi_1(1600)$~\cite{E852:1998mbq}, and $\pi_1(2015)$~\cite{E852:2004gpn}. They are possible single-gluon hybrid mesons, which consist of one quark-antiquark pair together with only one valence gluon. Their masses and widths were measured to be~\cite{pdg,E852:2004rfa}:
\begin{eqnarray}
\pi_1(1400) &:& M = 1354 \pm 25 {\rm~MeV} \, ,
\\ \nonumber && \Gamma = 330 \pm 35 {\rm~MeV} \, ;
\\ \pi_1(1600) &:& M = 1660 ^{+15}_{-11} {\rm~MeV} \, ,
\\ \nonumber && \Gamma = 257 \pm 60 {\rm~MeV} \, ;
\\ \pi_1(2015) &:& M = 2014\pm20\pm16 {\rm~MeV} \, ,
\\ \nonumber && \Gamma = 230 \pm 32 \pm 73 {\rm~MeV} \, .
\end{eqnarray}
The $\pi_1(1400)$ was observed in the $\eta \pi$ channel by several collaborations~\cite{IHEP-Brussels-LosAlamos-AnnecyLAPP:1988iqi,Aoyagi:1993kn,E852:1997gvf,VES:2001rwn,CrystalBarrel:1998cfz,CrystalBarrel:2019zqh}. Besides, it was observed in the $\rho \pi$ channel by OBELIX~\cite{OBELIX:2004oio}, but not observed in this channel by COMPASS~\cite{COMPASS:2018uzl}. The $\pi_1(1600)$ was observed in the $\rho \pi$, $\eta^\prime \pi$, $f_1(1285) \pi$, and $b_1(1235) \pi$ channels~\cite{E852:1998mbq,Khokhlov:2000tk,Baker:2003jh,COMPASS:2009xrl,CLEO:2011upl}. The $\pi_1(2015)$ was observed in the $f_1(1285) \pi$ and $b_1(1235) \pi$ channels only by the BNL E852 experiments~\cite{E852:2004gpn,E852:2004rfa}. However, their existence is still disputed~\cite{Meyer:2015eta,Ketzer:2019wmd,COMPASS:2021ogp,Jin:2021vct}.

\begin{figure}[hbtp]
\begin{center}
\subfigure[]{\includegraphics[width=0.45\textwidth]{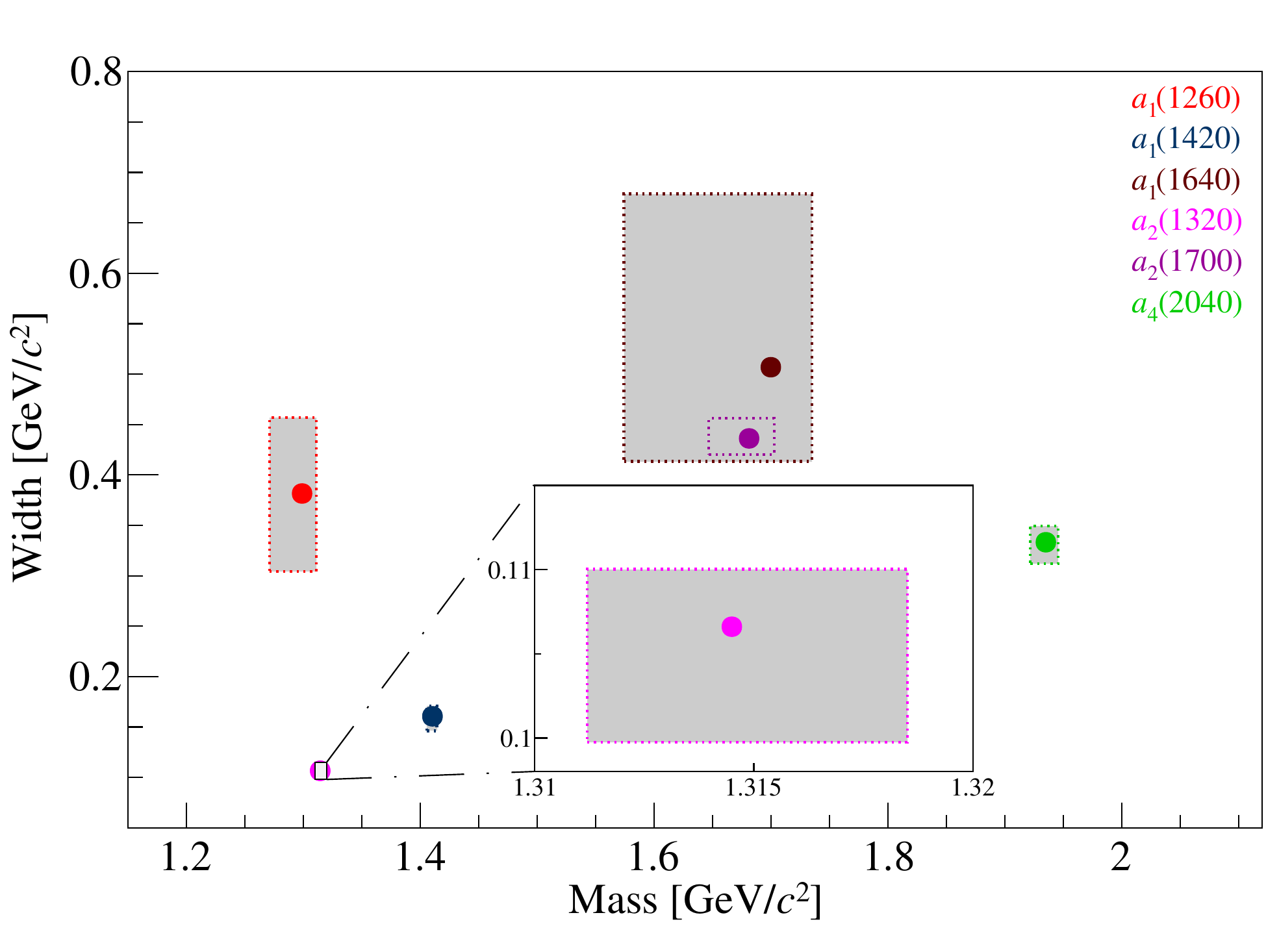}}
~~~~~
\subfigure[]{\includegraphics[width=0.45\textwidth]{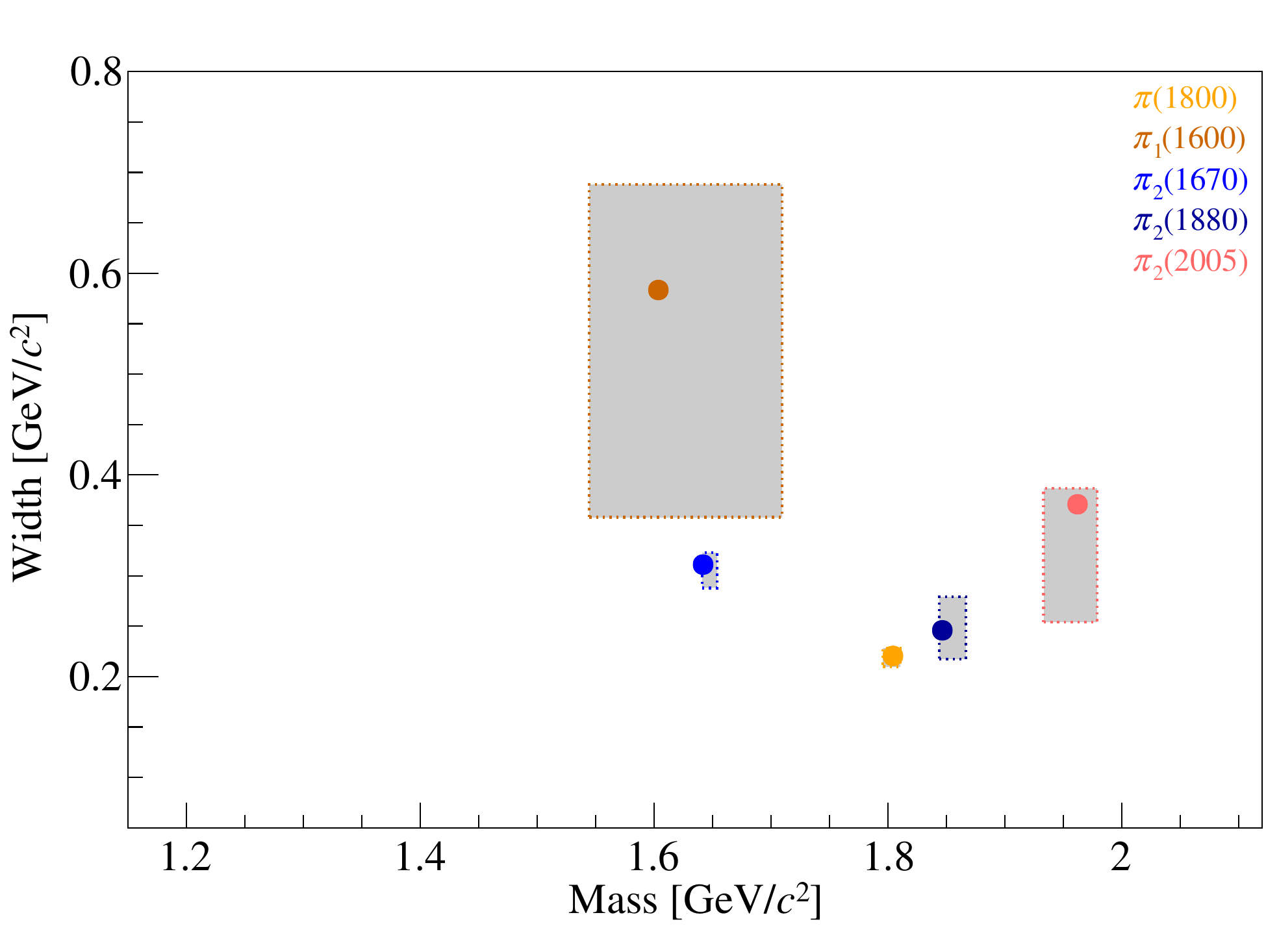}}
\end{center}
\caption{Masses and widths of (a) $a_J$ and (b) $\pi_J$ resonances measured in the COMPASS experiment. Source: Ref.~\cite{COMPASS:2018uzl}.}
\label{fig:pi1600}
\end{figure}

In 2018 the COMPASS collaboration performed a comprehensive resonance-model fit of the $\pi^-\pi^-\pi^+$ states produced in the reaction $\pi^- + p \to \pi^- \pi^- \pi^+ + p_{\rm recoil}$ with a $190$~GeV$/c$ pion beam~\cite{COMPASS:2018uzl}. They investigated altogether 11 isovector light mesons, including the $\pi_1(1600)$ in the $\rho \pi$ final state. As depicted in Fig.~\ref{fig:pi1600}, its mass and width were measured to be $1600^{+110}_{-~60}$~MeV and $580^{+100}_{-230}$~MeV, respectively.

\begin{figure}[hbtp]
\begin{center}
\subfigure[]{\includegraphics[width=0.3\textwidth]{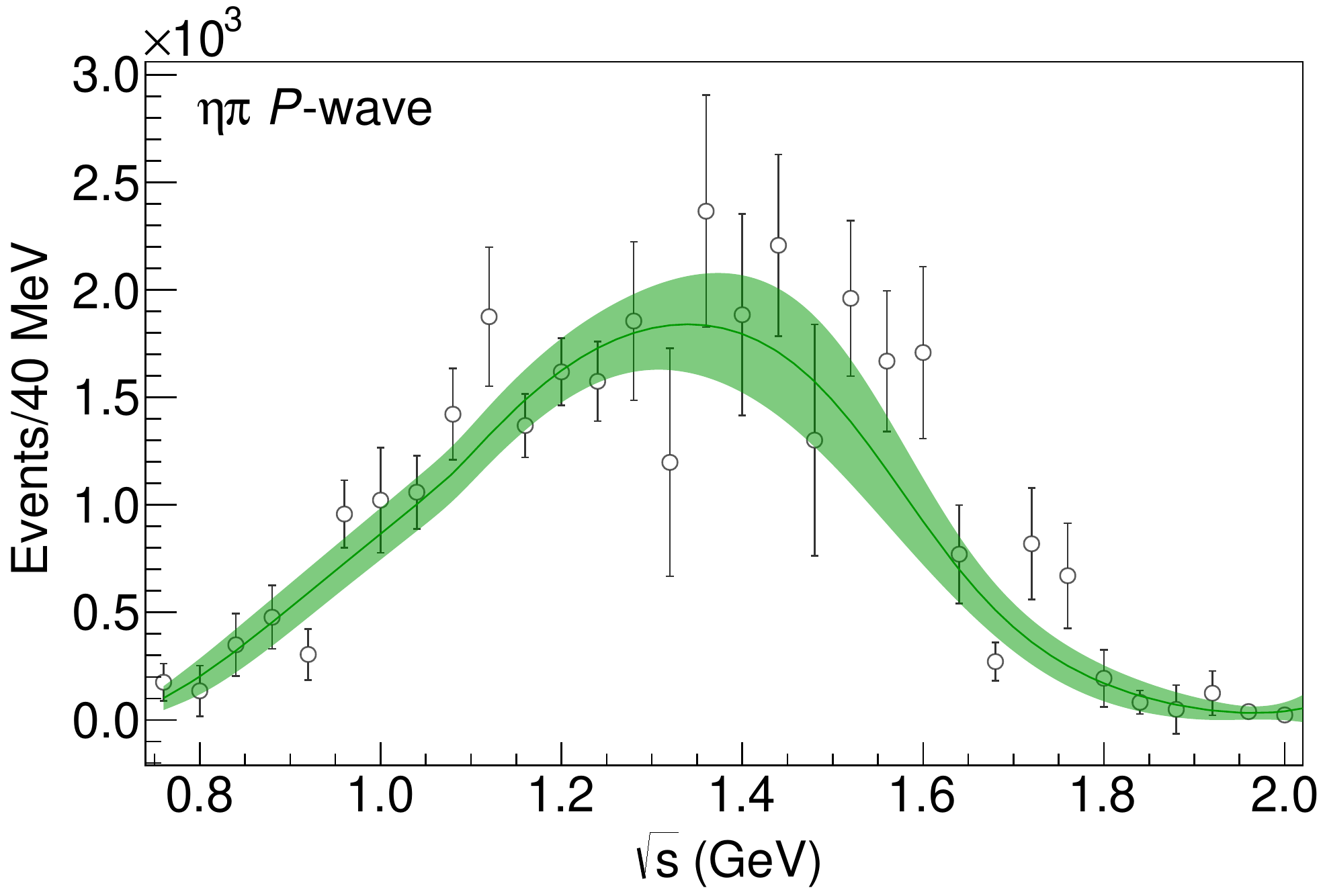}}
~~
\subfigure[]{\includegraphics[width=0.3\textwidth]{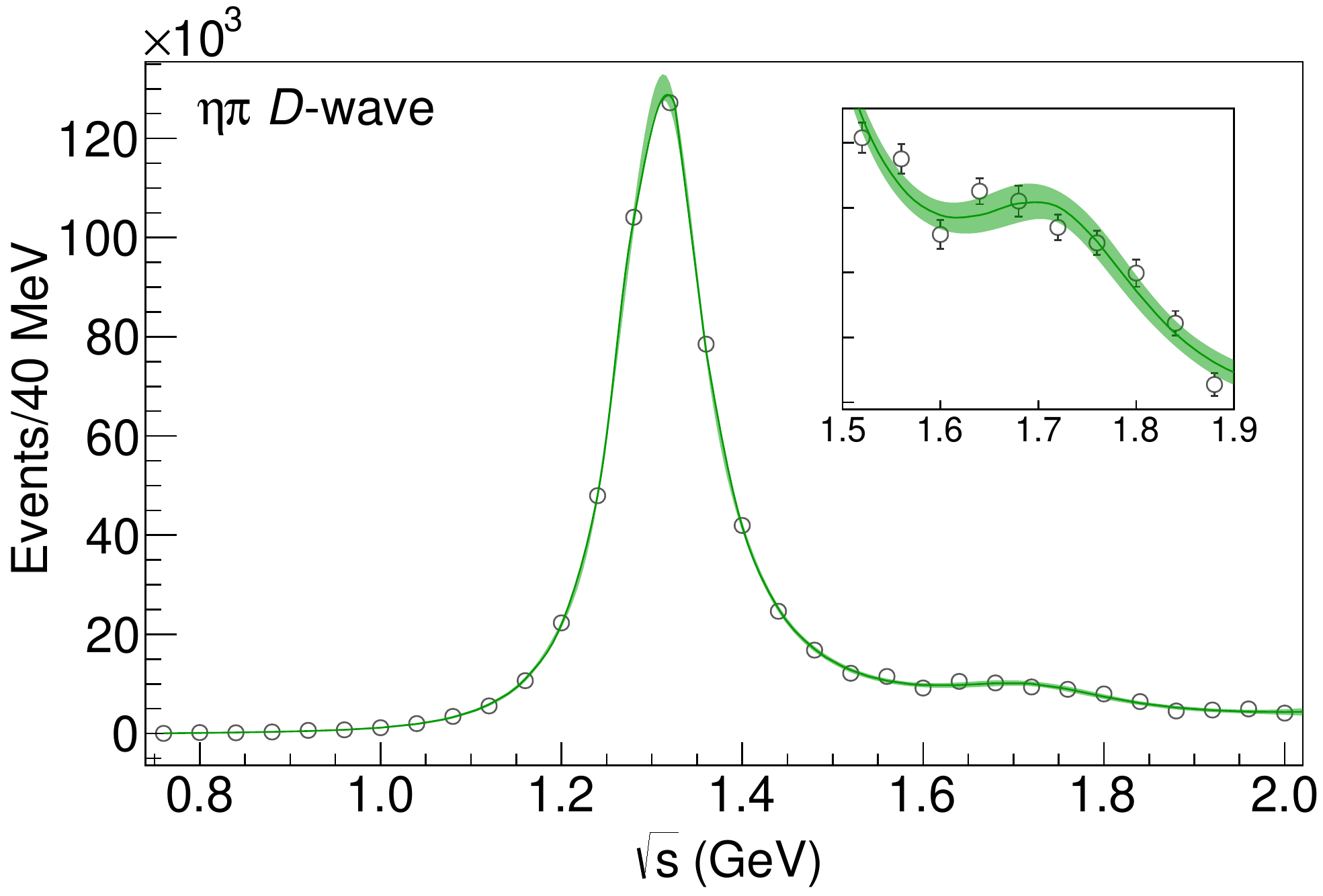}}
~~
\subfigure[]{\includegraphics[width=0.3\textwidth]{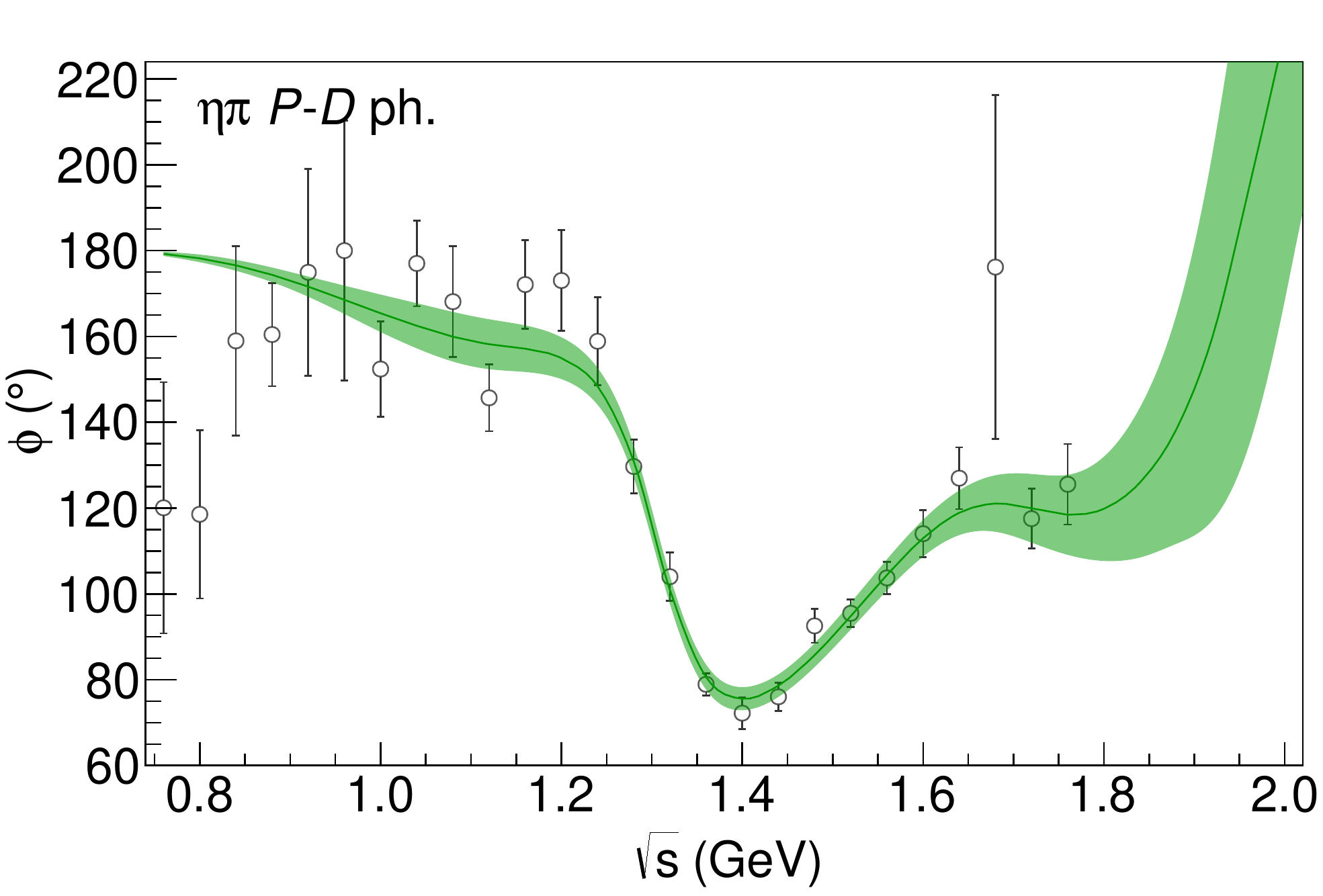}}
\\
\subfigure[]{\includegraphics[width=0.3\textwidth]{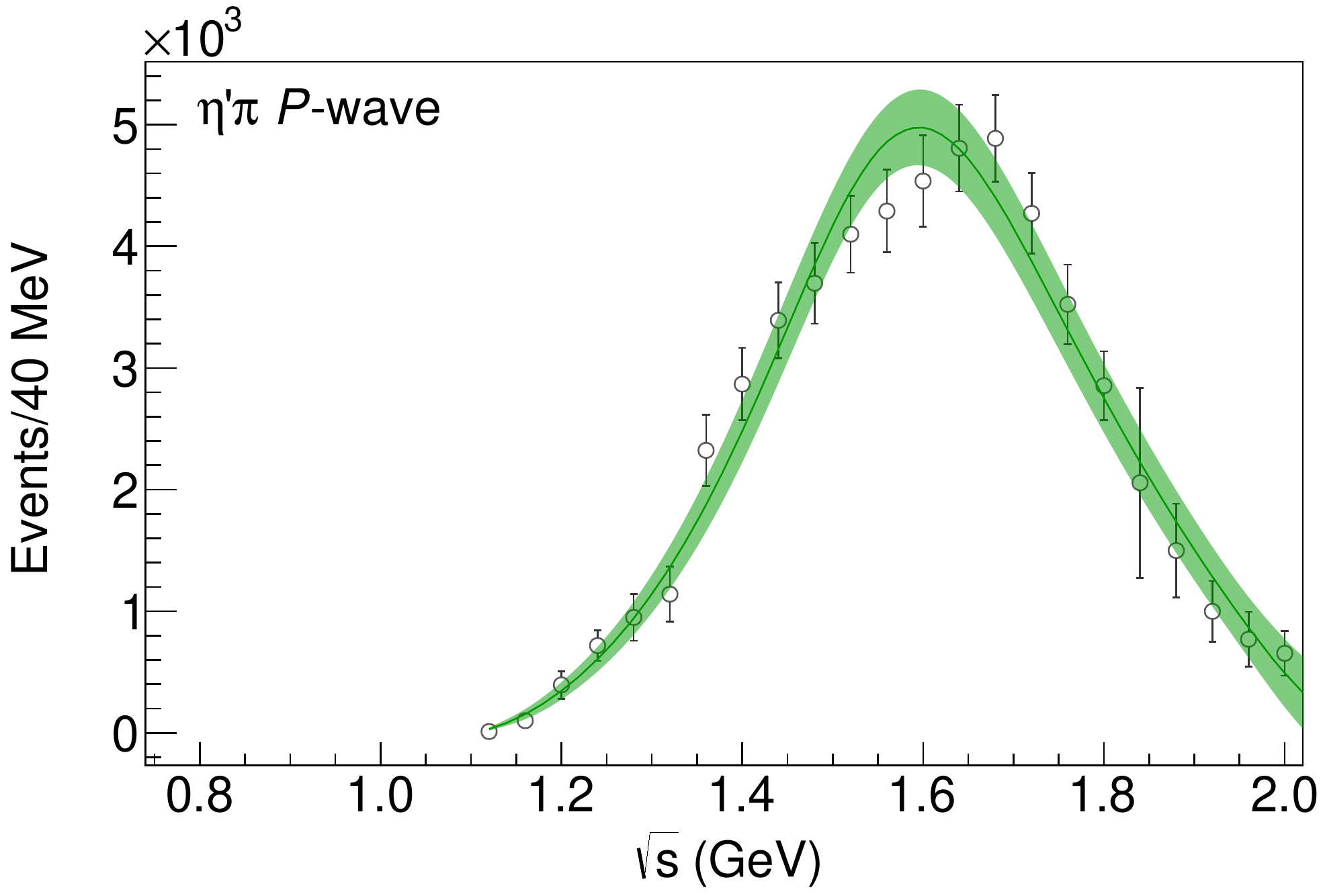}}
~~
\subfigure[]{\includegraphics[width=0.3\textwidth]{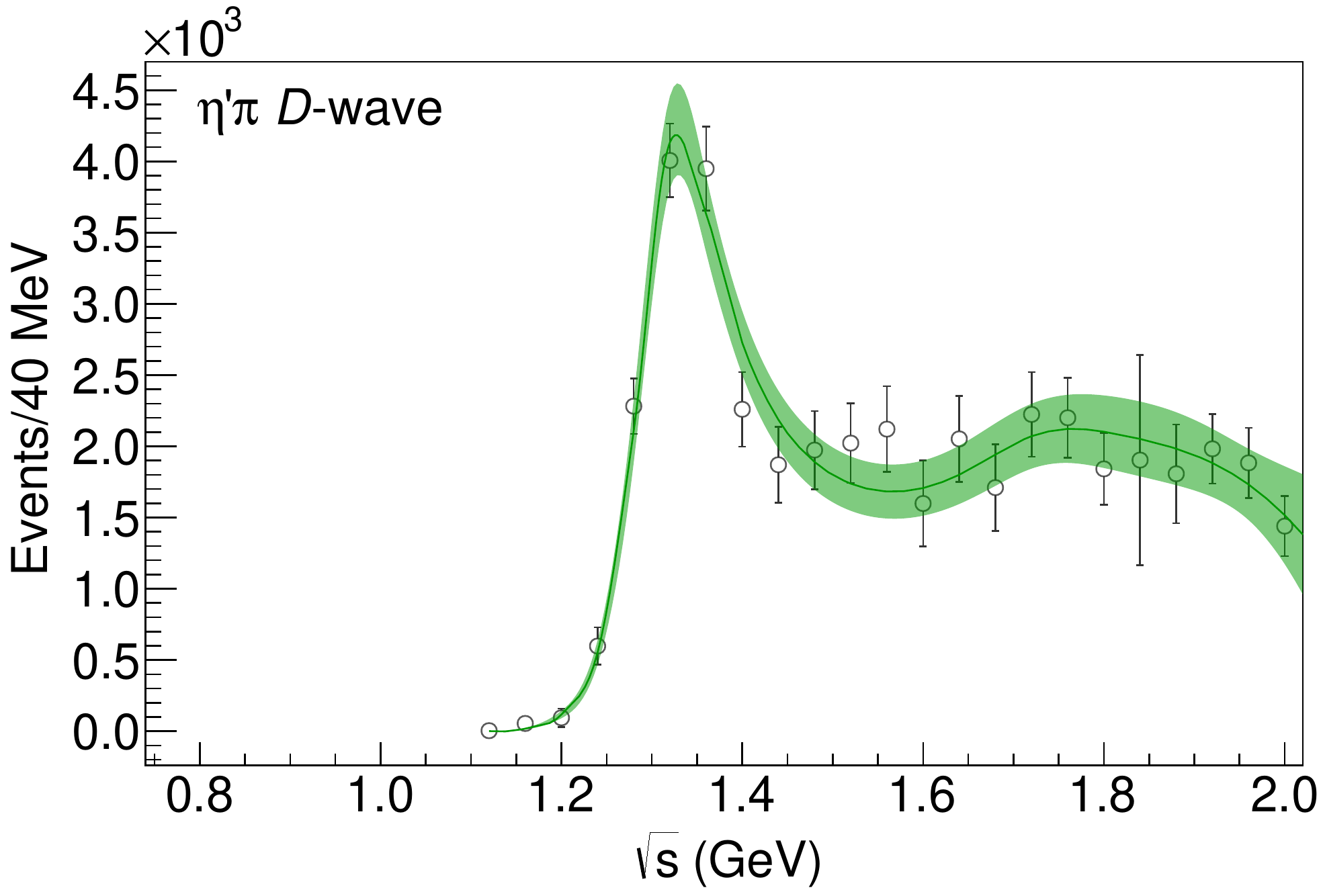}}
~~
\subfigure[]{\includegraphics[width=0.3\textwidth]{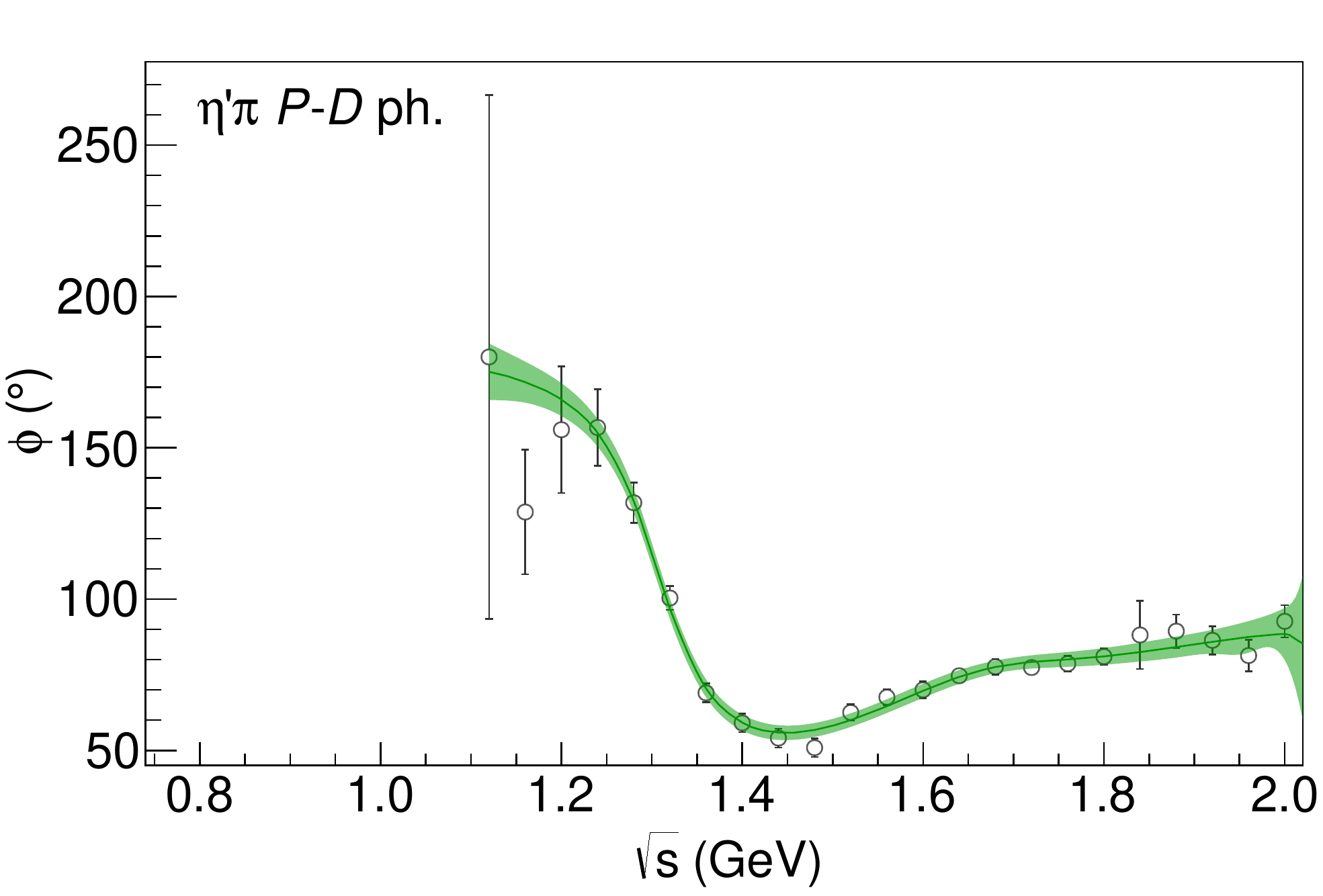}}
\end{center}
\caption{Fits to the COMPASS (a,b,c) $\eta \pi$ and (d,e,f) $\eta^\prime \pi$ data~\cite{COMPASS:2014vkj}. Source: Ref.~\cite{JPAC:2018zyd}.}
\label{fig:pi1}
\end{figure}

Still in 2018 the JPAC collaboration studied the $\eta \pi$ and $\eta^\prime \pi$ data collected by the COMPASS collaboration~\cite{COMPASS:2014vkj}, and fitted them with a coupled-channel amplitude that enforces the unitarity and analyticity of the $S$-matrix~\cite{JPAC:2018zyd}. Given that the $\pi_1(1400)$ and $\pi_1(1600)$ couple separately to the $\eta\pi$ and $\eta^\prime \pi$ channels~\cite{pdg}, their results suggest that there exists a single exotic $\pi_1$ resonant pole coupling to both $\eta \pi$ and $\eta^\prime \pi$, as depicted in Fig.~\ref{fig:pi1}. There is no evidence for a second exotic state. The mass and width of this single pole were determined to be $1564\pm24\pm86$~MeV and $492\pm54\pm102$~MeV respectively, consistent with the above COMPASS experiment.

There have been a lot of theoretical studies on the single-gluon hybrid mesons in the past fifty years, based on various methods and models that have also been applied to study the glueballs, such as the MIT bag model~\cite{Barnes:1977hg,Hasenfratz:1980jv,Chanowitz:1983ci,Karl:1999wq}, flux-tube model~\cite{Isgur:1983wj,Merlin:1986tz,Close:1994hc,Page:1996rj,Page:1998gz,Isgur:1999kx,Capstick:1999qq,Burns:2006wz}, constituent gluon model~\cite{Horn:1977rq,LeYaouanc:1984gh,Iddir:1988jd,Kalashnikova:1993xb,Swanson:1998kx,Szczepaniak:2001rg,Iddir:2002rf,Guo:2007sm}, AdS/QCD model~\cite{Andreev:2012hw,Andreev:2012mc,Bellantuono:2014lra}, lattice QCD~\cite{Griffiths:1983ah,Michael:1985ne,Perantonis:1990dy,Foster:1998wu,Juge:2002br,Bali:2003jq}, QCD sum rules~\cite{Frere:1988ac,Kisslinger:1995yw,Chen:2010ic,Huang:2010dc,Huang:2016upt,Li:2021fwk}, and others~\cite{Giles:1976hh,Buchmuller:1979gy,Burden:1996nh,Hilger:2015hka,Berwein:2015vca,Eshraim:2020ucw}.

\begin{figure}[hbtp]
\begin{center}
\includegraphics[width=1\textwidth]{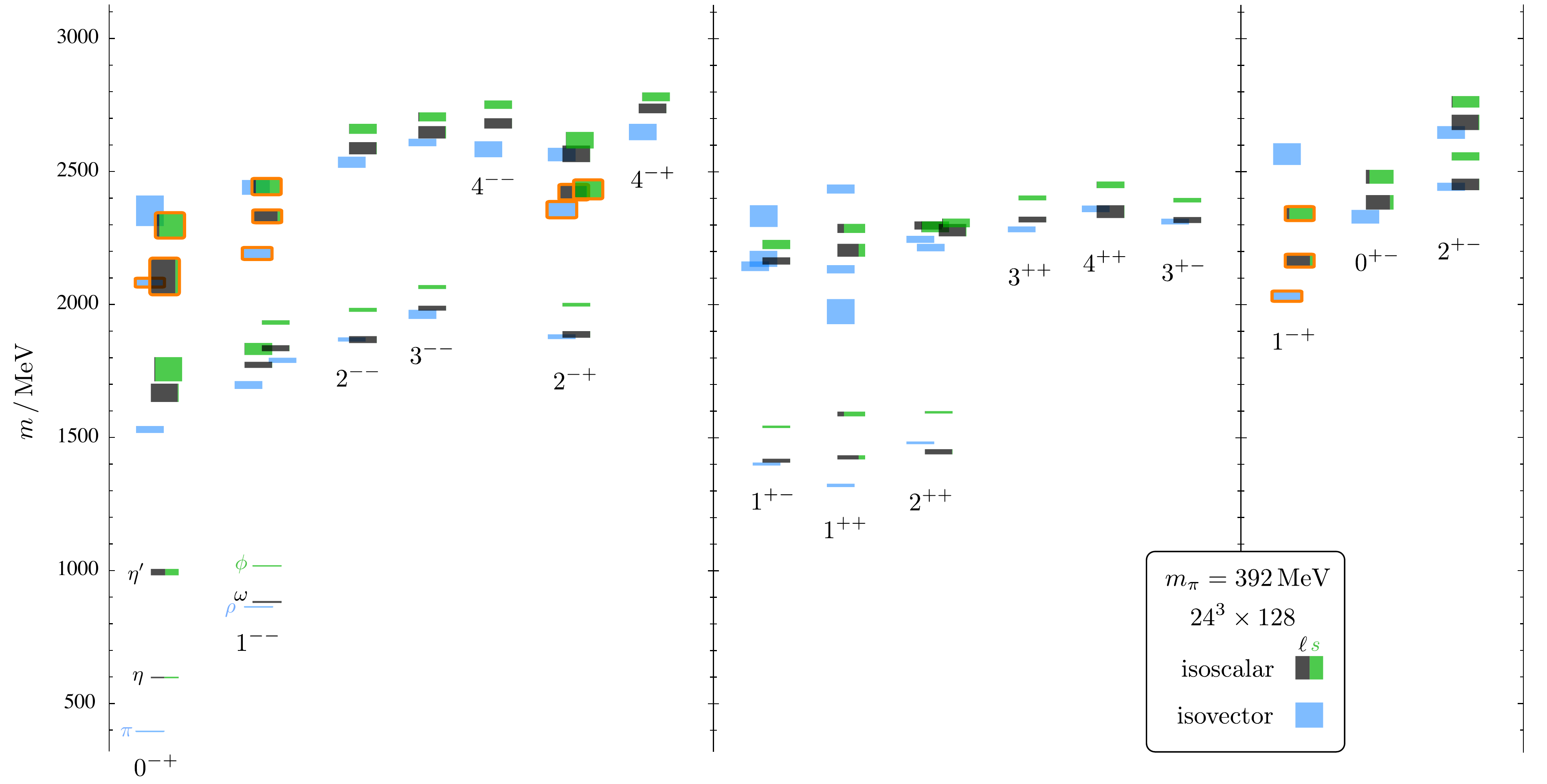}
\end{center}
\caption{Isoscalar (green/black) and isovector (blue) light meson spectrum on the $m_\pi = 392$~MeV, $24^3 \times 128$~lattice. Source: Ref.~\cite{Dudek:2013yja}.}
\label{fig:hybridlattice}
\end{figure}

\begin{figure}[hbtp]
\begin{center}
\includegraphics[width=0.6\textwidth]{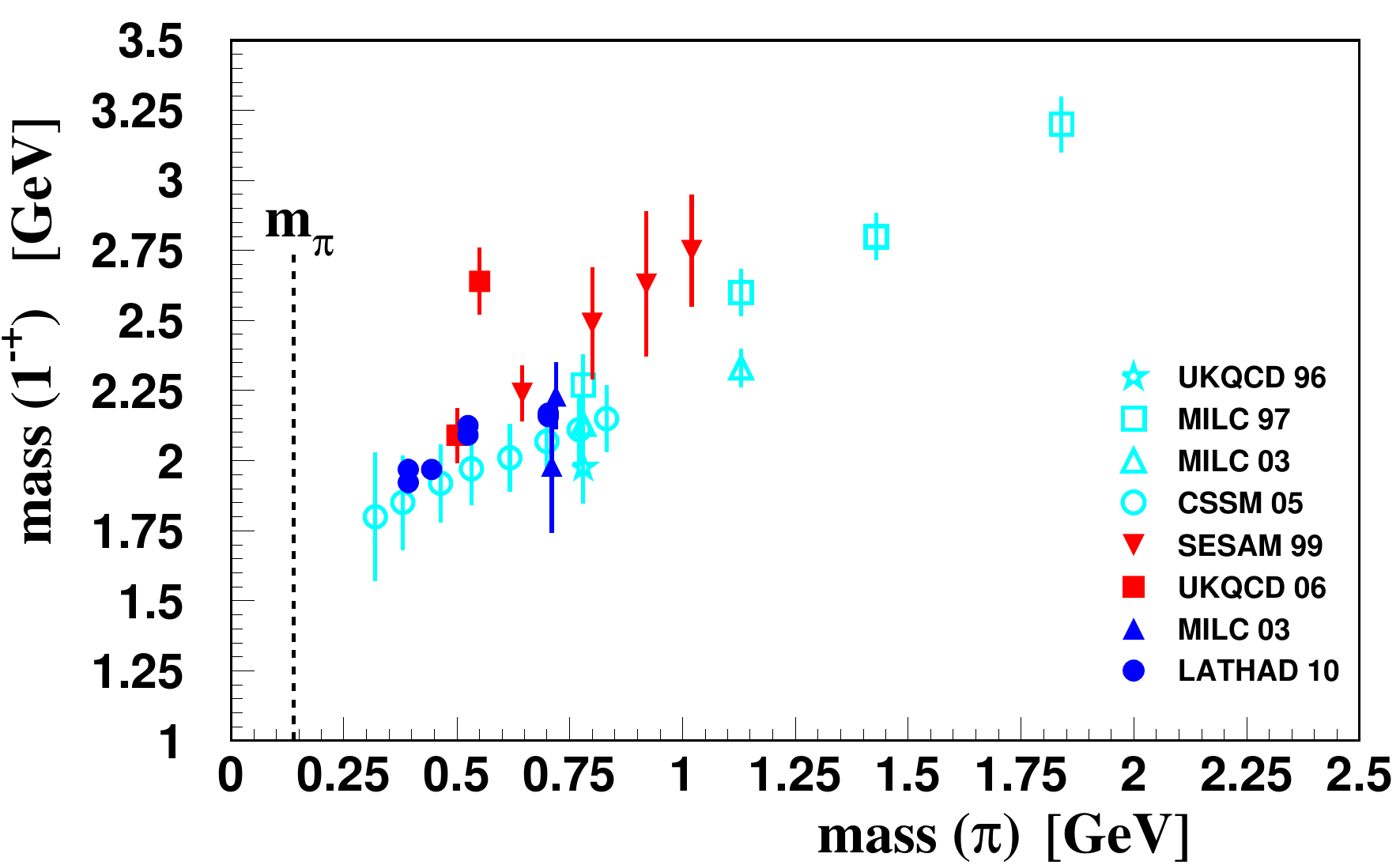}
\end{center}
\caption{The mass of the $J^{PC} = 1^{-+}$ hybrid meson as a function of the pion mass from various lattice QCD calculations. Source: Ref.~\cite{Meyer:2015eta}.}
\label{fig:latticepion}
\end{figure}

Lattice QCD calculations may provide the most accurate estimate of the masses of the hybrid mesons. In recent years the Hadron Spectrum collaboration performed exhaustive analyses on the light meson spectrum~\cite{Dudek:2009qf,Dudek:2010wm,Dudek:2011bn,HadronSpectrum:2012gic,Dudek:2013yja,Woss:2020ayi,Briceno:2017max}. We show their results in Fig.~\ref{fig:hybridlattice} where the masses of the isovector and isoscalar mesons were calculated using the pion mass $m_\pi = 392$~MeV. The lattice QCD calculations at the physical pion mass are not yet warranted, and it is useful to study the $m_\pi$ dependence. As shown in Fig.~\ref{fig:latticepion}, the authors of Ref.~\cite{Meyer:2015eta} summarized the mass of the $J^{PC} = 1^{-+}$ hybrid meson from various lattice QCD calculations, and its naive extrapolation to the physical pion mass turns out to be approximately 1.6~GeV.

\begin{figure}[hbtp]
\begin{center}
\includegraphics[width=1\textwidth]{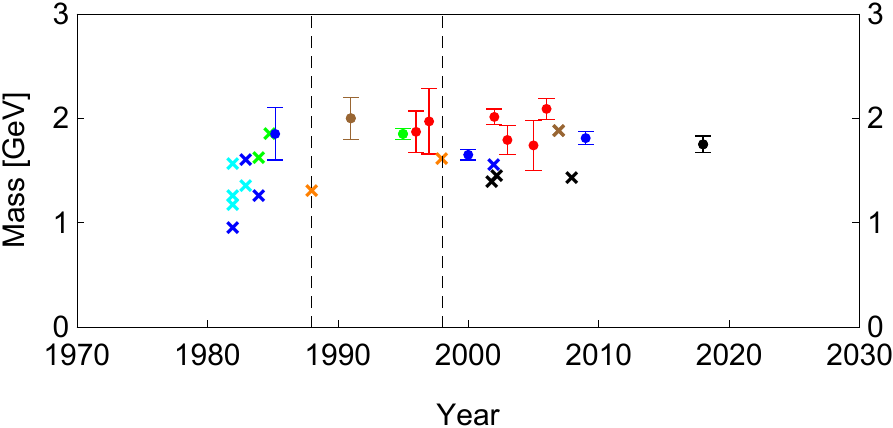}
\end{center}
\caption{Theoretical predictions of the $J^{PC} = 1^{-+}$ hybrid meson mass with uncertainties (error bars) and without uncertainties (crosses), calculated using the MIT bag model~\cite{Barnes:1982zs,Chanowitz:1982qj,Barnes:1982tx,Flensburg:1983ab} (cyan), constituent gluon model~\cite{Ishida:1991mx,Iddir:2007dq} (brown), flux-tube model~\cite{Isgur:1984bm,Isgur:1985vy,Barnes:1995hc} (green), lattice QCD~\cite{Lacock:1996ny,MILC:1997usn,Mei:2002ip,Bernard:2003jd,Hedditch:2005zf,McNeile:2006bz} (red), QCD sum rules~\cite{Balitsky:1982ps,Govaerts:1983ka,Govaerts:1984bk,Latorre:1985tg,Chetyrkin:2000tj,Jin:2002rw,Narison:2009vj} (blue), and others~\cite{Burden:2002ps,Kim:2008qh,Xu:2018cor} (black). See Fig.~\ref{fig:latticepion} for the dependence of lattice QCD calculations on the pion mass. The two dashed lines with orange crosses denote the $\pi_1(1400)$ first observed in 1988~\cite{IHEP-Brussels-LosAlamos-AnnecyLAPP:1988iqi} and $\pi_1(1600)$ first observed in 1998~\cite{E852:1998mbq}, whose masses were measured to be $1354 \pm 25$~MeV and $1660^{+15}_{-11}$~MeV, respectively~\cite{pdg}.}
\label{fig:hybrid}
\end{figure}

We collect as many theoretical predictions of the $J^{PC} = 1^{-+}$ hybrid meson mass as we can, and summarize them in Fig.~\ref{fig:hybrid}. Excluding the lattice QCD calculations, the average value of the mass predictions obtained after the year 1990 is
\begin{equation}
M_{| {\bar u u g/\bar d d g} ; 1^{-+} \rangle} \sim 1700{\rm~MeV} \, .
\end{equation}
This mass value is more or less consistent with the above COMPASS and JPAC results as well as the extrapolated lattice QCD result, suggesting the $\pi_1(1600)$ to be a very good candidate for the low-lying hybrid meson.

\subsubsection{$\eta_1(1855)$ of $I^GJ^{PC} = 0^+1^{-+}$.}
\label{sec6.2.2}

Very recently, the BESIII collaboration performed a partial wave analysis of the $J/\psi \to \gamma \eta \eta^\prime$ decay process, and reported the first observation of a nonconventional state with the exotic quantum number $I^GJ^{PC} = 0^+1^{-+}$~\cite{BESIII:2022riz,BESIII:2022qzu}. This state, labeled as $\eta_1(1855)$, was observed in the $\eta \eta^\prime$ invariant mass spectrum with a statistical significance larger than $19\sigma$. Its mass and width were measured to be
\begin{eqnarray}
\eta_1(1855) &:& M = 1855 \pm 9 ^{+6}_{-1} {\rm~MeV}/c^2 \, ,
\\ \nonumber && \Gamma = 188 \pm 18 ^{+3}_{-8} {\rm~MeV} \, .
\end{eqnarray}

\begin{figure*}[hbtp]
\begin{center}
\subfigure[]{\includegraphics[width=0.3\textwidth]{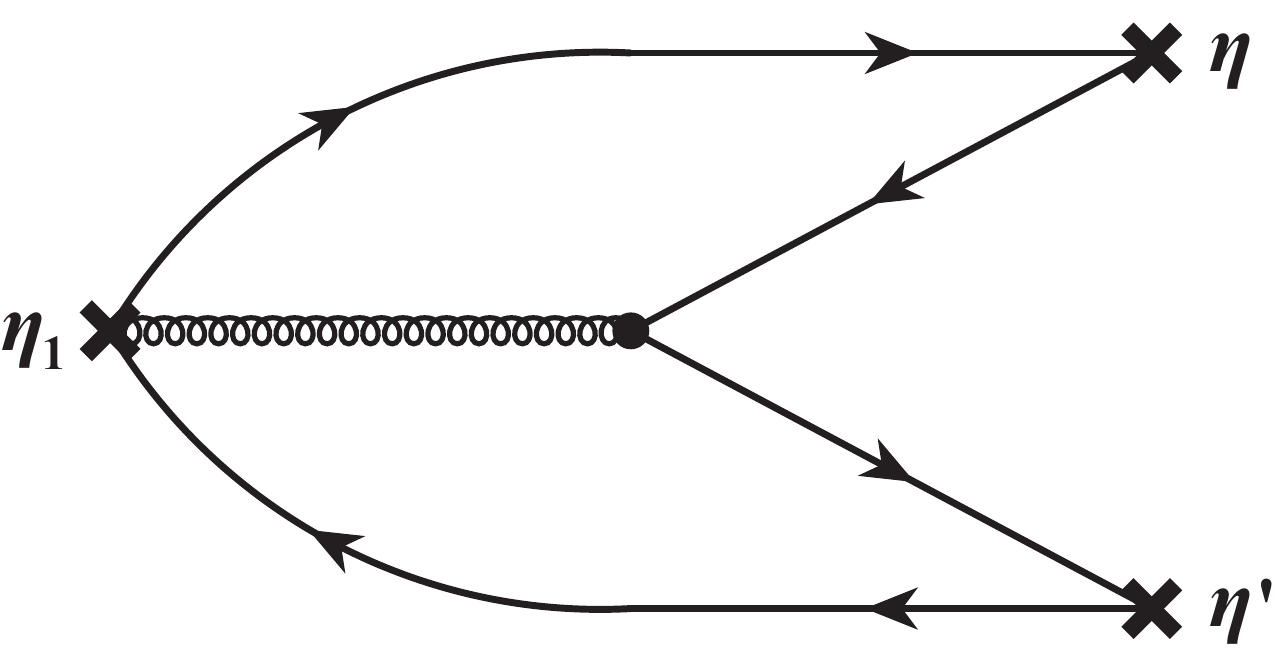}}
~~
\subfigure[]{\includegraphics[width=0.3\textwidth]{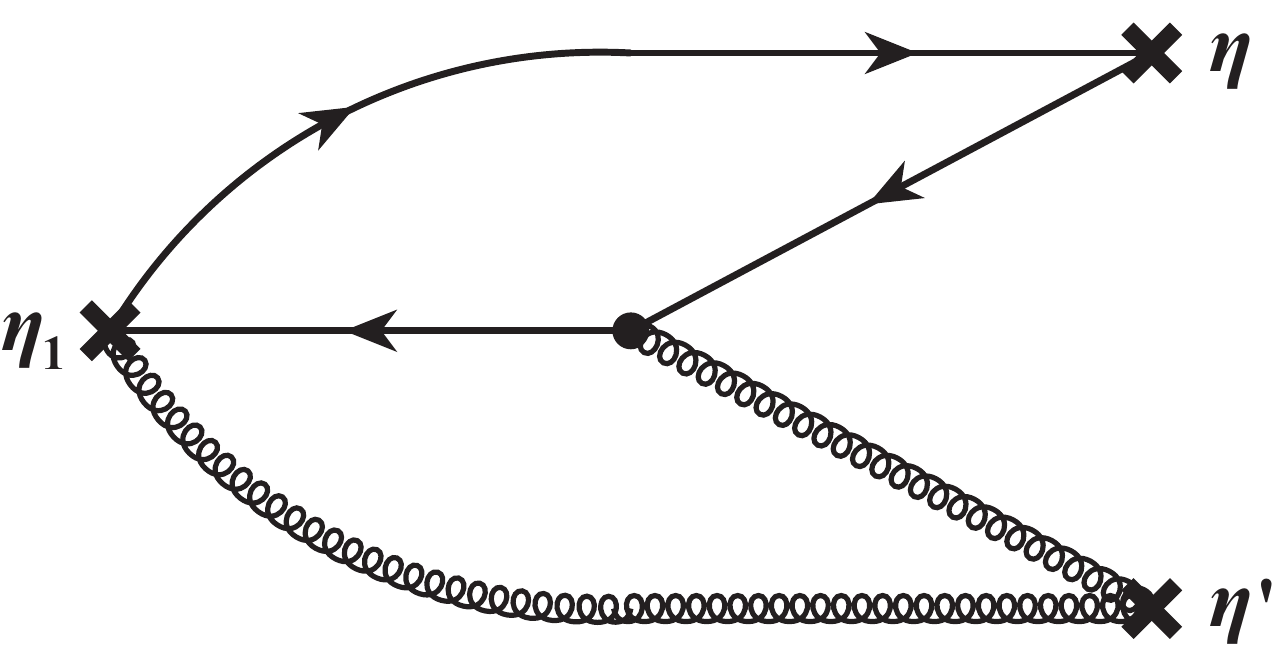}}
~~
\subfigure[]{\includegraphics[width=0.3\textwidth]{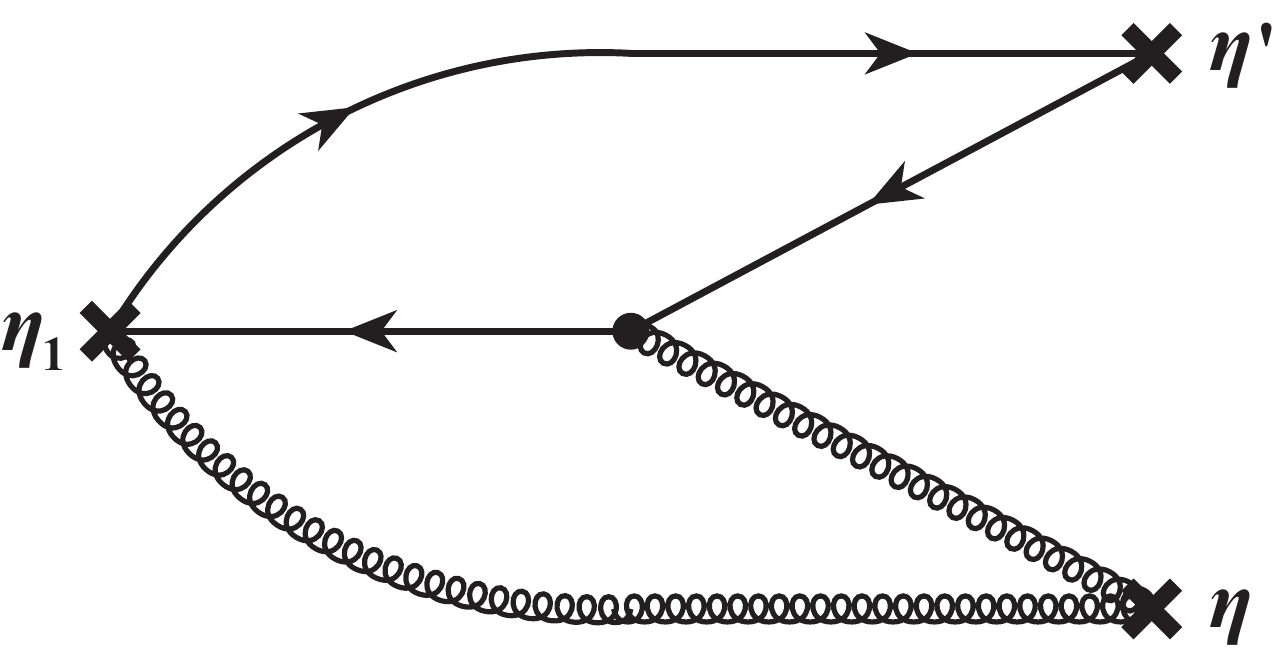}}
\end{center}
\caption{Possible decay mechanisms of the single-gluon hybrid meson through (a) the normal process with one quark-antiquark pair excited from the valence gluon as well as the abnormal processes with the (b) $\eta^\prime$ and (c) $\eta$ mesons produced by the QCD axial anomaly.}
\label{fig:anomaly}
\end{figure*}

There are many possible explanations for the $\eta_1(1855)$, one of which is the exotic hybrid meson. In 2010 the authors of Refs.~\cite{Chen:2010ic,Huang:2010dc} systematically studied the decay properties of the $\bar q q g$ ($q=u/d$) hybrid meson with $I^GJ^{PC} = 0^+1^{-+}$ using the methods of QCD sum rules and light-cone sum rules. They pointed out its $\eta \eta^\prime$ decay channel, which is just the discovery channel of the $\eta_1(1855)$. However, the partial width of this decay mode was estimated to be quite small, where the authors only considered the normal decay process depicted in Fig.~\ref{fig:anomaly}(a) with one quark-antiquark pair excited from the valence gluon.

Actually, there exists a long puzzle in the search of hybrid mesons. The two isovector states $\pi_1(1400)$ and $\pi_1(1600)$ were observed in the $\eta \pi$ and $\eta^\prime \pi$ channels, respectively. However, a ``selection rule'' was proposed in Refs.~\cite{Page:1996rj,Page:1998gz} through the flux tube model that the $J^{PC} = 1^{-+}$ hybrid meson does not decay into two ground-state mesons, {\it i.e.}, the $\eta \pi$ and $\eta^\prime \pi$ decay modes are strictly forbidden in the flux tube model. This is in strong contrast to the experimental fact that these two modes and the $\eta\eta^\prime$ mode are the discovery modes of the $\pi_1(1400)$, $\pi_1(1600)$, and $\eta_1(1885)$, respectively. This selection rule can be explained using the Feynman diagram depicted in Fig.~\ref{fig:anomaly}(a) to some extent, where the leading-order perturbative contribution vanishes in the chiral limit.

In a recent study~\cite{Chen:2022qpd} the authors further applied the QCD sum rule method to investigate the abnormal decay processes depicted in Fig.~\ref{fig:anomaly}(b,c) with the $\eta$ and $\eta^\prime$ mesons produced by the QCD axial anomaly. They found that the QCD axial anomaly significantly enhances the decay width of the $\eta \eta^\prime$ mode, and the obtained results support the interpretation of the $\eta_1(1855)$ as the $\bar s s g$ hybrid meson of $I^GJ^{PC}=0^+1^{-+}$. Especially, the QCD axial anomaly ensures the $\eta \eta^\prime$ decay mode to be a characteristic signal of the hybrid nature of the $\eta_1(1855)$. We refer to Refs.~\cite{Frere:1988ac,Chen:2022isv} for more QCD sum rule and lattice QCD studies. This mechanism was also investigated in Ref.~\cite{Eshraim:2020ucw}, where the authors applied the extended linear sigma model to derive the non-zero $\eta_1 \to \eta \eta^\prime$ decay mode, from a chirally symmetric interaction term that breaks explicitly the axial anomaly.

\begin{figure*}[hbtp]
\begin{center}
\subfigure[]{\includegraphics[width=0.45\textwidth]{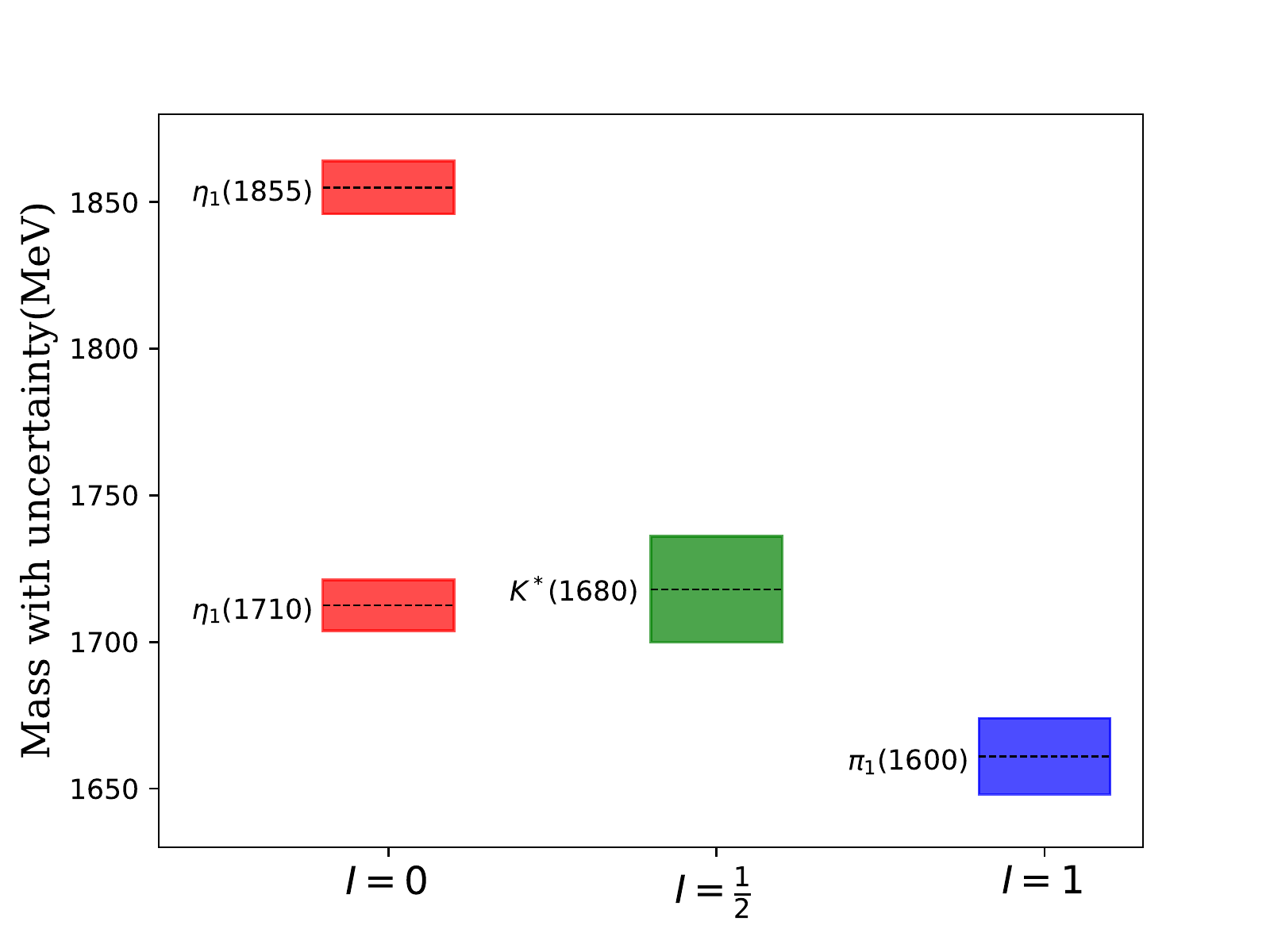}}
~~
\subfigure[]{\includegraphics[width=0.45\textwidth]{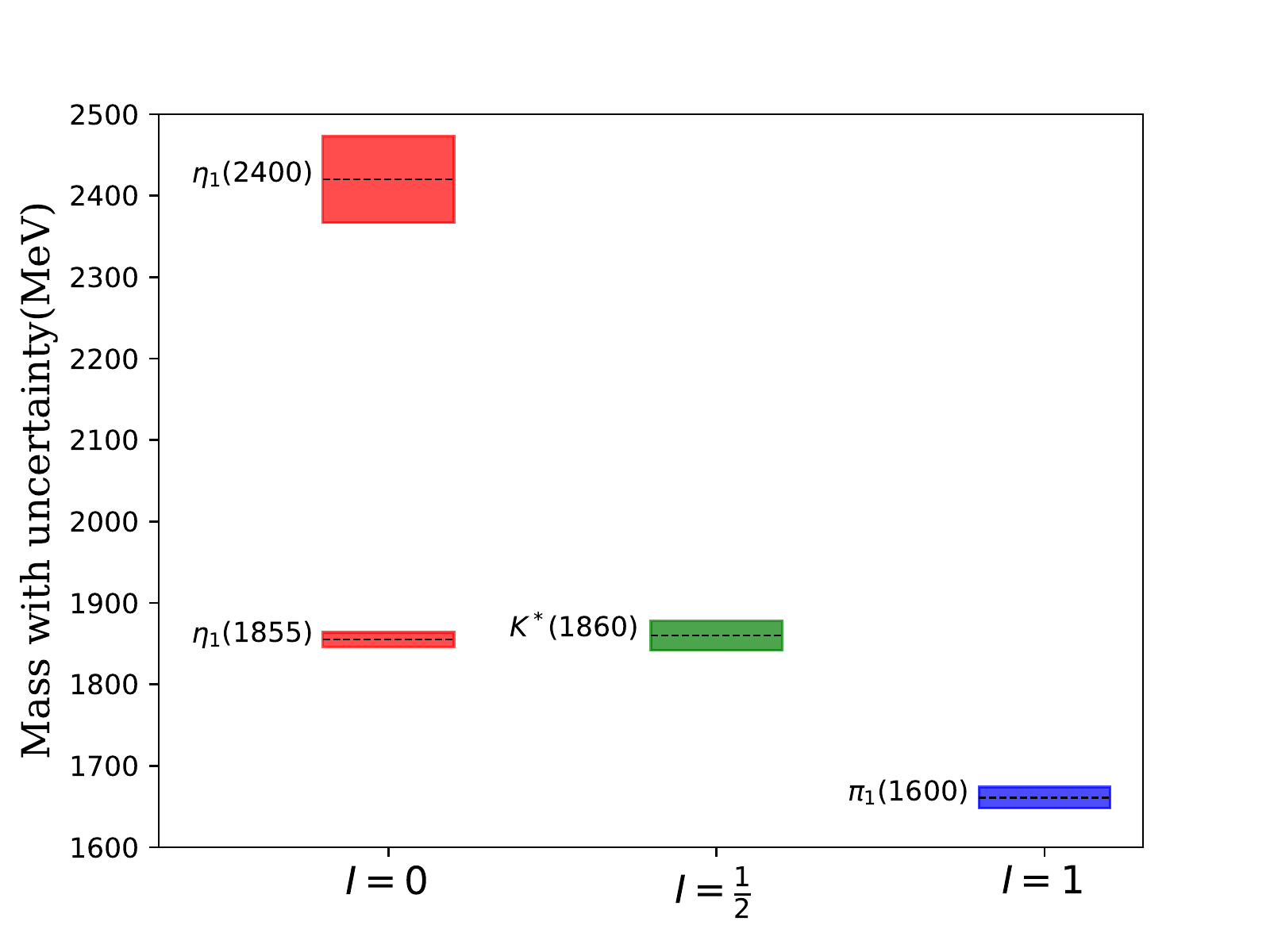}}
\end{center}
\caption{The $J^{P(C)} = 1^{-(+)}$ hybrid nonet with the $\eta_1(1855)$ assigned as either the (a) higher or (b) lower state. Source: Ref.~\cite{Qiu:2022ktc}.}
\label{fig:hybridnonet}
\end{figure*}

Based on the interpretation of the $\eta_1(1855)$ as the $I^GJ^{PC} = 0^+1^{-+}$ hybrid meson, the authors of Ref.~\cite{Qiu:2022ktc} investigated its flavor $SU(3)$ partners using a parametrization method based on the flux tube picture. They examined two schemes for the $J^{P(C)} = 1^{-(+)}$ hybrid nonet by assigning the $\eta_1(1855)$ to be either the higher or lower state, as shown in Fig.~\ref{fig:hybridnonet}. Possible channels for detecting the missing states were suggested. In a recent study~\cite{Shastry:2022mhk}, the authors studied the decays of this hybrid nonet using a Lagrangian invariant under the flavor symmetry, parity reversal, and charge conjugation. Their results suggest that the light isoscalar is significantly narrow, while the width of the heavy isoscalar can be matched to the recently observed $\eta_1(1855)$.

Besides the hybrid assignment, there exist some other possible explanations. In Ref.~\cite{Dong:2022cuw} the authors applied the one-meson-exchange model to interpret the $\eta_1(1855)$ as the $K \bar K_1(1400)$ hadronic molecule. They found that the attractive force between $K\bar K_1(1400)$ of $J^{PC}=1^{-+}$ is strong enough to form a bound state with the binding energy around 40~MeV, which was used to explain the $\eta_1(1855)$. They adopted a momentum cutoff larger than 2 GeV while the cutoff parameter is around 1~GeV for the deuteron and $T_{cc}$ within the same meson exchange framework. They further predicted the existence of its $J^{PC}=1^{--}$ partner with the binding energy around $10\sim30$~MeV. Their possible decay modes were extracted, as shown in Table~\ref{sec6:hybridwidth}. This assignment was supported by Ref.~\cite{Yang:2022lwq}, where the authors studied the radiative and strong decays of the $S$-wave $K \bar K_1(1400)$ molecular state within the effective Lagrangians approach. A similar $K \bar K_1(1270)$ molecular state was proposed in Ref.~\cite{Zhang:2019ykd} by investigating the $\eta \bar K K^*$ system through the Faddeev equation.

\begin{table}[htbp]
\centering
\renewcommand{\arraystretch}{1.2}
\scriptsize
\caption{Partial decay widths of the $K\bar K_1(1400)$ molecular states with $I^GJ^{PC}=0^-1^{--}$ and $0^+1^{-+}$, in unit of MeV. Source: Ref.~\cite{Dong:2022cuw}.}
\label{sec6:hybridwidth}
\begin{tabular}{c|*{6}{c}}
\hline
\multirow{2}*{Mode} & \multicolumn{2}{c}{Widths ($\mathrm{MeV}$)} \\
\cline{2-3}
& {$1^{--}~E_B=20~\mathrm{MeV}$} & {$1^{-+}~E_B=40~\mathrm{MeV}$} \\
\hline
$K^*\bar{K}^*$         & 38.1  & 26.3\\
$K \bar{K}$            & 0.5   & 0\\
$K \bar{K}^*$          & 1.0   & 0.9\\
$a_1 \pi$              & 0     & 9.2\\
$f_1 \eta$             & 0     & 0.2\\
$\eta \eta^{\prime}$   & 0     & 26.9\\
$\sigma \omega$        & 0.2   & 0\\
$\rho \rho$            & 0     & 0.04\\
$\pi \rho$             & 6.4   & 0\\
$\eta \omega$          & 0.4   & 0\\
$\omega \omega$        & 0     & 0.01\\
$\omega\phi$           & 0     & 0.4\\
$K \bar{K}^* \pi$      & 130.0 & 105.0 \\
\hline
2-body & 46.5  & 64.0\\
Total& 176.5  & 169.0\\
\hline
\end{tabular}
\end{table}

\begin{figure*}[hbtp]
\begin{center}
\includegraphics[width=0.6\textwidth]{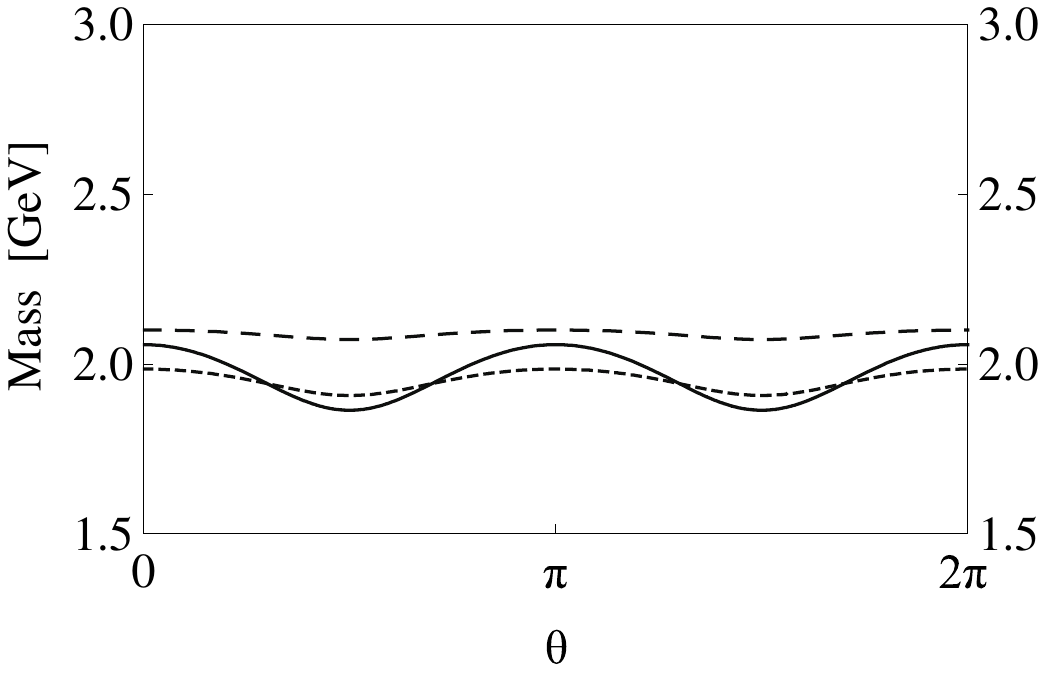}
\end{center}
\caption{Mass of the $qs\bar q \bar s$ ($q=u/d$) tetraquark state with $I^GJ^{PC} = 0^+1^{-+}$ as a function of the mixing parameter $\theta$. Source: Ref.~\cite{Chen:2008ne}.}
\label{fig:eta1mix}
\end{figure*}

In 2008 the QCD sum rule method was applied to study the $qs\bar q \bar s$ ($q=u/d$) tetraquark state of $I^GJ^{PC} = 0^+1^{-+}$ in Refs.~\cite{Chen:2008ne,Chen:2008qw}. As shown in Fig.~\ref{fig:eta1mix}, the authors evaluated its mass to be around $1.8$-$2.1$~GeV. They also estimated its decay width to be around 150~MeV and extracted its possible $\eta \eta^\prime$ decay mode. Accordingly, the $\eta_1(1855)$ may also be interpreted as a $qs\bar q \bar s$ tetraquark state. Its partner states, the $ss\bar s \bar s$ tetraquark states of $I^GJ^{PC} = 0^+1^{-+}$, were studied in Refs.~\cite{Wan:2022xkx,Su:2022eun} through the same QCD sum rule approach.

\subsubsection{Double-gluon hybrids.}
\label{sec6.2.3}

Very recently, the double-gluon hybrid meson consisting of one quark-antiquark pair together with two valence gluons was studied in Ref.~\cite{Chen:2021smz}. This work was motivated by the D0 and TOTEM experiment observing the evidence of a $C$-odd three-gluon glueball~\cite{TOTEM:2020zzr}, given that two-gluon glueballs have been studied intensively but are still difficult to be identified in experiments unambiguously. The authors concentrated on the double-gluon hybrid meson with the exotic quantum number $J^{PC} = 2^{+-}$ that conventional $\bar q q$ mesons can not reach, and applied the QCD sum rule method to evaluate its mass to be $2.26^{+0.11}_{-0.12}$~GeV. This mass value is accessible in the BESIII, Belle-II, GlueX, LHC, and PANDA experiments. Later in Ref.~\cite{Tang:2021zti} the authors studied the $J^{PC} = 1^{--}$ double-gluon charmonium and bottomonium hybrid mesons in terms of QCD sum rules, whose masses were evaluated to be about $5.5$~GeV and $11.5$~GeV, respectively.

To end this subsection, we briefly discuss possible decay patterns of the single and double-gluon hybrid mesons following Ref.~\cite{Chen:2021smz}. These decay patterns are similar to those of the two- and three-gluon glueballs discussed in Sec.~\ref{sec6.1.5}. A pure single-gluon hybrid meson, if it exists, can decay through exciting one quark-antiquark pair from the single gluon, and recombining two color-octet quark-antiquark pairs into two color-singlet mesons. However, it is rather difficult to differentiate it from conventional $\bar q q$ mesons and exotic tetraquark states.

\begin{figure}[hbtp]
\begin{center}
\subfigure[]{\includegraphics[width=0.45\textwidth]{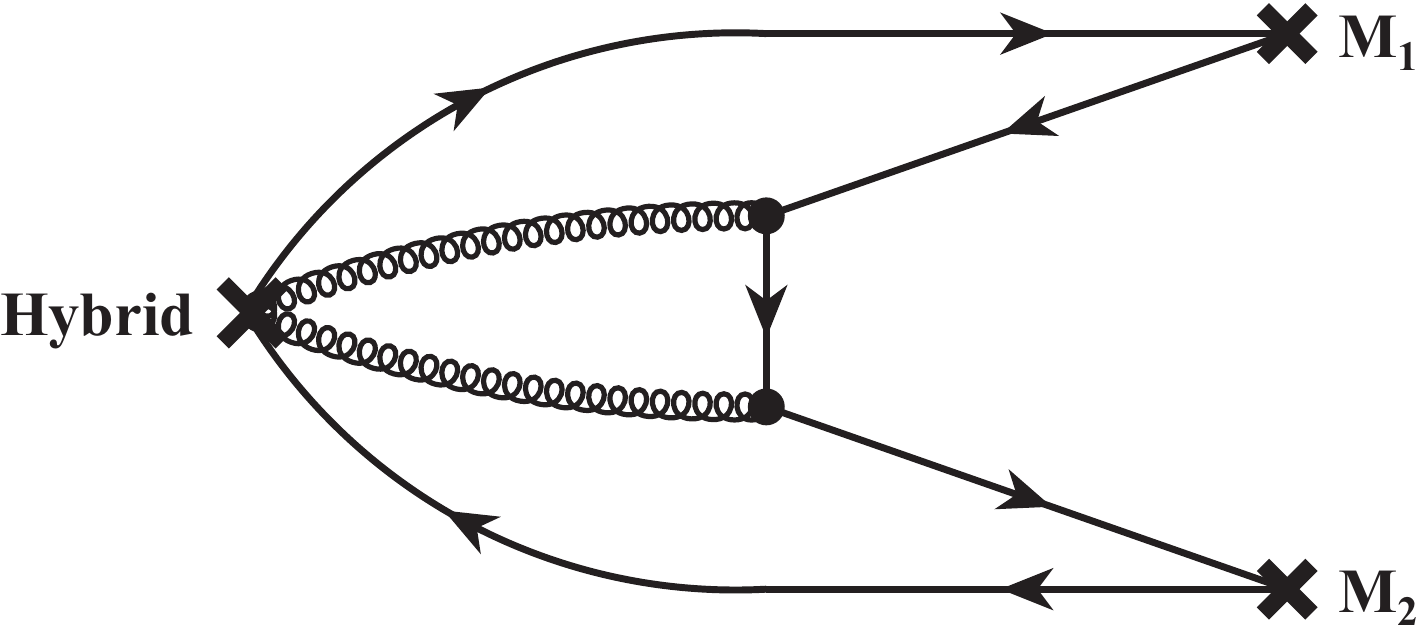}}
~~~~~
\subfigure[]{\includegraphics[width=0.45\textwidth]{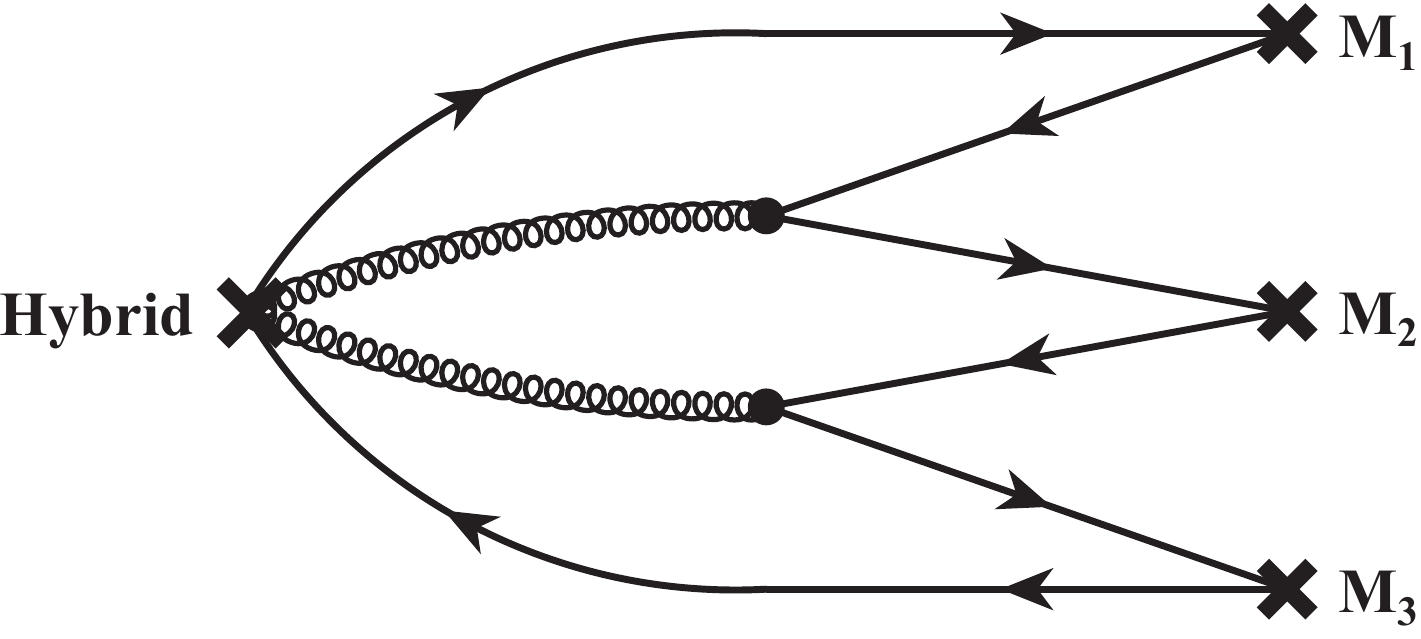}}
\end{center}
\caption{Two possible decay processes of the double-gluon hybrid meson, which decays through exciting two quark-antiquark pairs from two gluons, and recombining three color-octet quark-antiquark pairs into (a) two color-singlet mesons or (b) three color-singlet mesons.}
\label{fig:decayhybrid}
\end{figure}

A pure double-gluon hybrid meson, if it exists, can decay after exciting two quark-antiquark pairs from two gluons, followed by recombining three color-octet quark-antiquark pairs into two color-singlet mesons or three mesons, as depicted in Fig.~\ref{fig:decayhybrid}. The amplitudes of these two possible decay processes are both at the $\mathcal{O}(\alpha_s)$ order, so the three-meson decay patterns are generally not suppressed severely compared to the two-meson decay patterns, or even enhanced due to the quark-antiquark annihilation during the two-meson decay process. Especially, possible decay patterns of the $J^{PC} = 2^{+-}$ double-gluon hybrid meson were explicitly studied in Ref.~\cite{Chen:2021smz}, and their results suggest that the three-meson decay channels $f_1\omega\pi/f_1\rho\pi$ will be useful in identifying its nature, therefore, of particular importance to the direct test of QCD in the low energy sector.

\section{Summary and outlook}
\label{sec7}

The past decades have witnessed the golden era of the hadron physics. Based on our previous reviews~\cite{Chen:2016qju,Chen:2016spr,Liu:2019zoy}, in this paper we have provided an updated review of its experimental and theoretical progresses within the past five years. At the end, we would like to give a brief summary of the current status and our perspective for the future progresses in this active field.

\subsection{Open heavy flavor mesons and baryons}
\label{sec7.1}

The open heavy flavor mesons and baryons have been reviewed in Sec.~\ref{sec2} and Sec.~\ref{sec3}, respectively. Generally speaking, we have already understood them to some extents, from the viewpoint of hadron spectroscopy within the conventional quark model. Especially, the heavy quark symmetry plays an important role, which works quite well for the bottom system and does not work so well for the charm system, so one often needs to properly take into account the mixing effect.

To further study the open heavy flavor mesons and baryons, we would like to emphasize the fine structure observed in the heavy baryon system in recent years. In 2017 the LHCb collaboration observed as many as five excited $\Omega_c$ baryons in the narrow energy region $3.0\sim3.1$~GeV~\cite{LHCb:2017uwr}. Later, LHCb observed four excited $\Omega_b$ baryons and three excited $\Xi_c$ baryons in two other narrow energy regions~\cite{LHCb:2020iby,LHCb:2020tqd}.

These excited heavy baryons are all good candidates of the $P$-wave heavy baryons. Both the constitute quark model and the QCD sum rule method have been applied to systematically study their mass spectrum and decay properties. One can use three quark fields together with one derivative to construct a $P$-wave heavy baryon field with the definite internal structure, and it corresponds to a $P$-wave heavy baryon state with the same internal structure. The results obtained using the baryon field through QCD sum rules and those obtained using the baryon state through the constitute quark model are roughly consistent with each other, although not exactly the same.

There can be five $P$-wave excitations of the $\lambda$-mode, {\it e.g.}, two $\Omega_c(1P)$ baryons of $J^P = 1/2^-$, two $\Omega_c(1P)$ baryons of $J^P = 3/2^-$, and one $\Omega_c(1P)$ baryon of $J^P = 5/2^-$, as shown in Fig.~\ref{fig:Jacobi} and Fig.~\ref{fig:categorization}. It seems that these five $\lambda$-mode excitations can be used to explain the five excited $\Omega_c$ baryons observed by LHCb. However, the recent LHCb measurement~\cite{LHCb:2021ptx} tested several spin hypotheses, and their results disfavor this assignment. Besides, both the constitute quark model and QCD sum rule calculations suggest that one of the $\lambda$-mode excitations with $J^P = 1/2^-$ has a width too broad to be easily observed in experiments~\cite{Cheng:2021qpd}, {\it e.g.} see the constitute quark model calculations of Refs.~\cite{Wang:2017hej,Liang:2020hbo} and the QCD sum rule calculations of Refs.~\cite{Yang:2020zrh,Yang:2021lce}.

A natural question is how to explain the rest of the five excited $\Omega_c$ baryons observed by LHCb, given that at most four of them can be explained as the $\lambda$-mode excitations. There are two possible assignments: either the radial excitation ($2S$) or the $P$-wave excitation of the $\rho$-mode. The experimental measurements on the quantum numbers of these baryons are undoubtedly important and helpful. Besides, it will be helpful to examine the mass gaps existing in the excited heavy baryons, which show some universal behaviors~\cite{Chen:2021eyk,LHCb:2020iby,Arifi:2020yfp}.

Hopefully, these theoretical and experimental studies will significantly improve our understanding on the fine structure observed in the heavy baryon system, which is directly related to the rich internal structure of the excited heavy baryons. Recalling that the development of quantum physics is closely related to the better understanding of the gross, fine, and hyperfine structures of atomic spectra, one naturally expects that the current undergoing studies on the excited heavy baryons will not only improve our understanding on their internal structures, but also enrich our knowledge of the quantum physics.

Besides the fine structure of the excited heavy baryons, there still exist some problems to understand the $P$-wave charmed-strange mesons $D_{s0}^*(2317)$ and $D_{s1}(2460)$, probably related to their nearby $DK$ and $D^*K$ thresholds. A similar situation may exist for the $B_{sJ}(6064)$ and $B_{sJ}(6114)$, which are possible $D$-wave bottom-strange mesons. The experimental measurement of their $J^P$ quantum numbers will help us to understand whether they are candidates of multiquark states or molecular states closely related to the threshold effects.

\subsection{Open and hidden heavy flavor multiquark candidates}
\label{sec7.2}

The open heavy flavor multiquark candidates $X_{0,1}(2900)$ and $T_{cc}^+(3875)$ have been reviewed in Sec.~\ref{sec4}. The hidden heavy flavor multiquark candidates $X(6900)$, $P_{cs}(4459)^0$, $Z_{cs}(3985)^-$, $Z_{cs}(4000)^+$, and $Z_{cs}(4220)^+$ have been reviewed in Sec.~\ref{sec5}.

\begin{figure*}[hbtp]
\begin{center}
\subfigure[]{\includegraphics[width=0.4\textwidth]{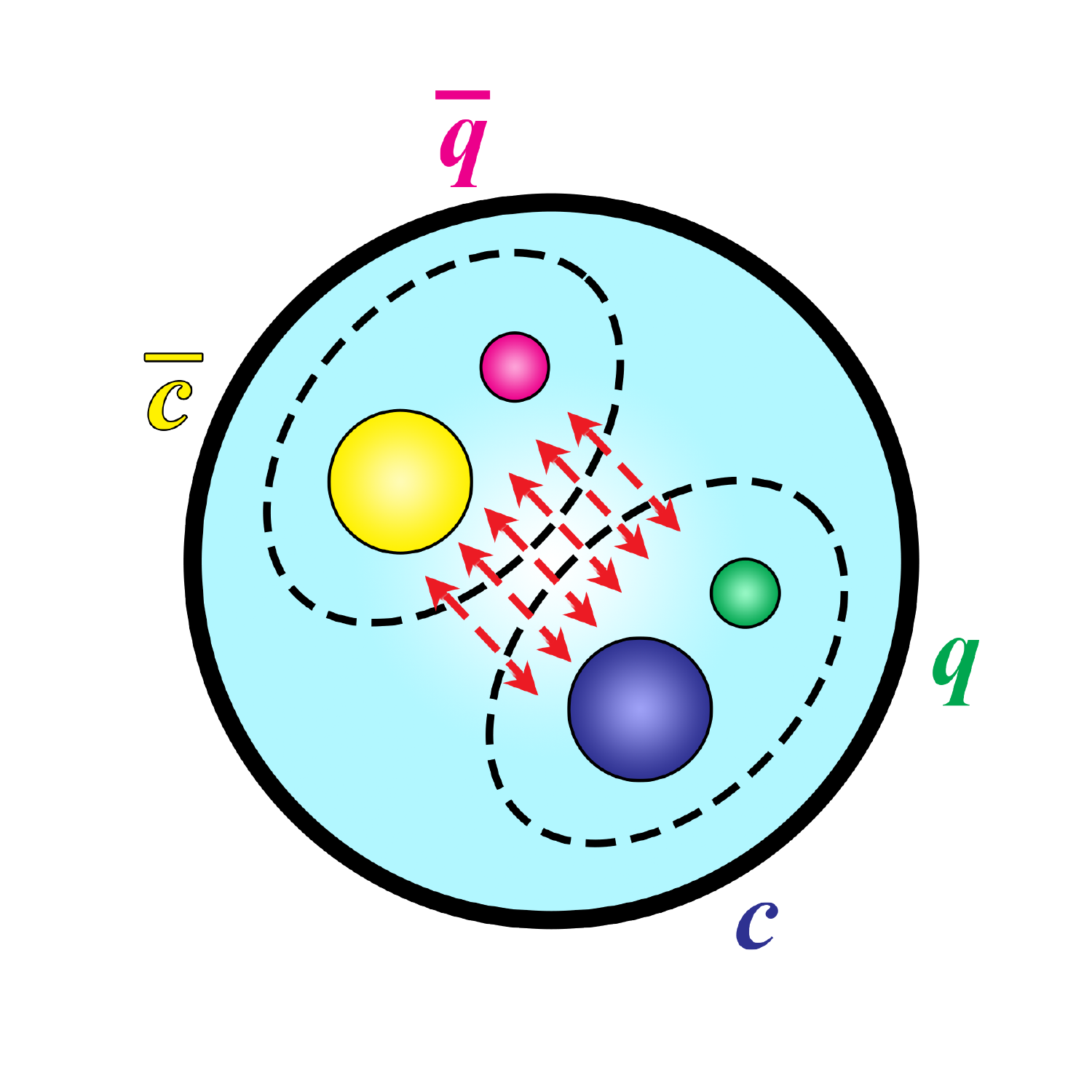}}
~~~~~
\subfigure[]{\includegraphics[width=0.4\textwidth]{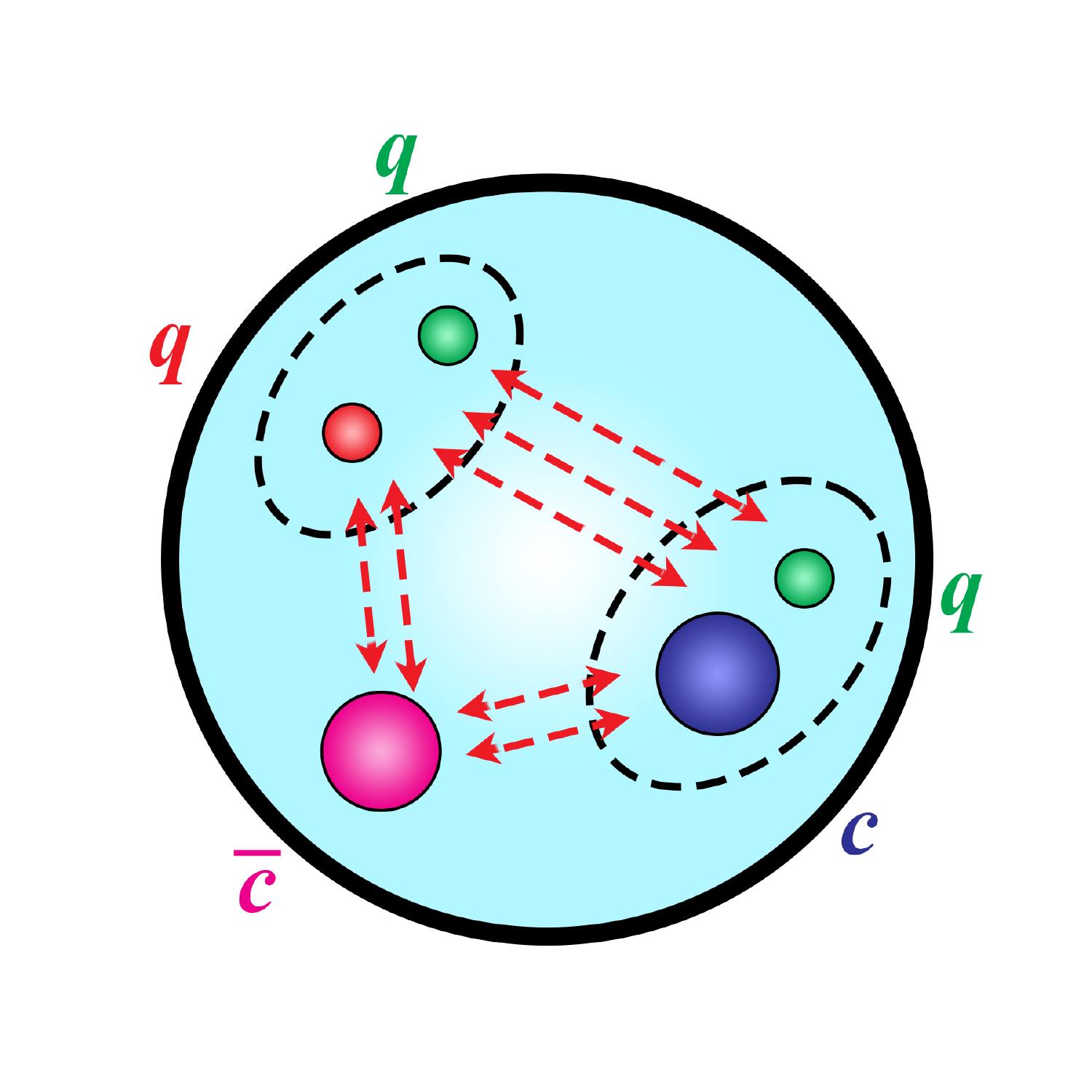}}
\end{center}
\caption{The compact (a) tetraquark and (b) pentaquark states, tightly bound by the strong interaction directly.}
\label{fig:compact}
\end{figure*}

\begin{figure*}[hbtp]
\begin{center}
\subfigure[]{\includegraphics[width=0.4\textwidth]{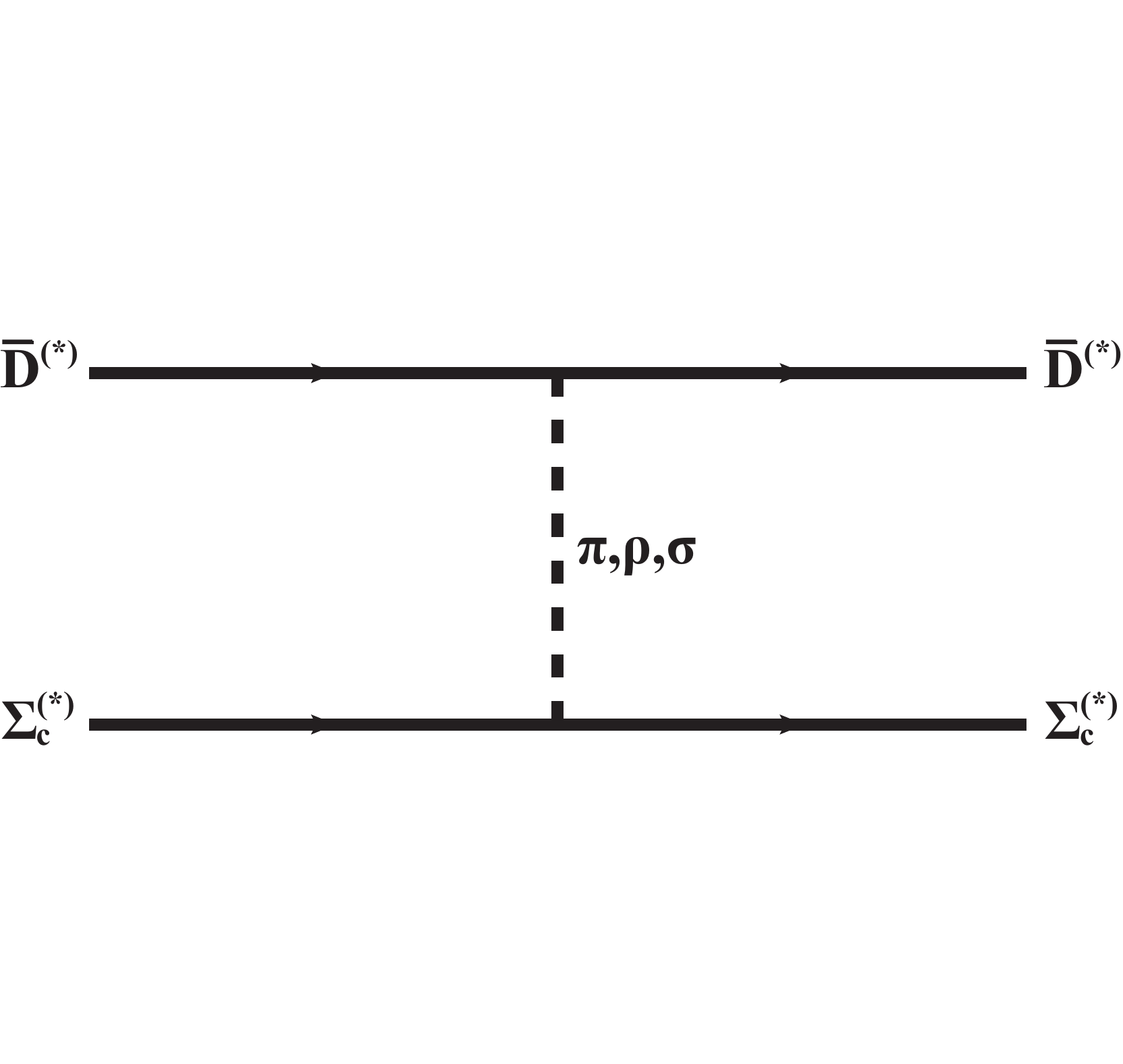}}
~~~~~
\subfigure[]{\includegraphics[width=0.4\textwidth]{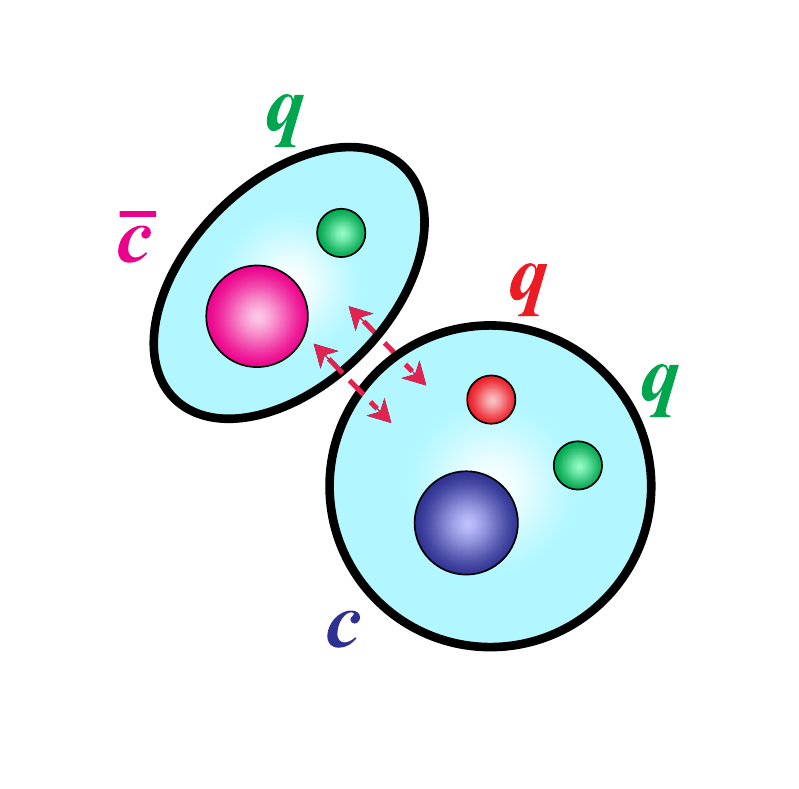}}
\end{center}
\caption{(a) Feynman diagram between $\bar D^{(*)}$ and $\Sigma_c^{(*)}$ describing the one-boson-exchange interaction, and (b) the $\bar D^{(*)} \Sigma_c^{(*)}$ hadronic molecules weakly bound by this residual strong interaction.}
\label{fig:OBE}
\end{figure*}

The interactions forming multiquarks are totally different within discrepant pictures, {\it e.g.}, the compact multiquark states are tightly bound directly by the strong interaction, while the hadronic molecular states are weakly bound by the residual strong interaction, as separately depicted in Fig.~\ref{fig:compact} and Fig.~\ref{fig:OBE}. Besides, some of the multiquark candidates may be caused by the kinematical effects, such as the threshold effects and triangle singularities, etc. When we were writing this review, we noticed that some theoretical studies have tried to describe many multiquark candidates simultaneously and accommodate them within one universal framework. Let us briefly discuss them as follows.

The compact multiquark states are tightly bound by the strong interaction directly, as depicted in Fig.~\ref{fig:compact} for the compact tetraquark and pentaquark states. In 1976 the diquark-antidiquark picture was proposed in Refs.~\cite{Jaffe:1976ig,Jaffe:1976ih} to explain the light scalar mesons as compact tetraquark states. Later in 2004 the authors of Ref.~\cite{Maiani:2004vq} proposed the ``type-I'' diquark-antidiquark model, where the $S$-wave hidden-charm tetraquark states are written in the spin basis as $|s_{cq}, \bar s_{\bar c \bar q} \rangle_J$, with $s_{cq}$ and $\bar s_{\bar c \bar q}$ the diquark and antidiquark spins, respectively. The ``type-II'' diquark-antidiquark model was developed in Ref.~\cite{Maiani:2014aja}. As discussed in Sec.~\ref{sec5.3.2}, this framework can explain many charmonium-like states as a whole~\cite{Maiani:2021tri}, including the $X(3872)$, $Z_c(3900)$, $Z_c(4020)$, $Z_{cs}(3985)$, $Z_{cs}(4000)$, $Z_{cs}(4220)$, and $X(4140)$, etc. Besides, the diquark-triquark picture was proposed in Ref.~\cite{Lebed:2015tna} to explain the compact pentaquark states.
\begin{itemize}

\item In Refs.~\cite{Ferretti:2020ewe,Ferretti:2021zis} the authors systematically calculated the mass spectra of the hidden-charm and hidden-bottom tetraquarks within the compact diquark-antidiquark model. Some of their results are shown in Fig.~\ref{fig:tetraquark} as two tetraquark nonets, which are similar to the results of Refs.~\cite{Maiani:2004vq,Maiani:2014aja,Maiani:2021tri}. Besides, they also applied the hadro-quarkonium model to systematically calculate the mass spectra of the hidden-charm and hidden-bottom tetraquarks and pentaquarks.

\begin{figure*}[hbtp]
\begin{center}
\subfigure[]{\includegraphics[width=0.4\textwidth]{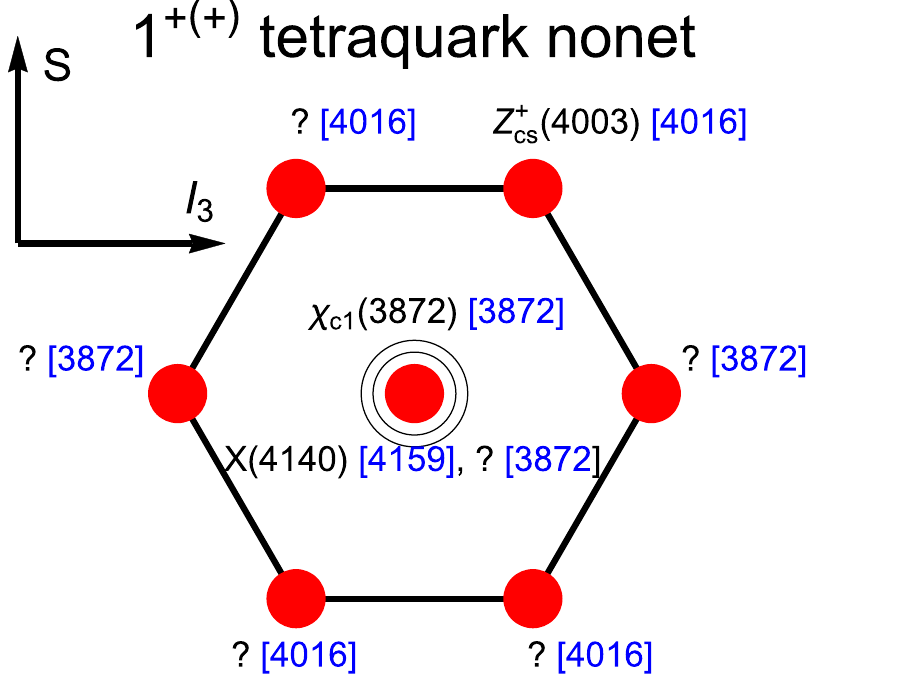}}
~~~~~
\subfigure[]{\includegraphics[width=0.4\textwidth]{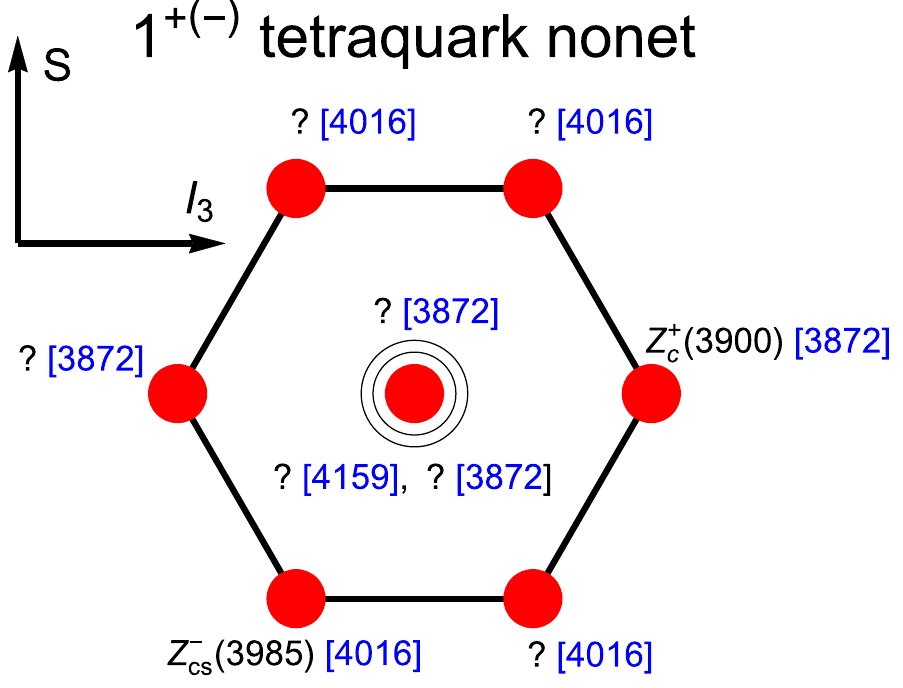}}
\end{center}
\caption{The (a) $J^{P(C)} = 1^{+(+)}$ and (b) $1^{+(-)}$ tetraquark nonets in the compact diquark-antidiquark model. Source: Ref.~\cite{Ferretti:2021zis}.}
\label{fig:tetraquark}
\end{figure*}

\item Within the compact diquark-antidiquark picture, the authors of Ref.~\cite{Faustov:2021hjs} reviewed their calculations of the mass spectrum of the hidden-charm tetraquark states through the relativistic quark model. Their results can describe the $X(3872)$, $Z_c(3900)$, $X(3940)$, $Z_{cs}(3985)$, $X(4140)$, $Y(4230)$, $Z_c(4240)$, $Z_c(4250)$, $Y(4260)$, $Y(4360)$, $Z_c(4430)$, $X(4500)$, $Y(4660)$, $X(4700)$, and $X(4740)$ as compact tetraquark states. They also calculated the mass spectrum of their bottom counterparts, but the obtained results can not explain the $Z_b(10610)$ and $Z_b(10650)$, which were suggested to be molecular states.

\end{itemize}

The hadronic molecular states are weakly bound by the residual strong interaction, as depicted in Fig.~\ref{fig:OBE} for the $\bar D^{(*)} \Sigma_c^{(*)}$ hadronic molecular states. Before the discovery of the $P_c(4380)$ and $P_c(4450)$ in 2015 by LHCb~\cite{LHCb:2015yax}, their existence had been predicted with various theoretical frameworks, such as the one-boson(pion)-exchange model~\cite{Yang:2011wz,Karliner:2015ina}, the chiral unitary model~\cite{Wu:2010jy,Wu:2010vk,Xiao:2013yca}, and the chiral quark model~\cite{Wang:2011rga}, etc. We refer to our previous review~\cite{Chen:2016qju} for their detailed discussions, together with a short introduction to the history of the hadronic molecular picture dating back to the study of the deuteron as the only stable hadronic molecule.
\begin{itemize}

\item Within the hadronic molecular picture, the authors of Ref.~\cite{Karliner:2015ina} predicted many states consisting of two heavy hadrons, which can couple through isospin exchange. These include $D \bar D^*$, $D^* \bar D^*$, $D^*B^*$, $\bar B B^*$, $\bar B^* B^*$, $\Sigma_c \bar D^*$, $\Sigma_c B^*$, $\Sigma_b \bar D^*$, $\Sigma_b B^*$, $\Sigma_c \bar \Sigma_c$, $\Sigma_c \bar \Lambda_c$, $\Sigma_c \bar \Lambda_b$, $\Sigma_b \bar \Sigma_b$, $\Sigma_b \bar \Lambda_b$, and $\Sigma_b \bar \Lambda_c$, as well as their corresponding doubly heavy molecular states.

\item In Ref.~\cite{Wang:2021aql} the authors systematically studied the interactions in the $D^{(*)}\bar{D}^{(*)}$, $\bar{D}^{(*)}\bar{D}_1$, $D^{(*)}\bar{D}_2^*$, $D_s^{(*)}\bar{D}_s^{(*)}$, ${D}_s^{(*)}\bar{D}_{s0}^*$, $D_s^{(*)}\bar{D}_{s1}^{\prime}$, ${D}_s^{(*)}\bar{D}_{s1}$, and $D_s^{(*)}\bar{D}_{s2}^*$ systems. After adopting the one-boson-exchange effective potentials, their results indicate that there exist a serial of isoscalar charmonium-like molecular states. Their results can fully exclude the charged charmonium-like $Z_c$ states as the isovector charmonium-like molecular states, some of which were suggested in Ref.~\cite{Wang:2020dmv} to be caused by kinematic effects.

\item In Ref.~\cite{Chen:2021cfl} the authors gave a unified description of the loosely bound molecular systems composed of the heavy flavor hadrons $(\bar{D},\bar{D}^*)$, $(\Lambda_c, \Sigma_c, \Sigma_c^*)$, and $(\Xi_c, \Xi_c^\prime,\Xi_c^*)$. They used a quark-level interaction, and their results indicate the same binding mechanism for the heavy flavor meson-meson and meson-baryon systems, {\it i.e.}, they are bound dominantly by the interactions from their light degrees of freedom. This study was recently extended in Ref.~\cite{Chen:2021spf} to study the loosely bound hadronic molecules composed of heavy flavor di-hadrons with strangeness.

\item In Ref.~\cite{Chen:2021xlu} the author systematically studied the $\bar D^{(*)} \Sigma_c^{(*)}$, $\bar D^{(*)} \Lambda_c$, $D^{(*)} \bar K^{*}$, and $D^{(*)} \bar D^{(*)}$ hadronic molecular states through the QCD sum rule method. As depicted in Fig.~\ref{fig:covalent}, the author proposed a binding mechanism induced by shared light quarks, together with a model-independent hypothesis: the light-quark-exchange interaction is attractive when the shared light quarks are totally antisymmetric and obey the Pauli principle. A unique feature of this framework was that the binding energies of the $(I)J^P = (0)1^+$ $D\bar B^*/D^* \bar B$ covalent hadronic molecules are much larger than those of the $(I)J^P = (0)1^+$ $DD^*/\bar B \bar B^*$ ones, while the $(I)J^P = (1/2)1/2^+$ $\bar D \Sigma_c/\bar D \Sigma_b/B \Sigma_c/B \Sigma_b$ covalent hadronic molecules have similar binding energies.

\begin{figure*}[hbtp]
\begin{center}
\subfigure[]{\includegraphics[width=0.4\textwidth]{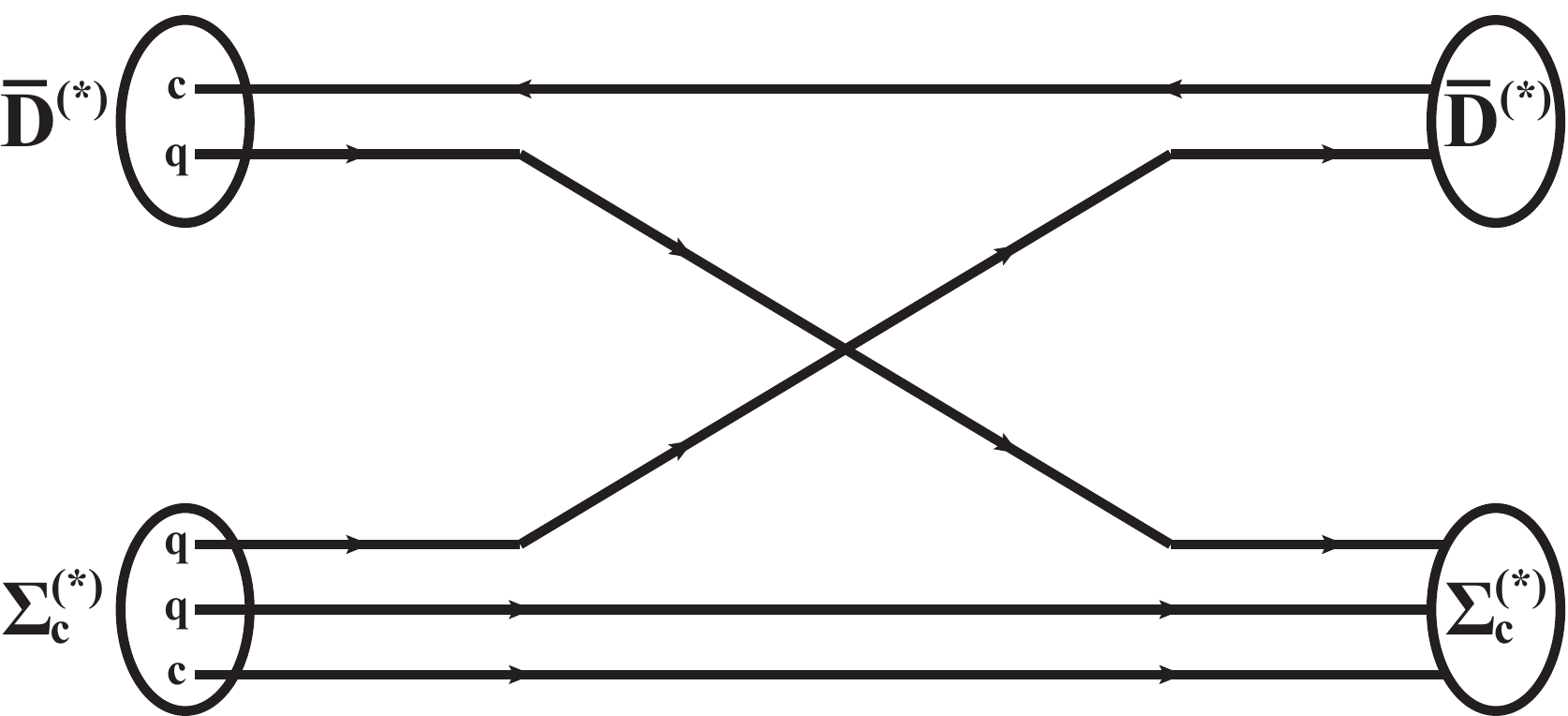}}
~~~~~
\subfigure[]{\includegraphics[width=0.43\textwidth]{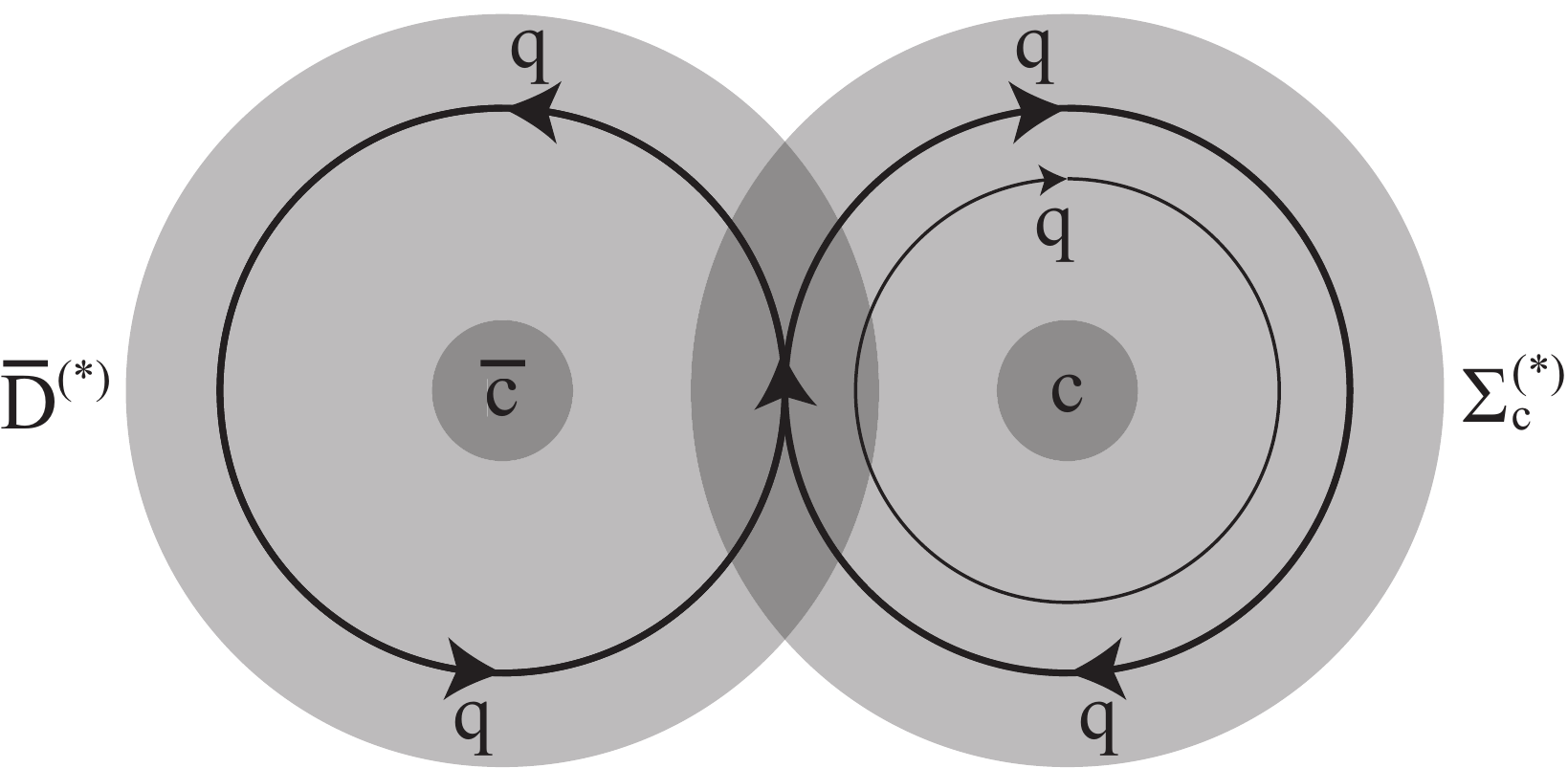}}
\end{center}
\caption{(a) Feynman diagram between $\bar D^{(*)}$ and $\Sigma_c^{(*)}$ describing the light-quark-exchange interaction, and (b) the $\bar D^{(*)} \Sigma_c^{(*)}$ covalent hadronic molecules weakly bound by this residual strong interaction.}
\label{fig:covalent}
\end{figure*}

\item The multiquark candidates are often located close to the thresholds of a pair of hadrons. The authors of Ref.~\cite{Dong:2020hxe} proposed a general argument that there should be a threshold cusp at any $S$-wave threshold, showing up as a peak only for the channels with attractive interactions, as depicted in Fig.~\ref{fig:threshold}. After carefully examining the threshold effects, the authors of Refs.~\cite{Dong:2021juy,Dong:2021bvy} systematically investigated the heavy-heavy and heavy-antiheavy hadronic molecules, and predicted 229 hidden-charm molecular states and 124 doubly charmed molecular states. Besides, several five-flavored molecular states with the quark content $udsc \bar b$ were predicted in Ref.~\cite{Shen:2022rpn} by investigating the coupled-channel effects of the $B^{(*)} \Xi_c$-$B^{(*)} \Xi_c^\prime$ system.

\begin{figure*}[hbtp]
\begin{center}
\includegraphics[width=0.6\textwidth]{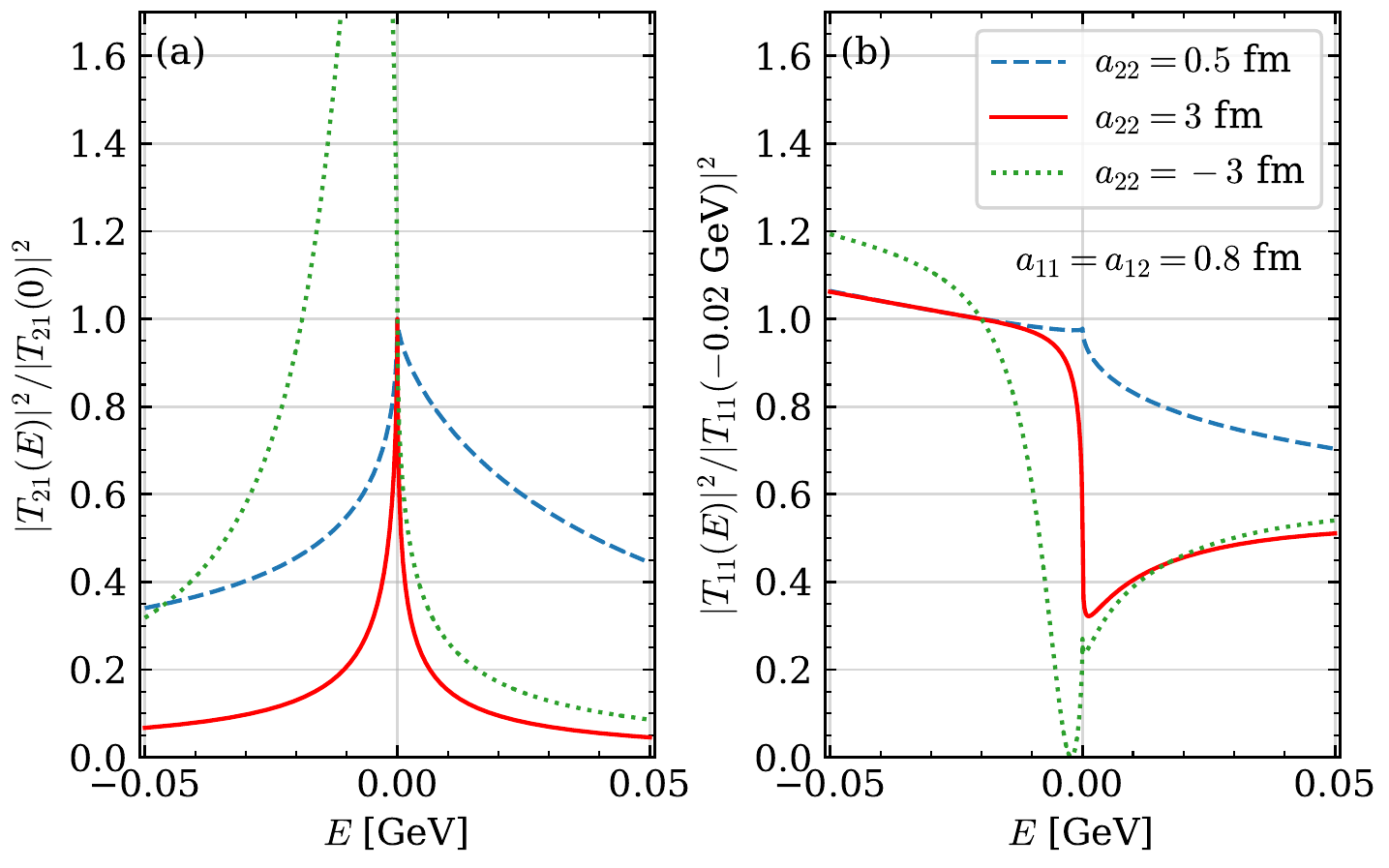}
\end{center}
\caption{Illustration of the threshold behaviors for a two-channel problem. Source: Ref.~\cite{Dong:2020hxe}.}
\label{fig:threshold}
\end{figure*}

\end{itemize}

Since the discovery of the $X(3872)$ by Belle in 2003~\cite{Belle:2003nnu}, many $XYZ$ and $P_c$ states have been discovered in the past twenty years~\cite{pdg}. And more multiquark states are expected in the coming future. These intriguing states bring us the renaissance of the hadron spectroscopy, although there is still a long long way to fully understand their internal structures and underlying non-perturbative QCD dynamics.

\subsection{Glueballs and hybrid mesons}
\label{sec7.3}

The glueballs and hybrid mesons have been reviewed in Sec.~\ref{sec6}. There have been various experimental and theoretical investigations in the past fifty years, and especially, the lattice QCD simulations have played a pivotal role~\cite{Morningstar:1999rf,Chen:2005mg,Meyer:2004gx,Gregory:2012hu,Athenodorou:2020ani,Dudek:2009qf}. However, there still lacks overwhelmingly definite and conclusive experimental evidence on the existence of the glueballs, and the academic community has not formed a common understanding on their nature. This is closely related to our poor understanding of the gluon degree of freedom in the low energy region, as discussed in Sec.~\ref{sec1.2} and at the beginning of Sec.~\ref{sec6}.

Experimentally, it is very difficult to differentiate the scalar glueball from the scalar $q \bar q$ mesons and scalar tetraquark states. Recently, the authors of Refs.~\cite{Sarantsev:2021ein,Klempt:2021wpg} proposed the existence of a distributed scalar glueball, with the mass around $1865$~MeV and width around $370$~MeV~\cite{Klempt:2021wpg}, which widely mixes with the scalar mesons $f_0(1370)$, $f_0(1500)$, $f_0(1710)$, $f_0(1770)$, $f_0(2020)$, and $f_0(2100)$, etc. Although these theoretical studies bring us a new viewpoint, the broad width and the mixing effect of this distributed glueball may hinder its experimental observation easily.

On the other hand, experimental signals with the exotic quantum numbers $J^{PC} =0^{--}/0^{+-}/1^{-+}/2^{+-}/3^{-+}/4^{+-}/\cdots$ are of particular interests, since these exotic quantum numbers can not be accessed by conventional $q \bar q$ mesons and may arise from the gluon degree of freedom, although we still need to differentiate them from either the exotic tetraquark or molecular states.

Up to now there are three candidates of the hybrid mesons with the exotic quantum number $I^GJ^{PC} = 1^-1^{-+}$, {\it i.e.}, the $\pi_1(1400)$, $\pi_1(1600)$, and $\pi_1(2015)$. The last one $\pi_1(2015)$ was observed only by the BNL E852 experiments~\cite{E852:2004gpn,E852:2004rfa}. Its existence might be questionable, and it was omitted from the summary table of PDG2020~\cite{pdg}. Besides, the results of Refs.~\cite{COMPASS:2018uzl,JPAC:2018zyd} suggest that there is only one exotic $\pi_1$ resonance with the mass around $1600$~MeV and width around $580$~MeV~\cite{COMPASS:2018uzl}. The $\eta_1(1855)$ of $I^GJ^{PC} = 0^+1^{-+}$ recently observed by BESIII~\cite{BESIII:2022riz,BESIII:2022qzu} is undoubtedly another interesting and important hybrid candidate.

The experimental observations, phenomenological analyses, and lattice QCD simulations have all contributed a lot to this old but still active, complicated but very important field. Hopefully, the interplay of these three approaches will continuously improve our understanding on the glueballs and hybrid mesons, the open heavy flavor mesons and baryons, the open and hidden heavy flavor multiquark states, as well as all the other conventional and exotic hadrons, and clarify their nature in the near future.

\vspace{0.2cm}

\section*{Acknowledgments}

We would like to express our gratitude to all the collaborators and colleagues who contributed to the investigations presented here, in particular to
Bing Chen, Dian-Yong Chen, Kan Chen, Rui Chen, Xiao-Lin Chen, Er-Liang Cui, Cheng-Rong Deng, Wei-Zhen Deng, Li-Sheng Geng, Jun He, Atsushi Hosaka, Ning Li, Xiao-Hai Liu, Li Ma, Lu Meng, T.~G.~Steele, Zhi-Feng Sun, Bo Wang, Guang-Juan Wang, Xin-Zhen Weng, Lu Zhao, Hai-Qing Zhou, and Zhi-Yong Zhou.
We thank Niu Su for drawing some of the figures in the paper, and thank Hui-Min Yang for helping prepare some relevant documents.
This review is dedicated to Professor Yuan-Ben Dai, Professor W-Y. P. Hwang, and Professor Takayuki Matsuki.
S.L.Z. also dedicates this review to his beloved mother.
This project is supported by
the National Natural Science Foundation of China under Grant No.~11722540, No.~11775132, No.~11975033, No.~12075019, and No.~12070131001,
the National Key R$\&$D Program of China under Contracts No.~2020YFA0406400,
the Jiangsu Provincial Double-Innovation Program under Grant No.~JSSCRC2021488,
and
the Fundamental Research Funds for the Central Universities.

\vspace{0.2cm}

\section*{References}
\bibliographystyle{elsarticle-num}
\bibliography{ref}

\begin{thebibliography}{1000}
\bibitem{Zweig:1964ruk}
G.~Zweig, {An SU(3) model for strong interaction symmetry and its breaking.
  Version 1}.
\end{thebibliography}

\end{document}